\documentclass[11pt,DIV11]{scrbook}

\usepackage[titletoc, title]{appendix}
\usepackage{float,amsmath,amstext,amsthm,amssymb,amsfonts,slashed,epigraph}
\usepackage[nouppercase]{scrpage2}      
\usepackage{makeidx}
\usepackage{makeidx}
\makeindex
\usepackage{hyperref}
\usepackage{graphicx}
\usepackage{caption}
\usepackage{feynmp}
\usepackage{subcaption}
\usepackage{bm}
\usepackage{enumerate}
\usepackage[utf8]{inputenc} 
\usepackage[T1]{fontenc}    
\usepackage[ngerman,english]{babel} 
\usepackage{changepage}
\usepackage{float}
\begin{document}
\setlength{\evensidemargin}{+0 cm}
\setlength{\oddsidemargin}{+0.6cm}

\bibliographystyle{unsrt}

\pagestyle{scrheadings}
    \clearscrheadfoot
    \ohead[\pagemark]{\pagemark}
    \renewcommand{\headfont}{\normalfont\sffamily}
    \renewcommand{\pnumfont}{\normalfont\sffamily} 
    \small
    \automark[section]{chapter}
    \chead{\headmark}
    \normalsize
    \setheadsepline{0.5pt}
    \pagenumbering{roman}\setcounter{page}{1}

\declareslashed{}{\not}{-0.4}{0}{\int}

\makeatletter
\newcommand{\contraction}[5][1ex]{%
  \mathchoice
    {\contraction@\displaystyle{#2}{#3}{#4}{#5}{#1}}%
    {\contraction@\textstyle{#2}{#3}{#4}{#5}{#1}}%
    {\contraction@\scriptstyle{#2}{#3}{#4}{#5}{#1}}%
    {\contraction@\scriptscriptstyle{#2}{#3}{#4}{#5}{#1}}}%
\newcommand{\contraction@}[6]{%
  \setbox0=\hbox{$#1#2$}%
  \setbox2=\hbox{$#1#3$}%
  \setbox4=\hbox{$#1#4$}%
  \setbox6=\hbox{$#1#5$}%
  \dimen0=\wd2%
  \advance\dimen0 by \wd6%
  \divide\dimen0 by 2%
  \advance\dimen0 by \wd4%
  \vbox{%
    \hbox to 0pt{%
      \kern \wd0%
      \kern 0.5\wd2%
      \contraction@@{\dimen0}{#6}%
      \hss}%
    \vskip 0.9ex
    \vskip\ht2}}
\newcommand{\contracted}[5][1ex]{%
  \contraction[#1]{#2}{#3}{#4}{#5}\ensuremath{#2#3#4#5}}
\newcommand{\contraction@@}[3][0.05em]{%
  \hbox{%
    \vrule width #1 height 0pt depth #3%
    \vrule width #2 height 0pt depth #1%
    \vrule width #1 height 0pt depth #3%
    \relax}}
\makeatother

\newcounter{saveeqn}

\newcommand{\alpheqn}{\setcounter{saveeqn}{\value{equation}}
  \stepcounter{saveeqn}\setcounter{equation}{0}
  \renewcommand{\theequation}
   {\thechapter.\mbox{\arabic{saveeqn}-\alph{equation}}}}

\newcommand{\reseteqn}
 {\setcounter{equation}{\value{saveeqn}}
 \renewcommand{\theequation}{\thechapter.\arabic{equation}}}

\setcounter{chapter}{0}



\begin{titlepage}
\vspace*{0cm}
\begin{center}
\Huge
\textsc{Phenomenology of a pseudoscalar Glueball and charmed mesons}\\
\vspace{2cm}
Dissertation \\
\LARGE
zur Erlangung des Doktorgrades\\
der Naturwissenschaften\\
\vspace{2cm}
vorgelegt beim Fachbereich Physik\\
der Johann Wolfgang Goethe-Universit\"at\\
in Frankfurt am Main\\
\vspace{2cm}
\LARGE
von\\
\LARGE
Walaa I. Eshraim\\
\LARGE
\vspace{1cm}
Frankfurt am Main 2015\\
(D30)
\vspace{5cm}
\end{center}
\large
\vfill
\newpage
~\\
\vspace{5cm}
~\\
vom Fachbereich Physik (13) \\
der Johann Wolfgang Goethe-Universit\"at als Dissertation angenommen.\\
\vspace{6cm}\\
Dekan: Prof. Dr. Rene Reifarth \\
\vspace{0.5cm}\\
1. Gutachter:  Prof. Dr. Dirk-Hermann Rischke \\
2. Gutachter:  Prof. Dr. Stefan Schramm\\
\vspace{0.5cm}\\
Datum der Disputation: 24.07.2015
\end{titlepage}
\newpage






\newpage

\vspace*{5cm}
\begin{center}
\textbf{
\textit{To}\\
\textit{the spirit that gives life to my body}\\
\textit{My Mother}\\
\textbf{\textit{Sanaa}}\\
\vspace{1cm}
\vspace{1cm}
\textit{To}\\
\textit{the blood that lets my heart beat}\\
\textit{My Father}\\
\textit{Ibrahim}\\
\vspace{1cm}
\textit{and}\\
\vspace{1cm}
\textit{To}\\
\textit{the world that has to be full of love,}\\
\textit{peace, safety, fidelity,}\\
\textit{probity, sincerity,}\\
\textit{justice, equality}\\
\textit{and liberty.}}\\
\end{center}

\chapter*{Zusammenfassung\label{abstract} }

Die Quantenchromodynamik (QCD) ist die Theorie, welche die Wechselwirkung zwischen Quarks und
Gluonen beschreibt. Die fundamentale Symmetrie, die der QCD zugrunde liegt,
ist die lokale $SU(3)_c$-Farbsymmetrie. Aufgrund vom Confinement der Quarks und Gluonen
werden im niederenergetischen Bereich die physikalischen
Freiheitsgrade durch Hadronen (Mesonen und Baryonen)
repr\"{a}sentiert. In den letzten Jahren wurden zahlreiche
effektive niederenergetische Modelle f\"{u}r die starke
Wechselwirkung entwickelt, denen eine chirale Symmetrie zugrunde
liegt. Die chirale Symmetrie ist eine weitere Symmetrie der QCD-Lagrangedichte, die im Limes verschwindender Quarkmassen (dem
sogenannten chiralen Limes) realisiert ist. Diese Symmetrie wird
durch nichtverschwindende Stromquarkmassen explizit gebrochen. Im
QCD-Vakuum ist die chirale Symmetrie spontan gebrochen. Als
Konsequenz entstehen pseudoskalare (Quasi-)Goldstone-Bosonen, die
f\"{u}r Up- und Down-Quarks (d.h. f\"{u}r $N_f=2$ Quarkflavors) den
Pionen entsprechen. F\"{u}r $N_f=3$, d.h. wenn auch das Strange-Quark betrachtet wird, entsprechen die Goldstone-Bosonen den
Pionen, Kaonen und dem Eta-Meson. Das $\eta'$-Meson ist kein
Goldstone-Boson wegen der chiralen Anomalie. Die chirale Symmetrie
kann in hadronischen Modellen in sogenannter linearer    oder
nichtlinearer Repr\"{a}sentation realisiert werden. Im
nichtlinearen Fall werden nur Goldstone-Bosonen betrachtet. In
neueren Modellen jedoch werden auch die Vektormesonen dazu
addiert. Im linearen Fall enthalten die Modelle auch die chiralen
Partner der Goldstone-Bosonen. Wenn man diese Modelle auf den
Vektorsektor erweitert, enthalten Sie sowohl Vektor-als
auch Axialvektor-Mesonen. In diesem Zusammenhang haben
aktuelle Bem\"{u}hungen zur Entwicklung des sogenannten
erweiterten linearen Sigma-Modells (eLSM) f\"{u}r $N_f=2$ und
f\"{u}r $N_f=3$ gef\"{u}hrt. Zus\"{a}tzlich zur chiralen Symmetrie
wird im eLSM die Symmetrie unter Dilatation (Skaleninvarianz) und
die anomale Brechung dieser Symmetrie (Spuranomalie)
ber\"{u}cksichtigt. F\"{u}r $N_f=2$ war es im Rahmen des eLSM zum
ersten Mal m\"{o}glich, (pseudo-)skalare sowie (axial-)vektorielle
Mesonen in einem chiralen Modell zu beschreiben: Die Massen und
Zerfallsbreiten stimmen gut mit den Resultaten der Particle Data
Group (PDG) \"{u}berein. Als Folge der nicht-abelschen Natur der
lokalen $SU(3)$-Farbsymmetrie tragen die Eichfelder der QCD, die
Gluonen, eine Farbladung. Daher wechselwirken sie stark
miteinander. Wegen des Confinements erwartet man, dass Gluonen auch
farblose, bzw. ``weiße'', Objekte bilden k\"{o}nnen. Diese
werden als Glueb\"{a}lle bezeichnet.

Die ersten Berechnungen der Glueball-Massen basierten auf dem
Bag-Modell. Sp\"{a}ter erlaubten numerische Gitterrechnungen die
Bestimmung des vollen Glueballspektrums. In voller QCD (d.h.
Gluonen plus Quarks) findet eine Mischung zwischen den
Glueb\"{a}llen und Quark-Antiquark-Konfigurationen mit denselben
Quantenzahlen statt, was die Identifikation der Resonanzen, die von
der PDG gelistet sind, zus\"{a}tzlich erschwert. Die Suche nach
Zust\"{a}nden, die vorrangig Glueb\"{a}lle sind, ist ein aktives
aktuelles Forschungsgebiet. Dadurch erhofft man sich ein besseres
Verst\"{a}ndnis f\"{u}r das nichtperturbative Verhalten der QCD.
Obwohl zurzeit einige Kandidaten f\"{u}r Glueb\"{a}lle existieren,
wurde noch kein Zustand eindeutig identifiziert, der vorrangig ein
Glueball ist. Im Allgemeinen sollten die Glueball-zustände zwei
Eigenschaften im Hinblick auf den Zerfall erf\"{u}llen. Erstens
ist ein Glueball flavorblind, da die Gluonen an alle Quarkflavors
mit derselben Stärke koppeln. Zweitens besitzen Glueb\"{a}lle
eine schmale Zerfallsbreite, die im Large-$N_c$-Limes wie $1/N_c^2$
skaliert. Im Vergleich dazu skaliert ein Quark-Antiquark-Zustand
wie $1/N_c$. Der leichteste Glueballzu-stand, den die
Gitterrechnungen vorhersagen, ist ein skalar-isoskalarer Zustand
($J^{PC}=0^{++}$) mit einer Masse von etwa $1.7$ GeV. Die Zerfallsbreite
der Resonanz $f_0(1500)$ ist flavorunabh\"{a}ngig und schmal. Aus
diesem Grund ist diese Resonanz ein guter Kandidat f\"{u}r einen Zustand, der
vorrangig ein skalarer Glueball ist. Zus\"{a}tzlich ist die
Resonanz $f_0(1700)$ ein Glueball-Kandidat, da ihre Masse in der
N\"{a}he der Vorhersagen der Gitterrechnungen liegt, und da sie in
den gluonreichen Zerf\"{a}llen des $J/\psi$-Mesons produziert wird.
Beide Szenarien wurden in vielen Arbeiten untersucht, in denen die
Mischung zwischen $f_0(1370)$, $f_0(1500)$ und $f_0(1710)$ betrachtet wird.
Der zweitleichteste Glueball, der von den Gitterrechnungen
vorhergesagt wird, ist ein Tensor-Zustand mit den Quantenzahlen
$2^{++}$ und einer Masse von etwa $2.2$ GeV. Ein guter Kandidat
daf\"{u}r k\"{o}nnte die sehr schmale Resonanz $f_0(2200)$ sein,
falls sich ihr Drehimpuls experimentell zu $J=2$ bestimmen
l\"{a}sst. Der drittleichteste Glueball ist ein pseudoskalarer
Zustand ($J^{PC}=0^{-+}$) mit einer Masse von etwa $2.6$ GeV.
Open-charm Mesonen bestehen aus einem Charm-Quark und einem Up-,
Down- oder Strange-Antiquark. Sie wurden im Jahre $1976$, zwei
Jahre sp\"{a}ter als das $J/\psi$-Meson ($c\overline{c}$-Zustand), entdeckt.
Seit dieser Zeit gab es signifikante experimentelle und
theoretische Fortschritte im Bereich der Spektroskopie und bei der
Bestimmung der Zerf\"{a}lle dieser Mesonen. In dieser Arbeit
zeigen wir, wie im Rahmen eines chiral-symmetrischen Modells,
welches das Charm-Quark als zus\"{a}tzlichen Freiheitsgrad
enth\"{a}lt, die urspr\"{u}ngliche $SU(3)$-Flavor-Symmetrie der
Hadronen zu einer $SU(4)$-Symmetrie erweitert werden kann. Die
chirale Symmetrie wird durch die große Strommasse des Charm-Quarks stark explizit
verletzt.\\

\textbf{Zerfall von pseudoskalaren Glueb\"{a}llen in skalare und
pseudoskalare
Mesonen}\\

In dieser Arbeit untersuchen wir die Zerf\"{a}lle des
pseudoskalaren Glueballs, dessen Masse laut Gitterrechnungen
zwischen $2$ und $3$ GeV liegt. Wir konstruieren eine effektive
chirale Lagrangedichte, die das pseudoskalare Glueballfeld $G$ an
skalare und pseudoskalare Mesonen mit $N_f=3$ koppelt. Danach
berechnen wir die Breiten f\"{u}r die Zerf\"{a}lle $G\rightarrow
PPP$ und $G\rightarrow PS$, wobei $P$ und $S$ pseudoskalare und
skalare Quark-Antiquark-Zust\"{a}nde kennzeichnet. Die
pseudoskalaren Zustände umfassen das Oktett der pseudo-Goldstone-Bosonen, während sich der skalare Zustand $S$
auf das Quark-Antiquark-Nonet oberhalb von $1$ GeV bezieht. Der Grund dafür besteht darin, dass
die chiralen Partner der pseudoskalaren Zust\"{a}nde nicht mit
Resonanzen unterhalb von $1$ GeV identifiziert werden sollten. Die
konstruierte chirale Lagrangedichte enth\"{a}lt eine unbekannte
Kopplungskonstante, die nur experimentell bestimmt werden kann.
Aus diesem Grund pr\"{a}sentieren wir die Resultate in Form von
Verzweigungsverh\"{a}ltnissen f\"{u}r die Zerf\"{a}lle des
pseudoskalaren Glueballs $G$ in drei pseudoskalare Mesonen oder
ein skalares und ein pseudoskalares Meson. Diese
Verzweigungsverh\"{a}ltnisse h\"{a}ngen von keinen weiteren
Parametern ab, sobald die Glueballmasse fixiert wird. Wir
betrachten zwei M\"{o}glichkeiten: i)  In \"{U}bereinstimmung mit
Gitterrechnungen w\"{a}hlen wir die Masse des pseudoskalaren
Glueballs zu etwa $2.6$ GeV. Die Existenz und die
Zerfallseigenschaften hypothetischer pseudoskalarer Resonanzen
k\"{o}nnen im zuk\"{u}nftigen PANDA-Experiment getestet werden.
(Das PANDA-Experiment mißt die Proton-Antiproton-Streuung.
Daher kann der pseudoskalare Glueball direkt als ein
Zwi-schenzustand produziert werden.) ii) Wir nehmen an, dass die
Resonanz $X(2370)$ (gemessen im Experiment BESIII) vorrangig ein
pseudoskalarer Glueball-Zustand ist. Daher benutzen wir daf\"{u}r
die Masse $2.37$ GeV. Unsere Ergebnisse sagen voraus, dass $K K
\pi$ der dominante Zerfallskanal ist, gefolgt von einem beinahe
gleich großem $\eta \pi \pi$-und $\eta' \pi \pi$-Zerfallskanal. Der Zerfallskanal in drei Pionen verschwindet. Beim
BESIII-Experiment w\"{a}re es m\"{o}glich, durch die Messung des
Verzweigungsverh\"{a}ltnisses f\"{u}r $\eta' \pi \pi$ und anderer
Zerfallskan\"{a}le zu bestimmen, ob $X(2370)$ vorrangig ein
pseudoskalarer Glueball ist. F\"{u}r das PANDA-Experiment liefern
unsere Resultate n\"{u}tzliche Hinweise f\"{u}r die
Suche nach pseudoskalaren Glueb\"{a}llen.\\

\textbf{Ph\"{a}nomenologie der Charm-Mesonen}\\

Wir vergr\"{o}$\beta$ern die globale Symmetrie des erweiterten
linearen Sigma Modells (eLSM) zu einer globalen $SU(4)_R\times
SU(4)_L$ Symmetrie, indem wir das Charm-Quark einbauen. Das eLSM
enth\"{a}lt zus\"{a}tzlich zu skalaren und pseudoskalaren Mesonen
auch Axialvektor- und Vektormesonen. Wir benutzen
die Parameter aus dem niederenergetischen Sektor der Mesonen. Die verbleibenden drei freien Parameter (die von der Strommasse
des Charm-Quarks abhängen) werden an Massen der Charmed Mesonen angepaßt. Die Resultate für
Open-Charm-Mesonen stimmen gut mit den experimentellen
Ergebnissen \"{u}berein (die Abweichung betr\"{a}gt etwa $150$
MeV). Für Charmonia weichen unsere Resultate für die
Massen stärker von der experimentellen Daten ab. Unser Modell
stellt dennoch ein n\"{u}tzliches Werkzeug dar, um einige
Eigenschaften der Charm-Zustände, wie zum Beispiel das chirale Kondensat, zu untersuchen.\\

Zusammenfassend bedeutet die Tatsache, dass eine (obwohl in diesem Stadium nur grobe) qualitative Beschreibung durch die
Verwendung eines chiralen Modells und
insbesondere der ermittelten Parameter durch die Untersuchung von $N_f=3$ Mesonen erzielt wurde, dass auch im Sektor der
Charm-Mesonen ein
Überrest der chiralen Symmetrie vorhanden ist. Die chirale Symmetrie ist immer noch präsent, da sich die Parameter des eLSM
als Funktion der Energieskala kaum verändern. Neben den Massentermen, die den Großteil der gegenwärtigen Charm-Quarkmasse
beschreiben,
sind alle Wechselwirkungsterme dieselben wie im niederenergetischen effektiven Modell, welches unter der Forderung nach
chiraler Symmetrie und Dilatationsinvarianz konstruiert wurde. Als Nebenprodukt unserer Arbeit haben wir das Charm-Kondensat auf
die gleiche Größenordnung wie die strange und non-strange Quarkkondensate bestimmt. Dies stimmt ebenfalls mit der auf
$U(4)_R\times U(4)_L$ erweiterten chiralen Dynamik überein.

Was die Zuweisung der skalaren und axial-vektoriellen Strange-Charm-Quarkonium-Zu-stände $D_{S0}$ und $D_{S1}$ betrifft, erhalten
wir das folgende: falls ihre Masse über dem jeweiligen Schwellenwert liegt, ist ihre Zerfallsbreite zu groß. Dies wiederum
bedeutet, dass sich diese Zustände, auch wenn sie existieren, der Detektion entzogen haben. In diesem Fall kann es sich bei
den Resonanzen $D_{S0}^{\ast}(2317)$ und $D_{S1}(2460)$ um dynamisch generierte Pole handeln (alternativ hierzu auch um
Tetraquarks oder molekulare Zustände). Unsere Ergebnisse implizieren auch, dass die Interpretation der Resonanz $D_{S1}(2536)$
als Mitglied des axial-vektoriellen Multiplets nicht favorisiert ist, da die experimentelle Breite zu schmal im Vergleich zur
theoretischen Breite eines Quarkonium-Zustands derselben Masse ist. Die Untersuchung dieser Resonanzen erfordert
die Berechnung von
Quantenfluktuationen und wird Thema zukünftiger Studien sein.\\

\textbf{Zerf\"{a}lle von Open-Charm-Mesonen}\\

Die Ergebnisse f\"{u}r die Konstanten aus den schwachen
Zerf\"{a}llen von pseudoskalaren Open-Charm-$D$- und $D_s$-Mesonen sind
in guter \"{U}bereinstimmung mit den experimentellen Werten. Wir
berechnen die OZI-dominierten Zerf\"{a}lle der Charmed-Mesonen.
Die Resultate f\"{u}r $D_0(2400)^+$, $D_0(2400)^0$, $D_0(2007)$,
$D(2010)$, $D(2420)^0$, und $D(2420)^+$ sind vergleichbar mit den
Ergebnissen f\"{u}r die Ober- und Untergrenzen der PDG, obwohl die
theoretischen Fehler ziemlich groß sind. In unserem Modell
ist es dennoch m\"{o}glich, gleichzeitig die Zerf\"{a}lle von
Open-Charm-Vektormesonen und ihren chiralen
Partnern, den Axialvektormesonen, zu beschreiben.\\

\textbf{Zerf\"{a}lle von Charmonium-Mesonen}\\

Wir erweitern unser $U(4)_R\times U(4)_L$-symmetrisches lineares
Sigma Modell mit Axialvektor- und Vektormesonen um ein Dilaton-Feld, welches ein skalarer Glueball ist. Zus\"{a}tzlich bauen wir
Wechselwirkungen eines pseudoskalaren Glueballs mit
(pseudo-)skalaren Mesonen ein, um die Eigenschaften der
OZI-unterdr\"{u}ckten Charmonia zu untersuchen. Wir berechnen die
OZI-unterdr\"{u}ckten Zerf\"{a}lle der skalaren und pseudoskalaren
Charmonium-Zust\"{a}nde, $\chi_{c0}(1P)$ und $\eta_c(1S)$. Wir
machen Vorhersagen f\"{u}r einen pseudoskalaren Glueball mit einer
Masse von etwa $2.6$ GeV, welcher im PANDA-Experiment bei FAIR
gemessen werden kann. Zus\"{a}tzlich geben wir Vorhersagen f\"{u}r
einen pseudoskalaren Glueball mit einer Masse von etwa $2.37$ GeV an.
Dieser Glueball entspricht der im BESIII-Experiment gemessenen
Resonanz $X(2370)$, die beim Zerfall vom Charmonium-Zustand
$\eta_c$ gemessen wird. Wir berechnen auch den Mischungswinkel
zwischen pseudoskalaren Glueb\"{a}llen mit einer Masse von $2.6$
GeV und dem Hidden-Charm-Meson $\eta_c$.

Die Tatsache, dass eine qualitative Beschreibung dieser
Zerf\"{a}lle im Rahmen eines chiralen Modells, dessen Parameter
f\"{u}r $N_f=3$ bestimmt wurden, m\"{o}glich ist, ist ein Indiz
daf\"{u}r, dass ein Teil der chiralen $SU(3)_R\times SU(3)_L$-Symmetrie im Charm-Sektor weiterhin erhalten ist. Ein weiterer
Hinweis f\"{u}r eine teilweise erhaltene chirale $SU(3)_R\times
SU(3)_L$-Symmetrie besteht darin, dass die Parameter des
erweiterten linearen Sigma-Modells keine starke
Energieabh\"{a}ngigkeit besitzen. Wir berechnen schließlich
das Charm-Kondensat, welches von derselben Gr\"{o}ßenordnung
ist wie das Non-strange- und das Strange-Quarkkondensat. Das ist
auch in \"{U}bereinstimmung mit der zu $U(4)_R\times U(4)_L$-vergr\"{o}ßerten chiralen Dynamik.

Darüber hinaus haben wir ein Dilatonfeld, ein skalares Glueballfeld und die Wechselwirkung eines pseudoskalaren
Glueballfeldes mit (pseudo-)skalaren Mesonen unter $U(4)_R\times U(4)_L$-Symmetrie einbezogen. Anschließend haben wir die Breite
des Zerfalls des Charm- onium-Mesons $\chi_{c0}$ in zwei oder drei Strange-oder Non-Strange-Mesonen und in einen skalaren Glueball $G$ berechnet. Letzterer ist eine Mischung der Resonanzen $f_{0}(1370)$ sowie $f_{0}(1500)$ und $f_0(1700)$.
Der Zerfall des Charmonium-Zustands in Open-Charm-Mesonen ist hingegen innerhalb des eLSM verboten. Ferner haben wir die Breite des Zerfalls des pseudoskalaren Charmonium-Zustands $\eta_C$ in leichte Mesonen und in einen pseudoskalaren Glueball $\widetilde{G}$ über den Kanal $\eta_C \rightarrow \pi\pi\widetilde{G}$ bestimmt.
Dies wurde mittels des Wechselwirkungsterms des pseudoskalaren Glueballs für zwei Fälle durchgeführt.
Zum einen für eine Masse von 2.6 GeV, wie sie von Gitter-QCD-Rechnungen in der Quenched-Näherung vorhergesagt wurde und
im bevorstehenden PANDA-Experiment an der FAIR-Anlage gemessen werden kann. Zum anderen für eine Glueballmasse von 2.37 GeV, die der
Masse der Resonanz $X(2370)$ entspricht und im BESIII-Experiment ermittelt wurde.
Der Mischungswinkel zwischen dem pseudoskalaren Glueball und $\eta_C$ wurde ausgewertet. Er ist sehr klein und beträgt lediglich
$-1^\circ$. Wir haben begründet, dass das eLSM keinerlei Zerfallskanal für (axial-)vektorielle Charmonium-Zustände aufweist, wobei
$\Gamma_{J/\psi}=0$ und $\Gamma_{\chi_{c1}}=0$.
Die Ergebnisse der Zerfallsbreiten $\chi_{c0}$ und $\eta_C$ stimmen gut mit experimentellen Daten überein.
Dies zeigt, wie erfolgreich das eLSM
im Bezug auf das Studium der Hidden-Charm- und Open-Charm-mesonischen Phänomenologie ist.
Die vier bestimmten Parameter im Falle von $N_f=3$, die zur erfolgreichen Auswertung der Massen von Open- und Hidden-Charm-Mesonen und der Zerfallsbreite des Open-Charm-Mesons dienten, sind:
(i) $\lambda_1$ und $h_1$, die gleich Null gesetzt werden im Falle von $N_f=3$, da sie so klein sind und die
vorherigen Resultate nicht beeinflussen. Dagegen hängt die Zerfallsbreite der Charmonium-Zustände $\chi_{c0}$ und $\eta_C$
von beiden ab. Deshalb
wurden diese beiden Parameter durch das Minimieren der Zerfallsbreite des $\chi_{c0}$ festgelegt, siehe Tabelle 8.2.

(ii) Der Parameter $c$, der im axialen Term vorkommt, wird auch durch
den Fit von Gl.(\ref{chi2}) bestimmt.

(iii) $c_{\widetilde{G}\Phi}$, das durch die Beziehung $c_{\widetilde{G}\Phi(N_f=3)}$ festgelegt wird.\\

\textbf{Ausblick}\\

In der modernen Hadronenphysik ist die Wiederherstellung der chiralen Symmetrie bei endlicher Temperatur und Dichte eine der
fundamentalsten Fragestellungen. Das eLSM hat es im Gegensatz zu alternativen Ansätzen geschafft, den
zwei-Flavor Fall bei einen chemischen Potential ungleich Null zu ergründen. All dies führt uns dazu, die Restauration
der chiralen Symmetrie bei nichtverschwindender Temperatur und Dichte für $N_f=3$ und $N_f=4$ mithilfe des eLSM zu untersuchen.
Dies bringt viele Herausforderungen mit vielen unbekannten Parametern in sich. In zukunft werden wir
die Vakuumphänomenologie des leichten Tetraquark-Nonets und dessen Erweiterung auf $N_f=4$ untersuchen.

\selectlanguage{english}


 \tableofcontents
    \cleardoublepage
      \pagenumbering{arabic}
      \setcounter{page}{1}



\chapter{\bigskip Introduction \label{ch1}}

\section{Historical Remarks}

``\textit{The most incomprehensible thing about the universe is that it's comprehensible at all...}''\\

$\,\,\,\,\,\,\,\,\,\,\,\,\,\,\,\,\,\,\,\,\,\,\,\,\,\,\,\,\,\,\,\,\,\,\,\,\,\,\,\,\,\,\,\,\,\,\,\,\,\,\,\,\,\,\,\,\,\,\,\,\,\,\,\,\,\,\,\,\,\,\,\,\,\,\,\,\,\,\,\,\,\,\,\,\,\,\,\,\,\,\,\,\,\,\,\,\,\,\,\,\,\,\,\,\,\,\,\,\,\,\,\,\,\,\,\,\,\,\,\,\,\,\,\,\,\,\,\,\,\,\,\,\,\,\,\,\,\,\,\,\,\,\,\,\,\,\,\,\,\,\,\,\,\,\,\,\,\,\,\,\,\,\,\,\,\,\,\,\,\,\,\,\,\,\,\,\,\,\,\,\,\,\,\,\,\,\,$ Albert Einstein\\

Billions of years ago, all of space was contained in a single
point which, exposed to an enormous and incomprehensible
explosion (the Big Bang), scattered the matter that constitutes
the Universe. At that time, it was hot and dense, but within the
first three minutes after the Big Bang the Universe became
sufficiently cool to consist of subatomic particles,
 including protons, neutrons, and electrons. More than ten billion years passed before the stars and galaxies formed. After some time, planets surrounded some
 stars... life formed...finally, after billions
 of years of changes, the human being was created with a complex brain which has a deep and insatiable curiosity about the world. Humans found that understanding the
 world is not easy
  and noticed that understanding the nature of matter is an important and complementary approach to understanding the nature of reality and answering the deep and
   pressing questions in their minds. To answer these questions,
   they used the observational method which creates a lot of
ideas. The Greek philosopher Empedocles surmised that everything
was made from a suitable mix of
  four basic elements: air, fire, water, and earth. These four elements were perceived as the fundamental elements in nature. Consequently, concentration moved
  towards understanding the nature of the elements'
  permanence. The ancient philosophers Leucippus and well-known Democritus of Greece are the earliest philosophers who conceived the idea that matter is composed
   entirely of various imperishable,
   indestructible, indivisible elements, always in motion, having empty space between them; called atoms. The name is derived from the Greek
    $\ddot{\alpha}\tau o \mu o \varsigma$ which means ``indivisible''. In 1661, Robert Boyle established the atomic idea (molecules).
However, this knowledge about the
   existence of atoms brings with it a lot of important questions: How do these atoms make molecules?... How do the molecules make gases, liquids and solids?...
   It must be forces that act on these atoms to keep them
   together in molecules, but what are these forces? They arrived at that time at the idea that the inter-atomic forces are gravity, static electricity, and
    magnetism. After that, there were a lot of efforts from philosophers and
scientists directed towards the fundamental building blocks of
matter. At the close of the 19$^{th}$ century, it was known
that more than 100 elements exist and that all matter is composed
of atoms which have an internal structure and are not indivisible,
which is opposite of
 what Democritus foresaw of the indivisible property of the atom.\\

At the beginning of the 20$^{th}$ century, Rutherford
presented the subatomic structure as a result of his experiments:
an atom is composed of a dense nucleus surrounded by a cloud of
electrons. Consequently, physicists found that the nucleus
decomposed into smaller particles, which they called protons and
neutrons, and in turn that protons and neutrons themselves contain
even smaller particles called quarks. Moreover, there are other
small ingredients making up the atom, which are called leptons.
These include the electron in the orbits of the nucleus, situated
at (relatively) large distance from the nucleus, but not inside
it. We conclude that quarks and leptons constitute all fundamental matter in the Universe \cite{Povh:2008jx}. This information is the
starting point for
 understanding the formation of the Universe.

\section{Standard Model}

``\textit{Daring ideas are like chessmen moved forward; they may be
detected, but they start a winning game}'' $\,\,\,\,\,\,\,\,\,\,\,\,\,\,\,\,\,\,\,\,\,\,\,\,\,\,\,\,\,\,\,\,\,\,\,\,\,\,\,\,\,\,\,\,\,\,\,\,\,\,\,\,\,\,\,\,\,\,\,\,\,\,\,\,\,\,\,\,\,\,\,\,\,\,\,\,\,\,\,\,\,\,\,\,\,\,\,\,\,\,\,\,\,\,\,\,\,\,\,\,\,\,\,\,\,\,\,\,\,\,\,\,\,\,\,\,\,\,\,\,\,\,\,\,\,\,\,\,\,\,\,\,\,\,\,\,\,\,\,\,\,\,\,\,\,\,\,\,\,\,\,\,\,\,\,\,\,\,\,\,\,\,\,\,\,\,\,\,$ Goethe\\

Quarks and leptons are the basic types of fundamental matter
particles. Each group consists of six types of flavour. The
combinations of these form the hundreds of particles discovered in
the 1950s and 1960s. Two flavours each can be classified as
a generation under the weak interaction, in which the first
generation consists of the lightest
 flavours which make the most stable particles in the Universe, whereas the second and third generations contain the heavier flavours which make the less stable
  particles which decay
 quickly to the next-most stable state belonging to the previous generation. The quark generations are: the ($ u \equiv up,$ and the $d\equiv down$)
 quark flavours (the first generation),
 followed by the ($ c \equiv charm,$ and the $s \equiv strange$) quark flavours as a second generation, and the third consists of the
 ($ t \equiv top,$ and the $b\equiv bottom$) quark flavours.
  Concerning the electric charges of quarks, quarks carry colour charge which corresponds to the electric charge of electrons. They also carry a fractional electric
   charge (the $u,c$,
  and $t$ quark flavours carry $(2/3)e$, whereas the $d,s,$ and $b$ quark flavours carry $(-1/3)e$). Each quark has its corresponding antiparticle with opposite charge.
   Similarly, there are three generations for
  the six lepton flavours: the $ e \equiv electron,$ and the $\nu_e \equiv electron\, neutrino$, the $ \mu\equiv muon,$ and the $\nu_\mu \equiv muon\,neutrino$,
  and the $ \tau \equiv tau
  ,$ and the $\nu_\tau \equiv tau \,neutrino$. The three lepton flavours (the electron, the muon and the tau) have a sizable mass with charge $-e$, whereas the
   other three -the neutrinos-
  are neutral and have a small mass. (See Table 1.1 for a compilation of quarks and leptons).\\

\begin{center}
\begin{tabular}
[c]{|c|c|c|c|c|}\hline Particle & Generation I & Generation II &
Generation III & charge\\\hline
& up ($u$)  & charm ($c$)& top ($t$)& +(2/3)e \\
& (0.0015-0.0033) & (1.5) & (172) & \\
Quarks ($q$)&&&&\\
  & down ($d$) & strange ($s$) & bottom ($b$)& -(1/3)e\\
  & (0.0035-0.006) & (0.1) & (4.5) &\\\hline
  & electron ($e$)& muon ($\mu$)& tau ($\tau$)& -1\\
  & (0.0005) & (0.1) & (1.7) &\\
 Leptons ($l$) &&&&\\
  & electron neutrino ($\nu_e$)& muon neutrino ($\nu_\mu$)& tau neutrino ($\nu_\tau$)& 0\\
  & (<0.000000015) & (<0.00017) & (<0.024)&\\\hline
\end{tabular}
Table 1.1:\thinspace\thinspace\ Summary of quarks and leptons. The numbers in parentheses are \\the masses in GeV.
\end{center}

\indent The central rule in creating the Universe depends on the
four fundamental forces. Physicists use these forces to describe
quantitatively all the phenomena from the small scale of quarks
and leptons to the large scale of the whole Universe. Then, what are the four fundamental forces that govern the Universe? Let us list them as follows:\\
(i) The gravitational force: Attracts any two pieces of matter. It
has an infinite range and is the weakest force. It is responsible
for keeping stars, galaxies, and planetary systems in order, but
it has no significance in the particle physics realm.\\
(ii) The electromagnetic force: causes electric and magnetic
effects. It also has infinite range, but is much stronger than
gravity. This force acts only between electrically charged matter. Therefore,
it governs the motion of electrons around the nucleus. Note that
the relations between the spatial- and time-dependences of the
electric and magnetic fields \cite{Maxwell:1865zz} were explained by James Maxwell in
1865 through his equations; the Maxwell Equations.\\
(iii) The nuclear force: As the name suggests, it acts only
between nucleons. It is a result of the strong force which has a
very short range, acting only over of a range of $10^{-13}$ cm. It
is the strongest force
 and has the responsibility of binding quarks together, keeping them inside protons and neutrons, and also binds the protons and neutrons together.
 Therefore, it is responsible
 for the stability of the nucleons.\\
(iv) The weak force: Dominates only at the level of subatomic
particles. It is effective also over a very short range (see Table 1.2), and it is
stronger than gravity and weaker than
 others.\\
\indent The electromagnetic, strong, and weak forces arise from the
exchange of force-carrier particles which are bosons called-gauge bosons. Each of these four fundamental forces has a
different type of carrier: the electromagnetic force is carried by
the massless photon ($\gamma$) which is chargeless and is known as
the particle of light. The strong force has a corresponding boson,
the gluon ($g$), which is massless and is not charged electrically,
just as the photon, but which carries a different sort of charge,
called colour which holds the quarks confined within nucleons.
The gluon is thus related to nucleon stability. The bosons
$W$ and $Z$ ($W^\pm,\,Z^0$) are the corresponding force-carrying
particles of the weak force, these carriers are massive,
having a mass about 100 times that of the proton mass. The properties of the interaction forces in the Standard Model are summarized in Table 1.2. Moreover,
gravity may be carried by the ``graviton'', but it has not yet been
found. Note that leptons carry electromagnetic charge and weak
isospin as quantum numbers, but quarks may experience all four fundamental interactions, and carry the strong charge which is also called colour charge.\\

\begin{center}
\begin{tabular}
[c]{|c|c|c|c|c|c|}\hline Interaction & Mediator & Spin & Mass
(GeV) & Range (m) & acts on \\\hline Electromagnetic & $\gamma$  & 1 & 0 & $\infty$ & Quarks, Leptons, \\
    &     &    &    &    & $W^\pm$ \\\hline
 Weak & $W^\pm$ & 1 & $80.398\pm0.025$ & $\leq 10^{-18}$ &  Quarks, Leptons\\
      &  $Z^0 $ & 1 & $91.1876\pm0.0021$  &  & \\\hline
 Strong& $g$ & 1 & 0 & $<10^{-15}$ &  Quarks, Gluons \\\hline
\end{tabular}
\\
Table 1.2:\thinspace\thinspace The properties of the interactions in the Standard Model.
\end{center}

Concerning quarks, leptons, and all their fundamental
interactions, theory and experiment together produced a gauge
theory called the Standard Model of elementary particles.
Recently the Higgs boson which is an essential component of the standard Model was discovered by the ATLAS \cite{Aad:2012tfa} and CMS \cite{Chatrchyan:2012ufa} experiments at the Large Hadronic Collider (LHC) in 2012.\\
The Standard Model is based on a Lagrangian density with fields as
degrees of freedom. The strong and weak interactions are described
by Quantum Chromodynamics (QCD) and the
Glashow-Weinberg-Salam Theory of the Weak Interaction (GWS) \cite{Bilenky:1982ms},
respectively. Quarks and gluons carry colour charge
and have never been seen to exist as single-particle states. They
couple to themselves which leads to confinement
and asymptotic freedom. These can be further distinguished into baryons
and mesons. Leptons do not interact by the strong force. The production of hadrons as
observed in the final state of high-energy collisions, which
arise due to how quarks and gluons arrange themselves, is
described by the theory called quantum chromodynamics
(QCD), described in the following.

\section{Quantum Chromodynamics (QCD)\label{ch1.b}}

``\textit{In modern physics, there is no such thing as `nothing'.
Even in a perfect vacuum, pairs of virtual particle are constantly
being created and destroyed. The existence of these particles is
no mathematical fiction. Though they cannot be directly observed,
the effects they create are quite real. The assumption that
they exist leads to predictions that have been confirmed by experiment to a high degree of accuracy}'' $\,\,\,\,\,\,\,\,\,\,\,\,\,\,\,\,\,\,\,\,\,\,\,\,\,\,\,\,\,\,\,\,\,\,\,\,\,\,\,\,\,\,\,\,\,\,\,\,\,\,\,\,\,\,\,\,\,\,\,\,\,\,\,\,\,\,\,\,$  Richard Morris\\

The dynamics of baryons and mesons (hadrons) are described by
the theory of the fundamental interactions of quarks and
gluons, i.e., quantum chromodynamics (QCD) \cite{Marciano:1977su}.\\
\indent The fundamental symmetry underlying QCD is an exact local
$SU(3)_c$ colour symmetry. The quarks are coloured objects: $q \in 3_c$. As a consequence of the non-Abelian
nature of the $SU(3)_c$ symmetry, the gauge fields of QCD -
the gluons - are also coloured objects:  $g\in 8_c$.
Therefore, quarks and gluons interact strongly with each other.
The dynamics of this interaction are described by the
 QCD Lagrangian, see Sec. 2.2, which implies \textit{asymptotic freedom} and \textit{confinement}. Perturbation theory works in the high-energy regime \cite{Muta:1987Pert,Hagiwara:1984jk}, due to the
 asymptotically-free nature of QCD, such as for deep inelastic lepton-hadron scattering (DIS). However, at low energy (energies comparable to the low-lying hadron masses $\sim 1 GeV$), perturbation theory fails due to confinement and the dynamical breaking of chiral symmetry.\\
\indent The development of an effective low-energy approach to the
strong interaction plays an important role in the description of
the masses and the interactions of low-lying hadron resonances
\cite{Amsler:2004ps, Klempt:2007cp}, which is done by imposing chiral symmetry. One of the basic
symmetries of the QCD Lagrangian in the limit of vanishing mass
(the so-called chiral limit) \cite{Gasiorowicz, Meissner:1987ge} is the chiral symmetry which is
explicitly broken by the nonzero current quark masses, but it is
also spontaneously broken by a nonzero quark condensate in the QCD
vacuum \cite{Vafa:1983tf, Giusti:2007cn}. As a consequence, pseudoscalar (quasi-)Goldstone
bosons emerge. In a world with only $u$ and $d$ quarks (i.e., for
$N_f=2$ quark flavours), these are the pions, while for $N_f=3$,
i.e., when also the strange quark $s$ is considered, these are the
pions, kaons, and the $\eta$ meson. (The $\eta'$ meson is not a
Goldstone boson because of the chiral anomaly \cite{Pisarski:1983ms, 'tHooft:1976fv,'tHooft:1976up, 'tHooft:1986nc}). In the
present work we study the vacuum phenomenology of mesons in the framework of the extended Linear Sigma Model (eLSM)
which is an effective chiral model that emulates the global
symmetries of the QCD Lagrangian (see the details in Sec. 2.3).
The fundamental features of QCD are described in the following
section.

\section{Features of QCD}

\subsection{Asymptotic freedom}

This feature was observed by Gross, Wilczek, and Politzer in 1973
\cite{Gross:1973id, Politzer:1973fx}. (They won the Nobel Prize in Physics in 2004). Quarks
behave quasi-free at small distance or at high energies (high
compared to the rest mass of the proton). That means the
coupling/interaction strength $\alpha_s=g^2/4\pi$ between quarks
becomes weaker or smaller with  increasing energy, increasing
momentum, and decreasing interparticle distance. This prediction
was confirmed experimentally by deep-inelastic scattering of
leptons by nucleons \cite{Povh:2008jx}. A
quark-gluon plasma was predicted for high temperature and/or
baryonic chemical potential based on asymptotic freedom.
Perturbation theory confirmed the existence
 of a quark-gluon plasma phase \cite{Kapusta:1979fh}. Furthermore, at high temperature and high density, colour charged particles are liberated from hadrons, they become deconfined.\\
\indent  The opposite effect occurs at low energies, which means
the interaction/coupling strength between quarks becomes stronger with increasing distance. This leads to the emergence of
confinement, which means that colour-charged particles are
confined in colour-neutral states (hadrons).

\indent QCD seems like an expanded version of quantum
electrodynamics (QED). Both have charges: QED has electric charge
and QCD has the colour charges (red, green, and blue). Therefore, just as one considers the
force between two electric charges to understand and study
electromagnetic physics, one can analogously consider the strong
force between two colour charges to understand the strong interaction.
This leads us to explain asymptotic freedom in a simple way by
referring firstly to electromagnetic physics as follows \cite{Aitchison:1982kj}: \\

Coulomb's law describes the force between two charges $q_1$ and
$q_2$ in vacuum as

\begin{equation}
F= \frac{1}{4\pi}\frac{q_1\,q_2}{r^2}\,.
\end{equation}
However, in a medium with
dielectric constant $\epsilon > 1$, the force between them
becomes
\begin{equation}
F=\frac{1}{4\pi\epsilon}\frac{q_1\,q_2}{r^2}\,,
\end{equation}
which has the same form as in vacuum with the effective charge
$\widetilde{q}_i=q_{1,2}/\sqrt{\epsilon}$.\\
In quantum field theory, the vacuum is the lowest energy state of
a system. In QED, it is not empty but filled with
electrons of negative energies. When the photon travels through the
vacuum, an electron can be induced to jump from a negative to a
positive energy state, which creates a virtual pair of an electron
and a positron (the hole in the negative-energy continuum). That is known as a vacuum fluctuation. For this
reason, the interaction force between two electrons in the vacuum
becomes
\begin{equation}
F=\frac{e^2_{eff}}{4\pi\,r^2}=\frac{\alpha_{em}(r)}{r^2}\,,
\end{equation}
where $\alpha_{em}$ is an effective fine structure constant and
depends on the distance $r$ or the momentum transfer $q \sim
\frac{1}{r}$. The interaction strength of a low-energy photon at $r \rightarrow
\infty$ or ($\equiv q \rightarrow 0$) is
$\alpha_{em}(q=0)=1/137.035$ \cite{Nakamura:2010zzi}.\\
\indent In QED, the coupling as a function of the momentum scale $\mu$ can be determined by the
following differential equation
\begin{equation}
\mu\frac{\partial\alpha(\mu)}{\partial\mu}=\beta(\alpha(\mu))\,.\label{betadiff}
\end{equation}
From perturbation theory, the
$\beta$-function can be obtained at one-loop order as
$$\beta=2\,\alpha^2_{em}/3\,\pi> 0\,.$$
Then the solution can be obtained as
\begin{equation}
\alpha_{em}(\mu)=\frac{\alpha_{em}(\mu_0)}{1-\frac{\alpha_{em}(\mu_0)}{3\pi}ln\frac{\mu^2}{\mu^2_0}}\,.
\end{equation}
Now it is clear that when the distance between the two electrons
becomes smaller, their interaction strength gets stronger.
Therefore, QED is a strong-coupling theory at very short
distance scales.\\

Now let us turn to QCD which has a classical scale symmetry (see the
details in the next chapter). At the quantum level, this symmetry is spontaneously broken due to
the energy scale which is introduced by the renormalization of
quantum fluctuations. Therefore, the strong coupling $g$ depends on
the energy scale $\mu$ \cite{Gross:1973ju, Politzer:1974sm, Politzer:1974fr}:
$$ g \,\,\,\underrightarrow{\textit{renormalization}}\,\,\, g(\mu)$$

The beta function $\beta(g(\mu))$ of the renormalization group
describes the variation of the strong coupling with energy scale
$\mu$, called running coupling. It has the same differential equation
(\ref{betadiff}) as in QED 
\begin{equation}
\beta(g(\mu))=\mu\frac{\partial
g(\mu)}{\partial\mu}\,.\label{betaQCD}
\end{equation}
At one-loop level in perturbation theory the beta function
of QCD has the following form \cite{Gross:1973id, Politzer:1973fx}
\begin{equation}
\beta(g(\mu))= \frac{-11\,N_c+2\,N_f}{48\,\pi^2}\,g^3\,,
\end{equation}
where $N_f$ is the number of active quark flavours and $N_c$ the
number of colours. In nature, there are six quark flavours and
three colours. As seen in Eq.(\ref{betaQCD}), if
the $\beta-$function is negative ($\beta(g(\mu))\textless 0)$, then the QCD coupling decreases with increasing energy scale $\mu$.\\
The coupling constant of QCD is obtained from the solution of the differential equation (\ref{betaQCD}) as
\begin{equation}
g^2(\mu)=\frac{24\,\pi^2}{(11N_c-2N_f)\,ln(\mu/\Lambda_{QCD})}\,,\label{betaQCD1}
\end{equation}
which describes clearly that, when the energy scale is
increasing ($\mu\rightarrow \infty$) or the distance is decreasing
($d\,\rightarrow\, 0$), the QCD coupling constant
is decreasing ($g\,\rightarrow\, 0$). This feature is called asymptotic freedom.

\subsection{Quark Confinement}

The fact that the strong coupling grows in the increasing distance 
leads to the confinement of quarks, which means that no isolated elementary excitations of QCD, quarks, exist in
nature. Experimentally, no one has observed an isolated quark.
Quarks usually clump together to form hadrons, such as
baryons and mesons. In QCD, the confinement hypothesis has not
been directly derived until now. Note that, at any finite order in perturbation theory, there is no confinement.
Therefore, it is a nonperturbative phenomenon. Confinement has a lot of meanings. Four different
meanings are considered \cite{Hartmann:1999}: \\

(i) \textit{`Quarks cannot leave a certain region
in space'} \cite{Hartmann:1999},
 which is called \textit{`Spatial Confinement'}. The MIT-bag model explores the consequences of spatial confinement,
whereas the Chromodielectric Soliton Model \cite{Wilets:1996kr} attempts
to understand the mechanisms producing this confinement.

(ii) \textit{`String confinement'}: it is especially for mesons which are produced from scattering processes.
 From the features of the meson spectrum, the attractive
force between quark and anti-quark increases linearly with the
distance of the quarks. Moreover, quark and antiquark are linked
together by something which expands with
increasing energy. For all of that, free quarks never appear. Note
that the string breaks and new particles are created when the
corresponding energy exceeds a certain critical value
and the separation becomes large enough.

(iii)\textit{`There are no poles in the quark propagator'}. This definition is
limited to the, a priori unknown, quark propagator. Asymptotic quark states
cannot appear when the full quark propagator has no poles. This
means no free quarks exist.

(iv)\textit{`Colour Confinement'} means any composite particle must be a colour singlet
under the strong interaction at zero
temperature and density, and at distance scales larger than
$1/\Lambda_{QCD}$. M. Gell-Mann is the first one who introduced this type of confinement to solve his original quark-model problem.\\

Finally, we have to conclude from all of this that there are no isolated quarks and gluons in nature.

\subsection{Chiral Symmetry}

Chiral symmetry and its dynamical breaking are very important features of QCD at low energy. From the dynamical breaking of chiral symmetry, an effective quark mass is generated. \\

\textit{What is the meaning of chirality?}\\

\textbf{Kelvin's definition of chirality:} ``I call any
geometrical figure, or group of points \textit{`chiral'}, and say
it has chirality, if its image in a plane mirror, ideally
realized,
cannot be brought to coincide with itself'' $\,\,\,\,\,\,\,\,\,\,\,\,\,\,\,\,\,\,\,\,\,\,\,\,\,\,\,\,\,\,\,\,\,\,\,\,$(Lord Kelvin, 1904, The Baltimore Lectures)\\

In general, \textit{chirality} is the property of having for the same object a left-form and a right-form which are mirror images of each other.\\
The property of \textit{chirality} (or ``handedness'') is well-known
of many physical, chemical, and biological systems. In theoretical
physics, it is demonstrated that a quantum field theory cannot be
chirally symmetric if its Lagrangian density has explicit
mass terms. However, comparing to the rest mass of the proton (about 1000 MeV), the current
quark masses of the relevant quarks are small (about 10 MeV)
in the low-energy domain of QCD which
leads to an approximate realization of chiral symmetry. In the end,
there is a chiral partner (with the same mass, but opposite
parity and G-parity) for every eigenstate of the interaction. This is not seen in experimental data, from which are concludes that the chiral symmetry is broken. Note that the confinement and dynamical chiral symmetry breaking cannot be obtained in a simple
  perturbative analysis of QCD because they are low-energy phenomena, and in this regime, perturbation theory breaks down (for more details of
  low-energy theorems see
   \cite{Gasiorowicz, Meissner:1987ge, Pisarski:1994yp}). For this reason, effective chiral models are widely used to study the phenomenology of hadrons.\\
More details of the chiral symmetry and its spontaneous and
explicit breaking are described in the next chapter.

\section{Evidence for colour}

There is experimental and theoretical evidence for the existence of colour in nature.\\

\textbf{\textit{Experimentally:}}\\

(i) \textbf{e$^+$e$^-$ annihilation experiments}\\

In 1967, the results of high-energy electron and positron
annihilation experiments at the Stanford Linear Accelerator (SLAC) supported the colour charge of quarks.\\
\indent The confirmation of the existence of the colour quantum
number can be obtained from a comparison of the cross section of
the following two processes:
\begin{equation}
e^+e^- \longrightarrow \mu^+\mu^-\,\,\,\,
and\,\,\,\,e^+e^- \longrightarrow
\textit{hadrons}\,.\label{had}
\end{equation}

 Note that hadron production occurs only when
quarks are in the final state as a result of confinement.
Therefore, the production of hadrons occurs through
$$e^+e^- \longrightarrow \gamma\,\,(or\,\,Z) \longrightarrow \overline{q}q \longrightarrow \textit{hadrons}.$$

\begin{figure}[H]
\begin{center}
\includegraphics[width=5cm]{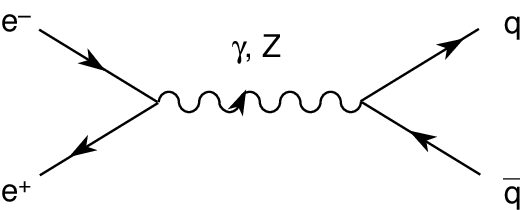}
\caption{Tree-level Feynman diagram for the \ $e^+e^-$
annihilation into hadrons.} \label{fig:eeqqhad}
\end{center}
\end{figure}

In this comparison of the cross sections, the weak
production factor involving the $Z$ for the previous process is neglected as well as
for the $e^+e^- \longrightarrow \mu^+\mu^-$ process as
seen in Eq.(\ref{had}), because of the dominance of the
cross section due to $\gamma$ exchange amplitude at the energies
below the $Z$ peak. The ratio of the cross sections for
the processes described in Eq.(\ref{had}) depends on the quark colour $N_c$ \cite{Pich:2005mk},

\begin{equation}
R \equiv \frac{\sigma(e^+e^- \longrightarrow
\textit{hadrons})}{\sigma(e^+e^- \longrightarrow
\mu^+\mu^-)}\simeq N_c\sum_{f=1}^{N_f} Q_f^2=\left\lbrace
\begin{array}
[c]{c}%
\frac{5}{9} N_c=\frac{5}{3},\,\,\,\,\text{for}\,\,\,(N_f=2:\,u,\,d)\,\,\,\,\,\,\,\,\,\,\\
\frac{2}{3} N_c=2,\,\,\,\,\text{for}\,\,\,(N_f=3:\,u,\,d,\,s)\,\,\,\,\,\\\
\frac{10}{9}N_c=\frac{10}{3},\,\,\text{for}\,\,(N_f=4:\,u,\,d,\,s,\,c)\,\,\,\,\,\\\
\frac{11}{9}N_c=\frac{11}{3},\,\,\text{for}\,\,(N_f=5:\,u,\,d,\,s,\,c,\,t)
\end{array}
\right.\,,
\end{equation}
where $Q_f$ denotes the electric charge of quark flavours $f$. The value of
the ratio, which corresponds to the experimental data of Fig. 1.2 \cite{Eidelman:2004wy} is obtained for $N_c=3$. Therefore, there are three quark colours in the physical world.\\

\begin{figure}[H]
\begin{center}
\includegraphics[width=11cm,height=5cm]{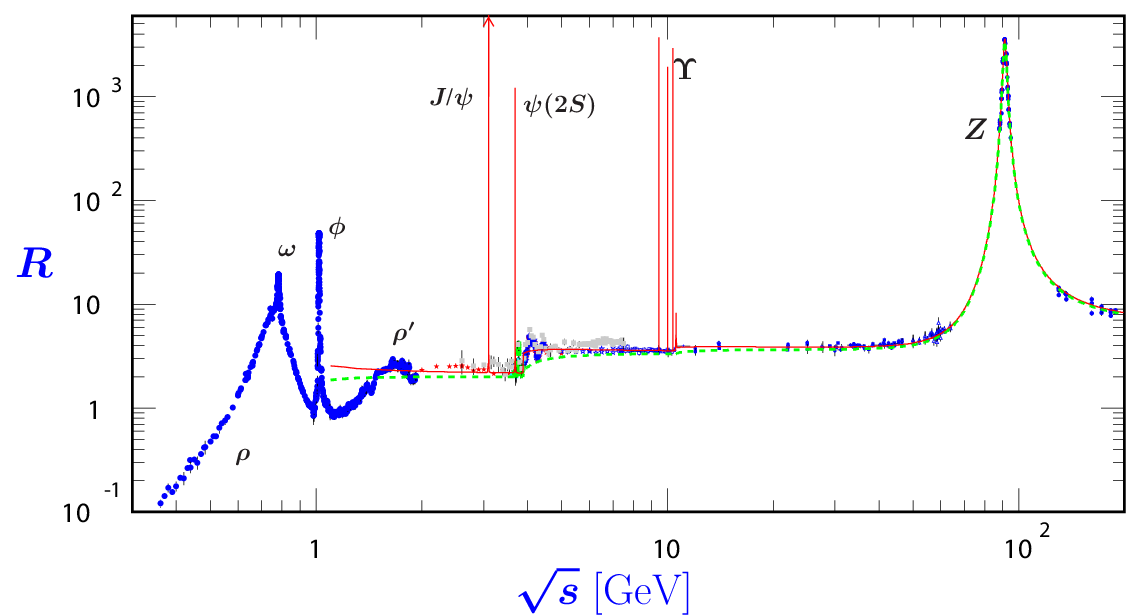}       
\caption{World data on the ratio $R_{e^+e^-}$.
The solid curve is the 3-loop perturbative QCD prediction. The broken lines show the naive quark model approximation with
$N_C=3$.}
\label{fig:Ree}
\end{center}
\end{figure}

(ii) \textbf{Decay of $\pi^0$ into $2\gamma$}\\

In 1967, Veltman tried to calculate the
$\pi^0$ decay rate and obtained that it is forbidden \cite{Veltmana1}. Following
this study, Adler, Bell, and Jackiw (1968-1970) \cite{Adler:1969gk, Bell:1969ts}, using the `fix'
field theory, which allowed $\pi^0$ to decay, found that its decay
rate is off by the factor of 9. During (1973-1974) many
physicists, notably Gell-Mann and Fritzsch, used QCD with three colours and arrived at the correct result of $\pi^0$ decay. Let us explain this in more detail:\\

The neutral pion $\pi^0$ is a meson composed of quarks,
$\pi^0=\frac{1}{\sqrt{2}}(\overline{u}u-\overline{d}d)$. The decay
width of the pion into two photons
 ($\pi^0\rightarrow 2\gamma$) is determined by a triangular quark loop in the Standard Model \cite{Pich:2005mk} as

\begin{equation}
\Gamma_{\pi^0\rightarrow
2\gamma}=\frac{\alpha^2\,m_\pi^2}{64\,\pi^3\,f_\pi^2}\left(\frac{N_c}{3}\right)^2\equiv
\left(\frac{N_c}{3}\right)^2 7.73\,eV\,,\label{TheGmapi}
\end{equation}
which depends on the number of colours and the decay constant of
the pion $f_\pi=92.4$ MeV. Experimental data give \cite{Nakamura:2010zzi}
\begin{equation}
\Gamma^{exp}_{\pi^0\rightarrow 2\gamma}=(7.83\pm0.37)eV\,.
\end{equation}
These experimental data is in very good agreement with the
Standard Model calculation (\ref{TheGmapi}) only when the number
of colours $N_c=3$. This is further evidence for the existence of
quark colour in nature.\\

\textbf{\textit{Theoretically}: solving the spin-statistics problem}\\

In 1965, the $\Delta-$baryon was discovered \cite{Roper:1964zza, Olsson:1966zza}, which is composed of
three quarks. Considering the baryon's charge $q=2e$, spin
$S=3/2$, and angular momentum $l=0$,
 there emerged a spin-statistics problem. When describing this particle in terms of $u$ and $d$ quarks, the spin-flavour wave function $\Delta^{++}$ had to be expressed as
\begin{equation}
|\Delta^{++}\rangle=|u_\uparrow u_\uparrow u_\uparrow \rangle\,,
\end{equation}

which describes an overall symmetric state. This violated
Fermi-Dirac statistics and the Pauli principle \cite{Pauli:1940zz} as well. This
paradox can be avoided by assuming the existence of colour as a degree of freedom for
quarks: a quark can carry three different colours which are
red ($r$), blue ($b$), green ($g$). Consequently, in the $\Delta^{++}$ the three $u$ quarks combine their colours in an antisymmetric
way as follows:

\begin{equation}
|\Delta^{++}\rangle_{
colour}=\frac{1}{\sqrt{6}}|r_1g_2b_3-g_1r_2b_3+b_1r_2g_3-b_1g_2r_3+g_1b_2r_3-r_1b_2g_3\rangle\,,
\end{equation}
which is in accordance with the Pauli principle.\\

\section{Baryons and Mesons}

Baryons and mesons are hadronic particles composed of
quarks and gluons bound together strongly and confined in colour
singlet states (colourless states).
They bear evidence for the existence of elementary (quark) constituents of matter because they come in many different forms in nature. \\

\textit{\textbf{Baryons}}: These are fermionic hadronic states with
half-integer spin and composed of three valence quark ($qqq$)
or three antiquarks ($\bar{q}\bar{q}\bar{q}$). A proton and a neutron are the lightest baryons, which
consist of $uud$ and $ddu$, respectively. Therefore, baryons are a
central part of nature
 and form the complex structure of the cores of atoms. The baryon number of a quark is $1/3$. Consequently, the baryon number for baryons is $1$, while for antibaryons it is $-1$. \\
The wave function of a baryon $B=qqq$ is antisymmetric under colour
exchange and can thus be described as
\begin{equation}
|B\rangle_{
colour}=\frac{1}{\sqrt{6}}|q_{1_r}q_{2_g}q_{3_b}-q_{1_g}q_{2_r}q_{3_b}+q_{1_b}q_{2_r}q_{3_g}-q_{1_b}q_{2_g}q_{3_r}+q_{1_g}q_{2_b}q_{3_r}-q_{1_r}q_{2_b}q_{3_g}\rangle\,,
\end{equation}
which can also be written as
\begin{equation}
|B\rangle_{
colour}=\frac{1}{\sqrt{6}}\varepsilon^{\alpha\beta\gamma}|q_{1_\alpha}q_{2_\beta}q_{3_\gamma}\rangle\,,
\end{equation}
where $\varepsilon^{\alpha\beta\gamma}$ is the totally
antisymmetric tensor and $\alpha,\,\beta,\,\gamma$ refers to the
three different colours ($r,\,g,\,b$). Furthermore, there is a
hypothetical
 ``exotic'' baryon with an extra quark-antiquark pair additional to the original three quarks, which is called a pentaquark ($qqqq\bar{q}$). The $qqqg$ states, bound states of three quark and a gluon,
 are hybrid states and called hermaphrodite baryons. There is also a hypothetical dibaryon state which consist of six quarks and has baryon number $+2$.\\

\textit{\textbf{Mesons}}: These are bosonic hadronic states with integer
spin. Many meson types are known in nature. Most states consist
of $q\overline{q}$ (a quark bound with its antiquark). The pion is
the lightest meson which has a mass of about $140$ MeV/c$^2$ and is the
first meson to have been discovered \cite{Lattes:1947mw, Lattes:1947mx, Lattes:1947my}. The colour wave
function for the $q\overline{q}$ state is antisymmetrised as

\begin{equation}
|M\rangle_{
colour}=\frac{1}{\sqrt{3}}|r\overline{r}+g\overline{g}+b\overline{b}\rangle\,,
\end{equation}
or,
\begin{equation}
|M\rangle_{
colour}=\frac{1}{\sqrt{3}}\delta^{\alpha\beta}|q_{\alpha}\overline{q}_{\beta}\rangle\,,
\end{equation}
where $\delta^{\alpha\beta}$ denotes the antisymmetric tensor and
$\alpha,\,\beta\,\in$ $\{r,\,g,\,b\}$. A meson may decay into
electrons, neutrons, and photons as seen in the previous section
with the decay of the pion $\pi$ into two photons $\gamma\gamma$.
According to the hypothesis of ``exotic'' mesons, there are
tetraquarks, consisting of a quark pair and an antiquark pair
$[qq][\overline{q}\,\overline{q}]$, and also glueballs, bound states of gluons ($gg$). The recently discovered $XYZ$ states are candidates for tetraquarks. Moreover there are hybrid states of mesons
consisting of $q\overline{q}g$, bound states of quark-antiquark and gluon,
called hermaphrodite mesons. In reality, there are many particles
still not known in nature and every day presents a possibility of
discovery for experimental physicists. In the recent past (2012),
the Higgs boson was discovered.

The main goal of the present work is the study of the vacuum properties of the pseudoscalar glueball and charmed mesons via the chirally symmetric eLSM.

\subsection{Charmed mesons}

The charm quark ($c$) is a special one in the quark family, as it
is heavier than the first three light quarks and does not belong
to the regular flavour $SU(3)$, but stands in a weak doublet
 with the light strange quark. Therefore, it can act as a bridge between the light and heavy flavours. There are two types of charmed mesons: (i) The heavy-light $Q\overline{q}$ and $\overline{Q}q$
  mesons called open charmed mesons, where $Q$ is a heavy quark (referring to the charm quark $c$) and $q$ is a light quark (referring to $u,\,d,$ and $s$ quarks).
   (ii) The heavy-heavy $Q\overline{Q}$ mesons, composite
   states of charm and anticharm quark, are called hidden charmed mesons (charmonia). The current charm quark mass ($m_c\sim 1.3$ GeV) is larger than the characteristic
   energy scale for the strong interaction ($\Lambda_{QCD}\sim 300 $ MeV), by which enter, the perturbative regime, $m_{Q}>>\Lambda_{QCD}$.   \\
\indent Since the discovery of the charmonium/hidden charmed state
$J/\psi$ in November 1974 at the Stanford Linear Accelerator
Center (SLAC) \cite{Augustin:1974xw} and Brookhaven National Laboratory (BNL)
\cite{Aubert:1974js}, and two years later (in 1976) the discovery of open
charmed states at SLAC, the study of charmed meson spectroscopy
and decays has made significant experimental \cite{Li:2012tr, Albrecht:1993jn, Anastassov:2001cw, Bartelt:1995rq}
and theoretical process \cite{Godfrey:1985xj, Godfrey:1986wj, Capstick:1986bm, Neubert:1993mb}. Therefore, we are interested to study the vacuum properties of open and hidden charmed mesons.\\
\indent In present work, we show how the original $SU(3)$ flavour
symmetry of hadrons can be extended to $SU(4)$ in the framework of
a chirally symmetric model with charm as an extra quantum number.
Twelve new charmed mesons are included in addition to the
nonstrange-strange sector. The new charmed mesons of lowest mass,
the $D,\,D_s$, and the higher mass $\eta_c$ are quark-antiquark
spin-singlet states with quantum number $J^{PC} \equiv 0^{-+}$, i.e.,
pseudoscalar mesons. The scalar mesons $D^\ast_0, D^\ast_{S0}$, and
$\chi_{C0}$ are spin-singlet states with $J^{PC}=0^{++}$. The
vector mesons $D^\ast,\,D^\ast_S,$ and $J/\psi$ are
quark-antiquark spin triplets with $J^{PC}=1^{--}$. The
axial-vector mesons $D_1$, $D_{S1}$, and $\chi_{C1}$ are
quark-antiquark spin triplets with $J^{PC}=1^{++}$. The additional
charmed fields $D^{\ast0},\,D^\ast,\,D^{\ast0}_0,\,\chi_{C1},\,
\chi_{C0}$, and $J/\psi$ are assigned to the physical resonances
$D^\ast(2007)^0,\, D^\ast(2010)^\pm, \, D^\ast_0(2400)^0,
\,D^\ast_0(2400)^\pm,\,\chi_{C1}(1P),\,\chi_{C0}(1P)$, and the
well known ground state $J/\psi(1P)$, respectively. The isospin
doublet $D^0_1$ is $D_1(2420)$. The $D$ is assigned to the well-established $D$
resonance. The isospin singlet $D_{S1}$ can be assigned to two
different physical resonances, $D_{S1}(2460)$ and $D_{S1}(2536)$
as listed by the Particle  Data Group PDG \cite{Beringer:1900zz}. The first
candidate can be interpreted as a molecular or a tetraquark state
as shown in Refs.\cite{Bartelt:1995rq, Guo:2009id, Liu:2012zya, Gutsche:2010zz, Gutsche:2010zza, Guo:2006rp, Cleven:2010aw}, which leads us assign $D_{S1}$ to
$D_{S1}(2536)$. Finally, the strange-charmed meson $D^\ast_{S0}$
is assigned to the only existing candidate $D^\ast_{S0}(2317)^\pm$
although it is also interpreted as a molecular or a tetraquark
\cite{Godfrey:1985xj, Godfrey:1986wj, Guo:2009id, Liu:2012zya, Guo:2009ct, Brambilla:2010cs} (For more details of the charmed meson assignment, see
Sec. 4.2). We compute charmed meson masses, weak decay constants, and
strong decay widths of
    (open and hidden) charmed mesons. Moreover, we calculate the decay width
      of a pseudoscalar ground state charmonium $\eta_c$ into a pseudoscalar glueball and the decay widths of a scalar charmonium $\chi_{C0}$ into a scalar glueball.
      The precise description of the decays of open
      charmed states is important for the CBM experiment at FAIR, while the description of
     hidden charmed states and the pseudoscalar glueball is vital for the PANDA experiment at the upcoming FAIR facility.

\subsection{Glueball}

The bound states of gluons form colourless, or `white', states
which are called glueballs. The first calculations of glueball
masses were based on the bag-model approach \cite{Jaffe:1975fd, Konoplich:1981ed, Jezabek:1982ic, StrohmeierPresicek:1999yv, Jaffe:1985qp}.
Later on, the rapid improvement of lattice QCD allowed for precise
simulations of Yang-Mills theory, leading to a determination of
the full glueball spectrum \cite{Morningstar:1999rf, Loan:2005ff, Chen:2005mg} (see Table 1.3). However, in full
QCD (i.e., gluons plus quarks) the mixing of glueball and
quark-antiquark configurations with the same quantum number
occurs, rendering the identification of the resonances listed by
the Particle Data Group (PDG) \cite{Nakamura:2010zzi} more difficult. The
search for states which are (predominantly) glueballs represents an
active experimental and theoretical area of research, see Refs.\
\cite{Klempt:2007cp, Close:1987er, Godfrey:1998pd, Giacosa:2009bj} and refs.\ therein. The reason for these efforts is
that a better understanding of the glueball properties would
represent an important step in the comprehension of the
non-perturbative behavior of QCD. However, although up to now some
glueball candidates exist (see below), no state which is
(predominantly) glueball has been unambiguously identified.

\begin{center}
\includegraphics[
height=3in, width=2.9in
]%
{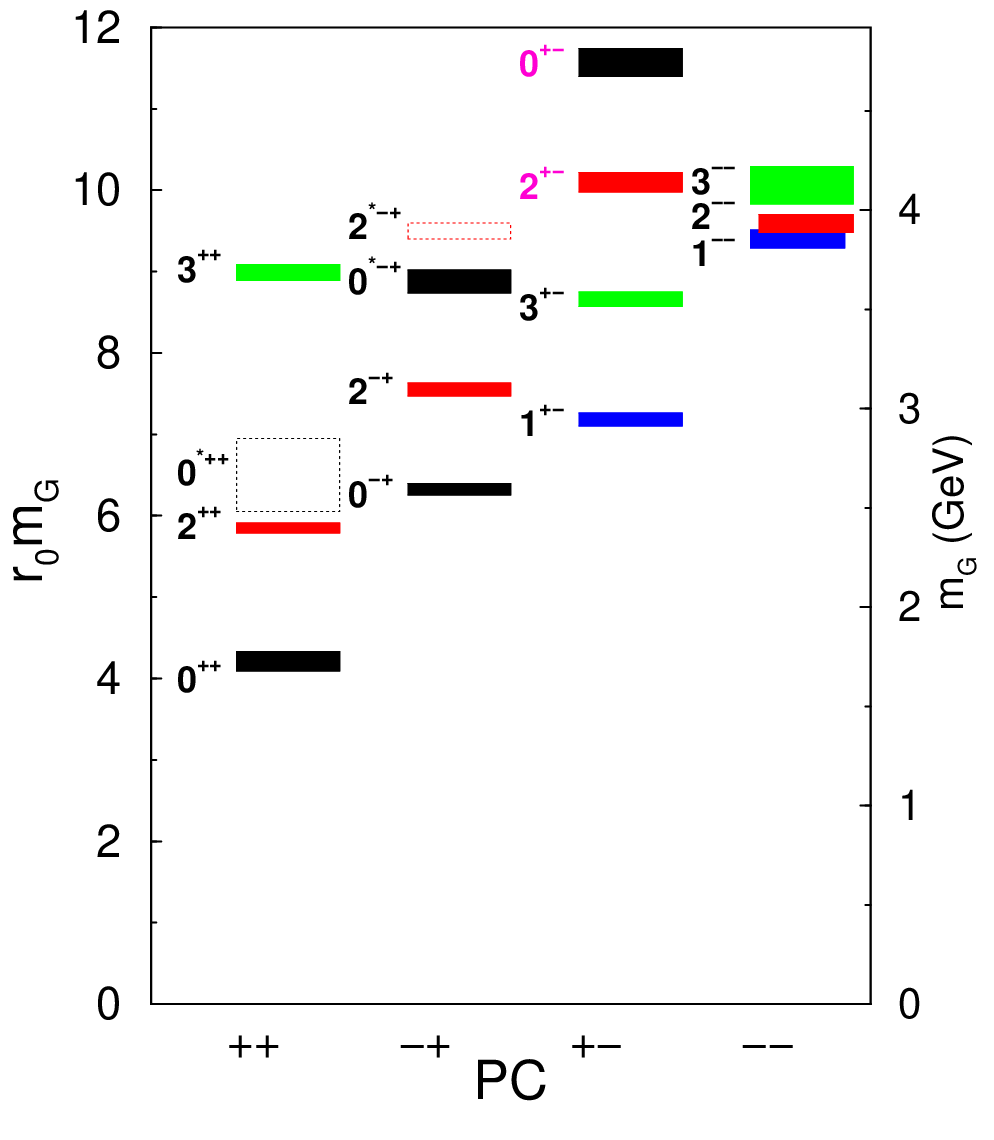}\\
Figure 1.3: The lightest six states in the spectrum of the $SU(3)$ Yang-Mills theory \cite{Morningstar:1999rf}.
\end{center}

In general, a glueball state should fulfill two properties
regarding its decays: it exhibits `flavour blindness', because the
gluons couple with the same strength to all quark flavours, and it
is narrow, because QCD in the
large-$N_{c}$ limit shows that all glueball decay widths scale as $N_{c}%
^{-2},$ which should be compared to the $N_{c}^{-1}$ scaling law
for a quark-antiquark state.

In Fig. 1.3, one can obtained that the lightest glueball state predicted by lattice QCD
simulations is a scalar-isoscalar state ($J^{PC}=0^{++}$) with a
mass of about $1.7$ GeV \cite{Morningstar:1999rf, Loan:2005ff, Chen:2005mg, Gregory:2012hu}. The resonance
$f_{0}(1500)$ shows a flavour-blind decay pattern and is narrow,
thus representing an optimal candidate to be (predominantly) a
scalar glueball. It has been investigated in a large variety of
works, e.g.\ Refs.\ \cite{Amsler:1995td, Lee:1999kv, Close:2001ga, Giacosa:2005qr, Giacosa:2004ug, Mathieu:2008me, Janowski:2011gt, Giacosa:2005zt, Cheng:2006hu, Chatzis:2011qz, Gutsche:2012zz, Janowski:2014ppa} and refs.\ therein, in
which mixing scenarios involving the scalar resonances
$f_{0}(1370),$ $f_{0}(1500)$, and $f_{0}(1710)$ are considered.
The second lightest lattice-predicted glueball state has tensor
quantum numbers ($J^{PC}=2^{++}$) and a mass of about $2.2$ GeV; a
good candidate could be the very narrow resonance $f_{J}(2200)$
\cite{Giacosa:2005bw, Burakovsky:1997ci}, if the total spin of the latter will be
experimentally confirmed to be $J=2$.

The third least massive glueball predicted by lattice QCD has
pseudoscalar quantum numbers ($J^{PC}=0^{-+}$) and a mass of about
$2.6$ GeV. Quite remarkably, most theoretical works investigating
the pseudoscalar glueball did not take into account this
prediction of Yang-Mills lattice studies, but concentrated their
search around $1.5$ GeV in connection with the
isoscalar-pseudoscalar resonances $\eta(1295),\eta(1405)$, and
$\eta(1475)$. A candidate for a predominantly light pseudoscalar
glueball is the middle-lying state $\eta(1405)$ due to the fact
that it is largely produced in (gluon-rich) $J/\psi$ radiative
decays and is missing in $\gamma\gamma$ reactions \cite{Masoni:2006rz, Gutsche:2009jh, Cheng:2008ss, Mathieu:2009sg, DiDonato:2011kr, Li:2009rk}.
In this framework the resonances $\eta(1295)$ and $\eta(1475)$
represent radial excitations of the resonances $\eta$ and
$\eta^{\prime}$. Indeed, in relation to $\eta$ and
$\eta^{\prime}$, a lot of work has been done in determining the
gluonic amount of their wave functions. The KLOE Collaboration
found that the pseudoscalar glueball fraction in the mixing of the
pseudoscalar-isoscalar states $\eta$ and $\eta^{\prime}$ can be
large ($\sim14$\%) \cite{Ambrosino:2009sc}. However, the theoretical
work of Ref.\ \cite{Escribano:2007cd} found that the glueball amount in
$\eta$ and $\eta^{\prime}$ is compatible with zero.

In this work we study the decay properties of a pseudoscalar
glueball state, (see chapter 6), whose mass lies, in agreement with lattice QCD,
between $2$ and $3$ GeV.

\section{Thesis Content}

\subsection{Second chapter}

We construct the QCD Lagrangian and discuss the symmetries of QCD
and their breaking. We then construct also a chirally invariant
Lagrangian for mesons, the so-called extended Linear Sigma Model
(eLSM) which satisfies two requirements: (i) global chiral
symmetry, and (ii) dilatation invariance. This model includes
scalar and pseudoscalar mesons, as well as vector and axial-vector mesons.

\subsection{Third chapter}

We present the extended linear sigma model (eLSM) for $N_f=2$
which describes the interaction of a scalar glueball and a
tetraquark with baryonic degrees of freedom which are
 the nucleon $N$ and its chiral partner $N^\ast$. We present the outline of extension of the eLSM from the
two-flavour case ($N_f=2$) to the three-flavour case ($N_f=3$) which
includes the strange sector as a new degree of freedom. We discuss
the fit parameters and present the results for the light meson masses.

\subsection{Fourth chapter}

We enlarge the so-called extended linear Sigma model (eLSM) by
including the charm quark to a global $U(4)_r\times
U(4)_l$ chiral symmetry. Most of the parameters of the model have been
determined in a previous work by fitting hadron
properties involving three quark flavours. Only three new parameters, all related to the current charm
quark mass, appear when introducing charmed mesons. Surprisingly, within the
accuracy expected from our approach, the masses of
open charmed mesons turn out to be in quantitative agreement with
experimental data. On the other hand, with the exception of $J/\psi$,
the masses of charmonia are underpredicted
by about 10\%. It is remarkable
that our approach correctly predicts (within errors)
the mass splitting between spin-0 and spin-1 
negative-parity open charm states. This indicates that, although
the charm quark mass breaks chiral symmetry quite strongly explicitly, 
this symmetry still seems to have some influence on the properties
of charmed mesons.


\subsection{Fifth chapter}

In our framework we study the decays of the pseudoscalar
glueball and charmed mesons. Therefore, we develop the two- and three-body decay formalisms
which are used in this study. Moreover, we present a simple method
for the calculation of the decay constants by using an axial transformation.

\subsection{Sixth chapter}

In this chapter we present a chirally invariant Lagrangian for
$N_f=2$ which describes the interaction of a pseudoscalar
glueball, $\tilde{G}$, with baryonic degrees of freedom which are
 the nucleon $N$ and its chiral partner $N^\ast$. Then we consider an $N_f=3$ chiral Lagrangian which describes the interaction between the pseudoscalar
 glueball, $J^{PC}=0^{-+}$, and scalar and pseudoscalar mesons. We calculate the mesonic and baryonic decays of the pseudoscalar glueball, where we fixed its
 mass to $2.6$ GeV, as predicted by lattice-QCD simulations, and take a closer look at the scalar-isoscalar decay channel. We present our results as branching
 ratios which are relevant for the future PANDA experiment at the FAIR facility. For completeness, we also repeat the calculation for a
glueball mass of $2.37$ GeV which corresponds to the mass of the
resonance $X(2370)$ measured in the BESIII experiment.

\subsection{Seventh chapter}

In this chapter we study the OZI-dominant decays of the heavy
open charmed states into light mesons within the chiral model at
tree level. We also obtain the value of the charm-anticharm
condensate and the values of the weak decay constants of open
charmed $D$ and $D_S$ mesons and the charmonium state $\eta_C$. Most
of the parameters in the model have been determined in the case of
$N_f=3$ by fitting hadron properties for three flavours. Only three new parameters, all related to the current charm
quark mass, are fixed in the fourth chapter.  The results are
compatible with the experimental data, although the theoretical
uncertainties are still large. The precise description of the
decays of open charmed states is important for the CBM experiment
at FAIR and the upcoming PANDA experiment.

\subsection{Eighth chapter}

In this chapter we expand our work of a $U(4)_r \times U(4)_l$
symmetric linear sigma model with (axial-)vector mesons by
including a dilaton field, a scalar glueball, and the interaction of a
pseudoscalar glueball with (pseudo)scalar mesons to study the
phenomenology of charmonium states. We compute the decay channels
of the scalar and pseudoscalar charmonium states
$\chi_{c0}(1P)$ and $\eta_c(1S)$, respectively. We calculate the
decays of $\chi_{c0}$ into the two scalar-isoscalar resonances
$f_0(1370)$ and $f_0(1500)$. We also study the decay of $\eta_C$ into a pseudoscalar
glueball. We compute the mixing angle between a
pseudoscalar glueball, with a mass of $2.6$ GeV, and the hidden
charmed meson $\eta_C$.

\chapter{Construction of mesonic Lagrangians}\label{chapterCh2 CSM}

\section{Introduction}

For many years, it has been known that chiral symmetry breaking in
QCD is responsible for the low mass of pions, which leads us to
describe the low-energy limit of QCD by the effective chiral
Lagrangian as in Ref. \cite{Meissner:1987ge, Pisarski:1994yp, Gasiorowicz:1969kn}. We have to note that effective
chiral models are widely used to study such phenomena because
perturbative QCD calculations cannot reproduce low-energy hadronic
properties due to confinement of colour charges.\\

Effective field theory (EFT) provides a fundamental framework to
describe physical systems with quantum field theory. The chiral
model is an effective field theory containing hadrons as degrees of freedom, which are colour-neutral because of the
confinement hypothesis. Moreover, in effective hadronic theories the
chiral symmetry of QCD can be realized in
the so-called nonlinear or in the linear representations. In the nonlinear
case, only the Goldstone bosons are considered
\cite{Schwinger:1957em, GellMann:1960np,Weinberg:1966fm,Gasser:1983yg, Scherer:2002tk} [in recent extensions vector mesons are
also added, see e.g.\ Refs.\ \cite{Bando:1987br, Ecker:1988te, Jenkins:1995vb, Terschlusen:2011pm}]. On the contrary, in
the linear case also the chiral partners of the Goldstone bosons,
the scalar mesons, are retained \cite{Schwinger:1967tc, Weinberg:1968de, Ko:1994en, Urban:2001ru}. When
extending this approach to the vector sector, both vector and
axial-vector mesons are present \cite{Ko:1994en, Urban:2001ru}. Along these
lines, recent efforts have led to the construction of the
so-called extended linear sigma model (eLSM), first for $N_{f}=2$
\cite{Janowski:2011gt, Parganlija:2010fz, Gallas:2009qp} and then for $N_{f}=3$ \cite{Parganlija:2012fy} (see
the details in chapter 3). In the eLSM, besides chiral symmetry, a
basic phenomenon of QCD in the chiral limit has been taken into
account: the symmetry under dilatation transformation and its
anomalous breaking (trace anomaly), see e.g.\ Ref.\
\cite{Rosenzweig:1981cu, Salomone:1980sp, Rosenzweig:1982cb, Gomm:1984zq, Gomm:1985ut}. For these reasons, we have used the linear sigma model (LSM) which has also other important motivations, summarized as follows:\\
(a) The LSM contains pseudoscalar states and their chiral partners from the outset.\\
(b) The LSM can be extended to include a large number of different
fields, i.e., quark-antiquarks with various flavours (two flavours \cite{Parganlija:2010fz, Ko:1994en, Janowski:2011gt, Parganlija:2012xj},
three flavours \cite{Lenaghan:2000ey, Ecker:1988te, Parganlija:2012fy, Parganlija:2012xj, Janowski:2014ppa}, four flavours \cite{Eshraim:2014afa, Eshraim:2014eka, Eshraim:2014vfa, Eshraim:2014gya, Eshraim:2014tla}), the nucleon and
its chiral partner \cite{Gallas:2009qp, Gallas:2013ipa}, a pseudoscalar glueball \cite{Eshraim:2012jv, Eshraim:2012ju, Eshraim:2012rb, Eshraim:2013dn},
tensor mesons, and
tetraquarks ($\overline{q}\,\overline{q}\,q\,q$ mesons) \cite{Giacosa:2006tf}.\\
(c) In the LSM the properties of fields at non-zero temperatures and
densities $(T \neq 0 \neq \mu)$ can be studied. As seen in Refs.
\cite{Kovacs:2006ym, Kovacs:2007sy, Heinz:2008cv} the critical point of QCD at
 finite densities is studied, as well as the chiral phase transition.\\

In this chapter we will construct the QCD Lagrangian and discuss
its symmetries. We then go further to construct an effective model,
the extended Linear Sigma Model (eLSM), based on
the global chiral symmetry and dilatation symmetry and containing
the (axial-)vector mesons as well as the (pseudo-)scalar mesons. The
spontaneous and explicit symmetry breaking allows us to
study the phenomenology of mesons and glueballs such as masses, decay
widths, and scattering lengths, etc.

\section{Construction of QCD Lagrangian}

In this section we construct the QCD Lagrangian which is required to
possess two kinds of symmetries: (i) a global chiral symmetry.
(ii) a local (gauge) $SU(N_c=3)$ symmetry \cite{Han:1965pf}. The QCD
Lagrangian contains quarks $q_f$ with $N_f$ flavours and
gluons. It can be constructed by gauging the colour degrees of
freedom
with an $SU(3)-$colour gauge transformation. \\
The full QCD Lagrangian density is the sum of quark and gluon
terms
\begin{equation}
\mathcal{L}_{QCD}=\mathcal{L}_{q}+\mathcal{L}_{g}\,.
\end{equation}

Firstly, let us describe the quark term $\mathcal{L}_{q}$:\\

\textbf{Quarks} are spin-$\frac{1}{2}$ fermions. The Dirac-conjugate spinor is
denoted as $\bar{q}\equiv q^\dagger \gamma_0$. Each quark of a particular
flavour $q_f$ is a triplet in colour space, which
 is given by the following 3-component quark vector
\begin{equation}
q_f =\left(
\begin{array}
[c]{c}%
q_{f,r}\\
q_{f,g}\\
q_{f,b}
\end{array}
\right)\,,
\end{equation}
where $f$ represents the flavour index $u,\,d,\,s,$, etc. whereas
$r,\, g,$ and $b$ refer to the three different fundamental
colour charges of quarks, $red,\, green,$ and $blue$,
respectively.\\

In the fundamental representation each quark flavour $q_f$
separately transforms under the local colour $SU(3)_c$
group as

\begin{equation}
q_f(x) \rightarrow q'_f(x)=  U(\theta(x))\,q_f(x)\,,\label{qt}
\end{equation}
where $U(\theta(x))$ is a special unitary matrix, $U(\theta(x))\in SU(3)$, acting on the colour index and
requires eight real parameters. It is usually written in the form

\begin{equation}
U(\theta(x))=
\exp\bigg(-i\sum_{a=1}^{8}\frac{\lambda_a}{2}\Theta^a(x)\bigg)\,,\,\,\,\,\,\,\,\,\,\,\,\,U^\dagger
U=UU^\dagger=\textbf{1},\,\,\, detU=\textbf{1}\,,\label{U(x)}
\end{equation}

where $\Theta^a(x)$ denotes the associated local parameters and
$\frac{\lambda_a}{2}$ are hermitian $3\times 3$ matrices; the
generators of the $SU(3)$ gauge group, while $\lambda_a$ are the
Gell-Mann matrices which can be found in the Appendix. \\

In general, the quarks are represented by $4N_cN_f-$ dimensional
Dirac spinors. The Dirac Lagrangian for a free fermion
$\psi$ takes the following form
\begin{equation}
\mathcal{L}_{Dirac}=\overline{\psi}(i\gamma^\mu
\partial_\mu-m_\psi)\psi\,,\label{LDirac}
\end{equation}
which leads us to construct a Lagrangian involving the quark
flavours, which is invariant under arbitrary local $SU(3)$
transformations (\ref{qt}) in colour space; with the same structure
as the Dirac Lagrangian (\ref{LDirac}) 

\begin{equation}
\mathcal{L}_{q}=\sum_{f=1}^{N_f}\overline{q}_f(i\gamma^\mu D_\mu-
M_f)q_f\,,\label{lq}
\end{equation}
 where $\gamma^\mu$ are the Dirac matrices and $M_f$ denotes the diagonal $N_f\times N_f$ quark mass matrix.
The $SU(3)$
covariant derivative in Eq. (\ref{lq}) is
\begin{equation}
D_\mu=\partial_\mu-ig\, A^a_\mu\,\frac{\lambda_a}{2}\,,
\end{equation}
where $A^a_\mu$ are the gauge fields (gluons), and $g$ is the strong coupling constant.\\

Now let us turn to describe the gluon term $\mathcal{L}_{g}$:\\

\textbf{Gluons} are massless gauge bosons with
spin-one and form an octet under the global colour $SU(3)$ group,
because the gluons are described by eight real-valued functions
$A^a_\mu$. They mediate the strong interaction between quarks and do not
make a distinction between different quark flavours. The gauge
fields (gluons), $A^a_\mu$, can be written as

\begin{equation}
A_\mu(x)=\sum_{a=1}^{8} A^a_\mu(x)\,\frac{\lambda^a}{2}\,,
\end{equation}

where $A_\mu$ is the matrix-valued vector potential of the non-Abelian gauge group $SU(3)$. It is a $3 \times 3$ matrix. The gluon fields transform
under local $SU(3)$ transformations as follows:

\begin{equation}
A_\mu (x) \rightarrow A'_\mu(x)=
U(\theta(x))\bigg(A_\mu(x)-\frac{i}{g}\partial_\mu\bigg)U^\dagger(\theta(x))\,.\label{At}
\end{equation}

The gluons carry colour charge, and their self-interactions are
responsible for many of the unique features of QCD. The
self-interactions of the gauge fields are described
by
\begin{equation}
G^a_{\mu\nu}(x)=\partial_\mu A^a_\nu(x)-\partial_\nu A^a_\mu(x)+g
f^{abc} A^b_\mu(x) A^c_\nu(x)\,,\label{GSInt}
\end{equation}
representing the gauge-invariant gluon field strength tensor. Here, $f^{abc}$ denote the
antisymmetric structure constants of the $SU(3)$ group.
\begin{equation}
[t^a,t^b]=i\,f^{abc}t^c, \,\,\, a,b,c=1,...,8\,.
\end{equation}
The field strength tensor (\ref{GSInt})
transform as follows under local $SU(N_c)$ transformation

\begin{equation}
G^a_{\mu\nu}(x)\, \frac{\lambda_a}{2}\, \rightarrow\,
\bigg(G_{\mu\nu}^a(x)\, \frac{\lambda_a}{2}\bigg)' = U(\theta(x))\, G^a_{\mu\nu}
\,\frac{\lambda_a}{2}\, U^\dagger(\theta(x))\,.
\end{equation}

We can then construct an additional gauge-invariant term involving
only gluons \cite{Yang:1954ek}, which is identical to the Yang-Mills Lagrangian
$\mathcal{L}_{YM}$

\begin{equation}
\mathcal{L}_{g}=-\frac{1}{4}\, G_{\mu\nu}^a(x)\,G^{\mu\nu}_a(x)\,.
\label{G1}
\end{equation}

Therefore, we construct the $SU(3)_c$-invariant Lagrangian of
Quantum Chromodynamics (QCD) from the sum of
Eq.(\ref{lq}) and Eq.(\ref{G1})

 \begin{equation}
\mathcal{L}_{QCD}=\sum_{f=1}^{N_f}\overline{q}_f(i\gamma^\mu
D_\mu- M_f)q_f-\frac{1}{4}\, G_{\mu\nu}^a
G^{\mu\nu}_a\,.\label{LQCD}
 \end{equation}

QCD has unique features, such as asymptotic freedom, quark
confinement, and chiral symmetry breaking, which are mentioned in detail in
the introduction. Beside these features there are other symmetry
features of the Lagrangian (\ref{LQCD}), which will be discussed in
the following section.

\section{Symmetry features of QCD}

In the previous section, the QCD Lagrangian has been constructed, which is the basis for all hadronic
models. Therefore, all models must implement the features of the
QCD Lagrangian, such as symmetries and their
spontaneous and explicit breaking. For this reason let us study
these features before constructing the so-called extended Linear
Sigma Model (eLSM) with vector and axial-vector mesons
\cite{Parganlija:2012xj}. \\

The features of the QCD Lagrangian are as follows:\\


\subsection{$Z(N_c)$ Symmetry}

The Abelian group $Z(N_c)$ is the center of the $SU(N_c)$ gauge group. The
special unitary $N_f \times N_f$ matrix, containing the center
elements $Z_n$, has the following general form
\begin{equation}
U(\theta(x))=Z_n
\exp\bigg(-i\frac{\lambda_a}{2}\Theta^a(x)\bigg),\,\,\,\,\,\,\,\,a=1,...,N_f^2-1\,,
\end{equation}
where
\begin{equation}
Z_n\equiv \exp\bigg(-i\frac{2\pi
n}{N_c}\,t^0\bigg),\,\,\,\,\,\,\,\,n=0,1,2,...,N_c-1\,,
\end{equation}
where $t^0=\lambda^0/2$. Quarks and gluon fields transform under
the $Z(N_c)$ group as
\begin{align}
q_f\,&\rightarrow\, q'_f=Z_n\,q_f\,,\label{qzn}\\
A_\mu\,&\rightarrow\,
A'_\mu=Z_n\,A_\mu\,Z_n^\dagger=A_\mu\,.\label{Azn}
\end{align}
Consequently, these transformations leave the QCD Lagrangian
$\mathcal{L}_{QCD}$ (\ref{LQCD}) invariant. At large temperature,
this symmetry is spontaneously broken in the gauge sector of QCD
(without quarks). The spontaneous breaking of this symmetry
indicates deconfinement of gluons. At nonzero temperature, the
$Z(N_c)$ symmetry is explicitly broken in the presence of
quarks, since the necessary antisymmetric boundary conditions are
not fulfilled for the fermion field. \\
This symmetry is important for much modern research in hadronic
physics at nonzero temperature and density. The order parameter of
the spontaneous symmetry breaking of the center group
$Z(N_c)$ at high $T$ is the Polyakov loop.\\

\subsection{Local $SU(3)_c$ colour symmetry}

The colour group $SU(3)$ corresponds to a local symmetry. As seen
in Eq.(\ref{At}), the QCD Lagrangian $\mathcal{L}_{QCD}$ is
invariant under local $SU(3)_c$ symmetry transformations. In hadronic
models, this local symmetry is satisfied automatically. Hadrons are
colour singlets. The $SU(N_c=3)$ group involves also the
transformation under the center elements $Z_n$, while
the gauge fields (i.e., gluons) transform under the $Z(N_c)$ group as
seen in Eq.
(\ref{Azn}).\\

\subsection{Scale invariance and trace anomaly}

Scale invariance (the so-called dilatation symmetry) is one of
the most important features of QCD because the classical scale
invariance is a profound phenomenon for the QCD Lagrangian. The
classical QCD Lagrangian (\ref{LQCD}) is invariant under space-time
dilatations in the limit of vanishing quark masses
($M_f\rightarrow 0$). That is clear, since no dimensionful
  parameters appear in the QCD Lagrangian density, which has a dimensionless coupling constant $g$ (discussed in chapter 1).\\
Note that the dilatation symmetry is broken by
quantum fluctuations. Let us consider the gauge sector
(no quarks), which is described by the
Yang-Mills (YM)
 Lagrangian (\ref{G1})
\begin{equation}
\mathcal{L}_{YM}=-\frac{1}{4}\,
G_{\mu\nu}^a(x)\,G^{\mu\nu}_a(x)\,.\label{LYM}
\end{equation}
The scalar (or dilatation) transformation is defined as
\begin{equation}
x^\mu \longrightarrow x'^\mu= \lambda^{-1}\,x^\mu\,.
\end{equation}
The gauge fields transform as
\begin{equation}
A^a_\mu(x) \longrightarrow {A^a_\mu}'(x')= \lambda\,A^a_\mu(x)\,,
\end{equation}
The action $S_{YM}=\int\,d^4x \mathcal{L}_{YM}$
is dimensionless and invariant under the scale transformations,\\

\begin{align}
S_{YM}^{\prime } &=\int d^{4}X^{\prime }\mathcal{L} _{YM}^{\prime }=-\frac{1%
}{4}\int d^{4}X^{\prime }G_{\mu \nu }^{a\prime }G_{a}^{\mu \nu \prime }
\notag \\
&=-\frac{1}{4}\int \lambda ^{-4}d^{4}X\lambda ^{2}G_{\mu \nu }^{a}\lambda
^{2}G_{a}^{\mu \nu }  \notag \\
&=-\frac{1}{4}\int d^{4}XG_{\mu \nu }^{a}G_{a}^{\mu \nu }  \notag \\
&=\int d^{4}X \mathcal{L}_{YM}=S_{YM}\,.\label{sproofspar}
\end{align}
Then, the scale invariance (dilatation symmetry) is fulfilled in
the limit $M_f=0$. Dilatation symmetry is continuous and
leads to the conserved Noether current
\begin{equation}
J_{scale}^\mu=x_\nu\,\,T^{\mu\nu}\,,
\end{equation}
where the energy-momentum tensor for the gauge field $T^{\mu\nu}$
reads
\begin{equation}
T^{\mu\nu}=\frac{\partial\mathcal{L}_{YM}}{\partial(\partial_\mu
A_{\xi})}\,\partial^\nu A_\xi - g^{\mu\nu} \mathcal{L}_{YM}\,.
\end{equation}
On the classical level, the current is conserved because the action is
invariant under a continuous scale transformation (as
discussed above)
\begin{equation}
\partial_\mu J_{scale}^\mu=0\,.
\end{equation}
Then, the divergence
\begin{align}
\partial_\mu J_{scale}^\mu&=\partial_\mu(x_\nu T^{\mu\nu}(x))=\partial_\mu (g_{\nu\rho}\,x^\rho\, T^{\mu\nu}(x))\nonumber\\
&=g_{\nu\rho}\, g^{\rho}_{\mu}\,T^{\mu\nu}(x)+g_{\nu\rho}\, x^\rho \partial_\mu T^{\mu\nu}(x)\nonumber\\
&=g_{\nu\mu}T^{\mu\nu}(x)=T^{\mu}_{\mu}(x)\,,
\end{align}
where the energy-momentum tensor is conserved for a time-translation invariant and homogeneous system
\begin{equation}
\partial_\mu T^{\mu\nu}(x)=0\,.
\end{equation}
A conserved scaling current leads to a vanishing trace of the
energy-momentum tensor, 
\begin{equation}
T^{\mu}_{\mu}(x)=0\,.
\end{equation}
Note that if all particles are massive, the
scaling current would be not conserved, having a nonvanishing trace of the
energy-momentum tensor.\\

At the quantum level, the scale current $J_{scale}^\mu$ is
anomalous and the classical scale invariance is broken. This
breaking is generate by a gluon condensate, i.e., a nonvanishing vacuum
expectation value of  $(
G_{\mu\nu}^a\,G^{\mu\nu}_a)$. From the renormalization
techniques \cite{Gross:1973id, Gross:1973ju, Politzer:1974sm, Politzer:1974fr} and perturbative QCD, one knows
\begin{equation}
\partial_\mu J_{scale}^\mu=T^{\mu}_{\mu}=\frac{\beta(g)}{4\,g}G_{\mu\nu}^a\,G^{\mu\nu}_a\neq0\,,
\end{equation}
where $\beta(g)$ is the $\beta-$function of QCD (\ref{betaQCD1}).\\
For massive quark flavours ($M_f \neq 0$), the quark fields transform unbder dilatation as
\begin{equation}
q_f \longrightarrow q'_f= \lambda^{3/2}\,q_f\,.
\end{equation}
The result of explicit breaking
of the scale invariance by nonvanishing quark masses gives
\begin{equation}
T^{\mu}_{\mu}(x)=\sum^{N_f}_{f=1}\overline{q}_f\,M_f\,q_f\,.
\end{equation}
Therefore, the divergence of the quantum current becomes
\begin{equation}
\partial_\mu J_{scale}^\mu=T^{\mu}_{\mu}=\frac{\beta(g)}{4\,g}G_{\mu\nu}^a\,G^{\mu\nu}_a+\sum^{N_f}_{f=1}\overline{q}_f\,M_f\,q_f\,.
\end{equation}

 \subsection{CP- symmetry}

The charge-conjugation and parity symmetry (CP-symmetry) is one of the fundamental properties of QCD, which describes the symmetry between matter and antimatter.\\
\indent CP-symmetry is the combination of a C-symmetry (charge-conjugation symmetry) and P-symmetry (parity symmetry). The strong
interaction as described by the QCD Lagrangian (\ref{LQCD}) is
 invariant under the combination of CP transformations, as is the electromagnetic interaction, while CP-symmetry is violated by the weak interaction.
 Now, let us prove the CP-symmetry of the QCD
Lagrangian (\ref{LQCD}):\\

i) \textit{C-symmetry}: is the transformation of a particle into
an antiparticle, without a change in the physical law. The charge
conjugation transformation $C$ maps matter into antimatter.
This symmetry is between positive and negative charge. \\

The charge conjugation of quarks is
\begin{equation}
q\,\,
\overset{C}{\longrightarrow}-i\,\gamma^2\,\gamma^0\,\overline{q}^t
= -i\,\gamma^2\,\gamma^0\,(\gamma^0)^t\,q^*\,,\label{qc}
\end{equation}
where the superscript $t$ is the transposition. Using the
Dirac notation, $\delta=-i\gamma^2\gamma^0$, the previous equation becomes
\begin{equation}
q\,\, \overset{C}{\longrightarrow}\,\delta\,\overline{q}^t =
\delta\,(\gamma^0)^t\,q^*\,,\label{qc1}
\end{equation}
and then,
\begin{equation}
q^{\dagger}\overset{C}{\rightarrow}[\delta(\gamma^{0})^{t}q^{\ast}]^{\dagger
}=q^{t}(\gamma^{0})^{\ast}\delta^{\dagger}\text{ .} \label{qaC}%
\end{equation}

Note that the properties of the Dirac notation are:\\
 $$\delta^{-1}=\delta^\dagger\,\,\,\text{(unitary\,\,transformation)}\,,$$
 $$\delta^{-1}\gamma^\mu\delta =\delta^\dagger\gamma^\mu\delta=(-\gamma^\mu)^t,\,\,\text{and}\,\,\,\, \delta^{\dagger}\gamma^\mu =(-\gamma^\mu)^t\,\delta^{-1}\,.$$

Moreover, useful properties for Dirac matrices are
$$\gamma^{0}(\gamma^{0})^{\dagger}=1,\,\,\,\,(\gamma^{0})^{\ast}(\gamma^{0})^{t}=[\gamma^{0}(\gamma^{0})^{\dagger}]^{t}=1^t = 1.$$

All of these properties are used to prove the C-symmetry of
$\mathcal{L}_{QCD}$ (\ref{LQCD}) (see below). Furthermore, the
gluon fields in the term $(D_\mu=\partial_\mu-igA_\mu)$ transforms odd under $C$ and the commutation of the quark fields which
are fermions leads to an additional minus sign.\\

\textbf{\textit{ The QCD Lagrangian is invariant under charge-conjugation transformations.}}\\

The quark part of the QCD Lagrangian (\ref{LQCD}) transforms under
charge conjugation as proven in Ref. \cite{Parganlija:2012xj}: 
\begin{align}%
\mathcal{L}%
_{q} &  =\bar{q}_{f}i\gamma^{\mu}D_{\mu}q_{f}-\bar{q}_{f}M_{f}q_{f}\nonumber\\
&  \overset{C}{\longrightarrow}iq_{f}^{t}(\gamma^{0})^{\ast}\delta^{\dagger}%
\gamma^{0}\gamma^{\mu}(\partial_{\mu}+ig\mathcal{A_{\mu}})\delta(\gamma
^{0})^{t}q_{f}^{\ast}-q_{f}^{t}(\gamma^{0})^{\ast}\delta^{\dagger}\gamma
^{0}M_{f}\delta(\gamma^{0})^{t}q_{f}^{\ast}\nonumber\\
&  =i\bar{q}_{f}\gamma^{\mu}D_{\mu}q_{f}-\bar{q}_{f}M_{f}q_{f}\text{ .} \label{QCDC}%
\end{align}

$\\$

ii) \textit{P-symmetry}: is the symmetry under reflection of spatial
coordinates. The parity transformation creates the reflection of
spatial coordinates (mirror image) of a physical system.

The parity transformation for quark fields (fermions) reads

\begin{equation}
q(t,\vec{x})\overset{P}{\longrightarrow}\gamma^{0}q(t,-\vec{x})\,, \label{qP}%
\end{equation}

and thus%

\begin{equation}
q^{\dagger}(t,\vec{x})\overset{P}{\longrightarrow}q^{\dagger}(t,-\vec{x}%
)\gamma^{0}\text{ .} \label{qaP}%
\end{equation}

The anticommutation formula of the Dirac matrices

\begin{equation}
\{\gamma_\mu,\,\gamma_\nu\}=2\,g_{\mu\nu},\,\,
\,\,\,\,\,\,\,\,\,g_{\mu\nu}=diag(1,-1,-1,-1)\,,
\end{equation}
is used below.\\

\textbf{\textit{ The QCD Lagrangian is invariant under parity transformations.}}

\begin{align}%
\mathcal{L}%
_{q}  &
=\bar{q}_{f}(t,\vec{x})i\gamma^{i}D_{i}q_{f}(t,\vec{x})-\bar
{q}_{f}(t,\vec{x})M_{f}q_{f}(t,\vec{x})\nonumber\\
&
\overset{P}{\longrightarrow}\bar{q}_{f}(t,-\vec{x})i\gamma^{i}D_{i}q_{f}(t,-\vec{x})-\bar{q}%
_{f}(t,-\vec{x})M_{f}q_{f}(t,-\vec{x})\text{\,,} \label{lgP}%
\end{align}

where $i=1,2,3$ and it is invariant when $\mu=0$ \cite{Parganlija:2012xj}. The gauge part of the QCD Lagrangian
(\ref{LQCD}) is also conserved under parity
transformations.\\

From Eq.(\ref{QCDC}) and Eq.(\ref{lgP}), we conclude that the QCD
Lagrangian (\ref{LQCD}) has a CP-symmetry (it is invariant under the
combined set of
 transformations CP and also separately under $C$  and $P$).\\

\subsection{Chiral symmetry and $U(1)_A$ anomaly}

The QCD Lagrangian (\ref{LQCD}) for $N_f$ flavours of massless quarks
possesses a large global symmetry, namely a global chiral
$U(N_f)_R \times U(N_f)_L$ symmetry \cite{Koch:1997ei, Halzen}. The notion of chirality allows us
to decompose a quark spinor into two-component spinors
corresponding to left- and right-handed components as

\begin{equation}
q_{f}=\left(  \mathcal{P}_{R}+\mathcal{P}_{L}\right)  q_{f}=q_{f,
\, R}+q_{f, \,
L}\text{ ,} \label{qRL}%
\end{equation}

and for the Dirac-conjugate spinors :%

\begin{equation}
\bar{q}_{f}=\bar{q}_{f}\left(
\mathcal{P}_{R}+\mathcal{P}_{L}\right)
=\bar{q}_{f, \, L}+\bar{q}_{f, \, R}\text{\,,} \label{aqRL}%
\end{equation}

where $ \mathcal{P}_{R,\,L}$ are the left- and right-handed projection
operators which are defined as

\begin{equation}
\mathcal{P}_{R}=\frac{1+\gamma_{5}}{2}\,,\,\,\,\,\,\,\,\,\mathcal{P}_{L}=\frac{1-\gamma_{5}}{2}\,, \label{PRL2}%
\end{equation}

including a Dirac matrix
$$\gamma_{5}  =i\gamma_{0}\gamma^{1}\gamma^{2}\gamma^{3}=\left(
\begin{array}
[c]{cc}%
1 & 0 \\
 0 & -1 \\
\end{array}
\right)\,.$$

The QCD Lagrangian (\ref{LQCD}) can be written in terms of ``\textit{right-handed quarks}'', $q_R=\mathcal{P}_{R}\,q$, and
the ``\textit{left-handed quarks}'', $q_R=\mathcal{P}_{R}\,q$, by
using decomposed quark fields (\ref{qRL}) and (\ref{aqRL}) as
\begin{align}
\mathcal{L}_{QCD}=&\sum_{f=1}^{N_f}i(\overline{q}_{f,L}\gamma^\mu
D_\mu q_{f,L}+ \overline{q}_{f,R}\gamma^\mu D_\mu q_{f,R})\nonumber\\
&-\sum_{f=1}^{N_f}(\overline{q}_{f,R} M_f q_{f,L}+
\overline{q}_{f,L} M_f q_{f,R})-\frac{1}{4}\sum_{a=1}^{8}
G_{\mu\nu}^a G^{\mu\nu}_a\,,\label{LQCDpq}
\end{align}
which is invariant under the following global chiral  $U(N_f)_R
\times U(N_f)_L$ transformations of right- and left-handed quark
spinors in the chiral limit (without the terms containing $M_f$)\\
\begin{align}
q_{f,\,L}  &  \longrightarrow
q_{f,\,L}^{\prime}=U_{L}q_{f,\,L}=\exp\left\{
-i\sum_{a=0}^{N_{f}^{2}-1}\Theta_{L}^{a}t^{a}\right\}  q_{f,\,L}\text{ ,}%
\label{qfLt}\\
q_{f,\,R}  &  \longrightarrow
q_{f,\,R}^{\prime}=U_{R}q_{f,\,R}=\exp\left\{
-i\sum_{a=0}^{N_{f}^{2}-1}\Theta_{R}^{a}t^{a}\right\}
q_{f,\,R}\text{ .}
\label{qfRt}%
\end{align}
This is the so-called \textit{chiral symmetry} which is exact only when
$M_f\rightarrow 0$, because the terms which contain masses ($M_f$)
break the chiral symmetry explicitly. Note that \textit{chiral
symmetry} is not exact in nature. However, on the hadronic mass
scale $\sim 1$ GeV, the current masses of up, down, and strange
quarks are very small, which leads one to approximate their
masses as nearly massless
 ($m_u=m_d=m_s\simeq0$). Then, chiral symmetry approximately holds. Moreover, the current mass of the charm quark is already of the order of the typical hadronic mass scale,
  and the masses of bottom
 and top quark exceed the hadronic mass scale. In the chiral limit $m_u=m_d=m_s\sim0$, the QCD Lagrangian (\ref{LQCDpq}) reads 
\begin{equation}
\mathcal{L}^0_{QCD}=\sum_{f=1}^{N_f}i(\overline{q}_{f,L}\gamma^\mu
D_\mu q_{f,L}+ \overline{q}_{f,R}\gamma^\mu D_\mu
q_{f,R})-\frac{1}{4}\sum_{a=1}^{8} G_{\mu\nu}^a G^{\mu\nu}_a\,,
\end{equation}
where the superscript $0$ denotes the chiral limit. As mentioned
above this Lagrangian is invariant under the chiral symmetry group
$U(N_f)_R\times U(N_f)_L$. Note that if the Lagrangian contains non-vanishing 
quark mass terms, some of the $2N_f^2$ chiral currents are not conserved, which reveals the
pattern of explicit chiral symmetry
  breaking, see below.\\
In addition, the transformation of the left- and right-handed
quarks under the symmetry group $U(N_f)_A \times U(N_f)_V$
whereby the subscript $A$ stands for ``\textit{axial vector}''
 and $V$ for ``\textit{vector}'' is defined as
\begin{equation}
q_{f,L} \rightarrow q'_{f,L}=U_V U^{\dagger}_A q_{f,L}=\exp(2i\theta^a
t^a) \exp(-2i\tilde{\theta}^a t^a)q_{f,L}\,,
\end{equation}
\begin{equation}
q_{f,R} \rightarrow q'_{f,R}=U_V U_A q_{f,R}=\exp(2i\theta^a
t^a) \exp(2i\tilde{\theta}^a t^a)q_{f,R}\,,
\end{equation}
where $U_V \in U(N_f)_V$ and $U_A \in U(N_f)_A$. This
transformation is equivalent to the transformation under $U(N_f)_L
\times U(N_f)_R$, if one sets
\begin{equation}
\theta^a_L=2(\theta^a-\tilde{\theta}^a),\,\,\,\,\,\,\,\,\,\theta^a_R=2(\theta^a+\tilde{\theta}^a)\,.
\end{equation}
Therefore,
\begin{equation}
U(N_f)_L \times U(N_f)_R \cong U(N_f)_A \times
U(N_f)_V\,.\label{ulequr}
\end{equation}
Then, in the chiral limit, the QCD Lagrangian (\ref{LQCD}) is invariant also
under the symmetry group $U(N_f)_A\times U(N_f)_V$. Note that, for
all $V\in U(n)$
 with $n\in {\mathbb N}$ there exists $U\in SU(n)$,
so that
\begin{equation}
V=\text{det}(V)^{\frac{1}{n}} U\,,
\end{equation}
and for all $n\in {\mathbb N} \Longrightarrow
\text{det}(V)^{\frac{1}{n}}\in U(1)$, yields
\begin{equation}
U(n)=U(1)\times SU(n)\,.
\end{equation}
Therefore, the unitary group can be represented as a product of a
special unitary group and a complex phase as
\begin{align}
U(N_f)_A \times U(N_f)_V&=[U(1)_A \times SU(N_f)_A] \times [U(1)_V \times SU(N_f)_V]\nonumber\\
&=SU(N_f)_A \times SU(N_f)_V \times U(1)_A \times U(1)_V\,.
\end{align}
Similarly,
\begin{equation}
U(N_f)_R \times U(N_f)_L=SU(N_f)_R \times SU(N_f)_L \times U(1)_R
\times U(1)_L\,.
\end{equation}
Using Eq.(\ref{ulequr}), we obtain
\begin{equation}
SU(N_f)_R \times SU(N_f)_L \times U(1)_R \times U(1)_L \cong
SU(N_f)_R \times SU(N_f)_L \times U(1)_V \times U(1)_A\,,
\end{equation}
which gives
\begin{equation}
U(N_f)_R \times U(N_f)_L \equiv SU(N_f)_R \times SU(N_f)_L \times
U(1)_V \times U(1)_A\, .
\end{equation}

In the quantized theory \cite{Gasiorowicz:1969kn}, QCD is not invariant
under $U(1)_A$ anymore, as a result of the explicit breaking of
axial $U(1)_A$ symmetry, which is known as the $U(1)_A$ anomaly
of QCD \cite{Pisarski:1983ms, 'tHooft:1976fv, 'tHooft:1976up}. Therefore the chiral symmetry is
reduced to $SU(N_f)_R \times SU(N_f)_L \times U(1)_V$. However, in
classical field theory, $\mathcal{L}_{QCD}$ is invariant under
$U(1)_A$. Therefore, one has to take
this symmetry breaking into account when constructing the effective chiral model. Moreover, this symmetry is broken at the classical level for massive quark.   \\

\indent According to the Noether
theorem \cite{Hung:1994eq}, the conserved Noether current is:
\begin{align}
\frac{\partial\mathcal{L}(\varphi(x),\,\partial_\mu\varphi(x))}{\partial(\partial_\mu\varphi(x))}\delta\varphi(x)
+\delta x^\mu\mathcal{L}(\varphi(x),\,
\partial_\mu\varphi(x))\,.
\end{align}

where the Lagrangian
$\mathcal{L}(\varphi(x),\,\partial_\mu\varphi(x))$ is invariant
under the transformation of the form $x\rightarrow
x'(x)= x+\delta x$ and $\varphi(x)\rightarrow
\varphi'(x)=\varphi(x) +\delta\varphi(x)$. This symmetry leads to the
conserved left-handed and right-handed currents denoted as $L^\mu_a$
and $R^\mu_a$, respectively,

\begin{equation}
R^\mu_a=\overline{q}_R\gamma^\mu t_a
q_R\,\,\,\,\,\Rightarrow\,\,\,\,\,
R^\mu=V^\mu-A^\mu\,,\label{Rconv}
\end{equation}
\begin{equation}
L^\mu_a=\overline{q}_L\gamma^\mu t_a
q_L\,\,\,\,\,\Rightarrow\,\,\,\,\,
L^\mu=V^\mu+A^\mu\,.\label{Lconv}
\end{equation}

$\bullet$ The vector $U(1)_V$ symmetry of the full
Lagrangian (\ref{LQCD}) coincides with quark number conservation. According to the Noether theorem \cite{Peskin:1995ev}, the conserved $U(1)_V$ current reads\\
\begin{equation}
V^\mu_0=\frac{\partial \mathcal{L}}{\partial(\partial_\mu
q_f)}\delta q_f=\overline{q}_f \gamma^\mu t_0\delta q_f\,,
\end{equation}
and its divergence is
\begin{equation}
\partial_\mu V^\mu_0=i\overline{q}_f [t_0,\,M_f]q_f=0\,.
\end{equation}
The integration over
the zero$^{th}$ component of $V^\mu_0$ yields the conserved baryon-number charge
\begin{equation}
Q=\int d^3x \,\overline{q}_f\,\gamma^0\,q_f\,.
\end{equation}

$\bullet$ The $SU(N_f)_V$ symmetry: the Lagrangian is symmetric
under $SU(N_f)_V$ transformations only when the quark masses of all
flavours are degenerate $m_1=m_2=...=m_{N_f}$. According to the Noether theorem \cite{Noether:1918zz}, the conserved vector
current is
\begin{equation}
V^{\mu a}=\overline{q}_f \gamma^\mu t^a q_f\,.\label{Vcurrent}
\end{equation}
Its divergence reads
\begin{equation}
\partial_\mu V^{\mu a}=i\overline{q}_f [t^a,\,M_f]q_f\,,
\end{equation}
which vanishes only for degenerate quark masses.\\

$\bullet$ The $SU(N_f)_A$ symmetry: This symmetry is broken
spontaneously. The axial-vector current and its divergence
(according to the Noether theorem), respectively, read

\begin{equation}
A^{\mu a}=\overline{q}_f \gamma^\mu \gamma^5 t^a
q_f\,,\label{Acuurent}
\end{equation}
and
\begin{equation}
\partial_\mu A^{\mu a}=\,i\overline{q}_f\,\{t^a,\,M_f\}\,q_f\,,
\end{equation}
 which is conserved only if all quarks are massless.\\

 From the linear combination of the left- and right-handed
 currents, described in Eq. (\ref{Rconv}) and Eq.
 (\ref{Lconv}), one can obtain the
 vector and axial vector currents as follows:
\begin{equation}
V^\mu=\frac{L^\mu+R^\mu}{2}\,,
\end{equation}
and
\begin{equation}
A^\mu=\frac{L^\mu-R^\mu}{2}\,.
\end{equation}
Using the definition of the transformation under parity
$$q(t,\textbf{x}) \overset{P}{\longrightarrow} \gamma_0 q(t,\textbf{x})\,,$$
the vector and axial-vector transform under parity transformations into
$+$ or $-$ themselves:
\begin{equation}
V^{\mu,0}(t,\textbf{x}) \overset{P}{\longrightarrow}
PV^{\mu,0}(t,\textbf{x})P^{-1}=+V^0_\mu(t,-\textbf{x})\,,
\end{equation}
\begin{equation}
V^{\mu,a}(t,\textbf{x}) \overset{P}{\longrightarrow}
PV^{\mu,a}(t,\textbf{x})P^{-1}=-V^a_\mu(t,-\textbf{x})\,,
\end{equation}
and
\begin{equation}
A^{\mu,0}(t,\textbf{x}) \overset{P}{\longrightarrow}
PA^{\mu,0}(t,\textbf{x})P^{-1}=-A^0_\mu(t,-\textbf{x})\,,
\end{equation}
\begin{equation}
A^{\mu,a}(t,\textbf{x}) \overset{P}{\longrightarrow}
PV^{\mu,a}(t,\textbf{x})P^{-1}=+A^a_\mu(t,-\textbf{x})\,,
\end{equation}
where $a$ denotes the spatial index.

\section{Chiral symmetry breaking}

\subsection{Explicit symmetry breaking}
The chiral symmetry of QCD is completely broken in the case of
non-vanishing quark masses $M_f\neq 0$, which enters the
QCD Lagrangian via the mass term (combining the left-
and right- handed components) as
\begin{equation}
\mathcal{L}_{mass}=\sum_{f=1}^{N_f} \overline{q}_f M_f
q_f=\sum_{f=1}^{N_f}(\overline{q}_{f,L} M_f
q_{f,R}+\overline{q}_{f,L} M_f q_{f,R})\,.
\end{equation}
 The mass term breaks the $SU(N_f)_A$ symmetry. The axial $U(N_f)_A$ symmetry of the QCD is explicitly broken even when all quark
 masses are equal and non-vanishing $m_1=m_2=...=M_f\neq 0$. This breaking leaves only the $SU(N_f)_V$ symmetry.
 Consequently, the $SU(N_f)_V$ of QCD is preserved, but only if the
 quark masses of all flavours are degenerate. In nature $m_u\approx
 m_d$ which leads to the so-called isospin symmetry. The
 $SU(3)_V$ flavour symmetry is also approximately preserved although it is
 explicitly broken due to a sizable mass of the $s$
 quark.\\

\subsection{Spontaneous symmetry breaking}

The central phenomenon in the low-energy hadronic realm is the spontaneous chiral symmetry breaking. This mechanism is the reason for the almost massless pions, and their weak interaction \cite{Koch:1995vp}. It has profound consequences for the hadron masses, especially the mass splitting of chiral partners (see below), and causes mass differences between multiplets and influences many strong decay modes. \\
As discussed previously, the QCD Lagrangian for massless quarks is invariant under chiral transformations. Consequently, one should expect that the approximate chiral symmetry should be evident in the mass spectrum of the lightest mesons. For $N_f=2$, the current masses of the up and down quark flavours are small compared to the typical hadronic scale which is about $1$ GeV, $m_u\simeq 0.002$ GeV and $m_d\simeq 0.005$ GeV. Therefore, these two lightest quark flavours can be considered to be approximately massless. As a consequence, the QCD Lagrangian is invariant under a global $SU(2)_R\times SU(2)_L$ transformation. One may write the lightest mesonic states composed of up and down quarks, $q\equiv(u,d)$, ($\sigma,\,\,\pi,\,\,\rho,\,\,a_1$) \cite{Parganlija:2012xj} as
\begin{align}
\,\,\,\,\,\,\,\,\,\,\,\,\,\,\,\,\,\,\,\,\,\,\,\,\,\,\,\,\,\,\,\,\,\,\,\,\,\,\,\,\,\,\,\,\,\,\,\,\,\,\,\,\,\,\,\,\,\,\,\,\,&scalar \,\,singlet\,\,state: & \sigma \equiv\bar{q}q\,,\,\,\,\,\,\,\,\,\,\,\,\,\,\,\,\,\,\,\,\,\,\,\,\,\,\,\,\,\,\,\,\,\,\,\,\,\,\,\,\,\,\,\,\,\,\,\,\,\,\,\,\,\,\,\,\,\,\,\,\,\,\,\,\,\,\,\,\,\,\,\,\,\,\,\,\,\,\,\,\,\,\,\,\,\,\,\,\,\,\,\,\,\,\,\,\,\,\,\,\,\,\,\,\,\,\,\,\,\,\,\,\,\,\,\,\,\nonumber\\
\,\,\,\,\,\,\,\,\,\,\,\,\,\,\,\,\,\,\,\,\,\,\,\,\,\,\,\,\,\,\,\,\,\,\,\,\,\,\,\,&pseudoscalar\,\, triplet\,\,state: &\vec{\pi}  \equiv i\bar{q}\vec{\tau}\gamma_{5}q\,,\,\,\,\,\,\,\,\,\,\,\,\,\,\,\,\,\,\,\,\,\,\,\,\,\,\,\,\,\,\,\,\,\,\,\,\,\,\,\,\,\,\,\,\,\,\,\,\,\,\,\,\,\,\,\,\,\,\,\,\,\,\,\,\,\,\,\,\,\,\,\,\,\,\,\,\,\,\,\,\,\,\,\,\,\,\,\,\,\,\,\,\,\,\,\,\,\,\,\,\,\,\,\,\,\,\,\nonumber\\
\,\,\,\,\,\,\,\,\,\,\,\,\,\,\,\,\,\,\,\,\,\,\,\,\,\,\,\,\,\,\,\,\,\,\,\,\,\,\,\,&vector\,\, triplet\,\,state : &\vec{\rho}^\mu  \equiv\bar{q}\vec{\tau}\gamma^{\mu}q\,,\,\,\,\,\,\,\,\,\,\,\,\,\,\,\,\,\,\,\,\,\,\,\,\,\,\,\,\,\,\,\,\,\,\,\,\,\,\,\,\,\,\,\,\,\,\,\,\,\,\,\,\,\,\,\,\,\,\,\,\,\,\,\,\,\,\,\,\,\,\,\,\,\,\,\,\,\,\,\,\,\,\,\,\,\,\,\,\,\,\,\,\,\,\,\,\,\,\,\,\,\,\,\,\,\,\,\,\nonumber\\
\,\,\,\,\,\,\,\,\,\,\,\,\,\,\,\,\,\,\,\,\,\,\,\,\,\,\,\,\,\,\,\,\,\,\,\,\,\,\,\,&axial-vector\,\, triplet \,\,state: &\vec{a}_1^\mu  \equiv \bar{q} \vec{\tau} \gamma^\mu \gamma_5 q\,.\,\,\,\,\,\,\,\,\,\,\,\,\,\,\,\,\,\,\,\,\,\,\,\,\,\,\,\,\,\,\,\,\,\,\,\,\,\,\,\,\,\,\,\,\,\,\,\,\,\,\,\,\,\,\,\,\,\,\,\,\,\,\,\,\,\,\,\,\,\,\,\,\,\,\,\,\,\,\,\,\,\,\,\,\,\,\,\,\,\,\,\,\,\,\,\,\,\,\,\,\,\,\,\,\,\,\,\label{states}
\end{align}
The quark flavour transforms under an axial-vector transformation as
\begin{equation}
U(2)_A: q=q'\rightarrow e^{-i\gamma_5\frac{\vec{\tau}}{2} \cdot \vec{\Theta}}q \simeq (1-i\gamma_5\frac{\vec{\tau}}{2} \cdot \vec{\Theta})\,q\,,
\end{equation}
where $\tau_i$ are the Pauli matrices. Consequently, the states (\ref{states}) transform under an axial-vector transformation as
\begin{align}\label{states1}
&U(2)_A: \sigma \rightarrow \sigma' = \sigma - \vec{\Theta}\cdot\vec{\pi}\,,\nonumber\\
&U(2)_A: \vec{\pi} \rightarrow \vec{\pi}' = \vec{\pi} + \vec{\Theta}\cdot\vec{\pi}\,,\nonumber\\
&U(2)_A: \vec{\rho}^\mu \rightarrow \vec{\rho}\,'^\mu =   \vec{\rho}^\mu+\vec{\Theta}\times \vec{a}_1^\mu\,,\nonumber\\
&U(2)_A: \vec{a}_1^\mu \rightarrow  \vec{a}_1'^\mu =   \vec{a}_1^\mu+ \vec{\Theta}\times\vec{\rho}^\mu\,,
\end{align}
which gives that the scalar state $\sigma$ is rotated to the pseudoscalar state $\pi$ and vice versa, i.e., they are chiral partners. Likewise, the vector state $\vec{\rho}$ is rotated to its partner, the axial-vector $\vec{a}_1$, and vice versa. The axial symmetry $SU(N_f)_A$ is still exact within the QCD Lagrangian. The explicit breaking of axial symmetry does not occur in the limit of small $u,d$ quark masses. The chiral partners have the same masses. In this case, the vector state $\rho$ is assigned to the $\rho(770)$ meson with a mass of $m_\rho=775.49$ MeV and the axial-vector state $a_1$ to the $a_1(1260)$ meson with a mass of $m_{a_1}\simeq 1230$ MeV \cite{Nakamura:2010zzi}. The mass difference of the chiral partners $\rho$ and $a_1$ is of the order of the $\rho$ mass itself and cannot be explained by the explicit symmetry breaking even if the nonvanishing masses of the up and down quark flavours are taken into account. However, the spontaneous symmetry breaking explains this phenomena successfully, i.e., the axial symmetry of the QCD Lagrangian is spontaneously broken in the ground state at zero temperature as
\begin{equation}
SU(N_f)_R \times SU(N_f)_L  \rightarrow SU(N_f)_V\,.
\end{equation}
In the case of nonvanishing quark masses, the chiral symmetry is explicitly broken by the mass term. However, even in the limit of $M_f\rightarrow 0$ the chiral symmetry is also broken, but this time spontaneously, when the ground state has a lower symmetry than the Lagrangian. The QCD vacuum has a nonvanishing expectation value for the quark condensate \cite{Muta:1998vi}
\begin{equation}
<\bar{q}q>_{vac}=<\bar{q}_R\,q_L+\bar{q}_L\,q_R>\neq0\,.
\end{equation}
According to Goldstone's theorem \cite{Goldstone:1961eq}, through the spontaneous breaking of a global symmetry there emerge massless Goldstone bosons, whose mumber is identical to the number of broken symmetries $(N_f^2-1)$ and which are indeed experimentally observed. In the $N_f=2$ case, three Goldstone bosons were observed and are identified with the pions \cite{Lattes:1947mw, Lattes:1947mx, Lattes:1947my}. Their mass of about $140$ MeV is small on a hadronic mass scale, but evidently they are not completely massless. The small nonvanishing mass arises from explicit chiral symmetry breaking, and thus they are named pseudo-Goldstone bosons. In the $N_f=3$ case, additionally five pseudo-Goldstone bosons have been experimentally observed, which are named the four kaons and the $\eta$ meson. In the $N_f=4$ case, there are 15 pseudoscalar Goldstone bosons comprising pions, kaons, $\eta$'s, and charmed mesons, which consist of the fourth quark flavour; the so-called charm quark. Charm thus strongly breaks the chiral symmetry explicitly.

\section{Construction of an effective model}

The main aim of the present work is to study low-energy
hadronic properties from an effective chiral model which is based on
QCD. Therefore, the effective model must possess all
features of the QCD Lagrangian, which are: The exact $SU(3)_c$
local gauge symmetry, the dilatation
 symmetry, the chiral $U(1)_A$ anomaly, CP symmetry, the global $U(N_f)_R\times U(N_f)_L$
 chiral symmetry for massless quark flavours, as well as the
 explicit and spontaneous breaking of chiral symmetry. The
 relationship between these symmetries gives us an opportunity to
 formulate an effective model: the so-called extended Linear Sigma
Model (eLSM).\\
In this section, we construct the eLSM \cite{Parganlija:2012xj, FGhabilit} for the
(pseudo)scalar and (axial-)vector mesons as well as a dilaton
field, which is
 valid for an arbitrary number of flavours $N_f$ and colours
 $N_c$. Hadrons are the degrees of freedom in the eLSM; they are colour neutral as a result of the confinement hypothesis. Therefore, in
 the construction of the eLSM, we do not have to take
 into account the $SU(3)_c$
 colour symmetry, it is automatically fulfilled. Note that we construct all terms of the eLSM with global chiral invariance up
 to naive scaling dimension four \cite{Pisarski:1994yp, Urban:2001ru, Boguta:1982wr, Kaymakcalan:1984bz}. As shown in Refs. \cite{Gasiorowicz:1969kn, Parganlija:2010fz, Parganlija:2012fy, Parganlija:2012xj}, the
 description of meson decay widths is quite reasonable.\\

 The first and fundamental step for the construction of the eLSM is to define the
 mesonic matrix $\Phi$ which contains bound quark-antiquark states. The matrix $\Phi$ is a non-perturbative object, and one uses it to
 build the multiplet of the scalar and pseudoscalar mesons as

 \begin{equation}
\Phi _{ij}\equiv\sqrt{2}\bar{q}_{j,R}q_{i,L}\,.\label{phide}
\end{equation}
According to the left- and right-handed transformations of quarks
(\ref{qfLt}) and (\ref{qfRt}), the mesonic matrix transforms under chiral
transformations as
\begin{equation}
\Phi_{ij} \longrightarrow
\sqrt{2}\bar{q}_{k,R}U^\dagger_{kj,R}U_{il,L}q_{l,L}\equiv U_{il,L}\Phi_{lk}
U_{kj,R}^{\dagger }, \label{tphi}
\end{equation}
thus,
\begin{equation}
\Phi \longrightarrow U_{L}\Phi U_{R}^{\dagger }\,, \label{tphi1}
\end{equation}
Using Eq.(\ref{PRL2}), Eq.(\ref{phide}) can be written as
\begin{eqnarray}
\Phi _{ij} &\equiv &\sqrt{2}\bar{q}_{j,R}q_{i,L}=\sqrt{2}\bar{q}_{j}\mathcal{P}_{L}%
\mathcal{P}_{L}q_{i}=\sqrt{2}\bar{q}_{j}\mathcal{P}_{L}q_{i}  \notag \\
&=&\frac{1}{\sqrt{2}}\left( \bar{q}_{j}q_{i}-\bar{q}_{j}\gamma
^{5}q_{i}\right) =\frac{1}{\sqrt{2}}\left( \bar{q}_{j}q_{i}+i\bar{q}%
_{j}i\gamma ^{5}q_{i}\right)\notag \\
&\equiv& S_{ij}+iP_{ij}\, ,
\end{eqnarray}
 where $S_{ij}$ and $P_{ij}$ are the scalar and the pseudoscalar
 quark-antiquark currents, respectively, which are defined by
\begin{equation}
S_{ij}\equiv\frac{1}{\sqrt{2}}\bar{q}_{j}q_{i}\,,
\end{equation}
\begin{equation}
P_{ij}\equiv\frac{1}{\sqrt{2}}\bar{q}_{j}i\gamma ^{5}q_{i}\,.
\end{equation}

Eventually, one can write the combination of scalar and
pseudoscalar currents via the $\Phi$ matrix as
\begin{equation}
\Phi =S+iP\,,
\end{equation}
The matrices $S$ and $P$ are Hermitian and can be expressed as
follows:
\begin{equation}
S=S_{a}t_{a},\,\,\,\,\,\,\,\,\,\,\,\,\,P=P_{a}t_{a}\,, \label{s}
\end{equation}%
with
\begin{equation}
S_{a}=\sqrt{2}\bar{q}t_{a}q,\,\,\,\,\,\,\,\,\,\,\,\,\,
P_{a}=\sqrt{2}\bar{q}i\gamma ^{5}t_{a}q\,.\label{SPEqm1}
\end{equation}
where $t^a$ denotes the generators of a unitary group $U(N_f)$
with $a=0,...,N_f^2-1$. \\
We summarize the transformation properties of the scalar
fields $S$, the pseudoscalar fields $P$, and $\Phi$ in Table 2.1.

\begin{center}
\begin{tabular}
[c]{|c|c|c|c|} \hline         &
$S=\frac{1}{\sqrt{2}}\sum_{a=0}^8 S^a\,\lambda_a$ &
$P=\frac{1}{\sqrt{2}}\sum_{a=0}^8 P^a\,\lambda_a$ &
$\Phi=S+iP$\\
\hline Elements  & $S_{ij}\equiv\bar{q}_j\,q_i$  & $P_{ij}\equiv\bar{q}_j i\gamma^5 q_i$ & $\Phi_{ij}\equiv \sqrt{2}\bar{q}_{j,R} q_{i,L}$\\
\hline Currents  & $S^a\equiv\bar{q}\frac{\lambda_a}{\sqrt{2}}q$  & $P^a\equiv\bar{q}i\gamma^5\frac{\lambda_a}{\sqrt{2}}q$ & $\Phi^a\equiv\sqrt{2}\bar{q}_R\frac{\lambda_a}{\sqrt{2}}q_L$\\
\hline P  & $S(t,-\textbf{x})$ & $-P(t,-\textbf{x})$ & $\Phi^\dagger(t,-\textbf{x})$ \\
\hline C  & $S^t$ & $P^t$ & $\Phi^t$\\
\hline $U(N_f)_V$ & $U_V S U_V^\dagger$ & $U_V P U_V^\dagger$ & $U_V \Phi U_V^\dagger$ \\
\hline $U(N_f)_A$ & $\frac{1}{2}(U_A\Phi U_A+U^\dagger_A\Phi^\dagger U^\dagger_A)$ & $\frac{1}{2i}(U_A\Phi U_A-U^\dagger_A\Phi^\dagger U^\dagger_A)$ & $U_A\Phi U_A$\\
\hline $U(N_f)_R \times U(N_f)_L$ & $\frac{1}{2}(U_L\Phi U_R^\dagger+U_R\Phi^\dagger U^\dagger_L)$ & $\frac{1}{2i}(U_L\Phi U^\dagger_R-U_R\Phi^\dagger U^\dagger_L)$ & $U_L\Phi U_R^\dagger$ \\
\hline
\end{tabular}\\
Table 2.1: The transformation properties of $S,\,P,\,and\,\,\Phi$ \cite{FGhabilit}.\\
\end{center}

The eLSM contains also the vector and axial-vector mesons, which
are the basic degrees of freedom for the construction of the right- and the
left-handed vector fields. Now let us define the $N_f
\times N_f$ right-handed $R^\mu$ and left-handed $L^\mu$ matrices,
respectively, as
\begin{equation}
R_{ij}^{\mu }\equiv\sqrt{2}\bar{q}_{j,R}\gamma ^{\mu }q_{i,R}=\frac{1}{\sqrt{2}}%
\left( \bar{q}_{j}\gamma ^{\mu }q_{i}-\bar{q}_{j}\gamma ^{5}\gamma
^{\mu }q_{i}\right)\equiv V_{ij}^{\mu }-A_{ij}^{\mu }\,,
\label{rhs}
\end{equation}%
\begin{equation}
L_{ij}^{\mu }\equiv\sqrt{2}\bar{q}_{j,L}\gamma ^{\mu }q_{i,L}=\frac{1}{\sqrt{2}}%
\left( \bar{q}_{j}\gamma ^{\mu }q_{i}+\bar{q}_{j}\gamma ^{5}\gamma
^{\mu }q_{i}\right)\equiv V_{ij}^{\mu }+A_{ij}^{\mu }\,,
\label{lhs}
\end{equation}
where the vector and axial-vector currents are defined,
respectively, as
\begin{equation}
V_{ij}^{\mu }\equiv\frac{1}{\sqrt{2}}\bar{q}_{j}\gamma ^{\mu
}q_{i}=V^{\mu,a }t^{a};\,\,V^{\mu,a}\equiv\sqrt{2}\bar{q}\gamma
^{\mu }t^{a}q\, ,
\end{equation}%
\begin{equation}
A_{ij}^{\mu }\equiv\frac{1}{\sqrt{2}}\bar{q}_{j}\gamma ^{5}\gamma
^{\mu }q_{i}=A^{\mu,a}t^{a}\,; A^{\mu,a}=\sqrt{2}\bar{q}\gamma
^{5}\gamma ^{\mu }t^{a}q\,,
\end{equation}%
which are also Hermitian matrices. The right-handed matrix and the
left-handed matrix transform under the chiral transformation as
\begin{equation}
R^{\mu }\longrightarrow R^{\mu \prime }=U_{R}R^{\mu
}U_{R}^{\dagger }\,, \label{chtrv}
\end{equation}%
and
\begin{equation}
L^{\mu }\longrightarrow L^{\mu \prime }=U_{L}L^{\mu
}U_{L}^{\dagger }\,. \label{chtlv}
\end{equation}
From $R^\mu$ and $L^\mu$, we construct the right- and
left-handed field-strength tensors, $R^{\mu\nu}$ and $L^{\mu\nu}$,
respectively, as
\begin{equation}
R^{\mu \nu }=\partial ^{\mu }R^{\nu }-\partial ^{\nu }R^{\mu }\,,
\label{rfst}
\end{equation}%
\begin{equation}
L^{\mu \nu }=\partial ^{\mu }L^{\nu }-\partial ^{\nu }L^{\mu }\,,
\label{lfst}
\end{equation}%
which transform under chiral transformations as%
\begin{equation}
R^{\mu \nu }\longrightarrow R^{\mu \nu \prime }=U_{R}R^{\mu \nu
}U_{R}^{\dagger }\,,\label{trRmunu}
\end{equation}%
\begin{equation}
L^{\mu \nu }\longrightarrow L^{\mu \nu \prime }=U_{L}L^{\mu \nu
}U_{L}^{\dagger }\,.\label{trLmunu}
\end{equation}
We present the transformation properties of the right- and left-handed ($R^\mu,\,\, L^\mu$) fields in Table 2.2 and the vector and
the axial-vector fields ($V^\mu,\,\,A^\mu$) in Table 2.3.

\begin{center}
\begin{tabular}
[c]{|c|c|c|} \hline         &
$R_\mu=\frac{1}{\sqrt{2}}\sum_{a=0}^8 R^a_\mu\,\lambda_a$ &
$L_\mu=\frac{1}{\sqrt{2}}\sum_{a=0}^8 L^a_\mu\,\lambda_a$ \\
\hline Elements  & $R_{ij}^\mu\equiv\sqrt{2}\bar{q}_{j,R}\gamma^\mu\,q_{i,R}$  & $L_{ij}^\mu\equiv\sqrt{2}\bar{q}_{j,L}\gamma^\mu q_{i,L}$ \\
\hline Currents  & $R_\mu^a\equiv\bar{q}_R\,\gamma^\mu\frac{\lambda_a}{\sqrt{2}}q_R$  & $L^a_\mu\equiv\bar{q}_Li\gamma^\mu\frac{\lambda_a}{\sqrt{2}}q_L$ \\
\hline P  & $g^{\mu\nu}L_\mu(t,-\textbf{x})$ & $g^{\mu\nu}R_\mu(t,-\textbf{x})$  \\
\hline C  & $-L_\mu^t$ & $R_\mu^t$ \\
\hline $U(N_f)_V$ & $U_V R_\mu U_V^\dagger$ & $U_V L_\mu U_V^\dagger$  \\
\hline $U(N_f)_A$ & $U_A R_\mu U_A^\dagger$ & $U_A^\dagger L_\mu U_A$\\
\hline $U(N_f)_R \times U(N_f)_L$ & $U_R R_\mu U_R^\dagger$ & $U_L R_\mu U^\dagger_L$ \\
\hline
\end{tabular}\\
Table 2.2: The transformation properties of $R_\mu\,\,and\,\,L_\mu$ \cite{FGhabilit}.\\
\end{center}

\begin{center}
\begin{tabular}
[c]{|c|c|c|} \hline         &
$V_\mu=\frac{1}{\sqrt{2}}\sum_{a=0}^8 V^a_\mu\,\lambda_a$ &
$A_\mu=\frac{1}{\sqrt{2}}\sum_{a=0}^8 A^a_\mu\,\lambda_a$ \\
\hline Elements  & $V_{ij}^\mu\equiv\sqrt{2}\bar{q}_{j}\gamma^\mu\,q_{i}$  & $A_{ij}^\mu\equiv\sqrt{2}\bar{q}_{j}\gamma^5\gamma^\mu q_{i}$ \\
\hline Currents  & $V^a\equiv\bar{q}\,\gamma^\mu\frac{\lambda_a}{\sqrt{2}}q$  & $A^a\equiv\bar{q}\gamma^5\frac{\lambda_a}{\sqrt{2}}q$ \\
\hline P  & $g^{\mu\nu}V_\mu(t,-\textbf{x})$ & $-g^{\mu\nu}A_\mu(t,-\textbf{x})$  \\
\hline C  & $-V_\mu^t$ & $A_\mu^t$ \\
\hline
\end{tabular}\\
Table 2.3: The transformation properties of $V_\mu\,\,and\,\,A_\mu$ \cite{FGhabilit}.\\
\end{center}

The basic construction of the mesonic Lagrangian of the effective model combines several terms:
\begin{equation}
\mathcal{L}_{mes}=\mathcal{L}_{\Phi,AV}+\mathcal{L}_{AV}+\mathcal{L}_{U(1)_A}+\mathcal{L}_{G}+\mathcal{L}_{ESB}\,.\label{BCL}
\end{equation}

Now let us construct every term in detail.\\

$\\$

\textbf{(i) The Lagrangian density $\mathcal{L}_{\Phi,AV}$}:\\

The chiral symmetry of QCD is exact in the chiral limit $M_f\rightarrow 0$. The Lagrangian density $\mathcal{L}_{\Phi,AV}$ fulfils the chiral symmetry exactly. It contains the (pseudo)scalar and the (axial-)vector degrees of freedom and describes the interaction between them. The covariant derivative for the coupling of the (pseudo)scalar degrees of freedom to the (axial-)vector ones has the following structure
\begin{equation}
D^{\mu }\Phi =\partial ^{\mu }\Phi +ig_{1}(\Phi R^{\mu }-L^{\mu }\Phi )\,,
\label{ka}
\end{equation}
then
\begin{equation}
\left( D^{\mu }\Phi \right) ^{\dagger }=\partial _{\mu }\Phi ^{\dagger
}-ig_{1}(R_{\mu }\Phi ^{\dagger }-\Phi ^{\dagger }L_{\mu })\,,  \label{kad}
\end{equation}
which are invariant under global $U(N_f)_L\times U(N_f)_R$ transformations.
\begin{eqnarray}
\left( D^{\mu }\Phi \right) \longrightarrow \left( D^{\mu }\Phi \right) ^{\prime } = U_{L}D^{\mu }\Phi U_{R}^{\dagger }\,.
\end{eqnarray}
and
\begin{equation}
\left( D^{\mu }\Phi \right) ^{\dagger} \longrightarrow \left( D^{\mu }\Phi \right) ^{\dagger \prime }=U_{R}\left( D^{\mu }\Phi
\right) ^{\dagger }U_{L}^{\dagger }\,.
\end{equation}
Therefore, the chirally invariant kinetic term can be constructed as
\begin{equation}
\text{\textrm{Tr}}\left[ (D^{\mu }\Phi )^{\dag }(D^{\mu }\Phi )\right] .
\end{equation}
The following self-interaction terms can be introduced up to naive scaling dimension four,
\begin{equation}
-\lambda _{1}( \text{\textrm{Tr}}[ \Phi ^{\dag }\Phi]) ^{2}\,,
\end{equation}%
\begin{equation}
-\lambda _{2}\text{\textrm{Tr}}(\Phi ^{\dag }\Phi) ^{2}\, ,
\end{equation}
which are also invariant under global chiral transformations. \\
\indent \textit{Proof:}
\begin{eqnarray}
-\lambda _{1}(\text{\textrm{Tr}}[ \Phi ^{\dag \prime }\Phi ^{\prime }])^2
&=&-\lambda _{1}(\text{\textrm{Tr}}[ U_{R}\Phi ^{\dagger }U_{L}^{\dagger
}U_{L}\Phi U_{R}^{\dagger }])^2  \notag \\
&=&-\lambda _{1}(\text{\textrm{Tr}}[ U_{R}\Phi ^{\dagger }\Phi
U_{R}^{\dagger }])^2  \notag \\
&=&-\lambda _{1}(\text{\textrm{Tr}}[ U_{R}^{\dagger }U_{R}\Phi ^{\dagger
}\Phi])^2  \notag \\
&=&-\lambda _{1}(\text{\textrm{Tr}}[ \Phi ^{\dag }\Phi])^2 .  \label{mt}
\end{eqnarray}
Similarly,
\begin{equation}
-\lambda _{2}\text{\textrm{Tr}}(\Phi ^{\dag\prime  }\Phi^\prime) ^{2}=-\lambda _{2}\text{\textrm{Tr}}(\Phi ^{\dag }\Phi) ^{2}\, ,
\end{equation}
In the Lagrangian density $\mathcal{L}_{\Phi,AV}$, the fourth chirally invariant term, describes a four-body coupling of the scalar, pseudoscalar, vector, and axial-vector degrees of freedom, is constructed in the form
\begin{equation}
\frac{h_{1}}{2}\text{\textrm{Tr}}[\Phi ^{\dag }\Phi ]\text{\textrm{Tr}}%
[L_{\mu }L^{\mu }+R_{\mu }R^{\mu }]\,.
\end{equation}
While the fifth term is constructed in the following form:
\begin{equation}
h_{2}\text{\textrm{Tr}}[\Phi ^{\dag }L_{\mu }L^{\mu }\Phi +\Phi R_{\mu
}R^{\mu }\Phi ^{\dag }]\,.
\end{equation}
Furthermore, one can construct an additional term as follows:
\begin{equation}
2h_{3}\text{\textrm{Tr}}[\Phi R_{\mu }\Phi ^{\dag }L^{\mu }]\,,
\end{equation}
which are invariant under global chiral transformations \cite{St.Di.Th}.\\

Finally, we obtain the full Lagrangian density $\mathcal{L}_{\Phi,AV}$ as
\begin{align}
\mathcal{L}_{\Phi,AV}=&\text{\textrm{Tr}}\left[ (D^{\mu }\Phi )^{\dag }(D^{\mu }\Phi )\right]-\lambda _{1}(\text{\textrm{Tr}}[\Phi ^{\dag }\Phi])^{2}-\lambda _{2}\text{\textrm{Tr}}( \Phi ^{\dag }\Phi) ^{2}\\\nonumber &+\frac{h_{1}}{2}\text{\textrm{Tr}}[\Phi ^{\dag }\Phi ]\text{\textrm{Tr}}%
[L_{\mu }L^{\mu }+R_{\mu }R^{\mu }]+h_{2}\text{\textrm{Tr}}[\Phi ^{\dag }L_{\mu }L^{\mu }\Phi +\Phi R_{\mu
}R^{\mu }\Phi ^{\dag }]\notag\\&+2h_{3}\text{\textrm{Tr}}[\Phi R_{\mu }\Phi ^{\dag }L^{\mu }]\,,\label{LPhiAV}
\end{align}
which contains terms up to order four in naive scaling dimension. In the Lagrangian density $\mathcal{L}_{\Phi,AV}$ (\ref{LPhiAV}), the parameters depend on the number of colours $N_c$ \cite{ FGhabilit, Lebed:1998st, 'tHooft:1973jz, Witten:1979kh} as follows\\
\begin{align}
g_1\,&\propto N_c^{-1/2},\nonumber\\ \lambda_1,\, h_1&\propto N_c^{-2},\nonumber\\ \lambda_2,\,h_2,\,h_3\,&\propto N_c^{-1}\,.
\end{align}
The quantities $\lambda_2,\,h_2,\,h_3$ scale as $N_c^{-1}$ because it describes a four-point interaction of the quark-antiquark states. The quantities $\lambda_1,\, h_1$ are suppressed by an additional $N_c$ and scale as $N_c^{-2}$ because these terms are the product of two separate traces. At the microscopic quark gluon level, one needs furthers large-$N_c$ suppressed transversal gluons to generate these terms. The quantity $g_1$ scales as $N_c^{-1/2}$.\\
 
\textbf{(ii) The term $\mathcal{L}_{AV}$}:\\

The Lagrangian density $\mathcal{L}_{AV}$ includes the (axial-)vector degrees of freedom. The construction of this Lagrangian follows the same principles
as in the (pseudo)scalar sector with additional terms of naive scaling dimension four. The mass term can be constructed as
\begin{equation}
\frac{m_{1}^{2}}{2}\text{\textrm{Tr}}\left[ (L^{\mu })^{2}+(R^{\mu })^{2}%
\right]\, .  \label{mtv}
\end{equation}
Using  Eq.(\ref{rfst}) and Eq.(\ref{lfst}), one can construct the keinetic term of the vector degrees of freedom as
\begin{equation}
-\frac{1}{4}\text{\textrm{Tr}}\left[ (L^{\mu \nu })^{2}+(R^{\mu \nu })^{2}%
\right]\,.\label{mtv1}
\end{equation}
From the transformation (\ref{chtrv}, \ref{chtlv}) and (\ref{trRmunu}, \ref{trLmunu}), one can obtain that the mass term (\ref{mtv}) and the keintic term (\ref{mtv1}) are invariant under chiral transformations. Moreover, the Lagrangian density $\mathcal{L}_{AV}$ involves also additional terms with 3- and 4-point vertices of the (axial-)vector degrees of freedom. The full Lagrangian ${L}_{AV}$ has the following form:
\begin{align}
\mathcal{L}_{AV}  &  =\frac{m_{1}^{2}}{2}\text{\textrm{Tr}}\left[ (L^{\mu })^{2}+(R^{\mu })^{2}%
\right]-\frac{1}{4}\text{\textrm{Tr}}\left[ (L^{\mu \nu })^{2}+(R^{\mu \nu })^{2}%
\right]\nonumber\\&-i\frac{g_{2}}{2}\{\mathrm{Tr}(L_{\mu\nu}[L^{\mu},L^{\nu}])+\mathrm{Tr}(R_{\mu\nu
}[R^{\mu},R^{\nu}])\} +g_{3}[\mathop{\mathrm{Tr}}(L_{\mu}L_{\nu}L^{\mu}L^{\nu}%
)+\mathop{\mathrm{Tr}}(R_{\mu}R_{\nu}R^{\mu}R^{\nu})]\nonumber\\&+g_{4}%
[\mathop{\mathrm{Tr}}\left(  L_{\mu}L^{\mu}L_{\nu}L^{\nu}\right)
+\mathop{\mathrm{Tr}}\left(  R_{\mu}R^{\mu}R_{\nu}R^{\nu}\right)]+g_{5}\mathop{\mathrm{Tr}}\left(
L_{\mu}L^{\mu}\right)
\,\mathop{\mathrm{Tr}}\left(  R_{\nu}R^{\nu}\right) \nonumber\\& +g_{6}%
[\mathop{\mathrm{Tr}}(L_{\mu}L^{\mu})\,\mathop{\mathrm{Tr}}(L_{\nu}L^{\nu
})+\mathop{\mathrm{Tr}}(R_{\mu}R^{\mu})\,\mathop{\mathrm{Tr}}(R_{\nu}R^{\nu
})]\text{ ,}  \label{fulllag}%
\end{align}
with the large-$N_c$ dependence of the parameters as
\begin{align}
g_2\,&\propto N_c^{-1/2},\nonumber\\ g_3,\,\,g_4\,&\propto N_c^{-1},\nonumber\\g_5,\,g_6\,&\propto N_c^{-2}\,.\label{scalg}
\end{align}

\textbf{(iii) The term $\mathcal{L}_{U(1)_A}$}:\\

The Lagrangian density $\mathcal{L}_{U(1)_A}$ contains only the chiral anomaly term \cite{'tHooft:1976fv, 'tHooft:1986nc},
\begin{equation}
\mathcal{L}_{U(1)_A}=c\left( \det \Phi ^{\dag }-\det \Phi \right)^2\, .  \label{det}
\end{equation}
which contributes to the mass of the isoscalar-pseudoscalar bosons. Their mass also does not disappear in the chiral limit. These fields are therefore no Goldstone bosons. This term is invariant under $SU(N_f)_R\times SU(N_f)_L$, but not under $U(1)_A$, as shown in the following :\\
\textit{Proof:}
\begin{eqnarray}
c\left( \det \Phi ^{\dag \prime }-\det \Phi ^{\prime }\right)^2
&=&c\left[ \det (U_{R}\Phi ^{\dag }U_{L}^{\dag })-\det (U_{L}\Phi
U_{R}^{\dag })\right]^2  \notag \\
&=&c\left[ \det (e^{i\theta _{R}^{a}t_{a}}\Phi ^{\dag }e^{-i\theta
_{L}^{a}t_{a}})-\det (e^{i\theta _{L}^{a}t_{a}}\Phi e^{-i\theta
_{R}^{a}t_{a}})\right]^2  \notag \\
&=&c\left[ \det (e^{-i\theta _{A}^{a}t_{a}}\Phi ^{\dag })-\det (e^{i\theta
_{A}^{a}t_{a}}\Phi )\right]^2  \notag \\
&=&c\left[ \det (e^{-i\sum_{a=1}^{N_{f}^{2}-1}\theta _{A}^{a}t_{a}})\det
(e^{-i\theta _{A}^{0}t_{0}})\det \Phi ^{\dag }\right.  \notag \\
&&\left. -\det (e^{i\sum_{a=1}^{N_{f}^{2}-1}\theta _{A}^{a}t_{a}})\det
(e^{i\theta _{A}^{0}t_{0}})\det \Phi \right]^2  \notag \\
&=&c\left[ \det (e^{-i\theta _{A}^{0}t_{0}})\det \Phi ^{\dag }-\det
(e^{i\theta _{A}^{0}t_{0}})\det \Phi \right]^2  \notag \\
&=&c\left[ e^{-i\theta _{A}^{0}N_{f}}\det \Phi ^{\dag }-e^{i\theta
_{A}^{0}N_{f}}\det \Phi \right]^2  \notag \\
&\neq &c\left[ \det \Phi ^{\dag }-\det \Phi \right]^2\, .  \label{proofdet}
\end{eqnarray}
The parameter $c$ scales in the large-$N_c$ limit \cite{FGhabilit} as
\begin{equation}
c\,\propto N_c^{-N_f/2}\,,
\end{equation}
i.e., it has a dependence on the number of quark flavours ($N_f$) in this model. For $N_f \geqslant 2$, the parameter $c$ vanishes which leads to neglect the anomaly for large $N_c$. The corresponding meson is then a Goldstone boson for $N_c \gg 1$. Note that for $N_f\neq 4$, the parameter $c$ is dimensionfull. This is an exception of the discussed rule. Which is possible since the anomaly also comes from the gauge sector.\\

\textbf{(iv) The term $\mathcal{L}_{G}$}:\\

The last field entering the model is the dilaton field /scalar
glueball $G$ through the Lagrangian density $\mathcal{L}_{G}$
which consists of the dilaton
Lagrangian $\mathcal{L}_{dil}$ and the coupling of the dilaton field with (pseudo)scalar and (axial-)vector degrees of freedom.\\

Firstly, let us discuss the dilaton Lagrangian\\

\indent As shown in Eq.(\ref{sproofspar}), the Yang-Mills (YM) sector
of QCD (which is described by $\mathcal{L}_{YM}$) is classically invariant
under dilatations. However, this symmetry is broken at
the quantum level. This scale invariance and its anomalous
breaking is one of the essential features of our effective model.
From the trace of the energy-momentum tensor $T^{\mu\nu}_{YM}$ of
the YM Lagrangian (\ref{LYM}), one can write the divergence of the
dilatation (Noether) current as follows:
\begin{equation}
\partial_\mu J^\mu_{YM,dil}=(T_{YM})_{\mu }^{\mu }=\frac{\beta (g)}{2g}\left( \frac{1}{2}G_{\mu \nu
}^{a}G_{a}^{\mu \nu }\right) \neq 0\,,  \label{speit}
\end{equation}
which does not vanish. The $\beta$-function is defined in
Eq.(\ref{betaQCD}), and $g=g(\mu)$ is the renormalised coupling
constant at the scale $\mu$. At the one-loop level,
\begin{equation}
\beta(g)=\frac{-11\,N_c}{48\pi^2}g^3\,.
\end{equation}
This implies
\begin{equation}
g^2(\mu) = \bigg[2\,\,\frac{11\,N_c}{48\pi^2}\,\,ln\bigg(\frac{\mu}{\Lambda_{YM}}\bigg)\bigg]^{-1}\,,
\end{equation}
where $\Lambda_{YM}$ is the YM scale and has a value of about
($\simeq 200$ MeV). The non-vanishing expectation value of the trace anomaly represents the gluon condensate
\begin{equation}
\left\langle T_{\mu }^{\mu }\right\rangle =\frac{-11\,N_c}{48}\left\langle
\frac{\alpha _{s}}{\pi }\,G_{\mu \nu
}^{a}\,G_{a}^{\mu \nu } \right\rangle =\frac{-11\,N_c}{48}C^4\,,  \label{vew}
\end{equation}%
where%
\begin{equation}
\left\langle \frac{\alpha _{s}}{\pi }\,G_{\mu \nu
}^{a}\,G_{a}^{\mu \nu } \right\rangle \equiv C^{4}\,.  \label{gk}
\end{equation}
The numerical values of $C^4$ have been computed from lattice-QCD simulations (higher range of the interval) \cite{Lashin:2003jv} and QCD sum rules
 (lower range of the interval) \cite{Shifman:1978rh}:
\begin{equation}
C^4\approx[(300-600)\, MeV]^4\,,
\end{equation}
whereas, in the lattice-QCD simulation of Ref. \cite{Migdal:1982jp}, its value has been found to be $C\approx 610$ MeV.\\
The effective theory of the YM sector of QCD can be built by
introducing a scalar dilaton/scalar field $G$ at the composite
level. The Lagrangian
density of the dilaton reads \cite{Rosenzweig:1981cu, Salomone:1980sp, Rosenzweig:1982cb, Gomm:1984zq, Gomm:1985ut, Migdal:1982jp}
\begin{equation}
\mathcal{L}_{dil}(G)=\frac{1}{2}(\partial ^{\mu }G)^{2}-V_{dil}(G)\,,  \label{dillag}
\end{equation}
where the dilaton potential is
\begin{equation}
V_{dil}(G)=\frac{1}{4}\frac{m_{G}^{2}}{\Lambda ^{2}}\left( G^{4}\ln
\left\vert \frac{G}{\Lambda }\right\vert -\frac{G^{4}}{4}\right)\, .
\label{potdil}
\end{equation}
The value $G_0=\Lambda$ corresponds to the minimum of the dilaton
potential $\mathcal{V}_{dil}(G)$. The particle mass $m_G$ emerges
upon shifting $G\rightarrow G_0+G$. This particle is interpreted
as the scalar glueball and its mass has been evaluated as $m_G
\approx (1500-1700)$ MeV by lattice QCD \cite{Morningstar:1999rf, Loan:2005ff, Gregory:2012hu}. The
scale invariance is broken explicitly by the logarithmic term of
the potential. The divergence of the dilatation current reads
\begin{equation}
\left\langle T_{dil,\,\mu }^{\mu }\right\rangle =\left\langle -\frac{1}{4}\frac{%
m_{G}^{2}}{\Lambda ^{2}}G^{4}\right\rangle =-\frac{1}{4}\frac{m_{G}^{2}}{%
\Lambda ^{2}}G_{0}^{4}=-\frac{1}{4}m_{G}^{2}\Lambda ^{2}\,,\label{div2}
\end{equation}
where $G$ is set to be equal to the minimum of the potential $G_0$. By
comparing Eq.(\ref{vew}) and Eq.(\ref{div2}), one obtains
\begin{equation}
\Lambda =\frac{\sqrt{11}}{2m_{G}}C^{2}\,.  \label{sp}
\end{equation}
When $m_G=1500$ MeV and $C \approx 610$ MeV \cite{DiGiacomo:2000va}, the parameter $\Lambda$ has the value $\Lambda=400$ MeV. Note that a narrow glueball is possible only if $\Lambda\gtrsim 1000$ MeV.\\

Now let us turn to couple the dilaton field/scalar
glueball with the (pseudo)scalar and (axial-)vector degrees of
freedom. This coupling must be scale invariance.\\
We assume that, a part from the $U(1)_A$ anomaly and terms related to quark masses, only the dilaton term breaks the
dilatation invariance and generates the scale anomaly in the
effective model. Note that the mass term for the (axial-)vector
mesons (\ref{mtv}) break the symmetry explicitly. It does not has
dilatation symmetry, since each scale with $
\lambda^{2}$. In order to achieve scale invariance, one should
write down the mass term of the scalar degrees of freedom as
follows
\begin{equation}
-aG^{2}\mathrm{Tr}\left[ \Phi ^{\dag }\Phi \right]\,, 
\end{equation}%
with the scalar mass parameter
\begin{equation}
m_{0}^{2}=aG_{0}^{2},
\end{equation}
where $a$ is a dimensionless constant larger than 0, which
represents the spontaneous symmetry breaking. The mass term for
the vector mesons should be written in the same way. Now let us
modify both mass terms by including the scalar glueball as
\begin{equation}
-m_{0}^{2}\mathrm{Tr}\left[ \Phi ^{\dag }\Phi \right]
\longrightarrow -m_{0}^{2}\left( \frac{G}{G_{0}}\right)
^{2}\mathrm{Tr}\left[ \Phi ^{\dag }\Phi \right]\,,\label{m0G}
\end{equation}%
and similarly,
\begin{equation}
\frac{m_{1}^{2}}{2}\text{\textrm{Tr}}\left[ (L^{\mu })^{2}+(R^{\mu })^{2}%
\right] \longrightarrow \frac{m_{1}^{2}}{2}\left( \frac{G}{G_{0}}\right) ^{2}%
\text{\textrm{Tr}}\left[ (L^{\mu })^{2}+(R^{\mu })^{2}\right]\,,\label{m1G}
\end{equation}
which implements the scale-invariance [as proven in Ref. \cite{St.Di.Th}]. 
From Eqs.(\ref{dillag}, \ref{potdil},
\ref{m0G}, \ref{m1G}), we get the full structure of the Lagrangian density $\mathcal{L}_{G}$ in the effective
model as
\begin{align}
\mathcal{L}(G)=&\frac{1}{2}(\partial ^{\mu }G)^{2}-\frac{1}{4}\frac{m_{G}^{2}}{\Lambda ^{2}}\left( G^{4}\ln
\left\vert \frac{G}{\Lambda }\right\vert -\frac{G^{4}}{4}\right)\nonumber\\ &-m_{0}^{2}\left( \frac{G}{G_{0}}\right)
^{2}\mathrm{Tr}\left[ \Phi ^{\dag }\Phi \right]+\frac{m_{1}^{2}}{2}\left( \frac{G}{G_{0}}\right) ^{2}%
\text{\textrm{Tr}}\left[ (L^{\mu })^{2}+(R^{\mu })^{2}\right]\,.\label{dillagL}
\end{align}
The Large-$N_c$ dependence of the parameter is given as
\begin{align}
m_G\,&\propto N_c^{0},\nonumber\\ \Lambda_G\,&\propto N_c\,.\label{scalLm}
\end{align}

\textbf{(v) The term $\mathcal{L}_{ESB}$}:\\

The chiral symmetry is explicitly broken by the quark masses. Two additional terms appear in the Lagrangian density $\mathcal{L}_{ESB}$,
to describe this breaking separately for the (pseudo)scalar and (axial-)vector fields. We discuss these separately: In the (pseudo)scalar sector, pions
are not really massless because of the explicit breaking of the
$SU(N_{f})_{V}\times SU(N_{f})_{A}$ -symmetry. Therefore the following term is introduced to break this symmetry explicitly

 \begin{equation}
\text{\textrm{Tr}}[H( \Phi ^{\dag }+\Phi)]\,,
\end{equation}
where
\begin{equation}
H=\text{diag}[h_1,\,h_2,\,...,\,h_{N_f}]\,,
\end{equation}
is the diagonal matrix with $h_0^{N}$ proportional the mass of the quark flavour number. For example: $N_f=1 \Rightarrow h_1\propto m_u$ and $N_f=2 \Rightarrow h_2\propto m_d$,...etc.\\
For the (axial-)vector sector, one can construct the following mass term which breaks the chiral symmetry explicitly
\begin{equation}
\text{\textrm{Tr}}[\Delta(L_\mu^2+ R_\mu^2)]\,,
\end{equation}
where
\begin{equation}
\Delta=\text{diag}[\delta_u,\,\delta_d,\,...,\,\delta_{N_f}]\,,
\end{equation}
which is also proportional to the quark mass terms as $$\delta_u,\,\delta_d,\,...,\,\delta_{N_f}\propto m^2_u,\,m^2_d,\,...,\,m^2_{N_f}$$ Thus, the Lagrangian density $\mathcal{L}_{ESB}$
is obtained as
\begin{equation}
 \mathcal{L}_{ESB}=\text{\textrm{Tr}}[H( \Phi ^{\dag }+\Phi)]+\text{\textrm{Tr}}[\Delta(L_\mu^2+ R_\mu^2)]\,.
\end{equation}
The large-$N_c$ dependence of the parameters in the previous Lagrangian
density is given as
\begin{align}
h_i\,&\propto N_c^{1/2},\nonumber\\ \delta_{i}\,&\propto N_c^{0}\,.\label{scald}
\end{align}

\textbf{Spontaneous symmetry breaking}\\

The spontaneous breaking of the chiral symmetry is an important requirement for all phenomena linked to hadrons in low-energy QCD. One can discuss this point from
the following potential of the mesonic Lagrangian $
\mathcal{L}_{mes}$ along the axis $\Phi=\sigma\,t^0$
\begin{equation}
V(G,\sigma)=V_{dil}(G)+m_0^2\sigma^2+(\lambda_1+\lambda_2)\sigma^4\,.
\end{equation}
The symmetry is broken spontaneously by non-trivial minima, which are in
the present case at
$$G_0\neq 0,\,\,\sigma_0\neq0,\,\,\text{for},\,m_0^2<0\,,$$
and
$$\sigma_0=0,\,\,G_0\neq0,\,\,\text{for}\,\,\,m_0^2>0\,,$$
which means that the vacuum is not invariant under $SU(N_f)_A$ transformations.
The dilatation symmetry is broken explicitly which is an
important source for the phenomenology in the vacuum. Note that
the conservation of parity and $SU(N_f)_V$ symmetry are required,
whereas the $U(1)_A$ anomaly is neglected here. The only state that can
condense in the vacuum is the scalar-isosinglet state, because this
state is the only one which has the same quantum numbers as the
vacuum.\\

\textit{\textbf{Now let us summarize the full Lagrangian of the effective model, the so-called extended Linear Sigma Model
(eLSM), for a generic number $N_f$ of flavours in the
following section.}}
\newpage

\section{The extended Linear Sigma Model}\label{sec-1}

In this subsection we present the chirally symmetric linear sigma
model Lagrangian which is essentially constructed with two
requirements stemming from the underlying theory QCD: (i) global
chiral symmetry $U(N)_R \times U(N)_L$. (ii) dilatation
invariance (with the exceptions of the scale anomaly, the $U(1)_A$
anomaly and terms proportional to quark masses). It is also
invariant under the discrete symmetries charge conjugation $C$,
parity $P$, and time reversal $T$. It has
the following form for a generic number $N_f$ of flavours \cite{Parganlija:2010fz, Parganlija:2012fy, Parganlija:2012xj}:
\begin{align}
\mathcal{L}  &  =\mathcal{L}_{dil}+\mathrm{Tr}[(D_{\mu}\Phi)^{\dagger}(D_{\mu}\Phi
)]-m_{0}^{2}\bigg(\frac{G}{G_0}\bigg)^2\mathrm{Tr}(\Phi^{\dagger}\Phi)-\lambda_{1}[\mathrm{Tr}%
(\Phi^{\dagger}\Phi)]^{2}-\lambda_{2}\mathrm{Tr}(\Phi^{\dagger}\Phi)^{2}\nonumber\\ &
-\frac{1}{4}\mathrm{Tr}[(L^{\mu\nu})^{2}+(R^{\mu\nu})^{2}]+\mathrm{Tr}\left\{  \left[  \left(  \frac{G}{G_{0}}\right)  ^{2}%
\frac{m_{1}^{2}}{2}+\Delta\right]  \left[  (L^{\mu})^{2}+(R^{\mu})^{2}\right]
\right\}+\mathrm{Tr}%
[H(\Phi+\Phi^{\dagger})] \nonumber\\
& +c(\mathrm{det}\Phi-\mathrm{det}\Phi^{\dagger})^{2}
+\frac{h_{1}}{2}\mathrm{Tr}(\Phi^{\dagger}\Phi)\mathrm{Tr}\left(
L_{\mu }^{2}+R_{\mu}^{2}\right)+h_{2}\mathrm{Tr}[\left\vert
L_{\mu}\Phi\right\vert
^{2}+\left\vert \Phi R_{\mu}\right\vert ^{2}]\nonumber\\
& +2h_{3}\mathrm{Tr}(L_{\mu}\Phi
R^{\mu}\Phi^{\dagger})+i\frac{g_{2}}{2}\{\mathrm{Tr}(L_{\mu\nu}[L^{\mu},L^{\nu}])+\mathrm{Tr}(R_{\mu\nu
}[R^{\mu},R^{\nu}])\}\nonumber\\
& +g_{3}[\mathop{\mathrm{Tr}}(L_{\mu}L_{\nu}L^{\mu}L^{\nu}%
)+\mathop{\mathrm{Tr}}(R_{\mu}R_{\nu}R^{\mu}R^{\nu})]+g_{4}%
[\mathop{\mathrm{Tr}}\left(  L_{\mu}L^{\mu}L_{\nu}L^{\nu}\right)
+\mathop{\mathrm{Tr}}\left(  R_{\mu}R^{\mu}R_{\nu}R^{\nu}\right)]
\nonumber\\& +g_{5}\mathop{\mathrm{Tr}}\left(
L_{\mu}L^{\mu}\right)
\,\mathop{\mathrm{Tr}}\left(  R_{\nu}R^{\nu}\right)  +g_{6}%
[\mathop{\mathrm{Tr}}(L_{\mu}L^{\mu})\,\mathop{\mathrm{Tr}}(L_{\nu}L^{\nu
})+\mathop{\mathrm{Tr}}(R_{\mu}R^{\mu})\,\mathop{\mathrm{Tr}}(R_{\nu}R^{\nu
})]\text{ .}  \label{fulllag}%
\end{align}

Here, $G$ is the dilaton field/scalar glueball and the dilaton Lagrangian
$\mathcal{L}_{dil}$ \cite{Salomone:1980sp, Gomm:1984zq, Migdal:1982jp} reads
\begin{equation}
\mathcal{L}_{dil}=\frac{1}{2}(\partial_\mu
G)^2-\frac{1}{4} \frac{m_G^2}{\Lambda^2}\bigg(G^4
\,\text{ln}\bigg|\frac{G}{\Lambda}\bigg|-\frac{G^4}{4}\bigg)\,,\label{Lial2}
\end{equation}

which mimics the trace anomaly of QCD \cite{Rosenzweig:1981cu, Parganlija:2012xj}. The dimensionful parameter $\Lambda_G \sim N_c
\Lambda_{QCD}$ sets the energy scale of low-energy QCD; in the
chiral limit it is the only dimensionful parameter besides the
coefficient of the term representing the axial anomaly. All other
interaction terms of the Lagrangian are described by dimensionless
coupling constants.  The minimum of the dilaton potential in
Eq.(\ref{potdil}) is given by $G_0=\Lambda$. A massive particle
will arise after shifting the dilaton field $G \rightarrow G_0+G$,
where the dilaton field $G$ is interpreted as the scalar glueball
which consists of two gluons ($G\equiv\left\vert
gg\right\rangle$). The value of $G_0$ is related to the gluon
condensate of QCD. According to lattice QCD the glueball mass
$m_{G}$, in the quenched approximation (no quarks), is about
1.5-1.7 GeV \cite{Morningstar:1999rf}. As mentioned above, the
identification of $G$ is still uncertain, the two most likely
candidates are $f_{0}(1500)$ and $f_{0}(1710)$ and/or admixtures
of them. Note that, we include the scalar glueball because
it is conceptually important to guarantee dilatation invariance of
the model (thus constraining the number of possible terms that the Lagrangian
can have). We do not make an assignment for the scalar glueball in
the framework of the strange and nonstrange ($N_f=2$ and
$N_f=3$) cases (see chapter 3) and it does not affect the results
of the study of the $N_f=4$ case for the masses of open and hidden
charmed mesons as well as the decay of open charmed  mesons.
Therefore, the scalar glueball is a frozen field in these
investigations, whereas it becomes a
dynamical field in the study of the decay of hidden charmed states
as we will see in chapter 8. The logarithmic term of the dilaton
potential breaks the dilatation symmetry explicitly, $x^\mu
\rightarrow \Lambda^{-1}\,x^\mu$, which leads to the divergence of
the corresponding current:
\begin{equation}
\partial_{\mu}J_{dil}^{\mu}=T_{dil,\,\mu}^{\;\mu} =-\frac{1}{4}m_{G}^{2}%
\Lambda^{2}\text{ }. \label{dil}%
\end{equation}

The model has also mesonic fields
described as quark-antiquark fields. We have to note that when working in the
so-called large-$N_c$ limit \cite{'tHooft:1973jz, Witten:1979kh}: (i) the glueball
self-interaction term vanishes, (ii) the glueball becomes a free
field, (iii) the masses are $N_c-$independent, (iv) the widths scale as $N_c^{-1}$.\\

Let us now turn to the question: \textbf{How can we introduce a pseudoscalar glueball into the chiral model (\ref{fulllag})?}\\
The structure of the chiral anomaly term,
$ic(\text{det}\Phi-\text{det}\Phi^\dagger)^2$, allows one to incorporate a
pseudoscalar glueball field $\tilde{G}$ into the model in a simple
way, of the form 
$$ic_{\tilde{G}\Phi}\tilde{G}\left(
\text{\textrm{det}}\Phi-\text{\textrm{det}}\Phi^{\dag}\right).$$
This form describes the interaction of the pseudoscalar field
$\tilde{G}$ with the scalar and pseudoscalar fields by a
dimensionless coupling constant $c_{\tilde{G}\Phi}$. Through this
term one can study the phenomenology of a pseudoscalar glueball.
The details of the introduction of the pseudoscalar glueball in
the extended Linear Sigma Model are presented in chapter 6. Furthermore, it is relevant in the decay of hidden charmed mesons, see chapter 8 below.


\chapter{The extended Linear Sigma Model for two- and three-flavours}\label{The eLSM cases}

\section{Introduction}

\indent In the last decades, effective low-energy approaches to
the strong interaction have been developed by imposing chiral
symmetry, one of the basic symmetries of the QCD Lagrangian in the
limit of vanishing quark masses (the so-called chiral limit)
\cite{Gasiorowicz, Meissner:1987ge}. Chiral symmetry is explicitly broken
by the nonzero current quark masses, but is also spontaneously
broken by a nonzero quark condensate in the QCD vacuum
\cite{Vafa:1983tf}. As a consequence, pseudoscalar (quasi-)Goldstone
bosons emerge, as discussed in details in the previous chapter. We develop the eLSM to study the vacuum properties of mesons and glueballs. In the case of $N_{f}=2$ quark flavours, there are only mesons made of $u$ and $d$ quarks. However, the nucleons can also be taken into account in the context of a
chiral model. The interaction of nucleons with mesons,
tetraquarks, and glueballs in the chiral model for $N_f=2$ will be described in the present chapter. Furthermore, in the case
$N_{f}=3$, mesons are made of up, down and strange quarks. It is for the first time possible to describe
(pseudo)scalar as well as (axial-)vector meson nonets in a chiral
framework: masses and decay widths turn out to be in
very good agreement with the results listed by the Particle Data
Group (PDG) \cite{Beringer:1900zz}. In this chapter, we thus present the extension of the eLSM from non-strange hadrons ($N_f=2$) \cite{Janowski:2011gt, Ko:1994en, Parganlija:2010fz, Gallas:2009qp, Parganlija:2012xj} to strange hadrons ($N_f=3$) \cite{Janowski:2014ppa, Ecker:1988te, Parganlija:2012fy, Parganlija:2012xj, Lenaghan:2000ey}. Consequently, we investigate the vacuum properties of the three-flavour case \cite{ Parganlija:2012fy, Parganlija:2012xj}.

\section{A $U(2)_R \times U(2)_L$ interaction with nucleons}

In this section we present the chirally symmetric linear sigma
model in the case of $N_f=2$ \cite{Gallas:2009qp}. It contains (pseudo)scalar and
(axial-)vector fields, as well as nucleons and their chiral
partners. Then we describe how the pseudoscalar glueball interacts
with the nucleon and its chiral partner. This allows us to compute
the decay widths of a pseudoscalar glueball into two nucleons (see Sec. 6.5).

\subsection{A chirally invariant mass term}

\ The mesonic Lagrangian for the Linear Sigma Model with global
chiral $U(2)_R\times U(2)_L$ symmetry has the same form of the
Lagrangian (\ref{fulllag}).\\ 



In this case, the matrix $\Phi$ reads

\begin{equation}
\label{scalars}\Phi= \sum_{a=0}^{3} \phi_{a} t_{a} =
(\sigma+i\eta_{N})\,t^{0} +(\vec{a}_{0}+i\vec{\pi})
\cdot\vec{t}\;,
\end{equation}

and includes scalar and pseudoscalar fields. The eta meson
$\eta_N$ contains only non-strange degrees of freedom. Under the
global $U(2)_R \times U(2)_L$ chiral symmetry, $\Phi$ transforms
as $\Phi\rightarrow U_L \Phi^\dagger U_R$. The vector and
axial-vector fields are described as

\begin{equation}
V^{\mu}  = \sum_{a=0}^{3} V_{a}^{\mu}t_{a} = \omega^{\mu}\, t^{0}
+\vec{\rho}^{\mu} \cdot\vec{t}\; ,
\end{equation}
and
\begin{equation}
 A^{\mu}  =
\sum_{a=0}^{3} A_{a}^{\mu}t_{a} = f_{1}^{\mu} \,t^{0} +\vec
{a}_{1}^{\mu} \cdot\vec{t}\;,
\end{equation}

respectively, where the generators of $U(2)$ are
$\overrightarrow{t}=\overrightarrow{\tau}/2$, with the vector of
Pauli matrices $\overrightarrow{\tau}$ and $t^0=\textbf{1}_{2}/2$.
Under global $U(2)_{R} \times U(2)_{L}$ transformations, these
fields behave as $R^{\mu }\rightarrow U_{R}
R^{\mu}U_{R}^{\dagger}\, , \; L^{\mu}\rightarrow U_{L}
L^{\mu}U_{L}^{\dagger}$.\\
The mesonic
Lagrangian (\ref{fulllag}) is
invariant under $U(2)_R \times U(2)_L$ transformations for $c=h_0=0$ whereas for $h_0 \neq 0$, this symmetry is explicitly
broken to the vectorial subgroup $U(2)_V$ \cite{Gallas:2009qp}, where $V=L+R$. Moreover,
the $U(1)_A$ symmetry, where $A=L-R$, is explicitly broken for
$c\neq 0$. The spontaneous chiral symmetry breaking is implemented by
shifting the scalar-isoscalar field $\sigma$ by its vacuum
expectation value $\varphi$ as $\sigma \rightarrow \sigma + \varphi$, where
the chiral condensate $\varphi= \left\langle 0\left\vert
\sigma\right\vert 0\right\rangle =Zf_{\pi}$. The parameter
$f_\pi= 92.4$ MeV is the pion decay constant and $Z$ is the wave function renormalization constant of the pseudoscalar
 fields \cite{Parganlija:2010fz, Struber:2007bm}.\\
The meson fields of the model (\ref{fulllag}) \cite{Parganlija:2010fz, Gallas:2009qp} are assigned to the following resonances listed by the PDG \cite{Beringer:1900zz}: \\
 (i) The pseudoscalar fields $\overrightarrow{\pi}$ and $\eta_N$ correspond to the pion and the $SU(2)$ counterpart of the $\eta$ meson,
   $\eta_{N} \equiv |\overline{u}u+\overline{d}d\rangle /\sqrt{2}$,
  with a mass of about $700$ MeV which can be obtained by ``unmixin'' the physical $\eta$ and $\eta'$ mesons. In the case of $N_f=3$ \cite{Parganlija:2012fy}
  this mixing is calculated and the results are presented in the
   next section, where the model contains also contributions from strange quarks.\\
  (ii) The vector fields $\omega^\mu$ and $\overrightarrow{\rho}^\mu$ represent the $\omega(782)$ and $\rho(770)$ vector mesons, respectively.\\
  (iii) The axial-vector fields $f_1^\mu$ and $\overrightarrow{a}_1^\mu$ represent the $f_1(1285)$ and $a_1(1260)$, respectively. The physical
  $\omega$ and $f_1$ states contain
   $\overline{s}s$ contributions which are negligibly small.\\
  (iv) The scalar fields $\sigma$ and $\overrightarrow{a}_0$ are assigned to the physical $f_0(1370)$ and $a_0(1450)$ resonances.\\

We now turn to the baryonic sector in the eLSM for two flavours.
The baryon sector involves two baryon doublets $\psi_1$ and
$\psi_2$, where $\psi_1$ has positive parity and $\psi_2$ has
negative parity. In the so-called mirror assignment \cite{Jido:1998av, Zschiesche:2006zj, Detar:1988kn} they
transform under chiral transformation as:
\begin{equation}
\Psi_{1R} \rightarrow U_R \Psi_{1R},\,\,\,\, \Psi_{1L} \rightarrow
U_L \Psi_{1L},\,\,\,\,\Psi_{2R} \rightarrow U_L \Psi_{2R},\,\,\,\,
\Psi_{2L} \rightarrow U_R \Psi_{2L}\,.
\end{equation}
While $\psi_1$ transforms as usual, $\psi_2$ transforms in a
``mirror way'' \cite{Detar:1988kn, Lee}. These field transformations allow
us to write the following baryonic Lagrangian for $N_f=2$ with a
chirally invariant mass term $\mathcal{L}_{\mathrm{mas}}$ for the
fermions \cite{Gallas:2009qp}:
\begin{align}
\mathcal{L}_{\mathrm{bar}}  &  =
\overline{\Psi}_{1L}i\gamma_{\mu}D_{1L}^{\mu }\Psi_{1L}
+\overline{\Psi}_{1R}i\gamma_{\mu}D_{1R}^{\mu}\Psi_{1R}
+\overline{\Psi}_{2L}i\gamma_{\mu}D_{2R}^{\mu}\Psi_{2L} +\overline{\Psi}%
_{2R}i\gamma_{\mu}D_{2L}^{\mu}\Psi_{2R}\nonumber\\
&  -\widehat{g}_{1} \left(  \overline{\Psi}_{1L}\Phi\Psi_{1R}
+\overline{\Psi }_{1R}\Phi^{\dagger}\Psi_{1L}\right)
-\widehat{g}_{2} \left(  \overline{\Psi
}_{2L}\Phi^{\dagger}\Psi_{2R}
+\overline{\Psi}_{2R}\Phi\Psi_{2L}\right)+\mathcal{L}_{\mathrm{mass}}\;,
\label{nucl lagra}%
\end{align}
where
$$D_{1R}^{\mu}=\partial^{\mu}-ic_{1}R^{\mu},\,\,\,D_{1L}^{\mu}=\partial
^{\mu}-ic_{1}L^{\mu}\,,$$ and
$$D_{2R}^{\mu}=\partial^{\mu}-ic_{2}R^{\mu},\,\,\,D_{2L}^{\mu}=\partial^{\mu}-ic_{2}L^{\mu}\,,$$
are the covariant derivatives for the nucleonic fields, with the
coupling constants $c_{1}$ and $c_{2}$. Note that the three
coupling constants  $c_{1},\,c_{2},\,\text{and}\,g_{1}$ are equal
in the case of local chiral symmetry. The interaction of the
baryonic fields with the scalar and (pseudo)scalar mesons is
parameterized by $\widehat{g}_{1}$ and $\widehat{g}_{2}$. The
chirally invariant mass term $\mathcal{L}_{mass}$ for
fermions parameterized by $\mu_0$, reads
\begin{align}
\mathcal{L}_{\mathrm{mass}} & =
-\mu_{0}(\overline{\Psi}_{1L}\Psi_{2R}
-\overline{\Psi}_{1R}\Psi_{2L}
-\overline{\Psi}_{2L}\Psi_{1R}+\overline{\Psi}_{2R}\Psi_{1L})\nonumber\\
& =
-\mu_0(\overline{\Psi}_2\gamma_5\Psi_1-\overline{\Psi}_1\gamma_5\Psi_2)\,,\label{mterm}
\end{align}
where $\mu_0$ has the dimension of mass. This mass term plays an
important role in generating the nucleon mass. The
physical fields are the nucleon $N$ and its chiral partner
$N^\ast$. They engender by diagonalizing the baryonic part of the
Lagrangian. As a result \cite{Gallas:2009qp} we have
\begin{equation}
\left(
\begin{array}
[c]{c}%
N\\
N^\ast
\end{array}
\right)=\widehat{M}\left(
\begin{array}
[c]{c}%
\Psi_1\\
\Psi_2
\end{array}
\right) =\frac{1}{\sqrt{2 \cosh\delta}}\left(
\begin{array}
[c]{cc}%
e^{\delta/2} & \gamma_5\,e^{-\delta/2} \\
\gamma_5\,e^{-\delta/2}  & -e^{\delta/2}
\end{array}
\right)\left(
\begin{array}
[c]{c}%
\Psi_{1}\\
\Psi_2
\end{array}
\right)  \text{ ,} \label{nucleonmixing}%
\end{equation}
where $\delta$ measures the intensity of the mixing and is
related to the parameter $\mu_0$ and the physical masses of $N$ and
$N^\ast$ by the expression:
\begin{equation}
\cosh\delta=\frac{m_N + m_{N^\ast}}{2\, m_0}\,.
\end{equation}
The masses of nucleon and its partner \cite{Gallas:2009qp} read
\begin{equation}
m_{N,N^{\ast}}= \sqrt{ m_{0}^{2} +\left[  \frac{1}{4} (\widehat{g}%
_{1}+\widehat{g}_{2}) \varphi\right]  ^{2}} \pm\frac{1}{4}(\widehat{g}%
_{1}-\widehat{g}_{2})\varphi\;. \label{nuclmasses}%
\end{equation}
The coupling constants $\widehat{g}_{1,2}$ are determined by the
masses of nucleon ($m_{N}$), its partner ($m_{N^{\ast}}$), and the parameter $m_0$,
\begin{equation}
\label{g12}\widehat{g}_{1,2}=\frac{1}{\varphi} \left[  \pm(m_{N}-m_{N^{\ast}%
})+\sqrt{(m_{N}+m_{N^{\ast}})^{2} - 4m_{0}^{2}} \right]  \;.
\end{equation}
The masses of the nucleon and its partner turn into be degenerate,
$m_N=m_{N^\ast}=m_0$, in the chirally restored phase where
$\varphi \rightarrow 0$
as observed from Eq. (\ref{nuclmasses}). The breaking of chiral symmetry, $\varphi \neq 0$, generates the mass splitting.\\
Note that the mass term (\ref{mterm}) is not dilatation
invariant but we can modify this term to restore the dilatation
symmetry by coupling it to the chirally invariant dilaton field
$G$ and a tetraquark field
$\chi\equiv[\overline{u},\overline{d}][u,d]$ in an
$U(2)_R\times U(2)_L$ invariant way. Then we obtain the dilatation invariant
mass term as follows:
\begin{equation}
\mathcal{L}_{mass}=-(\alpha\chi+\beta
G)(\overline{\Psi}_2\gamma_5\Psi_1-\overline{\Psi_1}\gamma_5\Psi_2)\,,\label{mterm2}
\end{equation}
where $\alpha$ and $\beta$ are dimensionless coupling constants.
The term $\mathcal{L}_{mass}$ would not be possible if the field
$\psi_2$ would transform as $\psi_1$. If both scalar fields are
shifted around their vacuum expectation values $\chi \rightarrow
\chi_0+\chi$ and $G\rightarrow G_0+G$, there emerges a
nonvanishing chiral mass
\begin{equation}
m_0=\alpha\,\chi_0+\beta G_0\,,
\end{equation}
where $\chi_0$ and $G_0$ are the tetraquark and gluon condensates,
respectively. $m_0$ is the mass contribution to the nucleon which
does not stem from the chiral ($q\overline{q}$) condensate
$\sigma=\phi$. In Ref. \cite{Gallas:2009qp} the quantitative value for
the parameter $m_0$ has been obtained by a fit to vacuum
properties as
\begin{equation}
m_0= (460 \pm 136)\,\,\, MeV.
\end{equation}
Under the simplifying assumption $\beta=0$, as seen in the Ref. \cite{Gallas:2011qp},
the parameter $m_0$ is saturated by the tetraquark condensate,
where $\chi$ is identified with the resonance $f_0(600)$ while for
$N^\ast$ there are two candidates with quantum numbers ($J^P={\frac{1}{2}^-}$), which are the lightest state
$N(1535)$ and the
heavier state $N(1650)$ \cite{Gallas:2013ipa}. But in the present section, we are
interested in studying the case $\alpha=0$, when the parameter
$m_0$ is saturated by the glueball condensate. The fermions
involved are represented by the spinors $\Psi_1$ and $\Psi_2$.
Then we obtain the following interaction term for the glueball
with a nucleon
\begin{equation}
\mathcal{L}_{G-baryons}=\beta\,G(\overline{\Psi}_2\,\gamma_5\,\Psi_1-\overline{\Psi}_1\,\gamma_5\,\Psi_2),
\end{equation}
where $\beta$ is a dimensionless coupling constant. The physical
fields $N$ and $N^\ast$ are related to the spinors $\Psi_1$ and
$\Psi_2$ according to Eq. (\ref{nucleonmixing}) by the following
relations:
\begin{equation}\label{psi1}
\Psi_1=\frac
1{\sqrt{2\cosh\delta}}\left(Ne^{\delta/2}+\gamma_5
N^*e^{-\delta/2}\right)\,,
\end{equation}

\begin{equation}\label{psi2}
\Psi_2=\frac 1{\sqrt{2\cosh\delta}}\left(\gamma_5
N e^{-\delta/2}- N^*e^{\delta/2}\right)\,,
\end{equation}

\begin{equation}
\overline{\Psi}_1=\frac 1{\sqrt{2\cosh\delta}}\left(\overline{N}
e^{\delta/2}- \overline{N}^*\gamma_5
e^{-\delta/2}\right)\,,
\end{equation}
and
\begin{equation}
\overline{\Psi}_2=\frac 1{\sqrt{2\cosh\delta}}\left(-
\overline{N}\gamma_5 e^{-\delta/2}- \overline{N}^*
e^{\delta/2}\right)\,.\label{psi4}
\end{equation}

\section{The $U(3)_R \times U(3)_L$ linear sigma model}\label{nf3lag}

In this section we present the extended linear sigma
model (eLSM) including the strange sector, $N_f=3$, and its
implications which have been
investigated in Refs. \cite{Parganlija:2012fy, Parganlija:2012xj}.\\

In the case $N_f=3$, all quark-antiquark mesons in the
Lagrangian (\ref{fulllag}) are assigned to the light (i.e., with mass $\lesssim 2$
GeV) resonances in the strange-nonstrange sector. The pseudoscalar fields
$P$ and the scalar fields $S$ read

\begin{equation}
P=\frac{1}{\sqrt{2}}\left(
\begin{array}
[c]{ccc}%
\frac{\eta_{N}+\pi^{0}}{\sqrt{2}} & \pi^{+} & K^{+}\\
\pi^{-} & \frac{\eta_{N}-\pi^{0}}{\sqrt{2}} & K^{0}\\
K^{-} & {\bar{K}^{0}} & \eta_{S}%
\end{array}
\right)  \text{}, \label{P}%
\end{equation}
and
\begin{equation}
S=\frac{1}{\sqrt{2}}\left(
\begin{array}
[c]{ccc}%
\frac{\sigma_{N}+a_{0}^{0}}{\sqrt{2}} & a_{0}^{+} & K_{0}^{\ast+}\\
a_{0}^{-} & \frac{\sigma_{N}-a_{0}^{0}}{\sqrt{2}} & K_{0}^{\ast0}\\
K_{0}^{\ast-} & {\bar{K}_{0}^{\ast0}} & \sigma_{S}%
\end{array}
\right) \text{}, \label{S}%
\end{equation}
which together form the matrix $\Phi$ describing the
multiplet of the scalar and pseudoscalar mesons, as follows
\begin{align}
\Phi & =\sum_{a=0}^{8}(S_{a}+iP_{a})T_{a}=\frac{1}{\sqrt{2}}\left(
\begin{array}
[c]{ccc}%
\frac{(\sigma_{N}+a_{0}^{0})+i(\eta_{N}+\pi^{0})}{\sqrt{2}} & a_{0}^{+}%
+i\pi^{+} & K_{0}^{\star+}+iK^{+}\\
a_{0}^{-}+i\pi^{-} & \frac{(\sigma_{N}-a_{0}^{0})+i(\eta_{N}-\pi^{0})}%
{\sqrt{2}} & K_{0}^{\star0}+iK^{0}\\
K_{0}^{\star-}+iK^{-} & {\bar{K}_{0}^{\star0}}+i{\bar{K}^{0}} &
\sigma
_{S}+i\eta_{S}%
\end{array}
\right)  \text{ ,}\label{eq:matrix_field_Phi}\\
\end{align}

and the adjoint matrix $\Phi^{\dagger}$ is%

\begin{equation}
\Phi^{\dagger}  =\sum_{a=0}^{8}(S_{a}-iP_{a})T_{a}=
\frac{1}{\sqrt{2}}\left(
\begin{array}
[c]{ccc}%
\frac{(\sigma_{N}+a_{0}^{0})-i(\eta_{N}+\pi^{0})}{\sqrt{2}} & a_{0}^{+}%
-i\pi^{+} & K_{0}^{\star+}-iK^{+}\\
a_{0}^{-}-i\pi^{-} & \frac{(\sigma_{N}-a_{0}^{0})-i(\eta_{N}-\pi^{0})}%
{\sqrt{2}} & K_{0}^{\star0}-iK^{0}\\
K_{0}^{\star-}-iK^{-} & {\bar{K}_{0}^{\star0}}-i{\bar{K}^{0}} &
\sigma
_{S}-i\eta_{S}%
\end{array}
\right)  \text{ ,}%
\end{equation}
where  $T_{a}\,(a=0,\ldots,8)$ denote the generators of $U(3)$. The assignment of the quark-antiquark fields is as follows: \\
(i) In the pseudoscalar sector the fields $\vec{\pi}$ and $K$
represent the pions
and the kaons, respectively \cite{Beringer:1900zz}. The bare fields $\eta_{N}%
\equiv\left\vert \bar{u}u+\bar{d}d\right\rangle /\sqrt{2}$ and $\eta_{S}%
\equiv\left\vert \bar{s}s\right\rangle $ are the non-strange and
strange
contributions of the physical states $\eta$ and $\eta^{\prime}$ \cite{Beringer:1900zz}:%
\begin{align}
\eta & =\eta_{N}\cos\varphi+\eta_{S}\sin\varphi,\text{ }\label{mixetas1}\\
\eta^{\prime} & =-\eta_{N}\sin\varphi+\eta_{S}\cos\varphi, \label{mixetas2}%
\end{align}
where $\varphi\simeq-44.6^{\circ}$ is the mixing angle \cite{Parganlija:2012fy}
between $\eta$ and $\eta^{\prime}$. There are other values for the
mixing angle, e.g.\ $\varphi=-36^{\circ}$ \cite{Giacosa:2007up}
or $\varphi=-41.4^{\circ}$, as determined by the KLOE
Collaboration \cite{Ambrosino:2009sc}, but using these
affects the presented results only
marginally.\\
(ii) In the scalar sector we assign the field $\vec{a}_{0}$ to the
physical isotriplet state $a_{0}(1450)$ and the scalar kaon fields
$K_{0}^{\star}$ to the resonance $K_{0}^{\star}(1430).$ Finally, the
non-strange and strange bare fields $\sigma_{N}\equiv\left\vert
\bar{u}u+\bar{d}d\right\rangle /\sqrt{2}$ and
$\sigma_{S}\equiv\left\vert \bar{s}s\right\rangle $ mix with a scalar glueball $G\equiv gg\rangle$ and generate the three physical isoscalar resonances $f_{0}(1370)$, $f_0(1500)$ and
$f_{0}(1710)$. As seen in Ref. \cite{Janowski:2014ppa} that $f_{0}(1370)$, $f_0(1500)$ and
$f_{0}(1710)$ are predominantly a $\sigma_N$, $\sigma_S$, and a glueball state, respectively. The mixing of the bare fields $\sigma_{N}$ and
$\sigma_{S}$ is small \cite{Parganlija:2012fy, Parganlija:2012xj} (in agreement
with large-$N_{c}$ arguments) and is
neglected in this work.\\
Now let us turn to the vector fields $V$ (with quantum numbers
$J^{PC}=1^{--}$) and the axial-vector fields $A$ (with quantum
numbers $J^{PC}=1^{++}$) which are summarized in the following $3
\times 3$ matrices, respectively:

\begin{equation}
V^{\mu}=\frac{1}{\sqrt{2}}\left(
\begin{array}
[c]{ccc}%
\frac{\omega_{N}^{\mu}+\rho^{\mu0}}{\sqrt{2}} & \rho^{\mu+} & K^{\star\mu+}\\
\rho^{\mu-} & \frac{\omega_{N}^{\mu}-\rho^{\mu0}}{\sqrt{2}} & K^{\star\mu0}\\
K^{\star\mu-} & {\bar{K}}^{\star\mu0} & \omega_{S}^{\mu}%
\end{array}
\right) \text{}, \label{Ve}%
\end{equation}

and

\begin{equation}
A^\mu=\frac{1}{\sqrt{2}}\left(
\begin{array}
[c]{ccc}%
\frac{f_{1N}^{\mu}+a_{1}^{\mu0}}{\sqrt{2}} & a_{1}^{\mu+} & K_{1}^{\mu+}\\
a_{1}^{\mu-} & \frac{f_{1N}^{\mu}-a_{1}^{\mu0}}{\sqrt{2}} & K_{1}^{\mu0}\\
K_{1}^{\mu-} & {\bar{K}}_{1}^{\mu0} & f_{1S}^{\mu}%
\end{array}
\right)  \text{ ,} \label{A}%
\end{equation}
 which are combined into right-handed and left-handed vector fields
 as follows:
\begin{align}
R^{\mu}  &  =\sum_{a=0}^{8}(V_{a}^{\mu}-A_{a}^{\mu})T_{a}=\frac{1}{\sqrt{2}%
}\left(
\begin{array}
[c]{ccc}%
\frac{\omega_{N}+\rho^{0}}{\sqrt{2}}-\frac{f_{1N}+a_{1}^{0}}{\sqrt{2}}
&
\rho^{+}-a_{1}^{+} & K^{\star+}-K_{1}^{+}\\
\rho^{-}-a_{1}^{-} & \frac{\omega_{N}-\rho^{0}}{\sqrt{2}}-\frac{f_{1N}%
-a_{1}^{0}}{\sqrt{2}} & K^{\star0}-K_{1}^{0}\\
K^{\star-}-K_{1}^{-} & {\bar{K}}^{\star0}-{\bar{K}}_{1}^{0} &
\omega
_{S}-f_{1S}%
\end{array}
\right)  ^{\mu}\text{ ,} \label{eq:matrix_field_R}\\
\nonumber \\
L^{\mu}  &  =\sum_{a=0}^{8}(V_{a}^{\mu}+A_{a}^{\mu})T_{a}=\frac{1}{\sqrt{2}%
}\left(
\begin{array}
[c]{ccc}%
\frac{\omega_{N}+\rho^{0}}{\sqrt{2}}+\frac{f_{1N}+a_{1}^{0}}{\sqrt{2}}
&
\rho^{+}+a_{1}^{+} & K^{\star+}+K_{1}^{+}\\
\rho^{-}+a_{1}^{-} & \frac{\omega_{N}-\rho^{0}}{\sqrt{2}}+\frac{f_{1N}%
-a_{1}^{0}}{\sqrt{2}} & K^{\star0}+K_{1}^{0}\\
K^{\star-}+K_{1}^{-} & {\bar{K}}^{\star0}+{\bar{K}}_{1}^{0} &
\omega
_{S}+f_{1S}%
\end{array}
\right)  ^{\mu}\text{ ,}\label{eq:matrix_field_L}
\end{align}
Note that the so-called strange-nonstrange basis in the $(0-8)$ vector is used \cite{Parganlija:2012fy}, which is defined as
\begin{align}
\varphi_{N}  &  =\frac{1}{\sqrt{3}}\left(
\sqrt{2}\;\varphi_{0}+\varphi
_{8}\right)  \;,\nonumber\\
\varphi_{S}  &  =\frac{1}{\sqrt{3}}\left(
\varphi_{0}-\sqrt{2}\;\varphi _{8}\right)
\;,\quad\quad\varphi\in(S_{a},P_{a},V_{a}^{\mu},A_{a}^{\mu})\;,
\label{eq:nsbase}%
\end{align}
The quark-antiquark (axial-)vector fields in the matrices
(\ref{Ve}, \ref{A}) are assigned as follows:\\
(i) In the vector sector the fields $\omega_N$ and $\rho$
represent the $\omega(782)$ and $\rho(770)$ vector mesons,
respectively, while the $\omega_S$ and $K^\ast$ fields correspond
to the physical $\phi(1020)$ and $K^\ast(892)$ resonances,
respectively. \\
(ii) In the axial-vector sector we assign the fields $f_{1N}^\mu$
and $\overrightarrow{a}_1^\mu$ to the physical resonances
$f_1(1285)$ and $a_1(1260)$ mesons, respectively. The strange
fields $f_{1S}$ and $K_1$ correspond to the $f_1(1420)$ and
$K_{1270}$ [or $K_1(1400)$] mesons, respectively. [The details of
this assignment are given in Ref. \cite{Parganlija:2012fy}].\\
The matrices $H$ and $\Delta$ are defined as
\begin{align}
H  &  =H_{0}T_{0}+H_{8}T_{8}=\left(
\begin{array}
[c]{ccc}%
\frac{h_{0N}}{2} & 0 & 0\\
0 & \frac{h_{0N}}{2} & 0\\
0 & 0 & \frac{h_{0S}}{\sqrt{2}}%
\end{array}
\right)  \;,\label{eq:expl_sym_br_epsilon}\\
\Delta &  =\Delta_{0}T_{0}+\Delta_{8}T_{8}=\left(
\begin{array}
[c]{ccc}%
\frac{\tilde{\delta}_{N}}{2} & 0 & 0\\
0 & \frac{\tilde{\delta}_{N}}{2} & 0\\
0 & 0 & \frac{\tilde{\delta}_{S}}{\sqrt{2}}%
\end{array}
\right)  \equiv\left(
\begin{array}
[c]{ccc}%
\delta_{N} & 0 & 0\\
0 & \delta_{N} & 0\\
0 & 0 & \delta_{S}%
\end{array}
\right)  \text{ ,} \label{eq:expl_sym_br_delta}%
\end{align}
where $h_{N}\sim m_{u}$, $h_{S}\sim m_{s}%
$, $\delta_{N}\sim m_{u}^{2}$, $\delta_{S}\sim m_{s}^{2}$. The
matrices $H$ (\ref{eq:expl_sym_br_epsilon}) and $\Delta$
(\ref{eq:expl_sym_br_delta}) enter the
terms Tr$[H(\Phi+\Phi^{\dagger})]$ and
Tr$[\Delta(L_\mu^2+R_\mu^2)]$ which explicitly break the global
symmetry, $U(3)_{R}\times U(3)_{L}$ [$=U(3)_{V}\times U(3)_{A}$],
in the (pseudo)scalar and (axial-)vector sectors due to different nonzero
values for the quark masses. They
break $U(3)_A$, if $H_{0},\Delta_{0}\neq0$, and $U(3)_{V}\rightarrow
SU(2)_{V}\times U(1)_{V}$, if $H_{8},\Delta_{8}\neq0$, for details see Ref. \cite{Lenaghan:2000ey}. \\
The spontaneous symmetry breaking of the chiral symmetry is
implemented by condensing the scalar-isosinglet states which are  $\sigma_{N}%
\equiv(\bar{u}u+\bar{d}d)/\sqrt{2}$ and $\sigma_{S}\equiv\bar{s}s$. We shift these fields by their vacuum expectation values
$\phi_N$ and $\phi_S$,
\begin{equation}
\sigma_{N}\rightarrow\sigma_{N}+\phi_{N}\text{ and
}\sigma_{S}\rightarrow
\sigma_{S}+\phi_{S}\text{ ,} \label{shift1}%
\end{equation}
where the condensates $\phi_N$ and $\phi_S$ are functions of the
pion decay constant $f_\pi$ and the kaon decay constant $f_K$,
respectively, (the detailed calculation is presented in
the Appendix)

\begin{align}
\phi_{N} & =Z_{\pi}f_{\pi}\text{ ,}\label{phin}\\
\phi_{S} & =\frac{2Z_{K}f_{K}-\phi_{N}}{\sqrt{2}}\,,\label{phis}
\end{align}

Thus leads to the mixing between (axial-)vector and (pseudo)scalar
states in the Lagrangian (\ref{fulllag}),
\begin{align}
&-g_1\phi_N(f_{1N}^\mu\partial_\mu\eta_N+\overrightarrow{a}_1^\mu\cdot\partial_\mu\overrightarrow{\pi})
-\sqrt{2}\,g_1\phi_Sf^\mu_{1S}\partial_\mu\eta_S\nonumber\\
&-i\frac{g_1}{2}(\phi_N-\phi_S)(\overline{K}\,^{*\mu0}\,\partial_\mu
K^{\ast0}_0+K^{*\mu-}\,\partial_\mu
K^{\ast+}_0)\nonumber\\
&+i\frac{g_1}{2}(\phi_N-\sqrt{2}\phi_s)(K^{*\mu0}\,\partial_\mu\overline{K}^{\ast0}_0+K^{*\mu+}\,
\partial_\mu K^{\ast-}_0)\nonumber\\
&-\frac{g_1}{2}(\phi_N+\sqrt{2}\,\phi_S)(K_1^{\mu 0}\,\partial_\mu
\overline{K}^0+K_1^{\mu+}\,\partial_\mu K^-+\overline{K}_1^{\mu
0}\,\partial_\mu K^0+K_1^{\mu-}\,\partial_\mu
K^+)\,.\label{mixing1}
\end{align}

In order to eliminate this
mixing, one performs shifts of the (axial-)vector fields as
follows
\begin{align}
f_{1N/S}^{\mu}  &  \longrightarrow f_{1N/S}^{\mu}+Z_{\eta_{N/S}}w_{f_{1N/S}%
}\partial^{\mu}\eta_{N/S}\text{ },\label{eq:shifts1}\\
{a_{1}^{\mu}}^{\pm,0} & \longrightarrow
{a_{1}^{\mu}}^{\pm,0}+Z_{\pi}w_{a_{1}}\partial^{\mu}\pi^{\pm,0},\label{eq:shifts2}\\
{K_{1}^{\mu}}^{\pm,0,\bar{0}}  &  \longrightarrow{K_{1}^{\mu}}^{\pm,0,\bar{0}%
}+Z_{K}w_{K_{1}}\partial^{\mu}K^{\pm0,\bar{0}}\text{ },\label{eq:shifts3}\\
{K^{\star\mu}}^{\pm,0,\bar{0}} & \longrightarrow{K^{\star\mu}}^{\pm,0,\bar{0}}+Z_{K^{\star}%
}w_{K^{\star}}\partial^{\mu}K_{0}^{\star\pm,0,\bar{0}}\text{
}.\text{ }\label{eq:shifts4}%
\end{align}
 which produce additional kinetic terms for the (pseudo)scalar
 fields. The shift (\ref{eq:shifts1}) was performed in the case $N_f=2$ in
Ref. \cite{Parganlija:2010fz}, and the shifts (\ref{eq:shifts2} - \ref{eq:shifts4}) were
performed in the case $N_f=3$ Ref. \cite{Parganlija:2012fy}. The wave-function renormalization constants have been
 introduced to retain the canonical normalization
\begin{align}
\pi^{\pm,0} & \rightarrow Z_{\pi}\pi^{\pm,0},\label{eq:shifts5}\\
K^{\pm,0,\bar{0}} & \rightarrow Z_{K}K^{\pm,0,\bar{0}}\text{ },\label{eq:shifts6}\\
\text{ }\eta_{N/S} & \rightarrow Z_{\eta_{N} /\eta_{S}}
\eta_{N/S}\text{ , }\label{eq:shifts7}\\
{K^{\star\mu}_0}^{\pm,0,\bar{0}} & \rightarrow
Z_{K^{\star}_0}{K_0^{\star\mu}}^{\pm,0,\bar{0}}\;.\label{eq:shifts8}
\end{align}
Note that for simplicity the isotriplet states have been grouped
together with the notation $\pi^{\pm,0},{a_{1}^{\mu}}^{\pm,0}$ and
the isodoublet states with the notation
$K^{\pm,0,\bar{0}},{K^{\star\mu}_0}^{\pm,0,\bar{0}}$, where
$\bar{0}$ refers to $\bar{K}^{0}$. The explicit expressions for
the coefficients $w_i$ are obtained after some
straightforward calculation (for details see Ref.\
\cite{Parganlija:2012xj}) as 
\begin{align}
w_{f_{1N}} & = w_{a_{1}}=\frac{g_{1}\phi_{N}}{m_{a_{1}}^{2}}\text{ ,}\label{wfn}\\
w_{f_{1S}} & =\frac{\sqrt{2}g_{1}\phi_{S}}{m_{f_{1S}}^{2}}\text{ ,}\label{wfs}\\%
w_{K^{\star}} & =\frac{ig_{1}(\phi_{N}-\sqrt{2}\phi_{S})}{2m_{K^{\star}}^{2}}\text{ ,}\label{wks}\\
w_{K_{1}} & =\frac{g_{1}(\phi_{N}+\sqrt{2}\phi_{S})}{2m_{K_{1}}^{2}}\text{ .}\label{wk}%
\end{align}%
The wave-function renormalization constants $Z_{i}$, introduced in
Eq.\ (\ref{eq:shifts1}-\ref{eq:shifts4}), are determined such that
one obtains the canonical normalization of the $\pi,\,\eta_N,\,
\eta_S, K$ and $K^\ast_0$. Their explicit expressions
read \cite{Parganlija:2012fy, Parganlija:2012xj}:%
\begin{align}
Z_{\pi} &
=Z_{\eta_{N}}=\frac{m_{a_{1}}}{\sqrt{m_{a_{1}}^{2}-g_{1}^{2}\phi
_{N}^{2}}}\;, \label{zpi}\\
Z_{K} &
=\frac{2m_{K_{1}}}{\sqrt{4m_{K_{1}}^{2}-g_{1}^{2}(\phi_{N}+\sqrt{2}
\phi_{S})^{2}}}\;, \label{zk}\\
Z_{K_{S}} &
=\frac{2m_{K^{\star}}}{\sqrt{4m_{K^{\star}}^{2}-g_{1}^{2}(\phi
_{N}-\sqrt{2}\phi_{S})^{2}}}\;, \label{zks}\\
Z_{\eta_{S}} & =\frac{m_{f_{1S}}}{\sqrt{m_{f_{1S}}^{2}-2g_{1}^{2}\phi_{S}^{2}}}\;, \label{zets}%
\end{align}
which are always larger than one. After some straightforward
calculation the tree-level (squared) masses for all nonets in the
chiral Lagrangian (\ref{fulllag}) are given\cite{Parganlija:2012fy, Parganlija:2012xj} by
\begin{align}
m_{\pi}^{2}  &  =Z_{\pi}^{2}\left[  m_{0}^{2}+\left(  \lambda_{1}%
+\frac{\lambda_{2}}{2}\right)
\phi_{N}^{2}+\lambda_{1}\phi_{S}^{2}\right]
\equiv\frac{Z_{\pi}^{2}h_{0N}}{\phi_{N}}\text{ ,}\label{m_pi}\\
m_{K}^{2}  &  =Z_{K}^{2}\left[  m_{0}^{2}+\left( \lambda_{1}+\frac
{\lambda_{2}}{2}\right)  \phi_{N}^{2}-\frac{\lambda_{2}}{\sqrt{2}}\phi_{N}%
\phi_{S}+\left(  \lambda_{1}+\lambda_{2}\right)
\phi_{S}^{2}\right]  \text{
,}\label{mkaon}\\
m_{\eta_{N}}^{2}  &  =Z_{\pi}^{2}\left[  m_{0}^{2}+\left(  \lambda_{1}%
+\frac{\lambda_{2}}{2}\right)  \phi_{N}^{2}+\lambda_{1}\phi_{S}^{2}%
+c_{1}\,\phi_{N}^{2}\phi_{S}^{2}\right]  \equiv Z_{\pi}^{2} \left(
\frac{h_{0N}}{\phi_{N}}+c_{1}\,\phi_{N}^{2}\phi_{S}^{2}\right)
\;,\label{eq:etaN}\\
m_{\eta_{S}}^{2}  &  =Z_{\eta_{S}}^{2}\left[
m_{0}^{2}+\lambda_{1}\phi
_{N}^{2}+\left(  \lambda_{1}+\lambda_{2}\right)  \phi_{S}^{2}+\frac{c_{1}}%
{4}\phi_{N}^{4}\right]  \equiv Z_{\eta_{S}}^{2} \left(
\frac{h_{0S}}{\phi
_{S}}+\frac{c_{1}}{4}\phi_{N}^{4}\right)  \text{ ,}\label{eq:etaS}\\
m_{\eta_{NS}}^{2}  &
=Z_{\pi}Z_{\pi_{S}}\frac{c_{1}}{2}\phi_{N}^{3}\phi
_{S}\text{ ,} \label{eq:etaNS}%
\end{align}
for the (squared) pseudoscalar masses, while
\begin{align}
m_{a_{0}}^{2}  &  =m_{0}^{2}+\left( \lambda_{1}+\frac{3}{2}\lambda
_{2}\right)  \phi_{N}^{2}+\lambda_{1}\phi_{S}^{2}\text{ ,}\label{m_a_0}\\
m_{K_{0}^{\star}}^{2}  &  =Z_{K_{0}^{\star}}^{2}\left[
m_{0}^{2}+\left(
\lambda_{1}+\frac{\lambda_{2}}{2}\right)  \phi_{N}^{2}+\frac{\lambda_{2}%
}{\sqrt{2}}\phi_{N}\phi_{S}+\left(  \lambda_{1}+\lambda_{2}\right)
\phi
_{S}^{2}\right]  \text{ ,}\\
m_{\sigma_{N}}^{2}  &  =m_{0}^{2}+3\left(  \lambda_{1}+\frac{\lambda_{2}}%
{2}\right)  \phi_{N}^{2}+\lambda_{1}\phi_{S}^{2}\text{ ,}\label{eq:sigN}\\
m_{\sigma_{S}}^{2}  &  =m_{0}^{2}+\lambda_{1}\phi_{N}^{2}+3\left(
\lambda
_{1}+\lambda_{2}\right)  \phi_{S}^{2}\text{ ,}\label{eq:sigS}\\
m_{\sigma_{NS}}^{2}  &  =2\lambda_{1}\phi_{N}\phi_{S}\text{ ,}
\label{eq:sigNS}%
\end{align}
are the (squared) scalar masses. Moreover, the (squared) vector
masses are obtained as
\begin{align}
m_{\rho}^{2}  &  =m_{1}^{2}+\frac{1}{2}(h_{1}+h_{2}+h_{3})\phi_{N}^{2}%
+\frac{h_{1}}{2}\phi_{S}^{2}+2\delta_{N}\;,\label{m_rho}\\
m_{K^{\star}}^{2}  &  =m_{1}^{2}+\frac{1}{4}\left(  g_{1}^{2}+2h_{1}%
+h_{2}\right)  \phi_{N}^{2}+\frac{1}{\sqrt{2}}\phi_{N}\phi_{S}(h_{3}-g_{1}%
^{2})+\frac{1}{2}(g_{1}^{2}+h_{1}+h_{2})\phi_{S}^{2}+\delta_{N}+\delta
_{S}\;,\label{m_K_star}\\
m_{\omega_{N}}^{2}  &  =m_{\rho}^{2}\;,\\
m_{\omega_{S}}^{2}  &
=m_{1}^{2}+\frac{h_{1}}{2}\phi_{N}^{2}+\left(
\frac{h_{1}}{2}+h_{2}+h_{3}\right)  \phi_{S}^{2}+2\delta_{S}\;,
\end{align}
while the (squared) axial-vector meson masses are
\begin{align}
m_{a_{1}}^{2}  &
=m_{1}^{2}+\frac{1}{2}(2g_{1}^{2}+h_{1}+h_{2}-h_{3})\phi
_{N}^{2}+\frac{h_{1}}{2}\phi_{S}^{2}+2\delta_{N}\;,\label{m_a_1}\\
m_{K_{1}}^{2}  &  =m_{1}^{2}+\frac{1}{4}\left(
g_{1}^{2}+2h_{1}+h_{2}\right)
\phi_{N}^{2}-\frac{1}{\sqrt{2}}\phi_{N}\phi_{S}(h_{3}-g_{1}^{2})+\frac{1}%
{2}\left(  g_{1}^{2}+h_{1}+h_{2}\right)
\phi_{S}^{2}+\delta_{N}+\delta
_{S}\;,\label{m_K_1}\\
m_{f_{1N}}^{2}  &  =m_{a_{1}}^{2}\;,\\
m_{f_{1S}}^{2}  &  =m_{1}^{2}+\frac{h_{1}}{2}\phi_{N}^{2}+\left(  2g_{1}%
^{2}+\frac{h_{1}}{2}+h_{2}-h_{3}\right)
\phi_{S}^{2}+2\delta_{S}\;.
\label{m_f1_S}%
\end{align}
All previous expressions coincide with Refs.\ \cite{Parganlija:2012fy, Parganlija:2012xj}.

\subsection{Model Parameters}

The chirally symmetric model Eq.(\ref{fulllag}) contains 18
parameters which are:
$m_{0}^{2}\text{, }m_{1}^{2}\text{, }c_{1}\text{, }\delta_{N}\text{, }$ \\
$\delta_{S},\,g_{1}\text{, }g_{2}\text{, }g_{3}\text{, }g_{4}\text{, }g_{5}\text{, }g_{6}\text{, }h_{0N}\text{, }h_{0S}\text{, }h_{1}\text{, }%
h_{2}\text{, }h_{3}\text{, }\lambda_{1}\text{, }\lambda_{2}$. Note
that the coupling of the glueball with the other mesons has been
neglected. The parameters $g_{3}\text{, }g_{4}\text{,
}g_{5}\text{, }$ and $g_{6}$ are not considered in the fit because
they do not influence any decay channels in the case of the
$N_f=3$ \cite{Parganlija:2012fy} investigation. The explicit symmetry breaking
in the vector and axial-vector channel is described by $\delta_N$
and $\delta_S$. The ESB arises from non-vanishing quark masses,
which leads us to the correspondence $\delta_N \propto
m_{u,d}^2$ and $\delta_S \propto m_S^2$. The linear combination
$m_{1}^{2}/2+\delta_{N/S}$ appears in the vector-meson mass term $\mathrm{Tr}[(m_{1}^{2}/2+\Delta)(L_{\mu}%
^{2}+R_{\mu}^{2})]$. We can redefine $m_{1}^{2}/2\rightarrow
m_{1}^{2}/2-\delta_{N}$ which leads to the appearance of only the
combination $\delta_{S}-\delta_{N}$ in the mass formulas. This
difference is determined by the fit of the (axial-)vector masses.
Without los of generally, we may take $\delta_N\equiv 0$. Then the unknown
parameters are decreased to $13$ in the chiral Lagrangian (or the
so-called the extended Linear Sigma Model) \cite{Parganlija:2012fy}:
$m_{0}^{2}\text{, }m_{1}^{2}\text{, }c_{1}\text{, }%
\delta_{S}\text{, }g_{1}\text{, }g_{2}\text{, }h_{0N}\text{, }h_{0S}\text{, }h_{1}\text{, }%
h_{2}\text{, }h_{3}\text{, }\lambda_{1}\text{, }\lambda_{2}\text{
.}$ \\

Moreover, the experimental quantities that are used in the fit do
not depend on all previous 13 parameters. The
following two linear combinations have been used in the fit rather
than the parameters $m_0,\,\lambda_1,\,m_1,\,h_1$ separately.

\begin{align}
&C_{1} = m_{0}^{2} + \lambda_{1} \left(  \phi_{N}^{2}
+\phi_{S}^{2}\right) \;,\nonumber\\
&C_{2} = m_{1}^{2} + \frac{h_{1}}{2}
\left(  \phi_{N}^{2} + \phi _{S}^{2}\right)  \;.
\end{align}
The condensates $\phi_N$ and $\phi_S$ are used instead of the
parameters $h_{0N}$ and $h_{0S}$ which are determined by the
masses of pion and $\eta_S$, as presented in Eqs. (\ref{m_pi}),
(\ref{eq:etaS}). Consequently, there are eleven parameters left:
$$C_{1}, ~C_{2} , ~c_{1}, ~\delta_{S}, ~g_{1}, ~g_{2}, ~\phi_{N}, ~\phi_{S},
~h_{2}, ~h_{3}, ~\lambda_{2}\,.$$ The eleven parameters are fitted
by 21 experimental quantities as seen in
Ref. \cite{Parganlija:2012xj}. The parameter values are obtained with
$\chi^2\simeq1$ \cite{Parganlija:2012fy} as summarized in the Table
\ref{Tab:param1}.

\begin{table}[H]
\centering
\begin{tabular}
[c]{|c|c|}\hline Parameter & Value\\\hline $C_{1}$  & $(-0.9183
\pm0.0006)\,\text{GeV}^{2} $\\\hline $C_{2}$  & $(0.4135
\pm0.0147)\,\text{GeV}^{2}$\\\hline $c_{1}$  & $(450.5420
\pm7.0339)\,\text{GeV}^{-2}$\\\hline $\delta_{S}$ & $(0.1511
\pm0.0038)\,\text{GeV}^{2}$\\\hline $g_{1}$ & $5.8433
\pm0.0176$\\\hline $g_{2}$ & $3.0250 \pm0.2329$\\\hline $\phi_{N}$
& $(0.1646 \pm0.0001)\,\text{GeV}$\\\hline $\phi_{S}$ & $(0.1262
\pm0.0001)\,\text{GeV}$\\\hline $h_{2}$ & $9.8796
\pm0.6627$\\\hline $h_{3}$ & $4.8667 \pm0.0864$\\\hline
$\lambda_{2}$ & $68.2972 \pm0.0435$\\\hline
\end{tabular}
\caption{Parameters and their errors.}%
\label{Tab:param1}%
\end{table}

Note that the parameters $\lambda_1$ and $h_1$ are set to zero
because the fit uses all scalar mass terms except $m_{\sigma_N}$
and $m_{\sigma_S}$, due
to the well-known ambiguities regarding the assignment of
scalar mesons. The parameter $\lambda_1$ is
expressed via the bare mass parameter $m_0^2$ which is allowed due
to the knowledge of the mentioned linear combination. Moreover the
parameter $h_1$ is suppressed in large-$N_c$ as well as
$\lambda_1$.

\subsection{Results}
The fit results for masses \cite{Parganlija:2012fy} are interesting because
they prove that a chiral framework is applicable for the study of hadron vacuum phenomenology up to 1.7 GeV, (as listed in Table
\ref{Tab:param1})

\begin{table}[H]
\centering
\begin{tabular}
[c]{|c|c|c|}\hline Observable & Fit [MeV] & Experiment
[MeV]\\\hline $m_{\pi}$ & $141.0 \pm5.8$ & $137.3 \pm6.9$\\\hline
$m_{K}$ & $485.6 \pm3.0$ & $495.6 \pm24.8$\\\hline $m_{\eta}$ &
$509.4 \pm3.0$ & $547.9 \pm27.4$\\\hline $m_{\eta^{\prime}}$ &
$962.5 \pm5.6$ & $957.8 \pm47.9$\\\hline $m_{\rho}$ & $783.1
\pm7.0$ & $775.5 \pm38.8$\\\hline $m_{K^{\star}}$ & $885.1 \pm6.3$
& $893.8 \pm44.7$\\\hline $m_{\phi}$ & $975.1 \pm6.4$ & $1019.5
\pm51.0$\\\hline $m_{a_{1}}$ & $1186 \pm6$ & $1230 \pm62$\\\hline
$m_{f_{1}(1420)}$ & $1372.5 \pm5.3$ & $1426.4 \pm71.3$\\\hline
$m_{a_{0}}$ & $1363 \pm1$ & $1474 \pm74$\\\hline
$m_{K_{0}^{\star}}$ & $1450 \pm1$ & $1425 \pm71$\\\hline
\end{tabular}
\caption{Best-fit results for masses compared with experiment
(from Ref. \cite{Parganlija:2012fy}).}%
\label{Comparmasli}%
\end{table}

The mass results of the (pseudo)scalar and (axial-)vector sectors
are in good agreement with experimental results as seen in Table \ref{Comparmasli}. The resonances
$a_{0}(1450)$ and $K_{0}^{\ast}(1430)$ are well described as
quark-antiquark fields. The scalar-isoscalar mesons are not
included in the fit. The decay pattern and the masses suggest that
$f_{0}(1370)$ and $f_{0}(1710)$ are (predominantly) the
non-strange and strange scalar-isoscalar fields. In the
Refs.\cite{Parganlija:2012fy, Parganlija:2012xj} there are further results for the
decay widths for light mesons in the chiral model (\ref{fulllag})
 which are also in a good agreement with experiment.\\

\chapter{Charmed mesons in the extended Linear Sigma Model}\label{ch5 chaptercharmed}

``\textit{Research is to see what everybody has seen and to think
what nobody else has thought}'' \\ 

$\,\,\,\,\,\,\,\,\,\,\,\,\,\,\,\,\,\,\,\,\,\,\,\,\,\,\,\,\,\,\,\,\,\,\,\,\,\,\,\,\,\,\,\,\,\,\,\,\,\,\,\,\,\,\,\,\,\,\,\,\,\,\,\,\,\,\,\,\,\,\,\,\,\,\,\,\,\,\,\,\,\,\,\,\,\,\,\,\,\,\,\,\,\,\,\,\,\,\,\,\,\,\,\,\,\,\,\,\,\,\,\,\,\,\,\,\,\,\,\,\,\,\,\,\,\,\,\,\,\,\,\,\,\,\,\,\,\,\,\,\,\,\,\,\,\,\,\,\,\,\,\,\,\,\,\,\,\,\,\,\,\,\,\,\,\,\,\,\,\,\,\,$ Albert Szent-Györgi\\
\section{Introduction}

In this chapter we investigate the eLSM model in the four-flavour case ($N_{f}%
=4$), i.e., by considering mesons which contain at least one charm quark. This chapter is based on Refs.\cite{Eshraim:2014afa, Eshraim:2014eka, Eshraim:2014vfa, Eshraim:2014gya, Eshraim:2014tla}. This
study is a straightforward extension of Sec 3.3 \cite{Parganlija:2012fy}: the Lagrangian has
the same structure as in the $N_{f}=3$ case, except that all (pseudo)scalar
and (axial-)vector meson fields are now parametrized in terms of $4\times4$
(instead of $3\times3$) matrices. These now also include the charmed degrees
of freedom. Since low-energy (i.e., nonstrange and strange) hadron
phenomenology was described very well \cite{Parganlija:2012fy}, we retain the values for
the parameters that already appear in the three-flavour sector. Then, extending
the model to $N_{f}=4$, three additional parameters enter, all of which are
related to the current charm quark mass (two of them in the (pseudo)scalar
sector and one in the (axial-)vector sector).

Considering that the explicit breaking of chiral and dilatation
symmetries by the current charm quark mass, $m_{c}\simeq1.275$ GeV, are quite
large, one may wonder whether it is at all justified to apply a model based on
chiral symmetry. Related to this, the charmed mesons entering our model have a
mass up to about $3.5$ GeV, i.e., they are strictly speaking no longer part of
the low-energy domain of the strong interaction. Naturally, we do not expect
to achieve the same precision as refined potential models \cite{Godfrey:1985xj, Godfrey:1986wj, Capstick:1986bm, Isgur:1991wq, Eichten:1993ub, Maki:1977ri, Segovia:2013kg, Cao:2012du}, lattice-QCD calculations \cite{Bali:2006xt, Donald:2012ga, Kalinowski:2012re}, and heavy-quark effective
theories \cite{Neubert:1993mb, Casalbuoni:1996pg, Georgi:1990um, Georgi:1991mr, DeFazio:2000up, Wise:1992hn, Kolomeitsev:2003ac, Bardeen:1993ae, Bardeen:2003kt, Nowak:1992um, Nowak:2003ra, Sasaki:2014asa, Lutz:2007sk} [see also the review of
Ref.\ \cite{Brambilla:2010cs} and refs.\ therein]. Nevertheless, it is still
interesting to see how a successful model for low-energy hadron phenomenology
based on chiral symmetry and dilatation invariance fares when extending it to
the high-energy charm sector. Quite surprisingly, a quantitative agreement with
experimental values for the open charmed meson masses is obtained by fitting
just the three additional parameters mentioned above (with deviations of the order of 150 
MeV, i.e., $\sim 5$\%). 
On the other hand, with the exception of $J/\psi$, the charmonium
states turn out to be about 10\% too light
when compared to experimental data. Nevertheless, the main conclusion
is that it is, to first approximation, not unreasonable to delegate the strong breaking
of chiral and dilatation symmetries to three mass terms only and
still have chirally and dilatation invariant interaction terms. Moreover, our
model correctly predicts the mass splitting between spin-0 and spin-1 negative-parity
open charm states, i.e., naturally incorporates the right amount of breaking of 
the heavy-quark spin symmetry.

\section{The $U(4)_r \times U(4)_l$ linear sigma model}
\label{Lagr}

\indent In this section we extend the eLSM \cite{Parganlija:2012fy, Parganlija:2012xj} to the four-flavour case. To this end, we introduce
$4\times4$ matrices which contain, in addition to the usual nonstrange and
strange mesons, also charmed states. The matrix of pseudoscalar fields $P$
(with quantum numbers $J^{PC}=0^{-+}$) reads
\begin{equation}
P=\frac{1}{\sqrt{2}}\left(
\begin{array}
[c]{cccc}%
\frac{1}{\sqrt{2}}(\eta_{N}+\pi^{0}) & \pi^{+} & K^{+} & D^{0}\\
\pi^{-} & \frac{1}{\sqrt{2}}(\eta_{N}-\pi^{0}) & K^{0} & D^{-}\\
K^{-} & \overline{K}^{0} & \eta_{S} & D_{S}^{-}\\
\overline{D}^{0} & D^{+} & D_{S}^{+} & \eta_{c}%
\end{array}
\right)  \sim\frac{1}{\sqrt{2}}\left(
\begin{array}
[c]{cccc}%
\bar{u}\Gamma u & \bar{d}\Gamma u & \bar{s}\Gamma u & \bar{c}\Gamma u\\
\bar{u}\Gamma d & \bar{d}\Gamma d & \bar{s}\Gamma d & \bar{c}\Gamma d\\
\bar{u}\Gamma s & \bar{d}\Gamma s & \bar{s}\Gamma s & \bar{c}\Gamma s\\
\bar{u}\Gamma c & \bar{d}\Gamma c & \bar{s}\Gamma c &
\bar{c}\Gamma c
\end{array}
\right)  \text{ ,} \label{p2}%
\end{equation}
where, for sake of clarity, we also show the quark-antiquark content of the
mesons (in the pseudoscalar channel $\Gamma=i\gamma^{5}$). In the
nonstrange-strange sector (the upper left $3\times3$ matrix) the matrix $P$
contains the pion triplet $\vec{\pi},$ the four kaon states $K^{+},$ $K^{-},$
$K^{0},$ $\bar{K}^{0},$ and the isoscalar fields $\eta_{N}=\sqrt{1/2}(\bar
{u}u+\bar{d}d)$ and $\eta_{S}=\bar{s}s$, see Eq. (\ref{P}). The latter two fields mix and
generate the physical fields $\eta$ (\ref{mixetas1}) and $\eta^{\prime}$ (\ref{mixetas2}) [see details in
Ref.\ \cite{Parganlija:2012fy}]. In the charm sector (fourth line and fourth column) the
matrix $P$ contains the open charmed states $D^{+},$ $D^{-},$ $D^{0},$
$\bar{D}^{0}$, which correspond to the well-established $D$ resonance, the
open strange-charmed states $D_{S}^{\pm}$, and, finally, the hidden charmed
state $\eta_{c},$ which represents the well-known pseudoscalar ground state
charmonium $\eta_{c}(1S)$.

The matrix of scalar fields $S$ (with quantum numbers $J^{PC}=0^{++}$) reads%
\begin{equation}
S=\frac{1}{\sqrt{2}}\left(
\begin{array}
[c]{cccc}%
\frac{1}{\sqrt{2}}(\sigma_{N}+a_{0}^{0}) & a_{0}^{+} &
K_{0}^{\ast+} &
D_{0}^{\ast0}\\
a_{0}^{-} & \frac{1}{\sqrt{2}}(\sigma_{N}-a_{0}^{0}) &
K_{0}^{\ast0} &
D_{0}^{\ast-}\\
K_{0}^{\ast-} & \overline{K}_{0}^{\ast0} & \sigma_{S} & D_{S0}^{\ast-}\\
\overline{D}_{0}^{\ast0} & D_{0}^{\ast+} & D_{S0}^{\ast+} & \chi_{c0}%
\end{array}
\right)  \text{ .}%
\end{equation}
The quark-antiquark content is the same as in Eq.\ (\ref{p2}), but
using $\Gamma=1_{4}$. A long debate about the correct assignment
of light scalar states has taken place in the last decades.
Present results \cite{Amsler:2004ps, Klempt:2007cp, Amsler:1995td, Lee:1999kv, Close:2001ga}, which have been
independently confirmed in the framework of the eLSM \cite{Parganlija:2012fy},
show that the scalar quarkonia have masses between 1-2 GeV. In
particular, the isotriplet $\vec{a}_{0}$ is assigned to the
resonance $a_{0}(1450)$ (and not to the lighter state
$a_{0}(980)$). Similarly, the kaonic states $K_{0}^{\ast+},$
$K_{0}^{\ast-},$ $K_{0}^{\ast+},$ $\overline {K}_{0}^{\ast0}$ are
assigned to the resonance $K_{0}^{\ast}(1430)$ (and not to the
$K_{0}^{\ast}(800)$ state). The situation in the scalar-isoscalar
sector is more complicated, due to the presence of a scalar
glueball state $G$, see Ref.\ \cite{Janowski:2011gt} and below. Then,
$\sigma_{N},$ $\sigma_{S}$, $G$ mix and generate the three
resonances $f_{0}(1370),$ $f_{0}(1500),$ and $f_{0}(1710).$ There
is evidence \cite{Janowski:2014ppa} that $f_{0}(1370)$ is predominantly a
$\sqrt{1/2}(\bar{u}u+\bar{d}d)$ state, while $f_{0}(1500)$ is
predominantly a $\bar{s}s$ state and $f_{0}(1710)$ predominantly a
glueball state. As a consequence, the light scalar states
$f_{0}(500)$ and $f_{0}(980)$ are not quarkonia (but, arguably,
tetraquark or molecular states) \cite{Giacosa:2006tf, Caprini:2005zr, Yndurain:2007qm, GarciaMartin:2011jx, Bugg:2007ja, Black:1999dx, Fariborz:2007ai, Fariborz:2009cq, Fariborz:2011in, Mukherjee:2012xn}. In the open charm sector, we assign the charmed
states $D_{0}^{\ast}$ to the resonances $D_{0}^{\ast}(2400)^{0}$
and $D_{0}^{\ast}(2400)^{\pm}$ (the latter state has not yet been
unambiguously established). In the strange-charm sector we assign
the state $D_{S0}^{\ast \pm}$ to the only existing candidate
$D_{S0}^{\ast}(2317)^{\pm}$; it should, however, be stressed that
the latter state has also been interpreted as a tetraquark or
molecular state because it is too light when compared to
quark-model predictions, see Refs.\
\cite{Godfrey:1985xj, Godfrey:1986wj, Guo:2009id, Liu:2012zya, Guo:2009ct, Brambilla:2010cs, Mohler:2011ke, Moir:2013ub, Ebert:2009ua}. In the next
section, we discuss in more detail the possibility that a heavier,
very broad (and therefore not yet discovered) scalar charmed state
exists. In the hidden charm sector the resonance $\chi_{c0}$
corresponds to the ground-state scalar charmonium $\chi_{c0}(1P)$.

The matrices $P$ and $S$ are used to construct the matrix $\Phi$
as follows,

\begin{equation}
\Phi=S+i\,P=\frac{1}{\sqrt{2}}
\left(%
\begin{array}{cccc}
  \frac{(\sigma_{N}+a^0_{0})+i(\eta_N +\pi^0)}{\sqrt{2}} & a^{+}_{0}+i \pi^{+} & K^{*+}_{0}+iK^{+} & D^{*0}_0+iD^0 \\
  a^{-}_{0}+i \pi^{-} & \frac{(\sigma_{N}-a^0_{0})+i(\eta_N -\pi^0)}{\sqrt{2}} & K^{*0}_{0}+iK^{0} & D^{*-}_0+iD^{-} \\
  K^{*-}_{0}+iK^{-} & \overline{K}^{*0}_{0}+i\overline{K}^{0} & \sigma_{S}+i\eta_{S} & D^{*-}_{S0}+iD^{-}_S\\
  \overline{D}^{*0}_0+i\overline{D}^0 & D^{*+}_0+iD^{+} & D^{*+}_{S0}+iD^{+}_S & \chi_{C0}+i\eta_C\\
\end{array}%
\right)\,,\label{Phi}
\end{equation}
and the adjoint matrix $\Phi^\dagger$ reads
\begin{equation}\label{phid}
\Phi^\dagger=S-i\,P=\frac{1}{\sqrt{2}}
\left(%
\begin{array}{cccc}
  \frac{(\sigma_{N}+a^0_{0})-i(\eta_N +\pi^0)}{\sqrt{2}} & a^{+}_{0}-i \pi^{+} & K^{*+}_{0}-iK^{+} & D^{*0}_0-iD^0 \\
  a^{-}_{0}-i \pi^{-} & \frac{(\sigma_{N}-a^0_{0})-i(\eta_N -\pi^0)}{\sqrt{2}} & K^{*0}_{0}-iK^{0} & D^{*-}_0-iD^{-} \\
  K^{*-}_{0}-iK^{-} & \overline{K}^{*0}_{0}-i\overline{K}^{0} & \sigma_{S}-i\eta_{S} & D^{*-}_{S0}-iD^{-}_S\\
  \overline{D}^{*0}_0-i\overline{D}^0 & D^{*+}_0-iD^{+} & D^{*+}_{S0}-iD^{+}_S & \chi_{C0}-i\eta_C\\
\end{array}%
\right)\,.
\end{equation}

The multiplet matrix $\Phi$ transforms as $\Phi\rightarrow
U_{L}\Phi U_{R}^{\dagger}$ under $U_{L}(4)\times U_{R}(4)$ chiral
transformations, where $U_{L(R)}=e^{-i\theta_{L(R)}^at^a}$ is an
element of $U(4)_{R(L)}$, under parity, 
$\Phi(t,\overrightarrow{x})\rightarrow\Phi^{\dagger}(t,-\overrightarrow{x})$,
and under charge conjugate $\Phi\rightarrow\Phi^{\dagger}$. The
determinant of $\Phi$ is invariant under $SU(4)_{L} \times
SU(4)_{R}$, but not under $U(1)_{A}$ because ${ det
\Phi}\rightarrow { det} U_{A}\Phi
U_A=e^{-i\theta_{A}^0\sqrt{2N_f}}{ det \Phi}\neq { det
\Phi}$.\\

We now turn to the vector sector. The matrix $V^{\mu}$ which
includes the
vector degrees of freedom is:%
\begin{equation}
V^{\mu}=\frac{1}{\sqrt{2}}\left(
\begin{array}
[c]{cccc}%
\frac{1}{\sqrt{2}}(\omega_{N}+\rho^{0}) & \rho^{+} &
K^{\ast}(892)^{+} &
D^{\ast0}\\
\rho^{-} & \frac{1}{\sqrt{2}}(\omega_{N}-\rho^{0}) &
K^{\ast}(892)^{0} &
D^{\ast-}\\
K^{\ast}(892)^{-} & \bar{K}^{\ast}(892)^{0} & \omega_{S} & D_{S}^{\ast-}\\
\overline{D}^{\ast0} & D^{\ast+} & D_{S}^{\ast+} & J/\psi
\end{array}
\right)  ^{\mu}\text{ .}%
\end{equation}

The quark-antiquark content is that shown in Eq.\ (\ref{p2}), setting
$\Gamma=\gamma^{\mu}.$ The isotriplet field $\vec{\rho}$ corresponds to the
$\rho$ meson, the four kaonic states correspond to the resonance $K^{\ast
}(892),$ the isoscalar states $\omega_{N}$ and $\omega_{S}$ correspond to the
$\omega$ and $\phi$ mesons, respectively. [No mixing between strange and
nonstrange isoscalars is present in the eLSM; this mixing is small anyway
\cite{Klempt:2004yz}.] In the charm sector, the fields $D^{\ast0},$ $\overline
{D}^{\ast0},D^{\ast+},$ and $D^{\ast-}$ correspond to the vector charmed
resonances $D^{\ast}(2007)^{0}$ and $D^{\ast}(2010)^{\pm}$, respectively,
while the strange-charmed $D_{S}^{\ast\pm}$ corresponds to the resonance
$D_{S}^{\ast\pm}$ (with mass $M_{D_{S}^{\ast\pm}}=(2112.3\pm0.5)$ MeV; note,
however, that the quantum numbers $J^{P}=1^{-}$ are not yet fully
established). Finally, $J/\psi$ is the very well-known lowest vector
charmonium state $J/\psi(1S).$\\

The matrix $A^{\mu}$ describing the axial-vector degrees of
freedom is given
by:%
\begin{equation}
A^{\mu}=\frac{1}{\sqrt{2}}\left(
\begin{array}
[c]{cccc}%
\frac{1}{\sqrt{2}}(f_{1,N}+a_{1}^{0}) & a_{1}^{+} & K_{1}^{+} & D_{1}^{0}\\
a_{1}^{-} & \frac{1}{\sqrt{2}}(f_{1,N}-a_{1}^{0}) & K_{1}^{0} & D_{1}^{-}\\
K_{1}^{-} & \bar{K}_{1}^{0} & f_{1,S} & D_{S1}^{-}\\
\bar{D}_{1}^{0} & D_{1}^{+} & D_{S1}^{+} & \chi_{c,1}%
\end{array}
\right)  ^{\mu}\,.
\end{equation}
The quark-antiquark content is that shown in Eq.\ (\ref{p2}),
setting $\Gamma=\gamma^{\mu}\gamma^{5}.$ The isotriplet field
$\vec{a}_{1}$ corresponds to the field $a_{1}(1260),$ the four
kaonic states $K_{1}$ correspond (predominantly) to the resonance
$K_{1}(1200)$ [but also to $K_{1}(1400),$ because of mixing
between axial-vector and pseudovector states, see Refs.\
\cite{Divotgey:2013jba, Cheng:2003bn, Hatanaka:2008gu, Ahmed:2011vr, Liu:2014dxa} and ref.\ therein]. The isoscalar fields
$f_{1,N}$ and $f_{1,S}$ correspond to $f_{1}(1285)$ and
$f_{1}(1420),$ respectively. In the charm sector, the $D_{1}$
field is chosen to correspond to the resonances $D_{1}(2420)^{0}$
and $D_{1}(2420)^{\pm}$. (Another possibility would be the not yet
very well established resonance $D_{1}(2430)^{0}$, or, due to
mixing between axial- and pseudovector states, to a mixture of
$D_{1}(2420)$ and $D_{1}(2430)$. Irrespective of this uncertainty,
the small mass difference between these states would leave our
results virtually unchanged.) The assignment of the
strange-charmed doublet $D_{S1}^{\pm}$ is not yet settled, the two
possibilities listed by the PDG are the resonances
$D_{S1}(2460)^{\pm}$ and $D_{S1}(2536)^{\pm}$ \cite{Beringer:1900zz}.
According to various studies, the latter option is favored, while
the former can be interpreted as a molecular or a tetraquark state
\cite{Bartelt:1995rq, Guo:2009id, Liu:2012zya, Guo:2006rp, Cleven:2010aw, Guo:2009ct, Gutsche:2010jf}. Thus, we assign our
quark-antiquark $D_{1}$ state to the resonance
$D_{S1}(2536)^{\pm}.$ Finally, the charm-anticharm state
$\chi_{c,1}$ can be unambiguously assigned to the charm-anticharm
resonance $\chi_{c,1}(1P)$.\\

From the matrices $V^{\mu}$ and $A^{\mu}$ we construct the
left-handed and right-handed vector fields as follows:

\begin{equation}\label{4}
L^\mu=V^\mu+A^\mu=\frac{1}{\sqrt{2}}
\left(%
\begin{array}{cccc}
  \frac{\omega_N+\rho^{0}}{\sqrt{2}}+ \frac{f_{1N}+a_1^{0}}{\sqrt{2}} & \rho^{+}+a^{+}_1 & K^{*+}+K^{+}_1 & D^{*0}+D^{0}_1 \\
  \rho^{-}+ a^{-}_1 &  \frac{\omega_N-\rho^{0}}{\sqrt{2}}+ \frac{f_{1N}-a_1^{0}}{\sqrt{2}} & K^{*0}+K^{0}_1 & D^{*-}+D^{-}_1 \\
  K^{*-}+K^{-}_1 & \overline{K}^{*0}+\overline{K}^{0}_1 & \omega_{S}+f_{1S} & D^{*-}_{S}+D^{-}_{S1}\\
  \overline{D}^{*0}+\overline{D}^{0}_1 & D^{*+}+D^{+}_1 & D^{*+}_{S}+D^{+}_{S1} & J/\psi+\chi_{C1}\\
\end{array}%
\right)^\mu,
\end{equation}
$$\\$$
\begin{equation}\label{5}
R^\mu=V^\mu-A^\mu=\frac{1}{\sqrt{2}}
\left(%
\begin{array}{cccc}
  \frac{\omega_N+\rho^{0}}{\sqrt{2}}- \frac{f_{1N}+a_1^{0}}{\sqrt{2}} & \rho^{+}-a^{+}_1 & K^{*+}-K^{+}_1 & D^{*0}-D^{0}_1 \\
  \rho^{-}- a^{-}_1 &  \frac{\omega_N-\rho^{0}}{\sqrt{2}}-\frac{f_{1N}-a_1^{0}}{\sqrt{2}} & K^{*0}-K^{0}_1 & D^{*-}-D^{-}_1 \\
  K^{*-}-K^{-}_1 & \overline{K}^{*0}-\overline{K}^{0}_1 & \omega_{S}-f_{1S} & D^{*-}_{S}-D^{-}_{S1}\\
  \overline{D}^{*0}-\overline{D}^{0}_1 & D^{*+}-D^{+}_1 & D^{*+}_{S}-D^{+}_{S1} & J/\psi-\chi_{C1}\\
\end{array}%
\right)^\mu\,,
\end{equation}

which transform under chiral transformations as
$L^{\mu}\rightarrow U_{L} L^{\mu}U_{L}^{\dag}$ and
$R^{\mu}\rightarrow U_{R} R^{\mu}U_{R}^{\dag}.$\\

The Lagrangian of the $N_f=4$ model with global chiral invariance
has an analogous form as the corresponding eLSM Lagrangian for
$N_f=3$ \cite{Parganlija:2012fy, Parganlija:2012xj}, which is discussed in Sec. 2.6 and described in Eq.(\ref{fulllag}), with the additional mass term $$-2\,
\text{Tr}[E\Phi^\dagger\Phi]$$ which has been added to
account for the mass of the charm quark, as well as to obtain a better fit to the masses.\\

The terms involving the matrices $H,\,E,$ and $\Delta$
break the dilatation symmetry explicitly, because
they involve dimensionful coupling constants, and chiral
symmetry due to nonzero current quark masses in the (pseudo)scalar
and (axial-)vector sectors. They are of
particular importance when the charmed mesons are considered,
because the charm quark mass is large. In the light sectors, these
terms are surely subleading when the quarks $u$ and $d$ are
considered (unless one is studying some particular
isospin-breaking processes), while the quark $s$ is somewhat on
the border between light and heavy. We describe these terms separately:\\

(i) The term $\mathrm{Tr}[H(\Phi+\Phi^{\dagger})]$ with%

\begin{equation}
H=\frac{1}{2}\left(
\begin{array}
[c]{cccc}%
h_{U} & 0 & 0 & 0\\
0 & h_{D} & 0 & 0\\
0 & 0 & \sqrt{2}h_{S} & 0\\
0 & 0 & 0 & \sqrt{2}h_{C}%
\end{array}
\right)  \, , \label{h}%
\end{equation}
describes the usual explicit symmetry breaking (tilting of the
Mexican-hat potential). The constants $h_{i}$ are proportional to
the current quark masses, $h_{i}\propto m_{i}.$ Here we work in
the isospin limit, $h_{U}=h_{D}=h_{N}.$ The pion mass, for
instance, turns out to be $m_{\pi}^{2}\propto m_{u},$ in agreement
with the Gell-Mann--Oakes--Renner (GOR) relation \cite{GellMann:1968rz}. The
parameter $h_{C}$ is one of the three new parameters entering the
$N_{f}=4$ version of the model when compared to the $N_{f}=3$ case
of Ref.\ \cite{Parganlija:2012fy}.\\

(ii) The term $-2\,\mathrm{Tr}[E\Phi^{\dagger}\Phi]$
with
\begin{equation}
E=\left(
\begin{array}
[c]{cccc}%
\varepsilon_{U} & 0 & 0 & 0\\
0 & \varepsilon_{D} & 0 & 0\\
0 & 0 & \varepsilon_{S} & 0\\
0 & 0 & 0 & \varepsilon_{C}%
\end{array}
\right)  \text{ ,} \label{8}%
\end{equation}
where $\varepsilon_{i}\propto m_{i}^{2}$, is the next-to-leading
order correction in the current quark-mass expansion. In the
isospin-symmetric limit
$\varepsilon_{U}=\varepsilon_{D}=\varepsilon_{N}$ one can subtract
from $\varepsilon$ a matrix proportional to the identity in such a
way that the parameter $\varepsilon_{N}$ can be absorbed in the
parameter $m_{0}^{2}.$ Thus, without loss of generality we can set
$\varepsilon_{N}=0.$ Following Ref.\ \cite{Parganlija:2012fy}, for the sake of
simplicity we shall here also set $\varepsilon_{S}=0,$ while we
keep $\varepsilon_{C}$ nonzero. This is the second additional
parameter with respect to Ref.\ \cite{Parganlija:2012fy}.\\

(iii) The term $\mathrm{Tr}\left[
\Delta(L^{\mu}{}^{2}+R^{\mu}{}^{2})\right] $ with
\begin{equation}
\Delta=\left(
\begin{array}
[c]{cccc}%
\delta_{U} & 0 & 0 & 0\\
0 & \delta_{D} & 0 & 0\\
0 & 0 & \delta_{S} & 0\\
0 & 0 & 0 & \delta_{C}%
\end{array}
\right)  \,, \label{7}%
\end{equation}
where $\delta_{i}\sim m_{i}^{2}$, describes the current quark-mass
contribution to the masses of the (axial-)vector mesons. Also in
this case, in the
isospin-symmetric limit it is possible to set $\delta_{U}=\delta_{D}%
=\delta_{N}=0$ because an identity matrix can be absorbed in the
term proportional to $m_{1}^{2}.$ The parameter $\delta_{S}$ is
taken from Ref.\ \cite{Parganlija:2012fy}. The third new parameter with respect
to Ref.\ \cite{Parganlija:2012fy} is $\delta_{C}.$ Note that in the present
effective model the mass parameters $\delta_{C}$ and
$\varepsilon_{C}$ are not to be regarded as the second-order
contribution in an expansion in powers of $m_{C}$. They simply
represent the direct, and in this case dominant, contribution
$\sim m^2_C$ of a charm quark to the masses of charmed
(pseudo)scalar and (axial-)vector mesons.

Another important term in the Lagrangian (\ref{fulllag}) is $c(\text{det}\Phi
-\text{det}\Phi^{\dagger})^{2},$ which is
responsible for the large $\eta^{\prime}$ mass. Care is needed,
because a determinant changes when the number of flavours changes.\\

We conclude this section with a few remarks on how to extend
the Lagrangian (\ref{fulllag}) in order to improve the description of hadron vacuum
properties. First of all, note that the requirement of dilatation invariance restricts 
the interaction terms in the Lagrangian to have naive scaling dimension four: 
higher-order dilatation-invariant terms would have to contain inverse powers of $G$, 
and thus would be non-analytic in this field. In this sense, our Lagrangian is
complete and cannot be systematically improved by the inclusion
of higher-order interaction terms, such as in theories with nonlinearly realized
chiral symmetry. However, one may add further terms that violate dilatation
invariance (which is already broken by the mass terms $\sim H, \varepsilon, \Delta$,
and the $U(1)_A$--violating term $\sim c$) to improve the model. 

Another possibility is to sacrifice chiral symmetry. For instance, since the
explicit breaking of chiral symmetry by the charm quark mass is large (which
is accounted for by the terms $\sim h_C, \varepsilon_C, \delta_C$), one could
also consider chiral-symmetry violating interaction terms, e.g.\ replace
\begin{equation} \label{U4symbreak}
\lambda_{2}\mathrm{Tr}(\Phi^{\dagger}  \Phi)^{2}
\longrightarrow \lambda_{2}\mathrm{Tr}(\Phi^{\dagger}  \Phi)^{2}
+ \delta \lambda_2 \mathrm{Tr}( \mathbb{P}_C \Phi^{\dagger} \Phi)^{2}
\end{equation}
where $\mathbb{P}_C = { diag}\{0,0,0,1\}$ is a projection operator onto the
charmed states. A value $\delta\lambda_{2}\neq 0$ 
explicitly breaks the symmetry of this interaction term from 
$U_{R}(4)\times U_{L}(4)$ to $U_{R}(3)\times U_{L}(3)$. 
One could modify the interaction terms proportional to $\lambda_1$,
$c$, $g_{1}$, $g_{2},$ $h_{1},$ $h_{2},$
and $h_{3}$ in Eq.\ (\ref{fulllag}) in a similar manner.

\section{Four-flavour linear sigma model implications}

The Lagrangian (\ref{fulllag}) induces spontaneous symmetry breaking if $m_{0}%
^{2}<0$ : as a consequence, the scalar-isoscalar fields $G,\sigma_{N}%
,\,\sigma_{S},$ and $\chi_{C0}$ develop nonzero vacuum expectation
values. One has to perform the shifts as
\begin{equation}
G_0\rightarrow G+G_0,\,\,\,
\sigma_N\rightarrow\sigma_N+\phi_N,\,\,\,
\sigma_S\rightarrow\sigma_S+\phi_S\;,
\end{equation}
as obtained in Eq.(\ref{shift1}) and in Refs.\ \cite{Parganlija:2012xj}, and
similarly for $\chi_{C0}$,
\begin{equation}\label{10}
\chi_{C0}\rightarrow\chi_{C0}+\phi_C\;,
\end{equation}
to implement this breaking. The quantity $G_0$ is proportional to the gluon condenstate \cite{Janowski:2011gt}, while the quantities $\phi_{N},$
$\phi_{S},$ and $\phi_{C}$ correspond to the nonstrange, strange
and charm quark-antiquark condensates.

The relations between the nonstrange, strange, and charm
condensates with the pion decay constant $f_\pi$ and the kaon decay
constant $f_K$ are presented in  Eq.(\ref{phin}) and  Eq. (\ref{phis}), respectively, whereas the decay constants of the pseudoscalar
$D,\,D_S,\,and\, \eta_C$ mesons, $f_{D},
\,f_{D_{s}},\,\text{and}\,f_{\eta_C}$ are
\begin{equation}
f_{D}=\frac{\phi_{N}+\sqrt{2}\phi_{C}}{\sqrt{2}Z_{D}}\text{ },\text{ }%
\end{equation}
\begin{equation}
f_{D_{S}}=\frac{\phi_{S}+\phi_{C}}{Z_{D_{S}}}\text{ },\text{ }
\end{equation}
\begin{equation}
f_{\eta_{c}}=\frac{2\phi_{C}}{Z_{\eta_{C}}}.
\end{equation}

where the detailed calculation is presented in\ Appendices A.3 and A.4. The chiral condensates $\phi_{N,S,C}$ lead to the mixing between
(axial-)vector and (pseudo)scalar states in the Lagrangian
(\ref{fulllag}), with the additional term $-2\,\mathrm{Tr}[E\Phi^{\dagger}\Phi]$, with bilinear mixing terms involving the light
mesons $\eta_N-f_{1N}$,
$\overrightarrow{\pi}-\overrightarrow{a}_1$ \cite{Parganlija:2012xj}, $\eta_S-f_{1S}$,
$K_S-K^*$, and $K-K_1$, which are presented in Eq.(\ref{mixing1}). In addition, for charmed mesons similar mixing terms of the type
$D-D_1$, $D^*_0-D^*$, $D_S-D_{S1}$, $D^*_{S0}-D^*_{S}$, and
$\eta_C-\chi_{C1}$ are present:
\begin{align}
&-g_{1}\phi_{C}\,\chi_{C1}^{\mu}\,\partial_{\mu}\eta_{C}-\frac{g_{1}}{\sqrt{2}}\,g_{1}
\phi_{S}(D_{S1}^{\mu-}\,\partial_{\mu}D_{S}^{+}+D_{S1}^{\mu+}\,\partial_{\mu
}D_{S}^{-})\nonumber\\
&+i\frac{g_{1}}{\sqrt{2}}\,\phi_{S}(D_{S}^{\ast\mu-}\,\partial_{\mu}%
D_{S0}^{\ast+}-D_{S}^{\ast\mu+}\,\partial_{\mu}D_{S0}^{\ast-})\nonumber\\
&+\,i\frac{g_1}{2}\,\phi_N(D^{*\mu-}\,\partial_\mu
D^{*+}_0-D^{*\mu+}\,
\partial_\mu D^{*-}_0+ D^{*\mu 0}\,\partial_\mu \overline{D}\,^{*0}_0-\overline{D}\,^{*\mu 0}\,\partial_\mu D^{*0}_0)\nonumber\\
&-\frac{g_1}{2}\,\phi_N(D^{0\mu}_1\,
\partial_\mu \overline{D}^0+\overline{D}\,^{\mu 0}_1\,
\partial_\mu D^0+D^{\mu
+}_1\, \partial_\mu D^-+D^{\mu -}_1\, \partial_\mu D^{+})\text{
.}\label{mixing2}
\end{align}

Note that the Lagrangian (\ref{fulllag}) is real despite the imaginary
$K_S-K^*$, $D^*_{S0}-D^*_{S1}$, $D_S-D_{S1}$ and $D^*_0-D^*$ coupling because
these mixing terms are equal to their Hermitian conjugates.

The mixing terms (\ref{mixing1}) are removed by performing field
transformations of the (axial-)vector states as presented in Eqs.(\ref{eq:shifts1}-\ref{eq:shifts4}). The mixing
terms (\ref{mixing2}) are removed by performing field
transformations of the (axial-)vector states as follows
\begin{align}
&\chi_{C1}^{\mu}\rightarrow\chi_{C1}^{\mu}+w_{\chi_{C1}}\,Z_{\eta_{C}}
\,\partial^{\mu}\eta_{C}\,,\label{shhh1}\\
&D_{S1}^{\mu\pm}\rightarrow D_{S1}^{\mu\pm }+w_{D_{S1}}\,Z_{D_{S}}
\,\partial^{\mu}D_{S}^{\pm}\>, \\
 &D^{*\mu-}_{S}\rightarrow
D^{*\mu-}_{S}+w_{D^{*}_{S}}\,Z_{D^{*}_{S0}}
\,\partial^{\mu}D^{*-}_{S0}\,,\\
&D^{*\mu+}_{S}\rightarrow D^{*\mu+}%
_{S}+w^{*}_{D^{*}_{S}}\,Z_{D^{*}_{S0}}
\,\partial^{\mu}D^{*+}_{S0}\>,\\
 &D^{*\mu+}\rightarrow
D^{*\mu+}+w^{*}_{D^{*}}\,Z_{D^{*}_{0}}
\,\partial^{\mu}D^{*+}_{0}\,,\\
&D^{*\mu-}\rightarrow D^{*\mu-}+w_{D^{*}%
}\,Z_{D^{*}_{0}} \,\partial^{\mu}D^{*-}_{0}\>,\\
&\overline{D}\,^{*\mu0}\rightarrow\overline{D}\,^{*\mu0}%
+w^{*}_{D^{*0}}\,Z_{D^{*0}_{0}}
\,\partial^{\mu}\overline{D}\,^{*0}_{0}\,,\\
&D^{*\mu0}\rightarrow D^{*\mu0}+w_{D^{*0}}\,Z_{D^{*0}_{0}}
\,\partial^{\mu}D^{*0}_{0}\>,\\
&D_{1}^{\mu\pm,0,\bar{0}}\rightarrow
D_{1}^{\mu\pm,0,\bar{0}}+w_{D_{1}}\,Z_{D} \,\partial^{\mu
}D^{\pm,0,\bar{0}}\>.
\end{align}

These shifts produce additional kinetic terms for the open and
hidden (pseudo)scalar charmed fields. Furthermore, one has to
rescale the strange-nonstrange (pseudo)scalar fields as
described in Eqs.(\ref{eq:shifts5}-\ref{eq:shifts5}) as well as
the open and hidden charmed pseudoscalar fields as
\begin{align}
& D^{\pm,0,\bar{0}}\rightarrow Z_{D}D^{\pm,0,\bar{0}}\,,\\
& D^{*\pm}_{0}\rightarrow Z_{D^{*}_{0}}D^{*\pm}_{0}\>,\\
& D_{0}^{\ast0,\bar{0}}\rightarrow Z_{D_{0}^{\ast0}}D_{0}^{\ast0,\bar{0}}\,,\\
& D_{S0}^{\ast\pm}\rightarrow Z_{D_{S0}^{\ast}}D_{S0}^{\ast\pm
}\text{ ,}\\
& \eta_{C}\rightarrow Z_{\eta_{C}}\eta_{C}\,.\label{shhhf}
\end{align}
Note that for the sake of simplicity we have grouped together the
isodoublet states with the notation $D^{\pm,0,\bar{0}},\,\text{and}\,
D_{0}^{\ast0,\bar{0}}$,  where $\bar{0}$ refers to $\bar{D}^{0}$
and $\bar{D}^{\ast 0}_0$. The coefficients $w_{i}$ and $Z_{i}$ are
determined in order to eliminate the mixing terms and to obtain
the canonical normalization of the $D$, $D_{0}^{\star}$, $D_{S0}^{\star}$, and $\eta_{C}$ fields. This is described next.\\
The quantities
$w_{f_{1N}},\,w_{a_{1}},$ $w_{f_{1S}},\,w_{K^{\ast}},\,w_{K_{1}}$, are given in Eqs. (\ref{eq:shifts1}-\ref{eq:shifts4}), while $w_{\chi_{C1}},$ 
$w_{D_{S1}},$ $w_{D_{S}^{\ast}},\,w_{D^{\ast}},$
$w_{D^{\ast0}},$ and $w_{D_{1}}$ are calculated from the condition
that the mixing terms (\ref{mixing2}) vanish once the shifts of
the (axial-)vectors have been implemented:
\begin{multline}
\bigg[-g_1\,\phi_N+\frac{1}{2}(2\,g_1^2+h_1+h_2-h_3)w_{f_{1N}}\,\phi_N^2+(m_1^2+2\delta_N)w_{f_{1N}}+\frac{h_1}{2}w_{f_{1N}}\,\phi_S^2+\frac{h_1}{2}w_{f_{1N}}\,\phi_C^2\bigg]\\
\times(f_{1N}^{\mu}\partial_{\mu}\eta_{N}+\overrightarrow{a}%
_{1}^{\mu}\cdot\partial_{\mu}\overrightarrow{\pi})=0\,,\,\,\,\,\,\,\,\,\,\,\,\,\,\,\,\,\,\,\,\,\,\,\,\,\,\,\,\,\,\,\,\,\,\,\,\,\,\,\,
\,\,\,\,\,\,\,\,\,\,\,\,\,\,\,\,\,\,\,\,\,\,\,\,\,\,\,\,\,\,\,\,\,\,\,\,\,\,\,
\,\,\,\,\,\,\,\,\,\,\,\,\,\,\,\,\,\,\,\,\,\,\,\,\,\,\,\,\,\,\,\,\,\,\,\,\,\,\,\,\,\,\,\,\,\,\,\,\,\,\,\,\,\,\,\,\,\,\,\,\,\,\,\,\,\label{SNSmix1}
\end{multline}
\begin{multline}
\bigg[-\sqrt{2}g_1\,\phi_S+(2\,g_1^2+\frac{h_1}{2}+h_2-h_3)w_{f_{1S}}\,\phi_S^2+(m_1^2+2\delta_S)w_{f_{1S}}+\frac{h_1}{2}w_{f_{1S}}\,\phi_N^2+\frac{h_1}{2}w_{f_{1S}}\,\phi_C^2\bigg]\\
\times f_{1S}^{\mu}\partial_{\mu}\eta_{S}=0\,,\,\,\,\,\,\,\,\,\,\,\,\,\,\,\,\,\,\,\,\,\,\,\,\,\,\,\,\,\,\,\,\,\,\,\,\,\,\,\,
\,\,\,\,\,\,\,\,\,\,\,\,\,\,\,\,\,\,\,\,\,\,\,\,\,\,\,\,\,\,\,\,\,\,\,\,\,\,\,
\,\,\,\,\,\,\,\,\,\,\,\,\,\,\,\,\,\,\,\,\,\,\,\,\,\,\,\,\,\,\,\,\,\,\,\,\,\,\,\,\,\,\,\,\,\,\,\,\,\,\,\,\,\,\,\,\,\,\,\,\,\,\,\,\,\,\,\,\,\,\,\,\,\,\,\,\,\,
\,\,\,\,\,
\end{multline}
\begin{multline}
\bigg[i\frac{g_1}{\sqrt{2}}\phi_S+\frac{h_3-g_1^2}{\sqrt{2}}w_{K^*}\phi_N\phi_S+\frac{g_1^2+h_1+h_2}{2}w_{K^*}\phi_S^2+(m_1^2+\delta_N+\delta_S)w_{K^*}-i\frac{g_1}{2}\phi_N\\+\frac{1}{4}
(g_1^2+2h_1+h_2)w_{K^*}\phi_N^2+\frac{h_1}{2}w_{K^*}\phi_C^2\bigg](\overline{K}\,^{\ast\mu0}\,\partial_{\mu}K_{S}^{0}+K^{\ast\mu-}\,\partial_{\mu}K_{S}^{+})=0\,,\,\,\,\,\,\,\,\,\,\,\,\,\,\,\,\,\,
\end{multline}
\begin{multline}
\bigg[-i\frac{g_1}{\sqrt{2}}\phi_S+\frac{h_3-g_1^2}{\sqrt{2}}w^*_{K^*}\phi_N\phi_S+\frac{g_1^2+h_1+h_2}{2}w^*_{K^*}\phi_S^2+(m_1^2+\delta_N+\delta_S)w^*_{K^*}+i\frac{g_1}{2}\phi_N\\+\frac{1}{4}
(g_1^2+2h_1+h_2)w^*_{K^*}\phi_N^2+\frac{h_1}{2}w^*_{K^*}\phi_C^2\bigg](K^{\ast\mu0}\,\partial_{\mu}\overline
{K}_{S}^{0}+K^{\ast\mu+}\,\partial_{\mu}K_{S}^{-})=0\,,\,\,\,\,\,\,\,\,\,\,\,\,\,\,\,\,
\end{multline}
\begin{multline}
\bigg[-\frac{g_1}{\sqrt{2}}\phi_S+\frac{g_1^2-h_3}{\sqrt{2}}w_{K_1}\phi_N\phi_S+\frac{g_1^2+h_1+h_2}{2}w_{K_1}\phi_S^2+(m_1^2+\delta_N+\delta_S)w_{K_1}-\frac{g_1}{2}\phi_N\\+\frac{1}{4}
(g_1^2+2h_1+h_2)w_{K_1}\phi_N^2+\frac{h_1}{2}w_{K_1}\phi_C^2\bigg](K_{1}^{\mu0}\,\partial_{\mu}\overline{K}^{0}+K_{1}^{\mu
+}\,\partial_{\mu}K^{-}+\overline{K}_{1}^{\mu0}\,\partial_{\mu
}K^{0}+K_{1}^{\mu-}\,\partial_{\mu}K^{+})=0\,,\label{SNSmixf}
\end{multline}
\begin{multline}
\bigg[-\frac{g_1}{\sqrt{2}}\phi_S+(g_1^2-h_3)w_{D_{S1}}\phi_S\phi_C+\frac{g_1^2+h_1+h_2}{2}w_{D_{S1}}\phi_S^2+(m_1^2+\delta_S+\delta_C)w_{D_{S1}}-\frac{g_1}{\sqrt{2}}\phi_C\\+\frac{g_1^2+h_1+h_2}{2}
w_{D_{S1}}\phi_C^2+\frac{h_1}{2}w_{D_{S1}}\phi_N^2\bigg](D_{S1}^{\mu-}\,\partial_{\mu}D_{S}^{+}+D_{S1}^{\mu+}\,\partial_{\mu
}D_{S}^{-})=0\,,\,\,\,\,\,\,\,\,\,\,\,\,\,\,\,\,\,\,\,\,\,\,\,\,\,\,\,
\end{multline}
\begin{multline}
\bigg[-i\frac{g_1}{\sqrt{2}}\phi_S+(h_3-g_1^2)w_{D^*_{S1}}\phi_S\phi_C+\frac{g_1^2+h_1+h_2}{2}w_{D^*_{S1}}\phi_S^2+(m_1^2+\delta_S+\delta_C)w_{D^*_{S1}}+i\frac{g_1}{\sqrt{2}}\phi_C\\
+\frac{g_1^2+h_1+h_2}{2}w_{D^*_{S1}}\phi_C^2+\frac{h_1}{2}w_{D^*_{S1}}\phi_N^2\bigg]D_{S1}^{\ast\mu+}\,\partial_{\mu}D_{S0}^{\ast-}=0\,,\,\,\,\,\,\,\,\,\,\,\,\,\,\,\,\,\,\,\,\,\,\,\,\,\,\,\,\,\,\,\,\,\,\,\,\,\,\,\,\,\,\,\,\,\,\,\,\,\,\,\,\,\,\,\,\,\,\,\,\,\,\,\,
\end{multline}
\begin{multline}
\bigg[i\frac{g_1}{\sqrt{2}}\phi_S+(h_3-g_1^2)w^\ast_{D^*_{S1}}\phi_S\phi_C+\frac{g_1^2+h_1+h_2}{2}w^\ast_{D^*_{S1}}\phi_S^2+(m_1^2+\delta_S+\delta_C)w^\ast_{D^*_{S1}}-i\frac{g_1}
{\sqrt{2}}\phi_C\\+\frac{g_1^2+h_1+h_2}{2}w^\ast_{D^*_{S1}}\phi_C^2+\frac{h_1}{2}w^\ast_{D^*_{S1}}\phi_N^2\bigg]D_{S1}^{\ast\mu-}\,\partial_{\mu}D_{S0}^{\ast+}=0\,,\,\,\,\,\,\,\,\,\,\,\,\,\,\,\,\,\,\,\,\,\,\,\,\,\,\,\,\,\,\,\,\,\,\,\,\,\,\,\,\,\,\,\,\,\,\,\,\,\,\,\,\,\,\,\,\,\,\,\,\,\,\,\,
\end{multline}
\begin{multline}
\bigg[-i\frac{g_1}{2}\phi_N+\frac{h_3-g_1^2}{\sqrt{2}}w_{D^*}\phi_N\phi_C+\frac{1}{4}(g_1^2+2h_1+h_2)w_{D^\ast}\phi_N^2+(m_1^2+\delta_N+\delta_C)w_{D^*}+i\frac{g_1}{\sqrt{2}}\phi_C\\
+\frac{g_1^2+h_1+h_2}{2}w_{D^*}\phi_C^2+\frac{h_1}{2}w_{D^*}\phi_S^2\bigg]D^{\ast\mu
+}\,\partial_{\mu}D_{0}^{\ast-}=0\,,\,\,\,\,\,\,\,\,\,\,\,\,\,\,\,\,\,\,\,\,\,\,\,\,\,\,\,\,\,\,\,\,\,\,\,\,\,\,\,\,\,\,\,\,\,\,\,\,\,\,\,\,\,\,\,\,\,\,\,\,\,\,\,\,\,\,
\end{multline}
\begin{multline}
\bigg[i\frac{g_1}{2}\phi_N+\frac{h_3-g_1^2}{\sqrt{2}}w^\ast_{D^*}\phi_N\phi_C+\frac{1}{4}(g_1^2+2h_1+h_2)w^\ast_{D^\ast}\phi_N^2+(m_1^2+\delta_N+\delta_C)w^\ast_{D^*}-i\frac{g_1}{\sqrt{2}}\phi_C\\
+\frac{g_1^2+h_1+h_2}{2}w^\ast_{D^*}\phi_C^2+\frac{h_1}{2}w^\ast_{D^*}\phi_S^2\bigg]D^{\ast\mu
-}\,\partial_{\mu}D_{0}^{\ast+}=0\,,\,\,\,\,\,\,\,\,\,\,\,\,\,\,\,\,\,\,\,\,\,\,\,\,\,\,\,\,\,\,\,\,\,\,\,\,\,\,\,\,\,\,\,\,\,\,\,\,\,\,\,\,\,\,\,\,\,\,\,\,\,\,\,\,\,\,
\end{multline}
\begin{multline}
\bigg[-i\frac{g_1}{2}\phi_N+\frac{h_3-g_1^2}{\sqrt{2}}w_{D^{\ast 0}}\phi_N\phi_C+\frac{1}{4}(g_1^2+2h_1+h_2)w_{D^{\ast 0}}\phi_N^2+(m_1^2+\delta_N+\delta_C)w_{D^{\ast0}}+i\frac{g_1}{\sqrt{2}}\phi_C\\
+\frac{g_1^2+h_1+h_2}{2}w_{D^{*0}}\phi_C^2+\frac{h_1}{2}w_{D^{*0}}\phi_S^2\bigg]\overline{D}\,^{\ast\mu0}%
\,\partial_{\mu}D_{0}^{\ast0}=0\,,\,\,\,\,\,\,\,\,\,\,\,\,\,\,\,\,\,\,\,\,\,\,\,\,\,\,\,\,\,\,\,\,\,\,\,\,\,\,\,\,\,\,\,\,\,\,\,\,\,\,\,\,\,\,\,\,\,\,\,\,\,\,\,\,\,\,
\end{multline}
\begin{multline}
\bigg[i\frac{g_1}{2}\phi_N+\frac{h_3-g_1^2}{\sqrt{2}}w^\ast_{D^{\ast
0}}\phi_N\phi_C+\frac{1}{4}(g_1^2+2h_1+h_2)w^\ast_{D^{\ast
0}}\phi_N^2+(m_1^2+\delta_N+\delta_C)w^\ast_{D^{\ast0}}
-i\frac{g_1}{\sqrt{2}}\phi_C\\+\frac{g_1^2+h_1+h_2}{2}w^\ast_{D^{*0}}\phi_C^2+\frac{h_1}{2}w^\ast_{D^{*0}}\phi_S^2\bigg]D^{\ast\mu0}%
\,\partial_{\mu}\overline{D}\,_{0}^{\ast0}=0\,,\,\,\,\,\,\,\,\,\,\,\,\,\,\,\,\,\,\,\,\,\,\,\,\,\,\,\,\,\,\,\,\,\,\,\,\,\,\,\,\,\,\,\,\,\,\,\,\,\,\,\,\,\,\,\,\,\,\,\,\,\,\,\,\,\,\,
\end{multline}
\begin{multline}
\bigg[-\frac{g_1}{2}\phi_N+\frac{g_1^2-h_3}{\sqrt{2}}w_{D_1}\phi_N\phi_C+\frac{1}{4}(g_1^2+2h_1+h_2)w_{D_1}\phi_N^2+(m_1^2+\delta_N+\delta_C)w_{D_1}-\frac{g_1}{\sqrt{2}}\phi_C\\
+\frac{g_1^2+h_1+h_2}{2}w_{D_1}\phi_C^2+\frac{h_1}{2}w_{D_1}\phi_S^2\bigg](D_{1}^{0\mu}\,\partial_{\mu}\overline{D}^{0}+\overline
{D}\,_{1}^{\mu0}\,\partial_{\mu}D^{0}+D_{1}^{\mu+}\,\partial_{\mu}D^{-}%
+D_{1}^{\mu-}\,\partial_{\mu}D^{+})=0 \,,
\end{multline}
\begin{multline}
\bigg[-\sqrt{2}\,g_1\,\phi_C+(2g_1^2+\frac{h_1}{2}+h_2-h_3)w_{\chi_{c1}}\phi_C^2+(m_1^2+2\delta_C)w_{\chi_{c1}}+\frac{h_1}{2}w_{\chi_{c1}}\phi_N^2+\frac{h_1}{2}w_{\chi_{c1}}\phi_S^2\bigg]\\
\times \chi_{C1}^{\mu}\,\partial_{\mu}\eta_{C}=0\,,\,\,\,\,\,\,\,\,\,\,\,\,\,\,\,\,\,\,\,\,\,\,\,\,\,\,\,\,\,\,\,\,\,\,\,\,\,\,\,\,\,\,\,\,\,\,\,\,\,\,\,\,\,\,\,\,\,\,\,\,\,\,\,\,\,\,\,\,\,\,\,\,\,\,\,\,\,\,\,\,\,\,\,\,\,\,\,\,\,\,\,\,\,\,\,\,\,\,\,\,\,\,\,\,\,\,\,\,\,\,\,\,\,\,\,\,\,\,\,\,\,\,\,\,\,\,\,\,\,\,\,\,\,\,\,\,\,\,\,\,\,\,\,\,\,\,\,\,\,\,\,\,\,\,\,\,\,\,\label{SNSmixDf}
\end{multline}
Equations (\ref{SNSmix1}) - (\ref{SNSmixf}) correspond to the mixing terms (\ref{mixing1}) obtained in the case of
strange-nonstrange investigation \cite{Parganlija:2012xj}. Equations
(\ref{SNSmix1})-(\ref{SNSmixDf}) are fulfilled only if we define
\begin{align}
& w_{\chi_{C1}}=\frac{\sqrt{2}g_{1}\phi_{C}}{m_{\chi_{C1}}^{2}}\,,\\
& w_{D_{S1}}=\frac{g_{1}(\phi_{S}+\phi_{C})}{\sqrt{2}m_{D_{S1}}^{2}}\>,\\
& w_{D_{S}^{\ast}}=\frac{ig_{1}(\phi_{S}-\phi_{C})}{\sqrt{2}m_{D_{S}^{\ast}}^{2}}\,,\\
& w_{D^{\ast}}=\frac{ig_{1}(\phi_{N}-\sqrt{2}\phi_{C})}{2\,m_{D^{\ast}}^{2}}\>,\\
& w_{D^{\ast0}}=\frac{ig_{1}(\phi_{N}-\sqrt{2}\phi_{C})}{2\,m_{D^{\ast0}}^{2}}\,,\\
& w_{D_{1}}=\frac{g_{1}(\phi_{N}+\sqrt{2}\phi_{C})}{2\,m_{D_{1}}^{2}}\>. %
\end{align}

The wave-function renormalization constants of the strange-nonstrange fields are given in Eqs.(\ref{zpi}-\ref{zets}) \cite{Parganlija:2012fy, Parganlija:2012xj}. For the charmed fields, they read
\begin{align}
& Z_{\eta_{C}}=\frac{m_{\chi_{C1}}}{\sqrt{m_{\chi_{C1}}^{2}-2g_{1}^{2}\phi_{C}^{2}}}\,,\label{zetc}\\
& Z_{D_{S}}=\frac{\sqrt{2}m_{D_{S1}}}{\sqrt{2m_{D_{S1}}^{2}-g_{1}^{2}(\phi_{S}+\phi_{C})^{2}}}\>, \\
& Z_{D_{S0}^{\ast}}=\frac{\sqrt{2}m_{D_{S}^{\ast}}}{\sqrt{2m_{D_{S}^{\ast}}^{2}-g_{1}^{2}(\phi_{S}-\phi_{C})^{2}}}\text{ ,}\,\\
& Z_{D_{0}^{\ast}}=\frac{2\,m_{D^{\ast}}}{\sqrt{4m_{D^{\ast}}^{2}-g_{1}^{2}(\phi_{N}-\sqrt{2}\phi_{C})^{2}}}\>, \\
& Z_{D_{0}^{\ast0}}=\frac{2\,m_{D^{\ast0}}}{\sqrt{4m_{D^{\ast0}}^{2}-g_{1}^{2}(\phi_{N}-\sqrt{2}\phi_{C})^{2}}}\,,\\
& Z_{D}=\frac{2\,m_{D_{1}}}{\sqrt{4m_{D_{1}}^{2}-g_{1}^{2}(\phi_{N}+\sqrt{2}\phi_{C})^{2}}}\,.\label{zf1}%
\end{align}
It is obvious from Eqs.(\ref{zpi}-\ref{zets}) and Eqs.(\ref{zetc})-(\ref{zf1}) that all the renormalization coefficients will have values larger than one.

\section{Tree-level masses}

In this section we present the squared meson masses of all mesons
in the model after having performed the transformation above as
well as transforming the unphysical fields to the physical fields, e.g. for the
pseudoscalar mesons $\eta_N,\, \eta_S$ to $\eta,\, \eta'$, and the
scalar mesons $\sigma_N,\, \sigma_S$ to $\sigma_1,\, \sigma_2$ due
to the mixing terms of these fields. \\
We obtain the tree-level masses of nonstrange-strange mesons in
the eLSM:

(i) Pseudoscalar mesons:
\begin{align}
m_{\pi}^{2}  & =Z_{\pi}^{2}\bigg[m_{0}^{2}+\bigg(\lambda_{1}+\frac
{\lambda_{2}}{2}\bigg)\,\phi_{N}^{2}+\lambda_{1}\,\phi_{S}^{2}+\lambda
_{1}\,\phi_{C}^{2}\bigg]\>,\label{45}\\
m_{K}^{2}  &  =Z_{K}^{2}\bigg[m_{0}^{2}+\bigg(\lambda_{1}+\frac{\lambda_{2}%
}{2}\bigg)\,\phi_{N}^{2}-\frac{\lambda_{2}}{\sqrt{2}}\,\phi_{N}\,\phi
_{S}+(\lambda_{1}+\lambda_{2})\,\phi_{S}^{2}+\lambda_{1}\,\phi_{C}%
^{2}\bigg]\>,\label{46}\\
m_{\eta_{N}}^{2}  &  =Z_{\pi}^{2}\bigg[m_{0}^{2}+\bigg(\lambda_{1}%
+\frac{\lambda_{2}}{2}\bigg)\,\phi_{N}^{2}+\lambda_{1}\,\phi_{S}^{2}%
+\lambda_{1}\,\phi_{C}^{2}+\frac{c}{2}\,\phi_{N}^{2}\,\phi_{S}^{2}\,\phi
_{C}^{2}\bigg]\>,\label{47}\\
m_{\eta_{S}}^{2}  &
=Z_{\eta_{S}}^{2}\bigg[m_{0}^{2}+\lambda_{1}\,\phi
_{N}^{2}+(\lambda_{1}+\lambda_{2})\,\phi_{S}^{2}+\lambda_{1}\,\phi_{C}%
^{2}+\frac{c}{8}\,\phi_{N}^{4}\,\phi_{C}^{2}\bigg]\>. \label{48}%
\end{align}

(ii) Scalar mesons:
\begin{align}
m_{a_{0}}^{2}  &  =m_{0}^{2}+\bigg(\lambda_{1}+\frac{3}{2}\lambda
_{2}\bigg)\phi_{N}^{2}+\lambda_{1}\,\phi_{S}^{2}+\lambda_{1}\,\phi_{C}%
^{2}\>,\label{41}\\
m_{K_{0}^{\ast}}^{2}  &
=Z_{K_{0}^{\ast}}^{2}\bigg[m_{0}^{2}+\bigg(\lambda
_{1}+\frac{\lambda_{2}}{2}\bigg)\,\phi_{N}^{2}+\frac{\lambda_{2}}{\sqrt{2}%
}\,\phi_{N}\,\phi_{S}+(\lambda_{1}+\lambda_{2})\,\phi_{S}^{2}+\lambda
_{1}\,\phi_{C}^{2}\bigg]\>,\label{42}\\
m_{\sigma_{N}}^{2}  &  =m_{0}^{2}+3\bigg(\lambda_{1}+\frac{\lambda_{2}}%
{2}\bigg)\,\phi_{N}^{2}+\lambda_{1}\,\phi_{S}^{2}+\lambda_{1}\,\phi_{C}%
^{2}\>,\label{43}\\
m_{\sigma_{S}}^{2}  & =m_{0}^{2}+\lambda_{1}\phi_{N}^{2}+3(\lambda
_{1}+\lambda_{2})\,\phi_{S}^{2}+\lambda_{1}\,\phi_{C}^{2}\>\text{
.}
\label{44}%
\end{align}

(iii) Vector mesons:%
\begin{align}
m_{\rho}^{2} = & m_{\omega_{N}}^{2}\;,\\
m_{\omega_{N}}^{2} = & m_{1}^{2}+2\,\delta_{N}+\frac{\phi_{N}^{2}}{2}%
\,(h_{1}+h_{2}+h_{3})+\frac{h_{1}}{2}\,\phi_{S}^{2}+\frac{h_{1}}{2}\,\phi
_{C}^{2}\;,\label{m_rho}\\
m_{\omega_{S}}^{2}  = & m_{1}^{2}+2\,\delta_{S}+\frac{h_{1}}{2}\,\phi_{N}%
^{2}+\bigg(\frac{h_{1}}{2}+h_{2}+h_{3}\bigg)\,\phi_{S}^{2}+\frac{h_{1}}%
{2}\,\phi_{C}^{2}\>,\label{34}\\
m_{K^{\ast}}^{2} = & m_{1}^{2}+\delta_{N}+\delta_{S}+\frac{\phi_{N}^{2}}%
{2}\bigg(\frac{g_{1}^{2}}{2}+h_{1}+\frac{h_{2}}{2}\bigg)+\frac{1}{\sqrt{2}%
}\,\phi_{N}\,\phi_{S}\,(h_{3}-g_{1}^{2})\nonumber \\&+\frac{\phi_{S}^{2}}{2}(g_{1}%
^{2}+h_{1}+h_{2})+\frac{h_{1}}{2}\,\phi_{C}^{2}\>\text{ .} \label{36}%
\end{align}

(iv) Axial-vector mesons:
\begin{align}
m_{f_{1N}}^{2} = & m_{a_{1}}^{2},\label{38}\\
m_{a_{1}}^{2}  = & m_{1}^{2}+2\delta_{N}+g_{1}^{2}\phi_{N}^{2}+\frac{\phi
_{N}^{2}}{2}(h_{1}+h_{2}-h_{3})+\frac{h_{1}}{2}\,\phi_{S}^{2}+\frac{h_{1}}%
{2}\,\phi_{C}^{2}\>,\label{37}\\
m_{f_{1S}}^{2} = & m_{1}^{2}+2\delta_{S}+\frac{h_{1}}{2}\,\phi_{N}^{2}%
+\frac{h_{1}}{2}\,\phi_{C}^{2}+2g_{1}^{2}\,\phi_{S}^{2}+\phi_{S}%
^{2}\bigg(\frac{h_{1}}{2}+\,h_{2}-h_{3}\bigg)\>,\label{39}\\
m_{K_{1}}^{2} = & m_{1}^{2}+\delta_{N}+\delta_{S}+\frac{\phi_{N}^{2}}%
{2}\bigg(\frac{g_{1}^{2}}{2}+h_{1}+\frac{h_{2}}{2}\bigg)+\frac{1}{\sqrt{2}%
}\,\phi_{N}\,\phi_{S}\,(g_{1}^{2}-h_{3})\nonumber \\&+\frac{\phi_{S}^{2}}{2}(g_{1}%
^{2}+h_{1}+h_{2})+\frac{h_{1}}{2}\,\phi_{C}^{2}\>. \label{40}%
\end{align}
Note that all the squared strange-nonstrange mesons masses have the same expressions obtained in the $N_f=3$ case as shown in Sec. 3.3, but with an additional term related to the charm sector. However, this term will not affect the results because it is multiplied by a vanishing parameter as we will see below in the results section.\\
The masses of (open and hidden) charmed mesons are as follows:\\
(i) Pseudoscalar charmed mesons:
\begin{align}
m_{\eta_{C}}^{2}  &  =Z_{\eta_{C}}^{2}[m_{0}^{2}+\lambda_{1}\phi_{N}%
^{2}+\lambda_{1}\phi_{S}^{2}+(\lambda_{1}+\lambda_{2})\phi_{C}^{2}+\frac{c}%
{8}\,\phi_{N}^{4}\,\phi_{S}^{2}+2\,\varepsilon_{C}]\>,\label{61}\\
m_{D}^{2}  &  =Z_{D}^{2}\bigg[m_{0}^{2}+\bigg(\lambda_{1}+\frac{\lambda_{2}%
}{2}\bigg)\phi_{N}^{2}+\lambda_{1}\phi_{S}^{2}-\frac{\lambda_{2}}{\sqrt{2}%
}\phi_{N}\phi_{C}+(\lambda_{1}+\lambda_{2})\phi_{C}^{2}+\varepsilon
_{C}\bigg]\>,\label{62}\\
m_{D_{S}}^{2}  &  =Z_{D_{S}}^{2}[m_{0}^{2}+\lambda_{1}\phi_{N}^{2}%
+(\lambda_{1}+\lambda_{2})\phi_{S}^{2}-\lambda_{2}\phi_{C}\phi_{S}%
+(\lambda_{1}+\lambda_{2})\phi_{C}^{2}+\varepsilon_{C}]\>\text{ .} \label{63}%
\end{align}

(ii) Scalar charmed mesons:%
\begin{align}
m_{\chi_{C0}}^{2}  &
=m_{0}^{2}+\lambda_{1}\phi_{N}^{2}+\lambda_{1}\phi
_{S}^{2}+3(\lambda_{1}+\lambda_{2})\phi_{C}^{2}+2\,\varepsilon_{C}\text{
},\label{57}\\
m_{D_{0}^{\ast}}^{2}  &
=Z_{D_{0}^{\ast}}^{2}\bigg[m_{0}^{2}+\bigg(\lambda
_{1}+\frac{\lambda_{2}}{2}\bigg)\phi_{N}^{2}+\lambda_{1}\phi_{S}^{2}%
+\frac{\lambda_{2}}{\sqrt{2}}\phi_{C}\phi_{N}+(\lambda_{1}+\lambda_{2}%
)\phi_{C}^{2}+\varepsilon_{C}\bigg]\>,\label{58}\\
m_{D_{0}^{\ast0}}^{2}  &
=Z_{D_{0}^{\ast0}}^{2}\bigg[m_{0}^{2}+\bigg(\lambda
_{1}+\frac{\lambda_{2}}{2}\bigg)\phi_{N}^{2}+\lambda_{1}\phi_{S}^{2}%
+\frac{\lambda_{2}}{\sqrt{2}}\phi_{N}\phi_{C}+(\lambda_{1}+\lambda_{2}%
)\phi_{C}^{2}+\varepsilon_{C}\bigg]\>,\label{59}\\
m_{D_{S0}^{\ast}}^{2}  &  =Z_{D_{S0}}^{2}[m_{0}^{2}+\lambda_{1}\phi_{N}%
^{2}+(\lambda_{1}+\lambda_{2})\phi_{S}^{2}+\lambda_{2}\phi_{C}\phi
_{S}+(\lambda_{1}+\lambda_{2})\phi_{C}^{2}+\varepsilon_{C}]\>\text{
.}
\label{60}%
\end{align}

(iii) Vector charmed mesons:
\begin{align}
m_{D^{\ast}}^{2}  = & m_{1}^{2}+\delta_{N}+\delta_{C}+\frac{\phi_{N}^{2}}%
{2}\bigg(\frac{g_{1}^{2}}{2}+h_{1}+\frac{h_{2}}{2}\bigg)+\frac{1}{\sqrt{2}%
}\,\phi_{N}\,\phi_{C}(h_{3}-\,g_{1}^{2})\nonumber\\ &+\frac{\phi_{C}^{2}}{2}(g_{1}%
^{2}+h_{1}+h_{2})+\frac{h_{1}}{2}\,\phi_{S}^{2}\>,\label{51}\\
m_{J/\psi}^{2}  = & m_{1}^{2}+2\delta_{C}+\frac{h_{1}}{2}\,\phi_{N}^{2}%
+\frac{h_{1}}{2}\,\phi_{S}^{2}+\bigg(\frac{h_{1}}{2}+h_{2}+h_{3}%
\bigg)\,\phi_{C}^{2}\>,\label{52}\\
m_{D_{S}^{\ast}}^{2} = & m_{1}^{2}+\delta_{S}+\delta_{C}+\frac{\phi_{S}^{2}%
}{2}(g_{1}^{2}+h_{1}+h_{2})+\phi_{S}\,\phi_{C}(h_{3}-g_{1}^{2})\nonumber\\ &+\frac{\phi
_{C}^{2}}{2}(g_{1}^{2}+h_{1}+h_{2})+\frac{h_{1}}{2}\,\phi_{N}^{2}\>\text{
.}
\label{53}%
\end{align}

(iv) Axial-vector charmed mesons:
\begin{align}
m_{D_{S1}}^{2}  = & m_{1}^{2}+\delta_{S}+\delta_{C}+\frac{\phi_{S}^{2}}%
{2}(g_{1}^{2}+h_{1}+h_{2})+\phi_{S}\,\phi_{C}(g_{1}^{2}-h_{3})\nonumber\\ &+\frac{\phi
_{C}^{2}}{2}(g_{1}^{2}+h_{1}+h_{2})+\frac{h_{1}}{2}\,\phi_{N}^{2}%
\>,\label{54}\\
m_{D_{1}}^{2} = & m_{1}^{2}+\delta_{N}+\delta_{C}+\frac{\phi_{N}^{2}}%
{2}\bigg(\frac{g_{1}^{2}}{2}+h_{1}+\frac{h_{2}}{2}\bigg)+\frac{1}{\sqrt{2}%
}\,\phi_{N}\,\phi_{C}(g_{1}^{2}-h_{3})\nonumber\\ & +\frac{\phi_{C}^{2}}{2}(g_{1}^{2}%
+h_{1}+h_{2})+\frac{h_{1}}{2}\,\phi_{S}^{2}\>,\label{55}\\
m_{\chi_{C1}}^{2}  &  =m_{1}^{2}+2\delta_{C}+\frac{h_{1}}{2}\,\phi_{N}%
^{2}+\frac{h_{1}}{2}\phi_{S}^{2}+2g_{1}^{2}\,\phi_{C}^{2}+\phi_{C}%
^{2}\bigg(\frac{h_{1}}{2}+h_{2}-h_{3}\bigg)\>\text{ .} \label{56}%
\end{align}

Further interesting quantities are also the following mass
differences, in which the explicit dependence on the parameters
$\varepsilon_{C}$ and $\delta_{C}$ cancels:
\begin{align}
m_{D_{1}}^{2}-m_{D^{\ast}}^{2} & =\sqrt{2}\,(g_{1}^{2}-h_{3})\phi
_{N}\phi_{C}\,,\label{diffmas1}\\
 m_{\chi_{C1}}^{2}-m_{J/\psi}^{2}&=2(g_{1}%
^{2}-h_{3})\phi_{C}^{2}\,,\label{diffmas2}\\
 m_{D_{S1}}^{2}-m_{D_{S}^{\ast}}^{2}%
&=2\,(g_{1}^{2}-h_{3})\phi_{S}\,\phi_{C}\,.\label{diffmas3}
\end{align}

\subsection{ \boldmath $\eta$ and \boldmath $\eta^{\prime}$ Masses} \label{sec.eta-eta}

From the Lagrangian (\ref{fulllag}), we obtain mixing between the pure non-strange and strange fields $\eta_{N}\equiv(\bar {u}u-\bar{d}d)/\sqrt{2}$ and
$\eta_{S}\equiv\bar{s}s$ as
\begin{equation}
\mathcal{L}_{\eta_{N}\eta_{S}}=-\frac{c}{4}Z_{\eta_{S}}Z_{\pi}\phi_{N}%
^{3}\phi_{S}\phi_{C}^{2}\eta_{N}\eta_{S}\text{ .} \label{eta-eta}%
\end{equation}
which has the same formula for $N_f=3$ case, as seen in Ref. \cite{Parganlija:2012xj}, but it includes charm quark-antiquark condensate ($\phi_C$).

The generate part of the $\eta_{N}$-$\eta_{S}$ Lagrangian \cite{Parganlija:2012xj} has the form%
\begin{equation}
\mathcal{L}_{\eta_{N}\eta_{S,}\,\mathrm{full}}=\frac{1}{2}(\partial_{\mu}%
\eta_{N})^{2}+\frac{1}{2}(\partial_{\mu}\eta_{S})^{2}-\frac{1}{2}m_{\eta_{N}%
}^{2}\eta_{N}{}^{2}-\frac{1}{2}m_{\eta_{S}}^{2}\eta_{S}{}^{2}+\Upsilon_{\eta} \eta
_{N}\eta_{S}\text{ ,} \label{eta-eta-1}%
\end{equation}

where $\Upsilon_{\eta}$ defines the mixing term of the pure states $\eta_{N}$ and $\eta_{S}$.\\
By comparing Eqs.\ (\ref{eta-eta}) and (\ref{eta-eta-1}) we obtain the mixing term $\Upsilon_{\eta}$ as follows%
\begin{equation}
\Upsilon_{\eta}=-\frac{c}{4}Z_{\eta_{S}}Z_{\pi}\phi_{N}^{3}\phi_{S}\phi_{C}^{2}\text{ ,}
\label{z1}%
\end{equation}

We can determine the physical states $\eta$ and $\eta^{\prime}$
as mixtures of the pure non-strange and strange fields $\eta_N$ and $\eta_S$, see the details in Ref. \cite{Parganlija:2012xj}, as
\begin{equation}
\left(
\begin{array}
[c]{c}%
\eta\\
\eta^{\prime}%
\end{array}
\right)  =\left(
\begin{array}
[c]{cc}%
\cos\varphi_{\eta} & \sin\varphi_{\eta}\\
-\sin\varphi_{\eta} & \cos\varphi_{\eta}%
\end{array}
\right)  \left(
\begin{array}
[c]{c}%
\eta_{N}\\
\eta_{S}%
\end{array}
\right)\,,
\end{equation}

which gives
\begin{align}
\eta &  =\cos\varphi_{\eta}\eta_{N}+\sin\varphi_{\eta}\eta_{S}\text{,} \label{eta}\\
\eta^{\prime}  &
=-\sin\varphi_{\eta}\eta_{N}+\cos\varphi_{\eta}\eta
_{S}\text{ ,} \label{etap}%
\end{align}

where $\varphi_{\eta}=44.6$ is the $\eta-\eta^{\prime}$ mixing angle \cite{Parganlija:2012fy}.

By overturning Eqs.\ (\ref{eta}) and (\ref{etap}), one obtain the pure states $\eta_{N}$ and $\eta_{S}$ as

\begin{align}
\eta_{N}  & =\cos\varphi_{\eta}\eta-\sin\varphi_{\eta}\eta^{\prime
}\text{ ,} \label{etaN}\\
\eta_{S}  &
=\sin\varphi_{\eta}\eta+\cos\varphi_{\eta}\eta^{\prime}\text{ ,}
\label{etaS}%
\end{align}

By substituting $\eta_{N}$ and $\eta_{S}$ by $\eta$ and
$\eta^{\prime}$ in the Lagrangian
Eq.\ (\ref{eta-eta-1}), we get \cite{Parganlija:2012xj}%
\begin{align}
\mathcal{L}_{\eta\eta^{\prime}}  &  =\frac{1}{2}[(\partial_{\mu}\eta)^{2}%
\cos^2\varphi_{\eta}+(\partial_{\mu}\eta^{\prime})^{2}\sin^2\varphi_{\eta
}-\sin(2\varphi_{\eta})\partial_{\mu}\eta\partial^{\mu}\eta^{\prime
}]\nonumber\\
&  +\frac{1}{2}[(\partial_{\mu}\eta)^{2}\sin^2\varphi_{\eta}%
+(\partial_{\mu}\eta^{\prime})^{2}\cos^2\varphi_{\eta}+\sin(2\varphi
_{\eta})\partial_{\mu}\eta\partial^{\mu}\eta^{\prime}]\nonumber\\
&
-\frac{1}{2}m_{\eta_{N}}^{2}[\eta^{2}\cos^2\varphi_{\eta}+(\eta
^{\prime})^{2}\sin^2\varphi_{\eta}-\sin(2\varphi_{\eta})\eta\eta^{\prime
}]\nonumber\\
&
-\frac{1}{2}m_{\eta_{S}}^{2}[\eta^{2}\sin^2\varphi_{\eta}+(\eta
^{\prime})^{2}\cos^2\varphi_{\eta}+\sin(2\varphi_{\eta})\eta\eta^{\prime
}]\nonumber\\
&  +\Upsilon_{\eta}\{[\eta^{2}-(\eta^{\prime})^{2}]\sin\varphi_{\eta}\cos
\varphi_{\eta}+\cos(2\varphi_{\eta})\eta\eta^{\prime}\}\nonumber\\
&
=\frac{1}{2}(\partial_{\mu}\eta)^{2}+\frac{1}{2}(\partial_{\mu}\eta
^{\prime})^{2}-\frac{1}{2}[m_{\eta_{N}}^{2}\cos^2\varphi_{\eta}%
+m_{\eta_{S}}^{2}\sin^2\varphi_{\eta}-\Upsilon_{\eta}\sin(2\varphi_{\eta}%
)]\eta^{2}\nonumber\\
&  -\frac{1}{2}[m_{\eta_{N}}^{2}\sin^2\varphi_{\eta}+m_{\eta_{S}}^{2}%
\cos^2\varphi_{\eta}+\Upsilon_{\eta}\sin (
2\varphi_{\eta})](\eta^{\prime
})^{2}\nonumber\\
&
-\frac{1}{2}[(m_{\eta_{S}}^{2}-m_{\eta_{N}}^{2})\sin(2\varphi_{\eta
})-2\Upsilon_{\eta}\cos(2\varphi_{\eta})]\eta\eta^{\prime}\text{ ,} \label{eta-eta-2}%
\end{align}

which gives the masses of the physical states $\eta$ and $\eta^{\prime}$, 
$m_{\eta}$ and $m_{\eta^{\prime}}$, in terms of the pure
non-strange and strange, $\eta_N$ and $\eta_S$, mass terms:

\begin{align}
m_{\eta}^{2}  &  =m_{\eta_{N}}^{2}\cos^{2}\varphi_{\eta}+m_{\eta_{S}}^{2}%
\sin^{2}\varphi_{\eta}-\Upsilon_{\eta}\sin(2\varphi_{\eta})\text{,} \label{m_eta}\\
m_{\eta^{\prime}}^{2}  &  =m_{\eta_{N}}^{2}\sin^{2}\varphi_{\eta}+m_{\eta_{S}%
}^{2}\cos^{2}\varphi_{\eta}+\Upsilon_{\eta}\sin(2\varphi_{\eta})\text{ .}
\label{m_eta'}%
\end{align}

where the mass terms $m_{\eta_{N}}$ and
$m_{\eta_{S}}$ are known from
Eqs.\ (\ref{47}) and (\ref{48}).

\subsection{Scalar-Isosinglet Masses}

There is a mixing between the pure states $\sigma_{N}\equiv(\bar {u}u+\bar{d}d)/\sqrt{2}$ 
and $\sigma_{S}\equiv\bar{s}s$ in the Lagrangian (\ref{fulllag}) with the mixing term given by

\begin{equation}
\mathcal{L}_{\sigma_{N}\sigma_{S}}=-2\lambda_{1}\phi_{N}\phi_{S}\sigma_{N}\sigma_{S}\text{ .}\label{mixtw41}
\end{equation}
Notice that the mixing term (\ref{mixtw41}) of $\sigma_{N}$
and $\sigma_{S}$ does not depend on the charm quark-antiquark condensate $\phi_C$. So it is the same mixing term in the case of $N_f=3$ \cite{Parganlija:2012xj}.\\
The generate part of the $\sigma_{N}$-$\sigma_{S}$\ Lagrangian has the form %
\begin{equation}
\mathcal{L}_{\sigma_{N}\sigma_{S},\,\mathrm{full}}=\frac{1}{2}(\partial_{\mu
}\sigma_{N})^{2}+\frac{1}{2}(\partial_{\mu}\sigma_{S})^{2}-\frac{1}%
{2}m_{\sigma_{N}}^{2}\sigma_{N}{}^{2}-\frac{1}{2}m_{\sigma_{S}}^{2}\sigma
_{S}{}^{2}+\Upsilon_{\sigma}\sigma_{N}\sigma_{S}\text{ ,}\label{Lms}
\end{equation}

where $\Upsilon_{\sigma}$ is the mixing term of the
$\sigma_{N}$ and $\sigma_{S}$ fields,

\begin{equation}
\Upsilon_{\sigma}=-2\lambda_{1}\phi_{N}\phi_{S}\text{
.}\label{sigma-sigma}
\end{equation}

The mixing between the states $\sigma_{N}$ and $\sigma_{S}$\
yields $\sigma_{1}$ and
$\sigma_{2}$ fields [see Ref. \cite{Parganlija:2012xj}]:

\begin{equation}
\left(
\begin{array}
[c]{c}%
\sigma_{1}\\
\sigma_{2}%
\end{array}
\right)  =\left(
\begin{array}
[c]{cc}%
\cos\varphi_{\sigma} & \sin\varphi_{\sigma}\\
-\sin\varphi_{\sigma} & \cos\varphi_{\sigma}%
\end{array}
\right)  \left(
\begin{array}
[c]{c}%
\sigma_{N}\\
\sigma_{S}%
\end{array}
\right)  \text{ .}\label{sigma-sigma_1}%
\end{equation}

which can be written as
\begin{align}
 \sigma_{1} &  =\cos\varphi_{\sigma}\sigma_{N}+\sin\varphi_{\sigma}\sigma_{S}\text{,} \label{sig1}\\
 \sigma_{2} &  =-\sin\varphi_{\sigma}\sigma_{N}+\cos\varphi_{\sigma}\sigma_{S}\text{ ,} \label{sig2}
\end{align}

where $\varphi_{\sigma}$ is the $\sigma_{N}$-$\sigma_{S}$ mixing angle.

By overturning Eqs.\ (\ref{sig1}) and (\ref{sig2}), one obtain the pure states $\sigma_{N}$ and $\sigma_{S}$ as

\begin{align}
\sigma_{N}  & =\cos\varphi_{\sigma}\sigma_1-\sin\varphi_{\sigma}\sigma_2\text{ ,} \label{sigN}\\
\sigma_{S}  &
=\sin\varphi_{\sigma}\sigma_1+\cos\varphi_{\sigma}\sigma_2\text{ ,}
\label{sigS}%
\end{align}

Substituting from Eqs. (\ref{sigN}) and (\ref{sigS}) into Eq. (\ref{Lms}), one obtain the $\sigma_1-\sigma_2$ Lagrangian as follows:

\begin{align}
\mathcal{L}_{\sigma_1\sigma_2}  &  =\frac{1}{2}[(\partial_{\mu}\sigma_1)^{2}%
\cos^2\varphi_{\sigma}+(\partial_{\mu}\sigma_2)^{2}\sin^2\varphi_{\sigma
}-\sin(2\varphi_{\sigma})\partial_{\mu}\sigma_1\partial^{\mu}\sigma_2]\nonumber\\
&  +\frac{1}{2}[(\partial_{\mu}\sigma_1)^{2}\sin^2\varphi_{\sigma}%
+(\partial_{\mu}\sigma_2)^{2}\cos^2\varphi_{\sigma}+\sin(2\varphi
_{\sigma})\partial_{\mu}\sigma_1\partial^{\mu}\sigma_2]\nonumber\\
&
-\frac{1}{2}m_{\sigma_{N}}^{2}[\sigma_1^{2}\cos^2\varphi_{\sigma}+(\sigma_2)^{2}\sin^2\varphi_{\sigma}-\sin(2\varphi_{\sigma})\sigma_1\sigma_2]\nonumber\\
&
-\frac{1}{2}m_{\sigma_{S}}^{2}[\sigma_1^{2}\sin^2\varphi_{\sigma}+(\sigma_2)^{2}\cos^2\varphi_{\sigma}+\sin(2\varphi_{\sigma})\sigma_1\sigma_2]\nonumber\\
&  +\Upsilon_{\sigma}\{[\sigma_1^{2}-(\sigma_2)^{2}]\sin\varphi_{\sigma}\cos
\varphi_{\sigma}+\cos(2\varphi_{\sigma})\sigma_1\sigma_2\}\nonumber\\
&
=\frac{1}{2}(\partial_{\mu}\sigma)^{2}+\frac{1}{2}(\partial_{\mu}\sigma
^{\prime})^{2}-\frac{1}{2}[m_{\sigma_{N}}^{2}\cos^2\varphi_{\sigma}%
+m_{\sigma_{S}}^{2}\sin^2\varphi_{\sigma}-\Upsilon_{\sigma}\sin(2\varphi_{\sigma}%
)]\sigma_1^{2}\nonumber\\
&  -\frac{1}{2}[m_{\sigma_{N}}^{2}\sin^2\varphi_{\sigma}+m_{\eta_{\sigma}}^{2}%
\cos^2\varphi_{\sigma}+\Upsilon_{\sigma}\sin (
2\varphi_{\sigma})](\sigma_2)^{2}\nonumber\\
&
-\frac{1}{2}[(m_{\sigma_{S}}^{2}-m_{\sigma_{N}}^{2})\sin(2\varphi_{\sigma})-2\Upsilon_{\sigma}\cos(2\varphi_{\sigma})]\sigma_1\sigma_2\text{ ,} \label{eta-eta-2}%
\end{align}

We then obtain the mass terms of $\sigma_1$ and $\sigma_2$ fields as

\begin{align}
m_{\sigma_{1}}^{2}  &  =m_{\sigma_{N}}^{2}\cos^{2}\varphi_{\sigma}%
+m_{\sigma_{S}}^{2}\sin^{2}\varphi_{\sigma}-\Upsilon_{\sigma}\sin(2\varphi_{\sigma
})\text{,} \label{m_sigma_1}\\
m_{\sigma_{2}}^{2}  &  =m_{\sigma_{N}}^{2}\sin^{2}\varphi_{\sigma}%
+m_{\sigma_{S}}^{2}\cos^{2}\varphi_{\sigma}+\Upsilon_{\sigma}\sin(2\varphi_{\sigma
})\,, \label{m_sigma_2}%
\end{align}

where $m_{\sigma_{N}}$ and $m_{\sigma_{S}}$ known from Eq.\
(\ref{43}) and 
Eq.\ (\ref{44}), respectively. While the mixing term $\Upsilon_{\sigma}$ is 

\begin{equation}
\Upsilon_{\sigma}=\frac{1}{2}(m_{\sigma_{S}}^{2}-m_{\sigma_{N}}^{2})\tan
(2\varphi_{\sigma})\text{ .} \label{Upsigma}%
\end{equation}

The resonances $\sigma_{1}$ and $\sigma_{2}$ will be assigned to the physical states  $f_0(1370)$ and $f_0(1710)$ \cite{Parganlija:2012xj}, respectively. \\

One can determine the $\sigma_1-\sigma_2$ mixing angle $\varphi_{\sigma}$ from Eqs.\ (\ref{sigma-sigma}) and (\ref{Upsigma}) \cite{Parganlija:2012xj} as 

\begin{align}
\varphi_{\sigma}  &=-\frac{1}{2}\arctan\left(
\frac{4\lambda_{1}\phi
_{N}\phi_{S}}{m_{\sigma_{S}}^{2}-m_{\sigma_{N}}^{2}}\right) \nonumber\\
& {=}\frac{1}%
{2}\arctan\left[  \frac{8\lambda_{1}\phi_{N}\phi_{S}}{(4\lambda_{1}%
+3\lambda_{2})\phi_{N}^{2}-(4\lambda_{1}+6\lambda_{2})\phi_{S}^{2}}\right]
\text{ ,} \label{phisigma1}%
\end{align}

In the large-$N_{c}$ limit, one sets $\lambda_{1}=0$, as shown in
Ref. \cite{Parganlija:2012xj} for the case $N_{f}=3$. Therefore we get from
Eq.(\ref{phisigma1}) that the mixing angle $\varphi_{\sigma}$
between $\sigma_1-\sigma_2$ is zero. Naturally the mixing angle
between $\sigma_1-\sigma_2$ is very small, which is in agreement
with our result. Then $\varphi_{\sigma}$ does not affect the
results in this framework.

\section{The Model Parameters}

The Lagrangian (\ref{fulllag}) contains the following 15 free
parameters:
$m_{0}^{2},\,\lambda_{1},\,\lambda_{2},\,m_{1},\,g_{1},\,c_{1},\,h_{1},$
$h_{2},\,h_{3},\,\delta_{S},\,\delta_{C},\,\varepsilon_{C},$
$h_{N},$ $h_{S},$ and $h_{C}$. For technical reasons, instead of
the parameters $h_{N},$ $h_{S},$ and $h_{C}$ entering Eq.\
(\ref{h}), it is easier to use the condensates $\phi_{N},$
$\phi_{S},$ $\phi_{C}$. This is obviously equivalent, because
$\phi_{N},$ $\phi_{S},$ and $\phi_{C}$ form linearly independent
combinations of the parameters.\\

We can obtain the relation between $c$ and its counterpart in the
three-flavour case $c_{N_{f}=3}$ of Ref. \cite{Parganlija:2012fy} as follows:\\
The axial-anomaly term as described in the Lagrangian (\ref{fulllag})
in the case of $N_f=3$ can be written as
\begin{equation}
\mathcal{L}_{ca3}=c_{N_f=3}(\text{det}\Phi_{N_f=3}-\text{det}\Phi^\dagger_{N_f=3})^2\,,\label{Lca3}
\end{equation}
and in the case of $N_f=4$
\begin{equation}
\mathcal{L}_{ca4}=c\,(\text{det}\Phi-\text{det}\Phi^\dagger)^2\,,\label{Lca4}
\end{equation}
The (pseudo)scalar multiplet matrix, $\Phi$, in the case of
$N_f=4$, which includes the $3\times3$ (pseudo)scalar multiplet
matrix, $\Phi_{N_f=3}$, can be written as
\begin{equation}
\Phi=
\left(%
\begin{array}{cccc}
  &  & & 0 \\
  & \Phi_{N_f=3} &  & 0 \\
  &  & & 0\\
 0 & 0 & 0 & \frac{\phi_C}{\sqrt{2}}\\
\end{array}%
\right)\>.\label{Phi4}
\end{equation}
By using Eq.(\ref{Phi4}) to calculate the determinant of $\Phi$,
we obtain
\begin{equation}
 \text{det} \Phi=\frac{\phi_C}{\sqrt{2}} \,\text{det} \Phi_{N_f=3},
\end{equation}
 The axial-anomaly term in the case of $N_f=3$, Eq. (\ref{Lca3}),
 can be transformed to the case of $N_f=4$ by using
 Eq.(\ref{Phi4}) as follows:
\begin{equation}
\mathcal{L}_{ca4}=\frac{2\,c_{N_f=3}}{\phi_C^2}(\text{det}\Phi-\text{det}\Phi^\dagger)^2\,,\label{lca4}
\end{equation}
Comparing between Eq.(\ref{Lca4}) and Eq.(\ref{lca4}), we get
\begin{equation}
c=\frac{2\,c_{N_{f}=3}}{\phi_{C}^{2}}\text{ .} \label{cQ2}%
\end{equation}
Thus, the parameter $c$ can be determined once the condensate
$\phi_{C}$ is obtained.\\

In the large-$N_{c}$ limit, one sets $h_{1}=\lambda_{1}=0.$ Then,
as shown in Ref.\ \cite{Parganlija:2012fy} for the case $N_{f}=3$, ten
parameters can be determined by a fit to masses and decay widths
of mesons below 1.5 GeV as shown in Table 1. In the following we
use these values for our numerical calculations. As a consequence,
the masses and the decay widths of the nonstrange-strange mesons
are -- by construction -- identical to the results of Ref.\
\cite{Parganlija:2012fy} (see Table 2 and Fig.\ 1 in that ref.). Note also
that, in virtue of Eq.\ (\ref{cQ2}), the parameter combination
$\phi_{C}^{2}c/2$ is determined by the fit of Ref.\ \cite{Parganlija:2012fy}.

\begin{table}[H]
\centering
\begin{tabular}
[c]{|c|c|c|c|}\hline Parameter & Value & Parameter & Value\\\hline
$m_{1}^{2}$ & $0.413\times10^{6}$ MeV$^{2}$ & $m_{0}^{2}$ &
$-0.918\times 10^{6}$ MeV$^{2}$\\\hline $\phi_{C}^{2}c/2$ &
$450\cdot10^{-6}$ MeV$^{-2}$ & $\delta_{S}$ &
$0.151\times10^{6}$MeV$^{2}$\\\hline $g_{1}$ & $5.84$ & $h_{1}$ &
$0$\\\hline $h_{2}$ & $9.88$ & $h_{3}$ & $3.87$\\\hline $\phi_{N}$
& $164.6$ MeV & $\phi_{S}$ & $126.2$ MeV\\\hline $\lambda_{1}$ &
$0$ & $\lambda_{2}$ & $68.3$\\\hline
\end{tabular}
\caption{ Values of the parameters (from
Ref. \cite{Parganlija:2012fy})}%
\label{Tab:param}
\end{table}

For the purposes of the present work, we are left with three
unknown parameters: $\phi_{C},\varepsilon_{C},$ $\delta_{C}.$ We
determine these by performing a fit to twelve experimental (hidden
and open) charmed meson masses listed by the PDG \cite{Beringer:1900zz}, minimizing
\begin{equation}
\chi^{2}\equiv\sum_{i}^{12}\bigg(\frac{M_{i}^{th}-M_{i}^{exp}}{\xi M_{i}%
^{exp}}\bigg)^{2}\text{ ,} \label{chich4}%
\end{equation}
where $\xi$ is a constant. We do not use the experimental errors
for the masses, because we do not expect to reach the same
precision with our effective model, which (besides other
effects) already neglects isospin breaking. In Ref.\ \cite{Parganlija:2012fy},
we required a minimum error of 5\% for experimental quantities
entering our fit, and obtained a reduced $\chi^{2}$ of about 1.23.
Here, we slightly change our fit strategy: we choose the parameter
$\xi$ such that the reduced $\chi^{2}$ takes the value
$\chi^{2}/(12-3)=1,$ which yields $\xi=0.07.$ This implies that we
enlarge the experimental errors to $7\%$ of the respective masses.

The parameters (together with their theoretical errors) are:

\begin{table}[H]
\centering
\begin{tabular}
[c]{|c|c|}\hline Parameter & Value\\\hline
$\phi_{C}$&$176\pm28\text{ MeV}$ \\\hline $\delta_{C}$
&$(3.91\pm0.36)\times10^{6}\text{ MeV}^{2}$ \\\hline
$\varepsilon_{C}$ & $(2.23\pm0.71)\times10^{6}\text{ MeV}$\\\hline
$c$ & $(2.91\pm 0.94)\times10^{-8}\text{ MeV}^{-4}$ \\\hline
\end{tabular}
\caption{ Values of the
unknown parameters.}%
\label{Tab:charmparam}
\end{table}

\section{Results}

In this section we present the $w_i$ and the wave-function renormalization constants $Z_i$ values which are important parameters for the
determination of the masses and the decay widths of mesons. We
present also the masses of light mesons as well as (open and
hidden) charmed mesons.

\subsection{The $w_i$ and the wave-function renormalization constants $Z_i$}

All 15 parameter have been determined as shown in the previous
section. Then, the parameters $w_i$ can be determined from Eqs.
(3.46 - 3.49) and (4.45 - 4.50) and the wave-function renormalization constants from
Eqs. (\ref{zpi}-\ref{zets}) as summarized in Table
\ref{Tab:ren1}:

\begin{table}[H]
\centering
\begin{tabular}
[c]{|c|c|c|c|c|} \hline parameter & value  & parameter & value\\
\hline $w_{a1}$      & 0.00068384 & $w_{f_{1N}}$ & 0.00068384\\
\hline $w_{f_{1S}}$ & 0.0005538   & $w_{K1}$ & 0.000609921\\
\hline $w_{K^\ast}$ & -0.0000523i   & $w_{D_{S1}}$ & 0.000203\\
\hline $w_{D^\ast}=w_{D^{\ast0}}$ & -0.0000523i  & $w_{D^\ast_0}$ & -0.0000423i\\
\hline $w_{D_1}$       & 0.00020   & $w_{\chi_{C1}}$  & 0.000138\\
\hline $Z_\pi=Z_{\eta_N}$ & 1.70927 & $Z_{\eta_S}$ & 1.53854\\
\hline $Z_K$  & 1.60406  & $Z_{K_{S}}$ & 1.00105\\
\hline $Z_{\eta_C}$ & 1.11892  & $Z_{D}$ & 1.15256\\
\hline $Z_{D_S}$    & 1.15716    & $Z_{D_{S0}^\ast}$ & 1.00437\\
\hline $Z_{D^\ast_0}=Z_{D^{\ast0}_0}$    & 1.00649  & $Z_{D^{\ast0}_0}$ & 1.00649\\
\hline
\end{tabular}
\caption{ $w_i$ and the wave-function renormalization\\ constants $Z_i$ values.}
 \label{Tab:ren1}
\end{table}

As seen in Table \ref{Tab:ren1}, all wave-function renormalization constants have values larger than one. The parameter $w_{D^\ast}$ is equal to 
$w_{D^{*0}}$ and  the parameter $w_{D^*_0}$ is equal to 
$w_{D^{*0}_{0}}$ for isospin symmetry reasons.\\

$$\\$$

\subsection{Masses of light mesons}

The results for the light meson masses are reported in Table
\ref{Tab:LM}.
 By construction, one finds the same values as in Refs.\ \cite{Parganlija:2012fy, Parganlija:2012xj}.
\begin{table}[H]
\centering
\begin{tabular}
[c]{|c|c|c|c|c|c|} \hline observable & $J^{P}$ &
theoretical value [MeV] & experimental value [MeV]\\
\hline $m_\pi$          &  $0^{-}$ & 141   &139.57018 $\pm$ 0.00035\\
\hline $m_{\eta}$       &  $0^{-}$ & 509   &547.853 $\pm$ 0.024\\
\hline $m_{\eta'}$      &  $0^{-}$ & 962   &957.78 $\pm$ 0.06\\
\hline $m_K$            &  $0^{-}$ & 485  &493.677 $\pm$ 0.016\\
\hline $m_{a_0}$        & $0^{+}$ & 1363  &1474 $\pm$ 19\\
\hline $m_{\sigma_1}$   & $0^{+}$ & 1362  &(1200-1500)-i(150-250)\\
\hline $m_{\sigma_2}$   & $0^{+}$ & 1531   &1720 $\pm$ 60\\
\hline $m_{K_0^\ast}$   & $0^{+}$ & 1449   &1425 $\pm$ 50\\
\hline $m_{\omega_N}$   & $1^{-}$ & 783   &782.65 $\pm$ 0.12\\
\hline $m_{\omega_S}$   & $1^{-}$ & 975   &1019.46 $\pm$ 0.020\\
\hline $m_\rho$         & $1^{-}$ & 783   &775.5 $\pm$ 38.8\\
\hline $m_{K^*}$        & $1^{-}$ & 885    &891.66 $\pm$ 0.26\\
\hline $m_{f_{1N}}$     & $1^{+}$ & 1186  &1281.8 $\pm$ 0.6\\
\hline $m_{a_{1}}$      & $1^{+}$ & 1185  &1230 $\pm$ 40\\
\hline $m_{f_{1S}}$     & $1^{+}$ & 1372  &1426.4 $\pm$ 0.9\\
\hline $m_{K_1}$        & $1^{+}$ & 1281  &1272 $\pm$ 7\\
\hline
\end{tabular}
\caption{ Light meson masses.}
 \label{Tab:LM}
\end{table}
Note that the values of the light mesons are the same as in
the case of $N_f=3$, shown in Table 3.2, as they are not
affected by the charm sector because $\lambda_1=h_1=0$.
\newpage

\subsection{Masses of charmed mesons}

In Table \ref{charmass} we present the results of our fit for the masses of the open and hidden charmed mesons, by comparing the theoretically
computed with the experimentally measured masses [see also
Ref.\ \cite{Eshraim:2014afa} for preliminary results]. For the nonstrange-charmed
states we use the masses of the neutral members of the multiplet in the fit,
because the corresponding resonances have been clearly identified and the
masses have been well determined for all quantum numbers. In view of the fact
that the employed model is built as a low-energy chiral model and that only
three parameters enter the fit, the masses are quite well described. The
mismatch grows for increasing masses because Eq.\ (\ref{chich4}) imposes, by
construction, a better precision for low masses. For comparison, in the right
column of Table \ref{charmass} we also show the value $0.07M_{i}^{exp}$ which represents
the `artificial experimental error' that we have used in our fit.

\begin{table}[H]
\centering
\begin{tabular}
[c]{|c|c|c|c|c|c|}\hline Resonance & Quark content & $J^{P}$ & Our
Value  & Experimental Value & 7\% of the exp.\\
   & &  &[MeV] & [MeV]& value [MeV]\\
\hline $D^{0}$ & $u\bar{c},\bar{u}c$ & $0^{-}$ &
$1981\pm73$ & $1864.86\pm0.13$ & $130$\\\hline $D_{S}^{\pm}$ &
$s\bar{c},\bar{s}c$ & $0^{-}$ & $2004\pm74$ & $1968.50\pm0.32$ &
$138$\\\hline $\eta_{c}(1S)$ & $c\bar{c}$ & $0^{-}$ & $2673\pm118$
& $2983.7\pm0.7$ & $209$\\\hline $D_{0}^{\ast}(2400)^{0}$ &
$u\bar{c},\bar{u}c$ & $0^{+}$ & $2414\pm77$ & $2318\pm29$ &
$162$\\\hline $D_{S0}^{\ast}(2317)^{\pm}$ & $s\bar{c},\bar{s}c$ &
$0^{+}$ & $2467\pm76$ & $2317.8\pm0.6$ & $162$\\\hline
$\chi_{c0}(1P)$ & $c\bar{c}$ & $0^{+}$ & $3144\pm128$ &
$3414.75\pm0.31$ & $239$\\\hline $D^{\ast}(2007)^{0}$ &
$u\bar{c},\bar{u}c$ & $1^{-}$ & $2168\pm70$ & $2006.99\pm0.15$ &
$140$\\\hline $D_{s}^{\ast}$ & $s\bar{c},\bar{s}c$ & $1^{-}$ &
$2203\pm69$ & $2112.3\pm0.5$ & $148$\\\hline $J/\psi(1S)$ &
$c\bar{c}$ & $1^{-}$ & $2947\pm109$ & $3096.916\pm0.011$ &
$217$\\\hline $D_{1}(2420)^{0}$ & $u\bar{c},\bar{u}c$ & $1^{+}$ &
$2429\pm63$ & $2421.4\pm0.6$ & $169$\\\hline $D_{S1}(2536)^{\pm}$
& $s\bar{c},\bar{s}c$ & $1^{+}$ & $2480\pm63$ & $2535.12\pm0.13$ &
$177$\\\hline $\chi_{c1}(1P)$ & $c\bar{c}$ & $1^{+}$ &
$3239\pm101$ & $3510.66\pm0.07$ & $246$\\\hline
\end{tabular}
\caption{Masses of charmed
mesons used
in the fit.}
\label{charmass}
\end{table}

The following remarks about our results are in order:\\

(i) Remembering that our model is a low-energy effective approach
to the strong interaction, it is quite surprising that the masses of the open charmed
states are in good quantitative agreement (within the theoretical error) with experiment data. In
particular, when taking into account the 7\% range (right column of Table 2),
almost all the results are within 1$\sigma$ or only slightly above it.
Clearly, our results cannot compete with the precision of other approaches,
but show that a connection to low-energy models is possible.

(ii) With the exception of $J/\psi$, the masses of the charmonia states are somewhat underestimated as is
particularly visible for the resonance $\eta_{c}(1S).$ On the one hand, this
is due to the way the fit has been performed; on the other hand, it points to
the fact that unique values of the parameters $h_{C},$ $\delta_{C}$, and
$\varepsilon_{C}$ are not sufficient for a precise description of both open
and hidden charmed states over the whole energy range. One way to
improve the fit of the charmonium masses
would be to include non-zero values for $\lambda_1, h_1$.
Another way is explicitly breaking the chiral symmetry as
discussed at the end of Sec.\ \ref{Lagr}. However, since the description
of open charm states is already reasonable, we only need to consider
the charmonium states. For the (pseudo)scalar charmonia, we would 
thus modify the
second term in Eq.\ (\ref{U4symbreak}) by introducing another projection
operator under the trace, ${ Tr} (\mathbb{P}_C \Phi^\dagger \Phi)^2
\rightarrow { Tr} (\mathbb{P}_C \Phi^\dagger \mathbb{P}_C \Phi)^2$.
A similar consideration could be done for the (axial-)vector charmonia.

(iii) The experimental value for the mass of the charged scalar state
$D_{0}^{\ast}(2400)^{\pm}$, which is $(2403\pm14\pm35)$ MeV, is in fair
agreement with our theoretical result, although the existence of this
resonance has not yet been unambiguously established.

(iv) The theoretically computed mass of the strange-charmed scalar state
$D_{S0}^{\ast\pm}$ turns out to be larger than that of the charmed state
$D_{0}^{\ast}(2400)^{0}.$ In this respect, the experimental result is puzzling
because the mass of $D_{S0}^{\ast}(2317)^{\pm}$ is smaller than that of
$D_{0}^{\ast}(2400)^{0}.$ A possibility is that the resonance $D_{S0}^{\ast
}(2317)$ is not a quarkonium, or that the current mass of the quarkonium field
is diminished by quantum fluctuations [see e.g.\ Ref.\ \cite{Achasov:2004uq, Giacosa:2007bn, Giacosa:2012de}.]

(v) The theoretical mass of the axial-vector strange-charmed state $D_{S1}$
reads $2480$ MeV, which lies in between the two physical states $D_{S1}%
(2460)^{\pm}$ and $D_{S1}(2536)^{\pm}.$ We shall re-analyze the scalar and the
axial-vector strange-charmed states in light of the results for the decay
widths, see chapter 7.

The theoretical results for the squared charmed vector and
axial-vector mass differences, in which the explicit dependence on
the parameters $\varepsilon_{C}$ and $\delta_{C}$ cancels and which were presented in
Eqs.(\ref{diffmas1}-\ref{diffmas3}) are

\begin{table}[H]
\centering
\begin{tabular}
[c]{|c|c|c|c|} \hline mass difference & theoretical value MeV$^{2}$& experimental value  MeV$^{2}$\\
\hline $m_{D_{1}}^{2}-m_{D^{\ast}}^{2}$ & $(1.2\pm0.6)\times10^{6}$ & $1.82\times10^{6}$ \\
\hline $m_{\chi_{C1}}^{2}-m_{J/\psi}^{2}$  & $(1.8\pm1.3)\times10^{6}$  & $2.73\times10^{6}$ \\
\hline $m_{D_{S1}}^{2}-m_{D_{S}^{\ast}}^{2}$ & $(1.2\pm0.6)\times10^{6}$  & $1.97\times10^{6}$ \\
\hline
\end{tabular}
\caption{ Mass differences.}
 \label{Tab:diff}
\end{table}

Here we compare them also with the experimental results ( the experimental
error is omitted because, being of the order of $10^{3}$ MeV$^{2},$ they 
are very small w.r.t. the theoretical ones). The agreement is
fairly good, which shows that our determination of the charm
condensate $\phi_{C}$ is compatible with the experiment, although
it still has a large uncertainty. Note that a similar determination of $\phi_{C}$ 
via the weak decay constants of
charmed mesons determined via the PCAC relations has been presented in Ref.\
\cite{Mishra:2003se}. Their result is $\phi_{C}/\phi_{N}\simeq1.35$, which is compatible to our ratio of
about $1.07\pm0.20$. Previously, Ref.\ \cite{Gottfried:1992bz}
determined $\phi_{C}/\phi_{N}\simeq1.08$ in the framework of the NJL model, which is
in perfect agreement with our result.\\

From a theoretical point of view, it is instructive to study the behavior of
the condensate $\phi_{C}$ as function of the heavy quark mass $m_{C}$. To this
end, we recall that the equation determining $\phi_{C}$ is of the third-order
type and reads (for $\lambda_{1}=0$)%
\begin{equation}
h_{0,C}=(m_{0}^{2}+2\varepsilon_{C})\phi_{C}+\lambda_{2}\phi_{C}^{3}\text{ .}%
\end{equation}
By imposing the scaling behaviors $h_{0,C}=m_{C}\tilde{h}_{0,C}$ and
$\varepsilon_{C}=\tilde{\varepsilon}_{C}m_{C}^{2}$, the solution for large
values of $m_{C}$ reads $\phi_{C}\simeq h_{0,C}/2\varepsilon_{C}\propto
1/m_{C}$, which shows that the mass differences of Eqs.\ (\ref{diffmas1}-\ref{diffmas3})
vanish in the heavy-quark limit. However, the fact that the value of the charm
condensate turns out to be quite large implies that the charm quark is not yet
that close to the heavy-quark limit. To show this aspect, we plot in Fig.\ 4.1
the condensate $\phi_{C}$ as function of $m_{c}$ by keeping all other
parameters fixed. Obviously, this is a simplifying assumption, but the main
point of this study is a qualitative demonstration that the chiral condensate
of charm-anticharm quarks is non-negligible.\\

\begin{figure}[H]
\begin{center}
\includegraphics[
height=2.9637in,
width=4.9191in
]%
{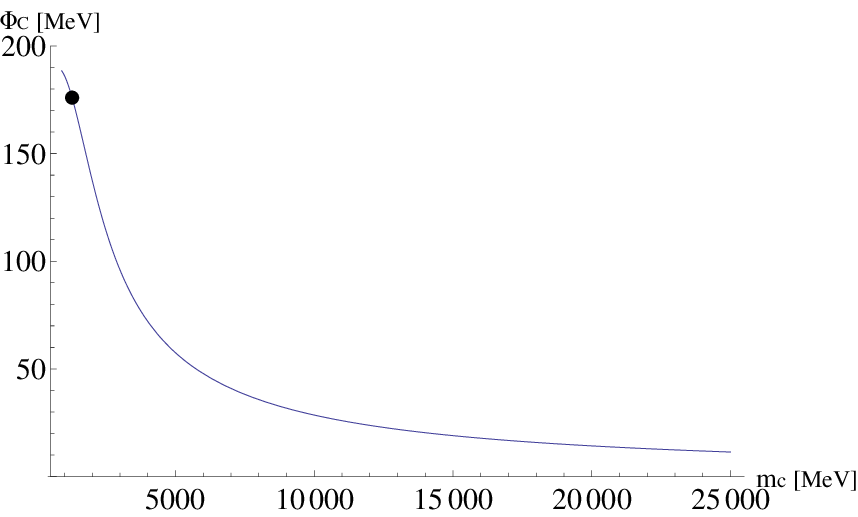}%
\caption{Condensate $\phi_{C}$ as function of the quark mass $m_{C}$. The dot
corresponds to the physical value $m_{C}=1.275$ GeV \cite{Beringer:1900zz}.}%
\end{center}
\end{figure}


Note that, in contrast to the demands of heavy-quark spin symmetry, our chiral approach does not necessarily imply an equal mass of the
vector and pseudoscalar states and of scalar and axial-vector states. Namely,
for a very large heavy quark mass one has $m_{D}^{2}\simeq\varepsilon_{C}$ and
$m_{D^{\ast}}^{2}\simeq\delta_{C}.$ Thus, in order to obtain a degeneracy of
these mesons, as predicted by heavy-quark symmetry \cite{Neubert:1993mb, Casalbuoni:1996pg, Georgi:1990um, Georgi:1991mr, DeFazio:2000up, Wise:1992hn}, we should
additionally impose that $\varepsilon_{C}=\delta_{C}.$ Numerically, the values
have indeed the same order of magnitude, but differ by a factor of two, see Table 4.2. Nevertheless, the differences of masses $m_{D^{\ast}}%
-m_{D}\simeq180$ MeV of the spin-0 and spin-1 negative-parity open charm
states turns out to be (at least within theoretical errors) in quantitative agreement with the experiment. 
Thus, at least for negative-parity mesons, our chiral approach seems 
to correctly predict the amount of breaking of the 
heavy-quark spin symmetry. For the mass difference $m_{D_1}-m_{D_0^*}$
of spin-0 and spin-1 mesons with positive parity, our model underpredicts
the experimental values by an order of magnitude, i.e., our approach based
on chiral symmetry follows the predictions of heavy-quark symmetry even more
closely than nature! \\
We conclude our discussion of the charmed meson masses by remarking
that chiral symmetry (and its breaking) may still have a sizable influence
in the charm sector. In this context it is interesting to note that in the theoretical works of
Refs.\ \cite{Bardeen:1993ae, Bardeen:2003kt, Nowak:1992um, Nowak:2003ra, Sasaki:2014asa} the degeneracy of vector and axial-vector
charmed states in the heavy-quark limit [see Eqs.\ (\ref{diffmas1}-\ref{diffmas3})] as well as
of scalar and pseudoscalar charmed states was obtained by combining the
heavy-quark and the (linear) chiral symmetries.

\newpage
\chapter{Particle decays}\label{Decay width}

\section{Introduction}

The study of decays of resonances plays a central role in
understanding hadron phenomenology. One can describe the behaviour and
the inner structure of subatomic
particles in mathematical expressions of decays based on Feynman diagrams, which are applications of quantum field theory. \\
\indent An unstable particle transforms spontaneously into other
more stable particles, with less mass. This can be followed
 by another transformation, until one arrives at the lightest particles. These transformations occur according
to some conservation rules and are called decays. There are three
types of decays:
 strong, weak, and electromagnetic. Strong decays occur when an unstable meson (quark-antiquark
state with gluons mediating the interaction) decay into lighter
mesons. Weak decays occur when a quark couples to the massive
bosons $W^\pm$ and $Z^0$ due to weak interactions. A famous
electromagnetic decay is that of the neutral pion into two photons ($\pi^0 \rightarrow 2\gamma$), but not into objects consisting of charged subcomponents.\\
\indent In the 1960s, Okubo, Zweig, and lizuka explained
independently the reason why certain decay modes appear less
frequently than otherwise might be expected. This is summarized in
their famous rule (OZI rule) \cite{Okubo:1963fa, Zweig:1964jf, Iizuka:1966fk} which
states that decays whose Feynman diagrams contain disconnected
quark lines
 (which occur by cutting internal gluon lines) are suppressed. \\
\indent The probability of a decay's occurrence can be computed.
For a particle with rest mass $M$ with energy and momentum $(E,\,\overrightarrow{P})$, the survival
probability ${\mathbb P}(t)$ (the probability that the particle survives for a
time $t$ before decaying) \cite{Beringer:1900zz} is
\begin{equation}
{\mathbb P}(t)=e^{-t/\gamma\tau}=e^{-Mt\Gamma/E}.
\end{equation}
where $\tau(\equiv 1/\Gamma)$ is the mean life time and $\Gamma$
is the decay width. After the particle moves a distance
$x$, the probability is
\begin{equation}
{\mathbb P}(t)=e^{-t/\gamma\tau}=e^{-Mx\Gamma/|\overrightarrow{P}|}.
\end{equation}
In this chapter, we will study two-body decays of mesons. Also three-body decays of mesons will
be analyzed. Moreover, the decay constant of mesons will be calculated. Therefore, in the present chapter, we develop formalisms for computing the corresponding decay constants and
for both the two- and three-body decays.

\newpage

\section{Decay constants}

In this section we develop the general formula for the decay
constant by using the transformation method. As seen in Sec. 2.5 the matrix $\Phi$ is a combination of scalar and pseudoscalar currents:

\begin{equation}
\Phi=S+iP\text{\,.} \label{PhiSP}%
\end{equation}

The hermitian conjugate is
\begin{equation}
\Phi^\dagger=S-iP\text{\,,} \label{PhidSP}%
\end{equation}
which transforms as

\begin{equation}
\Phi^{\dagger}\rightarrow U_{R}\Phi^{\dagger} U_{L}^{\dagger}\,. \label{Phidtr}%
\end{equation}

The pseudoscalar matrix can be written as
\begin{equation}
P=\frac{1}{2i}(\Phi-\Phi^\dagger)\text{ ,} \label{PhidSP}%
\end{equation}

which transforms as

\begin{equation}
P\rightarrow \frac{1}{2i}(U_{L}\Phi U_{R}^{\dagger}-U_{R}\Phi^{\dagger} U_{L}^{\dagger})\,, \label{Phidtr}%
\end{equation}

with
\begin{equation}
U_{L}\in U(N_f)_{L},\,\,U_{R}\in U(N_f)_{R}\,. \label{Phidtr}%
\end{equation}

The chiral symmetry group of QCD is

\begin{equation}
U(N_f)_{L}\times U(N_f)_{R}\equiv U(1)_V\times SU(N_f)_V\times U(1)_A\times SU(N_f)_A\,. \label{Phidtr}%
\end{equation}

The $U(1)_{V}$ transformation corresponds to

\begin{equation}
U_{1} = U_L = U_R =e^{i\theta t^0}\,,
\end{equation}
while an $SU(N_f)_V$ transformation corresponds to
\begin{equation}
U_{V} = U_L = U_R =e^{i\theta^V_a t^a}\,,\,\,\,\,\,
\end{equation}
and an $SU(N_f)_A$ transformation corresponds to
\begin{equation}
U_{A} = U_L = U_R^\dagger =e^{i\theta^A_a t^a}\,,\,\,\,\,\,
\end{equation}
 where $\theta^{V,A}_a $ are the parameters, and $t^a$ are the generators of the group, with $a=1,...,N_f^2-1$. For $N_f=2$,  $t_0=\frac{1}{2}1_2$ and
 $t_i=\frac{1}{2}\tau_i$ where $\tau_i$ are the Pauli matrices.
  Note that the matrices of the scalar mesons $S$ and of the pseudoscalar mesons in the model (\ref{SPEqm1}) are hermitian. Therefore they can be decomposed in terms of
  generators $t^a$ of a unitary group $U(N_f)$
  with  $a=0,...,N_f^2-1$. For small parameters
\begin{equation}
 U=1+i\theta_a\,t^a,\label{UA}%
\end{equation}
Under a $U_A(N_f)$ transformation, the pseudoscalar matrix becomes
\begin{equation}
P\rightarrow \frac{1}{2i}(U_{A}\Phi U_{A}-U_{A}^{\dagger}\Phi^{\dagger} U_{A}^{\dagger})\,. \label{Ptr}%
\end{equation}
Using Eqs.(\ref{PhiSP}), (\ref{PhidSP}), and (\ref{UA}), we obtain
the transformation of fields
\begin{equation}
P\rightarrow P+\theta^A_a \{t_a,S\}\,, \label{Ptr}%
\end{equation}
Introducing the wave-function renormalization of the fields
\begin{equation}
Z_i\,P\rightarrow Z_i\,P+\theta^A_a \{t_a,S\}\,,
\end{equation}
which can be written as 
\begin{equation}
P\rightarrow P+\theta^A_a\,\frac{\{t_a,S\}}{Z_i}\,,%
\end{equation}
The weak decay constants take the following form
\begin{equation}
f_i=\frac{\{t_a,S\}}{Z_i}\,.%
\end{equation}

\section{Two-body decay}

In this section, we investigate various aspects of the decay of a particle into a two-particle final state \cite{FGThesis}. \\

\begin{figure}[htb]
\begin{center}
\includegraphics[width=5cm]{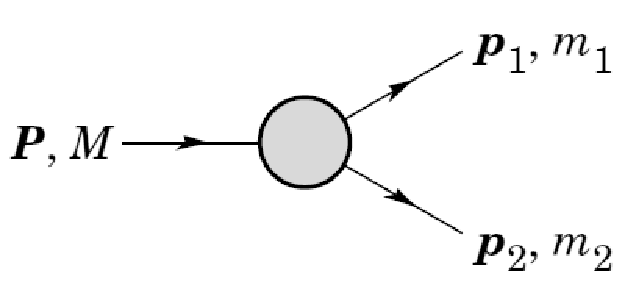}
\caption{Feynman diagram of a two-body decay \cite{Beringer:1900zz}.} \label{fig:2bodydecay}
\end{center}
\end{figure}

Consider a particle with four-momenta $P$ described by
$P(M,0)$, in its rest frame, which decays into two
particles with momenta $p_i$ and masses $m_i$, where $i=1,2$,
with $M> m_1+m_2$. This two-body decay is described by the Feynman diagram in Fig.
\ref{fig:2bodydecay}.\\

The initial state of the decaying particle and the final state can be defined as

\begin{align}
&|in> = b^\dagger_{P}\mid0>\,,\\
&|fin> = a^\dagger_{p_1}a^\dagger_{p_2}\mid0>\,.
\end{align}

Assume that the decaying particle and the decay products are
confined in a `box' with length $L$ and volume $V=L^3$. The
three-momenta are quantized in the box, as known from quantum
mechanics, as $P=2\pi n_{P}/L$ and $p_{1,2}=2\pi
n_{p_{1,2}}/L$. The corresponding energies of these particles
are described by $w_{P}=\sqrt{M^2+P^2}$
and $w_{p_1,p_2}=\sqrt{m^2_{1,2}+p^2_{1,2}}$.\\

The corresponding matrix element of the scattering matrix in terms
of the initial and final states is

\begin{equation}\label{apm1}
<fin|S^{(1)}|in> =
\frac{S}{V^{3/2}}\frac{-i\mathcal{M}}{\sqrt{2\omega_{p_1}2\omega_{p_2}2\omega_{P}}}(2\pi)^4\delta^4(P-p_1-p_2)\,,
\end{equation}
where $S$ refers to a symmetrization factor,
$\delta^4(P-p_1-p_2)$ denotes the delta distribution
which describes the energy-momentum conservation for each vertex,
and $\mathcal{M}$ is the corresponding tree-level
decay amplitude.\\
The squared modulus of the scattering matrix (\ref{apm1}) gives
the probability for the particle decaying from the initial
state $|in>$ state to the final $|fin>$
as:
\begin{equation}
|<fin|S^{(1)}|in>|^2 =
\frac{1}{V^3}\frac{|-i\mathcal{M}|^2}{\sqrt{2\omega_{p_1}2\omega_{p_2}2\omega_{P}}}(2\pi)^8(\delta^4(P-p_1-p_2))^2\,.\label{2pd1AB}
\end{equation}
Using the ``Fermi trick'' \cite{ONachtmann}, the
square of the delta distribution can be obtained as follows:
\begin{align}
(2\pi)^8(\delta^4(P-p_1-p_2))^2&=(2\pi)^4\delta^4(P-p_1-p_2)\int d^4x\,e^{ix(\textbf{P}-p_1-p_2)}\nonumber\\
\,\,\,\,\,\,\,\,\,\,\,\,\,\,\,\,\,\,&=(2\pi)^4\delta^4(P-p_1-p_2)\int d^4x\nonumber\\
\,\,\,\,\,\,\,\,\,\,\,\,\,\,\,\,\,\,&=(2\pi)^4\delta^4(P-p_1-p_2)\int_V d^3x\int^t_0dt\nonumber\\
\,\,\,\,\,\,\,\,\,\,\,\,\,\,\,\,\,\,&=(2\pi)^4\delta^4(P-p_1-p_2)Vt\,,\label{del1}
\end{align}
where $x$ is the Minkowski space-time vector. Consequently, the
probability of the two-body decay (\ref{2pd1AB}) can be written
as
\begin{equation}
|<fin|S^{(1)}|in>|^2 =
\frac{1}{V^3}\frac{|-i\mathcal{M}|^2}{\sqrt{2\omega_{p_1}2\omega_{p_2}2\omega_{P}}}(2\pi)^4\,\delta^4(P-p_1-p_2)Vt\,.\label{2pd1AB2}
\end{equation}
The number of final states is obtained as the factor
$V\,\frac{d^3p_1}{(2\pi)^3}\,V\,\frac{d^3p_2}{(2\pi)^3}$ when the
three-momenta of the outgoing particles lie between
$(p_1,\,p_1+d^3p_1)$ and $(p_2,\,p_2+d^3p_2)$. Consequently, the
probability for the decay in the momentum range becomes
\begin{align}
|<fin|S^{(1)}|in>|^2\, V\,
\frac{d^3p_1}{(2\pi)^3}\,V\,\frac{d^3p_2}{(2\pi)^3}=\frac{S|-i\mathcal{M}|^2}{2\omega_{p_1}2\omega_{p_2}2\omega_{P}}
\,\,\,\,\,\,\,\,\,\,\,\,\,\,\,\,\,\,\,\,\,\,\,\,\,\,\,\,\,\,\,\,\,\,\,\,\,\,\,\,\,\,\,\,\,\,\,\,\,\,\,\,\,\,\,\,\,\,\,\,\,\,\,\,\,\,\,\,\,\,\,\,\,\,\,\,\,\,\,\,\,\,\,\,\,\,\,\nonumber\\\times(2\pi)^4\delta^4(P-p_1-p_2)\times\frac{d^3p_1}{(2\pi)^3}\,\times\frac{d^3p_2}{(2\pi)^3}\,t\,.
\end{align}
By integrating over all possible final momenta ($p_1$ and $p_2$),
we obtain the definition of the decay rate $\Gamma$ of the two-body decay
as
\begin{equation}
\Gamma=S\int
\frac{d^3p_1}{(2\pi)^3}\,\frac{d^3p_2}{(2\pi)^3}\,\frac{|-i\mathcal{M}|^2}{2\omega_{p_1}\,2\omega_{p_2}2\omega_{P}}\,(2\pi)^4\,\delta^4(P-p_1-p_2)\,.\label{gamadp22}
\end{equation}
The probability to obtain the two particles as decay
products at any instant $t$ is
\begin{equation}
{\mathbb P}_d(t)=\Gamma t\,.
\end{equation}
Consequently, the probability of the initial particle surviving at
the same instant is
\begin{equation}
{\mathbb P}_s(t)=1-\Gamma t\,.
\end{equation}
which holds only when $t<<\Gamma^{-1}$. The mean-life time is
\begin{equation}
\tau=\Gamma^{-1}\,.
\end{equation}
Now, let us turn to the evaluation of the decay rate (\ref{gamadp22}) of the
two-body decay:
\begin{equation}
\Gamma=\frac{S}{2(2\pi)^2}\int
d^3p_1\,d^3p_2\,\frac{|-i\mathcal{M}|^2}{2\omega_{p_1}\,2\omega_{p_2}2\omega_{P}}\,\delta^4(P-p_1-p_2)\,.\label{gmdw2b}
\end{equation}
Consider the two outgoing particles to have the same mass
($m_1=m_2$) and note that, in the rest frame of the decaying particle
the four momentum is $P=(M,0)$. Therefore, the delta function
can be obtained as
\begin{align}
\delta^4(P-p_1-p_2)&=\delta^3(p_1+p_2)\,\delta(M-\omega_{p_1}-\omega_{p_2})\nonumber\\
&=\delta^3(p_1+p_2)\delta(M-2\omega_{p_1})\,.
\end{align}

Using the Dirac delta function to solve the integral over
$d^3p_2$, we obtain
\begin{equation}
\Gamma=\frac{S}{2(2\pi)^2}\int
d^3p_1\,\frac{|-i\mathcal{M}|^2}{(2\omega_{p_1})^2\,2M}\,\delta(M-2\omega_{p_1})\,.\label{gmdw2b}
\end{equation}

When we use the generic identity $\delta(g(x))=\sum_i
\delta(x-x_i)/|g'(x_i)|$, where $g(x_i)=0$, the $\delta$
distribution can be written as

\begin{equation}
\delta(M-2\omega_{p_1})=\frac{4M}{k_f}\delta(|p_1|-k_f)\,,
\end{equation}

 where the energy-momentum conservation gives

 \begin{equation}
 |p_1|=\sqrt{\frac{M^2}{4}-m_{1,2}^2}\equiv k_f,\,\,\,\,\,\,\,\,\,\,\,\,\,\text{for}\,\,\,\,\,\,(m_1=m_2)\,.
 \end{equation}

Using the spherical coordinates $d^3p_1\equiv p_1^2 d\varOmega
d|p_1|$ and integrating over $d|p_1|$, the decay rate
(\ref{gmdw2b}) becomes

\begin{equation}
\Gamma=\frac{S\,k_f}{32\,\pi^2\,M^2}\int
d\varOmega|-i\mathcal{M}|^2\,,\label{gmmdb2}
\end{equation}

When the decay amplitude does not depend on the angle, we obtain
the general formula \cite{Parganlija:2012xj, FGThesis} of the
two-body decay rate as

\begin{equation}
\label{B1}\Gamma_{A\rightarrow BC}=\textit{I}\frac{S k_f}{8 \pi
M^{2}}|-i\mathcal{M}|^{2}\,,
\end{equation}

with the center-of-mass momentum of the two decay particles

\begin{equation}
k_f=\frac{1}{2\,M}\sqrt{M^{4}+(m_{1}^{2}-m_{2}^{2})^{2}-2M^{2}\,(m_{1}^{2}+m_{2}^{2})}\theta(M-m_{1}-m_{2})\,,
\label{B2}%
\end{equation}

and the symmetrization factor $S$ equals $1$ if the outgoing
particles are different, and $2$ for two identical
outgoing particles (because of the inter change ability of outgoing particles lines). The isospin factor $\textit{I}$ considers all subchannels of a particular decay channel. The decay
threshold is encoded by the $\theta$ function.

\section{Three-body decay}

In this section we expand our investigation to various aspects of
the decay into a three-body final state
 which is more complicated than the two-body decay, as we will see in the following.\\

Consider a particle $A$ with four-momentum $P$ in its rest frame,
with $P=(M,0)$, decaying into three particles $(B_1,\,B_2,\,B_3)$
with momenta $p_i$ and mass $m_i$, where $i=1,2,3$, with
$M>m_1+m_2+m_3$. This decay is confined in a ``box'' that has length
$L$ with volume $V=L^3$. The momenta of the decaying particle and
the three outgoing particles are quantized in the box as $P=2\pi
n_P/L$ and $p_{1,2,3}=2\pi n_{p_{1,2,3}}/L$, respectively. The
corresponding energies for the decaying particle are
$\omega_P=\sqrt{M^2+P^2}$ whereas those for the produced particles are
$\omega_{p_{1,2,3}}=\sqrt{m_{1,2,3}^2+p_{1,2,3}^2}$. The Feynman
diagram that describes the three-body decay is presented in Fig. 5.2.

\begin{figure}[H]
\begin{center}
\includegraphics[width=5cm]{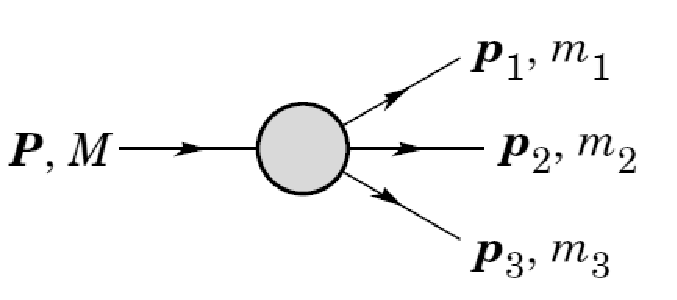}
\caption{Feynman diagram of the three body decay \cite{Beringer:1900zz}.} \label{fig:3bodydecay}
\end{center}
\end{figure}

The initial and final states read
\begin{align}
&|in> = b^\dagger_P\mid0>\,,\\
&|fin> = a^\dagger_{p_1}a^\dagger_{p_2}a^\dagger_{p_3}\mid0>\,,
\end{align}

and the corresponding element of the scattering matrix is

\begin{align}\label{amp}
<fin|S^{(1)}|in> =&
\frac{1}{V^{3/2}}\frac{1}{\sqrt{2\omega_{p_1}2\omega_{p_2}2\omega_{p_3}2\omega_{P}}}(2\pi)^4\nonumber\\&\times\delta^4(P-p_1-p_2-p_3)(-i\mathcal{M}_{A\rightarrow
B_{1}B_{2}B_{3}})\,,
\end{align}

where $-iM_1$ is the invariant amplitude of the vertex function
that enters into the Feynman rule for the vertex $A-B_1B_2B_3$.
The probability for the process $A\rightarrow B_1B_2B_3$ can be
computed by squaring the amplitude (\ref{amp})

\begin{align}
|<fin|S^{(1)}|in>|^2 =&
\frac{1}{V^3}\frac{1}{\sqrt{2\omega_{p_1}2\omega_{p_2}2\omega_{p_3}2\omega_{P}}}(2\pi)^8\nonumber\\&\times(\delta^4(P-p_1-p_2-p_3))^2|-i\mathcal{M}_{A\rightarrow
B_{1}B_{2}B_{3}}|^2\,.\label{p1ABBB}
\end{align}
Due to the Fermi Trick \cite{ONachtmann}, we simplify the delta
squared as
\begin{align}
(2\pi)^8(\delta^4(P-p_1-p_2-p_3))^2&=(2\pi)^4\delta^4(P-p_1-p_2-p_3)\int d^4x\,e^{ix(P-p_1-p_2-p_3)}\nonumber\\
\,\,\,\,\,\,\,\,\,\,\,\,\,\,\,\,\,\,&=(2\pi)^4\delta^4(P-p_1-p_2-p_3)\int d^4x\nonumber\\
\,\,\,\,\,\,\,\,\,\,\,\,\,\,\,\,\,\,&=(2\pi)^4\delta^4(P-p_1-p_2-p_3)\int_V d^3x\int^t_0dt\nonumber\\
\,\,\,\,\,\,\,\,\,\,\,\,\,\,\,\,\,\,&=(2\pi)^4\delta^4(P-p_1-p_2-p_3)Vt\,,\label{del1}
\end{align}
where $x=(t,x_1,x_2,x_3)$. The probability for the decay
$A\rightarrow B_1B_2B_3$ when the three particles
$B_1,\,B_2,\,{ and}\, B_3$ have momenta between
$(\textbf{p}_1,\,\textbf{p}_1+d^3p_1)$,
$(\textbf{p}_2,\,\textbf{p}_2+d^3p_2)$ and
$(\textbf{p}_3,\,\textbf{p}_3+d^3p_3)$, respectively, is given by
\begin{equation}
|<fin|S^{(1)}|in>|^2\, V\,
\frac{d^3p_1}{(2\pi)^3}\,V\,\frac{d^3p_2}{(2\pi)^3}\,V\,\frac{d^3p_3}{(2\pi)^3}\,,\label{pABBB1}
\end{equation}
where $Vd^3p_1/(2\pi)^3$ describes the number of states with four
momenta between $(\textbf{p}_1,\,\textbf{p}_1+d^3p_1)$. Using Eqs.
(\ref{p1ABBB}) and (\ref{del1}), the probability of the three-body
decay Eq.(\ref{pABBB1}) becomes
\begin{align}
|<fin|S^{(1)}|in>|^2\, V\,
&\frac{d^3p_1}{(2\pi)^3}\,V\,\frac{d^3p_2}{(2\pi)^3}\,V\,\frac{d^2p_3}{(2\pi)^3}&\nonumber\\
&=\frac{1}{2\omega_{p_1}2\omega_{p_2}2\omega_{p_3}2\omega_{P}}\times(2\pi)^4\delta^4(P-p_1-p_2-p_3)\nonumber\\&\times\frac{d^3p_1}{(2\pi)^3}\,\frac{d^3p_2}{(2\pi)^3}\,\frac{d^2p_3}{(2\pi)^3}|-i\mathcal{M}_{A\rightarrow
B_{1}B_{2}B_{3}}|^2t\,,
\end{align}
which does not depend on the normalization volume $V$. By
integrating over all possible final momenta, we can obtain the
decay rate $\Gamma$ as:
\begin{equation}
\Gamma=\int
\frac{d^3p_1}{(2\pi)^3}\,\frac{d^3p_2}{(2\pi)^3}\,\frac{d^3p_3}{(2\pi)^3}\frac{|-i\mathcal{M}_{A\rightarrow
B_{1}B_{2}B_{3}}|^2}{2\omega_{p_1}2\omega_{p_2}2\omega_{p_3}2\omega_{P}}(2\pi)^4\delta^4(P-p_1-p_2-p_3)\,,
\end{equation}
which can be written as
\begin{equation}
\Gamma=\frac{1}{(2\pi)^5}\int
d^3p_1\,d^3p_2\,d^3p_3\frac{|-i\mathcal{M}_{A\rightarrow
B_{1}B_{2}B_{3}}|^2}{2\omega_{p_1}2\omega_{p_2}2\omega_{p_3}2\omega_{P}}\delta^4(P-p_1-p_2-p_3)\,.\label{Gama}
\end{equation}
In the rest frame of the decaying particle with $P=(M,0)$ we have
\begin{equation}
\delta^4(P-p_1-p_2-p_3)=\delta^3(p_1+p_2+p_3)\delta(M-\omega_{p_1}-\omega_{p_2}-\omega_{p_3})\,.\label{del2}
\end{equation}
Substituting Eq.(\ref{del2}) into Eq.(\ref{Gama}), we obtain
\begin{align}
\Gamma=\frac{1}{16(2\pi)^5}&\int
d^3p_1\,d^3p_2\,d^3p_3\frac{|-i\mathcal{M}_{A\rightarrow
B_{1}B_{2}B_{3}}(p_1,\,p_2,\,p_3)|^2}{\omega_{p_1}\omega_{p_2}\omega_{p_3}M}\nonumber\\&\times\delta^3(p_1+p_2+p_3)\delta(M-\omega_{p_1}-\omega_{p_2}-\omega_{p_3})\,,
\end{align}
solving the integral over $d^3p_3$ (by use of the Dirac delta
function), we get
\begin{equation}
\Gamma=\frac{1}{16(2\pi)^5}\int
d^3p_1\,d^3p_2\,\frac{|-i\mathcal{M}_{A\rightarrow
B_{1}B_{2}B_{3}}(p_1,\,p_2,\,-(p_1+p_2)|^2}{\omega_{p_1}\omega_{p_2}\omega_{p_3}M}\,\delta(M-\omega_{p_1}-\omega_{p_2}-\omega_{p_3})\,,\label{gama4}
\end{equation}
where $\omega_{p_3}=(p_3^2+m_{B_3}^2)^{1/2}$ and now $p_3$ is a
notation for $-(p_1+p_2)$. The matrix element depends on the angle
$\theta$ between $p_1$ and $p_2$, and so Eq. (\ref{gama4}) becomes
\begin{align}
\Gamma=\frac{1}{8(2\pi)^3}\int
\frac{p_1\,dp_1.\,p_2dp_2}{\omega_{p_1}\omega_{p_2}\omega_{p_3}M}\,p_1\,p_2\,d\cos\theta\,&|-i\mathcal{M}_{A\rightarrow
B_{1}B_{2}B_{3}}\{p_1,\,p_2,\,-(p_1+p_2)\}|^2\,\nonumber\\&\times\delta(M-\omega_{p_1}-\omega_{p_2}-\omega_{p_3})\,,\label{gama5}
\end{align}
but
$$\omega_{p_1}^2=p_1^2+m_{B_1}^2\,\,\Rightarrow\,\,2\omega_{p_1}d\omega_{p_1}=2p_1dp_1,$$
\begin{equation}
\therefore\,\,\omega_{p_1}d\omega_{p_1}=p_1dp_1\,.\label{omg1}
\end{equation}
Similarly,
\begin{equation}
\omega_{p_2}d\omega_{p_2}=p_2dp_2\,,\label{omg2}
\end{equation}
$$\omega^2_{p_3}=(\overrightarrow{p_1}+\overrightarrow{p_2})^2+m_{B_3}^2=p_1^2+p_2^2+2p_1p_2\cos\theta+m_{B_3}^2,$$
At $p_1$ and $p_2$ fixed, we have
\begin{equation}
2\omega_{p_3}d\omega_{p_3}=2p_1\,p_2\,d\cos\theta\,,\label{omg3}
\end{equation}
Substituting Eqs.(\ref{omg1} - \ref{omg3}) into
Eq.(\ref{gama5}), we obtain
\begin{equation}
\Gamma=\frac{1}{8(2\pi)^3}\int
\frac{d\omega_{p_1}\,d\omega_{p_2}\,d\omega_{p_3}}{M}\,|-i\mathcal{M}_{A\rightarrow
B_{1}B_{2}B_{3}}|^2\,\delta(M-\omega_{p_1}-\omega_{p_2}-\omega_{p_3})\,,\label{gama6}
\end{equation}
Using the Dirac delta to eliminate $\omega_{p_3}$
\begin{align}
\Gamma&=\frac{1}{(2\pi)^3}\frac{1}{8M}\int |-i\mathcal{M}_{A\rightarrow
B_{1}B_{2}B_{3}}|^2\,d\omega_{p_1}\,d\omega_{p_2}\,,\nonumber\\
&=\frac{1}{(2\pi)^3}\frac{1}{32M^3}\int
|-i\mathcal{M}_{A\rightarrow
B_{1}B_{2}B_{3}}|^2\,dm_{12}^2\,dm^2_{23}\,,
\end{align}
which is the standard form for the Dalitz plot \cite{Beringer:1900zz}.

\begin{figure}[H]
\begin{center}
\includegraphics[width=8cm]{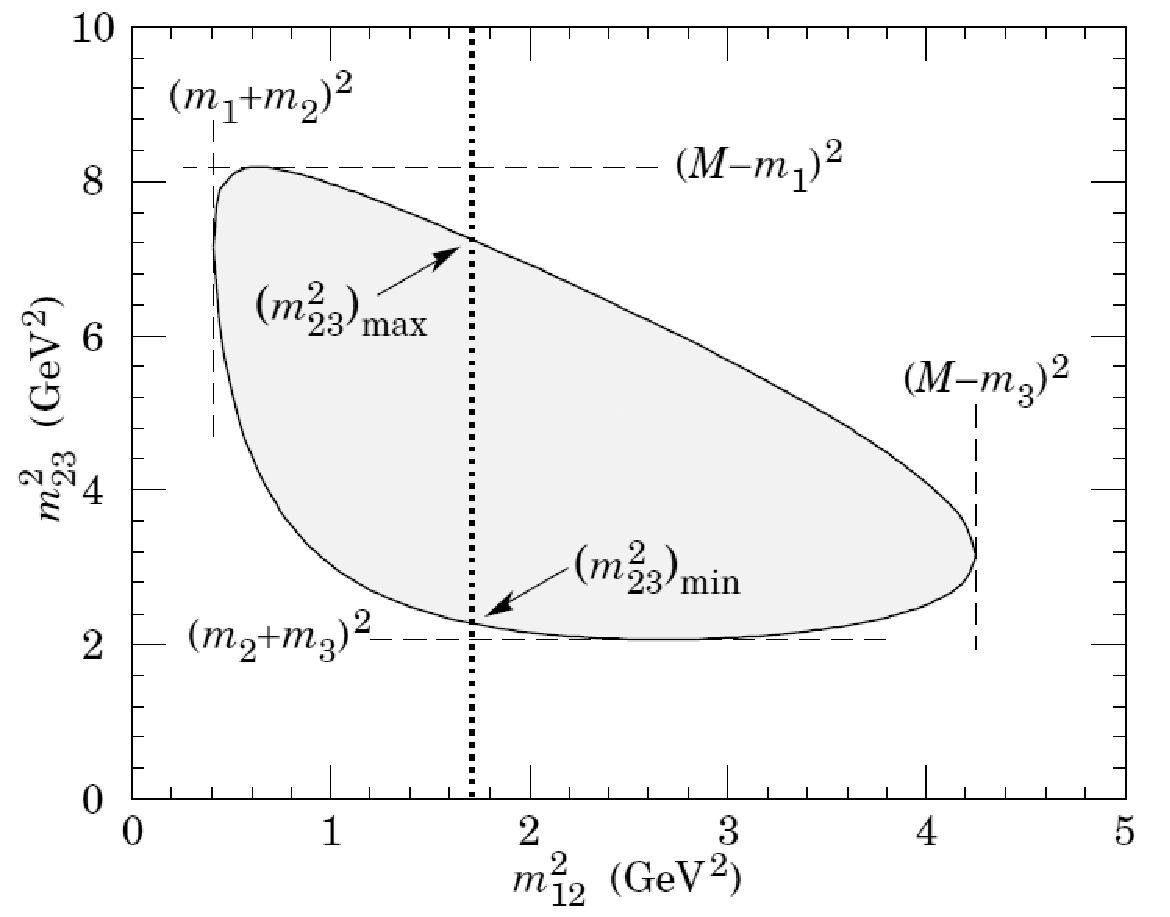}
\caption{Dalitz plot for a three-body final state \cite{Beringer:1900zz}.}
\label{fig:Dalitz plot}
\end{center}
\end{figure}

The limits of integration can be determined from the Dalitz plot
\cite{Beringer:1900zz}, which leads to the following formula for the decay
rate of A into $B_1,\,B_2,\,B_3$
\begin{equation}
\Gamma_{A\rightarrow B_{1}B_{2}B_{3}}=\frac{S_{A\rightarrow B_{1}B_{2}B_{3}}%
}{32(2\pi)^{3}M^{3}}\int_{(m_{1}+m_{2})^{2}}^{(M-m_{3})^{2}}%
\int_{(m_{23}^2)_{\min}}^{(m_{23}^2)_{\max}}|-i\mathcal{M}_{A\rightarrow
B_{1}B_{2}B_{3}}|^{2}\,dm_{23}^{2}\,dm_{12}^{2}\,. \label{B3}%
\end{equation}
Integrating over $m_{23}^2$, we obtain
\begin{align}
\Gamma_{A\rightarrow
B_{1}B_{2}B_{3}}=\frac{S_{A\rightarrow B_{1}B_{2}B_{3}}}{32(2\pi)^3\,M^3}\,\int_{(m_1+m_2)^2}^{(M-m_3)^2}|-i\mathcal{M}_{A\rightarrow
B_{1}B_{2}B_{3}}|^{2}\,[(m_{23}^2)_{ max}-(m_{23}^2)_{
min}]\,dm_{12}^2\,.
\end{align}
The range of $m_{23}^2$ is determined by the value of $m_{12}^2$
when $p_2$ is parallel or antiparallel to $p_3$ as follows \cite{Beringer:1900zz}
\begin{align}
(m^2_{23})_{\min}  &  =(E_{2}^{\ast}+E_{3}^{\ast})^{2}-\left(  \sqrt{E_{2}%
^{\ast2}-m_{2}^{2}}+\sqrt{E_{3}^{\ast2}-m_{3}^{2}}\right)  ^{2}\text{ ,}\\
(m^2_{23})_{\max}  &  =(E_{2}^{\ast}+E_{3}^{\ast})^{2}-\left(  \sqrt{E_{2}%
^{\ast2}-m_{2}^{2}}-\sqrt{E_{3}^{\ast2}-m_{3}^{2}}\right)  ^{2}\text{ ,}%
\end{align}
where $E^\ast_2$ and $E^\ast_3$ are the energies of particles
$B_2$ and $B_3$, respectively, in the $m_{12}$ rest frame,
\begin{align}
&E_{2}^{\ast}=\frac{m_{12}^{2}-m_{1}^{2}+m_{2}^{2}}{2m_{12}}\text{ , }\nonumber\\
&E_{3}^{\ast}=\frac{M^{2}-m_{12}^{2}-m_{3}^{2}}{2m_{12}}\text{
. }
\label{eneries}%
\end{align}
Finally, the explicit expression for the three-body decay
rate for the process $A\rightarrow B_{1}B_{2}B_{3}$ is:%
\begin{align}
\Gamma_{A\rightarrow B_{1}B_{2}B_{3}}&=\frac{S_{A\rightarrow
B_{1}B_{2}B_{3}}}{32(2\pi)^{3}M^{3}}\int_{(m_{1}+m_{2})^{2}%
}^{(M-m_{3})^{2}}|-i\mathcal{M}_{A\rightarrow B_{1}B_{2}B_{3}}%
|^{2}\nonumber\\&\times\sqrt{\frac{(-m_{1}+m_{12}-m_{2})(m_{1}+m_{12}-m_{2})(-m_{1}%
+m_{12}+m_{2})(m_{1}+m_{12}+m_{2})}{m_{12}^{2}}}\nonumber\\&\times
\sqrt{\frac{(-M+m_{12}-m_{3})(M+m_{12}-m_{3}%
)(-M+m_{12}+m_{3})(M+m_{12}+m_{3})}{m_{12}^{2}}%
}dm_{12}^{2}\text{ ,}\label{3bodydecay}
\end{align}
where $\mathcal{M}_{A\rightarrow B_{1}B_{2}B_{3}}$ is the
corresponding tree-level decay amplitude, and $S_{A\rightarrow
B_{1}B_{2}B_{3}}$ is a symmetrization factor (it is equal to $1$ if $B_{1},$ $B_{2}$, and $B_{3}$ are all different, and equal to $2$ for two
identical particles in the final state, and equal to $6$ for three
identical particles in the final state).


\chapter{Decay of the pseudoscalar glueball into scalar and pseudoscalar mesons}\label{Pseudoscalar glueball}

``\textit{The most exciting phrase to hear in science, the one that
heralds new discoveries, is not `Eureka' (I found it), but `that's
funny'...}'' \\

$\,\,\,\,\,\,\,\,\,\,\,\,\,\,\,\,\,\,\,\,\,\,\,\,\,\,\,\,\,\,\,\,\,\,\,\,\,\,\,\,\,\,\,\,\,\,\,\,\,\,\,\,\,\,\,\,\,\,\,\,\,\,\,\,\,\,\,\,\,\,\,\,\,\,\,\,\,\,\,\,\,\,\,\,\,\,\,\,\,\,\,\,\,\,\,\,\,\,\,\,\,\,\,\,\,\,\,\,\,\,\,\,\,\,\,\,\,\,\,\,\,\,\,\,\,\,\,\,\,\,\,\,\,\,\,\,\,\,\,\,\,\,\,\,\,\,\,\,\,\,\,\,\,\,\,\,\,\,\,\,\,\,\,\,\,\,\,\,\,\,\,\,\,\,\,\,\,\,\,\,\,\,\,\,\,\,\,\,\,\,\,\,\,\,\,\,\,\,\,$ Isaac Asimov

\section{Introduction}

The fundamental symmetry underlying Quantum Chromodynamics (QCD),
the theory of strong interactions, is the exact local $SU(3)_{c}$
colour symmetry. As a consequence of the non-Abelian nature of
this symmetry the gauge fields of QCD (the gluons) are coloured
objects and therefore interact strongly with each other. Because
of confinement, one expects that gluons can also form colourless,
or `white', states which are called glueballs.

In this chapter we study the decay properties of a pseudoscalar
glueball state whose mass lies, in agreement with lattice QCD,
between $2$ and $3$ GeV. Following Ref.\ \cite{Rosenzweig:1981cu, Rosenzweig:1982cb, Rosenzweig:1979ay, Kawarabayashi:1980dp} we write
down an effective chiral Lagrangian which couples the pseudoscalar
glueball field (denoted as $\tilde{G}$) to scalar and pseudoscalar
mesons. We can thus evaluate the widths for the decays
$\tilde{G}\rightarrow PPP$ and $\tilde{G}\rightarrow PS,$ where
$P$ and $S$ stand for pseudoscalar and scalar quark-antiquark
states. The pseudoscalar state $P$ refers to the well-known light
pseudoscalars $\{\pi,K,\eta ,\eta^{\prime}\}$, while the scalar
state $S$ refers to the quark-antiquark
nonet of scalars above 1 GeV: $\{a_{0}(1450),K_{0}^{\ast}(1430),f_{0}%
(1370),f_{0}(1710)\}$. The reason for the latter assignment is a
growing consensus that the chiral partners of the pseudoscalar
states should not be identified with the resonances below 1 GeV,
see Refs.\ \cite{Janowski:2011gt, Parganlija:2010fz, Parganlija:2012fy} for results within the extended
linear sigma model, see also other theoretical works in Refs.\
\cite{Amsler:1995td, Lee:1999kv, Close:2001ga, Giacosa:2005qr, Giacosa:2004ug, Mathieu:2008me, Heinz:2008cv, Maiani:2004uc, Giacosa:2006rg, Fariborz:2005gm, Fariborz:2003uj, Napsuciale:2004au, Giacosa:2009qh} (and refs.\ therein).

The chiral Lagrangian that we construct contains one unknown
coupling constant which cannot be determined without experimental
data. However, the branching ratios can be unambiguously
calculated and may represent a useful guideline for an
experimental search of the pseudoscalar glueball in the energy
region between $2$ to $3$ GeV. In this respect, the planned PANDA
experiment at the FAIR facility \cite{Lutz:2009ff} will prove fruitful,
since it will be capable of scanning the mass region above 2.5
GeV. The experiment is based on proton-antiproton scattering, thus
the pseudoscalar glueball $\tilde{G}$ can be directly produced as
an intermediate state. We shall therefore present our results for
the branching ratios for a putative pseudoscalar glueball with a
mass of 2.6 GeV.

In addition to the vacuum properties of a pseudoscalar glueball,
we describe (to our know-ledge \cite{Eshraim:2012jv, Eshraim:2012ju, Eshraim:2012rb, Eshraim:2013dn} for the first time) the interaction
of $\tilde{G}$ with baryons: we introduce the chiral effective
Lagrangian which couples $\tilde{G}$ to the nucleon field and its
chiral partner. This Lagrangian describes also the
proton-antiproton conversion process $\bar
{p}p\rightarrow\tilde{G}$, which allow us to study the decay of a
pseudoscalar glueball into two nucleons.

Additionally, it is possible that the pseudoscalar glueball
$\tilde{G}$ has a mass that is slightly lower than the lattice-QCD
prediction and that it has been already observed in the BESIII
experiment where pseudoscalar resonances have been investigated in
$J/\psi$ decays \cite{Ablikim:2005um, Kochelev:2005vd, Ablikim:2010au}. In particular, the resonance $X(2370)$
which has been clearly observed in the
$\pi^{+}\pi^{-}\eta^{\prime}$ channel represents a good candidate,
because it is quite narrow ($\sim80$ MeV) and its mass lies just
below the lattice-QCD prediction. For this reason we repeat our
calculation for a pseudoscalar glueball mass of $2.37$ GeV, and
thus make predictions for the resonance $X(2370)$, which can be
tested in the near future.


\section{The effective Lagrangian with a pseudoscalar glueball}

In this section we consider an $N_f=3$ chiral Lagrangian which
describes the interaction between the pseudoscalar glueball and
(pseudo)scalar mesons. We calculate the decay widths of the
pseudoscalar glueball, where we fixed its mass to 2.6 GeV, as
predicted by lattice-QCD simulations, and take a closer look at
the scalar-isoscalar decay channel. We present our results as
branching ratios which are relevant for the future PANDA
experiment at the FAIR facility.

We introduce a chiral Lagrangian which couples the pseudoscalar
glueball $\tilde{G}\equiv\left\vert gg\right\rangle$ with quantum
numbers $J^{PC}=0^{-+}$ to scalar and pseudoscalar mesons as
in Refs. \cite{Rosenzweig:1981cu, Eshraim:2012jv, Eshraim:2012ju, Eshraim:2012rb, Eshraim:2013dn, Kawarabayashi:1980dp}
\begin{equation}
\mathcal{L}_{\tilde{G}-mesons}^{int}=ic_{\tilde{G}\Phi}\tilde{G}\left(
\text{\textrm{det}}\Phi-\text{\textrm{det}}\Phi^{\dag}\right)
\text{ ,}
\label{intlag}%
\end{equation}
where $c_{\tilde{G}\phi}$ is a (unknown) dimensionless coupling
constant.
\begin{equation}
\Phi=(S^{a}+iP^{a})t^{a} \label{phimat}%
\end{equation}
represents the multiplet of scalar and pseudoscalar
quark-antiquark states, and $t^{a}$ are the generators of the
group $U(N_{f})$. In the present case $N_{f}=3$ the explicit
representation of the scalar and
pseudoscalar mesons reads \cite{Parganlija:2012fy, Parganlija:2012xj}:%
\begin{equation}
\Phi=\frac{1}{\sqrt{2}}\left(
\begin{array}
[c]{ccc}%
\frac{(\sigma_{N}+a_{0}^{0})+i(\eta_{N}+\pi^{0})}{\sqrt{2}} & a_{0}^{+}%
+i\pi^{+} & K_{0}^{\ast +}+iK^{+}\\
a_{0}^{-}+i\pi^{-} & \frac{(\sigma_{N}-a_{0}^{0})+i(\eta_{N}-\pi^{0})}%
{\sqrt{2}} & K_{0}^{\ast0}+iK^{0}\\
K_{0}^{\ast-}+iK^{-} & \bar{K}_{0}^{\ast 0}+i\bar{K}^{0} & \sigma_{S}+i\eta_{S}%
\end{array}
\right)  \; . \label{phimatex}%
\end{equation}
Let us consider the symmetry properties \cite{Amsler:2004ps, Klempt:2007cp} of the effective
Lagrangian (\ref{intlag}). The pseudoscalar glueball $\tilde{G}$
consists of gluons and is therefore a chirally invariant object.
Under $U(3)_{L}\times U(3)_{R}$ chiral transformations the
multiplet $\Phi$ transforms as $\Phi\rightarrow U_{L}\Phi
U_{R}^{\dagger}$ where $U_{L(R)}%
=e^{-i\theta_{L(R)}^{a}t^{a}}$ is an element of the group of $U(3)_{R(L)}$
matrices. Performing these transformations on the determinant of
$\Phi$ it is easy to prove that this object is invariant under
$SU(3)_{L} \times SU(3)_{R}$, but not under the axial $U(1)_{A}$
transformation,
\[
\mathrm{det}\Phi\rightarrow\mathrm{det}(U_{A}\Phi U_{A})=e^{-i\theta_{A}%
^{0}\sqrt{2N_{f}}}\mathrm{det}\Phi\neq\mathrm{det}\Phi\text{ .}%
\]

This is in agreement with the so-called axial anomaly. Consequently the effective
Lagrangian (\ref{intlag}) possesses only an $SU(3)_{L} \times
SU(3)_{R}$ symmetry. Further essential symmetries of the strong
interaction are the parity $P$ and charge
conjugation $C$. The pseudoscalar glueball and the
multiplet $\Phi$ transform under parity as
$$\tilde{G} (t,\vec{x}) \rightarrow -
\tilde{G}(t,-\vec{x})\,,$$
$$\Phi(t,\vec{x})\rightarrow\Phi^{\dagger}(t,-\vec{x})\,.$$
Performing the discrete transformations $P$ and
$C$ on the effective Lagrangian (\ref{intlag}) leave it
unchanged. In conclusion, one can say that the symmetries of the
effective Lagrangian (\ref{intlag}) are in agreement with the
symmetries of the QCD Lagrangian. The rest of the mesonic
Lagrangian which describes the interactions of $\Phi$ and also
includes (axial-)vector degrees of freedom is presented in Sec.\
\ref{nf3lag}.\\

\subsection{Implications of the interaction Lagrangian}

We have to consider that when $m_0^2<0$ spontaneous chiral
symmetry breaking occurs and the scalar-isoscalar fields condense.
When this breaking takes place, we need to shift the
scalar-isoscalar fields by their vacuum expectation values
$\phi_{N}$ and $\phi_{S}$,

\begin{equation}
\sigma_{N}\rightarrow\sigma_{N}+\phi_{N}\,,
\end{equation}
and 
\begin{equation}
\sigma_{S}\rightarrow
\sigma_{S}+\phi_{S}\text{ .} \label{shift}%
\end{equation}
In addition, when (axial-)vector mesons are present in the
Lagrangian, one also has to `shift' the (axial-)vector fields and to
define the wave-function
renormalization constants of the (pseudo)scalar fields:%
\begin{align}
& \vec{\pi}\rightarrow Z_{\pi}\vec{\pi}\text{, }\nonumber\\ &K^{i}\rightarrow Z_{K}%
K^{i}\text{, }\nonumber\\ & \eta_{j}\rightarrow Z_{\eta_{j}}\eta_{j}\text{, }\nonumber\\ & K^{\ast\,i}_0\rightarrow Z_{K^{\ast}_0} K^{\ast\,i}_0   \;, \label{psz}%
\end{align}
where $i=1,2,3,4$ runs over the four kaonic fields and $j=N,S.$
Once the field transformations in Eqs.\ (\ref{shift}) and
(\ref{psz}) have been performed, the Lagrangian
(\ref{intlag}) contains the relevant tree-level vertices for the
decay processes of $\tilde{G}$, and takes the form \cite{Eshraim:2012jv}:
\begin{align}
\mathcal{L}_{\tilde{G}-mesons}^{int}  =\frac{c_{\tilde{G}\Phi}}{2\sqrt{2}}%
&
\tilde{G}(\sqrt{2}Z_{K^{\ast}_{0}}Z_{K}a_{0}^{0}K_{0}^{\ast0}\overline{K}^{0}+\sqrt
{2}Z_{K}Z_{K^\ast_{0}}a_{0}^{0}K^{0}\overline{K}_{0}^{\ast}-2Z_{K_{0}^\ast}Z_{K}a_{0}%
^{+}K_{0}^{\ast0}K^{-}\nonumber\\
& 
 -2Z_{K_{0}^\ast}Z_{K}a_{0}^{+}K_{0}^{\ast-}K^{0}-2Z_{K_{0}\ast}Z_{K}a_{0}^{-}%
\overline{K}_{0}^{\ast 0}K^{+}-\sqrt{2}Z_{K_{0}^\ast}Z_{K}a_{0}^{0}K_{0}^{\ast-}K^{+}\nonumber\\
&
-\sqrt{2}Z_{K}^{2}Z_{\eta_{N}}K^{0}\overline{K}^{0}\eta_{N}
+\sqrt{2}Z_{K_{S}}^{2}Z_{\eta_{N}}K_{S}^{0}\overline{K}_{S}^{0}\eta
_{N}+2Z_{\eta_{S}}a_{0}^{-}a_{0}^{+}\eta_{S}\nonumber\\
&
-\sqrt{2}Z_{K}^{2}Z_{\eta_{N}}K^{-}K^{+}\eta_{N}-\sqrt{2}Z_{K}^{2}Z_{\pi
}K^{0}\overline{K}^{0}\pi^{0}+\sqrt{2}Z_{K_{0}^\ast}^{2}Z_{\pi}K_{0}^{\ast0}%
\overline{K}_{0}^{\ast0}\pi^{0}\nonumber\\
& 
+\sqrt{2}Z_{K}^{2}Z_{\pi}K^{-}K^{+}\pi
^{0}+Z_{\eta_{S}}{a_{0}^{0}%
}^{2}\eta_{S}
+Z_{\eta_{N}}^{2}Z_{\eta_{S}}\eta_{N}^{2}\eta_{S}-Z_{\eta_{S}}Z_{\pi}^{2}\eta_{S}\,{\pi^{0}}^{2}\nonumber\\
&
+2Z_{K}^{2}Z_{\pi}\overline{K}^{0}K^{+}\pi^{-}+2Z_{K}^{2}Z_{\pi}K^{0}K^{-}\pi^{+}-2Z_{K_{0}^\ast}%
^{2}Z_{\pi}K_{0}^{\ast0}K_{0}^{\ast-}\pi^{+}\nonumber\\
&  
-2Z_{\eta_{S}}Z_{\pi}^{2}\eta_{S}\pi^{-}\pi^{+}-2Z_{K_{0}^\ast}Z_{K}a_{0}%
^{-}K_{0}^{\ast +}\overline{K}^{0}+\sqrt{2}Z_{K_{0}^\ast}^{2}Z_{\eta_{N}}K_{0}^{\ast +}%
K_{0}^{\ast -}\eta_{N}\nonumber\\
&
-\sqrt{2}Z_{K_{0}^\ast}^{2}Z_{\pi}K_{0}^{\ast+}K_{0}^{\ast-}\pi
^{0} -2Z_{K_{0}^\ast}^{2}Z_{\pi}K_{0}^{\ast +}\overline{K}_{0}^{\ast 0}\pi^{-}
+2Z_{\pi}a_{0}^{0}\pi^{0}\phi_{S}\nonumber\\
&
+2Z_{\pi}a_{0}^{0}\pi^{0}\sigma
_{S}
-\sqrt{2}%
Z_{K_{0}^\ast}Z_{K}a_{0}^{0}K_{0}^{\ast+}K^{-}
+\sqrt{2}Z_{K}Z_{K_{0}^\ast}K^{-}K_{0}^{\ast +}%
\phi_{N} \nonumber\\
&+\sqrt{2}Z_{K}Z_{K_{0}^\ast}K^{-}K_{0}^{\ast +}\sigma_{N} +\sqrt{2}Z_{K_{0}^\ast}Z_{K}K_{0}^{\ast 0}\overline{K}^{0}\phi_{N}-2Z_{\eta_{N}}\eta_{N}\phi_{N}\phi_{S}\nonumber\\
&
+\sqrt{2}Z_{K_{0}^\ast%
}Z_{K}K_{0}^{\ast 0}\overline{K}^{0}\sigma_{N}+\sqrt{2}Z_{K}Z_{K_{0}^\ast}K^{0}%
\overline{K}_{0}^{\ast 0}\phi_{N}-Z_{\eta_{S}}\eta_{S}\phi_{N}^{2}\nonumber\\
&
+\sqrt{2}Z_{K}Z_{K_{0}^\ast}K^{0}\overline{K}_{0}%
^{\ast 0}\sigma_{N} +\sqrt{2}Z_{K_{0}^\ast}Z_{K}K_{0}^{\ast -}K^{+}\phi_{N}\nonumber\\
&
-Z_{\eta_{S}}\eta
_{S}\sigma_{N}^{2}-2Z_{\eta_{S}}\eta_{S}\phi_{N}\sigma_{N}+\sqrt{2}Z_{K_{0}^\ast}Z_{K}%
K_{0}^{\ast-}K^{+}\sigma_{N}\nonumber\\
& 
+2Z_{\pi}a_{0}^{+}\pi^{-}\phi_{S}+2Z_{\pi}a_{0}^{+}\pi^{-}\sigma
_{S}+2Z_{\pi}a_{0}^{-}\pi^{+}\phi_{S}+2Z_{\pi}a_{0}^{-}\pi^{+}\sigma
_{S}\nonumber\\
&  
-2Z_{\eta_{N}}\eta_{N}\phi_{N}%
\sigma_{S}-2Z_{\eta_{N}}\eta_{N}\sigma_{N}\phi_{S}-2Z_{\eta_{N}}\eta_{N}%
\sigma_{N}\sigma_{S})\text{ .}\label{explag}%
\end{align}

which is used to determine the coupling of the field $\tilde{G}$
to the scalar and pseudoscalar mesons.

\section{Field assignments}

\indent The assignment of the quark-antiquark fields in Eq.
(\ref{intlag}) or (\ref{phimatex}) is as follows: \\
(i) In the pseudoscalar sector the fields $\vec{\pi}$ and $K$
represent the pions
and the kaons, respectively \cite{Nakamura:2010zzi}. The bare fields $\eta_{N}%
\equiv\left\vert \bar{u}u+\bar{d}d\right\rangle /\sqrt{2}$ and $\eta_{S}%
\equiv\left\vert \bar{s}s\right\rangle $ are the non-strange and
strange
contributions of the physical states $\eta$ and $\eta^{\prime}$ \cite{Nakamura:2010zzi}:%
\begin{align}
&\eta=\eta_{N}\cos\varphi+\eta_{S}\sin\varphi,\text{
}\nonumber\\&\eta^{\prime}=-\eta
_{N}\sin\varphi+\eta_{S}\cos\varphi, \label{mixetas}%
\end{align}
where $\varphi\simeq-44.6^{\circ}$ is the mixing angle
\cite{Parganlija:2012fy}. Using other values for the mixing angle, e.g.\
$\varphi=-36^{\circ}$ \cite{Giacosa:2007up} or
$\varphi=-41.4^{\circ}$, as determined by the KLOE Collaboration
\cite{Ambrosino:2009sc}, affects the presented results only marginally.
In the effective Lagrangian (\ref{intlag}) there exists a mixing
between the bare pseudoscalar glueball $\widetilde{G}$ and both
bare fields $\eta_N$ and $\eta_S$, but due to the large mass
difference between the pseudoscalar glueball and the pseudoscalar
quark-antiquark fields, it turns out that this mixing is very small
and is therefore negligible.\\
(ii) In the scalar sector the field $\vec{a}_{0}$ corresponds to
the physical isotriplet state $a_{0}(1450)$ and the scalar kaon
field $K_{0}^\ast$ to the physical isodoublet state
$K_{0}^{\star}(1430)$ \cite{Nakamura:2010zzi}. The field
$\sigma_{N}\equiv(\bar{u}u+\bar{d}d)/\sqrt{2}$ is the bare
nonstrange isoscalar field and it corresponds to the resonance
$f_{0}(1370)$ \cite{Parganlija:2012fy, Cheng:2006hu}. The field
$\sigma_{S}\equiv\bar{s}s$ is the bare strange isoscalar field and
the debate about its assignment to a physical state is still
ongoing; in a first approximation it can be assigned to the
resonance $f_{0}(1710)$ \cite{Parganlija:2012fy} or $f_{0}(1500)$ \cite{Cheng:2006hu}.
(Scalars below 1 GeV are predominantly tetraquarks or mesonic
molecular states, as seen in Refs. \cite{Maiani:2004uc, Giacosa:2006rg, Fariborz:2005gm, Jaffe:1976ig, Napsuciale:2004au, vanBeveren:1986ea, Tornqvist:1995kr, Boglione:2002vv, vanBeveren:2006ua, Pelaez:2003dy, Oller:1997ti},
which are not considered here). In order to properly take into
account mixing effects in the scalar-isoscalar sector, we have
also used the results of Refs. \cite{Giacosa:2005zt, Cheng:2006hu}. The mixing takes
the form:

\begin{equation}
\left(
\begin{array}
[c]{c}%
\,f_{0}(1370)\\
\,f_{0}(1500)\\
\,f_{0}(1710)
\end{array}
\right)  =B\cdot\left(
\begin{array}
[c]{c}%
\sigma_{N}\equiv \bar{n}n =(\bar{u}u+\bar{d}d)/\sqrt{2}\\
G\equiv gg\\
\sigma_{S}\equiv\bar{s}s
\end{array}
\right)  \label{isomix} ,%
\end{equation}%
where $G\equiv gg$ is a scalar glueball field which is absent in
this study and $B$ is an orthogonal (3 $\times$ 3) matrix which
has three solutions. The solution 1 and 2 from Ref.
\cite{Giacosa:2005zt} is:

\begin{equation}
B_{1}= \left(
\begin{array}
[c]{ccc}%
0.86 & 0.45 & 0.24\\
-0.45 & 0.89 & -0.06\\
-0.24 & -0.06 & 0.97
\end{array}
\right)  \label{s1} ,
\end{equation}

\begin{equation}
B_{2}= \left(
\begin{array}
[c]{ccc}%
0.81 & 0.54 & 0.19\\
-0.49 & 0.49 & 0.72\\
0.30 & -0.68 & 0.67
\end{array}
\right)\,,  \label{s2}
\end{equation}
and the solution 3 from Ref. \cite{Cheng:2006hu}:%

\begin{equation}
B_{3}=\left(
\begin{array}
[c]{ccc}%
0.78 & -0.36 & 0.51\\
-0.54 & 0.03 & 0.84\\
0.32 & 0.93 & 0.18
\end{array}
\right) \,. \label{s3} %
\end{equation}%

In the solution 1 of Ref. \cite{Giacosa:2005zt} the resonance $f_{0}(1370)$
is predominantly an $\bar{n}n$ state, the resonance $f_{0}(1500)$ is
predominantly a glueball, and $f_{0}(1710)$ is predominantly a
strange $\bar {s}s$ state. In the solution 2 of Ref. \cite{Giacosa:2005zt}
and in the solution of Ref. \cite{Cheng:2006hu} the resonance
$f_{0}(1370)$ is still predominantly a nonstrange $\bar{n}n$ state,
but $f_{0}(1710)$ is now predominantly a glueball, and
$f_{0}(1500)$ predominantly a strange $\bar{s}s$ state. \\

Note that the experimental values of the fields are used,
which are summarized with the numerical values of the
renormalization constants $Z_i$ \cite{Parganlija:2012fy},  the vacuum
expectation values of $\sigma_N$ and $\sigma_S$ which are
$\phi_{N}$ (\ref{phin}) and $\phi_{S}$ (\ref{phis}), respectively,
and the decay constants of pion ($f_{\pi}$ ) and kaon ($f_{K}$)
\cite{Beringer:1900zz} in the following Table \ref{Comparison1}

\begin{table}[H]
\centering
\begin{tabular}
[c]{|c|c|c|c|}\hline Observable & Experiment(\cite{Beringer:1900zz})[MeV] &
constants & Value\\\hline $m_{\pi}$  & $137.3 \pm6.9$ & $Z_{\pi}$
& 1.709\\\hline $m_{K}$  & $495.6 \pm24.8$ & $Z_{K}$ &
1.604\\\hline $m_{\eta}$ &  $547.9 \pm27.4$ & $Z_{K_{S}}$ &
1.001\\\hline $m_{\eta^{\prime}}$ & $957.8 \pm47.9$
&$Z_{\eta_{N}}$ & 1.709\\\hline
 $m_{a_{0}}$ & $1474 \pm74$ &$Z_{\eta_{S}}$ & 1.539\\\hline
$m_{K_S}$ & $893.8 \pm44.7$ &$\phi_{N}$ & 158 MeV \\\hline
$m_{f_{1}(1370)}$ & $(1200-1500)-i(150-250)$ & $\phi_{S}$ & 138
MeV\\\hline $m_{f_{1}(1500)}$ & $1505 \pm 6$ & $f_{\pi}$ & 92.2
MeV\\\hline $m_{f_{1}(1710)}$ & $1722_{-5}^{+6}$ & $f_{K}$ & 110
MeV\\\hline
\end{tabular}
\caption{Masses and wave-function renormalization constants.}%
\label{Comparison1}%
\end{table}

\section{Decay widths of a pseudoscalar glueball into (pseudo)scalar mesons}

The chiral Lagrangian (\ref{explag}) describes the two- and three-body decays of a pseudoscalar glueball, $\tilde{G}$, into scalar
and pseudoscalar mesons. All the decay rates depend on the unknown
coupling constant $c_{\tilde{G}\Phi}$. The decay widths
for the two- and three-body decays of a pseudoscalar glueball are listed in the following:\\

Firstly, let us list the two-body decay widths for a
pseudoscalar glueball, $\tilde{G}$. Performing the two-body decay-width calculation Eq.(\ref{B1}), the decay widths for every channel
can be obtain as follows:
\begin{align}
\Gamma_{\tilde{G}\rightarrow{KK_0^\ast}} &
=\Gamma_{\tilde{G}\rightarrow{K^-K_0^{\ast+}}}+\Gamma_{\tilde{G}\rightarrow{\overline{K}^0K_0^{\ast0}}}+\Gamma_{\tilde{G}\rightarrow{K^0\overline{K}_0^{\ast0}}}
+\Gamma_{\tilde{G}\rightarrow{K^+K_0^{\ast-}}}\nonumber\\
 & = \frac{Z_{K}^2\,\phi_N^2\,c_{\tilde{G}\phi}^2}{16\,\pi\,m_{\tilde{G}}^3}\big[m_{\tilde{G}}^4+(m_K^2-m_{K_0^\ast}^2)^2-2(m_K^2+m_{K_0^\ast}^2)\,m_{\tilde{G}}^2\,\big]^{1/2}\,,
\end{align}

\begin{align}
\Gamma_{\tilde{G}\rightarrow{a_0\,\pi}} & =\Gamma_{\tilde{G}\rightarrow{a^0_0\,\pi^0}}+\Gamma_{\tilde{G}\rightarrow{a_0^+\,\pi^-}}+\Gamma_{\tilde{G}\rightarrow{a_0^-\,\pi^+}}\nonumber\\
 & = \frac{3\,Z_{\pi}^2\,\phi_S^2\,c_{\tilde{G}\phi}^2}{32\,\pi\,m_{\tilde{G}}^3}\,\big[m_{\tilde{G}}^4+(m_{a_0}^2-m_{\pi}^2)^2-2(m_{a_0}^2+m_{\pi}^2)\,m_{\tilde{G}}^2\,\big]^{1/2}\,,
\end{align}

\begin{align}
\Gamma_{\tilde{G}\rightarrow{\eta\,\sigma_N}} & =
\frac{Z_{\eta_S}^2\,\phi_N^2\,c_{\tilde{G}\phi}^2
\sin^2\varphi}{32\,\pi\,m_{\tilde{G}}^3}\,\big[m_{\tilde{G}}^4+(m_\eta^2-m_{\sigma_N}^2)^2-2(m_\eta^2+m_{\sigma_N}^2)\,m_{\tilde{G}}^2\,\big]^{1/2}\,,
\end{align}

\begin{align}
\Gamma_{\tilde{G}\rightarrow{\eta'\,\sigma_N}} & =
\frac{Z_{\eta_S}^2\,\phi_N^2\,c_{\tilde{G}\phi}^2
\cos^2\varphi}{32\,\pi\,m_{\tilde{G}}^3}\,\big[m_{\tilde{G}}^4+(m_{\eta'}^2-m_{\sigma_N}^2)^2-2(m_{\eta'}^2+m_{\sigma_N}^2)\,m_{\tilde{G}}^2\,\big]^{1/2}\,,
\end{align}

\begin{equation}
\Gamma_{\tilde{G}\rightarrow{\eta\,\sigma_S}} =
\frac{Z_{\eta_N}^2\,\phi_N^2\,c_{\tilde{G}\phi}^2
\cos^2\varphi}{32\,\pi\,m_{\tilde{G}}^3}\,\big[m_{\tilde{G}}^4+(m_\eta^2-m_{\sigma_S}^2)^2-2(m_\eta^2+m_{\sigma_S}^2)\,m_{\tilde{G}}^2\,\big]^{1/2}\,,
\end{equation}

\begin{equation}
\Gamma_{\tilde{G}\rightarrow{\eta'\,\sigma_S}} =
\frac{Z_{\eta_N}^2\,\phi_N^2\,c_{\tilde{G}\phi}^2
\sin^2\varphi}{32\,\pi\,m_{\tilde{G}}^3}\,\big[m_{\tilde{G}}^4+(m_{\eta'}^2-m_{\sigma_S}^2)^2-2(m_{\eta'}^2+m_{\sigma_S}^2)\,m_{\tilde{G}}^2\,\big]^{1/2}\,.
\end{equation}

Secondly, let us turn to the three-body decay widths for a
pseudoscalar glueball, $\tilde{G}$. We use the three-body
decay-width expression Eq.(\ref{3bodydecay}), which reads in the present case for the decay of $\tilde{G}$ into three pseudoscalar mesons $\bar{P}_{1},$
$\bar{P}_{2},$ and $\bar{P}_{3}$,
\begin{align}
\Gamma&_{\tilde{G}\rightarrow
\bar{P}_{1}\bar{P}_{2}\bar{P}_{3}}=\frac{S_{\tilde{G}\rightarrow
\bar{P}_{1}\bar{P}_{2}\bar{P}_{3}}}
{32(2\pi)^{3}m_{\tilde{G}}^{3}}\int_{(m_{1}+m_{2})^{2}}^{(m_{\tilde{G}}-m_{3})^{2}}
|-i\mathcal{M}_{\tilde{G}\rightarrow
\bar{P}_{1}\bar{P}_{2}\bar{P}_{3}}|^{2} \,dm_{12}^{2}\nonumber\\& \times %
\sqrt{\frac{(-m_1+m_{12}-m_{2})(m_1+m_{12}-m_2)(-m_1+m_{12+m_2})(m_1+m_{12}+m_2)}{m_{12}^2}}\nonumber\\
&\times\sqrt{\frac{(-m_{\tilde{G}}+m_{12}-m_{3})(m_{\tilde{G}}+m_{12}-m_{3})(-m_{\tilde{G}}+m_{12}+m_{3})(m_{\tilde{G}}+m_{12}+m_{3})}{m_{12}^2}}\,, \label{B3}%
\end{align}

The quantities $m_{1},$ $m_{2},$ $m_{3}$ are the masses of
$\bar{P}_{1},$ $\bar{P}_{2},$ and $\bar{P}_{3}$,
$\mathcal{M}_{\tilde{G}\rightarrow
\bar{P}_{1}\bar{P}_{2}\bar{P}_{3}}$ is the tree-level decay
amplitude, and $S_{\tilde{G}\rightarrow
\bar{P}_{1}\bar{P}_{2}\bar{P}_{3}}$ is a symmetrization factor (it
is equal to $1$ if $\bar{P}_{1}, \bar{P}_{2}$, and
$\bar{P}_{3}$ are all different, equal to $2$ for two identical
particles in the final state, and equal to $6$ for three identical particles in the final state). The decay channels can be obtained as follows:\\

(1) The full decay width into the channel $KK\eta$ results from the sum%
\begin{equation}
\Gamma_{\tilde{G}\rightarrow{KK\eta}} =\Gamma_{\tilde{G}\rightarrow{K^{0}\bar{K}^{0}\eta}}%
+\Gamma_{\tilde{G}\rightarrow{K^{-}%
K^{+}\eta}}=2\Gamma_{\tilde{G}\rightarrow{K^{-}K^{+}%
\eta}}\text{ ,}%
\end{equation}
with the modulus squared decay amplitude
$$|-i\mathcal{M}_{\tilde{G}\rightarrow{K^{-}K^{+}\eta}}|^{2}=\frac{1}%
{4}c_{\tilde{G}\Phi}^{2}Z_{K}^{2}Z_{\eta_N}^{2}\cos^2\varphi,$$
where $m_{1}=m_{2}=m_{K}$ and $m_{3}=m_{\eta}$.\\

(2) The full decay width into the channel $KK\eta'$ results from the sum%
\begin{equation}
\Gamma_{\tilde{G}\rightarrow{KK\eta'}} =\Gamma_{\tilde{G}\rightarrow{K^{0}\bar{K}^{0}\eta'}}%
+\Gamma_{\tilde{G}\rightarrow{K^{-}%
K^{+}\eta}} =2\Gamma_{\tilde{G}\rightarrow{K^{-}K^{+}%
\eta'}}\text{ ,}%
\end{equation}
with the modulus squared decay amplitude
$$|-i\mathcal{M}_{\tilde{G}\rightarrow{K^{-}K^{+}\eta'}}|^{2}=\frac{1}%
{4}c_{\tilde{G}\Phi}^{2}Z_{K}^{2}Z_{\eta_N}^{2}\sin^2\varphi,$$\\
where $m_{1}=m_{2}=m_{K}$, and $m_{3}=m_{\eta}$.\\

(3) The decay width into the channel $\eta\eta\eta$ has as modulus squared decay amplitude $$|-i\mathcal{M}_{\tilde{G}\rightarrow{\eta\eta\eta}}|^{2}=\frac{1}%
{8}c_{\tilde{G}\Phi}^{2}Z_{\eta_N}^{4}Z_{\eta_S}^{2}\cos^4\varphi\,\sin^2\varphi,$$\\
 where $m_{1}=m_{2}=m_{3}=m_{\eta}$ and the
symmetrization factor $S_{\tilde{G}\rightarrow
\eta\eta\eta}=6$.\\

(4) The decay width into the channel $\eta\eta\eta'$ has as modulus squared decay amplitude $$|-i\mathcal{M}_{\tilde{G}\rightarrow{\eta\eta\eta'}}|^{2}=\frac{1}%
{8}c_{\tilde{G}\Phi}^{2}Z_{\eta_N}^{4}Z_{\eta_S}^{2}(\cos^3\varphi-2\cos\varphi\sin^2\varphi)^2,$$
where $m_{1}=m_{2}=m_{\eta}$, $m_{3}=m_{\eta'}$ and the
symmetrization factor $S_{\tilde{G}\rightarrow
\eta\eta\eta'}=2$.\\

(5) The decay width into the channel $\eta\eta'\eta'$ has as modulus squared decay amplitude $$|-i\mathcal{M}_{\tilde{G}\rightarrow{\eta\eta'\eta'}}|^{2}=\frac{1}%
{8}c_{\tilde{G}\Phi}^{2}Z_{\eta_N}^{4}Z_{\eta_S}^{2}(\sin^3\varphi-2\cos^2\varphi\sin\varphi)^2,$$
where $m_{1}=m_{\eta}$, $m_{2}=m_{3}=m_{\eta'}$ and the
symmetrization factor $S_{\tilde{G}\rightarrow
\eta\eta'\eta'}=2$.\\

(6) The full decay width into the channel $\eta\pi\pi$ is computed from the sum%
\begin{equation}
\Gamma_{\tilde{G}\rightarrow{\eta\pi\pi}} =\Gamma_{\tilde{G}\rightarrow{\eta\pi^0\pi^0}}%
+\Gamma_{\tilde{G}\rightarrow{\eta\pi^-\pi^+}}=\frac{3}{2}\Gamma_{\tilde{G}\rightarrow{\eta\pi^-\pi^+}}\text{ ,}%
\end{equation}
with the modulus squared decay amplitude
$$|-i\mathcal{M}_{\tilde{G}\rightarrow{\eta\pi^-\pi^+}}|^{2}=\frac{1}%
{8}c_{\tilde{G}\Phi}^{2}Z_{\pi}^{4}Z_{\eta_S}^{2}\sin^2\varphi\,,$$
where $m_{1}=m_{\eta}$ and $m_{2}=m_{3}=m_{\pi}$.\\

(7) The full decay width into the channel $\eta'\pi\pi$ is computed from the sum%
\begin{equation}
\Gamma_{\tilde{G}\rightarrow{\eta'\pi\pi}} =\Gamma_{\tilde{G}\rightarrow{\eta'\pi^0\pi^0}}%
+\Gamma_{\tilde{G}\rightarrow{\eta'\pi^-\pi^+}} =\frac{3}{2}\Gamma_{\tilde{G}\rightarrow{\eta'\pi^-\pi^+}}\text{ ,}%
\end{equation}
with the modulus squared decay amplitude
$$|-i\mathcal{M}_{\tilde{G}\rightarrow{\eta\pi^-\pi^+}}|^{2}=\frac{1}%
{8}c_{\tilde{G}\Phi}^{2}Z_{\pi}^{4}Z_{\eta_S}^{2}\cos^2\varphi,$$
where $m_{1}=m_{\eta'}$ and $m_{2}=m_{3}=m_{\pi}$.\\

(8) In the case $\tilde{G}\rightarrow{K^{-}K^{+}\pi^{0}}$ one has:
$$|-i\mathcal{M}_{\tilde{G}\rightarrow{K^{-}K^{+}\pi^{0}}}|^{2}=\frac{1}%
{4}c_{\tilde{G}\Phi}^{2}Z_{K}^{4}Z_{\pi}^{2},$$\\
 where $m_{1}=m_{2}=m_{K}$ and $m_{3}=m_{\pi^{0}}$. Then:
\begin{equation}
\Gamma_{\tilde{G}\rightarrow{K^{-}K^{+}\pi^{0}}}=0.00041\,c_{\tilde{G}\Phi
}^{2}\text{ [GeV]}\;. \label{4}%
\end{equation}
The full decay width into the channel $KK\pi$ results from the sum%
\begin{align}
\Gamma_{\tilde{G}\rightarrow{KK\pi}}&=\Gamma_{\tilde{G}\rightarrow{K^{-}%
K^{+}\pi^{0}}}+\Gamma_{\tilde{G}\rightarrow{K^{0}\bar{K}^{0}\pi^{0}}}%
+\Gamma_{\tilde{G}\rightarrow{\bar{K}^{0}K^{+}\pi^{-}}}+\Gamma_{\tilde
{G}\rightarrow{K^{0}K^{-}\pi^{+}}}\nonumber\\&=6\Gamma_{\tilde{G}\rightarrow{K^{-}K^{+}%
\pi^{0}}}\text{ .}%
\end{align}

There exists an interesting and subtle issue: a decay channel
which involves some scalar states which decay further into two
pseudoscalar ones. For instance, $K_{0}^\ast\equiv$ $K_{0}^{\ast
}(1430)$ decays into $K\pi$. There are then two possible decay
amplitudes for the process $\tilde{G}\rightarrow KK\pi$: one is
the direct decay mechanism reported in Table 6.2, and the other is
the decay chain $\tilde{G}\rightarrow KK_{0}^\ast\rightarrow KK\pi$.
The immediate question is, if interference effects emerge which
spoil the results presented in Table \ref{3pG} and \ref{2pG}. Namely, simply performing the sum of the direct three-body decay
(Table \ref{3pG}) and the corresponding two-body decay (Table \ref{2pG}) is
not correct.\\
\indent We now describe this point in more detail using the neutral
channel $\tilde {G}\rightarrow K^{0}\bar{K}^{0}\pi$ as an
illustrative case. To this end, we describe the coupling
$K_{0}^{\ast}$ to $K\pi$ via the Lagrangian
\begin{equation}
\mathcal{L}_{K_{0}^\ast K\pi}=gK_{0}^{\ast}\bar{K}_{0}\pi^{0}+\sqrt{2}gK_{0}^{\ast
}K^{-}\pi^{+}+h.c.\text{ .}%
\end{equation}
The coupling constant $g=2.73$ GeV is obtained by using the
experimental value for the total decay width
$\Gamma_{K_{0}^{\ast}}=270$ MeV \cite{Nakamura:2010zzi}. The full amplitude for
the process $\tilde{G}\rightarrow K^{0}\bar{K}^{0}\pi^{0}$ will
result from the sum
\begin{equation}
\mathcal{M}_{\tilde{G}\rightarrow K^{0}\bar{K}^{0}\pi^{0}}^{\text{full}%
}=\mathcal{M}_{\tilde{G}\rightarrow K^{0}\bar{K}^{0}\pi^{0}}^{\text{direct}%
}+\mathcal{M}_{\tilde{G}\rightarrow\bar{K}^{0}K_{0}^{\ast 0}\rightarrow K^{0}%
\bar{K}^{0}\pi^{0}}^{\text{via}K_{0}^\ast}+\mathcal{M}_{\tilde{G}\rightarrow
K^{0}\bar{K}_{0}^{\ast 0}\rightarrow
K^{0}\bar{K}^{0}\pi^{0}}^{\text{via}\bar
{K}_{0}^\ast}\text{ .}%
\end{equation}
Thus for the decay width we obtain
\begin{align}
\Gamma_{\tilde{G}\rightarrow
K^{0}\bar{K}^{0}\pi^{0}}^{\text{full}} &
=\Gamma_{\tilde{G}\rightarrow K^{0}\bar{K}^{0}\pi^{0}}^{\text{direct}}%
+\Gamma_{\tilde{G}\rightarrow K^{0}K_{0}^{\ast0}\rightarrow K^{0}\bar{K}^{0}%
\pi^{0}}^{\text{via}K_{0}^\ast}+\nonumber\\
&  \Gamma_{\tilde{G}\rightarrow K^{0}\bar{K}_{0}^{\ast 0}\rightarrow
K^{0}\bar
{K}^{0}\pi^{0}}^{\text{via}\bar{K}_{0}^\ast}+\Gamma_{\tilde{G}\rightarrow K^{0}%
\bar{K}^{0}\pi^{0}}^{\text{mix}}\,,%
\end{align}
where $\Gamma_{\tilde{G}\rightarrow
K^{0}\bar{K}^{0}\pi^{0}}^{\text{mix}}$ is the sum of all
interference terms. We can then investigate the magnitude of the
mixing term $\Gamma_{\text{mix}}$, and thus the error incurred
when it is neglected. The explicit calculation for the
$K^{0}\bar{K}^{0}\pi^{0}$ case gives a
relative error of%
\begin{equation}
\left\vert \frac{\Gamma_{\tilde{G}\rightarrow K^{0}\bar{K}^{0}\pi^{0}%
}^{\text{mix}}}{\Gamma_{\tilde{G}\rightarrow K^{0}\bar{K}^{0}\pi^{0}%
}^{\text{direct}}+\Gamma_{\tilde{G}\rightarrow
K^{0}K_{0}^{\ast0}\rightarrow
K^{0}\bar{K}^{0}\pi^{0}}^{\text{via}K_{0}^\ast}+\Gamma_{\tilde{G}\rightarrow
K^{0}\bar{K}_{0}^{\ast0}\rightarrow
K^{0}\bar{K}^{0}\pi^{0}}^{\text{via}\bar
{K}_{0}^\ast}}\right\vert \approx%
\begin{array}
[c]{c}%
7.3\text{ \% (}g>0\text{)}\\
2.2\text{ \% (}g<0\text{)}%
\end{array}
\text{ .}%
\end{equation}
Present results from the model in Ref. \cite{Parganlija:2012fy} show that
$g<0$: the estimates presented in Ref. \cite{Eshraim:2012jv} may be regarded
as upper limits. We thus conclude that the total error for the
channel $\tilde{G}\rightarrow K^{0}\bar{K}^{0}\pi^{0}$ is not
large and can be neglected at this stage. However, in any future,
more detailed and precise theoretical calculation, these
interference effects should also be taken into account.

\newpage
\subsection{Results}

The branching ratios of $\tilde{G}$ for the decays into three
pseudoscalar mesons are reported in Table \ref{3pG} for both choices of
the pseudoscalar masses, $2.6$ and $2.37$ GeV (relevant for PANDA
and BESIII experiments, respectively). The branching ratios are
presented relative to the total decay width of the pseudoscalar
glueball $\Gamma_{\tilde{G}}^{tot}$.

\begin{center}%
\begin{table}[h] \centering
\begin{tabular}
[c]{|c|c|c|}\hline Quantity & Case (i): $m_{\tilde{G}}=2.6$ GeV &
Case (ii): $m_{\tilde{G}}=2.37$ GeV\\\hline
$\Gamma_{\tilde{G}\rightarrow KK\eta}/\Gamma_{\tilde{G}}^{tot}$ &
$0.049$ & $0.043$\\\hline $\Gamma_{\tilde{G}\rightarrow
KK\eta^{\prime}}/\Gamma_{\tilde{G}}^{tot}$ & $0.019$ &
$0.011$\\\hline
$\Gamma_{\tilde{G}\rightarrow\eta\eta\eta}/\Gamma_{\tilde{G}}^{tot}$
& $0.016$ & $0.013$\\\hline
$\Gamma_{\tilde{G}\rightarrow\eta\eta\eta^{\prime}}/\Gamma_{\tilde{G}}^{tot}$
& $0.0017$ & $0.00082$\\\hline
$\Gamma_{\tilde{G}\rightarrow\eta\eta^{\prime}\eta^{\prime}}/\Gamma_{\tilde
{G}}^{tot}$ & $0.00013$ & $0$\\\hline
$\Gamma_{\tilde{G}\rightarrow KK\pi}/\Gamma_{\tilde{G}}^{tot}$ &
$0.47$ & $0.47$\\\hline
$\Gamma_{\tilde{G}\rightarrow\eta\pi\pi}/\Gamma_{\tilde{G}}^{tot}$
& $0.16$ & $0.17$\\\hline
$\Gamma_{\tilde{G}\rightarrow\eta^{\prime}\pi\pi}/\Gamma_{\tilde{G}}^{tot}$
& $0.095$ & $0.090$\\\hline
\end{tabular}%
\caption{Branching ratios for the decay of the pseudoscalar
glueball $\tilde
{G}$ into three pseudoscalar mesons.}
\label{3pG}%
\end{table}%

\end{center}

Next we turn to the decay process $\tilde{G}\rightarrow PS.$ The
results, for both choices of $m_{\tilde{G}},$ are reported in
Table \ref{2pG} for the cases in which the bare resonance $\sigma_{S}$ is
assigned to $f_{0}(1710)$ or to $f_{0}(1500).$

\begin{center}%
\begin{table}[h] \centering
\begin{tabular}
[c]{|c|c|c|}\hline Quantity & Case (i): $m_{\tilde{G}}=2.6$ GeV &
Case (ii): $m_{\tilde{G}}=2.37$ GeV\\\hline
$\Gamma_{\tilde{G}\rightarrow KK_{S}}/\Gamma_{\tilde{G}}^{tot}$ &
$0.060$ & $0.070$\\\hline $\Gamma_{\tilde{G}\rightarrow
a_{0}\pi}/\Gamma_{\tilde{G}}^{tot}$ & $0.083$ & $0.10$\\\hline
$\Gamma_{\tilde{G}\rightarrow\eta\sigma_{N}}/\Gamma_{\tilde{G}}^{tot}$
& $0.0000026$ & $0.0000030$\\\hline
$\Gamma_{\tilde{G}\rightarrow\eta^{\prime}\sigma_{N}}/\Gamma_{\tilde{G}}%
^{tot}$ & $0.039$ & $0.026$\\\hline
$\Gamma_{\tilde{G}\rightarrow\eta\sigma_{S}}/\Gamma_{\tilde{G}}^{tot}$
& $0.012$ $(0.015)$ & $0.0094$ $(0.017)$\\\hline
$\Gamma_{\tilde{G}\rightarrow\eta^{\prime}\sigma_{S}}/\Gamma_{\tilde{G}}%
^{tot}$ & $0$ $(0.0082)$ & $0$ $(0)$\\\hline
\end{tabular}%
\caption{Branching ratios for the decay of the pseudoscalar
glueball $\tilde {G}$ into a scalar and a pseudoscalar meson. In
the last two rows $\sigma_{S}$ is assigned to $f_{0}(1710)$ or to
$f_{0}(1500)$
(values in parentheses).}%
\label{2pG}
\end{table}%
\end{center}

Note that the results are presented as branching ratios because of the as of yet
undetermined coupling constant $c_{\tilde{G}\Phi}$. Concerning
the decays involving scalar-isoscalar mesons, one should go beyond
the results of Table \ref{2pG} by including the full mixing pattern above
1 GeV, in which the resonances $f_{0}(1370),$ $f_{0}(1500),$ and
$f_{0}(1710)$ are mixed states of the bare quark-antiquark
contributions $\sigma_{N}\equiv\left\vert
\bar{u}u+\bar{d}d\right\rangle /\sqrt{2}$, $\sigma_{S}=\bar{s}s\rangle$, and a
bare scalar glueball field $G.$ This mixing is described by an
orthogonal $(3\times3)$ matrix, see Eq. (\ref{isomix})
\cite{Amsler:1995td, Lee:1999kv, Close:2001ga, Giacosa:2005qr, Giacosa:2004ug, Mathieu:2008me, Janowski:2011gt, Giacosa:2005zt, Cheng:2006hu, Chatzis:2011qz, Gutsche:2012zz}. In view of the fact
that a complete evaluation of this mixing in the framework of our
chiral approach has not yet been done, we use the two solutions
for the mixing matrix of Ref.\ \cite{Giacosa:2005zt} and the solution of
Ref.\ \cite{Cheng:2006hu} in order to evaluate the decays of the
pseudoscalar glueball into the three scalar-isoscalar resonances
$f_{0}(1370),$ $f_{0}(1500),$ and $f_{0}(1710)$. In all three
solutions $f_{0}(1370)$ is predominantly described by the bare
configuration $\sigma_{N}\equiv\left\vert
\bar{u}u+\bar{d}d\right\rangle /\sqrt{2}$, but the assignments for
the other resonances vary: in the first solution of Ref.\
\cite{Giacosa:2005zt} the resonance $f_{0}(1500)$ is predominantly
gluonic, while in the second solution of Ref.\ \cite{Giacosa:2005zt} and
the solution of Ref.\ \cite{Cheng:2006hu} the resonance $f_{0}(1710)$ has
the largest gluonic content. The results for the decay of the
pseudoscalar glueball into
scalar-isoscalar resonances are reported in Table \ref{solde}.%

\begin{table}[h] \centering
\begin{tabular}
[c]{|c|c|c|c|}\hline Quantity & Sol. 1 of Ref. \cite{Giacosa:2005zt} &
Sol. 2 of Ref. \cite{Giacosa:2005zt} & Sol. of Ref. \cite{Cheng:2006hu}\\\hline
$\Gamma_{\tilde{G}\rightarrow\eta
f_{0}(1370)}/\Gamma_{\tilde{G}}^{tot}$ &
\multicolumn{1}{|l|}{$0.00093$ $(0.0011)$} &
\multicolumn{1}{|l|}{$0.00058$ $(0.00068)$} &
\multicolumn{1}{|l|}{$0.0044$ $(0.0052)$}\\\hline
$\Gamma_{\tilde{G}\rightarrow\eta
f_{0}(1500)}/\Gamma_{\tilde{G}}^{tot}$ &
\multicolumn{1}{|l|}{$0.000046$ $(0.000051)$} &
\multicolumn{1}{|l|}{$0.0082$ $(0.0090)$} &
\multicolumn{1}{|l|}{$0.011$ $(0.012)$}\\\hline
$\Gamma_{\tilde{G}\rightarrow\eta
f_{0}(1710)}/\Gamma_{\tilde{G}}^{tot}$ &
\multicolumn{1}{|l|}{$0.011$ $(0.0089)$} &
\multicolumn{1}{|l|}{$0.0053$ $(0.0042)$} &
\multicolumn{1}{|l|}{$0.00037$ $(0.00029)$}\\\hline
$\Gamma_{\tilde{G}\rightarrow\eta^{\prime}f_{0}(1370)}/\Gamma_{\tilde{G}%
}^{tot}$ & \multicolumn{1}{|l|}{$0.038$ $(0.026)$} &
\multicolumn{1}{|l|}{$0.033$ $(0.022)$} &
\multicolumn{1}{|l|}{$0.043$ $(0.029)$}\\\hline
$\Gamma_{\tilde{G}\rightarrow\eta^{\prime}f_{0}(1500)}/\Gamma_{\tilde{G}%
}^{tot}$ & \multicolumn{1}{|l|}{$0.0062$ $(0)$} &
\multicolumn{1}{|l|}{$0.00020$ $(0)$} & \multicolumn{1}{|l|}{$0.00013$ $(0)$%
}\\\hline
$\Gamma_{\tilde{G}\rightarrow\eta^{\prime}f_{0}(1710)}/\Gamma_{\tilde{G}%
}^{tot}$ & \multicolumn{1}{|l|}{$0$ $(0)$} &
\multicolumn{1}{|l|}{$0$ $(0)$} & \multicolumn{1}{|l|}{$0$
$(0)$}\\\hline
\end{tabular}%
\caption{Branching ratios for the decays of the pseudoscalar
glueball $\tilde {G}$ into $\eta$ and $\eta '$, respectively, and
one of the scalar-isoscalar states: $f_0(1370), f_0(1500)$, and $
f_0(1710)$ by using three different mixing scenarios of these
scalar-isoscalar states reported in Refs \cite {Giacosa:2005zt, Cheng:2006hu}.
The mass of the pseudoscalar glueball is $m_{\tilde{G}}%
=2.6$ GeV and $m_{\tilde{G}}=2.37$ GeV (values in parentheses),
respectively.}%
\label{solde}
\end{table}%

\begin{figure}[htb]
\begin{center}
\includegraphics[
height=2.1958in, width=3.1099in
]%
{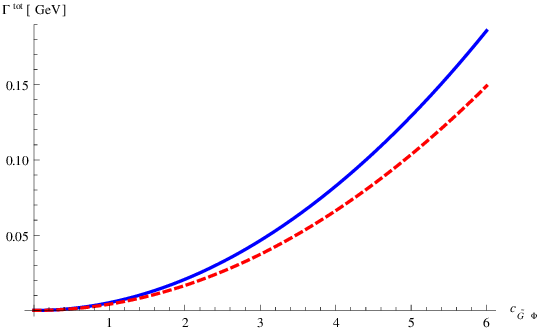}%
\caption{{\small Solid (blue) line: Total decay width of the
pseudoscalar glueball with the bare mass }${\small
M}_{\tilde{G}}{\small =2.6}${\small \ GeV as function of the
coupling }$c_{\tilde{G}\Phi}${\small. Dashed (red)
line: Same curve for }${\small M}_{\tilde{G}}{\small =2.37}$ {\small GeV.}}%
\label{fig1}%
\end{center}
\end{figure}

In Fig.\ \ref{fig1} we show the behavior of the total decay width
$\Gamma_{\tilde{G}}^{tot}=\Gamma_{\tilde{G}\rightarrow
PPP}+\Gamma_{\tilde {G}\rightarrow PS}$ as function of the
coupling constant $c_{\tilde{G}\Phi}$ for both choices of the
pseudoscalar glueball mass. (We assume here that other decay
channels, such as decays into vector mesons or baryons are
negligible.) In the case of $m_{\tilde{G}}=2.6$ GeV, one expects
from large-$N_{c}$
considerations that the total decay width $\Gamma_{\tilde{G}}^{tot}%
\lesssim100$ MeV. In fact, as discussed in the Introduction, the
scalar glueball candidate $f_{0}(1500)$ is roughly $100$ MeV broad
and the tensor candidate $f_{J}(2220)$ is even narrower. In the
present work, the condition $\Gamma_{\tilde{G}}^{tot}\lesssim100$
MeV implies that $c_{\tilde{G}\Phi }\lesssim$ $5$. Moreover, in
the case of $m_{\tilde{G}}=2.37$ GeV for which the identification
$\tilde{G}\equiv X(2370)$ has been made, we can indeed use the
experimental knowledge of the full decay width
[$\Gamma_{X(2370)}=83\pm17$ MeV \cite{Ablikim:2005um, Kochelev:2005vd, Ablikim:2010au}] to determine the
coupling constant to be $c_{\tilde{G}\Phi }=4.48\pm0.46$.
(However, we also refer to the recent work of
Ref.\ \cite{Bugg:2012ys}, where the possibility of a broad pseudoscalar glueball is discussed.)%

Some comments are in order:

(i) The results depend only slightly on the glueball mass. Thus,
the two columns of Table \ref{3pG} and \ref{2pG} are similar. It turns out that
the channel $KK\pi$
is the dominant one (almost 50\%), and also that the $\eta\pi\pi$ and $\eta^{\prime}%
\pi\pi$ channels are sizable. On the other hand, the two-body decays
are subdominant and reach only 20\% of the full mesonic decay
width.

(ii) The decay of the pseudoscalar glueball into three pions
vanishes:
\begin{equation}
\Gamma_{\tilde{G}\rightarrow\pi\pi\pi}=0\text{ .}%
\end{equation}
This result represents a further testable prediction of our
approach.

(iii) The decays of the pseudoscalar glueball into a
scalar-isoscalar meson amount only to $5\%$ of the total decay
width. Moreover, the mixing pattern in the scalar-isoscalar sector
has a negligible influence on the total decay width of
$\tilde{G}.$ Nevertheless, in the future it may represent an
interesting and additional test for scalar-isoscalar states.

(iv) Once the shifts of the scalar fields have been performed,
there are also bilinear mixing terms of the form
$\tilde{G}\eta_{N}$ and $\tilde{G}\eta_{S}$ which lead to a
non-diagonal mass matrix. In principle, one should take these
terms into account (in addition to the already-mentioned
$\eta_{N}\eta_{S}$ mixing) and solve a three-state mixing problem
in order to determine the masses of the pseudoscalar particles.
This will also affect the calculation of the decay widths.
However, due to the large mass difference of the bare glueball
fields $\tilde{G}$ in contrast to the other quark-antiquark
pseudoscalar fields, the mixing of $\tilde{G}$ turns out to be
very small in the present work, and can be safely neglected. For
instance, it turns out that the mass of the mixed state which is
predominantly glueball is (at most) just $0.002$ GeV larger than
the bare mass $m_{\tilde{G}}=2.6$ GeV.

(v) If a standard linear sigma model without (axial-)vector mesons
is studied, the replacements
$Z_{\pi}=Z_{K}=Z_{\eta_{N}}=Z_{\eta_{S}}=1$ need to be performed.
Most of the results of the branching ratios for the three-body
decay are qualitatively (but not quantitatively) similar to the
values of Table \ref{3pG} (variations of about $25$-$30\%$). However, the
branching ratios for the two-body decay change sizably w.r.t. the
results of Table \ref{2pG}. This fact shows once more that the inclusion
of (axial-)vector degrees of freedom has sizable effects also
concerning the decays of the pseudoscalar glueball.

(vi) In principle, the three-body final states for the decays
shown in Table \ref{3pG} can also be reached through a sequential decay
from the two-body final states shown in Table \ref{2pG}, where the scalar
particle $S$ further decays into $PP$, for instance,
$K_{0}^{\ast}(1430)\rightarrow K\pi$. There are then two possible
decay amplitudes: One from the direct three-body decay, and one
from the sequential decay, which have to be added coherently
before taking the modulus square to obtain the total three-body
decay width. Summing the results shown in Table \ref{3pG} and \ref{2pG} gives a
first estimate (which neglects interference terms) for the
magnitude of the total three-body decay width. We have verified
that the correction from the interference term to this total
three-body decay width in a given channel is at most of the order
of $10$\% for $m_{\tilde{G}}=2.6$ GeV and $15\%$ for
$m_{\tilde{G}}=2.37$ GeV. For a full understanding of the
contribution of the various decay amplitudes to the final
three-body state, one needs to perform a detailed study of the
Dalitz plot for the three-body decay.

\

\section{Interaction of a pseudoscalar glueball with nucleons}

In this section we compute the decay width of the pseudoscalar
glueball into a nucleon and an antinucleon and we present the result as a
branching
ratio to remove the effects of the undetermined coupling constant.\\

A $U(2)_R \times U(2)_L$ Lagrangian of the interaction of a
pseudoscalar glueball $\widetilde{G}$ with the baryon fields
$\Psi_1$ and $\Psi_2$ \cite{Eshraim:2012rb, Antje} is
\begin{equation}
\mathcal{L}^{int}_{\widetilde{G}-baryons}=i\,c_{\widetilde{G}\Psi}\,\widetilde{G}(\overline{\Psi}_2\,\Psi_1-\overline{\Psi}_1\,\Psi_2)\,.\label{LG}
\end{equation}
This interaction Lagrangian (\ref{LG}) describes the fusion of a
proton and an antiproton, which is hermitian and invariant under
$SU(2)_R \times SU(2)_L$, parity ($\widetilde{G}\rightarrow
-\widetilde{G}$), and charge conjugation as proved in Ref. \cite{Antje}.
Now let us write the interaction Lagrangian with the
physical fields $N$ and $N^\ast$ which refer to the nucleon and its
partner \cite{Antje} , respectively. By substituting Eqs.
(\ref{psi1}-\ref{psi4}) into Eq.(\ref{LG}), we obtain

\begin{align}
\mathcal{L}^{int}_{\widetilde{G}-nucleons} &= \frac
{c_{\widetilde{G}\Psi}}{2\cosh\delta}\,\widetilde{G}\big(-i\overline{N}\gamma_5N-i\overline{N}^*
N e^{\delta} -i\overline{N}N^* e^{-\delta}
- i\overline{N}^*\gamma_5 N^*\nonumber\\&\,\,\,\,\,\,\,\,\, \,\,\,\,\,\,\,\,\, \,\,\,\,\,\,\,\,\, \,\,\,\,\,\,\,\,\, \,\,\, -i\overline{N}\gamma_5 N+ i\overline{N} N^* e^{\delta} + i\overline{N}^* N e^{-\delta} - i\overline{N}^*\gamma_5 N^*\big)\nonumber\\
&=\frac
{c_{\widetilde{G}\Psi}}{2\cosh\delta}\,\widetilde{G}\big(-2i\overline{N}\gamma_5
N +  i\overline{N} N^*\left[e^{\delta}- e^{-\delta}\right]\nonumber\\
&\,\,\,\,\,\,\,\, \,\,\,\,\,\,\,\,\, \,\,\,\,\,\,\,\,\, \,\,\,\,\,\,\,\,\, \,\,\,+
i\overline {N}^* N \left[e^{-\delta}
- e^{\delta}\right] - 2i\overline{N}^*\gamma_5 N^* \big)\nonumber\\
&=\frac{-ic_{\widetilde{G}\Psi}}{\cosh\delta}\,\widetilde{G}\left(\overline{N}\gamma_5
N +  \sinh\delta\overline{N}^* N - \sinh\delta\overline{N} N^*
+\overline{N}^* \gamma_5 N^* \right)\>.\label{LG1}
\end{align}

 We consider first the pseudoscalar field $\widetilde{G}$ and the nucleon fields
 $N, \, N^\ast,\,\overline{N},$ and $\overline{N}^\ast$ as confined in a cube of length $L$ and volume
 $V=L^3$. The four-momenta of $\widetilde{G}$, $N$, and
 $\overline{N}$ are denoted as $p$, $k_1$, and $k_2$,
 respectively: From Quantum Mechanics it is known that their
 3-momenta are quantized as $\mathbf{p} = 2\pi \mathbf{n}_p /L,$  $\mathbf{k_1} = 2\pi
 \mathbf{n}_{k_1}
 /L,$ and $\mathbf{k_2} = 2\pi
 \mathbf{n}_{k_2}$. Using a Fourier transformation the field
 operators \cite{Antje} can be obtained as

\begin{equation}
\tilde{G}\left(X\right)=\frac 1{\sqrt V}\sum_{\vec{p}} \frac
1{\sqrt{2E_p}}\left(a_p e^{-iP\cdot X}+{a_p}^{\dagger} e^{iP\cdot
X}\right)\,,\label{G}
\end{equation}

\begin{equation}
\overline{N}\left(X\right)=\frac 1{\sqrt
V}\sum_{\vec{k_1},s}\sqrt{\frac {m_N}{E_{k_1}}}\left(d_{\vec
{k_1},s}\overline{v}\left(\vec {p}, s\right)e^{-iK_1\cdot X} +
b^{\dagger}_{\vec {k_1},s} \overline{u}\left(\vec
{k_1},s\right)e^{iK_1\cdot X}\right)\,,\label{Nbar}
\end{equation}
and
\begin{equation}
N\left(X\right)=\frac 1{\sqrt V}\sum_{\vec{k_2},r}\sqrt{\frac
{m_{N}}{E_{k_2}}}\left(b_{\vec {k_2},r} u\left(\vec {k_2},
r\right)e^{-iK_2\cdot X} + d^{\dagger}_{\vec {k_2},r} v\left(\vec
{k_2},r\right)e^{iK_2\cdot X}\right)\,.\label{N}
\end{equation}

where, as known from Quantum Field Theory, the glueball and the two fermionic fields were
decomposed in terms of annihilation and creation operators, $a_p$,
$b,\,\,d$, and ${a_p}^{\dagger}$, $b^{\dagger},\,\,d^{\dagger}$,
respectively.\\
In Eq. (\ref{LG1}) the coupling constant $c_{\widetilde{G}\Psi}$
cannot be determined, but it is easy to calculate the ratio of the
decay of the pseudoscalar glueball $\widetilde{G}$ into a
nucleon and an antinucleon and of the decay into $\overline{
N}^\ast$ and $N$ \cite{Antje},

\begin{equation}
Q=\frac{\Gamma_{\tilde{G}\rightarrow
\overline{N}N}}{\Gamma_{\tilde{G}\rightarrow \overline{N^*} N
+h.c.}}=\frac{\Gamma_{\tilde{G}\rightarrow
\overline{N}N}}{2\Gamma_{\tilde{G}\rightarrow \overline{N^*}
N}}\,.\label{Qration}
\end{equation}

\subsection{Decay of a pseudoscalar glueball into two nucleons}

Let us calculate the decay process of $\Gamma_{\widetilde{G}
\rightarrow \overline{N}N}$ which is described by the first term in Eq.(\ref{LG1}),
\begin{equation}
\mathcal{L}_{1}=\frac{-ig}{\cosh\delta}\tilde{G}\overline{N}\gamma_5
N\,.\label{l1GNNbar}
\end{equation}

 The $\widetilde{G}$ resonance represents the initial state $|i
\rangle$

\begin{equation}
|i\rangle = a^{\dagger}_{{\vec p}^{\prime}} | 0\rangle\,,
\end{equation}
whereas the final state is
\begin{equation}
| f \rangle= b^{\dagger}_{{\vec{k_1}}^{\prime},s^{\prime}}
d^{\dagger}_{{\vec{k_2}}^{\prime},r^{\prime}}| 0\rangle\,,
\end{equation}

The corresponding matrix element of the scattering matrix reads

\begin{equation}
\langle f|S|i\rangle \,,
\end{equation}
We now calculate the expectation value $S_{fi}$ in terms of the
initial and final states:
\begin{equation}
S_{fi} =\langle f | S | i \rangle=\langle f | i\int
d^4X\mathcal{L}_1| i \rangle\,,\label{sfi1}
\end{equation}
Inserting Eqs. (\ref{G}), (\ref{Nbar}), (\ref{N}), and
(\ref{l1GNNbar}) into (\ref{sfi1}) and performing a time-ordered
product of creation and annihilation operators \cite{Antje} we obtain
\begin{align}
\langle f | S | i \rangle & =\frac {-ig\sqrt{ {m_N}^2}}{\cosh
\delta V^{\frac 3 2}\sqrt{2E_{k_2} E_pE_{k_1}}} \langle 0|
b_{\vec{k_1}^{\prime},s^{\prime}}
d_{\vec{k_2}^{\prime},r^{\prime}}
 \int d^4X \nonumber\\
 & \times\sum_{\vec{p}}\left(a_{\vec{p}}e^{-iP\cdot X} +{a_{\vec{p}}}^{\dagger} e^{iP\cdot X}\right)\nonumber\\
      & \times\sum_{\vec{k_1},s}\left(d_{\vec {k_1},s}\overline{v}\left(\vec {k_1}, s\right)e^{-iK_1\cdot X} + b^{\dagger}_{\vec k_1,s} \overline{u}\left(\vec {k_1},s\right)e^{iK_1\cdot X}\right)\gamma_5\nonumber\\
\nonumber&\times\sum_{\vec{k_2},r}\left(b_{\vec {k_2},r}
u\left(\vec {k_2}, r\right)e^{-iK_2\cdot X} + d^{\dagger}_{\vec
{k_2},r} v\left(\vec {k_2},r\right)e^{iK_2\cdot X}\right)
a^{\dagger}_{\vec{p^{\prime}}}
| 0\rangle \nonumber\\
  & \propto\langle 0| b_{\vec{{k_1}}^{\prime},s^{\prime}}d_{\vec{{k_2}}^{\prime},r^{\prime}}b^{\dagger}_{\vec{k_1},s}d^{\dagger}_{\vec {k_2},r}a_{\vec{p}}a^{\dagger}_{\vec{p^{\prime}}}e^{-i\left(P-K_2-K_1\right)X}\overline{u}\gamma_5 v
| 0\rangle \nonumber\\
     & \propto \langle 0 | b_{{\vec{k_1}}^{\prime},s^{\prime}}d_{{\vec{k_2}}^{\prime},r^{\prime}}b^{\dagger}_{\vec{k_1},s}d^{\dagger}_{\vec {k_2},r}\left(\delta_{\vec p {\vec p}^{\prime}} + a^{\dagger}_{{\vec p}^{\prime}}a_{\vec {p}}\right) | 0 \rangle e^{-i\left(P-K_2-K_1\right)X}\overline{u}\gamma_5 v \nonumber \\
    & =\langle 0| \left(\delta_{\vec {k_1} {\vec {k_1}}^{\prime}}\delta_{s s^{\prime}} - b^{\dagger}_{{\vec {k_1}},s}b_{{\vec {k_1}}^{\prime},s^{\prime}}\right)d_{\vec{{k_2}}^{\prime}} d^{\dagger}_{\vec{k_2}}\delta_{\vec p {\vec p}^{\prime}}| 0\rangle e^{-i\left(P-K_2-K_1\right)X}\overline{u}\gamma_5 v \nonumber\\
   & =\langle 0| \delta_{\vec {k_1} {\vec {k_1}}^{\prime}}\delta_{s s^{\prime}}\delta_{\vec p {\vec p}^{\prime}}\left(\delta_{\vec {k_2} {\vec {k_2}}^{\prime}}\delta_{r r^{\prime}} - d^{\dagger}_{{\vec {k_2}}^{\prime},r^{\prime}}d_{\vec {k_2},r}\right) | 0\rangle e^{-i\left(P-K_2-K_1\right)X}\overline{u}\gamma_5 v \nonumber\\
  & = \delta_{\vec {k_1} {\vec {k_1}}^{\prime}}\delta_{s s^{\prime}}\delta_{\vec p {\vec p}^{\prime}}\delta_{\vec {k_2} {\vec {k_2}}^{\prime}}\delta_{r r^{\prime}} \,e^{-i\left(P-K_2-K_1\right)X}\overline{u}\gamma_5 v\,.
\end{align}

We therefore obtain
\begin{align}
\langle f | S | i \rangle & =\frac{m_N}{V^{\frac 32}\,\sqrt{2E_{k_2} E_pE_{k_1}}}\int d^4Xi\mathcal{M}e^{-i\left(P-K_2-K_1\right)X}\nonumber\\
 & =\frac{m_N}{V^{\frac 32}\,\sqrt{2E_{k_2} E_pE_{k_1}}}\,i\mathcal{M}\,(2\pi)^4\,\delta^4(K_1+K_2-p)\,,
\end{align}
where $V$ is the volume of the `box' which contains the fields and
$i\mathcal{M}$ is the invariant amplitude which is given by
\begin{equation}\label{M-Matrix2}
i\mathcal{M}\equiv \frac
g{\cosh\delta}\overline{u}\left(\vec{k_1},s\right)\gamma_5
v\left(\vec{k_2},r\right)\,.
\end{equation}
To find the lifetime of $\tilde{G}$ we have to take the square of the amplitude, which is the probability for the process,
\begin{align}
\langle f | S | i \rangle ^2 & =\frac{m_N^2}{V^{3}\,2E_{k_2} E_pE_{k_1}}\,|i\mathcal{M}|^2\,(2\pi)^8\,(\delta^4(K_1+K_2-p))^2\nonumber\\
 &  =\frac{m_N^2}{V^{3}\,2  E_p E_{k_1} E_{k_2}}\,|i\mathcal{M}|^2\,(2\pi)^4\,(\delta^4(p-K_1-K_2))^2 Vt\,,
\end{align}
where
$$(2\pi)^8\,(\delta^4(K_1+K_2-p))^2=(2\pi)^4\,(\delta^4(p-K_1-K_2))^2 V\,t\,.$$
This is proved in Sec.5.3. The probability for the decay, when the two
particles $\overline{N},\,N$ have momenta between
$(\textbf{k}_1,\,\textbf{k}_1+d^3k_1)$ and
$(\textbf{k}_2,\,\textbf{k}_2+d^3k_2)$, is given by
\begin{equation}
|<f|S|i>|^2\, V\,
\frac{d^3k_1}{(2\pi)^3}\,V\,\frac{d^3k_2}{(2\pi)^3}\,.\label{pABBB}
\end{equation}
By integrating over all possible final momenta, summing over all
final spin orientations and division by $t$, one can calculate
the decay rate as
\begin{equation}
\Gamma=\frac{{m_N}^2}{(2\pi)^2}\sum_{r,s} \int d^3k_1 \,\int
d^3k_2 \frac{|i\mathcal{M}|^2}{2\,E_p\, E_{k_1}\, E_{k_2}}
\delta^4\left(P-K_1+K_2\right)\,.
\end{equation}
In the rest frame of the decaying particle,
$p=(m_{\widetilde{G}},\textbf{0})$, one finds
\begin{equation}
\delta^4\left(P-K_1+K_2\right)=\delta^3(\textbf{k}_1+\textbf{k}_2)\delta(m_{\widetilde{G}}-E_{k_1}-E_{k_2})\,.
\end{equation}
Solving the integral over $d^3k_2$ (by using the Dirac delta
function) we are left with
\begin{equation}
\Gamma=\frac{{m_N}^2}{(2\pi)^2}\sum_{r,s} \int d^3k_1 \,
\frac{|i\mathcal{M}|^2}{2\, E_{k_1}^2\,m_{\widetilde{G}} }
\delta\left(m_{\widetilde{G}}-2 E_{k_1}\right)\,.
\end{equation}
We can write
\begin{equation}
\delta\left(m_{\widetilde{G}}-2
E_{k_1}\right)=\frac{m_{\widetilde{G}}}{4k_f}\delta(|k_1|-k_f)\,,
\end{equation}
where
\begin{equation}\label{k_f2}
k_f=\sqrt{\frac{m_{\widetilde{G}}^2}{4}-{m_N}^2}\,
\end{equation}
is the modulus of the momenta of the outgoing particles. Using the
spherical coordinates, $d^3 k_1= d\Omega |k_1|^2 d|k_1|$ and
solving the integral over $dk_1$, one obtains the decay rate as

\begin{equation}
\Gamma=\frac{{m_N}^2}{2\pi
m_{\widetilde{G}}^2}\sum_{r,s}|i\mathcal{M}|^2 k_f\,.
\end{equation}

For the computation of $\sum_{r,s}|i\mathcal{M}|^2$, one should
use the following two identities \cite{Antje} :

\begin{equation}
\sum_{s}u_{\alpha}(\vec{k},s)\overline{u}_{\beta}(\vec{k},s)=\bigg(\frac{\gamma^\mu
K_\mu+m_N}{2m_N}\bigg)_{\alpha \beta}\,,
\end{equation}

and

\begin{equation}
\sum_{s}v_{\alpha\beta}(\vec{k},s)\overline{v}_{\beta}(\vec{k},s)=\bigg(\frac{-\gamma^\mu
K_\mu+m_N}{2m_N}\bigg)_{\alpha \beta}\,.
\end{equation}
The averaged modulus squared amplitude will thus be
\begin{align}
 \sum_{r,s}|i\mathcal{M}|^2\nonumber & =\frac {g^2}{\cosh^2\delta} \times  \sum_{r,s}\overline{u}\left(\vec{k_1},s\right)\gamma_5 v\left(\vec{k_2},r\right)\left[\overline{u}\left(\vec{k_1},s\right)\gamma_5 v\left(\vec{k_2},r\right)\right]^{\dagger}\\
                                  \nonumber &=\frac {g^2}{\cosh^2\delta}\times -\sum_{r}{\left(\gamma_5\right)}_{\alpha \beta}v_{\beta}\left(\vec{k_2},r\right){\overline{v}}_{\mu}\left(\vec{k_2},r\right)\sum_{s}{\left(\gamma_5\right)}_{\mu \nu}u_{\nu}\left(\vec{k_1},s\right){\overline{u}}_{\alpha}\left(\vec{k_1},s\right)\\
                                 \nonumber  &=\frac {g^2}{\cosh^2\delta}\times {\left(\gamma_5\right)}_{\alpha \beta}{\left(\frac{-\gamma^{\mu}{K_2}_{\mu}+m_N}{2m_N}\right)}_{\beta\mu}{\left(\gamma_5\right)}_{\mu \nu}{\left(\frac{\gamma^{\mu}{K_1}_{\mu}+m_{N}}{2m_{N}}\right)}_{\nu\alpha}\\
                                \nonumber   &=\frac {g^2}{\cosh^2\delta}\times \text{Tr}\left[\gamma_5\left(\frac{-\gamma^{\mu}{K_2}_{\mu}+m_N}{2m_N}\right)\gamma_5\left(\frac{\gamma^{\mu}{K_1}_{\mu}+m_{N}}{2m_{N}}\right)\right]\\
                                 \nonumber  &=\frac {g^2}{\cosh^2\delta}\times \frac 1{4{m_N}^2}\left(4K_1\cdot K_2+4{m_N}^2\right)\\
                             \nonumber      &=\frac {g^2}{\cosh^2\delta}\times \frac {m_{\widetilde{G}}^2-2{m_N}^2+2m_N^2}{2\,m_N^2}\\
                               \nonumber    &=\frac {g^2}{\cosh^2\delta}\times \left(\frac {m_{\widetilde{G}}^2}{2{m_N}^2}\right)\,,
\end{align}

where

$$K_1 \cdot K_2= \frac{m_{\widetilde{G}}^2-2\,m_N^2}{2}\,.$$

Then we obtain the final result of the decay rate
$\Gamma_{\tilde{G}\rightarrow \overline{N}N}$ as follows
\begin{equation}
\Gamma_{\tilde{G}\rightarrow \overline{N}N}=\frac
{g^2}{4\pi\cosh^2\delta}\,k_f\,.\label{dGtoN1}
\end{equation}

Similarly, we now proceed to calculate the rate for the decay process $\tilde{G}\rightarrow
\overline{N}^*N$, which is described in Eq.(\ref{LG1}) by the term
\begin{equation}
\mathcal{L}_2 =\frac{-i\sinh
\delta}{\cosh\delta}\,g\,\tilde{G}\overline{N}^*N = -\tanh \delta
\,g\, \tilde{G}\overline{N}^*N\,.
\end{equation}

The corresponding matrix element can be obtained by
\begin{align}
S_{fi} & =\langle f| S | i \rangle =\langle f |i\int d^4X\mathcal{L}_2| i\rangle \nonumber\\
       & =\frac {-ig\,\tanh\delta \,\sqrt{ {m_N}{m_{N^*}}}}{ V^{\frac 3 2}\sqrt{2E_{k_2} E_pE_{k_1}}} \langle 0 |\tilde{b}_{\vec{k_1}^{\prime},s^{\prime}} d_{\vec{k_2}^{\prime},r^{\prime}} \int d^4X \nonumber\\         \nonumber&\times \sum_{\vec{p}}\left(a_{\vec{p}}e^{-iP\cdot X} +{a_{\vec{p}}}^{\dagger} e^{iP\cdot X}\right)\\
\nonumber& \times\sum_{\vec{k_1},s}\left(\tilde{d}_{\vec {k_1},s}\overline{\tilde{v}}\left(\vec {k_1}, s\right)e^{-iK_1\cdot X} + \tilde{b}^{\dagger}_{\vec k_1,s} \overline{\tilde{u}}\left(\vec {k_1},s\right)e^{iK_1\cdot X}\right)\\
\nonumber&\times\sum_{\vec{k_2},r}\left(b_{\vec {k_2},r}
u\left(\vec {k_2}, r\right)e^{-iK_2\cdot X} + d^{\dagger}_{\vec
{k_2},r} v\left(\vec {k_2},r\right)e^{iK_2\cdot X}\right)
a^{\dagger}_{\vec{p^{\prime}}}
| 0\rangle\\
&\propto
 \delta_{\vec {k_1} {\vec {k_1}}^{\prime}} \delta_{s s^{\prime}}\delta_{\vec p {\vec p}^{\prime}}\delta_{\vec {k_2} {\vec {k_2}}^{\prime}}\delta_{r r^{\prime}}\overline{u}v e^{-i\left(P-K_1-K_2\right)X}\,,
\end{align}

which can be written as:

\begin{equation}
\langle f| S | i \rangle=\frac {\sqrt{ {m_N}{m_{N^*}}}}{V^{\frac 3
2}\sqrt{2E_{k_2} E_pE_{k_1}}}\int
d^4Xi\mathcal{M}e^{-i\left(P-K_2-K_1\right)X}\,,
\end{equation}
where
\begin{equation}\label{M-Matrix3}
i\mathcal{M}\equiv
g\,\tanh\delta\,\,\overline{\tilde{u}}\left(\vec{k_1},s\right)
v\left(\vec{k_2},r\right)\,.
\end{equation}

The decay rate of the pseudoscalar glueball into
$\overline{N}^*\, N$ is obtained as

\begin{equation}
\Gamma_{\tilde{G}\rightarrow \overline{N}^*\,
N}=\frac{m_{N^*}m_N}{2\pi M^2}\, \sum_{r,s}|i\mathcal{M}|^2
k_f^{\prime}\,,
\end{equation}

where
\begin{equation}
k_f^{\prime}= \pm\sqrt{\frac{{m_{N^*}}^4 + {m_N}^4 +
m_{\widetilde{G}}^4 -2{m_N}^2{m_{N^*}}^2
-2{m_{N^*}}^2\,m_{\widetilde{G}}^2
-2{m_N}^2m_{\widetilde{G}}^2}{4m_{\widetilde{G}}^2}}\,,
\end{equation}

which is proved in Ref. \cite{Antje}. The averaged modulus squared decay amplitude is

\begin{align}
\sum_{r,s}{|i\mathcal{M}|}^2\nonumber &= g^2\,\tanh^2\delta\, \sum_{r,s} \overline{\tilde{u}}\left(\vec {k_1},s\right)v\left(\vec {k_2},r\right)\left[\overline{\tilde{u}}\left(\vec {k_1},s\right)v\left(\vec {k_2},r\right)\right]^{\dagger}\\
                             \nonumber   &=-{\left(\frac{\gamma^{\mu}{K_1}_{\mu}+m_{N^*}}{2m_{N^*}}\right)}_{\beta \alpha}{\left(\frac{-\gamma^{\mu}{K_2}_{\mu}+m_{N}}{2m_{N}}\right)}_{\alpha
\beta}\,g^2\,\tanh^2\delta\\
\nonumber&=- \text{Tr}\left[\left(\frac{\gamma^{\mu}{K_1}_{\mu}+m_{N^*}}{2m_{N^*}}\right)\left(\frac{-\gamma^{\mu}{K_2}_{\mu}+m_{N}}{2m_{N}}\right)\right]\,g^2\,\tanh^2\delta\\
             \nonumber&=\frac 1{4\,m_N m_{N^*}} \left(4K_1\cdot K_2-4m_N m_{N^*}\right)\,g^2\,\tanh^2\delta\\
            \nonumber &= \frac 1{2m_N m_{N^*}}   \left[m_{\widetilde{G}}^2-(m_N+ m_{N^*})^2 \right]\,g^2\,\tanh^2\delta\,.\\
\end{align}

We obtain the final result of the decay width
$\Gamma_{\tilde{G}\rightarrow \overline{N}^*\, N}$ as

\begin{equation}
\Gamma_{\tilde{G}\rightarrow \overline{N}^*\,
N}=\frac{g^2\,\tanh^2\delta}{4\pi\,m_{\widetilde{G}}^2}\left[m_{\widetilde{G}}^2-(m_N+
m_{N^*})^2\right]k_f^{\prime}\,.\label{dGtoN2}
\end{equation}

The mass of the nucleon and its partner are $m_N=938$ MeV and
$m_{N^*}=1535$ MeV, respectively
\cite{Nakamura:2010zzi}, whereas the mass of the lightest pseudoscalar glueball is predicted from lattice calculations to be about $2.6$
GeV. The moduli of the momenta of the outgoing particles are
$k_f=900$ MeV and ${k_f}{\prime}= 390.6$ MeV. From
Eq.(\ref{dGtoN1}) and Eq.(\ref{dGtoN2}) we obtain the branching
ratio (\ref{Qration}) of the pseudoscalar glueball decay processes
$\tilde{G}\rightarrow \overline{N}N$  and
$\tilde{G}\rightarrow \overline{N}^\ast N$ \cite{Eshraim:2012rb, Antje}:

\begin{align}
Q\nonumber & = \frac{\Gamma_{\tilde{G}\rightarrow \overline{N}N}}{2\Gamma_{\tilde{G}\rightarrow \overline{N}^*N}}\\
 \nonumber & = \frac{k_f\,m_{\tilde{G}}^2}{2\, \text{sinh}^2\,\delta[m_{\tilde{G}}^2-(m_N+m_{N^\ast})]\,k_f'}\\
   & =1.94\,,
\end{align}

which can be tested by the upcoming PANDA experiment at the FAIR
facility in Darmstadt.

\chapter{Decay of open charmed mesons}\label{ch6decayopencharmed}

\section{Introduction}

Charm physics is an experimentally and theoretically active field
of hadronic physics \cite{Brambilla:2004jw}. The study of strong decays of the heavy
mesons into light pseudoscalar mesons is useful to classify the
charmed states. In this chapter, which is based in the paper \cite{Eshraim:2014eka}, we study the strong OZI-dominant
decays of the open heavy charmed states into light mesons. In this
way, our model acts as a bridge between the high- and low-energy
sector of the strong interaction. It turns out that the
OZI-dominant decays are in agreement with current experimental
results, although the theoretical uncertainties for some of them
are still very large. Nevertheless, since our decay amplitudes
depend on the parameters of the low-energy sector of the theory,
there seems to be an important influence of chiral symmetry in the
determination of the decay widths of charmed states. As a
by-product of our analysis, we also obtain the value of the
charm-anticharm condensate and the values of the weak decay
constants of $D$ mesons. Moreover, in the light of our results we
shall discuss the interpretation of the enigmatic scalar
strange-charmed meson $D_{S0}^{\ast}(2317)$ and the axial-vector
strange-charmed mesons $D_{S1}(2460)$ and $D_{S1}(2536)$
\cite{Brambilla:2010cs, Mohler:2011ke, Moir:2013ub, Ebert:2009ua}. We show the explicit form of
open-charmed decay widths which were obtained from the Lagrangian (\ref{fulllag}) at
tree level. The formulas are organized according to the type of
decaying particle. The
precise description of the decays of open charmed states is important for the CBM experiment at
FAIR.

\section{Decay widths of open-charmed scalar mesons}

In this section we study the phenomenology of the open charmed
mesons in the scalar sector. The Lagrangian (\ref{fulllag})
contains two open charmed scalar mesons which are $D^*_0$ and
$D^*_{S0}$. The neutral scalar state $D_{0}^{\ast0}$ and the
charged $D_{0}^{\ast\overline{0},\pm}$ decay into $D\pi$, while the
strange-charmed state $D^{*\pm}_{S0}$ decays into $DK$. The corresponding
interaction Lagrangian from Eq.(\ref{fulllag}) for the
nonstrange-charmed meson $D^\ast_0$ with $D\pi$ and the
strange-charmed meson with $DK$ has the same structure, as we
shall see below. Then, we shall calculate the general decay width
of a scalar
state in this case.\\
We consider a generic decay process of a scalar state $S$ into two
pseudoscalar states $\widetilde{P}$, i.e., $S \rightarrow
\widetilde{P}_1\widetilde{P}_2$. The interaction Lagrangian for
the neutral scalar state will be given in the following general
simple form:

\begin{align}
\mathcal{L}_{SPP} = &A_{SPP} S^0 \widetilde{P}^0_1
\widetilde{P}^0_2+ B_{SPP} S^0 \partial_\mu \widetilde{P}_1^0
\partial^\mu \widetilde{P}^0_2\nonumber\\&+ C_{SPP}\, \partial_\mu S^0\,\partial^\mu \widetilde{P}^0_1\,\widetilde{P}_2^0
+ E_{SPP} \,\partial_\mu S^0\,\widetilde{P}^0_1\,\partial^\mu
\widetilde{P}_2^0\,.\label{SPP}
\end{align}


Firstly, to calculate the decay amplitude for this process  we
denote the momenta of
$S,\,\widetilde{P}_1,\, and\,\widetilde{P}_2 $ as
$P,\,P_1,\,\text{and}\,P_2,$ respectively. Then, (upon substituting
$\partial^{\mu}\rightarrow - i P^{\mu}$ for the decaying particles
and $\partial^{\mu}\rightarrow
 i \widetilde{P}^{\mu}_{1,2}$ for the decay products) we obtain
the Lorentz-invariant $S\widetilde{P}_1\widetilde{P}_2$ decay
amplitude $-iM_{S \rightarrow \widetilde{P}_1\widetilde{P}_2}$ as

\begin{equation}
-i\mathcal{M}_{S\rightarrow\widetilde{P}^0_1\widetilde{P}^0_2}=i(A_{SPP}-B_{SPP}%
P_{1}\cdot P_{2}+C_{SPP}P\cdot P_{1}+E_{SPP}P\cdot
P_{2})\text{ .}%
\end{equation}

Using energy momentum conservation on the vertex, $P=P_{1}+P_{2}$, we obtain%

\begin{equation}
-i\mathcal{M}_{S\rightarrow\widetilde{P}^0_1\widetilde{P}^0_2}=i[A_{SPP}-B_{SPP}%
P_{1}\cdot P_{2}+C_{SPP}(P_{1}^{2}+P_{1}\cdot
P_{2})+E_{SPP}(P_{2}^{2}+P_{1}\cdot
P_{2})]\text{ .}\label{DADstar}%
\end{equation}

In the decay process, the two pseudoscalar mesons
$\widetilde{P}^0_1$ and $\widetilde{P}^0_2$ are on-shell;
therefore $P_{1}^{2}=m_{\widetilde{P}^0_1}^{2}$ and
$P_{2}^{2}=m_{\widetilde{P}^0_2}^{2}$. Moreover,
\begin{equation}
P_{1}\cdot
P_{2}=\frac{P^{2}-P_{1}^{2}-P_{2}^{2}}{2}\equiv\frac{m_{S}^{2}-m_{\widetilde{P}_1}^{2}-m_{\widetilde{P}_2}%
^{2}}{2}\,.
\end{equation}
Therefore, the decay amplitude (\ref{DADstar}) can be written as
\begin{align}
 -i\mathcal{M}_{S\rightarrow\widetilde{P}^0_1\widetilde{P}^0_2} 
=i\big[ A_{SPP} &+(C_{SPP}+E_{SPP}-B_{SPP})\frac{m_{S}^{2}-m_{\widetilde{P}_1}%
^{2}-m_{\widetilde{P}_2}^{2}}{2} \nonumber\\&+
C_{SPP}\,m_{\widetilde{P}_1}^{2}+E_{SPP}\,m_{\widetilde{P}_2}^{2}\big]
\text{ .}\label{MDStDp}
\end{align}

The decay width $\Gamma_{S
\rightarrow\widetilde{P}^0_1\widetilde{P}^0_2}$ then reads%
\begin{equation} \Gamma_{S \rightarrow
\widetilde{P}^0_1\widetilde{P}^0_2}
=\frac{k(m_{S},m_{\widetilde{P}_1},m_{\widetilde{P}_2})}{8\pi
m_{S}^{2}}\,|-i\mathcal{M}_{S \rightarrow
 \widetilde{P}^0_1\widetilde{P}^0_2 }|^{2}%
\text{ .}\label{GDSP0P0}
\end{equation}
 A non-singlet scalar field will posses also charged decay
 channels. We then have to consider the contribution of the charged
 modes from the process
 $S\rightarrow\widetilde{P}^{\pm}_1\widetilde{P}^{\mp}_2$to the
 full decay width. By multiplying the neutral-mode decay width of
 Eq.(\ref{GDSP0P0}) with an isospin factor $I$, we obtain
\begin{equation}
\Gamma_{S \rightarrow \widetilde{P}_1\widetilde{P}_2}=\Gamma_{S
\rightarrow \widetilde{P}^0_1\widetilde{P}^0_2}+\Gamma_{S
\rightarrow \widetilde{P}^{\pm}_1\widetilde{P}^{\mp}_2}\equiv
I\,\Gamma_{S \rightarrow \widetilde{P}^0_1\widetilde{P}^0_2}\,.
\end{equation}
The full decay width can be written as

\begin{equation}
\Gamma_{S \rightarrow\widetilde{P}_1\widetilde{P}_2}
=I\,\frac{k(m_{S},m_{\widetilde{P}_1},m_{\widetilde{P}_2})}{8\pi
m_{S}^{2}}\,|-i\mathcal{M}_{S \rightarrow
 \widetilde{P}^0_1\widetilde{P}^0_2 }|^{2}%
\text{ ,}\label{DSppG}
\end{equation}
where $I$ is determined from isospin deliberations, or from the interaction Lagrangian of a given decay process.\\
Note that usually the contribution of the charged modes, is twice
the contribution of the neutral modes, which, as we will see in
the explicit interaction Lagrangian below, leads us to write the
general decay width of $S$ into charged modes,
$\widetilde{P}^{\pm}_1\widetilde{P}^{\mp}_2$, as follows:

\begin{equation}
\Gamma_{S \rightarrow\widetilde{P}^{\pm}_1\widetilde{P}^\mp_2}
=2\,\frac{k(m_{S},m_{\widetilde{P}_1},m_{\widetilde{P}_2})}{8\pi
m_{S}^{2}}\,|-i\mathcal{M}_{S \rightarrow
 \widetilde{P}^0_1\widetilde{P}^0_2 }|^{2}%
\text{  .}\label{DSppG2}
\end{equation}

Using this general structures of the decay process of a
 scalar into two pseudoscalar states, $S \rightarrow \widetilde{P}_1
\widetilde{P}_2$, in the
following we compute the decay width of the
nonstrange-charmed scalar state $D_{0}^{\ast0,\pm}$ into $D\pi$ and the
strange-charmed scalar state $D^{*\pm}_{S0}$ into $DK$.

\subsection{Decay Width $D_{0}^{\ast0,\pm}\rightarrow D\pi$}

\label{app4} Firstly, we study the phenomenology of the scalar
doublet charmed mesons $D_{0}^{\ast0,\pm}$ which are assigned to
$D^{*}_0(2400)^{0,\pm}$. These open-charmed mesons are known to
decay into $D\pi$ \cite{Eshraim:2014eka, Eshraim:2014vfa, Eshraim:2014tla}. The corresponding interaction
Lagrangian from Eq.(\ref{fulllag}), for only the neutral and
positively-charged components (the other ones like
$D_{0}^{\ast\overline{0},-}$ possess analogous forms) reads

\begin{align}
\mathcal{L}_{D_{0}^{\ast}D\pi}  &  =A_{D^{\ast0}_{0}D\pi}D^{\ast0}_{0}(\bar{D}^{0}\pi^{0}%
+\sqrt{2}D^{+}\pi^{-})+B_{D^{\ast0}_{0}D\pi}D^{\ast0}_{0}(\partial_{\mu}\bar{D}%
^{0}\partial^{\mu}\pi^{0}+\sqrt{2}\partial_{\mu}D^{+}\partial^{\mu}\pi
^{-})\nonumber\\
&  +C_{D^{\ast0}_{0}D\pi}\partial_{\mu}D^{\ast0}_{0}(\pi^{0}\partial^{\mu}\bar{D}%
^{0}+\sqrt{2}\pi^{-}\partial_{\mu}D^{+})+E_{D^{\ast0}_{0}D\pi}\partial_{\mu}D%
^{\ast0}_{0}(\bar{D}^{0}\partial^{\mu}\pi^{0}+\sqrt{2}D^{+}\partial^{\mu}\pi^{-})\nonumber\\
& +A_{D^{\ast}_{0}D\pi}D^{\ast+}_{0}(D^{-}\pi^{0}%
-\sqrt{2}D^{0}\pi^{-})+B_{D^{\ast}_{0}D\pi}D^{\ast+}_{0}(\partial_{\mu}D^{-}%
\partial^{\mu}\pi^{0}-\sqrt{2}\partial_{\mu}D^{0}\partial^{\mu}\pi
^{-})\nonumber\\
&  +C_{D^{\ast}_{0}D\pi}\partial_{\mu}D^{\ast+}_{0}(\pi^{0}\partial^{\mu}D%
^{-}-\sqrt{2}\pi^{-}\partial_{\mu}D^{0})+E_{D^{\ast}_{0}D\pi}\partial_{\mu}D%
^{\ast+}_{0}(D^{-}\partial^{\mu}\pi^{0}-\sqrt{2}D^{0}\partial^{\mu}\pi^{-})\,,
\label{IntDsDpi}%
\end{align}
where the coefficients read

\begin{align}
A_{D_{0}^{\ast0}D\pi}  &  =-\frac{Z_{\pi}Z_{D}Z_{D_{0}^{\ast0}}}{\sqrt{2}%
}\,\lambda_{2}\phi_{C},\\
B_{D_{0}^{\ast0}D\pi}  &  =\frac{Z_{\pi}Z_{D}Z_{D_{0}^{\ast0}}}{4}\,w_{a_{1}%
}w_{D_{1}}\big[  g_{1}^{2}(3\phi_{N}+\sqrt{2}\phi_{C})-2g_{1}\frac{w_{a_{1}%
}+w_{D_{1}}}{w_{a_{1}}w_{D_{1}}}\nonumber\\&\,\,\,\,\,\,\,\,\,\,\,\,\,\,\,\,\,\,\,\,\,\,\,\,\,\,\,\,\,\,\,\,\,\,\,\,\,\,\,\,\,\,\,\,\,\,\,\,\,\,\,\,\,\,\,\,+h_{2}(\phi_{N}+\sqrt{2}\phi_{C})-2h_{3}%
\phi_{N}\big]  ,\\
\nonumber\\
C_{D_{0}^{\ast0}D\pi}  &  =-\frac{Z_{\pi}Z_{D}Z_{D_{0}^{\ast0}}}{2}%
w_{D^{\star0}}w_{D_{1}}\big[  \sqrt{2}ig_{1}^{2}\phi_{C}-g_{1}\frac{w_{D_{1}%
}+iw_{D^{\star0}}}{w_{D^{\star0}}w_{D_{1}}}-\sqrt{2}ih_{3}\phi_{C}\big]\,  ,\\
E_{D_{0}^{\ast0}D\pi}  &  =\frac{Z_{\pi}Z_{D}Z_{D_{0}^{\ast0}}}{4}%
w_{D^{\star0}}w_{a_{1}}\big[  ig_{1}^{2}(3\phi_{N}-\sqrt{2}\phi_{C}%
)+2g_{1} \frac{w_{a_{1}}-iw_{D^{\star0}}}{w_{D^{\star0}}w_{a_{1}}%
}\nonumber\\&\,\,\,\,\,\,\,\,\,\,\,\,\,\, \,\,\,\,\,\,\,\,\,\,\,\,\,\,\,\,\,\,\,\,\,\,\,\,\,\,\,\,\,\,\,\,\,\,\,\,\,\,\,\,\,\,\,\,\,\,\,\,\,\,\,\,+ih_{2}(\phi_{N}-\sqrt{2}\phi_{C})-2ih_{3}\phi_{N}\big]
\,,\\
\nonumber\\
 A_{D_{0}^{\ast}D\pi}  & =
\frac{Z_{\pi}Z_{D}Z_{D^{\ast}_{0}}}{\sqrt{2}}\,\lambda_{2}\phi
_{C},\\
B_{D_{0}^{\ast}D\pi}  &
=-\frac{Z_{\pi}Z_{D}Z_{D^{\ast}_{0}}}{4}\,w_{a_{1}}w_{D_{1}}\big[
g_{1}^{2}(3\phi_N+\sqrt{2}\phi_{C})-2g_{1}\frac{w_{a_{1}}+w_{D_{1}}}{w_{a_{1}}w_{D_{1}}}%
\nonumber\\&\,\,\,\,\,\,\,\,\,\,\,\,\,\, \,\,\,\,\,\,\,\,\,\,\,\,\,\,\,\,\,\,\,\,\,\,\,\,\,\,\,\,\,\,\,\,\,\,\,\,\,\,\,\,\,\,\,\,\,\,\,\,\,\,\,\,+h_{2}(\phi_N+\sqrt{2}\phi_{C})-2h_{3}\phi_{N}\big]  ,\\
\nonumber\\
C_{D_{0}^{\ast}D\pi}  &  =\frac{Z_{\pi}Z_{D}Z_{D^{\ast}_{0}}}{2}w_{D^{\star}}w_{D_{1}}%
\left[\sqrt{2}ig_{1}^{2}\phi_C-g_{1}\frac{w_{D_{1}}+iw_{D^{\star}}}{w_{D^{\star}}w_{D_{1}}}-\sqrt{2}ih_{3}\phi_{C}\right],\\
E_{D_{0}^{\ast}D\pi}  &  =-\frac{Z_{\pi}Z_{D}Z_{D^{\ast}_{0}}}{4} w_{D^{\star}}w_{a_{1}}%
\big[ i g_{1}^2 (3\phi_{N}-\sqrt{2}\phi_C) + 2g_{1}\frac{w_{a_{1}}-%
iw_{D^{\star}}}{w_{D^{\star}}w_{a_{1}}}\nonumber\\&\,\,\,\,\,\,\,\,\,\,\,\,\,\, \,\,\,\,\,\,\,\,\,\,\,\,\,\,\,\,\,\,\,\,\,\,\,\,\,\,\,\,\,\,\,\,\,\,\,\,\,\,\,\,\,\,\,\,\,\,\,\,\,\,\,\,+ih_{2}(\phi_{N}-\sqrt{2}\phi_{C})-2ih_{3}\phi_{N} \big
]\,.
\end{align}
Note that the parameters $w_{D^\ast}$ and $w_{D^{\ast0}}$ are
 imaginary as shown in Eqs. (4.48) and (4.49),
 respectively, which means that the coefficients containing the imaginary unit are
 real. Furthermore the wave-function renormalisation factors $Z_{D^{\ast0}_0}$ and $Z_{D^{\ast}_0}$ are equal, as well as the parameters $w_{D^{\ast0}}$ and $w_{D^\ast}$
 (for isospin symmetry reasons), which leads to 

 $$ A_{D_{0}^{\ast}D\pi} = A_{D_{0}^{\ast0}D\pi}^\ast,\,\, B_{D_{0}^{\ast}D\pi} = B_{D_{0}^{\ast0}D\pi}^\ast,\,\, C_{D_{0}^{\ast}D\pi} =C_{D_{0}^{\ast0}D\pi}^\ast ,\,\, E_{D_{0}^{\ast}D\pi}=E_{D_{0}^{\ast0}D\pi}^\ast\,.$$

In the interaction Lagrangian (\ref{IntDsDpi}), the considered
decays are that of the neutral state $D^{\ast0}_0$ and the positively
charged state $D^{\ast+}_0$. Both have two relevant decay
channels .

Firstly, let us focus on the decay of $D^{\ast0}_0$ which decays
into neutral modes $D^0 \pi^0$ and charged modes $D^+ \pi^-$. The
explicit expression for the decay process $D^{\ast0}_0 \rightarrow
D^0\pi^0$ is similar to Eq.(\ref{SPP}) as seen in
Eq.(\ref{IntDsDpi}); when using Eq.(\ref{GDSP0P0}) upon
identifying the mesons $S,\,\widetilde{P}_1, \text{and}
\,\widetilde{P}_2$ with the scalar meson $D^{\ast 0}_0$,
 and the pseudoscalar mesons as $D^0$ and $\pi^0$, respectively. The coefficients $A_{SPP},\,B_{SPP},\,C_{SPP},$ and $D_{SPP}$ refer to
  $A_{D^{\ast0}_0 D \pi},$ $B_{D^{\ast0}_0 D \pi},\,C_{D^{\ast0}_0 D \pi},$ and $E_{D^{\ast0}_0 D \pi}$, respectively. We then obtain
   $\Gamma_{D^{\ast0}_0 \rightarrow D^0\pi^0}$ as follows:
\begin{align}%
\Gamma_{D^{\ast0}_0 \rightarrow D^0\pi^0}=&\frac{1}{8\pi m_{D^{\ast0}_0 }}\left[  \frac{(m_{D^{\ast0}_0 }%
^{2}-m_{D^0}^{2}-m_{\pi^0}^{2})^{2}-4m_{\pi^0}^{2}m_{D^0}^{2}}{4m_{D^{\ast0}_0 }^{4}%
}\right]  ^{1/2}  \nonumber\\ & \times \big[
A_{D_{0}^{\ast0}D\pi}+(C_{D_{0}^{\ast0}D\pi
}+E_{D_{0}^{\ast0}D\pi}-B_{D_{0}^{\ast0}D\pi})\frac{m_{D^{\ast0}_0}^{2}-m_{D^0}^{2}-m_{\pi^0}^{2}}{2}
  \nonumber\\ & +C_{D_{0}^{\ast0}D\pi}m_{D^0}^{2}+E_{D_{0}^{\ast0}D\pi}m_{\pi^0}^{2}\big]
^{2}\;.
\label{dscalar1}%
\end{align}
\ The decay width for $D^{\ast0}_0 \rightarrow D^+\pi^-$ has the
same expression as (\ref{dscalar1}) but is multiplied by an
isospin factor 2, which can be seen in Eq.(\ref{DSppG2}).  All the
parameters entering Eqs.(\ref{dscalar1}) have
been fixed as presented in Tables 4.1, 4.2, and 4.3.
The value of $\Gamma_{D^{\ast0}_0 \rightarrow D\pi}$ is then determined as follows:

\begin{align}
\Gamma_{D^{\ast}_0 (2400)^0\rightarrow D\pi} &
=\Gamma_{D^{\ast0}_0 \rightarrow D^0\pi^0}+\Gamma_{D^{\ast0}_0
\rightarrow D^+\pi^-}\\ & = 139_{-114}^{+243}\,\,\,MeV\,.
\end{align}

Now let us turn to the positively charged scalar state
$D_{0}^{\ast+}$ which decays into $D^{+}\pi^{0}$ and
$D^{0}\pi^{+}$. The explicit expression for the decay process
$D^{\ast0}_0 \rightarrow D^0\pi^0$ is also similar to
Eq.(\ref{SPP}) as seen in Eq.(\ref{IntDsDpi}), when using
Eq.(\ref{GDSP0P0}) (upon identifying the mesons
$S,\,\widetilde{P}_1, \text{and} \,\widetilde{P}_2$ with
$D_{0}^{\ast+},\,D^{+},\,$ and $\pi^{0}$, respectively). One may
proceed in a similar manner for the decay width of
$D_{0}^{\ast+}\rightarrow D^{0}\pi^{+}$, where an overall isospin
factor $2$ is also present. Then we obtain the value for  $\Gamma
_{D_{0}^{\ast+}\rightarrow D\pi}$:
\begin{align}
\Gamma_{D_{0}^{\ast}(2400)^{+}\rightarrow
D\pi} & =\Gamma_{D^{+}\pi^{0}}+\Gamma_{D^{0}\pi^{+}}\\
 & = 51_{-51}^{+182}\text{ MeV .} \label{GDStDp1}%
\end{align}

\subsection{Decay Width $D_{S0}^{\ast\pm}\rightarrow DK$}

We turn here to the phenomenology of the scalar state
$D_{S0}^{\ast\pm}$ which is assigned to $D_{S0}^{\ast}(2317)^{\pm}$.
This open strange-charmed meson decays into $D^+K^0$ and $D^0K^+$
\cite{Eshraim:2014eka} as known from \cite{Beringer:1900zz}. The corresponding
interaction Lagrangian from Eq.(\ref{fulllag}) reads

\begin{align}
\mathcal{L}_{D_{S0}^{\ast}DK}  &  =A_{D^{\ast}_{S0}DK}\,D^{\ast+}_{S0}(D^{-}\bar{K}^{0}+D^{0}K^{-})%
+B_{D^{\ast}_{S0}DK}\,D^{\ast+}_{S0}(\partial_{\mu}D%
^{-}\partial^{\mu}\bar{K}^{0}+\partial_{\mu}D^{0}\partial^{\mu}K
^{-})\nonumber\\
& +C_{D^{\ast}_{S0}DK}\,\partial_{\mu}D^{\ast+}_{S0}(\bar{K}^{0}\partial^{\mu}D^{-}%
+\partial_{\mu}D^{0}K^{-})+E_{D^{\ast}_{S0}DK}\,\partial_{\mu}D^{\ast+}_{S0}%
(D^{-}\partial^{\mu}\bar{K}^{0}+D^{0}\partial^{\mu}K^{-})\nonumber\\
& + A_{D^{\ast}_{S0}DK}\,D^{\ast-}_{S0}(D^{+}K^{0}+\bar{D}^{0}K^{+})%
+B_{D^{\ast}_{S0}DK}\,D^{\ast-}_{S0}(\partial_{\mu}D%
^{+}\partial^{\mu}K^{0}+\partial_{\mu}\bar{D}^{0}\partial^{\mu}K
^{+})\nonumber\\
& +C_{D^{\ast}_{S0}DK}\,\partial_{\mu}D^{\ast-}_{S0}(K^{0}\partial^{\mu}D^{+}%
+\partial_{\mu}\bar{D}^{0}K^{+})+E_{D^{\ast}_{S0}DK}\,\partial_{\mu}D^{\ast-}_{S0}%
(D^{+}\partial^{\mu}K^{0}+\bar{D}^{0}\partial^{\mu}K^{+})\,,
\label{Ds00DK}%
\end{align}
with the following coefficients

\begin{equation}
A_{D_{S0}^{\ast}DK}   =\frac{Z_{K}Z_{D}Z_{D_{S0}^{\ast}}}{\sqrt{2}}%
\lambda_{2}\left[  \phi_{N}+\sqrt{2}(\phi_{S}-\phi_{C})\right]\,,\,\,\,\,\,\,\,\,\,\,\,\,\,\,\,\,\,\,\,\,\,\,\,\,\,\,\,\,\,\,\,\,\,\,\,\,\,\,\,\,\,\,\,\,\,\,\,\,\,\,\,\,\,\,\,\,\,\,\,\,\,\,\,
\end{equation}
\begin{align}
B_{D_{S0}^{\ast}DK} =\frac{Z_{K}Z_{D}Z_{D_{S0}^{\ast}}}{2}\,w_{K_{1}%
}w_{D_{1}}\bigg[ & -\sqrt{2}g_{1}\frac{w_{K_{1}}+w_{D_{1}}}{w_{K_{1}}w_{D_{1}}%
}\nonumber\\&+\sqrt{2}(g_{1}^{2}-h_{3})\phi_{N}+(g_{1}^{2}+h_{2})(\phi_{S}+\phi
_{C})\bigg]\,,
\end{align}
\begin{align}
C_{D_{S0}^{\ast}DK}  =\frac{Z_{K}Z_{D}Z_{D_{S0}^{\ast}}}{2}\,w_{D_{1}%
}w_{D_{S}^{\ast}}\bigg[\sqrt{2}g_{1}&\frac{w_{D_{1}}+iw_{D_{S}^{\star}}%
}{w_{D_{1}}w_{D_{S}^{\star}}}-\frac{i}{\sqrt{2}}(g_{1}^{2}+h_{2})\phi
_{N}\nonumber\\&+i(g_{1}^{2}+h_{2})\phi_{S}+2i(h_{3}-g_{1}^{2})\phi_{C}\bigg]\,,
\end{align}
\begin{align}
E_{D_{S0}^{\ast}DK}=\frac{Z_{K}Z_{D}Z_{D_{S0}^{\ast}}}{2}\,w_{K_{1}%
}w_{D_{S}^{\ast}}\bigg[  \sqrt{2}g_{1}&\frac{w_{K_{1}}-iw_{D_{S}^{\star}}%
}{w_{K_{1}}w_{D_{S}^{\star}}}+\frac{i}{\sqrt{2}}(g_{1}^{2}+h_{2})\phi
_{N}\nonumber\\&-+2i(g_{1}^{2}-h_{3})\phi_{S}i(g_{1}^{2}+h_{2})\phi_{C}\bigg]  \text{ .}%
\end{align}
As seen in Eq.(\ref{Ds00DK}) the explicit expression of
$D^{\ast-}_{S0}$ has a form analogous to that of $D^{\ast+}_{S0}$
for isospin symmetry reasons. The decay width of the process
$D^{\ast+}_{S0}\rightarrow D^+ K^0$ is obtained from
Eq.(\ref{DSppG}) when using Eq.(\ref{MDStDp}) upon identifying
$S,\,\widetilde{P}_{1},\,$and $\widetilde{P}_{2}$ with $D_{S0}^{\ast+},\,D^{+},$ and
$K^{0}$ and upon replacing $A_{SPP}\rightarrow A_{D_{S0}^{\ast}D K
},\,B_{SPP}\rightarrow B_{D_{S0}^{\ast}D K},\,C_{SPP}\rightarrow
C_{D_{S0}^{\ast}D K},\,E_{SPP}\rightarrow E_{D_{S0}^{\ast}D K}$.
The decay width for the process $D^{\ast+}_{S0}\rightarrow D^0
K^+$ has an analogous analytic expression. The full decay width of
$D^{\ast}_{S0}$ with a mass of about $2467$ MeV into $DK$ \cite{Eshraim:2014eka}, see Table \ref{charmass}, is then
$$\Gamma_{D_{S0}^{\ast}\rightarrow DK}\simeq3\, GeV\,.$$

\newpage
\section{Decay widths of open-charmed vector mesons}

In this section we describe the phenomenology of open-charmed
vector mesons $D^{\ast0,+}$. The neutral state $D^{\ast0}$ is
assigned to $D^{\ast}(2007)^0$ while the positively charged state
$D^{\ast+}$ to $D^\ast(2010)^+$. These resonances decay into two
pseudoscalar states $D\pi$ \cite{Beringer:1900zz}. Let us now consider the
decay of a vector state $V$ into two pseudoscalar states
$\widetilde{P}$, i.e., $V \rightarrow \widetilde{P}_1
\widetilde{P}_2$, in general. The general simple form of the
interaction Lagrangian, as obtained from Eq.(\ref{fulllag}),
reads
\begin{align}
\mathcal{L}_{VPP}= A_{VPP} V^0_{\mu} \widetilde{P}_1^0
\partial^\mu \widetilde{P}_2^0 + B_{VPP} V^0_{\mu} \partial^\mu
\widetilde{P}_1^0 \widetilde{P}_2^0 + C_{VPP} \,\partial_\nu
V^0_\mu (\partial^\nu \widetilde{P}_1^0 \partial^\mu
\widetilde{P}_2^0 -
\partial^\mu\widetilde{P}_1^0 \, \partial^\nu \widetilde{P}_2^0)\,.\label{VPPL}
\end{align}

\begin{figure}[H]
\begin{center}
\includegraphics[
height=1.5in, width=2.5in]
{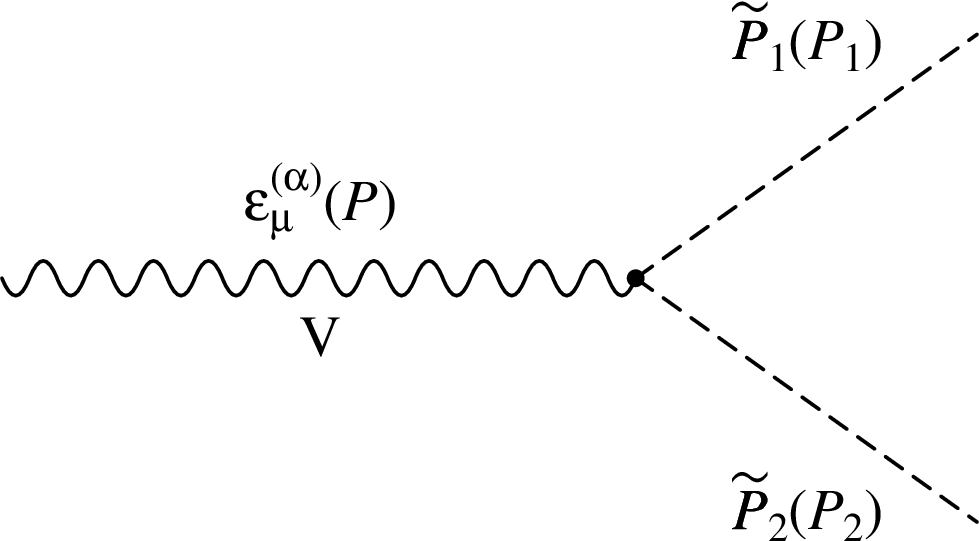}
\caption{Decay process $V \rightarrow \widetilde{P}_1
\widetilde{P}_2$ }
\label{figspp}
\end{center}
\end{figure}

We denote the momenta of $V$, $\widetilde{P}_1$, and
$\widetilde{P}_2$ as $P$, $P_{1}$, and $P_{2}$, respectively, and
compute the decay amplitude for this process. We
consider the polarization vector $\varepsilon
_{\mu}^{(\alpha)}(P)$ for the vector state. Then, upon
substituting $\partial^{\mu}\rightarrow -iP^{\mu}$\ for the
decaying particle and $\partial^{\mu}\rightarrow iP_{1,2}^{\mu}$
for the decay products, we obtain the following Lorentz-invariant
$V\widetilde{P}_1\,\widetilde{P}_2$ scattering amplitude
$-i\mathcal{M}_{V\rightarrow
\widetilde{P}_1 \widetilde{P}_2}$ from the Lagrangian (\ref{VPPL}):%
\begin{align}
-i\mathcal{M}^{(\alpha)}_{V\rightarrow \widetilde{P}_1
\widetilde{P}_2}=\varepsilon _{\mu}^{(\alpha)}(P)h_{VPP}^{\mu}\,,  \label{VPPA}%
\end{align}
with
\begin{equation}
h_{VPP}^{\mu}=-\left\{
A_{VPP}P_{2}^{\mu}+B_{VPP}P_{1}^{\mu}+C_{VPP}[P_{2}^{\mu}(P\cdot
P_{1})-P_{1}^{\mu
}(P\cdot P_{2})]\right\}\text{,}  \label{hVPP}%
\end{equation}

where $h_{VPP}^{\mu}$ denotes the
$V\widetilde{P}_1\,\widetilde{P}_2$ vertex.\newline

The calculation of the decay width requires the determination of
the modulus square of the scattering amplitude. The scattering amplitude
in Eq. (\ref{VPPA})
 depends on the polarization vector $\varepsilon_{\mu }^{(\alpha)}(P)$. \\
For a general scattering amplitude, one has to calculate
$|\overline{-i\mathcal{M}_{V\rightarrow \widetilde{P}_1
\widetilde{P}_2}}|^2$ of a process containing one vector state with
mass $m_V$. The calculation reads as follows:

\begin{align}
-i\mathcal{M}^{(\alpha)}_{V\rightarrow \widetilde{P}_1
\widetilde{P}_2}=\varepsilon_{\mu}^{(\alpha)}(P)h_{VPP}^{\mu}
&\Rightarrow\\\Rightarrow |\overline{-i\mathcal{M}_{V\rightarrow
\widetilde{P}_1
\widetilde{P}_2}}|^2&=\frac{1}{3}\sum_{\alpha=1}^{3}|-i\mathcal{M}^{(\alpha)}_{V\rightarrow
\widetilde{P}_1
\widetilde{P}_2}|^2\nonumber\\&=\frac{1}{3}\sum_{\alpha=1}^{3}\varepsilon_{\mu}^{(\alpha)}(P)\varepsilon_{\nu}^{(\alpha)}(P)h_{VPP}^{\mu}h_{VPP}^{\nu}\nonumber\\
&=\frac{1}{3}\bigg(-g_{\mu\nu}+\frac{P_\mu P_\nu}{m^2}\bigg)
h_{VPP}^\mu
h^\nu_{VPP}\nonumber\\
&=\frac{1}{3}\bigg[-(h^\mu_{VPP})^2+\frac{(P_\mu
h^\mu_{VPP})^2}{m_V^2}\bigg]\,,\label{VPPASQ}
\end{align}
where
\begin{equation}
\sum_{\alpha=1}^{3}\varepsilon_{\mu}^{(\alpha)}(P)\varepsilon_{\nu}^{(\alpha)}(P)=-g_{\mu\,\nu}+\frac{P_\mu\,P_\nu}{m_V^2}\,.
\end{equation}
Eq.(\ref{VPPASQ}) contains the metric tensor
$g_{\mu\nu}=diag(1,-1,-1,-1)$. Note that, if the vector
particle decays, then $P_\mu=(m_V,0)$ in the rest frame of the
decaying particle and thus
\begin{equation}
\frac{(P_\mu\,h^\mu)^2}{m_V^2} \equiv \frac{(m_V\,h^0)^2}{m_V^2}=(h^0)^2\,.\label{h0}
\end{equation}
Therefore, we have to calculate the squared vertex
$(h_{VPP}^\mu)^2$ using Eq.(\ref{hVPP}):

\begin{align}
(h_{VPP}^{\mu})^{2}  &
=A_{VPP}^{2}m_{\widetilde{P}_2}^{2}+B_{VPP}^{2}m_{\widetilde{P}_1}^{2}+C_{VPP}^{2}[P_{2}^{\mu}(P\cdot P_{1}%
)-P_{1}^{\mu}(P\cdot P_{2})]^{2}\nonumber\\
&  +2A_{VPP}B_{VPP}P_{1}\cdot P_{2}+2A_{VPP }C_{VPP}[(P\cdot
P_{1})m_{\widetilde{P}_2}^{2}-(P_{1}\cdot P_{2})(P\cdot
P_{2})]\nonumber\\
&  +2B_{VPP}C_{VPP}[(P_{1}\cdot P_{2})(P\cdot
P_{1})-(P\cdot P_{2})m_{\widetilde{P}_1}^{2}]\text{ .} \label{hVPP2}%
\end{align}

Now let us compute the squared vertex at rest (also using Eq.\ (\ref{hVPP})):%

\begin{align}
(h_{VPP}^{0})^{2}  &
=A_{VPP}^{2}E_{\widetilde{P}_2}^{2}+B_{VPP}^{2}E_{\widetilde{P}_1}^{2}+C_{VPP}^{2}[E_{\widetilde{P}_2}(P\cdot
P_{1})-E_{\widetilde{P}_1}(P\cdot
P_{2})]^{2}\nonumber\\
&
+2A_{VPP}B_{VPP}E_{\widetilde{P}_1}E_{\widetilde{P}_2}+2A_{VPP}C_{VPP}[(P\cdot P_{1})E_{\widetilde{P}_2}^{2}-E_{\widetilde{P}_1}E_{\widetilde{P}_2}(P\cdot P_{2}%
)]\nonumber\\
&  +2B_{VPP}C_{VPP}[E_{\widetilde{P}_1}E_{\widetilde{P}_2}(P\cdot
P_{1})-(P\cdot
P_{2})E_{\widetilde{P}_1}^{2}]\text{ .} \label{hVPP3}%
\end{align}

From Eqs.\ (\ref{VPPASQ}), (\ref{hVPP2}), and (\ref{hVPP3}) we obtain%

\begin{align}
|\overline{-i\mathcal{M}_{V\rightarrow PP}}|^{2}  & =\frac
{1}{3}\{(A_{VPP}^{2}+B_{VPP}^{2})k^{2}(m_{V}%
,m_{\widetilde{P}_2},m_{\widetilde{P}_1})\nonumber\\
&
+C_{VPP}^{2}\{k^{2}(m_{V},m_{\widetilde{P}_1},m_{\widetilde{P}_1})[(P\cdot
P_{1})^{2}+(P\cdot P_{2})^{2}]\nonumber\\
&  -2(P\cdot P_{1})(P\cdot P_{2})(E_{\widetilde{P}_1}E_{\widetilde{P}_2}-P_{1}\cdot P_{2})\}\nonumber\\
&  +2A_{VPP}B_{VPP}(E_{\widetilde{P}_1}E_{\widetilde{P}_2}-P_{1}\cdot P_{2}%
)\nonumber\\
&
+2A_{VPP}C_{VPP}[k^{2}(m_{V},m_{\widetilde{P}_2},m_{\widetilde{P}_2
})P\cdot P_{1}-(P\cdot P_{2})(E_{\widetilde{P}_1}E_{\widetilde{P}_2}-P_{1}\cdot P_{2})]\nonumber\\
&  +2B_{VPP}C_{VPP}[(P\cdot
P_{1})(E_{\widetilde{P}_1}E_{\widetilde{P}_2}-P_{1}\cdot
P_{2})-k^{2}(m_{V},m_{\widetilde{P}_2},m_{\widetilde{P}_1})(P\cdot P_{2})]\}\nonumber\\
&  =\frac{1}{3}\left\{  \{A_{VPP}^{2}+B_{VPP}%
^{2}+C_{VPP}^{2}[(P\cdot P_{1})^{2}+(P\cdot P_{2})^{2}]\right.
\nonumber\\
&  \left.  +2C_{VPP}[A_{VPP}(P\cdot P_{1})-B_{VPP}(P\cdot P_{2})]\}k^{2}(m_{V},m_{\widetilde{P}_2},m_{\widetilde{P}_1})\right. \nonumber\\
&  \left. +2\{A_{VPP}B_{VPP}-C_{VPP}^{2}(P\cdot P_{1})(P\cdot
P_{2})+C_{VPP}(B_{VPP}P\cdot P_{1}\right.
\nonumber\\
&  \left.  -A_{VPP}P\cdot
P_{2})\}(E_{\widetilde{P}_1}E_{\widetilde{P}_2}-P_{1}\cdot
P_{2})\right\}\text{ .}  \label{MVPPsq}%
\end{align}
The vertex $h_{VPP}^{\mu}$ from Eq.(\ref{hVPP}) can be transformed
as
\begin{align}
h_{VPP}^{\mu}  & =-[A_{VPP}P_{2}^{\mu}+B_{VPP}P_{1}^{\mu}+C_{VPP}(m_{V}E_{\widetilde{P}_1}P_{2}^{\mu}-m_{V}E_{\widetilde{P}_2}P_{1}^{\mu})]\nonumber\\
& =-(B_{VPP}-C_{VPP}m_{V}E_{\widetilde{P}_2})P_{1}^{\mu
}-(A_{VPP}+C_{VPP}m_{V}E_{\widetilde{P}_1})P_{2}^{\mu}\text{ ,}
\label{hVPP4}%
\end{align}
where
\begin{equation}
P_1\cdot P=m_{V}\,E_{\widetilde{P}_1} \,\,\,\,\text{and}\,\,\,\,
P_2\cdot P=m_{V}\,E_{\widetilde{P}_2}\,.\label{pem}
\end{equation}

We can then write Eq.(\ref{MVPPsq}) in a slightly different (but
equivalent) form, by inserting Eq.(\ref{hVPP4}) into Eq.\
(\ref{VPPASQ}) as follows:

\begin{align}
|\overline{-i\mathcal{M}_{V \rightarrow PP}}|^{2}  &
=\frac{1}{3}\{-[(B_{VPP}-C_{VPP} m_{V} E_{\widetilde{P}_2})P_{1}^{\mu}+(A_{VPP}+C_{VPP} m_{\widetilde{P}_2} E_{\widetilde{P}_1})P_{2}^{\mu}]^{2}\nonumber\\
& +\frac{1}{m_{V}^{2}}[(B_{VPP}-C_{VPP} m_{V} E_{\widetilde{P}_2})P_{1\mu}P^{\mu} \nonumber \\
& +(A_{VPP}+C_{VPP}m_{V}E_{\widetilde{P}_1})P_{2\mu}P^{\mu}]^{2}\}\text{ .} \label{MVPP2sq}%
\end{align}

Using Eq.(\ref{pem}),
$k^{2}(m_{V},m_{\widetilde{P}_2},m_{\widetilde{P}_1})=E_{\widetilde{P}_2}^2-m_{\widetilde{P}_2}^2$,
and
$k^{2}(m_{V},m_{\widetilde{P}_2},m_{\widetilde{P}_1})=E_{\widetilde{P}_1}^2-m_{\widetilde{P}_1}^2$,
we obtain that Eq.(\ref{MVPP2sq}) can be written as

\begin{align}
|\overline{-i\mathcal{M}_{V\rightarrow \widetilde{P}_1
\widetilde{P}_2}}|^{2}  &
=\frac{1}{3}[(B_{VPP}-C_{VPP}m_{V}E_{\widetilde{P}_2})^{2}+(A_{VPP}+C_{VPP}m_{V}E_{\widetilde{P}_1})^{2}\nonumber\\
&  -2(A_{VPP}+C_{VPP}m_{V}E_{\widetilde{P}_1})
(B_{VPP}-C_{VPP}m_{V}E_{\widetilde{P}_2})]k^{2}(m_{V},m_{\widetilde{P}_2},m_{\widetilde{P}_1})\nonumber\\
&
=\frac{1}{3}[A_{VPP}-B_{VPP}+C_{VPP}m_{V}(E_{\widetilde{P}_1}+E_{\widetilde{P}_2})]^{2}k^{2}(m_{V},m_{\widetilde{P}_2},m_{\widetilde{P}_1}) \nonumber\\
&
=\frac{1}{3}(A_{VPP}-B_{VPP}+C_{VPP}m_{V}^{2})^{2}k^{2}(m_{V},m_{\widetilde{P}_2},m_{\widetilde{P}_1})\text{
.} \label{MVPPsqf}
\end{align}
The decay width for a vector meson decaying into two pseudoscalar
mesons $\Gamma_{V \rightarrow \widetilde{P}_1\widetilde{P}_2}$
(as obtained from Eq. (\ref{VPPL}) can be computed with the
 following formula:
\begin{equation}
\Gamma_{V \rightarrow  \widetilde{P}_1\widetilde{P}_2 }=
I\frac{k(m_{V},m_{\widetilde{P}_2 },m_{\widetilde{P}_1})}{8\pi
m_{V}^{2}}|\overline{-i\mathcal{M}_{V \rightarrow \widetilde{P}_1
\widetilde{P}_2}}|^{2}\text{ ,}
\label{GVPP}%
\end{equation}
 where $I$ is an isospin factor and the modulus square of the decay amplitude $|-i\mathcal{\bar{M}}_{V \rightarrow \widetilde{P}_1
\widetilde{P}_2}|^{2}$ is given in Eq.\ (\ref{MVPPsqf}) or,
equivalently, Eq.\ (\ref{MVPPsq}). Now let us apply this result to
calculate the decay width of the vector meson $D^{\ast0,+}$ into
$D\pi$ from the Lagrangian (\ref{fulllag}).

\subsection{Decay Width $D^{\ast0,\pm}\rightarrow D\pi$}

The $D^\ast D \pi$ interaction Lagrangian from Eq.(\ref{fulllag})
reads

\begin{align}
\mathcal{L}_{D^{\ast}D\pi}  &
=A_{D^{\star}D\pi}D_{\mu}^{\star0}(\pi
^{0}\partial^{\mu}\bar{D}^{0}+\sqrt{2}\pi^{-}\partial^{\mu}D^{+})+B_{D^{\star
}D\pi}D_{\mu}^{\star0}(\bar{D}^{0}\partial^{\mu}\pi^{0}+\sqrt{2}D^{+}%
\partial^{\mu}\pi^{-})\nonumber\\
&  +C_{D^{\star}D\pi}\partial_{\nu}D_{\mu}^{\star0}(\partial^{\mu}\bar{D}%
^{0}\partial^{\nu}\pi^{0}+\sqrt{2}\partial^{\mu}D^{+}\partial^{\nu}\pi
^{-}) \nonumber \\
& + C_{D^{\star}D\pi}^{\ast}\partial_{\nu}D_{\mu}^{\star0}(\partial^{\mu}%
\pi^{0}\partial^{\nu}\bar{D}^{0}+\sqrt{2}\partial^{\mu}\pi^{-}\partial^{\nu
}D^{+})\nonumber\\
&  +A_{D^{\star}D\pi}^{\ast}\bar{D}_{\mu}^{\star0}(\pi^{0}\partial^{\mu}%
D^{0}+\sqrt{2}\pi^{+}\partial^{\mu}D^{-})+B_{D^{\star}D\pi}^{\ast}\bar{D}%
_{\mu}^{\star0}(D^{0}\partial^{\mu}\pi^{0}+\sqrt{2}D^{-}\partial^{\mu}\pi
^{+})\nonumber\\
&
+C_{D^{\star}D\pi}^{\ast}\partial_{\nu}\bar{D}_{\mu}^{\star0}(\partial
^{\mu}D^{0}\partial^{\nu}\pi^{0}+\sqrt{2}\partial^{\mu}D^{-}\partial^{\nu}%
\pi^{+}) \nonumber \\
& +
C_{D^{\star}D\pi}\partial_{\nu}\bar{D}_{\mu}^{\star0}(\partial^{\mu
}\pi^{0}\partial^{\nu}D^{0}+\sqrt{2}\partial^{\mu}\pi^{+}\partial^{\nu}D^{-})\nonumber \\
& + A_{D^{\star}D\pi}D_{\mu}^{\star+}(\pi
^{0}\partial^{\mu}D^{-}-\sqrt{2}\pi^{-}\partial^{\mu}D^{0})+B_{D^{\star
}D\pi}D_{\mu}^{\star+}(D^{-}\partial^{\mu}\pi^{0}-\sqrt{2}D^{0}%
\partial^{\mu}\pi^{-})\nonumber\\
&  +C_{D^{\star}D\pi}\partial_{\nu}D_{\mu}^{\star+}(\partial^{\mu}D%
^{-}\partial^{\nu}\pi^{0}-\sqrt{2}\partial^{\mu}D^{0}\partial^{\nu}\pi
^{-}) \nonumber \\
& + C_{D^{\star}D\pi}^{\ast}\partial_{\nu}D_{\mu}^{\star+}(\partial^{\mu}%
\pi^{0}\partial^{\nu}D^{-}-\sqrt{2}\partial^{\mu}\pi^{-}\partial^{\nu
}D^{0})\nonumber\\
&  +A_{D^{\star}D\pi}^{\ast}D_{\mu}^{\star-}(\pi^{0}\partial^{\mu}%
D^{+}-\sqrt{2}\pi^{+}\partial^{\mu}\bar{D}^{0})+B_{D^{\star}D\pi}^{\ast}D%
_{\mu}^{\star-}(D^{+}\partial^{\mu}\pi^{0}-\sqrt{2}\bar{D}^{0}\partial^{\mu}\pi
^{+})\nonumber\\
& +C_{D^{\star}D\pi}^{\ast}\partial_{\nu}D_{\mu}^{\star-}(\partial
^{\mu}D^{+}\partial^{\nu}\pi^{0}-\sqrt{2}\partial^{\mu}\bar{D}^{0}\partial^{\nu}%
\pi^{+}) \nonumber \\
& + C_{D^{\star}D\pi}\partial_{\nu}D_{\mu}^{\star-}(\partial^{\mu
}\pi^{0}\partial^{\nu}D^{+}-\sqrt{2}\partial^{\mu}\pi^{+}\partial^{\nu}\bar{D}^{0})\,,
\label{DstarDpionL}%
\end{align}
with the following coefficients
\begin{align}
A_{D^{\ast}D\pi}  &  =\frac{i}{2}Z_{\pi}Z_{D}\left[  g_{1}+\sqrt{2}w_{D_{1}%
}(h_{3}-g_{1}^{2})\phi_{C}\right]  \;,\\
B_{D^{\ast}D\pi}  &  =-\frac{i}{4}Z_{\pi}Z_{D}\left[  2g_{1}-w_{a_{1}}%
(3g_{1}^{2}+h_{2}-2h_{3})\phi_{N}+\sqrt{2}w_{a_{1}}(g_{1}^{2}+h_{2})\phi
_{C}\right]  \;,\\
C_{D^{\ast}D\pi}  &
=\frac{i}{2}Z_{\pi}Z_{D}w_{a_{1}}w_{D_{1}}g_{2}\;.
\end{align}
Note that the Lagrangian in Eq.\ (\ref{DstarDpionL}) contains the
parameter combinations $A_{D^{\star}D\pi}$, $B_{D^{\star}D\pi}$,
and $C_{D^{\star}D\pi}$ and their complex conjugates. Thus, it is
certain that the Lagrangian is hermitian; so we obtain
$\mathcal{L}_{D^{\star}D\pi}^{\dagger}=\mathcal{L}_{D^{\star
}D\pi}$. \\
In the following we will focus only on the decays
$D^{\star0}\rightarrow D\pi$ and $D^{\star+}\rightarrow D\pi$,
where the corresponding decay of $\bar{D}^{\star0}$ and
$D^{\star-}\rightarrow D\pi$ yields the same result due to isospin
symmetry.\\

The interaction Lagrangian (\ref{DstarDpionL}) has two decay
channels for the neutral charmed-vector meson $D^{\ast0}$, which
are $D^{\ast0}\rightarrow D^0 \pi^0$ and $D^{\ast0}\rightarrow D^+
\pi^-$. We have to note that the decay of $D^{\ast0}$ into $D^+
\pi^-$ is impossible because, in this situation, the sum of
the decay product masses is larger than the mass of the decay
particle, so
$$\Gamma_{D^{\ast0}
\rightarrow D^+ \pi^-}=0\,.$$ This result shows very good agreement
with the experiment result: experimentally it is not seen, as
listed by the Particle Data Group \cite{Beringer:1900zz}. Then the decay of the
neutral state $D^{\ast0}$ into the neutral modes $D^{0}\pi^{0}$ is
the only possible decay in the eLSM (\ref{fulllag}). To compute
$\Gamma_{D^{\ast0}\rightarrow D^0 \pi^0}$, we use Eqs. (\ref{MVPPsqf}) and (\ref{GVPP}) upon identifying
$V,\,\widetilde{P}_1,\,\text{and}\,\widetilde{P}_2$ with
$D^{\ast0},\,\pi^0,\text{and}\, D^0$ and the coefficients
$A_{VPP},\,B_{VPP},\,\text{and},\,C_{VPP}$ with
$A_{D^{\ast}D\pi},\,B_{D^{\ast}D\pi},\,\text{and}$ $C_{D^{\ast}D\pi}$,
respectively. We then obtain the corresponding expression
\begin{align}
\Gamma_{D^{\ast0}\rightarrow D^0 \pi^0}=\frac{1}{24\pi}&\left[  \frac{(m_{D^{\ast0}}%
^{2}-m_{\pi^0}^{2}-m_{D^{0}}^{2})^{2}-4m_{\pi^0}^{2}m_{D^{0}}^{2}}{4m_{D^{\ast0}}^{4}%
}\right]  ^{3/2}\nonumber\\&\times(A_{D^{\ast}D\pi}-B_{D^{\ast}D\pi}+C_{D^{\star}D\pi}%
\,m_{D^{\ast0}}^{2})^{2}\;. \label{DsDpiQc}%
\end{align}
The parameters entering Eq.(\ref{DstarDpionL}) have been
determined uniquely from the fit (see Tables 4.1, 4.2, and 4.3). Therefore
we can calculate the value of the decay width immediately:

\begin{equation}
\Gamma_{D^{\ast}(2007)^{0}\rightarrow D^{0}\pi^{0}}=
(0.025\pm0.003)MeV\;,%
\end{equation}
where the experimental value reads
$\Gamma_{D^{\ast}(2007)^{0}\rightarrow
D^{0}\pi^{0}}^{exp}<1.3$ MeV as listed in Ref. \cite{Beringer:1900zz}.\\

Now let us turn to the phenomenology of the positively charged
state $D^{\ast}(2010)^+$, which can decay into $D^0\pi^+$ and $D^+\pi^0$.
The decay width of the channel $D^{\ast}(2010)^+ \rightarrow
D^0\pi^+$ has an analogous analytic expression as
$\Gamma_{D^{\ast0}\rightarrow D^0 \pi^0}$, which is described in
Eq.(\ref{DsDpiQc}), whereas concerning the decay $D^{\ast}(2010)^+
\rightarrow D^+\pi^0$ Eq.\ (\ref{DsDpiQc}) holds upon
multiplication with an isospin factor $2$. We then obtain the
following results:
\begin{equation}
\Gamma_{D^{\ast}(2010)^{+}\rightarrow D^{+}\pi^{0}}=
(0.018_{-0.003}^{+0.002})\, MeV\;,%
\end{equation}

\begin{equation}
\Gamma_{D^{\ast}(2010)^{+}\rightarrow D^{0}\pi^{+}}=
(0.038_{-0.004}^{+0.005})\, MeV\;.%
\end{equation}

The experimental values \cite{Beringer:1900zz} read
$$\Gamma_{D^{\ast}(2010)^{+}\rightarrow D^{+}\pi^{0}}^{\exp} =
(0.029\pm0.008) MeV$$ and $$\Gamma_{D^{\ast}(2010)^{+}\rightarrow
D^{0}\pi^{+}}^{\exp} = (0.065\pm0.017) MeV\,.$$

\newpage

\section{Decay widths of open charmed axial-vector mesons}

In this section we turn to the study of the phenomenology of
axial-vector charmed mesons, i.e., the two- and three-body decays of $D_1$ and the two-body decays of the strange $D_{S1}$.
The nonstrange-charmed field $D_1$ corresponds to the resonances
$D_1(2420)^{0,\pm}$ while the strange-charmed doublet $D_{S1}^\pm$ is
assigned to the resonance $D_{S1}(2536)^\pm$. Firstly let us focus
on the two-body decays of the axial-vector charmed mesons
$D_1,\,D_{S1}$. The nonstrange-charmed $D_1$ decays into
vector-charmed and light pseudoscalar mesons,
$D_1\rightarrow D^\ast\pi$, while the strange-charmed doublet
$D_{S1}$ decays into $D^\ast K$, which also contains a
vector-charmed meson and a light pseudoscalar meson. Therefore,
both decays have the same structure of the corresponding
expression for the decay amplitude. This leads us to consider the general case of the
two-body decay process of an axial-vector meson $A$ into a vector
$V$ and a pseudoscalar state $\widetilde{P}$, i.e., $A \rightarrow V
\widetilde{P}$. The following interaction Lagrangian
describes the decay of the axial-vector state into neutral modes:
\begin{align}
\mathcal{L}_{AV\widetilde{P}}  &  =A_{AV\widetilde{P}}A^{\mu0}V_{\mu}^{0}{\bar{P}}%
^{0}\nonumber\\
&  +B_{AV\widetilde{P}}\left[ A^{\mu0}\left(  \partial_{\nu}V_{\mu}^{0}%
-\partial_{\mu}V_{\nu}^{0}\right)
\partial^{\nu}{\bar{P}}^{0}+\partial^{\nu }A^{\mu0}\left(
V_{\nu}^{0}\partial_{\mu}{\bar{P}}^{0}-V_{\mu}
^{0}\partial_{\nu}{\bar{P}}^{0}\right)  \right]\text{ .}
\label{K12}
\end{align}

\begin{figure}[H]
\begin{center}
\includegraphics[
height=1.5in, width=2.5in]%
{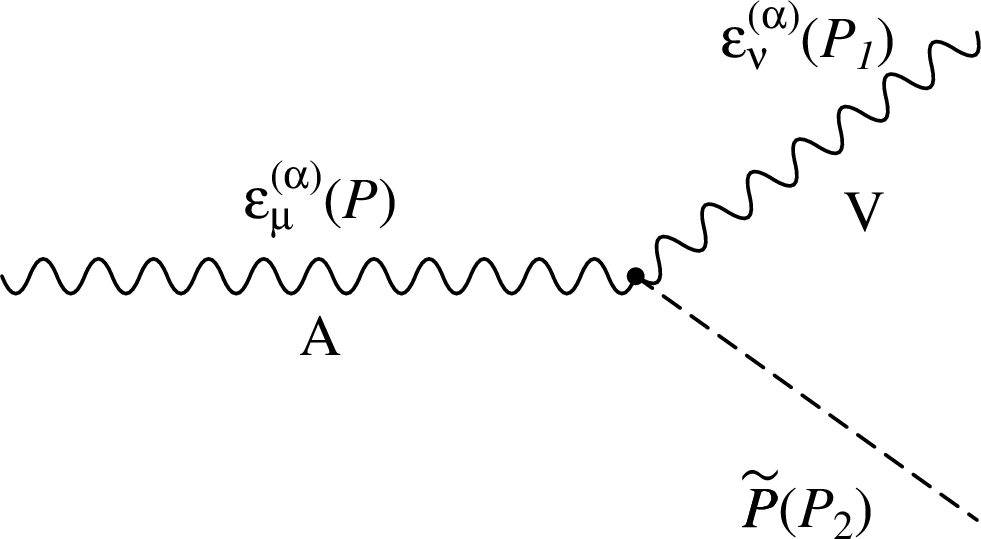}%
\caption{Decay process $A\rightarrow V{\tilde{P}}$.}%
\label{figAVP}%
\end{center}
\end{figure}

Let us study a generic decay process of the form $A\rightarrow V^0
\widetilde{P}^0$. The 4-momenta of $A,\,V$, and $\widetilde{P}$ are
denoted as $P,\,P_1$, and $P_2$, respectively. We have to consider
the corresponding polarization vectors because there are two
vector states involved in the decay process $A\rightarrow V^0
\widetilde{P}^0$, i.e., $A$ and $V$. These we denote as
$\varepsilon_\mu^{(\alpha)}(P)$ for $A$ and
$\varepsilon_\nu^{(\beta)}(P_1)$ for $V$. Using the
substitutions $\partial^\mu\rightarrow -iP^\mu$ for the decaying
particle and $\partial^\mu\rightarrow iP^\mu_{1,2}$ for the decay
products, we obtain the Lorentz-invariant $AV\widetilde{P}$
scattering amplitude
$-i\mathcal{M}_{A\rightarrow{V^0\widetilde{P}^0}}^{(\alpha,\beta)}$ as
follows:
\begin{align}
-i\mathcal{M}_{A\rightarrow V^{0}{\tilde{P}}^{0}}^{(\alpha,\beta)}
&
=\varepsilon_{\mu}^{(\alpha)}(P)\varepsilon_{\nu}^{(\beta)}(P_{1}
)h_{AV{\tilde{P}}}^{\mu\nu}\,,\label{iMAVP}
\end{align}

with

\begin{align}
h_{AV{\tilde{P}}}^{\mu\nu}=i\left\{  A_{AV{\tilde{P}}}g^{\mu\nu}
+B_{AV{\tilde{P}}}[P_{1}^{\mu}P_{2}^{\nu}+P_{2}^{\mu}P^{\nu}-(P_{1}\cdot
P_{2})g^{\mu\nu}-(P\cdot P_{2})g^{\mu\nu}]\right\}\text{ ,}
\label{hAVP}
\end{align}
$\,$\\
where $h_{AV{\tilde{P}}}^{\mu\nu}$ denotes the $AV{\tilde{P}}$
vertex.\newline

It will be necessary to determine the modulus square of the scattering
amplitude in order to calculate the decay width. We note that the
scattering amplitude in Eq.\ (\ref{iMAVP}) depends on the
polarization vectors $\varepsilon_{\mu }^{(\alpha)}(P)$ and
$\varepsilon_{\nu}^{(\beta)}(P_{1})$. Therefore, it is necessary
to calculate the average of the modulus squared amplitude for all
polarization values. Let us denote the masses of the vectors
states $A$ and $V$\ as $m_{A}$ and $m_{V}$, respectively. Then the
averaged modulus squared amplitude $|\overline{-i\mathcal{M}}|^{2}$ is
determined as follows:

\begin{align}
\left\vert \overline{-i\mathcal{M}_{A\rightarrow V^{0}{\tilde{P}}^{0}}}\right\vert ^{2}&=\frac{1}{3}%
\sum\limits_{\alpha,\beta=1}^{3}\left\vert
-i\mathcal{M}_{A\rightarrow
V^{0}{\tilde{P}}^{0}}^{(\alpha,\beta)}\right\vert ^{2}\nonumber\\
&
=\frac{1}{3}\sum\limits_{\alpha,\beta=1}^{3}\varepsilon_{\mu}^{(\alpha
)}(P)\varepsilon_{\nu}^{(\beta)}(P_{1})h_{AV{P}}^{\mu\nu}\varepsilon_{\kappa
}^{(\alpha)}(P)\varepsilon_{\lambda}^{(\beta)}(P_{1})h_{AV{\tilde{P}}}%
^{\ast\kappa\lambda} \text{ .} \label{iMK11}%
\end{align}

Given that

\begin{equation}
\sum\limits_{\alpha=1}^{3}\varepsilon_{\mu}^{(\alpha)}(P)\varepsilon_{\kappa
}^{(\alpha)}(P)=
-g_{\mu\kappa}+\frac{P_{\mu}P_{\kappa}}{m_{A}^{2} }\,,
\label{epsilon}
\end{equation}

[an analogous equation holds for $\varepsilon^{(\beta)}$], we
obtain from Eq.\ (\ref{iMK11}):

\begin{align}
|\overline{-i\mathcal{M}_{A\rightarrow V^{0}{\tilde{P}}^{0}}}|^{2}  &  =\frac{1}%
{3}\left(
-g_{\mu\kappa}+\frac{P_{\mu}P_{\kappa}}{m_{A}^{2}}\right)  \left(
-g_{\nu\lambda}+\frac{P_{1\nu}P_{1\lambda}}{m_{V}^{2}}\right)
h_{AV{\tilde
{P}}}^{\mu\nu}h_{AV{\tilde{P}}}^{\ast\kappa\lambda}\nonumber\\
&  =\frac{1}{3}\left[  h_{\mu\nu
AV{\tilde{P}}}h_{AV{\tilde{P}}}^{\ast\mu\nu }-\frac{
h_{AV{\tilde{P}}}^{\mu\nu}P_{\mu} \,\, h_{\nu
AV{\tilde{P}}}^{\ast\kappa}P_{\kappa}
}{m_{A}^{2}}-\frac{
h_{AV{\tilde{P}}}^{\mu\nu}P_{1\nu} \,\, h_{\mu AV{\tilde{P}}}%
^{\ast\lambda}P_{\lambda}  }{m_{V}^{2}}\right. \nonumber\\
&  \left.  +\frac{
h_{AV{\tilde{P}}}^{\mu\nu}P_{\mu}P_{1\nu}\,\,
 h_{AV{\tilde{P}}}^{\ast x \lambda}P_{x}P_{1\lambda} }{m_{V}^{2}%
m_{A}^{2}}\right] \nonumber\\
&  =\frac{1}{3}\left[  \left\vert
h_{AV{\tilde{P}}}^{\mu\nu}\right\vert
^{2}-\frac{\left\vert h_{AV{\tilde{P}}}^{\mu\nu}P_{\mu}\right\vert ^{2}}%
{m_{A}^{2}}-\frac{\left\vert
h_{AV{\tilde{P}}}^{\mu\nu}P_{1\nu}\right\vert
^{2}}{m_{V}^{2}}+\frac{\left\vert
h_{AV{\tilde{P}}}^{\mu\nu}P_{\mu}P_{1\nu
}\right\vert ^{2}}{m_{V}^{2}m_{A}^{2}}\right]\text{ ,} \label{iMAVP2}%
\end{align}

which contains the metric tensor $g_{\mu\nu}=\mathrm{diag}%
(1,-1-1,-1)$. The decay width for the process $A\rightarrow V^{0}{\tilde{P}%
}^{0}$ then reads

\begin{equation}
\Gamma_{A\rightarrow
V^{0}{\tilde{P}}^{0}}=\frac{k(m_{A},m_{V},m_{{\tilde{P}} })}{8\pi
m_{A}^{2}}|\overline{-i\mathcal{M}_{A\rightarrow V^{0}{\tilde{P}}^{0}
}}|^{2}\text{ .} \label{GAV0P0}
\end{equation}
 Note that a non-singlet axial-vector field will in general also possess
charged decay channels. Therefore, in addition to the decay
process considered in Eq.\ (\ref{GAV0P0}),\ we must consider the
contribution of the charged modes from the process $A\rightarrow
V^{\pm}{\tilde{P}}^{\mp}$ to the full decay width. Then the full
decay width is obtained as $$\Gamma
_{A\rightarrow V{\tilde{P}}}=\Gamma_{A\rightarrow V^{0}{\tilde{P}}^{0}}%
+\Gamma_{A\rightarrow V^{\pm}{\tilde{P}}^{\mp}}\equiv
I\Gamma_{A\rightarrow V^{0}{\tilde{P}}^{0}}\,.$$

Using Eq.(\ref{iMAVP2}) we obtain the two-body decay width of
vector meson, Eq.(\ref{GAV0P0}), as follows:
\begin{equation}
\Gamma_{A\rightarrow
V\tilde{P}}=\frac{K(m_{A},m_{V},m_{\tilde{P}})}{12\pi
m_{A}^{2}}\left[\left\vert h_{AV{\tilde{P}}}^{\mu\nu}\right\vert
^{2}-\frac{\left\vert h_{AV{\tilde{P}}}^{\mu\nu}P_{\mu}\right\vert ^{2}}%
{m_{A}^{2}}-\frac{\left\vert
h_{AV{\tilde{P}}}^{\mu\nu}P_{1\nu}\right\vert
^{2}}{m_{V}^{2}}+\frac{\left\vert
h_{AV{\tilde{P}}}^{\mu\nu}P_{\mu}P_{1\nu }\right\vert
^{2}}{m_{V}^{2}m_{A}^{2}}\right]  \text{ ,}
\label{d1dspionQ}%
\end{equation}
 where the quantities $P^{\mu}=(m_{A},\mathbf{0})$, $P_{1}^{\mu}=(E_{V}%
,\mathbf{k})$, and $P_{2}^{\mu}=(E_{\tilde{P}},-\mathbf{k})$ are
the four-momenta of $A$, $V$, and $\tilde{P}$ in the rest frame of
$A$, respectively. The following kinematic relations hold:
\begin{align*}
&  P_{V}\cdot P_{\tilde{P}}=\frac{m_{A}^{2}-m_{V}^{2}-m_{\tilde{P}}^{2}}{2}\,,{}\\
&  P_{A}\cdot P_{V}=m_{A}E_{V}=\frac{m_{A}^{2}+m_{V}-m_{\tilde{P}}^{2}}{2}\,,\\
&  P_{A}\cdot
P_{\tilde{P}}=m_{A}E_{\tilde{P}}=\frac{m_{A}^{2}-m_{V}+m_{\tilde{P}}^{2}}{2}\,.
\end{align*}
The terms entering in\ Eq. (\ref{d1dspionQ}) are given by%

\begin{align}
\left\vert h_{AV\tilde{P}}^{\mu\nu}\right\vert ^{2}  &  =4A_{AV\tilde{P}}^{2}+B_{AV\tilde{P}}^{2}\left[  m_{V}^{2}m_{\tilde{P}}^{2}+m_{A}%
^{2}m_{\tilde{P}}^{2}+2(P_{1}\cdot P_{2})^{2}+2(P\cdot
P_2)^{2}+6(P_{1}\cdot
P_2)(P\cdot P_2)\right] \nonumber\\
&-6A_{AV\tilde{P}}B_{AV\tilde{P}}(P_{1}\cdot P_2+ P \cdot
P_2)\text{ ,} \label{hD1dspQ1}%
\end{align}%
\begin{align}
\left\vert h_{AV\tilde{P}}^{\mu\nu}P_{\mu}\right\vert ^{2} &
=A_{AV\tilde{P}}^{2}m_{A}^{2}+B_{AV\tilde{P}}^{2}\left[ (P\cdot
P_{1})^{2}m_{\tilde{P}}^{2}+(P_{1}\cdot P_2)^{2}m_{A}^{2}-2(P\cdot P_{1}%
)(P\cdot P_2)(P_{1}\cdot P_2)\right] \nonumber\\
&  +2A_{AV\tilde{P}}B_{AV\tilde{P}}\left[  (P\cdot P_{1}%
)(P\cdot P_2)-(P_{1}\cdot P_2)m_{A}^{2}\right]  \text{ ,}%
\end{align}%
\begin{align}
\left\vert h_{AV\tilde{P}}^{\mu\nu}P_{1,\nu}\right\vert ^{2} &
=A_{AV\tilde{P}}^{2}m_{V}^{2}+B_{AV\tilde{P}}^{2}\left[ (P\cdot
P_{1})^{2}m_{\tilde{P}}^{2}+(P\cdot P_2)^{2}m_{V}^{2}-2(P\cdot P_{1}%
)(P\cdot P_2)(P_{1}\cdot P_2)\right] \nonumber\\
&  +2A_{AV\tilde{P}}B_{AV\tilde{P}}\left[  (P\cdot P_{1}%
)(P_{1}\cdot P_2)-(P\cdot P_2)m_{V}^{2}\right]  \text{ ,}%
\end{align}%
\[
\left\vert h_{AV\tilde{P}}^{\mu\nu}P_{\mu}P_{1,\nu}\right\vert
^{2}=[A_{AV\tilde{P}}(P\cdot P_{1})]^{2}%
\]
\newline with $E_{V}=\sqrt{K^{2}(m_{A},m_{V},m_{\tilde{P}})+m_{V}^{2}}$ and
$E_{\tilde{P}}=\sqrt{K^{2}(m_{A},m_{V},m_{\tilde{P}})+m_{\tilde{P}}^{2}}$.

\subsection{Two-body decay of $D_{1}$}

We present the relevant interaction Lagrangian for the nonstrange
axial-vector meson $D_1$ which is a light-heavy quark
$Q\overline{q}$ and is assigned to $D_1(2420)^{0,\pm}$. This
Lagrangian describes only the two-body decays of this state, and is given as:

\begin{align}
\mathcal{L}_{D_{1}D^\star\pi} &
=A_{D_{1}D^{\star}\pi}D_{1}^{\mu0}\left( \bar{D}_{\mu
}^{\star0}\pi^{0}+\sqrt{2}D_{\mu}^{\star+}\pi^{-}\right)  \nonumber\\
&  +B_{D_{1}D^{\star}\pi}\left\{  D_{1}^{\mu0}\left[  \left(
\partial_{\nu
}\bar{D}_{\mu}^{\star0}-\partial_{\mu}\bar{D}_{\nu}^{\star0}\right)
\partial^{\nu}\pi^{0}+\sqrt{2}\left(  \partial_{\nu}D_{\mu}^{\star+}%
-\partial_{\mu}D_{\nu}^{\star+}\right)
\partial^{\nu}\pi^{-}\right]  \right.
\nonumber\\
&  \left.  +\partial^{\nu}D_{1}^{\mu0}\left[  \left(
\bar{D}_{\nu}^{\star
0}\partial_{\mu}\pi^{0}-\bar{D}_{\mu}^{\star0}\partial_{\nu}\pi^{0}\right)
+\sqrt{2}\left(
D_{\nu}^{\star+}\partial_{\mu}\pi^{-}-D_{\mu}^{\star
+}\partial_{\nu}\pi^{-}\right)  \right]  \right\}  \nonumber \\
& + A_{D_{1}D^{\star}\pi}^{\star}\bar{D}_{1}^{\mu0}\left( D_{\mu
}^{\star0}\pi^{0}+\sqrt{2}D_{\mu}^{\star-}\pi^{+}\right)  \nonumber\\
&  +B_{D_{1}D^{\star}\pi}^{\star}\left\{ \bar{D}_{1}^{\mu0}\left[
\left(
\partial_{\nu
}D_{\mu}^{\star0}-\partial_{\mu}\bar{D}_{\nu}^{\star0}\right)
\partial^{\nu}\pi^{0}+\sqrt{2}\left(  \partial_{\nu}D_{\mu}^{\star-}%
-\partial_{\mu}D_{\nu}^{\star-}\right)
\partial^{\nu}\pi^{+}\right]  \right.
\nonumber\\
&  \left.  +\partial^{\nu}\bar{D}_{1}^{\mu0}\left[  \left(
D_{\nu}^{\star
0}\partial_{\mu}\pi^{0}-D_{\mu}^{\star0}\partial_{\nu}\pi^{0}\right)
+\sqrt{2}\left(
D_{\nu}^{\star-}\partial_{\mu}\pi^{+}-D_{\mu}^{\star
-}\partial_{\nu}\pi^{+}\right)  \right]  \right\}  \nonumber \\
& + A_{D_{1}D^{\star}\pi}D_{1}^{\mu+}\left( D_{\mu
}^{\star-}\pi^{0}-\sqrt{2}D_{\mu}^{\star0}\pi^{-}\right)  \nonumber\\
&  +B_{D_{1}D^{\star}\pi}\left\{  D_{1}^{\mu+}\left[  \left(
\partial_{\nu
}D_{\mu}^{\star-}-\partial_{\mu}D_{\nu}^{\star-}\right)
\partial^{\nu}\pi^{0}-\sqrt{2}\left(  \partial_{\nu}D_{\mu}^{\star0}%
-\partial_{\mu}D_{\nu}^{\star0}\right)
\partial^{\nu}\pi^{-}\right]  \right.
\nonumber\\
&  \left.  +\partial^{\nu}D_{1}^{\mu+}\left[  \left(
D_{\nu}^{\star
-}\partial_{\mu}\pi^{0}-D_{\mu}^{\star-}\partial_{\nu}\pi^{0}\right)
-\sqrt{2}\left(
D_{\nu}^{\star0}\partial_{\mu}\pi^{-}-D_{\mu}^{\star
0}\partial_{\nu}\pi^{-}\right)  \right]  \right\}  \nonumber \\
& + A_{D_{1}D^{\star}\pi}^{\star}D_{1}^{\mu-}\left( D_{\mu
}^{\star+}\pi^{0}-\sqrt{2}\bar{D}_{\mu}^{\star0}\pi^{+}\right)  \nonumber\\
&  +B_{D_{1}D^{\star}\pi}^{\star}\left\{  D_{1}^{\mu-}\left[
\left(
\partial_{\nu
}D_{\mu}^{\star+}-\partial_{\mu}D_{\nu}^{\star+}\right)
\partial^{\nu}\pi^{0}-\sqrt{2}\left(  \partial_{\nu}\bar{D}_{\mu}^{\star0}%
-\partial_{\mu}\bar{D}_{\nu}^{\star0}\right)
\partial^{\nu}\pi^{+}\right]  \right.
\nonumber\\
&  \left.  +\partial^{\nu}D_{1}^{\mu-}\left[  \left(
D_{\nu}^{\star
+}\partial_{\mu}\pi^{0}-D_{\mu}^{\star+}\partial_{\nu}\pi^{0}\right)
-\sqrt{2}\left(
\bar{D}_{\nu}^{\star0}\partial_{\mu}\pi^{+}-\bar{D}_{\mu}^{\star
0}\partial_{\nu}\pi^{+}\right)  \right]
\right\}\>,\label{D1Dstarpi}
\end{align}

with the following coefficients

\begin{align}
A_{D_{1}D^{\star}\pi} &
=\frac{i}{\sqrt{2}}Z_{\pi}(g_{1}^{2}-h_{3})\phi_{C}\text{ ,} \label{AD1Dstarpi}\\
B_{D_{1}D^{\star}\pi} &  =\frac{i}{2}Z_{\pi}g_{2}w_{a_{1}}\text{ .} \label{BD1dstarK}
\end{align}

 From the interaction Lagrangian (\ref{D1Dstarpi}) we obtain that the neutral
 state of
$D_{1}$ decays into $D^{\ast0}\pi^{0},\,D^{\ast+}\pi^{-}$ and that
the positively charged state decays into
$D^{\ast+}\pi^{0},\,D^{\ast0}\pi^{+}$ (see below). The interesting
point is that the decays $D_1^0 \rightarrow D^+ \pi^-$ and $D_1^+
\rightarrow D^0\pi^+$ (although kinematically allowed) do not
occur in our model; because there is no respective tree-level
coupling. This is in agreement with the small experimental upper
bound. Improvements in the decay channels of $D_1(2420)$ could be made by
taking into account also the multiplet of pseudovector
quark-antiquark states. In this way, one will be able to evaluate
the mixing of these configurations and describe at the same time
the resonances $D_1(2420)$ and $D_1(2430)$.

\subsubsection{Decay Width \boldmath$D^0_{1}\rightarrow D^{\star}\pi$}

 The decay width of $D_{1}^{0}$ into $D^{\ast0}\pi^{0}$ is given by Eq.\ (\ref{d1dspionQ}) upon substituting the
 fields $A,\,V^0,$ and $\tilde{P}^0$ with $D_{1}^{0},\,D^{\ast0},$ and
 $\pi^{0}$ respectively. One may do likewise for the decay width of $D_{1}^{0}$ into
 $D^{\ast+}\pi^{-}$, but in this case it is necessary to multiply the expression by an isospin factor $2.$ Given that all
parameters entering Eqs.\ (\ref{d1dspionQ}, \ref{AD1Dstarpi}, \ref{BD1dstarK})
are known from Tables 4.1, 4.2, and 4.3, we consequently obtain the following
value of the decay width of $D_1^0$ into $D^\ast\pi$
\begin{equation}
\Gamma_{D_{1}(2420)^{0}\rightarrow
D^{\ast}\pi}=\Gamma_{D_{1}(2420)^{0}\rightarrow
D^{\ast+}\pi^{-}}+\Gamma_{D_{1}(2420)^{0}\rightarrow
D^{\ast0}\pi^{0}}= 65_{-37}^{+51}\text{ MeV.} \label{GD10Dstarp}
\end{equation}
 Experimentally the decay width of $D_{1}(2420)^{0}$ into
 $D^{\ast+}\pi^{-}$ has been observed. 

\subsubsection{Decay Width \boldmath$D^+_{1}\rightarrow D^{\star}\pi$}

One can proceed in a similar manner for the $D^+_1$ state, but in this
case we set $A\equiv D^{+}_1,\,V^{0}\equiv D^{\star}$, and
${\bar{P}^0\equiv}\pi$ in Eq.\ (\ref{d1dspionQ}). We obtain the
following value of the decay width

\begin{equation}
\Gamma_{D^+_{1}\rightarrow
D^{\star}\pi}=\Gamma_{D_{1}(2420)^{+}\rightarrow
D^{\ast+}\pi^{0}}+\Gamma_{D_{1}(2420)^{+}\rightarrow
D^{\ast0}\pi^{+}}=65_{-36}^{+51}\text{ MeV.} \label{GD1pDstarp}
\end{equation}



\subsection{Decay Width $D_{1}\rightarrow D\pi\pi$}

Now we turn to the three-body decays of the open axial-vector nonstrange charmed
meson $D_1(2420)^{0,+}$ which decays into three pseudoscalar mesons,
$D\pi\pi$. The relevant interaction Lagrangian can be written in a
single equation as follows:

\begin{align}
\mathcal{L}_{D_{1}D\pi\pi} &  =A_{D_{1}D\pi\pi}D_{1}^{\mu0}\partial_{\mu}\bar{D}^{0}%
\left({\pi^{0}}^2+2\pi^{-}\pi^{+}\right)  \nonumber\\
&  +B_{D_{1}D\pi\pi} D_{1}^{\mu0}\bar{D}^{0}\partial_{\mu}\pi^{0}\pi^{0}%
+C_{D_{1}D\pi\pi} D_{1}^{\mu0}\bar{D}^{0}\partial_{\mu}\pi^{-}\pi^{+} \nonumber\\%
& + E_{D_{1}D\pi\pi} D_{1}^{\mu0}\bar{D}^{0}\pi^{-}\partial_{\mu}\pi^{+}%
+ F_{D_{1}D\pi\pi} D_{1}^{\mu0}D^{+}(\pi^{0}\partial_{\mu}\pi^{-}-\partial_{\mu}\pi^{0}\pi^{-})\nonumber\\%
& +
A_{D_{1}D\pi\pi}\bar{D}_{1}^{\mu0}\partial_{\mu}D^{0}%
\left({\pi^{0}}^2+2\pi^{-}\pi^{+}\right)  \nonumber\\
&  +B_{D_{1}D\pi\pi}\bar{D}_{1}^{\mu0}D^{0}\partial_{\mu}\pi^{0}\pi^{0}%
+C_{D_{1}D\pi\pi}\bar{D}_{1}^{\mu0}D^{0}\partial_{\mu}\pi^{+}\pi^{-} \nonumber\\%
& + E_{D_{1}D\pi\pi} \bar{D}_{1}^{\mu0}D^{0}\pi^{+}\partial_{\mu}\pi^{-}%
+ F_{D_{1}D\pi\pi}\bar{D}_{1}^{\mu0}D^{-}(\pi^{0}\partial_{\mu}\pi^{+}-\partial_{\mu}\pi^{0}\pi^{+})\nonumber\\%
& +
A_{D_{1}D\pi\pi}D_{1}^{\mu+}\partial_{\mu}D^{-}%
\left({\pi^{0}}^2+2\pi^{-}\pi^{+}\right)  \nonumber\\
&  +B_{D_{1}D\pi\pi} D_{1}^{\mu+}D^{-}\partial_{\mu}\pi^{0}\pi^{0}%
+C_{D_{1}D\pi\pi} D_{1}^{\mu+}D^{-}\partial_{\mu}\pi^{-}\pi^{+} \nonumber\\%
& + E_{D_{1}D\pi\pi} D_{1}^{\mu+}D^{-}\pi^{-}\partial_{\mu}\pi^{+}%
+ F_{D_{1}D\pi\pi} D_{1}^{\mu+}D^{0}(\pi^{-}\partial_{\mu}\pi^{0}-\partial_{\mu}\pi^{-}\pi^{0})\nonumber\\%
& +
A_{D_{1}D\pi\pi}D_{1}^{\mu-}\partial_{\mu}D^{+}%
\left({\pi^{0}}^2+2\pi^{-}\pi^{+}\right)  \nonumber\\
&  +B_{D_{1}D\pi\pi} D_{1}^{\mu-}D^{+}\partial_{\mu}\pi^{0}\pi^{0}%
+C_{D_{1}D\pi\pi} D_{1}^{\mu-}D^{+}\partial_{\mu}\pi^{+}\pi^{-} \nonumber\\%
& + E_{D_{1}D\pi\pi} D_{1}^{\mu-}D^{+}\pi^{+}\partial_{\mu}\pi^{-}%
+ F_{D_{1}D\pi\pi}
D_{1}^{\mu-}\bar{D}^{0}(\pi^{+}\partial_{\mu}\pi^{0}-\partial_{\mu}\pi^{+}\pi^{0})\>,\label{intD1pipi}
\end{align}

with the coefficients 
\begin{align}
A_{D_{1}D\pi\pi}  &  =\frac{1}{4}Z_{\pi}^{2}\,Z_{D}\,w_{D_{1}}(g_{1}%
^{2}+2h_{1}+h_{2})\;,\label{AD1Dpi}\\
B_{D_{1}D\pi\pi}  &  =\frac{1}{4}Z_{\pi}^{2}\,Z_{D}\,w_{a_{1}}(3g_{1}%
^{2}+h_{2}-2h_{3})\text{ ,} \label{BD1Dpi}\\%
C_{D_{1}D\pi\pi}  &  =\frac{1}{2}Z_{\pi}^{2}\,Z_{D}\,w_{a_{1}}(g_{1}^{2}+h_{2})\;,\label{AD1Dpi}\\
E_{D_{1}D\pi\pi}  &  =Z_{\pi}^{2}\,Z_{D}\,w_{a_{1}}(g_{1}^{2}-h_{3})\text{ ,} \label{BD1Dpi}\\%
F_{D_{1}D\pi\pi}  &  =\frac{\sqrt{2}}{4}Z_{\pi}^{2}\,Z_{D}\,w_{a_{1}}(g_{1}^{2}-h_{2}-2h_{3})\;.\label{AD1Dpi}\\
\end{align}

 As seen in the Lagrangian (\ref{intD1pipi}) there are
three relevant channels for the neutral state $D^0_1$ which are
 $D^0_1 \rightarrow D^{0}\pi^{0}\pi^{0}$, $D^0_1 \rightarrow D^{0}\pi^{+}\pi^{-}$, and $D^0_1 \rightarrow
 D^{+}\pi^{-}\pi^{0}$. The positive state also has three
 relevant channels due to  $D^+_1 \rightarrow D^{+}\pi^{+}\pi^{-}$, $D^+_1 \rightarrow D^{+}\pi^{0}\pi^{0}$, and $D^+_1 \rightarrow
 D^{0}\pi^{0}\pi^{+}$.

\subsubsection{Decay Width $D_{1}^{0,+}\rightarrow D^{0,+}\pi\pi$}

The Lagrangian (\ref{intD1pipi}) allows us to calculate the decay
width for the process $D_1 \rightarrow D \pi \pi$. Firstly let us
focus on the three body-decay of the neutral state $D_{1}^{0}$
into $D^0\pi^0\pi^0$. We denote the momenta of $D_1^0, \,
D^0,\,\pi^0$ and $\pi^0$ as $P,\, P_1,\,P_2$, and $P_3$. A vector
state $D_1$ is involved in this decay process. We then consider
the corresponding polarization vector
$\varepsilon_\mu^{(\alpha)}(P)$ for $D_1$, and substitute
$\partial^\mu \rightarrow -iP^\mu$ for the decaying particle and
$\partial^\mu \rightarrow iP_{1,2,3}^\mu$ for the decay products.
We straightforwardly obtain the following $D^0_1 D^0 \pi^0 \pi^0$
scattering amplitude $-iM_{D_1^0\rightarrow D^0 \pi^0
\pi^0}^{(\alpha)}$ from the Lagrangian (\ref{intD1pipi}):

\begin{align}
-iM_{D_1^0\rightarrow D^0 \pi^0
\pi^0}^{(\alpha)}& = \varepsilon _\mu^{(\alpha)}(P)\,h_{D_{1}D\pi\pi}^{\mu} \text{
,}\label{MD1Dpipi}
\end{align}
where $h_{D_{1}D\pi\pi}^{\mu}$ is the vertex following from the
relevant part of the Lagrangian, 
\begin{equation}
h_{D_{1}D\pi\pi}^{\mu}=-\left[  A_{D_{1}D\pi\pi}P_{1}^{\mu}%
+B_{D_{1}D\pi\pi}P_{2}^{\mu}\right]  \text{ .} \label{hD1Dpipi}%
\end{equation}
In order to calculate the decay width of $D_1^0$ we need to
calculate the averaged modulus squared decay amplitude
$|\overline{-iM_{D_1^0\rightarrow D^0 \pi^0 \pi^00}}|^2$ from Eq.(\ref{MD1Dpipi}):
\begin{align}
|\overline{-iM_{D_1^0\rightarrow D^0 \pi^0 \pi^0}}|^2 &=
\frac{1}{3}\sum_{\alpha=1}^3 |-iM_{D_1^0\rightarrow D^0 \pi^0
\pi^0}^{(\alpha)}|^2 \nonumber \\
& =\frac{1}{3}\sum_{\alpha=1}^3 \varepsilon _\mu^{(\alpha)}(P)
\varepsilon _\nu^{(\alpha)}(P)
h_{D_{1}D\pi\pi}^{\mu}\,h_{D_{1}D\pi\pi}^{\ast\nu}\nonumber\\
&=\frac{1}{3}\bigg(-g_{\mu\nu}+\frac{P_\mu
P_\nu}{M^2}\bigg)h_{D_{1}D\pi\pi}^{\mu}\,h_{D_{1}D\pi\pi}^{\ast\nu}\nonumber\\
&=\frac{1}{3} \left[ -|h_{D_{1}D\pi\pi}^{\mu}|^2
+\frac{|P_\mu\,h_{D_{1}D\pi\pi}^{\mu}|^2}{M^2}\right]\text{
,}\label{MD1Dpipisq}
\end{align}
 where
\begin{equation}
|h_{D_{1}D\pi\pi}^{\mu}|^2  =A_{D_{1}D\pi\pi}^2P_{1}^2%
+2 A_{D_{1}D\pi\pi}\,B_{D_{1}D\pi\pi}(P_1\cdot P_{2})+B_{D_{1}D\pi\pi}^2\,P_2^2  \text{ ,} \label{hD1Dpipisq}%
\end{equation}
\begin{equation}
|P_\mu\,h_{D_{1}D\pi\pi}^{\mu}|^2
=|-\left[A_{D_{1}D\pi\pi}(P\cdot \,P_{1})+B_{D_{1}D\pi\pi}(P\cdot P_{2})\right]|^2 \text{ ,} \label{PhD1Dpipi}%
\end{equation}
where $P^{\mu}=(M,\mathbf{0})$,
$P_{1}^{\mu}=(E_{P_{1}},\mathbf{p})$,
$P_{2}^{\mu}=(E_{P_{2}},-\mathbf{p})$, and $P_{3}^{\mu}=(E_{P_{3}}%
,-\mathbf{p})$ are the four-momenta of $D_{1}^{0}$, $D^{0}$,
$\pi^{0}$, and $\pi^{0}$ in the rest frame of $D_{1}^{0}$,
respectively. In this frame,
\begin{align*}
&  P_{1}\cdot P_{2}=\frac{m_{12}^{2}-m_{1}^{2}-m_{2}^{2}}{2}\,,\\
&  P\cdot P_{1}=m_{1}^{2}+\frac{m_{12}^{2}-m_{1}^{2}-m_{2}^{2}}{2}%
+\frac{m_{13}^{2}-m_{1}^{2}-m_{3}^{2}}{2}\,,\\
&  P\cdot P_{2}=m_{2}^{2}+\frac{m_{12}^{2}-m_{1}^{2}-m_{2}^{2}}{2}%
+\frac{m_{23}^{2}-m_{2}^{2}-m_{3}^{2}}{2}\,.
\end{align*}
The quantities $M,\,m_{1},\,m_{2},\,m_{3}$ refer to the mass of
$D_{1}^{0},\,D^{0},\,\pi^{0},\,\pi^{0}$, respectively. Using
Eq.(\ref{B3}) the decay width for the process $D_{1}^{0}\rightarrow
D^{0}\pi^{0}\pi^{0}$ can be obtained as
\begin{align}
\Gamma_{A\rightarrow P_{1}P_{2}P_{3}}=\frac{2}{96\,(2\pi)^{3}\,M^{3}%
}\int_{(m_{1}+m_{2})^{2}}^{(M-m_{3})^{2}}&\int_{(m^2_{23})_{\min}}^{(m^2_{23}%
)_{\max}}\nonumber\\
\times &\left[  -\left\vert h_{D_{1}D\pi\pi}^{\mu}\right\vert ^{2}%
+\frac{\left\vert h_{D_{1}D\pi\pi}^{\mu}P_{\mu}\right\vert ^{2}}{M^{2}%
}\right]  \,dm_{23}^{2}\,dm_{12}^{2}\>\,. \label{d1d2pionQ}%
\end{align}
The decay of the neutral states for the processes
$D_{1}^{0}\rightarrow D^{0}\pi ^{+}\pi^{-}$ and
$D_{1}^{0}\rightarrow D^{+}\pi ^{-}\pi^{0}$ has the same formula
as in Eq.\ (\ref{d1d2pionQ}), only in this case it is
multiplied by an isospin factor $2$ as given by the Lagrangian
(\ref{intD1pipi}). We then get
\begin{equation}
\Gamma_{D^0_{1}\rightarrow D^0
\pi\pi}=\Gamma_{D_{1}(2420)^{0}\rightarrow
D^{0}\pi^{0}\pi^{0}}+\Gamma_{D_{1}(2420)^{0}\rightarrow
D^{0}\pi^{+}\pi^{0}}=(0.59\pm 0.02)\text{ MeV.} \label{GD10D0pipi}
\end{equation}

For the positive state $D_1(2420)^+$ the decay process
$D_{1}^{+}\rightarrow D^{+}\pi^{0}\pi^{0}$ has an analogous
analytic expression as for the process $D_{1}^{0}\rightarrow
D^{0}\pi^{0}\pi^{0}$ (as presented in Eq.\ (\ref{d1d2pionQ})),
while $D_{1}^{+}\rightarrow D^{+}\pi^{+}\pi^{-}$ has the same
formula, multiplied by an isospin factor $2$ as given from the
Lagrangian (\ref{intD1pipi}).
\newline
\begin{equation}
\Gamma_{D^0_{1}\rightarrow D^+
\pi\pi}=\Gamma_{D_{1}(2420)^{+}\rightarrow
D^{+}\pi^{0}\pi^{0}}+\Gamma_{D_{1}(2420)^{+}\rightarrow
D^{+}\pi^{+}\pi^{-}}=(0.56\pm0.02)\text{ MeV.} \label{GD10D0pipi}
\end{equation}
This has been observed experimentally.

\subsubsection{Decay Width $D^{0,+}_{1}\rightarrow D\pi^{+}\pi^{0}$}

The scattering amplitudes for the processes $D^{0}_{1}\rightarrow
D^-\pi^{+}\pi^{0}$ and $D^{+}_{1}\rightarrow D^0\pi^{+}\pi^{0}$
differ markedly from the previous processes for the three-body
decay of $D^{0,+}_1$, as can be seen from the interaction Lagrangian
(\ref{intD1pipi}). Thus, we study the decay widths for these
channels differently. Let us firstly focus on the decay width of
the neutral state $D^{0}_{1}$ into $ D^-\pi^{+}\pi^{0}$, which has the same expression as for
the decay process $D^{+}_{1}\rightarrow D^0\pi^{+}\pi^{0}$. To write the scattering amplitude
for $D^{0}_{1}\rightarrow D^-\pi^{+}\pi^{0}$ we denote the momenta
of $D_1^0, \, D^-,\,\pi^+$, and $\pi^0$ as $P,\, P_1,\,P_2$, and
$P_3$. For the meson $D^0_1$, we consider the
polarization vector $\varepsilon_\mu^{(\alpha)}(P)$ and substitute $\partial^\mu \rightarrow -iP^\mu$ for
the decaying particle and $\partial^\mu \rightarrow
iP_{1,2,3}^\mu$ for the outgoing particles. We then obtain the
following decay amplitude from the Lagrangian (\ref{intD1pipi}):

\begin{align}
-iM_{D_1^0\rightarrow D^0- \pi^+
\pi^0}^{(\alpha)}= \varepsilon _\mu^{(\alpha)}(P)\,X_{D_{1}D\pi\pi}^{\mu} \text{
,}\label{MD1Dpipi2}
\end{align}
where the vertex $X_{D_{1}D\pi\pi}^{\mu}$ is
\begin{equation}
X_{D_{1}D\pi\pi}^{\mu}= - F_{D_{1}D\pi\pi}(P_{2}^{\mu}%
-P_{3}^{\mu}) \text{ .} \label{xD1Dpipi}%
\end{equation}
 The averaged modulus squared amplitude $|\overline{-i\,M_{D_1^0\rightarrow D^0- \pi^+
\pi^0}}|^2$ has the same general formula as in Eq.(\ref{MD1Dpipisq}),
with substituting the vertices  $h_{D_{1}D\pi\pi}^{\mu}\rightarrow
X_{D_{1}D\pi\pi}^{\mu}$,

\begin{align}
|\overline{-i M_{D_1^0\rightarrow D^0- \pi^+ \pi^0}}|^2 &=\frac{1}{3}
\left[ -|X_{D_{1}D\pi\pi}^{\mu}|^2
+\frac{|P_\mu\,X_{D_{1}D\pi\pi}^{\mu}|^2}{M^2}\right]\nonumber\\
&= \frac{F_{D_{1}D\pi\pi}^2}{3}\left[
\frac{1}{M^2}(P\cdot P_2-P\cdot P_3)^2-(m_2^2 + m_3^2 + 2\,
P_2\cdot P_3)\right]\,.
\end{align}
The decay width $D_{1}^{0}\rightarrow D^{+}\pi^{-}\pi^{0}$ reads
\begin{align}
\Gamma_{A\rightarrow P_{1}P_{2}P_{3}}=\frac{F_{D_{1}D\pi\pi}^{2}}%
{3\times\,32\,(2\pi)^{3}\,M^{3}} & \int_{(m_{1}+m_{2})^{2}}^{(M-m_{3})^{2}}%
 \int_{(m^2_{23})_{\min}}^{(m^2_{23})_{\max}}
\,dm_{23}^{2}\,dm_{12}^{2}\nonumber\\
&\times \left[
\frac{1}{M^2}(P\cdot P_2-P\cdot P_3)^2-(m_2^2 + m_3^2 + 2\, P_2\cdot P_3)\right]\>,\label{GADpp}
\end{align}
where the quantities $M,\,m_{1},\,m_{2},\,$ and $m_{3}$ refer to
the masses for the fields $D_{1}^{0},\,D^{+},\,\pi^{-},$ and
$\pi^{0}$, respectively, and the kinematic relations
\begin{align*}
&  P_{2}\cdot P_{3}=\frac{m_{23}^{2}-m_{2}^{2}-m_{3}^{2}}{2}\,,\\
&  P\cdot P_{3}=m_{3}^{2}+\frac{m_{13}^{2}-m_{1}^{2}-m_{3}^{2}}{2}%
+\frac{m_{23}^{2}-m_{2}^{2}-m_{3}^{2}}{2}\,.
\end{align*}
All the parameters entering Eq.(\ref{GADpp}) are fixed and listed
in the Tables 4.1 and 4.3. Consequently we obtain the decay width of
$D_{1}(2420)^{0}$ into $D^+ \pi^-\pi^0$:
\begin{equation}
\Gamma_{D_{1}(2420)^{0}\rightarrow D^+
\pi^-\pi^0}=0.21_{-0.015}^{+0.01}\text{MeV.} \label{GD10Dpimpi}
\end{equation}

The decay of the charged state $D_{1}^{+}$ into
$D^{0}\pi^{0}\pi^{+}$ has an analogous expression and is 
\begin{equation}
\Gamma_{D_{1}(2420)^{+}\rightarrow D^0 \pi^0\pi^+}=(0.22 \pm
0.01)\text{ MeV.} \label{GD1+D0pi0pi}
\end{equation}
This decay is observed experimentally.

\subsection{Decay Width $D_{S1}\rightarrow D^{\ast}K$}

As a last step we turn to the strange-charmed axial-vector state
$D_{S1}^+$ which is assigned to $D_{S1}(2536)$. This resonance
decays into $D^\ast(2010)^+K^0$ and $D^\ast(2007)^0K^+$ in the
model (\ref{fulllag}), whereas kinematically the decay to $DK$ is not allowed. This is in agreement with experimental data in which the
decays $D_{S1}(2536)^+\rightarrow D^+K^0$ and
$D_{S1}(2536)^+\rightarrow D^0K^+$ are not seen (as stated by the PDG \cite{Beringer:1900zz}). The $D_{S1}D^{\star}K$ interaction Lagrangian from
Eq.\ (\ref{fulllag}) reads

\begin{align}
\mathcal{L}_{D_{S1}D^{\star}K} & =
A_{D_{S1}D^{\ast}K}D_{S1}^{\mu+}(D_{\mu
}^{\star-}\bar{K}^{0}+D_{\mu}^{\star0}K^{-})\nonumber\\%
&  +B_{D_{S1}D^{\ast}K}\{D_{S1}^{\mu+}[(\partial_{\nu}D_{\mu}^{\star-}%
-\partial_{\mu}D_{\nu}^{\star-})\partial^{\nu}\bar{K}^{0}+(\partial_{\nu
}D_{\mu}^{\star0}-\partial_{\mu}D_{\nu}^{\star0})\partial^{\nu}K^{-}]\nonumber\\%
& +\partial^{\nu}D_{S1}^{\mu+}[D_{\nu}^{\star-}\partial_{\mu}\bar{K}^{0}%
-D_{\mu}^{\star-}\partial_{\nu}\bar{K}^{0}+D_{\nu}^{\star0}\partial_{\mu}K^{-}-D_{\mu}^{\star0}\partial_{\nu}K^{-}]\}\nonumber\\%
& + A_{D_{S1}D^{\ast}K}^{\star}D_{S1}^{\mu-}(D_{\mu
}^{\star+}K^{0}+\bar{D}_{\mu}^{\star0}K^{+})\nonumber\\%
&  + B_{D_{S1}D^{\ast}K}^{\star}\{D_{S1}^{\mu-}[(\partial_{\nu}D_{\mu}^{\star+}%
-\partial_{\mu}D_{\nu}^{\star+})\partial^{\nu}K^{0}+(\partial_{\nu
}\bar{D}_{\mu}^{\star0}-\partial_{\mu}\bar{D}_{\nu}^{\star0})\partial^{\nu}K^{+}]\nonumber\\%
& +\partial^{\nu}D_{S1}^{\mu-}[D_{\nu}^{\star+}\partial_{\mu}K^{0}%
-D_{\mu}^{\star+}\partial_{\nu}K^{0}+\bar{D}_{\nu}^{\star0}\partial_{\mu}K^{+}-
\bar{D}_{\mu}^{\star0}\partial_{\nu}K^{+}]\}\>,\label{LDS1DsK}
\end{align}

with the following coefficients

\begin{align}
&  A_{D_{S1}D^{\ast}K}=\frac{i}{4}Z_{K}\left[
g_{1}^{2}(\sqrt{2}\phi
_{N}-2\phi_{S}-4\phi_{C})+h_{2}(\sqrt{2}\phi_{N}-2\phi_{S})+4h_{3}\phi
_{C}\right]  ,\\
&  B_{D_{S1}D^{\ast}K}=-\frac{i}{\sqrt{2}}Z_{K}\,g_{2}w_{K_{1}}\,.
\end{align}

According to the interaction Lagrangian, $D_{S1}$ decays into the channels $D^{\ast+}K^{0}$ and $D^{\ast0}%
K^{+}$. The formula for the decay widths
$\Gamma_{D_{S1}^{+}\rightarrow D^{\ast+}K^{0}}$ and
$\Gamma_{D_{S1}^{+}\rightarrow D^{\ast0}K^{0}}$ is as in Eq.\
(\ref{d1dspionQ}), when setting $A \equiv D_{S1},\, V \equiv
D^\ast, \bar{P}\equiv K$ and replacing the vertex
$h_{AV\bar{P}}^{\mu\nu}$ by the following vertex
$h_{D_{S1}D^{\ast}K}^{\mu\nu}$:
\begin{equation}
h_{D_{S1}D^{\ast}K}^{\mu\nu}=i\left\{  A_{D_{S1}D^{\ast}K}g^{\mu\nu}%
+B_{D_{S1}D^{\ast}K}[P_{1}^{\mu}P_{2}^{\nu}+P_{2}^{\mu}P^{\nu}-(P\cdot
P_{2})g^{\mu\nu}-(P_{1}\cdot P_{2})g^{\mu\nu}]\right\}  \text{ ,}
\label{hDs1Dsk}%
\end{equation}
The parameters are fixed and presented in Tables 4.1, 4.2, and 4.3. We
then obtain the decay width into $D^\ast K$ as
\begin{align}
\Gamma_{D_{S1}(2536)^{+}\rightarrow D^\ast K}
&=\Gamma_{D_{S1}^{+}\rightarrow D^{\ast0}
K^+}+\Gamma_{D_{S1}^{+}\rightarrow D^{\ast+}
K^0}\nonumber\\
& = 25_{-15}^{+22}\text{ MeV.} \label{GDs1DsK}
\end{align}
 whereas $\Gamma_{D_{S1}(2536)^{+}\rightarrow D^\ast
 K}^{exp}=(0.92\pm0.03\pm0.04)$ MeV.

\section{Weak decay constants of Charmed mesons}

In this subsection we evaluate the weak decay constants of the
pseudoscalar mesons $D$, $D_{S},$ and $\eta_C$. Their analytic expressions
read [see Appendix A and also
Ref.\ \cite{Eshraim:2014afa, Eshraim:2014eka, Eshraim:2014tla}]%

\begin{align}
f_{D}=&\frac{\phi_{N}+\sqrt{2}\phi_{C}}{\sqrt{2}Z_{D}}\text{ },\\
f_{D_{S}}=&\frac{\phi_{S}+\phi_{C}}{Z_{D_{S}}}\text{ },\\ f_{\eta_{C}
}=&\frac{2\phi_{C}}{Z_{\eta_{C}}}\text{ }.
\end{align}
Using the parameters of the fit we obtain
\begin{align}
f_{D}=&(254\pm17)\text{ MeV },\text{ }\\f_{D_{S}}=&(261\pm17)\text{
MeV },\text{
}\\f_{\eta_{C}}=&(314\pm39)\text{ MeV.}%
\end{align}
The experimental values $f_{D}=(206.7\pm8.9)$ MeV and $f_{D_{s}}%
=(260.5\pm5.4)$ MeV \cite{Beringer:1900zz} show a good agreement for
$f_{D_{s}}$ and a slightly too large theoretical result for
$f_{D}$. The quantity $f_{\eta_{C}}$ is in fair agreement with the
experimental value $f_{\eta_{C}}=(335\pm75)$ MeV \cite{Edwards:2000bb} as
well as with the theoretical result $f_{\eta_{C}}=(300\pm50)$ MeV
obtained in Ref.\ \cite{Deshpande:1994mk}. These results show that our
determination of the condensate $\phi_{C}$ is reliable (even if
the theoretical uncertainty is still large).

\newpage

\section{Summary}

We summarize the results of the (OZI-dominant) strong decay widths
of the resonances $D_{0},$ $D^{\ast},$ $D_{1}$, and $D_{S1},$ in
Table 7.1. For the calculation of the decay widths we
have used the physical masses listed by the PDG \cite{Beringer:1900zz}. This is necessary
in order to have the correct phase space. With the exception of
$D_{S1}(2536)^{+}\rightarrow D^{\ast}K,$ all values are in reasonable
agreement with the mean experimental values and upper bounds.
Although the theoretical uncertainties are still large and the
experimental results are not yet well known, the qualitative
agreement is anyhow interesting if one considers that the decay
amplitudes depend on the parameters of the three-flavour version of
the model determined in Ref.\ \cite{Parganlija:2012fy}. Note that the
theoretical errors have been calculated by
taking into account the uncertainty in the charm condensate $\phi_{C}%
=(176\pm28)$ MeV. The lower theoretical value corresponds to
$\phi_{C}=\left( 176-28\right)  $ MeV, while the upper one to
$\phi_{C}=\left(  176+28\right) $ MeV. The explicit expressions
for the decay widths are reported in the previous.\\
Here we do not study the decay of other (hidden and open) charmed
states because we restrict ourselves to OZI-dominant processes.
The study of OZI-suppressed decays which involve the large-$N_{c}$
suppressed parameters $\lambda_{1}$ and $h_{1}$ is left for the next chapter. There, also the decays of the well-known charmonium states
(such as $\chi_{c0}$ and $\eta_{c}$) will be investigated.
\begin{center}
\begin{tabular}
[c]{|c|c|c|}\hline Decay Channel & Theoretical result [MeV] &
Experimental result [MeV]\\\hline
$D_{0}^{\ast}(2400)^{0}\rightarrow D\pi$
& $139_{-114}^{+243}$ & $D^{+}\pi^{-}$ seen; full width
$\Gamma=267\pm 40$\\\hline $D_{0}^{\ast}(2400)^{+}\rightarrow
D\pi$ & $51_{-51}^{+182}$ &
$D^{+}\pi^{0}$ seen; full width: $\Gamma=283\pm24\pm 34$\\\hline
$D^{\ast}(2007)^{0}\rightarrow D^{0}\pi^{0}$ & $0.025\pm0.003$ &
seen; $<1.3$\\\hline $D^{\ast}(2007)^{0}\rightarrow D^{+}\pi^{-}$
& $0$ & not seen\\\hline $D^{\ast}(2010)^{+}\rightarrow
D^{+}\pi^{0}$ & $0.018_{-0.003}^{+0.002}$ &
$0.029\pm0.008$\\\hline $D^{\ast}(2010)^{+}\rightarrow
D^{0}\pi^{+}$ & $0.038_{-0.004}^{+0.005}$ &
$0.065\pm0.017$\\\hline $D_{1}(2420)^{0}\rightarrow
D^{\ast}\pi$ & $65_{-37}^{+51}$
& $D^{\ast+}\pi^{-}$ seen; full width: $\Gamma=27.4\pm
2.5$\\\hline
$D_{1}(2420)^{0}\rightarrow D^{0}\pi\pi$ & $0.59\pm0.02$ & seen\\\hline
$D_{1}(2420)^{0}\rightarrow D^{+}\pi^{-}\pi^{0}$ &
$0.21_{-0.015}^{+0.01}$ & seen\\\hline
$D_{1}(2420)^{0}\rightarrow D^{+}\pi^{-}$ & $0$ & not seen; $\Gamma(D^{+}%
\pi^{-})/\Gamma(D^{\ast+}\pi^{-})<0.24$\\\hline
$D_{1}(2420)^{+}\rightarrow
D^{\ast}\pi$ & $65_{-36}^{+51}$
& $D^{\ast0}\pi^{+}$ seen; full width: $\Gamma=25\pm 6$\\\hline
$D_{1}(2420)^{+}\rightarrow D^{+}\pi\pi$ & $0.56\pm0.02$ & seen\\\hline
$D_{1}(2420)^{+}\rightarrow D^{0}\pi^{0}\pi^{+}$ & $0.22\pm0.01$ &
seen\\\hline
$D_{1}(2420)^{+}\rightarrow D^{0}\pi^{+}$ & $0$ & not seen; $\Gamma(D^{0}%
\pi^{+})/\Gamma(D^{\ast0}\pi^{+})<0.18$\\\hline
$D_{S1}(2536)^{+}\rightarrow
D^{\ast}K$ & $25_{-15}^{+22}$ &
seen; full width $\Gamma=0.92\pm0.03\pm0.04$\\\hline
$D_{S1}(2536)^{+}\rightarrow D^{+}K^{0}$ & $0$ & not seen\\\hline
$\,D_{S1}(2536)^{+}\rightarrow+D^{0}K^{+}$ & $0$ & not
seen\\\hline
\end{tabular}
Table 7.1: Decay widths of charmed
mesons.
\end{center}

The following comments are in order:

(i) The decay of $D_{0}^{\ast}(2400)^{0}$ into $D\pi$ has a very
large theoretical error due to the imprecise determination of
$\phi_{C}$. A qualitative statement is, however, possible: the
decay channel $D_{0}^{\ast }(2400)^{0}\rightarrow D\pi$ is large
and is the only OZI-dominant decay predicted by our model. This
decay channel is also the only one seen in experiment (although
the branching ratio is not yet known). A similar discussion holds
for the charged counterpart $D_{0}^{\ast}(2400)^{+}$.

(ii) The decay widths of the vector charmed states
$D^{\ast}(2007)^{0}$ and $D^{\ast}(2010)^{+}$ are slightly smaller than the experimental results, but
close to the lower bounds of the latter.

(iii) The results for the axial-vector charmed states
$D_{1}(2420)^{0}$ and $D_{1}(2420)^{+}$ are compatible with
experiment. Note that the decay into $D^{\ast}\pi$ is the only one
which is experimentally seen. Moreover, the
decays $D_{1}(2420)^{0}\rightarrow D^{+}\pi^{-}$ and $D_{1}(2420)^{+}%
\rightarrow D^{0}\pi^{+}$, although kinematically allowed, do not
occur in our model because there is no respective tree-level
coupling; this is in agreement with the small experimental upper
bound. Improvements in the decay channels of $D_{1}(2420)$ are
possible by taking into account also the multiplet of pseudovector
quark-antiquark states. In this way, one will be able to evaluate
the mixing of these configurations and describe at the same time
the resonances $D_{1}(2420)$ and $D_{1}(2430)$.

(iv) It is interesting that the decays of the vector states
$D^{\ast }(2007)^{0}$ and $D^{\ast}(2010)^{\pm}$ and of the
axial-vector states $D_{1}(2420)^{0}$ and $D_{1}(2420)^{+}$ can be
simultaneously described with the same set of (low-energy)
parameters. Namely, in the low-energy language these states are
chiral partners and the results (even if the experimental
knowledge is not yet conclusive and the theoretical uncertainties
are still large) show that chiral symmetry is still important in
the energy range relevant for charmed mesons.

(v) The decay of the axial-vector strange-charmed
$D_{S1}(2536)^{+}\rightarrow D^{\ast}K$ is too large in our model
when compared to the experimental data of about $1$ MeV. This
result is robust upon variation of the parameters, as the error
shows. We thus conclude that the resonance $D_{S1}(2536)^{\pm}$ is
not favored to be (predominantly) a member of the axial-vector
multiplet (it can be, however, a member of the pseudovector
multiplet). Then, we discuss two possible solutions to the problem
of identifying the axial-vector strange-charmed quarkonium:

\textit{Solution 1}: There is a `seed' quark-antiquark
axial-vector state $D_{S1}$ above the $D^{\ast}K$ threshold, which
is, however, very broad and for this reason has \emph{not yet}
been detected. Quantum corrections generate the state
$D_{S1}(2460)^{\pm}$ through pole doubling \cite{vanBeveren:1986ea, Boglione:2002vv, vanBeveren:2006ua}. In
this scenario, $D_{S1}(2460)$ is dynamically generated but is
still related to a broad quark-antiquark seed state. In this way,
the low mass of $D_{S1}(2460)$ in comparison to the quark-model
prediction \cite{Godfrey:1985xj, Godfrey:1986wj} is due to quantum corrections
\cite{Lutz:2007sk, Achasov:2004uq, Giacosa:2007bn, Giacosa:2012de, Coito:2011qn, Rupp:2012py}. Then, the state $D_{S1}(2460)$, being
below threshold, has a very small decay width.

\textit{Solution 2}: Also in this case, there is still a broad and
not yet detected quark-antiquark field above threshold, but
solution 1 is assumed not to apply (loops are not sufficient to
generate $D_{S1}(2460)$). The resonance $D_{S1}(2460)^{\pm}$ is
not a quark-antiquark field, but a tetraquark or a loosely bound
molecular state and its existence is not related to the
quark-antiquark state of the axial-vector multiplet.

(vi) For the state $D_{S0}^{\ast}(2317)$ similar arguments apply.
If the mass of this state is above the $DK$ threshold, we predict
a very large ($\gtrsim 1$ GeV) decay width into $DK$ (for example:
$\Gamma_{D_{S0}^{\ast}\rightarrow DK}\simeq3$ GeV for a
$D_{S0}^{\ast}$ mass of $2467$ MeV as determined in Table 4.5).
Then, the two solutions mentioned above are applicable also here:

\textit{Solution 1}: A quark-antiquark state with a mass above the
$DK$ threshold exists, but it is too broad to be seen in
experiment. The state $D_{S0}(2317)$ arises through the
pole-doubling mechanism.

\textit{Solution 2}: Loops are not sufficient to dynamically
generate $D_{S0}^{\ast}(2317).$ The latter is not a quarkonium but
either a tetraquark or a molecular state.

In conclusion, a detailed study of loops in the axial-vector and
scalar strange-charm sector needs to be performed. In the
axial-vector strange-charm sector mixing with a pseudovector
quark-antiquark state should also be included. These tasks go
beyond the tree-level analysis of our work but are an interesting
subject for the future. \newline


\chapter{Decay of hidden charmed mesons}

\section{Introduction}

\indent Charmonia exhibit a spectrum of resonances and play the
same role for understanding hadronic dynamics as the hydrogen atom
\cite{Novikov:1977dq}. The properties of charmonia are determined by the
strong interaction which is undoubtedly one of the most
challenging tasks. Since the discovery of the charmonium state
$(J/\psi)$ with quantum numbers $J^{PC}=1^{--}$ in November 1974 at BNL \cite{Aubert:1974js}
and at SLAC \cite{Augustin:1974xw}, a significant experimental progress has
been achieved in charmonium spectroscopy. As an example of this,
the hadronic and electromagnetic transitions between charmonium
states and their decays have been measured with high precision
with the BESIII spectrometer at the electron-positron collider at
IHEP Beijing. Moreover, unconventional narrow charmonium-rich
states have been recently discovered in an energy regime above the
open-charm threshold by Belle \cite{Choi:2002na, Vinokurova:2011dy, Uehara:2007vb, Liventsev:2011ks} and BaBar \cite{Lees:2010de},
which potentially initiates a new area in charmonium spectroscopy.
The upcoming PANDA experiment at the research facility FAIR will
exploit the annihilation of cooled anti-protons with protons to
perform charmonium spectroscopy with an incredible precision.

Recent theoretical developments such as nonrelativistic QCD
\cite{Sun:2014gca, Khan:2013ixa, Kniehl:2012ffa} and heavy-quark effective theory \cite{Mannel:1997ky},
potential models \cite{Cao:2012du, Segovia:2013kg}, lattice gauge theory
\cite{Kawanai:2011jt, DeTar:2011nn, Liu:2012ze}, and light front quantization have shown the direct
connection of charmonium properties with QCD. More details of the
experimental and theoretical situation is given in the Ref.
\cite{Brambilla:2010cs}. Therefore, we highlight the study of the decays of
charmonium states and their mixing with glueballs in the eLSM. The
charm quark has been included in the eLSM by its extension
from the case $N_f=3$ \cite{Parganlija:2012fy} to the case $N_f=4$ \cite{Eshraim:2014eka}. The eLSM has four charmonium states, which are the
(pseudo-)scalar ground states $\eta_c(1S)$ and $\chi_{c0}(1P)$
with quantum numbers $J^{PC}=0^{-+}$ and  $J^{PC}=0^{++}$ as well
as the ground-state (axial-)vector $J/\psi(1S)$  and $\chi_{c1}(1P)$
with quantum numbers $J^{PC}=1^{-+}$ and $J^{PC}=1^{++}$
\cite{Eshraim:2014afa, Eshraim:2014eka}, respectively. It also includes two
glueballs: a scalar glueball (denoted as $G$) and a pseudoscalar
glueball (denoted as $\widetilde{G}$), composed of two glouns.
There are two candidates for the scalar glueball which are the
resonance $f_0(1500)$ (which shows a flavour-blind decay
pattern) and the resonance $f_0(1710)$, because its mass is very
close to lattice-QCD predictions, and because it is produced in
the gluon-rich decay of the $J/\psi$, as seen in Refs.
\cite{Amsler:1995td, Lee:1999kv, Close:2001ga, Giacosa:2005qr, Giacosa:2004ug, Mathieu:2008me, Janowski:2011gt, Giacosa:2005zt, Cheng:2006hu, Chatzis:2011qz, Gutsche:2012zz, Giacosa:2009qh}. The latter is a
mixing between three bare fields: the nonstrange
$\sigma_N\equiv(u\overline{u}+d\overline{d})/\sqrt{2}$, the
hidden-strange $\sigma_S\equiv s\overline{s}$, and the scalar glueball
$G\equiv gg$. This three-body mixing in the scalar-isoscalar channel was
solved in Ref.\cite{Janowski:2014ppa} and generated the physical resonances
$f_0(1370),\,f_0(1500)$, and $f_0(1710)$. The last field that was
introduced in the eLSM is a pseudoscalar glueball via a term
describing the interaction between the pseudoscalar glueball with
scalar and pseudoscalar mesons; as seen in chapter 6. The decay
channels of the pseudoscalar glueball into scalar and pseudoscalar
mesons \cite{Eshraim:2012jv, Eshraim:2012ju, Eshraim:2012rb, Eshraim:2013dn} could potentially be
measured in the upcoming PANDA experiment at the FAIR facility
\cite{Lutz:2009ff}, which is based on proton-antiproton scattering and
has the ability to produce the pseudoscalar glueball in an
intermediate state with mass above $2.5$ GeV. The mass of a
pseudoscalar glueball predicted by lattice QCD (in the quenched
approximation) is about $2.6$ GeV \cite{Morningstar:1999rf, Loan:2005ff, Chen:2005mg, Gregory:2012hu}.

\indent In the present chapter we study the decay properties of the
charmonium states $\chi_{c0}$ and $\eta_c$ within the eLSM. The
charmonia region is a good one to look for exotics \cite{Peters:2007zza, Close:2005iz}. As seen in Ref. \cite{Bali:1993fb, Bali:2000vr, Morningstar:2003ew}, lattice QCD predicts the
existence of various glueball states in the charmonium mass
region, some of them with exotic quantum numbers which are not
possible in a $q\overline{q}$ state. We then obtain the scalar
glueball (due to the decay of $\chi_{c0}$) and the pseudoscalar
glueball (due to the decay of $\eta_c$). We calculate the decay
width of the charmonium state $\eta_c$ into a pseudoscalar
glueball (this decay is allowed in the channel $\eta_c \rightarrow
\pi\pi\widetilde{G}$), with a mass of $2.6$ GeV predicted by
lattice QCD and repeat it with a mass of $2.37$ GeV, as observed
in the BESIII experiment where pseudoscalar resonances have been
investigated in $J/\psi$ decays \cite{Ablikim:2005um, Kochelev:2005vd, Ablikim:2010au}. Particularly, we
consider the $X(2370)$ resonance, since its mass lies just below
the lattice-QCD prediction. Mixing phenomena, although believed to
be smaller than in the light mesonic sector, can occur also
between charmonia and glueballs with the same quantum numbers
\cite{Suzuki:2002bz, Chan:1999px}. The parameters have been used in the strange-nonstrange
investigation \cite{Parganlija:2012fy} (which are discussed in chapter 3),
whereas the three parameters related to the charm sector have been
fixed in chapter 4 through a fit including the masses of charmed
mesons. There are two parameters, $\lambda_1,\,h_1$, which had
zero values in the previous case ($N_f=4$), but in the present chapter this
should be re-evaluated, as the decay widths of $\chi_{c0}$ depend
on their values. For instance, there is a mixing
between a scalar glueball $G$ with the charmonium state $\chi_{c0}$,
but we neglect it in our evaluation because it is expected to be small. We
compute instead the mixing angle between the pseudoscalar glueball
(with a mass of $2.6$ GeV) with $\eta_c$.

\section{Decay of the scalar charmonium state $\chi_{c0}$}

\indent \indent In this section we study the decay properties of
the scalar ground-state charmonium $\chi_{c0}(1P)$ in the eLSM,
via computing the decay width of this charmonium state into
(axial-)vector and
(pseudo)scalar mesons and a scalar glueball as well. As a result of the study discussed in Ref.\cite{Janowski:2014ppa}, the resonance $f_0(1710)$ is predominantly a scalar glueball.\\

The terms in the eLSM relevant for the decay of $\chi_{C0}$ are

\begin{align}
\mathcal{L}_{\chi_{C0}}=&\mathcal{L}_{dil}-m_{0}^{2}\left(
\frac{G}{G_{0}}\right)  ^{2}\text{Tr}(\Phi^{\dagger}\Phi
)+\text{Tr}\left[  \left(  \left(  \frac{G}{G_{0}}\right)  ^{2}\frac{m_{1}^{2}}%
{2}+\Delta\right)  (L^{\mu}{}^{2}+R^{\mu}{}^{2})\right]\,\nonumber\\&-\lambda_{1}[\text{Tr}(\Phi^{\dagger}\Phi)]^{2}+\frac{h_{1}}{2}\text{Tr}(\Phi^{\dagger}%
\Phi)\text{Tr}[(L^{\mu})^{2}+(R^{\mu})^{2}]+c(\text{det}\Phi
-\text{det}\Phi^{\dagger})^{2},... \text{ .}
\label{lag}%
\end{align}
These terms contain the decay channels of the charmonium state
$\chi_{c0}$ into (pseudo)scalar and (axial-)vector mesons as well
as a scalar glueball $G$. The full Lagrangian (\ref{fulllag}) is
presented in chapter 2. The term denoted as
$\mathcal{L}_{dil}$ is the dilaton Lagrangian which describes the
scalar dilaton field which is represented by a scalar glueball
$G\equiv|gg\rangle$ with quantum numbers $J^{PC}=0^{++}$, and
emulates the trace anomaly of the pure Yang-Mills sector of QCD
\cite{Parganlija:2012fy, Rosenzweig:1981cu, Salomone:1980sp, Rosenzweig:1982cb, Migdal:1982jp, Gomm:1984zq, Gomm:1985ut}:
\begin{equation}
\mathcal{L}_{dil}=\frac{1}{2}(\partial_{\mu}G)^{2}-\frac{1}{4}\frac{m_{G}^{2}%
}{\Lambda^{2}}\left(
G^{4}\,\ln\left\vert\frac{G}{\Lambda}\right\vert-\frac{G^{4}}{4}\right)\,.
\label{dil}%
\end{equation}
 The energy scale of low-energy QCD is described by the dimensionful parameter $\Lambda$ which is identical to the minimum $G_0$ of the dilaton potential, ($G_0=\Lambda$).
 The scalar glueball mass
 $m_G$ has been evaluated by lattice QCD which gives a mass of about (1.5-1.7) GeV \cite{Morningstar:1999rf, Loan:2005ff, Chen:2005mg, Gregory:2012hu, Gregory:2005yr}. The
dilatation symmetry or scale invariance,
$x^\mu\rightarrow\lambda^{-1}x^{\mu}$, is realized at the
classical level of the Yang-Mills sector of QCD and explicitly
broken due to the logarithmic term of the potential in Eq. (\ref{dil}). This breaking
leads to a non-vanishing divergence of the corresponding current:
\begin{equation}
\partial_\mu
J^\mu_{dil}=T^\mu_{dil,\,\mu}=-\frac{1}{4}m_G^2\Lambda^2\,.
\end{equation}
The importance of including the scalar glueball in the eLSM is to
incorporate dilatation invariance meson mass terms. Note that the scalar
glueball state was
frozen in the previous discussion but here it is elevated to be a dynamical degree of freedom. The details of the terms and the field assignments are presented and
 discussed for the eLSM in the case $N_f=4$ in chapter 4.\\
\indent In our framework, $\Phi$
represents the $4\times4$ (pseudo)scalar multiplets, as seen in Sec.(4.2), as follows:

\begin{equation}\label{4}
\Phi=(S^{a}+iP^{a})t^{a}=\frac{1}{\sqrt{2}}
\left(%
\begin{array}{cccc}
  \frac{(\sigma_{N}+a^0_{0})+i(\eta_N +\pi^0)}{\sqrt{2}} & a^{+}_{0}+i \pi^{+} & K^{*+}_{0}+iK^{+} & D^{*0}_0+iD^0 \\
  a^{-}_{0}+i \pi^{-} & \frac{(\sigma_{N}-a^0_{0})+i(\eta_N -\pi^0)}{\sqrt{2}} & K^{*0}_{0}+iK^{0} & D^{*-}_0+iD^{-} \\
  K^{*-}_{0}+iK^{-} & \overline{K}^{*0}_{0}+i\overline{K}^{0} & \sigma_{S}+i\eta_{S} & D^{*-}_{S0}+iD^{-}_S\\
  \overline{D}^{*0}_0+i\overline{D}^0 & D^{*+}_0+iD^{+} & D^{*+}_{S0}+iD^{+}_S & \chi_{C0}+i\eta_C\\
\end{array}%
\right)\,,
\end{equation}

where $t^{a}$ are the generators of the group $U(N_{f})$. The
multiplet $\Phi$ transforms as $\Phi\rightarrow U_{L}\Phi
U_{R}^{\dagger}$ under $U_{L}(4)\times U_{R}(4)$ chiral
transformations, where $U_{L(R)}=e^{-i\theta_{L(R)}^at^a}$ is an
element of $U(4)_{R(L)}$. Under parity
$\Phi(t,\overrightarrow{x})\rightarrow\Phi^{\dagger}(t,-\overrightarrow{x})$,
and under charge conjugation $\Phi\rightarrow\Phi^{\dagger}$. The
determinant of $\Phi$ is invariant under $SU(4)_{L} \times
SU(4)_{R}$, but not under $U(1)_{A}$ because ${ det
\Phi}\rightarrow { det} U_{A}\Phi
U_A=e^{-i\theta_{A}^0\sqrt{2N_f}}{ det \Phi}\neq { det
\Phi}$. Note that Eq.\ (\ref{intlag}) is not invariant under
$U_{A}(1)$ which is in agreement with the so-called axial anomaly.\\
Now we present the left-handed and right-handed matrices
containing the vector, $V^{a}_\mu$, and axial-vector, $A^{a}_\mu$, degrees
of freedom \cite{Eshraim:2014eka}:
\begin{equation}\label{4}
L^\mu=(V^a+\,A^a)^{\mu}\,t^a=\frac{1}{\sqrt{2}}
\left(%
\begin{array}{cccc}
  \frac{\omega_N+\rho^{0}}{\sqrt{2}}+ \frac{f_{1N}+a_1^{0}}{\sqrt{2}} & \rho^{+}+a^{+}_1 & K^{*+}+K^{+}_1 & D^{*0}+D^{0}_1 \\
  \rho^{-}+ a^{-}_1 &  \frac{\omega_N-\rho^{0}}{\sqrt{2}}+ \frac{f_{1N}-a_1^{0}}{\sqrt{2}} & K^{*0}+K^{0}_1 & D^{*-}+D^{-}_1 \\
  K^{*-}+K^{-}_1 & \overline{K}^{*0}+\overline{K}^{0}_1 & \omega_{S}+f_{1S} & D^{*-}_{S}+D^{-}_{S1}\\
  \overline{D}^{*0}+\overline{D}^{0}_1 & D^{*+}+D^{+}_1 & D^{*+}_{S}+D^{+}_{S1} & J/\psi+\chi_{C1}\\
\end{array}%
\right)^\mu\,,
\end{equation}
$$\\$$
\begin{equation}\label{5}
R^\mu=(V^a-\,A^a)^\mu\,t^a=\frac{1}{\sqrt{2}}
\left(%
\begin{array}{cccc}
  \frac{\omega_N+\rho^{0}}{\sqrt{2}}- \frac{f_{1N}+a_1^{0}}{\sqrt{2}} & \rho^{+}-a^{+}_1 & K^{*+}-K^{+}_1 & D^{*0}-D^{0}_1 \\
  \rho^{-}- a^{-}_1 &  \frac{\omega_N-\rho^{0}}{\sqrt{2}}-\frac{f_{1N}-a_1^{0}}{\sqrt{2}} & K^{*0}-K^{0}_1 & D^{*-}-D^{-}_1 \\
  K^{*-}-K^{-}_1 & \overline{K}^{*0}-\overline{K}^{0}_1 & \omega_{S}-f_{1S} & D^{*-}_{S}-D^{-}_{S1}\\
  \overline{D}^{*0}-\overline{D}^{0}_1 & D^{*+}-D^{+}_1 & D^{*+}_{S}-D^{+}_{S1} & J/\psi-\chi_{C1}\\
\end{array}%
\right)^\mu\,,
\end{equation}

which transform as $L^{\mu}\rightarrow U_{L}%
L^{\mu}U_{L}^{\dag}$ and $R^{\mu}\rightarrow
U_{R}L^{\mu}U_{R}^{\dag}$ under chiral transformations. These
transformation properties of $\Phi,\, L^\mu$, and $R^\mu$ have
been used to build the chirally invariant Lagrangian (\ref{lag}).
The matrix $\Delta$ is defined as
\begin{equation}
\Delta=\left(
\begin{array}
[c]{cccc}%
0 & 0 & 0 & 0\\
0 & 0 & 0 & 0\\
0 & 0 & \delta_{S} & 0\\
0 & 0 & 0 & \delta_{C}%
\end{array}
\right)  \,, \label{delta}%
\end{equation}
where $\delta_S\sim m_S^2\,\,{ and} \,\,\delta_C\sim m_C^2$.

If $m_0^2<0$, the Lagrangian (\ref{lag}) features spontaneous
symmetry breaking. To implement this breaking we have to shift the
scalar-isoscalar fields $G,\sigma_{N},\,\sigma_{S},$ and
$\chi_{C0}$ by their vacuum expectation values
$G_{0},\,\phi_{N},\,\phi_{S},$ and $\phi_{C}$ \cite{Janowski:2011gt, Eshraim:2014eka}
\begin{align}
G\rightarrow & G+G_{0}\,,\nonumber\\
\sigma_N\rightarrow &\sigma_N+\phi_N\,,\nonumber\\
\sigma_S\rightarrow & \sigma_S+\phi_S\;,\nonumber\\
\chi_{C0}\rightarrow & \chi_{C0}+\phi_C\;.
\end{align}

The identification of the scalar glueball $G$ is still uncertain,
the two most likely candidates are $f_{0}(1500)$ and $f_{0}(1710)$
and/or admixtures of them. We assign the scalar glueball ($G$),
$\sigma_N$, and $\sigma_S$ from the following mixing matrix which
is constructed in Ref. \cite{Janowski:2014ppa}:
\begin{equation}
\left(
\begin{array}
[c]{c}%
f_0(1370)\\
f_0(1500)\\
f_0(1710)
\end{array}
\right) =\left(
\begin{array}
[c]{ccc}%
0.94 & -0.17 & 0.29 \\
0.21 & 0.97 & -0.12 \\
-0.26 & 0.18 & 0.95
\end{array}
\right)\left(
\begin{array}
[c]{c}%
\sigma_N\equiv(\overline{u}u+\overline{d}d)/\sqrt{2}\\
\sigma_S\equiv\overline{s}s\\
G\equiv gg\\
\end{array}
\right)  \text{ .} \label{scalmixmat}%
\end{equation}

Note that we used different mixing matrices in the study of the decay of the pseudoscalar
glueball into scalar mesons as seen in
chapter 5. From Eq.(\ref{scalmixmat}), one obtains

\begin{align}
G=-0.33\,f_0(1370)-0.172\,f_0(1500)+0.93\,f_0(1710)\,,\\
\sigma_N=-0.92\,f_0(1370)+0.31\,f_0(1500)-0.27\,f_0(1710)\,,\\
\sigma_S=0.23\,f_0(1370)+0.93\,f_0(1500)+0.25\,f_0(1710)\,.
\end{align}

These relations are used in the calculation of the decay widths of $\chi_{c0}$.

\subsection{Parameters and results}

All the parameters in the Lagrangian (\ref{lag}) have been fixed
in the case of $N_f=3$, see Ref. \cite{Parganlija:2012fy} for more details, and
the three additional parameters related to the charm sector
($\varepsilon_C,\,\delta_C$, and $\phi_C$), in the
case of $N_f=4$, have been determined in chapter 4. The values of
the parameters and the wave-function renormalization constants are
summarized in the following
Table \ref{Tab:ren}:\\

\begin{table}[H]
\centering
\begin{tabular}
[c]{|c|c|c|c|c|} \hline parameter & value  & renormalization factor & value\\
\hline $m_{1}^{2}$  & $0.413\times10^{6}$ MeV$^{2}$   & $Z_\pi=Z_{\eta_N}$ & 1.70927\\
\hline $m_{0}^{2}$  & $-0.918\times 10^{6}$ MeV$^{2}$ &  $Z_K$             & 1.60406  \\
\hline $\delta_{S}$ & $0.151\times10^{6}$MeV$^{2}$    &  $Z_{\eta_C}$      & 1.11892  \\
\hline $\delta_{C}$ & $3.91\times10^{6}$MeV$^{2}$     &  $Z_{D_S}$    & 1.15716    \\
\hline $\varepsilon_{C}$ & $2.23\times10^{6}$MeV$^{2}$ &  $Z_{D^\ast_0}=Z_{D^{\ast0}_0}$& 1.00649 \\
\hline  $g_{1}$ & $5.84$ & $Z_{\eta_S}$ & 1.53854\\
\hline $h_{2}$ & $9.88$ & $Z_{K_{S}}$ & 1.00105\\
\hline $\lambda_{2}$ & $68.3$ & $Z_{D}$ & 1.15256\\
\hline $h_{3}$ & $3.87$  & $Z_{D_{S0}^\ast}$ & 1.00437\\
\hline
\end{tabular}
\caption{ Parameters and wave-function renormalization constants.}
 \label{Tab:ren}
\end{table}

The wave-function renormalization constants for $\pi$ and $\eta_N$ are equal because of isospin symmetry, and for $D^\ast_0$ and
$D^{\ast0}$ as well. The gluon condensate $G_0$ is equal to $\Lambda\approx 3.3$ GeV \cite{Janowski:2014ppa} in pure YM theory, which is used in the present discussion.

Furthermore, we set the values of the parameters $\lambda_1$ and
$h_1$ to zero in eLSM in all cases studied in this framework
(as seen also in the previous chapters, in the case of $N_f=2$
\cite{Parganlija:2010fz, Janowski:2011gt, Janowski:2011gt}, $N_f=3$ \cite{Parganlija:2012fy}, and $N_f=4$ for
the masses of charmed mesons and the (OZI-dominant) strong decays of
open charmed mesons \cite{Eshraim:2014afa, Eshraim:2014eka, Eshraim:2014vfa, Eshraim:2014gya, Eshraim:2014tla}), because they are
expected to be
small, which is in agreement with large-${N_c}$ expectations. However, in the OZI-suppressed decays of the
charmonium state $\chi_{c0}$, non-zero values of $\lambda_1$ and $h_1$ will become important. To understand the reason for this let us explain in more detail.\\
Indeed, the decay of the charmonium state $\chi_{c0}$ into hadrons is mediated by gluon annihilation. This annihilation must proceed through a three-gluon exchange for the following two reasons:\\
(i) Gluons carry colour, but mesons in the final state are colour singlets (colourless). This leads to the fact that the annihilation must be mediated by more than one gluon.\\
(ii) The combination of gluons involved in the decay must be such
that it conserves all strong-interaction quantum numbers.
Consequently, vector particles,
 with charge-conjugation quantum number -1, cannot decay into two vectors through two-gluon exchange.
  The charge conjugation quantum number for a two-gluon state is +1, but for a three-gluon state it is -1. Therefore, vector mesons decays only through three-gluon annihilation.\\

\begin{figure}[H]
\begin{center}
\includegraphics[height=1in, width=2.3in]{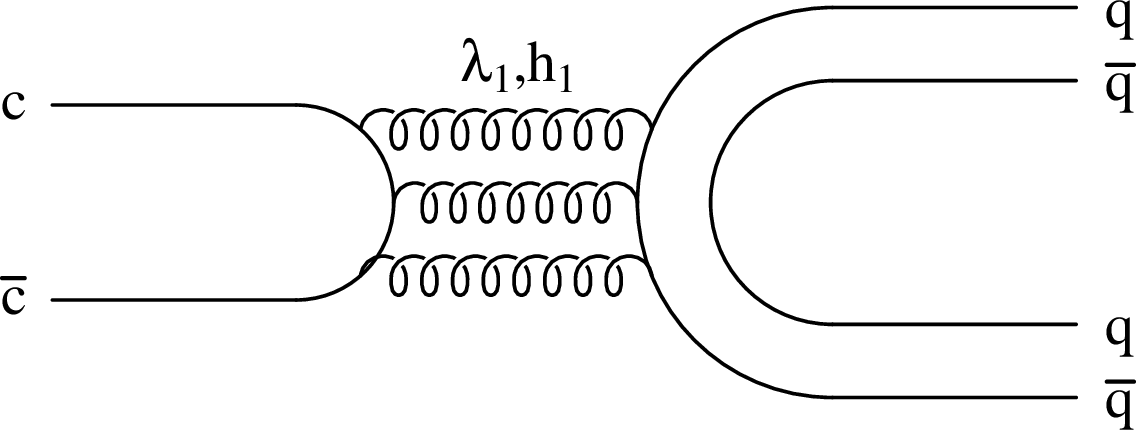}
\caption{Decay of charmonium state into two mesons. $q$ refers to the up (u), down (d), and strange (s) quark flavours.}
\label{figAVP}%
\end{center}
\end{figure}

\begin{figure}[H]
\begin{center}
\includegraphics[
height=1in, width=2.5in
]%
{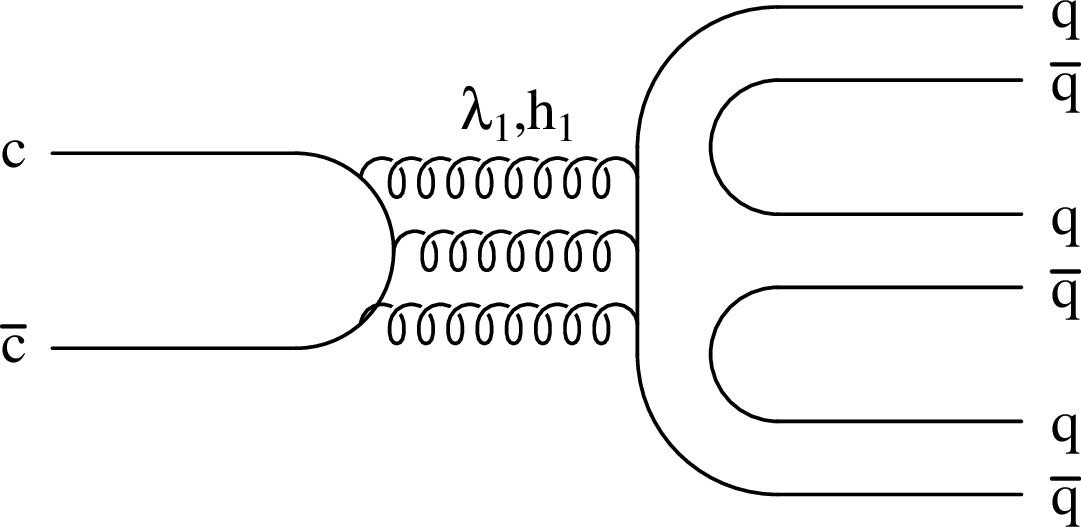}%
\caption{Decay of charmonium state into three mesons.}%
\label{figAVP}%
\end{center}
\end{figure}

In our case, gluons carry all the energy. Therefore, the
interaction is relatively weak due to asymptotic freedom, which leads to OZI-suppression. As a
consequence, the decay of the charmonium state $\chi_{C0}$ into
(axial-)vector and (pseudo)scalar mesons and a scalar glueball is
dynamically suppressed due to annihilation into three `hard'
gluons (see Fig. 8.1 and Fig. 8.2). In the eLSM, this is incorporated by small non-zero values of the
large-$N_c$ suppressed parameters $\lambda_1$ and $h_1$, as seen in Fig. 8.1 and Fig. 8.2. If we set them to zero, all the
decay channels of $\chi_{C0}$ which have been found in the eLSM
(\ref{lag}) are zero, which is an unacceptable result. This is
important evidence for these parameters being nonzero in the
OZI-suppressed case. For this reason we determine them using the
experimental decay widths of $\chi_{C0}$ listed by the PDG
\cite{Beringer:1900zz} via a $\chi^2$ fit, 

\begin{equation}
\chi^{2}(\lambda_1, h_1)\equiv\sum_{i}^{6}\bigg(\frac{\Gamma_{i}^{th}-\Gamma_{i}^{exp}}{\xi \Gamma_{i}%
^{exp}}\bigg)^{2}\text{ ,} \label{chi2,1}%
\end{equation}
where $\xi$ is a constant. We choose $\xi=1$ which leads to
$\chi^2/d.o.f= 0.7$. We then obtain

\begin{equation}
\lambda_1=-0.16\,,\label{lam}
\end{equation}
and
\begin{equation}
h_1=0.046.\label{h1}
\end{equation}

These values of the parameters $\lambda_1$ and $h_1$ are indeed very small, as mentioned before. So a
posteriori
 we justify the results of Refs.
\cite{Eshraim:2014eka, Eshraim:2014tla}. The partial decay widths of various decay
channels which we used in the fit (\ref{chi2,1}) are summarized in the
Table \ref{Tab:chi1} 

\begin{table}[H]
\centering
\begin{tabular}
[c]{|c|c|c|c|} \hline Decay Channel &  theoretical result [MeV] &
Experimental result [MeV]\\
\hline $\Gamma_{\chi_{c0}\rightarrow\overline{K}^{*0}_0K^{*0}_0}$   & 0.01  & 0.01$\pm$0.0047 \\
\hline $\Gamma_{\chi_{c0}\rightarrow K^-K^+}$ & 0.059  & 0.061$\pm$0.007\\
\hline $\Gamma_{\chi_{c0}\rightarrow\pi\pi}$      & 0.089  & 0.088$\pm$0.0092\\
\hline $\Gamma_{\chi_{c0}\rightarrow\overline{K}^{*0}K^{*0}}$ & 0.014  &0.017$\pm$0.0072\\
\hline $\Gamma_{\chi_{c0}\rightarrow \text{w w}}$    & 0.01  & 0.0099$\pm$0.0017\\
\hline $\Gamma_{\chi_{c0}\rightarrow \phi\phi}$    & 0.004  & 0.0081$\pm$0.0013\\
\hline
\end{tabular}
\caption{The partial decay widths of $\chi_{c0}$.}
 \label{Tab:chi1}
\end{table}

Furthermore, we have to change the value of the parameter $c$, which is the coefficient of the axial
anomaly term, to fit the results of the decay widths of $\chi_{c0}$. Therefore, for the determination of $c$, we use the decay widths of $\chi_{c0}$ into
$\eta\eta$ and $\eta' \eta'$, which are \cite{Beringer:1900zz}.
$$\Gamma_{\chi_{c0}\rightarrow \eta\eta}^{exp}= (0.031\pm0.0039) \text{MeV}\,,$$
 and 
$$\Gamma_{\chi_{c0}\rightarrow\eta'
\eta'}^{exp}=(0.02\pm0.0035) \text{MeV}\,.$$
We then perform a fit by minimizing the $\chi^2$-function,
\begin{equation}
\chi^{2}(c)\equiv\bigg(\frac{\Gamma_{\chi_{c0}\rightarrow \eta\eta}^{th}(c)-\Gamma_{\chi_{c0}\rightarrow \eta\eta}^{exp}}{\xi \Gamma_{\chi_{c0}\rightarrow \eta\eta}%
^{exp}}\bigg)^{2}+\bigg(\frac{\Gamma_{\chi_{c0}\rightarrow \eta'\eta'}^{th}(c)-\Gamma_{\chi_{c0}\rightarrow \eta'\eta'}^{exp}}{\xi \Gamma_{\chi_{c0}\rightarrow \eta'\eta'}%
^{exp}}\bigg)^{2}\text{ ,} \label{chi2}%
\end{equation}
which gives $c=7.178\times 10^{-10}$ MeV$^{-4}$ with $\chi^2/{ d.o.f}=
0.18$ where $\xi=1$. \\
The mixing between the hidden-charmed scalar meson $\chi_{c0}$ and the scalar glueball $G$ is neglected because it is small. \\
The two- and three-body decays of the hidden-charmed meson $\chi_{c0}$ into scalar glueballs and scalar mesons are reported in Table 8.3.\\

\begin{table}[th]
\centering
\begin{tabular}
[c]{|c|c|c|c|} \hline Decay Channel &  theoretical result [MeV] &
Experimental result [MeV]\\
\hline $\Gamma_{\chi_{c0}\rightarrow f_0(1370)f_0(1370)}$  & 5.10$^{-3}$  & $<$3.10$^{-3}$\\
\hline $\Gamma_{\chi_{c0}\rightarrow f_0(1500)f_0(1500)}$  & 4.10$^{-3}$  & $<$5.10$^{-4}$\\
\hline $\Gamma_{\chi_{c0}\rightarrow f_0(1370)f_0(1500)}$  & 2.10$^{-6}$  & $<$1.10$^{-3}$\\
\hline $\Gamma_{\chi_{c0}\rightarrow f_0(1370)f_0(1710)}$  & 1.10$^{-4}$  &    0.0069$\pm$0.004\\
\hline $\Gamma_{\chi_{c0}\rightarrow f_0(1500)f_0(1710)}$  & 2.10$^{-5}$ &  $<$7.10$^{-4}$\\
\hline $\Gamma_{\chi_{c0}\rightarrow f_0(1370) \eta\eta}$  & 4.10$^{-4}$  & -\\
\hline $\Gamma_{\chi_{c0}\rightarrow f_0(1500) \eta\eta}$   & 3.10$^{-3}$  &   -\\
\hline $\Gamma_{\chi_{c0}\rightarrow f_0(1370) \eta' \eta'}$ & 27.10$^{-4}$  &   -\\
\hline $\Gamma_{\chi_{c0}\rightarrow f_0(1370) \eta \eta'}$  & 89.10$^{-6}$  & - \\
\hline $\Gamma_{\chi_{c0}\rightarrow f_0(1500) \eta\eta'}$   & 11.10$^{-3}$  & - \\
\hline $\Gamma_{\chi_{c0}\rightarrow f_0(1710) \eta \eta}$   & 8.10$^{-5}$  & - \\
\hline $\Gamma_{\chi_{c0}\rightarrow f_0(1710) \eta \eta'}$  & 3.10$^{-5}$  & - \\

\hline
\end{tabular}\\
\caption{The partial decay widths of $\chi_{c0}$.}
\end{table}

Additionally, the two- and three-body decays of the hidden-charmed
meson $\chi_{c0}$ into (axial-)vector and (pseudo)scalar mesons are
reported in Table 8.4.
\begin{table}[H]
\centering
\begin{tabular}
[c]{|c|c|c|c|} \hline Decay Channel &  theoretical result [MeV] &
Experimental result [MeV]\\
\hline $\Gamma_{\chi_{c0}\rightarrow a_0 a_0}$   & 0.004  &   -\\
\hline $\Gamma_{\chi_{c0}\rightarrow K_{1} \overline{K}_{1}}$      & 0.005  &   -\\
\hline $\Gamma_{\chi_{c0}\rightarrow K^+_1 K^-}$     & 0.005  & 0.063$\pm$0.0233\\
\hline $\Gamma_{\chi_{c0}\rightarrow \eta\eta}$ & 0.022  & 0.031$\pm$0.0039 \\
\hline $\Gamma_{\chi_{c0}\rightarrow\eta' \eta'}$    & 0.02  & 0.02$\pm$0.0035\\
\hline $\Gamma_{\chi_{c0}\rightarrow \eta \eta'}$       & 0.004  & $<$0.0024 \\
\hline $\Gamma_{\chi_{c0}\rightarrow K^{*} K^{*}_0}$  & 0.00007 &   - \\
\hline $\Gamma_{\chi_{c0}\rightarrow \rho\rho}$  & 0.01  &   -\\
\hline $\Gamma_{\chi_{c0}\rightarrow K^*_0 K \eta}$   & 0.008  & - \\
\hline $\Gamma_{\chi_{c0}\rightarrow K^*_0 K \eta'}$ & 0.004  & - \\
\hline
\end{tabular}\\
\caption{The partial decay widths of $\chi_{c0}$.}
\end{table}

The results are in good agreement with experimental data
\cite{Beringer:1900zz}. All the relevant expressions for the two- and
three-body decay processes of $\chi_{c0}$ are presented in the
Appendix along with computational details.

\section{Decay of the pseudoscalar charmonium state $\eta_C$}

\indent \indent In this section we compute and discuss the decay
widths of the pseudoscalar charmonium state $\eta_C (1P)$
into (pseudo)scalar mesons and a pseudoscalar glueball ($\eta_c
\rightarrow \pi \pi \widetilde{G}$) in the
eLSM.\\
\indent Two terms in the eLSM are relevant for the decay of the
pseudoscalar hidden-charmed meson $\eta_C$ into (pseudo-)scalar
mesons and a pseudoscalar glueball $\widetilde{G}$. The first has
the form $c(\text{det}\Phi -\text{det}\Phi^{\dagger})^{2}$, which describes the
axial anomaly and represents a further breaking of dilatation and
chiral symmetry and is additionally responsible for the mass and
decays of the $\eta$'s. The other term (see Eq. (\ref{intlag})) describes the interactions of the
pseudoscalar glueball $\widetilde{G}\equiv|gg\rangle$, with
quantum numbers $J^{PC}=0^{-+}$, with scalar and pseudoscalar
mesons.

\subsection{Decay of $\eta_C$ into a pseudoscalar glueball}

The effective Lagrangian which describes the interaction of the
pseudoscalar glueball $\widetilde{G}$ with the (pseudo)scalar
mesons (which is described in detail for the case of $N_f=3$ (\ref{intlag}) in
chapter 6) reads
\begin{equation}
\mathcal{L}^{\textit{int}}_{\widetilde{G}}=ic_{\tilde{G}\Phi}\tilde{G}\left(
\text{\textrm{det}}\Phi-\text{\textrm{det}}\Phi^{\dag}\right)
\text{ ,}
\label{intlag2}%
\end{equation}
where $c_{\tilde{G}\phi}$ is a dimensionless coupling constant.
The pseudoscalar glueball $\tilde{G}$ is invariant under $U(4)_{L}
\times U(4)_{R}$ chiral transformations, while under parity,
$\tilde{G}(t,\overrightarrow{x}) \rightarrow-
\tilde{G}(t,-\overrightarrow{x})$, and under charge conjugation
$\tilde{G}\rightarrow \tilde{G}$. These considerations lead to the
interaction Lagrangian $\mathcal{L}_{\tilde{G}}^{int}$ of
Eq.\ (\ref{intlag}) which is invariant under $SU(4)_{L} \times
SU(4)_{R}$, parity, and charge conjugation.

To determine the value of the coupling constant
$c_{\tilde{G}\phi}$, one can relate it to its counterpart in the
three-flavour case $c_{\widetilde{G}\Phi(N_f=3)}$, which was
computed in the study of the decay of a pseudoscalar
glueball $\tilde{G}$ into scalar and pseudoscalar
mesons (as seen in chapter 6), with the result $c_{\widetilde{G}\Phi(N_f=3)}= 4.48\pm 0.46$ \cite{Eshraim:2012jv}.\\

The Lagrangian which describes the coupling of the pseudoscalar
glueball and (pseudo)scalar mesons was given in Eq. (\ref{intlag}) for the three-flavour case 
\begin{equation}
\mathcal{L}^{\textit{int}}_{\widetilde{G}(N_f=3)}=ic_{\tilde{G}\Phi(N_f=3)}\tilde{G}\left(
\text{\textrm{det}}\Phi_{(N_f=3)}-\text{\textrm{det}}\Phi^{\dag}_{(N_f=3)}\right)
\text{ .}\label{Lca3,2}
\end{equation} 
By using the relation of the multiplet matrix of (pseudo)scalar
mesons ($\Phi$) for the four-flavour case and for the three-flavour case (which is presented in Eq. (4.123)), we can transform the
interaction Lagrangian $\mathcal{L}^{\textit{int}}_{\widetilde{G}}$, for the case of $N_f=4$ (\ref{intlag2}), to be as
\begin{equation}
\mathcal{L}^{\textit{int}}_{\widetilde{G}}=i\frac{\sqrt{2}}{\phi_C}\,c_{\widetilde{G}\Phi(N_f=3)}\tilde{G}\left(
\text{\textrm{det}}\Phi-\text{\textrm{det}}\Phi^{\dag}\right)
\text{ ,}\label{lca4,b}
\end{equation}
Comparing Eq.(\ref{intlag2}) with Eq.(\ref{lca4,b}), we get
\begin{equation}
c_{\widetilde{G}\Phi}=\frac{\sqrt{2}\,c_{\widetilde{G}\Phi(N_f=3)}}{\phi_C}\,.\label{cGphi}
\end{equation}
We then get $c_{\widetilde{G}\Phi}=0.036$ in the present case ($N_f=4$). \\

The interaction Lagrangian (\ref{intlag2}) contains only one decay
process which describes the decay of the pseudoscalar charmonium
$\eta_c$ into a pseudoscalar glueball $\widetilde{G}$ through the
channel $\eta_C\rightarrow \widetilde{G}\pi\pi$. The tree-level
vertices of this process have the form
\begin{equation}
\mathcal{L}_{\eta_{C}\widetilde{G}\pi\pi}=-\frac{1}{4}c_{\widetilde{G}\Phi}\phi_S\,Z_{\eta_C}Z_\pi^2\,\eta_C\widetilde{G}{\pi^0}^2-\frac{1}{2}
c_{\widetilde{G}\Phi}\phi_S\,Z_{\eta_C}Z_\pi^2\eta_C\widetilde{G}\pi^-\pi^+\,.\label{veretacG}
\end{equation}
One can compute the full decay  width $\Gamma_{\eta_C\rightarrow \widetilde{G}\pi\pi}$ from
\begin{align}
\Gamma_{\eta_C\rightarrow \widetilde{G}\pi\pi}&= \Gamma_{\eta_C\rightarrow \widetilde{G}\pi^0\pi^0}+\Gamma_{\eta_C\rightarrow \widetilde{G}\pi^-\pi^+}\nonumber\\
&=\Gamma_{\eta_C\rightarrow \widetilde{G}\pi^0\pi^0}+2\Gamma_{\eta_C\rightarrow \widetilde{G}\pi^0\pi^0}\nonumber\\
&=3\Gamma_{\eta_C\rightarrow \widetilde{G}\pi^0\pi^0}\,.
\end{align}
The decay amplitude is
\begin{equation}
-iM= \frac{-i}{4}c_{\widetilde{G}\Phi}\phi_S\,Z_{\eta_C}Z_\pi^2\,.
\end{equation}
one also uses the corresponding decay width for the three-body case, Eq. (\ref{3bodydecay}). The decay width of the pseudoscalar
charmonium state $\eta_c$ into a pseudoscalar glueball with a mass
of $2.6$ GeV (as predicted by lattice QCD in the quenched
approximation \cite{Morningstar:1999rf, Loan:2005ff, Chen:2005mg, Gregory:2012hu, Gregory:2005yr}) is
\begin{equation}
\Gamma_{\eta_C\rightarrow \pi\pi\widetilde{G}(2600)}=0.124\,\,
{ MeV},
\end{equation}
and for a mass of the charmonium state $\eta_c$ which is about of $2.37$ GeV (corresponding to the mass of the resonance $X(2370)$
measured in the BESIII experiment \cite{Ablikim:2005um, Kochelev:2005vd, Ablikim:2010au})
\begin{equation}
\Gamma_{\eta_C\rightarrow \pi\pi\widetilde{G}(2370)}=0.16\,\, { MeV}\text{ .} \label{etadpg}%
\end{equation}
These results could be tested in the PANDA experiment at the upcoming FAIR facility.

\subsection{Decay of $\eta_C$ into (pseudo)scalar mesons}

The chiral Lagrangian contains the tree-level vertices for the
decay processes of the pseudoscalar $\eta_C$ into (pseudo)scalar
mesons, through the chiral anomaly term
\begin{equation}
\mathcal{L}_{U(1)_A}=c(\text{det}\Phi
-\text{det}\Phi^{\dagger})^{2}
\text{ ,}
\label{intlag3}%
\end{equation}
where $c$ is a dimensionful constant and has been determined in Sec. 8.2. After the field transformations Eqs.
(\ref{shhh1} - \ref{shhhf}) have been performed in Sec. 4.3, the terms in the Lagrangian (\ref{intlag3}) which correspond to decay processes of $\eta_C$ read
\begin{align}
\mathcal{L}_{\eta_{C}}=&\frac{c}{8}\phi_N^2\phi_C\,Z_{\eta_C}\eta_C\{\sqrt{2}\phi_S\phi_N Z_K
Z_{K^\ast_0}(K^{\ast0}_0\overline{K}^0+\overline{K}^{\ast0}_0 K^0+K^{\ast-}_0 K^{+}+K^{\ast+}_0 K^-)\nonumber\\
&+2 Z_\pi \phi_S^2 (a_0^{0} \pi^{0 }+ a_0^{+} \pi^{-} + a_0^{-}\pi^{+})-4 \phi_N \phi_S (Z_{\eta_S} \eta_S \sigma_N + Z_{\eta_N} \eta_N \sigma_S)\nonumber\\
&-6\phi_S^2\,Z_{\eta_N}\eta_N\,\sigma_N-\phi_N^2\eta_S\sigma_S\,Z_{\eta_S}+2\phi_N\,Z^2_{\eta_S}\,Z_{\eta_N}\eta_S^2\eta_N +6\phi_S\,Z_{\eta_S}\,Z_{\eta_N}^2\eta_N^2\eta_S \nonumber\\
&-\sqrt{2}\phi_N\,Z_{\eta_S}\,Z_K^2(\overline{K}^0K^0+K^-K^+)\,\eta_S
-3\sqrt{2}\phi_S\,Z_{\eta_N}Z_K^2(\overline{K}^0K^0+K^-K^+)\eta_N\nonumber\\
&+\sqrt{2}\phi_S\,Z_\pi\,Z_K^2\bigg[\sqrt{2}(\overline{K}^0 K^+ \pi^- + K^0 K^- \pi^+)-(K^0 \overline{K}^0 -K^{-}K^{+})\pi^0\bigg]\nonumber\\
&-2\phi_S\eta_S\,Z_{\eta_S}Z_\pi^2({\pi^0}^2+2\pi^-\pi^+)\}\,.\label{intLagofetadecay}
\end{align}

The decay widths of the pseudoscalar hidden charmed meson $\eta_C$
into scalar and pseudoscalar charmed mesons are presented
in Table 8.5. The relevant expressions for these decay processes are
presented in the Appendix.\\

\begin{table}[H]
\centering
\begin{tabular}
[c]{|c|c|c|c|} \hline Decay Channel &  theoretical result [MeV] &
Experimental result [MeV]\\
\hline $\Gamma_{\eta_{c}\rightarrow\overline{K}^{*}_0K}$   & 0.01  & - \\
\hline $\Gamma_{\eta_{c}\rightarrow a_0\pi}$      & 0.01  & -\\
\hline $\Gamma_{\eta_{c}\rightarrow f_0(1370) \eta}$  & 0.00018  & - \\
\hline $\Gamma_{\eta_{c}\rightarrow f_0(1500) \eta}$  & 0.006  & - \\
\hline $\Gamma_{\eta_{c}\rightarrow f_0(1710) \eta}$  & 0.000032  &  - \\
\hline $\Gamma_{\eta_{c}\rightarrow f_0(1370) \eta'}$  & 0.027  & - \\
\hline $\Gamma_{\eta_{c}\rightarrow f_0(1500) \eta'}$  & 0.024  & - \\
\hline $\Gamma_{\eta_{c}\rightarrow f_0(1710) \eta'}$  & 0.0006  &  - \\
\hline $\Gamma_{\eta_{c}\rightarrow \eta \eta \eta}$   & 0.052  &   -\\
\hline $\Gamma_{\eta_{c}\rightarrow \eta' \eta' \eta'}$      & 0.0023  &   -\\
\hline $\Gamma_{\eta_{c}\rightarrow \eta'\eta \eta}$ & 0.44  & - \\
\hline $\Gamma_{\eta_{c}\rightarrow \eta' \eta' \eta}$    & 0.0034  & \\
\hline $\Gamma_{\eta_{c}\rightarrow \eta K \overline{K}}$       & 0.15  &  0.32$\pm$0.17\\
\hline $\Gamma_{\eta_{c}\rightarrow \eta' K K}$      & 0.41  &  \\
\hline $\Gamma_{\eta_{c}\rightarrow \eta \pi \pi}$   & 0.12  &  0.54$\pm$0.18\\
\hline $\Gamma_{\eta_{c}\rightarrow \eta' \pi \pi}$  & 0.08  &  1.3$\pm$0.6\\
\hline $\Gamma_{\eta_{c}\rightarrow K K \pi}$     & 0.095  & - \\
\hline
\end{tabular}\\
\caption{The partial decay widths of $\eta_{c}$.}
\end{table}

There are experimental data for these decays firm by the PDG with which we can compare. The three
measured decay widths are in reasonably good agreement
with experimental data. \\

\subsection{Mixing of a pseudoscalar glueball and $\eta_C$ }

The mixing between the pseudoscalar glueball $\widetilde{G}$ with the
pseudoscalar charm-anticharm meson $\eta_c$ is described by the non-interacting Lagrangian as follows:

\begin{equation}\label{70}
 \mathcal{L}_{\widetilde{G}, \, \eta_C}= \frac{1}{2}(\partial_\mu\widetilde{G})^2+\frac{1}{2}(\partial_\mu\eta_C)^2-\frac{1}{2}m_{\widetilde{G}}^2\widetilde{G}^2
-\frac{1}{2}m_{\eta_c}^2\eta_c^2+Z_{\widetilde{G}\eta_C}\widetilde{G}\,\eta_C\,,
\end{equation}
where
\begin{equation}\label{70}
 Z_{\widetilde{G}\eta_C}=\frac{-1}{4}\,c_{\widetilde{G}\Phi}Z_{\eta_C}\,\phi_N^2\,\phi_S\,.
\end{equation}
The physical fields $\eta_C$ and $\widetilde{G}$ can be obtained
through an $SO(2)$ rotation
\begin{equation}\label{8}
\left(%
\begin{array}{c}
 \widetilde{G}' \\
  \eta_C'  \\
\end{array}%
\right)=\left(%
\begin{array}{cccc}
 \cos\phi & \sin\phi  \\
 -\sin\phi & \cos\phi \\
\end{array}%
\right)=\left(%
\begin{array}{c}
 \widetilde{G}\\
  \eta_C  \\
\end{array}%
\right)\,,
\end{equation}
with
\begin{equation}\label{70}
m^2_{\eta_C'}=m^2_{\widetilde{G}}\,
\sin^2\phi+m^2_{\eta_C}\cos^2\phi+Z_{\widetilde{G}\eta_C}\sin(2\phi)\,,
\end{equation}
\begin{equation}\label{70}
m^2_{\widetilde{G}'}=m^2_{\widetilde{G}}\,
\cos^2\phi+m^2_{\eta_C}\sin^2\phi-Z_{\widetilde{G}\eta_C}\sin(2\phi)\,,
\end{equation}
where the mixing angle $\phi$ reads
\begin{equation}\label{70}
\phi=\frac{1}{2}\arctan\bigg[\frac{-c_{\widetilde{G}\Phi}\,Z_{\eta_c}\phi_N^2\phi_S}{2(m^2_{\eta_C}-m^2_{\widetilde{G}})}\bigg]\,,
\end{equation}
where $c_{\widetilde{G}\Phi}$ is a dimensionless coupling constant
between $\widetilde{G}\Phi$ which was determined in Eq.(\ref{cGphi}).
We then obtain the mixing angle of the pseudoscalar glueball
$\widetilde{G}$ and the pseudoscalar charm-anticharm meson $\eta_c$
to be $-1^{\circ}$, for a mass of the pseudoscalar glueball which is $2.6$ GeV, as predicted
by lattice-QCD simulations \cite{Morningstar:1999rf, Loan:2005ff, Gregory:2012hu, Chen:2005mg, Gregory:2005yr}.

\chapter{Conclusions and Outlook \label{ch9}}


In this work, we have developed a four-flavour extended linear
sigma model with vector and axial-vector degrees of freedom. Within this model, we have calculated masses and
decay widths of charmed mesons.

For the coupling constants of the model, 
we have used the values determined in the low-energy study of
mesons in Ref.\ \cite{Parganlija:2012fy} and listed in Table 4.1. The three
remaining parameters related to the current charm quark mass were determined in
a fit to twelve masses of hidden and open charmed mesons. The results are shown in Table 4.5:
the open charmed mesons agree within theoretical errors with the
experimental values, while the masses of charmonia are (with the exception of $J/\psi$) underestimated by about 10\%.
The precision of our approach cannot compete with
methods based on heavy-quark symmetry, but it admits a
perspective on charmed states from a low-energy approach based on chiral symmetry 
and dilatation invariance. The level of agreement with experimental data 
proves that these symmetries are, at least to some degree, still
relevant for the charm sector. In this respect,
our approach is a useful tool to investigate the assignment of some charmed
states (see below) and to obtain an independent determination of quantities
such as the chiral condensate of charm-anticharm quarks. The latter turns out
to be sizable, showing that the charm quark, although heavy, is indeed still
connected to nontrivial vacuum dynamics.


We have also presented a chirally invariant effective
Lagrangian describing the interaction of a pseudoscalar glueball
with scalar and pseudoscalar mesons for the three-flavour case
$N_{f}=3$. We have studied the decays of the pseudoscalar glueball
into three pseudoscalar and into a scalar and pseudoscalar
quark-antiquark fields.

The branching ratios are parameter-free once the mass of the
glueball has been fixed. To this end, we have considered two
possibilities: (i) in agreement with lattice QCD, we have chosen
$m_{\tilde{G}}=2.6$ GeV. The existence and the decay properties of
such a hypothetical pseudoscalar resonance can be tested in the
upcoming PANDA experiment \cite{Lutz:2009ff}. (ii) We assumed that the
resonance $X(2370),$ measured in the experiment BESIII, is
(predominantly) a pseudoscalar glueball state, and thus we have
also used a mass of $2.37$ GeV \cite{Ablikim:2005um, Kochelev:2005vd, Ablikim:2010au}. The results for both
possibilities have been summarized in Tables 6.2 and 6.3: we predict
that $KK\pi$ is the dominant decay channel, followed by (almost
equally large) $\eta\pi\pi$ and $\eta^{\prime}\pi\pi$ decay
channels. In the case of BESIII, with a measurement of the
branching ratio for other decay channels than the measured
$\eta^{\prime}\pi\pi$ one could ascertain if $X(2370)$ is
(predominantly) a pseudoscalar glueball. In the case of PANDA, our
results may represent a useful guideline for the search of the
pseudoscalar glueball.


Then, we have calculated the weak-decay constants of the
pseudoscalar states $D,$ $D_{S},$ and $\eta_{c}$, which are in
fair agreement with the experimental values and, as a last step,
we have evaluated the OZI-dominant decays of charmed mesons (Table
7.1). The result for $D_{0}^{\ast }(2400)^{0},$
$D_{0}^{\ast}(2400)^{+},$ $D^{\ast}(2007)^{0},$ $D^{\ast
}(2010)^{+},$ $D_{1}(2420)^{0},$ and $D_{1}(2420)^{+}$ are
compatible with the results and the upper bounds listed by the PDG
\cite{Beringer:1900zz}, although the theoretical errors are still quite large.
Nevertheless, we could simultaneously describe the decays of open-charmed vector and axial-vector states which are chiral partners
within our theoretical treatment.

Concerning the assignment of the scalar and axial-vector
strange-charmed quarkonium states $D_{S0}$ and $D_{S1},$ we obtain
the following: If the masses of these quarkonia are above the
respective thresholds, we find that their decay widths are too
large, which probably means that these states, even if they exist,
have escaped detection. In this case, the resonances
$D_{S0}^{\ast}(2317)$ and $D_{S1}(2460)$ can emerge as dynamically
generated companion poles (alternatively, they can be tetraquark
or molecular states). Our results imply also that the
interpretation of the resonance $D_{S1}(2536)$ as a member of the
axial-vector multiplet is not favored because the experimental
width is too narrow when compared to the theoretical width of a
quarkonium state with the same mass. An investigation of these
resonances
necessitates the calculation of quantum fluctuations. \\
\indent In summary, the fact that a (although at this stage only rough)
qualitative description is obtained by using a chiral model and,
more remarkably, by using the parameters determined by a study of
$N_{f}=3$ mesons, means that a remnant of chiral symmetry is
present also in the sector of charmed mesons. Chiral symmetry is
(to a large extent) still valid because the parameters of the eLSM
do not vary too much as a function of the energy at which they are
probed. Besides mass terms which describe the large contribution
of the current charm quark mass, all interaction terms are the
same as in the low-energy effective model of Refs.\
\cite{Parganlija:2010fz, Janowski:2011gt, Parganlija:2012fy} which was built under the requirements of
chiral symmetry and dilatation invariance. As a by-product of our
work we also evaluate the charm condensate which is of the same
order as the nonstrange and strange quark condensates. This is
also in accord with chiral dynamics enlarged to the group
$U(4)_{R}\times U(4)_{L}$.\\


In the present work we have also represented a chirally invariant
linear sigma model with (axial-)vector mesons in the four-flavour
case, $N_f=4$, by including a dilaton field and a scalar glueball
field, and describing the interaction of the pseudoscalar glueball
with (pseudo-)scalar mesons. We have calculated the decay widths
of the hidden-charmed meson $\chi_{c0}$ into two and three strange and nonstrange mesons (Tables 8.2 and 8.4) as well as into a scalar
glueball $G$, which is an admixture of two resonances $f_0(1370)$ and
$f_0(1500)$ (Table 8.3). Note here that the decay of charmonium
states to open-charmed mesons is forbidden in the eLSM, as
discussed also in Ref.\cite{Bettoni:2005bb}. We have also computed the
decay widths of the pseudoscalar charmonium state $\eta_C$ into
light mesons (Table 8.5) and into a pseudoscalar glueball
$\widetilde{G}$, through the channel $\eta_C \rightarrow
\pi\pi\widetilde{G}$. The latter is obtained from the interaction
term of the pseudoscalar glueball. We have also
evaluated the mixing angle between the pseudoscalar glueball and
$\eta_c$, which is very small and equal to $-1^\circ$. We have
additionally found that the extended linear sigma model
(\ref{fulllag}) offers no decay channels for the (axial-)vector
charmonium states where $\Gamma_{J/\psi}=0$ and
$\Gamma_{\chi_{c1}}=0$. The results of the decay widths of
$\chi_{c0}$ are in good agreement with experimental data
\cite{Beringer:1900zz} and of $\eta_c$ are in reasonably good agreement with experiment
\cite{Beringer:1900zz}, which indicates to what extent the eLSM is a successful
and appropriate model to study the phenomenology of hidden-charmed
mesons and open-charmed mesons (chapter 7).\\
The parameters were determined in the case of $N_f=4$ (see chapter 4).
However, there are four parameters that we need to fix: (i) $\lambda_1$
and $h_1$; which are assume to have zero values in all the previous
investigation for $N_f=3$ case, (see chapter 3), and $N_f=4$ case,
(see chapter 4), because their values are numerically small and do not
affect the previous results. However, the decay widths of
charmonium states $\chi_{c0}$ and $\eta_C$ depend on both, and for
this reason we determined them by a $\chi^2$ fit to the decay
 widths of $\chi_{c0}$, see Table 8.2. (ii) The $c$ parameter; which is in the axial anomaly term. This
 parameter was determined by the fit (\ref{chi2}).
  (iii) $c_{\widetilde{G}\Phi}$; which was fixed by the relation $c_{\widetilde{G}\Phi(N_f=3)}$ \cite{Eshraim:2012jv}.\\
Note that the decay widths for the (axial-)vector charmonium states $J/\psi$ and $\chi_{c1}$ are zero for kinematical reasons.\\

\indent The restoration of chiral symmetry at nonzero temperature
and density is one of the fundamental questions of modern hadronic
physics \cite{Asner:1999kj, Bromberg:1980bk, Dionisi:1980hi, Harada:2003jx}. The two-flavour version of the eLSM has been
successful in a study at nonzero density
\cite{Struber:2007bm}. This leads us to consider the restoration of chiral
symmetry at nonzero temperature and density for $N_{f}=3$ and
$N_{f}=4$ with the eLSM, which offers many challenges for future work.


\appendix
 \cleardoublepage
\addappheadtotoc
 \appendixpage

\chapter{Determination of the weak decay constants}

We compute the decay constants of pion, kaon, the pseudoscalar open-charmed mesons $D$ and $D_S$, and the pseudoscalar hidden-charmed
meson $\eta_C$, which are denoted as $f_\pi,\,f_D,\,f_{D_S},$ and
$f_{\eta_C}$, by using the formula (\ref{Ptr}),
\begin{equation}
P\rightarrow P+\theta_a (t_a\, S+S\,t_a)\,,\label{Ptr1}
\end{equation}
which is discussed in chapter 5 in details. In the case $N_f=4$,
the pseudoscalar mesons are ordered in a $4\times 4$ matrix as
follows:
\begin{equation}
P=\frac{1}{\sqrt{2}}\left(
\begin{array}
[c]{cccc}%
\frac{1}{\sqrt{2}}(\eta_{N}+\pi^{0}) & \pi^{+} & K^{+} & D^{0}\\
\pi^{-} & \frac{1}{\sqrt{2}}(\eta_{N}-\pi^{0}) & K^{0} & D^{-}\\
K^{-} & \overline{K}^{0} & \eta_{S} & D_{S}^{-}\\
\overline{D}^{0} & D^{+} & D_{S}^{+} & \eta_{c}%
\end{array}
\right)  \text{ ,} \label{p}%
\end{equation}
where
\begin{equation}
\pi^-=\frac{\pi^1+i\pi^2}{\sqrt{2}}\,,\,\,\,\,\,\,\,\,\,\,\,\pi^+=\frac{\pi^1-i\pi^2}{\sqrt{2}}\,,
\end{equation}
\begin{equation}
K^-=\frac{K^1+i\,K^2}{\sqrt{2}}\,,\,\,\,\,\,\,\,\,\,\,\,K^+=\frac{K^1-iK^2}{\sqrt{2}}\,,
\end{equation}
\begin{equation}
\overline{D}^0=\frac{D^1+i\,D^2}{\sqrt{2}}\,,\,\,\,\,\,\,\,\,\,\,\,D^0=\frac{D^1-iD^2}{\sqrt{2}}\,.
\end{equation}
\begin{equation}
D^+_S=\frac{D^1_S+i\,D^2_S}{\sqrt{2}}\,,\,\,\,\,\,\,\,\,\,\,\,D^-_S=\frac{D^1_S-iD^2_S}{\sqrt{2}}\,.
\end{equation}
The vacuum expectation values
$\phi_N,\,\phi_S,$ and $\phi_C $ are contained in the following diagonal
matrix:
\begin{equation}
\langle\Phi\rangle=\left(
\begin{array}
[c]{cccc}%
\frac{\phi_N}{\sqrt{2}} & 0 & 0 & 0\\
0 & \frac{\phi_N}{\sqrt{2}} & 0 & 0\\
0 & 0 & \phi_S & 0\\
0 & 0 & 0 & \phi_C%
\end{array}
\right)  \text{ .}%
\end{equation}
In Eq.(\ref{Ptr1}), $t_a=\frac{\lambda_a}{2}$ are the generators with
$a=0,1,..., N_f^2-1$, where $\lambda_a$ are the Gell-Mann matrices
and chosen to satisfy $\text{Tr}(\lambda_a\,\lambda_b)=2\,\delta_{a\,b}$.
In the case $N_f=4$, the Gell-Mann matrices are rank-4 tensors
and there are 16 generators ($a=0,...,15$). For $a=0$,
$\lambda_0$ is a special unitary $SU(4)$ matrix but
it corresponds to a unitary  $U(1)$ matrix, 
\begin{equation}
t_0=\frac{1}{2\sqrt{2}}\left(
\begin{array}
[c]{cccc}%
1 & 0 & 0 & 0\\
0 & 1 & 0 & 0\\
0 & 0 & 1 & 0\\
0 & 0 & 0 & 1%
\end{array}
\right)  \text{ .}%
\end{equation}

The canonical form of the $4\times 4$ Gell-Mann matrices is
\begin{equation}
\lambda_1=\left(
\begin{array}
[c]{cccc}%
0 & 1 & 0 & 0\\
1 & 0 & 0 & 0\\
0 & 0 & 0 & 0\\
0 & 0 & 0 & 0%
\end{array}
\right)\,,\,\,\,\,\, \lambda_2=\left(
\begin{array}
[c]{cccc}%
0 & -i & 0 & 0\\
i & 0 & 0 & 0\\
0 & 0 & 0 & 0\\
0 & 0 & 0 & 0%
\end{array}
\right)\,,\,\,\,\,\, \lambda_3=\left(
\begin{array}
[c]{cccc}%
1 & 0  & 0 & 0\\
0 & -1 & 0 & 0\\
0 & 0  & 0 & 0\\
0 & 0  & 0 & 0 %
\end{array}
\right)  \text{ ,}\nonumber%
\end{equation}

\begin{equation}
\lambda_4=\left(
\begin{array}
[c]{cccc}%
0 & 0 & 1 & 0\\
0 & 0 & 0 & 0\\
1 & 0 & 0 & 0\\
0 & 0 & 0 & 0%
\end{array}
\right)\,,\,\,\,\,\, \lambda_5=\left(
\begin{array}
[c]{cccc}%
0 & 0 & -i & 0\\
0 & 0 & 0 & 0\\
i & 0 & 0 & 0\\
0 & 0 & 0 & 0%
\end{array}
\right)\,,\,\,\,\,\, \lambda_6=\left(
\begin{array}
[c]{cccc}%
0 & 0 & 0 & 0\\
0 & 0 & 1 & 0\\
0 & 1 & 0 & 0\\
0 & 0 & 0 & 0%
\end{array}
\right)  \text{ ,}\nonumber%
\end{equation}

\begin{equation}
\lambda_7=\left(
\begin{array}
[c]{cccc}%
0 & 0 & 0 & 0\\
0 & 0 & -i & 0\\
0 & i & 0 & 0\\
0 & 0 & 0 & 0%
\end{array}
\right)\,,\,\,\,\,\, \lambda_8=\left(
\begin{array}
[c]{cccc}%
1 & 0 & 0 & 0\\
0 & 1 & 0 & 0\\
0 & 0 & -2 & 0\\
0 & 0 & 0 & 0%
\end{array}
\right)\,,\,\,\,\,\, \lambda_9=\left(
\begin{array}
[c]{cccc}%
0 & 0 & 0 & 1\\
0 & 0 & 0 & 0\\
0 & 0 & 0 & 0\\
1 & 0 & 0 & 0%
\end{array}
\right)  \text{ ,}\nonumber%
\end{equation}

\begin{equation}
\lambda_{10}=\left(
\begin{array}
[c]{cccc}%
0 & 0 & 0 & -i\\
0 & 0 & 0 & 0\\
0 & 0 & 0 & 0\\
i & 0 & 0 & 0%
\end{array}
\right)\,,\,\,\,\,\, \lambda_{11}=\left(
\begin{array}
[c]{cccc}%
0 & 0 & 0 & 0\\
0 & 0 & 0 & 1\\
0 & 0 & 0 & 0\\
0 & 1 & 0 & 0%
\end{array}
\right)\,,\,\,\,\,\, \lambda_{12}=\left(
\begin{array}
[c]{cccc}%
0 & 0 & 0 & 0\\
0 & 0 & 0 & -i\\
0 & 0 & 0 & 0\\
0 & i & 0 & 0%
\end{array}
\right)  \text{ ,}%
\end{equation}

\begin{equation}
\lambda_{13}=\left(
\begin{array}
[c]{cccc}%
0 & 0 & 0 & 0\\
0 & 0 & 0 & 0\\
0 & 0 & 0 & 1\\
0 & 0 & 1 & 0%
\end{array}
\right)\,,\,\,\,\,\, \lambda_{14}=\left(
\begin{array}
[c]{cccc}%
0 & 0 & 0 & 0\\
0 & 0 & 0 & 0\\
0 & 0 & 0 & -i\\
0 & 0 & i & 0%
\end{array}
\right) \,,\,\,\,\,\, \lambda_{15}=\frac{1}{\sqrt{6}}\left(
\begin{array}
[c]{cccc}%
1 & 0 & 0 & 0\\
0 & 1 & 0 & 0\\
0 & 0 & 1 & 0\\
0 & 0 & 0 & -3%
\end{array}
\right)   \text{ .}\nonumber%
\end{equation}

Note that the rank-3 Gell-Mann matrices of the $SU(3)$ group
are described by the first eight matrices \cite{Tilma:2002kf},
whereas transitions between $SU(3)$ and $SU(4)$ elements
are generated by the matrices $\lambda_9-\lambda_{15}$ \cite{WGreiner}.
We now use Eq.(\ref{Ptr1}) to determine the decay constants. \\

\section{Pion decay constant}

In order to determine the pion decay constant, it suffices to take the corresponding direction in a-space, for instance $a=1$. \\

Firstly, for $a=1$, the generator
$t_1=\frac{1}{2}\left(\begin{array}
[c]{cccc}%
0 & 1 & 0 & 0\\
1 & 0 & 0 & 0\\
0 & 0 & 0 & 0\\
0 & 0 & 0 & 0%
\end{array}
\right) $, and Eq.(\ref{Ptr1}) takes the following form

\begin{equation}
P\rightarrow P+\theta_1 (t_1\, \langle S \rangle+\langle S \rangle\,t_1).%
\end{equation}

Explicitly,

\begin{align}
\frac{1}{2}\left(\begin{array}
[c]{cccc}%
0 & \pi^1 & 0 & 0\\
\pi^1 & 0 & 0 & 0\\
0 & 0 & 0 & 0\\
0 & 0 & 0 & 0%
\end{array}
\right) \mapsto & \frac{1}{2} \left(\begin{array}
[c]{cccc}%
0 & \pi^1 & 0 & 0\\
\pi^1 & 0 & 0 & 0\\
0 & 0 & 0 & 0\\
0 & 0 & 0 & 0%
\end{array}
\right)\nonumber\\ & +\frac{1}{2}\theta_1\bigg[
\left(\begin{array}
[c]{cccc}%
0 & 1 & 0 & 0\\
1 & 0 & 0 & 0\\
0 & 0 & 0 & 0\\
0 & 0 & 0 & 0%
\end{array}
\right) \left(
\begin{array}
[c]{cccc}%
\frac{\phi_N}{\sqrt{2}} & 0 & 0 & 0\\
0 & \frac{\phi_N}{\sqrt{2}} & 0 & 0\\
0 & 0 & \phi_S & 0\\
0 & 0 & 0 & \phi_C%
\end{array}
\right)\nonumber\\ & +\left(
\begin{array}
[c]{cccc}%
\frac{\phi_N}{\sqrt{2}} & 0 & 0 & 0\\
0 & \frac{\phi_N}{\sqrt{2}} & 0 & 0\\
0 & 0 & \phi_S & 0\\
0 & 0 & 0 & \phi_C%
\end{array}
\right)  \left(\begin{array}
[c]{cccc}%
0 & 1 & 0 & 0\\
1 & 0 & 0 & 0\\
0 & 0 & 0 & 0\\
0 & 0 & 0 & 0%
\end{array}
\right)\bigg]\,,\nonumber%
\end{align}

\begin{equation}
\frac{1}{2}\left(\begin{array}
[c]{cccc}%
0 & \pi^1 & 0 & 0\\
\pi^1 & 0 & 0 & 0\\
0 & 0 & 0 & 0\\
0 & 0 & 0 & 0%
\end{array}
\right) \mapsto \frac{1}{2} \left(\begin{array}
[c]{cccc}%
0 & \pi^1 & 0 & 0\\
\pi^1 & 0 & 0 & 0\\
0 & 0 & 0 & 0\\
0 & 0 & 0 & 0%
\end{array}
\right) +\theta_1 \left(\begin{array}
[c]{cccc}%
0 & \frac{\phi_N}{\sqrt{2}} & 0 & 0\\
\frac{\phi_N}{\sqrt{2}} & 0 & 0 & 0\\
0 & 0 & 0& 0\\
0 & 0 & 0 & 0%
\end{array}
\right)\,. %
\end{equation}
We obtain
\begin{equation}
\pi^1 \mapsto \pi^1+\sqrt{2}\phi_N \theta_1\,.
\end{equation}
Similarly, for $a=2,3$, we obtain
\begin{align}
&\pi^2 \mapsto \pi^2+\sqrt{2}\phi_N \theta_2, \\ &\pi^0 \mapsto
\pi^0+\sqrt{2}\phi_N \theta_3.
\end{align}
After introducing the wave-function renormalization for the pion, we get
\begin{align}
&Z_\pi\,\pi^0 \mapsto Z_\pi\,\pi^0+\sqrt{2}\phi_N \theta_3,\\&
Z_\pi\,\pi^1 \mapsto Z_\pi\,\pi^1+\sqrt{2}\phi_N \theta_1,\\&
Z_\pi\,\pi^2 \mapsto Z_\pi\,\pi^2+\sqrt{2}\phi_N \theta_2\,,
\end{align}
which can be written
\begin{align}
&\pi^0 \mapsto \pi^0+\frac{\sqrt{2}\phi_N}{Z_\pi} \theta_3\,,\\ 
&\pi^1 \mapsto \pi^1+\frac{\sqrt{2}\phi_N}{Z_\pi} \theta_1\,,\\
&\pi^2 \mapsto
\pi^2+\frac{\sqrt{2}\phi_N}{Z_\pi}\theta_2\,.
\end{align}
This gives the decay constant of the pion as
\begin{equation}
 f_\pi=\frac{\sqrt{2}\phi_N}{Z_\pi}\,.
\end{equation}

\section{Kaon decay constant}

Let us determine the decay constant $f_K$ of the kaon.\\

 In the case of $a=4$, the generator $t_4=\frac{1}{2}\left(\begin{array}
[c]{cccc}%
0 & 0 & 1 & 0\\
0 & 0 & 0 & 0\\
1 & 0 & 0 & 0\\
0 & 0 & 0 & 0%
\end{array}
\right) $, then Eq.(\ref{Ptr1}) takes the following form $\\$
\begin{equation}
P\rightarrow P+\theta_4 (t_4\, \langle S \rangle+ \langle S \rangle \,t_4).%
\end{equation}

\begin{align}
\frac{1}{\sqrt{2}}\left(\begin{array}
[c]{cccc}%
0 & 0 & \frac{K^1}{\sqrt{2}} & 0\\
0 & 0 & 0 & 0\\
\frac{K^1}{\sqrt{2}} & 0 & 0 & 0\\
0 & 0 & 0 & 0%
\end{array}
\right) \mapsto & \frac{1}{\sqrt{2}} \left(\begin{array}
[c]{cccc}%
0 & 0 & \frac{K^1}{\sqrt{2}} & 0\\
0 & 0 & 0 & 0\\
\frac{K^1}{\sqrt{2}} & 0 & 0 & 0\\
0 & 0 & 0 & 0%
\end{array}
\right)\nonumber\\ & +\frac{1}{2}\theta_4\bigg[
\left(\begin{array}
[c]{cccc}%
0 & 0 & 1 & 0\\
0 & 0 & 0 & 0\\
1 & 0 & 0 & 0\\
0 & 0 & 0 & 0%
\end{array}
\right) \left(
\begin{array}
[c]{cccc}%
\frac{\phi_N}{\sqrt{2}} & 0 & 0 & 0\\
0 & \frac{\phi_N}{\sqrt{2}} & 0 & 0\\
0 & 0 & \phi_S & 0\\
0 & 0 & 0 & \phi_C%
\end{array}
\right)\nonumber\\ & +\left(
\begin{array}
[c]{cccc}%
\frac{\phi_N}{\sqrt{2}} & 0 & 0 & 0\\
0 & \frac{\phi_N}{\sqrt{2}} & 0 & 0\\
0 & 0 & \phi_S & 0\\
0 & 0 & 0 & \phi_C%
\end{array}
\right)  \left(\begin{array}
[c]{cccc}%
0 & 0 & 1 & 0\\
0 & 0 & 0 & 0\\
1 & 0 & 0 & 0\\
0 & 0 & 0 & 0%
\end{array}
\right)\bigg]\,,\nonumber%
\end{align}

\begin{equation}
\frac{1}{\sqrt{2}}\left(\begin{array}
[c]{cccc}%
0 & 0 &  \frac{K^1}{\sqrt{2}} & 0\\
0 & 0 & 0 & 0\\
\frac{K^1}{\sqrt{2}} & 0 & 0 & 0\\
0 & 0 & 0 & 0%
\end{array}
\right) \mapsto \frac{1}{\sqrt{2}} \left(\begin{array}
[c]{cccc}%
0 & 0 &  \frac{K^1}{\sqrt{2}} & 0\\
0 & 0 & 0 & 0\\
\frac{K^1}{\sqrt{2}} & 0 & 0 & 0\\
0 & 0 & 0 & 0%
\end{array}
\right) +\frac{1}{2}\theta_4 \left(\begin{array}
[c]{cccc}%
0 & 0 & \phi_S+\frac{\phi_N}{\sqrt{2}} & 0\\
0 & 0 & 0 & 0\\
\phi_S+\frac{\phi_N}{\sqrt{2}}  & 0 & 0& 0\\
0 & 0 & 0 & 0%
\end{array}
\right)\,, %
\end{equation}
\begin{equation}
\frac{1}{2}\,K^1 \mapsto
\frac{1}{2}\,K^1+\frac{1}{2}(\phi_S+\frac{\phi_N}{\sqrt{2}})
\theta_4\,,
\end{equation}
\begin{equation}
K^1 \mapsto K^1+\frac{\sqrt{2}\,\phi_S+\phi_N}{\sqrt{2}}
\theta_4\,.
\end{equation}
After introducing the wave-function renormalization of the kaons,
\begin{align}
& Z_K\,K^1 \mapsto
Z_K\,K^1+\frac{(\sqrt{2}\,\phi_S+\phi_N)}{\sqrt{2}} \theta_4\nonumber\\
& \Rightarrow \,\,\,\,\,K^1 \mapsto
K^1+\frac{(\sqrt{2}\,\phi_S+\phi_N)}{\sqrt{2}\,Z_K} \theta_4\,.
\end{align}
Similarly for $a=5,\,6,\,7,8$, which can be written in general as
\begin{equation}
K \mapsto K+\frac{\sqrt{2}\,\phi_S+\phi_N}{\sqrt{2}\,Z_K}\,\,
\theta_{4,5,6,8} \,,
\end{equation}
Then we can obtain the kaon decay constant as

\begin{equation}
f_K=\frac{\sqrt{2}\,\phi_S+\phi_N}{\sqrt{2}\,Z_K} \,.
\end{equation}
Note that, in the case $N_f=3$, we used the
formulas for the weak decay constants of pion and kaon divided by
$\sqrt{2}$ as follows:
\begin{equation}
f_\pi=
\frac{\phi_N}{Z_\pi}\,,
\end{equation}
and
\begin{equation}
f_K=\frac{\sqrt{2}\,\phi_S+\phi_N}{2\,Z_K}\,,
\end{equation}
 because these theoretical formulas of the weak decay constants have been used in the fit \cite{Parganlija:2012fy}, which are compared to the experimental data as
 listed by the PDG \cite{Beringer:1900zz}, where they are divided by
 $\sqrt{2}$.\\

\section{Decay constant of $D$ and $D_S$}

Now let us turn to determine the weak decay constants of the open charmed mesons $D$ and $D_S$:\\

For $a=9$, the generator $t_9=\frac{1}{2}\left(\begin{array}
[c]{cccc}%
0 & 0 & 0 & 1\\
0 & 0 & 0 & 0\\
0 & 0 & 0 & 0\\
1 & 0 & 0 & 0%
\end{array}
\right) $, and Eq.(\ref{Ptr}) reads\\

\begin{equation}
P\rightarrow P+\theta_9 (t_9\, \langle S \rangle+\langle S \rangle\,t_9).%
\end{equation}

Then,

\begin{align}
\frac{1}{2}\left(\begin{array}
[c]{cccc}%
0 & 0 & 0 & D^1\\
0 & 0 & 0 & 0\\
0 & 0 & 0 & 0\\
D^1 & 0 & 0 & 0%
\end{array}
\right) \mapsto & \frac{1}{2} \left(\begin{array}
[c]{cccc}%
0 & 0 & 0 & D^1\\
0 & 0 & 0 & 0\\
0 & 0 & 0 & 0\\
D^1 & 0 & 0 & 0%
\end{array}
\right)\nonumber\\ & +\frac{1}{2}\theta_9\bigg[
\left(\begin{array}
[c]{cccc}%
0 & 0 & 0 & 1\\
0 & 0 & 0 & 0\\
0 & 0 & 0 & 0\\
1 & 0 & 0 & 0%
\end{array}
\right) \left(
\begin{array}
[c]{cccc}%
\frac{\phi_N}{\sqrt{2}} & 0 & 0 & 0\\
0 & \frac{\phi_N}{\sqrt{2}} & 0 & 0\\
0 & 0 & \phi_S & 0\\
0 & 0 & 0 & \phi_C%
\end{array}
\right)\nonumber\\ & +\left(
\begin{array}
[c]{cccc}%
\frac{\phi_N}{\sqrt{2}} & 0 & 0 & 0\\
0 & \frac{\phi_N}{\sqrt{2}} & 0 & 0\\
0 & 0 & \phi_S & 0\\
0 & 0 & 0 & \phi_C%
\end{array}
\right)  \left(\begin{array}
[c]{cccc}%
0 & 0 & 0 & 1\\
0 & 0 & 0 & 0\\
0 & 0 & 0 & 0\\
1 & 0 & 0 & 0%
\end{array}
\right)\bigg]\,,\nonumber%
\end{align}

\begin{equation}
\left(\begin{array}
[c]{cccc}%
0 & 0 & 0 & D^1\\
0 & 0 & 0 & 0\\
0 & 0 & 0 & 0\\
D^1 & 0 & 0 & 0%
\end{array}
\right) \mapsto \left(\begin{array}
[c]{cccc}%
0 & 0 &  0 & D^1\\
0 & 0 & 0 & 0\\
0 & 0 & 0 & 0\\
D^1 & 0 & 0 & 0%
\end{array}
\right) +\theta_9 \left(\begin{array}
[c]{cccc}%
0 & \,\,\,0 & \,\,\,0  & (\frac{\phi_N}{\sqrt{2}}+\phi_C)\\
0 & \,\,\,0 & \,\,\,0  & 0\\
0 & \,\,\,0 & \,\,\,0 & 0\\
(\frac{\phi_N}{\sqrt{2}}+\phi_C) & \,\,\,0 & \,\,\,0 & 0%
\end{array}
\right)\,, %
\end{equation}
\begin{equation}
D^1 \mapsto D^1+(\frac{\phi_N}{\sqrt{2}}+\phi_C) \theta_9\,.
\end{equation}
After introducing the wave-function renormalization of the $D$ mesons

\begin{align}
&Z_D\,D^1 \mapsto Z_D\,D^1+(\frac{\phi_N}{\sqrt{2}}+\phi_C)
\theta_9\,,\nonumber\\ &\Rightarrow \,\,\,\,\,D^1 \mapsto
D^1+\frac{\phi_N+\sqrt{2}\phi_C}{\sqrt{2}}\,\, \theta_9\,.
\end{align}
Similarly for $a=10,\,11,\,12$, which can be written in general as
\begin{equation}
D \mapsto D+\frac{\phi_N+\sqrt{2}\,\phi_C}{\sqrt{2}\,Z_D}\,
\theta_{9,\,10,\,11,\,12} \,.
\end{equation}
Then we can obtain the decay constant of $D$ as
\begin{equation}
f_D=\frac{\phi_N+\sqrt{2}\,\phi_C}{\sqrt{2}\,Z_D} \,.
\end{equation}

For $a=13$, the generator $t_{13}=\frac{1}{2}\left(\begin{array}
[c]{cccc}%
0 & 0 & 0 & 0\\
0 & 0 & 0 & 0\\
0 & 0 & 0 & 1\\
0 & 0 & 1 & 0%
\end{array}
\right) $, and Eq.(\ref{Ptr}) is\\

\begin{equation}
P\rightarrow P+\theta_{13} (t_{13}\, \langle S \rangle+\langle S\rangle \,t_{13})\,.%
\end{equation}

\begin{align}
\frac{1}{2}\left(\begin{array}
[c]{cccc}%
0 & 0 & 0 & 0\\
0 & 0 & 0 & 0\\
0 & 0 & 0 & D^1_S\\
0 & 0 & D^1_S & 0%
\end{array}
\right) \mapsto & \frac{1}{2} \left(\begin{array}
[c]{cccc}%
0 & 0 & 0 & 0\\
0 & 0 & 0 & 0\\
0 & 0 & 0 & D^1_S\\
0 & 0 & D^1_S & 0%
\end{array}
\right)\nonumber\\ & +\frac{1}{2}\theta_{13}\bigg[
\left(\begin{array}
[c]{cccc}%
0 & 0 & 0 & 0\\
0 & 0 & 0 & 0\\
0 & 0 & 0 & 1\\
0 & 0 & 1 & 0%
\end{array}
\right) \left(
\begin{array}
[c]{cccc}%
\frac{\phi_N}{\sqrt{2}} & 0 & 0 & 0\\
0 & \frac{\phi_N}{\sqrt{2}} & 0 & 0\\
0 & 0 & \phi_S & 0\\
0 & 0 & 0 & \phi_C%
\end{array}
\right)\nonumber\\ & +\left(
\begin{array}
[c]{cccc}%
\frac{\phi_N}{\sqrt{2}} & 0 & 0 & 0\\
0 & \frac{\phi_N}{\sqrt{2}} & 0 & 0\\
0 & 0 & \phi_S & 0\\
0 & 0 & 0 & \phi_C%
\end{array}
\right)  \left(\begin{array}
[c]{cccc}%
0 & 0 & 0 & 0\\
0 & 0 & 0 & 0\\
0 & 0 & 0 & 1\\
0 & 0 & 1 & 0%
\end{array}
\right)\bigg]\,,\nonumber\\%
\end{align}

\begin{equation}
\left(\begin{array}
[c]{cccc}%
0 & 0 & 0 & 0\\
0 & 0 & 0 & 0\\
0 & 0 & 0 & D^1_S\\
0 & 0 & D^1_S & 0%
\end{array}
\right) \mapsto \left(\begin{array}
[c]{cccc}%
0 & 0 & 0 & 0\\
0 & 0 & 0 & 0\\
0 & 0 & 0 & D^1_S\\
0 & 0 & D^1_S & 0%
\end{array}
\right) +\theta_{13} \left(\begin{array}
[c]{cccc}%
0 & \,\,\,0 & \,\,\,0  & 0\\
0 & \,\,\,0 & \,\,\,0  & 0\\
0 & \,\,\,0 & \,\,\,0 & \phi_S+\phi_C\\
0 & \,\,\,0 & \phi_S+\phi_C & 0%
\end{array}
\right)\,, %
\end{equation}
\begin{equation}
D^1_S \mapsto D^1_S+(\phi_S+\phi_C) \theta_{13}\,,
\end{equation}
we get

\begin{align}
&Z_{D_S}\,D^1_S \mapsto Z_{D_S}\,D^1_S+(\phi_S+\phi_C)
\theta_{13}\nonumber\\ &\Rightarrow \,\,\,\,\,D^1_S \mapsto
D^1_S+\frac{\phi_S+\phi_C}{ Z_{D_S}} \theta_{13}\,.
\end{align}
Similarly for $a=14$, which can be written in general as
\begin{equation}
D_S \mapsto D_S+\frac{\phi_S+\phi_C}{Z_{D_S}}
\theta_{13,14} \,,
\end{equation}
Then we obtain the decay constant of $D_S$ as
\begin{equation}
f_{D_S}=\frac{\phi_N+\phi_C}{Z_{D_S}} \,,
\end{equation}

\section{Decay constant of $\eta_C$}

Finally, the decay constant of the charmonium state $\eta_C$ can be
determined from the combination of two generators, $a=0$ and $a=15$, as follows: \\

For $a=0$, the generator
$t_{0}=\frac{\lambda_0}{2}=\frac{1}{2\sqrt{2}}\left(\begin{array}
[c]{cccc}%
1 & 0 & 0 & 0\\
0 & 1 & 0 & 0\\
0 & 0 & 1 & 0\\
0 & 0 & 0 & 1%
\end{array}
\right) $, and Eq.(\ref{Ptr}) reads\\

\begin{equation}
P\rightarrow P+\theta_{0} (t_{0}\, \langle S \rangle+\langle S \rangle \,t_{0})\,.%
\end{equation}

\begin{align}
\frac{1}{\sqrt{2}}\left(\begin{array}
[c]{cccc}%
\frac{\eta_N}{\sqrt{2}} & 0 & 0 & 0\\
0 & \frac{\eta_N}{\sqrt{2}} & 0 & 0\\
0 & 0 & \eta_S & 0\\
0 & 0 & 0 & \eta_C%
\end{array}
\right)& \mapsto  \frac{1}{\sqrt{2}} \left(\begin{array}
[c]{cccc}%
\frac{\eta_N}{\sqrt{2}} & 0 & 0 & 0\\
0 & \frac{\eta_N}{\sqrt{2}} & 0 & 0\\
0 & 0 & \eta_S & 0\\
0 & 0 & 0 & \eta_C%
\end{array}
\right)\nonumber\\ & +\frac{1}{2\sqrt{2}}\theta_{0}\bigg[
\left(\begin{array}
[c]{cccc}%
1 & 0 & 0 & 0\\
0 & 1 & 0 & 0\\
0 & 0 & 1 & 0\\
0 & 0 & 0 & 1%
\end{array}
\right) \left(
\begin{array}
[c]{cccc}%
\frac{\phi_N}{\sqrt{2}} & 0 & 0 & 0\\
0 & \frac{\phi_N}{\sqrt{2}} & 0 & 0\\
0 & 0 & \phi_S & 0\\
0 & 0 & 0 & \phi_C%
\end{array}
\right)\nonumber\\ &+\left(
\begin{array}
[c]{cccc}%
\frac{\phi_N}{\sqrt{2}} & 0 & 0 & 0\\
0 & \frac{\phi_N}{\sqrt{2}} & 0 & 0\\
0 & 0 & \phi_S & 0\\
0 & 0 & 0 & \phi_C%
\end{array}
\right)  \left(\begin{array}
[c]{cccc}%
1 & 0 & 0 & 0\\
0 & 1 & 0 & 0\\
0 & 0 & 1 & 0\\
0 & 0 & 0 & 1%
\end{array}
\right)\bigg]\,,\nonumber%
\end{align}

\begin{equation}
\left(\begin{array}
[c]{cccc}%
\frac{\eta_N}{\sqrt{2}} & 0 & 0 & 0\\
0 & \frac{\eta_N}{\sqrt{2}} & 0 & 0\\
0 & 0 & \eta_S & 0\\
0 & 0 & 0 & \eta_C%
\end{array}
\right) \mapsto \left(\begin{array}
[c]{cccc}%
\frac{\eta_N}{\sqrt{2}} & 0 & 0 & 0\\
0 & \frac{\eta_N}{\sqrt{2}} & 0 & 0\\
0 & 0 & \phi_S & 0\\
0 & 0 & 0 & \phi_C%
\end{array}
\right) +\theta_{0} \left(\begin{array}
[c]{cccc}%
\frac{\phi_N}{\sqrt{2}} & 0 & 0 & 0\\
0 & \frac{\phi_N}{\sqrt{2}} & 0 & 0\\
0 & 0 & \phi_S & 0\\
0 & 0 & 0 & \phi_C%
\end{array}
\right)\,. \label{Trans0} %
\end{equation}

For $a=15$, the generator
$t_{15}=\frac{1}{2\sqrt{6}}\left(\begin{array}
[c]{cccc}%
1 & 0 & 0 & 0\\
0 & 1 & 0 & 0\\
0 & 0 & 1 & 0\\
0 & 0 & 0 & -3%
\end{array}
\right) $, and Eq.(\ref{Ptr1}) reads\\

\begin{equation}
P\rightarrow P+\theta_{15} (t_{15}\, \langle S\rangle +\langle S \rangle\,t_{15})\,.%
\end{equation}
Then,
\begin{align}
\frac{1}{\sqrt{2}}\left(\begin{array}
[c]{cccc}%
\frac{\eta_N}{\sqrt{2}} & 0 & 0 & 0\\
0 & \frac{\eta_N}{\sqrt{2}} & 0 & 0\\
0 & 0 & \eta_S & 0\\
0 & 0 & 0 & \eta_C%
\end{array}
\right)& \mapsto  \frac{1}{\sqrt{2}} \left(\begin{array}
[c]{cccc}%
\frac{\eta_N}{\sqrt{2}} & 0 & 0 & 0\\
0 & \frac{\eta_N}{\sqrt{2}} & 0 & 0\\
0 & 0 & \eta_S & 0\\
0 & 0 & 0 & \eta_C%
\end{array}
\right)\nonumber\\ & +\frac{1}{2\sqrt{6}}\theta_{15}\bigg[
\left(\begin{array}
[c]{cccc}%
1 & 0 & 0 & 0\\
0 & 1 & 0 & 0\\
0 & 0 & 1 & 0\\
0 & 0 & 0 & -3%
\end{array}
\right) \left(
\begin{array}
[c]{cccc}%
\frac{\phi_N}{\sqrt{2}} & 0 & 0 & 0\\
0 & \frac{\phi_N}{\sqrt{2}} & 0 & 0\\
0 & 0 & \phi_S & 0\\
0 & 0 & 0 & \phi_C%
\end{array}
\right)\nonumber\\ &+\left(
\begin{array}
[c]{cccc}%
\frac{\phi_N}{\sqrt{2}} & 0 & 0 & 0\\
0 & \frac{\phi_N}{\sqrt{2}} & 0 & 0\\
0 & 0 & \phi_S & 0\\
0 & 0 & 0 & \phi_C%
\end{array}
\right)  \left(\begin{array}
[c]{cccc}%
1 & 0 & 0 & 0\\
0 & 1 & 0 & 0\\
0 & 0 & 1 & 0\\
0 & 0 & 0 & -3%
\end{array}
\right)\bigg]\,,\nonumber%
\end{align}

\begin{equation}
\left(\begin{array}
[c]{cccc}%
\frac{\eta_N}{\sqrt{2}} & 0 & 0 & 0\\
0 & \frac{\eta_N}{\sqrt{2}} & 0 & 0\\
0 & 0 & \eta_S & 0\\
0 & 0 & 0 & \eta_C%
\end{array}
\right) \mapsto \left(\begin{array}
[c]{cccc}%
\frac{\eta_N}{\sqrt{2}} & 0 & 0 & 0\\
0 & \frac{\eta_N}{\sqrt{2}} & 0 & 0\\
0 & 0 & \eta_S & 0\\
0 & 0 & 0 & \eta_C%
\end{array}
\right) +\frac{\theta_{15}}{\sqrt{3}} \left(\begin{array}
[c]{cccc}%
\frac{\phi_N}{\sqrt{2}} & 0 & 0 & 0\\
0 & \frac{\phi_N}{\sqrt{2}} & 0 & 0\\
0 & 0 & \phi_S & 0\\
0 & 0 & 0 & -3\phi_C%
\end{array}
\right)\,.\label{trans15} %
\end{equation}

To find a combination of diagonal Lambdas which has only a nonzero entry in the fourth column and in the fourth row, we consider the relation between the singlet angle of transformation
$\theta_0$ and the multiplet angle transformation $\theta_{15}$ as
follows:
$$\theta_{15}= -\sqrt{3}\theta_0\,.$$

Therefore, Eq.(\ref{trans15}), can be written as

\begin{equation}
\left(\begin{array}
[c]{cccc}%
\frac{\eta_N}{\sqrt{2}} & 0 & 0 & 0\\
0 & \frac{\eta_N}{\sqrt{2}} & 0 & 0\\
0 & 0 & \eta_S & 0\\
0 & 0 & 0 & \eta_C%
\end{array}
\right) \mapsto \left(\begin{array}
[c]{cccc}%
\frac{\eta_N}{\sqrt{2}} & 0 & 0 & 0\\
0 & \frac{\eta_N}{\sqrt{2}} & 0 & 0\\
0 & 0 & \eta_S & 0\\
0 & 0 & 0 & \eta_C%
\end{array}
\right) -\theta_{0} \left(\begin{array}
[c]{cccc}%
\frac{\phi_N}{\sqrt{2}} & 0 & 0 & 0\\
0 & \frac{\phi_N}{\sqrt{2}} & 0 & 0\\
0 & 0 & \phi_S & 0\\
0 & 0 & 0 & -3\phi_C%
\end{array}
\right)\,.\label{Trans152} %
\end{equation}
Adding Eq.(\ref{Trans0}) and Eq.(\ref{Trans152}), we obtain
\begin{equation}
2\left(\begin{array}
[c]{cccc}%
\frac{\eta_N}{\sqrt{2}} & 0 & 0 & 0\\
0 & \frac{\eta_N}{\sqrt{2}} & 0 & 0\\
0 & 0 & \eta_S & 0\\
0 & 0 & 0 & \eta_C%
\end{array}
\right) \mapsto 2\left(\begin{array}
[c]{cccc}%
\frac{\eta_N}{\sqrt{2}} & 0 & 0 & 0\\
0 & \frac{\eta_N}{\sqrt{2}} & 0 & 0\\
0 & 0 & \eta_S & 0\\
0 & 0 & 0 & \eta_C%
\end{array}
\right) +\theta_{0} \left(\begin{array}
[c]{cccc}%
0 & 0 & 0 & 0\\
0 & 0 & 0 & 0\\
0 & 0 & 0 & 0\\
0 & 0 & 0 & 4\phi_C%
\end{array}
\right)\,.%
\end{equation}
Then,
\begin{equation}
2\eta_C \mapsto 2\eta_C+ 4\phi_C \theta_{0}\,,
\end{equation}
After introducing the wave-function renormalization of the $\eta_C$ meson

\begin{equation}
Z_{\eta_C}\,\eta_C \mapsto Z_{\eta_C}\,\eta_C+2\phi_C
\theta_{0}\,.
\end{equation}
Therefore, the weak decay constant of $\eta_C$ is
\begin{equation}
f_{\eta_C}=\frac{2\,\phi_C}{Z_{\eta_C}} \,.
\end{equation}

\chapter{Decay rates for $\chi_{c0}$ \label{Sec2}}

We show the explicit expressions for the two- and three-body decay
rates for the scalar hidden-charmed meson $\chi_{c0}$, which
are extracted from the Lagrangian (\ref{lag}) at tree level. The
results are listed in Tables 8.2, 8.3, and Table 8.4 in Sec.8.2.

\section{Two-body decay rates for $\chi_{C0}$}

The explicit expression for the two-body decay rates of
$\chi_{c0}$ are extracted from the
Lagrangian (\ref{lag}), and are
presented in the following.\\

\textbf{Decay channel
$\chi_{C0}\rightarrow\overline{K}^{\ast0}_{0}K^{\ast0}_{0}$ }\\

The corresponding interaction Lagrangian from the Lagrangian
(\ref{lag}) reads
\begin{align}
\mathcal{L}_{\chi_{C0}\overline{K}^{\ast}_{0}K^{\ast}_{0}}=&-2\lambda_1
\,
Z_{K^\ast_0}^2\,\phi_C\,\chi_{C0}(\overline{K}^{\ast0}_{0}K^{\ast0}_{0}+
K^{\ast-}_{0}K^{\ast+}_{0})\nonumber\\&-h_1\,\phi_C\,Z_{K^\ast_0}^2
\omega^2_{K^\ast} \chi_{C0}(\partial_\mu K^{\ast0}_0\partial^\mu
\overline{K}^{\ast0}_0 + \partial_\mu K^{\ast-}_0\partial^\mu
K^{\ast+}_0 )\,.
\end{align}
Consider only the
$\chi_{C0}\rightarrow\overline{K}^{\ast0}_{0}K^{\ast0}_{0}$ decay
channel, the $\chi_{C0}\rightarrow
K^{\ast-}_{0}K^{\ast+}_{0}$ will give the same contribution due
to isospin symmetry,

\begin{align}
\mathcal{L}_{\chi_{C0}\overline{K}^{\ast0}_{0}K^{\ast0}_{0}}=&-2\lambda_1
\,
Z_{K^\ast_0}^2\,\phi_C\,\chi_{C0}\overline{K}^{\ast0}_{0}K^{\ast0}_{0}-h_1\,\phi_C\,Z_{K^\ast_0}^2
\omega^2_{K^\ast} \chi_{C0}\partial_\mu K^{\ast0}_0\partial^\mu
\overline{K}^{\ast0}_0 \,.
\end{align}
Let us denote the momenta of ${K}^{\ast0}_{0}$ and
$\overline{K}^{\ast0}_{0}$ as $P_1$ and $P_2$, respectively. The
energy-momentum conservation on the vertex implies $P=P_1+P_2$, where $P$
denotes the momenta of the decaying particle $\chi_{C0}$. Given
that our particles are on-shell, we obtain
\begin{equation}
P_1 \cdot P_2=\frac{P^2-P_1^2-P_2^2}{2}=\frac{m_{\chi_{C0}}^2-2\,m_{K^{\ast0}_0}}{2}\,.
\end{equation}

Upon substituting $\partial_\mu\rightarrow -iP^\mu$ for the decay
particle and $\partial_\mu\rightarrow +iP_{1,2}^\mu$ for the outgoing
particles, one obtains
\begin{align}
\mathcal{L}_{\chi_{C0}\overline{K}^{\ast0}_{0}K^{\ast0}_{0}}=\phi_C\,Z_{K^\ast_0}^2\bigg[-2\lambda_1+h_1\omega^2_{K^\ast}\frac{m_{\chi_{c0}}^2-2m_{K^\ast_0}^2}{2}\bigg]\chi_{C0}
K^{\ast0}_0\overline{K}^{\ast0}_0\,.
\end{align}

Consequently, the decay amplitude is given by
\begin{equation}
-i\mathcal{M}_{\chi_{C0}\rightarrow\overline{K}^{\ast0}_{0}K^{\ast0}_{0}}=i
\phi_C\,Z_{K^\ast_0}^2\bigg[2\lambda_1-h_1\omega^2_{K^\ast} \frac{m_{\chi_{c0}}^2-2m_{K^\ast_0}^2}{2}\bigg]\,.
\end{equation}
The decay width is obtained as
\begin{equation}
\Gamma_{\chi_{C0}\rightarrow\overline{K}^{\ast0}_{0}K^{\ast0}_{0}}=\frac{|\overrightarrow{k}_1|}{8\pi\,m_{\chi_{C0}}^2}|-i\mathcal{M}_{\chi_{C0}\rightarrow\overline{K}^{\ast0}_{0}K^{\ast0}_{0}}|^2\,.
\end{equation}
where
\begin{equation}
|\overrightarrow{k}_1|=\frac{1}{2m_{\chi_{C0}}}\bigg[m_{\chi_{C0}}^4+(m_{\overline{K}^{\ast0}_{0}}^2-m_{K^{\ast0}_{0}}^2)^2-2(m_{\overline{K}^{\ast0}_{0}}^2+m_{K^{\ast0}_{0}}^2)m_{\chi_{C0}}^2\bigg]^{1/2}\,.
\end{equation}
$$\\$$
\textbf{Decay channel $\chi_{C0}\rightarrow K^-K^+$ }\\

 The corresponding interaction Lagrangian from the Lagrangian (\ref{lag})
has the following form
\begin{align}
\mathcal{L}_{\chi_{C0}KK}=&-2\lambda_1 \,
Z_{K}^2\,\phi_C\,\chi_{C0}(\overline{K}^{0}K^{0}+
K^{-}K^{+}_{0})\nonumber\\&+h_1\,\phi_C\,Z_{K}^2 \omega^2_{K_1}
\chi_{C0}(\partial_\mu K^{0}\partial^\mu \overline{K}^{0} +
\partial_\mu K^{-}\partial^\mu K^{+})\,.
\end{align}
In a similar way as the previous case, one can obtain the decay
width for the channel $\chi_{C0}\rightarrow K^-K^+$ as
\begin{equation}
\Gamma_{\chi_{C0}\rightarrow K^-K^+
}=\frac{|\overrightarrow{k}_1|}{8\pi\,m_{\chi_{C0}}^2}\,\phi_C^2\,Z_{K}^4
\bigg[2\lambda_1+h_1\omega^2_{K_1}\bigg(\frac{m_{\chi_{c0}}^2-2m_{K}^2}{2}\bigg)\bigg]^2\,,
\end{equation}
where
\begin{equation}
|\overrightarrow{k}_1|=\frac{1}{2m_{\chi_{C0}}}\bigg[m_{\chi_{C0}}^4-4m_{K}^2\,m_{\chi_{C0}}^2\bigg]^{1/2}\,.
\end{equation}
$$\\$$
$$\\$$

\textbf{Decay channel $\chi_{C0}\rightarrow\pi\pi$}\\

The corresponding interaction Lagrangian is extracted from the
Lagrangian (\ref{lag}) as
\begin{align}
\mathcal{L}_{\chi_{C0}\pi\pi}=-\lambda_1\,\phi_C\,
Z_{\pi}^2\,\chi_{C0}({\pi^0}^2+2
\pi^{-}\pi^{+})+\frac{1}{2}h_1\,\phi_C\,Z_{\pi}^2 \omega^2_{a_1}
\chi_{C0}[(\partial_\mu\pi^0)^2+2\partial_\mu \pi^-\partial^\mu
\pi^+]\,.
\end{align}
The decay width for the channel $\chi_{C0}\rightarrow \pi \pi$ can be obtained in complete analogous as the previous cases
\begin{equation}
\Gamma_{\chi_{C0}\rightarrow\pi\pi}=\frac{3}{2}\frac{|\overrightarrow{k}_1|}{8\pi\,m_{\chi_{C0}}^2}\,\phi_C^2\,Z_{\pi}^4
\bigg[2\lambda_1+h_1\omega^2_{a_1}\bigg(\frac{m_{\chi_{c0}}^2-2m_{\pi}^2}{2}\bigg)\bigg]^2\,,
\end{equation}
where
\begin{equation}
|\overrightarrow{k}_1|=\frac{(m_{\chi_{C0}}^4-4m_\pi^2
m_{\chi_{C0}}^2)^{1/2}}{2m_{\chi_{C0}}}\,.
\end{equation}
$$\\$$

\textbf{Decay channel
$\chi_{C0}\rightarrow\overline{K}^{\ast0}K^{\ast0}$}\\

The corresponding interaction Lagrangian is extracted as
\begin{align}
\mathcal{L}_{\chi_{C0}\overline{K}^{\ast0}K^{\ast0}}=h_1\,\phi_C\,\chi_{C0}(
K^{\ast-}_\mu
K^{\ast+\mu}+K^{\ast0}_\mu\overline{K}^{\ast0\mu})\,.
\end{align}
Consider only the $K^{\ast0}_\mu\overline{K}^{\ast0\mu}$ decay
channel, then
\begin{align}
\mathcal{L}_{\chi_{C0}\overline{K}^{\ast0}K^{\ast0}}=h_1\,\phi_C\,\chi_{C0}K^{\ast0}_\mu\overline{K}^{\ast0\mu}\,.
\end{align}
Put
\begin{equation}
 A_{\chi_{C0}\overline{K}^{\ast0}K^{\ast0}}= h_1\,\phi_C\,.
\end{equation}

Let us denote the momenta of $\chi_{C0}$, $\overline{K}^{\ast0}$,
and $K^{\ast0}$ as $P$, $P_{1}$, and $P_{2}$, respectively, while
the polarisation vectors are denoted as
$\varepsilon_{\mu}^{(\alpha)}(P_{1})$ and
$\varepsilon_{\nu}^{(\beta )}(P_{2})$. Then, upon substituting
$\partial^{\mu}\rightarrow iP_{1,2}^{\mu}$ for the outgoing
particles, we obtain the following Lorentz-invariant
$\chi_{C0}\overline{K}^{\ast0}K^{\ast0}$ scattering amplitude
$-i\mathcal{M}^{(\alpha,\beta)}_{\chi_{C0}\rightarrow\overline{K}^{\ast0}K^{\ast0}}$:\\

\begin{equation}
-i\mathcal{M}_{\chi_{C0}\rightarrow\overline{K}^{\ast0}K^{\ast0}}^{(\alpha,\beta)}=\varepsilon_{\mu
}^{(\alpha)}(P_{1})\varepsilon_{\nu}^{(\beta)}(P_{2})h_{\chi_{C0}\overline{K}^{\ast0}K^{\ast0}}^{\mu\nu
}\,,\label{iMSVV}
\end{equation}

with

\begin{equation}
h_{\chi_{C0}\overline{K}^{\ast0}K^{\ast0}}^{\mu\nu}=iA_{\chi_{C0}\overline{K}^{\ast0}K^{\ast0}}g^{\mu\nu}\text{,}
\label{hSVV}
\end{equation}
$$\\$$
where $h_{\chi_{C0}\overline{K}^{\ast0}K^{\ast0}}^{\mu\nu}$
denotes the $\chi_{C0}\overline{K}^{\ast0}K^{\ast0}$
vertex.\newline

The averaged squared amplitude $|\overline{-i\mathcal{M}}|^{2}$ is
determined as follows:

\begin{align}
\left\vert
\overline{-i\mathcal{M}_{\chi_{C0}\rightarrow\overline{K}^{\ast0}K^{\ast0}}}\right\vert
^{2}
=&\frac{1}{3}\sum\limits_{\alpha,\beta=1}^{3}\left\vert
-i\mathcal{M}_{\chi_{C0}\rightarrow\overline{K}^{\ast0}K^{\ast0}}^{(\alpha,\beta)}\right\vert
^{2}\nonumber\\
=&\frac{1}{3}\sum\limits_{\alpha,\beta=1}^{3}\varepsilon_{\mu}^{(\alpha
)}(P_{1})\varepsilon_{\nu}^{(\beta)}(P_{2})h_{\chi_{C0}\overline{K}^{\ast0}K^{\ast0}}^{\mu\nu}\varepsilon
_{\kappa}^{(\alpha)}(P_{1})\nonumber\\
&\,\,\,\,\,\,\,\,\,\,\,\,\,\,\,\,\,\,\,\,\,\times \varepsilon_{\lambda}^{(\beta)}(P_{2})h_{\chi_{C0}\overline{K}^{\ast0}K^{\ast0}
}^{\ast\kappa\lambda}\text{ .}\label{iMSVV1}
\end{align}

Equation (\ref{iMSVV1}) then yields the same expression as the
one presented in Eq.\ (\ref{iMAVP2}):

\begin{align}
|\overline{-i\mathcal{M}_{\chi_{C0}\rightarrow\overline{K}^{\ast0}K^{\ast0}}}|^{2}=\frac{1}{3}\bigg[
\left\vert
h_{\chi_{C0}\overline{K}^{\ast0}K^{\ast0}}^{\mu\nu}\right\vert
^{2}&-\frac{\left\vert h_{\chi_{C0}\overline{K}^{\ast0}K^{\ast0}}
^{\mu\nu}P_{1\mu}\right\vert ^{2}}{m_{V_{1}}^{2}}-\frac{\left\vert
h_{\chi_{C0}\overline{K}^{\ast0}K^{\ast0}
}^{\mu\nu}P_{2\nu}\right\vert
^{2}}{m_{V_{2}}^{2}}\nonumber\\
&+\frac{\left\vert
h_{\chi_{C0}\overline{K}^{\ast0}K^{\ast0}
}^{\mu\nu}P_{1\mu}P_{2\nu}\right\vert
^{2}}{m_{V_{1}}^{2}m_{V_{2}}^{2} }\bigg]  \text{ .}\label{iMSVV2}
\end{align}

From Eq.\ (\ref{hSVV}) we obtain 

$$h_{\chi_{C0}\overline{K}^{\ast0}K^{\ast0}}^{\mu\nu}P_{1\mu}=iA_{\chi_{C0}\overline{K}^{\ast0}K^{\ast0}%
}P_{1}^{\nu}\,,$$
$$h_{\chi_{C0}\overline{K}^{\ast0}K^{\ast0}}^{\mu\nu}P_{2\nu}=iA_{\chi_{C0}\overline{K}^{\ast0}K^{\ast0}}P_{2}^{\mu}\,,$$
and
$$h_{\chi_{C0}\overline{K}^{\ast0}K^{\ast0}}^{\mu\nu}P_{1\mu}P_{2\nu}=iA_{\chi_{C0}\overline{K}^{\ast0}K^{\ast0}}P_{1}\cdot P_{2}\,,$$ and
consequently

\begin{equation}
|\overline{-i\mathcal{M}_{\chi_{C0}\rightarrow\overline{K}^{\ast0}K^{\ast0}}}|^{2}=\frac{1}{3}\left[
4-\frac{P_{1}^{2}}{m_{K^{\ast0}}^{2}}-\frac{P_{2}^{2}}{m_{\overline{K}^{\ast0}}^{2}}+\frac
{(P_{1}\cdot
P_{2})^{2}}{m_{K^{\ast0}}^{2}m_{\overline{K}^{\ast0}}^{2}}\right]
A_{\chi_{C0}\overline{K}^{\ast0}K^{\ast0}}
^{2}\text{.}\label{iMSVV3}
\end{equation}

For on-shell states,
$P_{1,2}^{2}=m_{\overline{K}^{\ast0},K^{\ast0}}^{2}$ and Eq.\
(\ref{iMSVV3}) reduces to

\begin{align}
|\overline{-i\mathcal{M}_{\chi_{C0}\rightarrow\overline{K}^{\ast0}K^{\ast0}}}|^{2}& =\frac{1}{3}\left[
2+\frac{(P_{1}\cdot
P_{2})^{2}}{m_{\overline{K}^{\ast0}}^{2}m_{K^{\ast0}}^{2}}\right]
A_{\chi_{C0}\overline{K}^{\ast0}K^{\ast0} }^{2}\nonumber\\
&=\frac{1}{3}\left[
2+\frac{(m_{\chi_{C0}}^{2}-m_{\overline{K}^{\ast0}}^{2}-m_{K^{\ast0}}^{2})^{2}
}{4m_{\overline{K}^{\ast0}}^{2}m_{K^{\ast0}}^{2}}\right]
A_{\chi_{C0}\overline{K}^{\ast0}K^{\ast0}}^{2}\text{.}\label{iMSVV4}
\end{align}

Consequently, the decay width is

\begin{equation}
\Gamma_{\chi_{C0}\rightarrow\overline{K}^{\ast0}K^{\ast0}}=\frac{|\overrightarrow{k}_1|}{8\pi
m_{\chi_{C0}}^{2}}|\overline{-i\mathcal{M}_{\chi_{C0}\rightarrow\overline{K}^{\ast0}K^{\ast0}}}|^{2}\,,\label{GSVV}
\end{equation}
where
\begin{equation}
|\overrightarrow{k}_1|=\frac{1}{2m_{\chi_{C0}}}[m_{\chi_{C0}}^4+(m^2_{\overline{K}^{\ast0}}-m^2_{{K}^{\ast0}})^2-2(m^2_{\overline{K}^{\ast0}}+m^2_{{K}^{\ast0}})^2
m_{\chi_{C0}}^2]^{1/2}\,.
\end{equation}
$$\\$$

\textbf{Decay channel $\chi_{C0}\rightarrow K_1\overline{K}_1$}\\

The corresponding interaction Lagrangian is extracted as
\begin{align}
\mathcal{L}_{\chi_{C0}\overline{K}_1K_1}=h_1\,\phi_C\,\chi_{C0}(
K^{+\mu}_1 K^{-}_{1\mu}+K^{0\mu}_1\overline{K}^{0}_{1\mu})\,.
\end{align}
Consider only the $\chi_{C0}\rightarrow K^{0}_1\overline{K}^{0}_1$
decay channel, which gives the same contribution as the
$\chi_{C0}\rightarrow K^{+\mu}_1 K^{-}_{1\mu}$ decay channel due
to the isospin symmetry
\begin{align}
\mathcal{L}_{\chi_{C0}K^0_1\overline{K}^0_1}=h_1\,\phi_C\,\chi_{C0}K^{0\mu}_1\overline{K}^{0}_{1\mu}\,,
\end{align}
which has the same form of the interaction Lagrangian
$\mathcal{L}_{\chi_{C0}\overline{K}^{\ast0}K^{\ast0}}$. Therefore,
in a similar way as was discussed in the previous case for the decay width
of $\chi_{C0}$ into $\overline{K}^{\ast0}K^{\ast0}$, one can
obviously obtain the decay width of the channel
$\chi_{C0}\rightarrow K^{0}_1\overline{K}^{0}_1$ as
\begin{equation}
\Gamma_{\chi_{C0}\rightarrow
K^{0}_1\overline{K}^{0}_1}=\frac{|\overrightarrow{k}_1|}{8\pi
m_{\chi_{C0}}^{2}}\,
\frac{1}{3}h_1^2\,\phi_C^2\bigg[2+\frac{(m_{\chi_{C0}}^2-m_{K^{0}_1}^2-m_{\overline{K}^{0}_1}^2)^2}{4m_{K^{0}_1}^2\,m_{\overline{K}^{0}_1}^2}\bigg]\,,\label{GSk1k1}
\end{equation}
where
\begin{equation}
|\overrightarrow{k}_1|=\frac{1}{2m_{\chi_{C0}}}[m_{\chi_{C0}}^4+(m^2_{K^{0}_1}-m^2_{\overline{K}^{0}_1})^2-2(m^2_{K^{0}_1}+m^2_{\overline{K}^{0}_1})^2
m_{\chi_{C0}}^2]^{1/2}\,.
\end{equation}
$$\\$$

\textbf{Decay channel $\chi_{C0}\rightarrow \omega\omega$}\\

The corresponding interaction Lagrangian is extracted as

\begin{align}
\mathcal{L}_{\chi_{C0}\omega\omega}=\frac{1}{2}h_1\,\phi_C\,\chi_{C0}\omega_N^\mu\omega_{N\mu}\,,
\end{align}
which also has the same form as the interaction Lagrangian
$\mathcal{L}_{\chi_{C0}\rightarrow\overline{K}^{\ast0}K^{\ast0}}$. Thus one can obtain the decay width of $\chi_{C0}\rightarrow
\omega\omega$ as
\begin{equation}
\Gamma_{\chi_{C0}\rightarrow
\omega\omega}=2\frac{[m_{\chi_{C0}}^{4}-4m_\omega^2m_{\chi_{C0}}^{2}]^{1/2}}{16\pi m_{\chi_{C0}}^{3}}\,\, \frac{1}{12}h_1^2\,\phi_C^2\bigg[2+\frac{(m_{\chi_{C0}}^2-2m_{\omega}^2)^2}{4m_{\omega}^4}\bigg]\,.\label{dw}
\end{equation}
$$\\$$

\textbf{Decay channel $\chi_{C0}\rightarrow \phi\phi$}\\

The corresponding interaction Lagrangian is extracted as
\begin{align}
\mathcal{L}_{\chi_{C0}
\omega\omega}=\frac{1}{2}h_1\,\phi_C\,\chi_{C0}\omega_S^\mu\omega_{S\mu}\,.
\end{align}
Similar to the decay width for $\chi_{C0}\rightarrow
\omega\omega$, the decay width for $\chi_{C0}\rightarrow \phi\phi$
is
\begin{equation}
\Gamma_{\chi_{C0}\rightarrow
\phi\phi}=2\frac{[m_{\chi_{C0}}^{4}-4m_\phi^2m_{\chi_{C0}}^{2}]^{1/2}}{16 \pi m_{\chi_{C0}}^{3}}\,\, \frac{1}{12}h_1^2\,\phi_C^2\bigg[2+\frac{(m_{\chi_{C0}}^2-2m_{\phi}^2)^2}{4m_{\phi}^4}\bigg]\,.\label{dw}
\end{equation}
$$\\$$

\textbf{Decay channel $\chi_{C0}\rightarrow \rho\rho$}\\

The corresponding interaction Lagrangian is extracted as
\begin{align}
\mathcal{L}_{\chi_{C0}\rho\rho}=\frac{1}{2}h_1\,\phi_C\,\chi_{C0}(\rho^{0\mu}\rho^0_\mu+2\rho^{-\mu}\rho^+_\mu)\,,
\end{align}
which also has the same form as
$\mathcal{L}_{\chi_{C0}\omega\omega}$. We thus obtain the decay
width as
\begin{align}
\Gamma_{\chi_{C0}\rightarrow\rho\rho}&=3\Gamma_{\chi_{C0}\rightarrow\rho^0\rho^0}\nonumber\\&
=3\frac{[m_{\chi_{C0}}^{4}-4m_{\rho^0}^2m_{\chi_{C0}}^{2}]^{1/2}}{16\pi m_{\chi_{C0}}^{3}}\times \frac{1}{12}h_1^2\,\phi_C^2\bigg[2+\frac{(m_{\chi_{C0}}^2-2m_{\rho^0}^2)^2}{4m_{\rho^0}^4}\bigg]\,.\label{dw}
\end{align}
$$\\$$

\textbf{Decay channel $\chi_{C0}\rightarrow a_0 a_0$}\\

The corresponding interaction Lagrangian has the form
\begin{align}
\mathcal{L}_{\chi_{C0}a_0a_0}=&-\lambda_1
\,\phi_C\,\chi_{C0}({a_0^0}^2+2a_0^-\,a_0^+)\,.
\end{align}
The decay width of $\chi_{C0}$ into $a_0a_0$ can be obtained as
\begin{align}
\Gamma_{\chi_{C0}\rightarrow
a_0\,a_0}&=3\Gamma_{\chi_{C0}\rightarrow a^0_0\,a^0_0}\nonumber\\&
=3\frac{\lambda_1^2\,\phi_C^2}{8\pi\,m_{\chi_{C0}}^2}\,\,\frac{[m_{\chi_{C0}}^4-4m_{a_0^0}^2\,m_{\chi_{C0}}^2]^{1/2}}{2m_{\chi_{C0}}}\,.
\end{align}
$$\\$$

\textbf{Decay channel
$\chi_{C0}\rightarrow K_1^+K^-$ }\\

The corresponding interaction Lagrangian from the Lagrangian
(\ref{lag}) reads
\begin{align}
\mathcal{L}_{\chi_{C0}K_1K}=
Z_{K}\,w_{K_1}\,h_1\,\phi_C\,\chi_{C0}({K}^{0\mu}_{1}\partial_\mu\overline{K}^{0}+\partial_\mu
K^+ K_1^\mu +K_1^{+\mu}\partial_\mu
K^-+\overline{K}_1^{0\mu}\partial_\mu K^0)\,.
\end{align}
Let us consider only the decay channel $\chi_{C0}\rightarrow
K_1^+K^-$; the decay channels $\chi_{C0}\rightarrow
K^{0}_{1}\overline{K}^{0},$ $\overline{K}^0_1K^0,\,K_1^-K^+$ will
give the same contribution as a result of the isospin symmetry,
so consider only
\begin{align}
\mathcal{L}_{\chi_{C0}K_1^+K^-}=Z_{K}\,w_{K_1}\,h_1\,\phi_C\,\chi_{C0}K_1^{+\mu}\partial_\mu
K^-\,.
\end{align}
We denote the momenta of $K^-$ and $K_1^+$ as $P_1$ and $P_2$,
respectively. The energy-momentum conservation on the vertex implies
$P=P_1+P_2$, where $P$ denotes the momentum of the decaying
particle $\chi_{C0}$. Given that our particles are on-shell, we
obtain
\begin{equation}
P_1 \cdot P_2=\frac{P^2-P_1^2-P_2^2}{2}=\frac{m_{\chi_{C0}}^2-m_{K^-}-m_{K_1^+}}{2}\,.
\end{equation}

Note that the scattering amplitude depends on
the polarisation vector $\varepsilon_\mu^{(\alpha)}(P_2)$, Upon substituting $\partial_\mu\rightarrow +iP^\mu$ for the
outgoing particles, one obtains

\begin{equation}
-i\mathcal{M}^{(\alpha)}_{\chi_{C0}\rightarrow
K_1^+K^-}=iA_{\chi_{c0}K_1K}\,iP_1^\mu\varepsilon_\mu^{(\alpha)}(P_2)\,,
\end{equation}
The average modulus squared amplitude reads
\begin{equation}
\left\vert \overline{-i\mathcal{M}_{\chi_{C0}\rightarrow
K_1^+K^-}}\right\vert^{2}=\frac{1}{3}A^2_{\chi_{c0}K_1K}\bigg[-P_1^2+\frac{(P_1\cdot P_2)^2}{m_{K_1^+}^2}\bigg]\,,
\end{equation}
where
\begin{equation}
A_{\chi_{c0}K_1K}=Z_{K}\,w_{K_1}\,h_1\,\phi_C\,,
\end{equation}
and
$$P_1^2=m_{K^-}^2\,.$$
$$\\$$

Then, the decay width $\Gamma_{\chi_{C0}\rightarrow K_1^+K^-}$ is

\begin{equation}
\Gamma_{\chi_{C0}\rightarrow
K_1^+K^-}=\frac{|\overrightarrow{k}_1|}{8\pi\,m_{\chi_{C0}}^2}\left\vert
\overline{-i\mathcal{M}_{\chi_{C0}\rightarrow
K_1^+K^-}}\right\vert^{2}\,,
\end{equation}
where
\begin{equation}
|\overrightarrow{k}_1|=\frac{1}{2m_{\chi_{C0}}}\bigg[m_{\chi_{C0}}^4+(m_{K^-}^2-m_{K_1^+}^2)^2-2(m_{K^-}^2+m_{K_1^+}^2)m_{\chi_{C0}}^2\bigg]^{1/2}\,.
\end{equation}

Similarly, one can obtain the decay width of $\chi_{C0}\rightarrow
K^\ast K^\ast_0$ as follows.\\

\textbf{Decay channel $\chi_{C0}\rightarrow K^{\ast}
\overline{K}^{\ast}_{0} $}\\

From the interaction Lagrangian
\begin{align}
\mathcal{L}_{\chi_{C0}K^{\ast0}
\overline{K}^{\ast0}_{0}}=Z_{K_0^\ast}\,w_{K^\ast}\,h_1\,\phi_C\,\chi_{C0}\overline{K}^{\ast0\mu}_{0}\partial_\mu
K^{\ast0}_0\,,
\end{align}
which obtain from the corresponding interaction Lagrangian
\begin{align}
\mathcal{L}_{\chi_{C0}K^{\ast}
\overline{K}^{\ast}_{0}}=Z_{K_0^\ast}\,w_{K^\ast}\,h_1\,\phi_C\,\chi_{C0}
(\overline{K}^{\ast0\mu}_{0}\partial_\mu K^{\ast0}_0-\partial_\mu
K^{\ast-}_0 K^{\ast+\mu} +\partial_\mu K^{\ast+}_0
K^{\ast-\mu}-\partial_\mu\overline{K}^{\ast0\mu}_{0}
K^{\ast0\mu})\,.
\end{align}
We compute the decay width as
\begin{equation}
\Gamma_{\chi_{C0}\rightarrow K^{\ast0}
\overline{K}^{\ast0}_{0}}=\frac{|\overrightarrow{k}_1|}{8\pi\,m_{\chi_{C0}}^2}w_{K^\ast}^2\,Z_{K^\ast}h_1^2\phi_C^2\left\vert-m_{K^{\ast0}}^2+\frac{(m_{\chi_{C0}}^2-m_{K^{\ast0}}^2-m^2_{\overline{K}^{\ast0}_{0}})^2}{4m^2_{\overline{K}^{\ast0}_{0}}}
\right\vert\,,
\end{equation}
where
\begin{equation}
|\overrightarrow{k}_1|=\frac{1}{2m_{\chi_{C0}}}\bigg[m_{\chi_{C0}}^4+(m_{K^{\ast0}}^2-m_{\overline{K}^{\ast0}_{0}}^2)^2-2(m_{K^{\ast0}}^2+m_{\overline{K}^{\ast0}_{0}}^2)m_{\chi_{C0}}^2\bigg]^{1/2}\,.
\end{equation}

Note that we considered only the decay channel
$\chi_{C0}K^{\ast0} \overline{K}^{\ast0}_{0}$ because the other
decay channels contribute the same of isospin symmetry
reasons. Thus,
\begin{align}
\Gamma_{\chi_{C0}\rightarrow K^{\ast}
\overline{K}^{\ast}_{0}}=\Gamma_{\chi_{C0}\rightarrow K^{\ast0}
\overline{K}^{\ast0}_{0}}+\Gamma_{\chi_{C0}\rightarrow K^{\ast+}
K^{\ast-}_{0}}+\Gamma_{\chi_{C0}\rightarrow K^{\ast-}
K^{\ast+}_{0}}+\Gamma_{\chi_{C0}\rightarrow K^{\ast0}_0
\overline{K}^{\ast0}}\,.
\end{align}

\textbf{Decay channels $\chi_{C0}\rightarrow \eta,\,\eta'$}\\

The corresponding interaction Lagrangian of $\chi_{C0}$ with the
$\eta'$ and the $\eta$ resonances reads

\begin{align}
\mathcal{L}_{\chi_{C0}\eta_N\eta_N,\eta_S\eta_S,\eta_N\eta_S}=&
(-\lambda_1-\frac{1}{2}c\phi_N^2\phi_S^2)Z^2_{\eta_N}\,\phi_C\,\chi_{C0}\eta_N^2+\frac{1}{2}h_1\,w_{f_{1N}}^2\,Z_{\eta_N}^2\chi_{C0}\,\partial_\mu\eta_N\partial^\mu\eta_N\nonumber\\&
+
(-\lambda_1-\frac{1}{8}c\phi_N^4)Z^2_{\eta_S}\,\phi_C\,\chi_{C0}\eta_S^2+\frac{1}{2}h_1\,w_{f_{1S}}^2Z_{\eta_S}^2\chi_{C0}\partial_\mu\eta_S\partial^\mu\eta_S\nonumber\\&
-\frac{1}{2}\phi_N^3\phi_C\phi_S\,Z_{\eta_N}\,Z_{\eta_S}\,\eta_N\eta_S\,.\label{chietas}
\end{align}
Using Eq.(4.101) and Eq.(4.102), the interaction Lagrangian
(\ref{chietas}) will transform to a Lagrangian which describes the interaction of $\chi_{C0}$
with $\eta$ and $\eta'$,
\begin{align}
\mathcal{L}_{\chi_{C0}\eta^2,\eta'^2,\eta\eta'}&=
[-\lambda_1(Z^2_{\eta_N}\cos^2\varphi_\eta+Z^2_{\eta_S}\sin^2\varphi_\eta)-\frac{1}{2}c\phi_N^2(\phi_S^2\,Z^2_{\eta_N}\,\cos^2\varphi_\eta\nonumber\\
&+\frac{1}{4}\phi_N^2Z^2_{\eta_S}\sin^2\varphi_\eta+\phi_N\phi_S\,Z_{\eta_N}Z_{\eta_S}
\sin\varphi_\eta \cos\varphi_\eta)]\phi_C\chi_{C0}\eta^2\nonumber\\
&+\bigg[\frac{1}{2}h_1\phi_C(w^2_{f_{1N}}Z_{\eta_N}^2\cos^2\varphi_\eta+w^2_{f_{1S}}Z_{\eta_S}^2\sin^2\varphi_\eta)\bigg]\chi_{C0}\partial_\mu
\eta\partial^\mu\eta\nonumber\\
&[-\lambda_1(Z^2_{\eta_N}\sin^2\varphi_\eta+Z^2_{\eta_S}\cos^2\varphi_\eta)-\frac{1}{2}c\phi_N^2(\phi_S^2\,Z^2_{\eta_N}\,\sin^2\varphi_\eta\nonumber\\
&+\frac{1}{4}\phi_N^2Z^2_{\eta_S}\cos^2\varphi_\eta-\phi_N\phi_S\,Z_{\eta_N}Z_{\eta_S}
\sin\varphi_\eta \cos\varphi_\eta)]\phi_C\chi_{C0}\eta'^2\nonumber\\
&+\bigg[\frac{1}{2}h_1\phi_C(w^2_{f_{1N}}Z_{\eta_N}^2\sin^2\varphi_\eta+w^2_{f_{1S}}Z_{\eta_S}^2\cos^2\varphi_\eta)\bigg]\chi_{C0}\partial_\mu
\eta'\partial^\mu\eta'\nonumber\\
&+[-2\lambda_1(-Z^2_{\eta_N}+Z^2_{\eta_S})\sin\varphi_\eta
\cos\varphi_\eta)\nonumber\\
&+\frac{1}{2}c\phi_N^2(2\phi_S^2\,Z^2_{\eta_N}-\frac{1}{2}\phi_N^2\,Z_{\eta_S}^2)\cos\varphi_\eta
\sin\varphi_\eta)\nonumber\\&-
 \frac{1}{2}c\phi_N^3\phi_S\,Z_{\eta_N}Z_{\eta_S}(\cos^2\varphi_\eta-
\sin^2\varphi_\eta \cos\varphi_\eta)] \phi_C\chi_{C0}\eta\eta'\nonumber\\
&+h_1\phi_C\,\cos\varphi_\eta
\sin\varphi_\eta(w^2_{f_{1S}}Z_{\eta_S}^2-w^2_{f_{1N}}Z_{\eta_N}^2)\chi_{C0}\partial_\mu
\eta\partial^\mu\eta'\,,
\end{align}
which contains three different decay channels,
$\chi_{C0}\rightarrow\eta\eta$, $\chi_{C0}\rightarrow\eta'\eta'$,
and $\chi_{C0}\rightarrow\eta\eta'$, with the following vertices
\begin{align}
A_{\chi_{C0}\eta\eta}=&-\lambda_1\phi_C(Z^2_{\eta_N}\cos^2\varphi_\eta+Z^2_{\eta_S}\sin^2\varphi_\eta)\nonumber\\
&-\frac{1}{2}c\phi_N^2\phi_C(\phi_S^2\,Z^2_{\eta_N}\,\cos^2\varphi_\eta+\frac{1}{4}\phi_N^2Z^2_{\eta_S}\sin^2\varphi_\eta+\phi_N\phi_S\,Z_{\eta_N}Z_{\eta_S}
\sin\varphi_\eta \cos\varphi_\eta)\,,\\
B_{\chi_{C0}\eta\eta}&=\frac{1}{2}h_1\phi_C(w^2_{f_{1N}}Z_{\eta_N}^2\cos^2\varphi_\eta+w^2_{f_{1S}}Z_{\eta_S}^2\sin^2\varphi_\eta)\,,
\end{align}
\begin{align}
A_{\chi_{C0}\eta'\eta'}=&-\lambda_1\phi_C(Z^2_{\eta_N}\sin^2\varphi_\eta+Z^2_{\eta_S}\cos^2\varphi_\eta)\nonumber\\
&-\frac{1}{2}c\phi_N^2\phi_C(\phi_S^2\,Z^2_{\eta_N}\,\sin^2\varphi_\eta+\frac{1}{4}\phi_N^2Z^2_{\eta_S}\cos^2\varphi_\eta-\phi_N\phi_S\,Z_{\eta_N}Z_{\eta_S}
\sin\varphi_\eta \cos\varphi_\eta)\,,\\
B_{\chi_{C0}\eta'\eta'}&=\frac{1}{2}h_1\phi_C(w^2_{f_{1N}}Z_{\eta_N}^2\sin^2\varphi_\eta+w^2_{f_{1S}}Z_{\eta_S}^2\cos^2\varphi_\eta)\,,
\end{align}
\begin{align}
A_{\chi_{C0}\eta\eta'}=&-2\lambda_1\phi_C(-Z^2_{\eta_N}+Z^2_{\eta_S})\sin\varphi_\eta
\cos\varphi_\eta+\frac{1}{2}c\phi_N^2\phi_C(2\phi_S^2\,Z^2_{\eta_N}-\frac{1}{2}\phi_N^2\,Z_{\eta_S}^2)\cos\varphi_\eta
\sin\varphi_\eta\nonumber\\&-
 \frac{1}{2}c\phi_N^3\phi_S\phi_C\,Z_{\eta_N}Z_{\eta_S}(\cos^2\varphi_\eta-
\sin^2\varphi_\eta \cos\varphi_\eta)\,,\\
B_{\chi_{C0}\eta\eta'}&=h_1\phi_C\,\cos\varphi_\eta
\sin\varphi_\eta(w^2_{f_{1S}}Z_{\eta_S}^2-w^2_{f_{1N}}Z_{\eta_N}^2)\,.
\end{align}
Let us firstly consider the channel
$\chi_{C0}\rightarrow\eta\eta$. We denote the momenta of the two
outgoing $\eta$ particles as $P_1$ and $P_2$, and $P$ denotes
the momentum of the decaying $\chi_{C0}$ particle. Given that our
particles are on shell, we obtain
\begin{equation}
P_1\cdot P_2=\frac{P^2-P_1^2-P_2^2}{2}=\frac{m_{\chi_{C0}}^2-2m_{\eta}^2}{2}\,.
\end{equation}
After replacing $\partial_\mu\rightarrow+iP^\mu$ for the
outgoing particles, one obtains the decay amplitude as
\begin{equation}
-iM_{\chi_{C0}\rightarrow\eta\eta}=i\bigg[A_{\chi_{C0}\eta\eta}-B_{\chi_{C0}\eta\eta}\frac{m_{\chi_{C0}}^2-2m_{\eta}^2}{2}\bigg]\,.
\end{equation}
Then the decay width is
\begin{equation}
\Gamma_{\chi_{C0}\rightarrow
\eta\eta}=2\frac{|\overrightarrow{k}_1|}{8\pi\,m_{\chi_{C0}}^2}\left\vert\overline{-i\mathcal{M}_{\chi_{C0}\rightarrow\eta\eta}}\right\vert^{2}\,,
\end{equation}
where
\begin{equation}
|\overrightarrow{k}_1|=\frac{1}{2m_{\chi_{C0}}}\bigg[m_{\chi_{C0}}^4-4m_{\eta}^2m_{\chi_{C0}}^2\bigg]^{1/2}\,.
\end{equation}
Similarly, the decay width of $\chi_{C0}$ into $\eta'\eta'$ is obtained
as
\begin{equation}
\Gamma_{\chi_{C0}\rightarrow
\eta'\eta'}=2\frac{|\overrightarrow{k}_1|}{8\pi\,m_{\chi_{C0}}^2}\left\vert
A_{\chi_{C0}\eta'\eta'}-B_{\chi_{C0}\eta'\eta'}\frac{m_{\chi_{C0}}^2-2m_{\eta'}^2}{2}\right\vert^{2}\,,
\end{equation}
where
\begin{equation}
|\overrightarrow{k}_1|=\frac{1}{2m_{\chi_{C0}}}\bigg[m_{\chi_{C0}}^4-4m_{\eta'}^2m_{\chi_{C0}}^2\bigg]^{1/2}\,.
\end{equation}
In a similar way, the decay width of $\chi_{C0}$ into
$\eta\eta'$ can be obtained as
\begin{equation}
\Gamma_{\chi_{C0}\rightarrow
\eta\eta'}=\frac{|\overrightarrow{k}_1|}{8\pi\,m_{\chi_{C0}}^2}\left\vert
A_{\chi_{C0}\eta\eta'}-B_{\chi_{C0}\eta\eta'}\frac{m_{\chi_{C0}}^2-m_\eta^2-m_{\eta'}^2}{2}\right\vert^{2}\,,
\end{equation}
where
\begin{equation}
|\overrightarrow{k}_1|=\frac{1}{2m_{\chi_{C0}}}\bigg[m_{\chi_{C0}}^4+(m_\eta^2+m_{\eta'}^2)^2-2(m_\eta^2+m_{\eta'}^2)m_{\chi_{C0}}^2\bigg]^{1/2}\,.
\end{equation}
$$\\$$

\textbf{Decay channels $\chi_{C0}\rightarrow f_0f_0$}\\

The corresponding interaction Lagrangian is extracted from the
Lagrangian (\ref{lag}) 

\begin{align}
\mathcal{L}_{\chi_{C0}
f_0'}=-\lambda_1\,\phi_C\,\chi_{C0}(\sigma_N^2+\sigma_S^2)-\frac{m_0^2}{G_0^2}\phi_C\chi_{C0}G^2\,.
\end{align}
Using the mixing matrix (\ref{scalmixmat}), the interaction Lagrangian
becomes
\begin{align}
\mathcal{L}_{\chi_{C0}
f_0f_0}=&-\bigg[0.9125\lambda_1+0.0841\,\frac{m_0^2}{G_0^2}\bigg]\phi_C\chi_{C0}f_0(1370)^2\nonumber\\
&-\bigg[0.985\lambda_1+0.0144\,\frac{m_0^2}{G_0^2}\bigg]\phi_C\chi_{C0}f_0(1500)^2\nonumber\\
&-\bigg[0.065\lambda_1-0.0696\,\frac{m_0^2}{G_0^2}\bigg]\phi_C\chi_{C0}f_0(1370)f_0(1500)\nonumber\\
&-\bigg[0.55\lambda_1-0.551\,\frac{m_0^2}{G_0^2}\bigg]\phi_C\chi_{C0}f_0(1370)f_0(1710)\nonumber\\
&-\bigg[0.24\lambda_1-0.228\,\frac{m_0^2}{G_0^2}\bigg]\phi_C\chi_{C0}f_0(1500)f_0(1710)\,.\label{f0decay}
\end{align}

Therefore, we obtain the decay widths for all channels
represented in the interaction Lagrangian (\ref{f0decay}) as
follows:
$$\\$$

\textbf{Decay channels $\chi_{C0}\rightarrow f_0(1370)f_0(1370)$}\\
\begin{equation}
\Gamma_{\chi_{C0}\rightarrow
f_0(1370)^2}=2\frac{[m_{\chi_{C0}}^2-4m^2_{f_0(1370)}m^2_{\chi_{C0}}]^{1/2}}{2\times8\pi
m_{\chi_{C0}}^{3}}\,\bigg|-0.9125\lambda_1\phi_C-0.0841\,\frac{m_0^2}{G_0^2}\phi_C\bigg|^{2}\,,\label{df1}
\end{equation}
$$\\$$
$$\\$$
\textbf{Decay channels $\chi_{C0}\rightarrow f_0(1500)f_0(1500)$}\\

\begin{equation}
\Gamma_{\chi_{C0}\rightarrow
f_0(1500)^2}=2\frac{[m_{\chi_{C0}}^2-4m^2_{f_0(1500)}m^2_{\chi_{C0}}]^{1/2}}{2\times8\pi
m_{\chi_{C0}}^{3}}\,\bigg|-0.985\lambda_1\phi_C-0.0144\,\frac{m_0^2}{G_0^2}\phi_C\bigg|^{2}\,,\label{df2}
\end{equation}
$$\\$$
$$\\$$
\textbf{Decay channels $\chi_{C0}\rightarrow f_0(1370)f_0(1500)$}\\
\begin{align}
\Gamma_{\chi_{C0}\rightarrow f_0(1370)f_0(1500)}=&\frac{1}{8\pi
m_{\chi_{C0}}^{2}}\bigg|-0.065\lambda_1\phi_C+0.0696\,\frac{m_0^2}{G_0^2}\phi_C\bigg|^{2}\,\nonumber\\
&\times\frac{[m_{\chi_{C0}}^2+(m^2_{f_0(1370)}-m^2_{f_0(1500)})^2-2(m^2_{f_0(1370)}+m^2_{f_0(1500)})m^2_{\chi_{C0}}]^{1/2}}{2
m_{\chi_{C0}}}\,\,,\label{df3}
\end{align}
$$\\$$

\textbf{Decay channels $\chi_{C0}\rightarrow f_0(1370)f_0(1710)$}\\
\begin{align}
\Gamma_{\chi_{C0}\rightarrow f_0(1370)f_0(1710)}=&\frac{1}{8\pi
m_{\chi_{C0}}^{2}}\bigg|-0.55\lambda_1\phi_C+0.551\,\frac{m_0^2}{G_0^2}\phi_C\bigg|^{2}\,\nonumber\\
&\times\frac{[m_{\chi_{C0}}^2+(m^2_{f_0(1370)}-m^2_{f_0(1710)})^2-2(m^2_{f_0(1370)}+m^2_{f_0(1710)})m^2_{\chi_{C0}}]^{1/2}}{2
m_{\chi_{C0}}}\,\,,\label{df3}
\end{align}
$$\\$$

\textbf{Decay channels $\chi_{C0}\rightarrow f_0(1500)f_0(1710)$}\\
\begin{align}
\Gamma_{\chi_{C0}\rightarrow f_0(1500)f_0(1710)}=&\frac{1}{8\pi
m_{\chi_{C0}}^{2}}\bigg|-0.228\lambda_1\phi_C-0.24\,\frac{m_0^2}{G_0^2}\phi_C\bigg|^{2}\,\nonumber\\
&\times\frac{[m_{\chi_{C0}}^2+(m^2_{f_0(1500)}-m^2_{f_0(1710)})^2-2(m^2_{f_0(1500)}+m^2_{f_0(1710)})m^2_{\chi_{C0}}]^{1/2}}{2
m_{\chi_{C0}}}\,.\label{df3}
\end{align}

\section{Three-body decay rates for $\chi_{C0}$}

The general formula for the three-body decay
width for $\chi_{C0}$, which is proved in chapter 5, takes the following form

\begin{align}
\Gamma&_{\chi_{C0}\rightarrow B_{1}B_{2}B_{3}}=\frac{S}{32(2\pi)^{3}m_{\chi_{C0}}^{3}}\int_{(m_{1}+m_{2})^{2}%
}^{(m_{\chi_{C0}}-m_{3})^{2}}|-i\mathcal{M}_{A\rightarrow B_{1}B_{2}B_{3}}%
|^{2}dm_{12}^{2}\nonumber\\&
\sqrt{\frac{(-m_{1}+m_{12}-m_{2})(m_{1}+m_{12}-m_{2})(-m_{1}%
+m_{12}+m_{2})(m_{1}+m_{12}+m_{2})}{m_{12}^{2}}}\nonumber\\&
\sqrt{\frac{(-m_{\eta_C}+m_{12}-m_{3})(m_{\chi_{C0}}+m_{12}-m_{3}%
)(-m_{\chi{C0}}+m_{12}+m_{3})(m_{\chi_{C0}}+m_{12}+m_{3})}{m_{12}^{2}}%
}\text{ ,}\label{3chidec}%
\end{align}

The quantities $m_{1},$ $m_{2},$ $m_{3}$ refer to the masses of the three outgoing particles ($P_{1},$ $P_{2}$, and $P_{3})$, which are (pseudo)scalar mesons in the present case,
 $\mathcal{M}_{\eta_C\rightarrow P_{1}P_{2}P_{3}}$ is the corresponding tree-level decay
amplitude, and $S$ is a symmetrization factor (it equals $1$ if all $P_{1},$ $P_{2}$, and $P_{3}$ are different, it
equals $2$ for two identical particles in the final state, and it equals $6$
for three identical particles in the final state)

Now let us list the corresponding tree-level decay amplitudes for $\chi_{C0}$ which are presented in Tables 8.3 and 8.4 as follows:\\
$$\\$$
\textbf{Decay channel $\chi_{C0}\rightarrow K^\ast_0K\eta,\eta'$}\\

The corresponding interaction Lagrangian can be obtained from the Lagrangian (\ref{lag}) as
\begin{align}
\mathcal{L}_{\chi_{C0}K^\ast_0K\eta_N,\eta_S}=&\frac{1}{\sqrt{2}}cZ_K\,Z_{K^\ast_0}\,Z_{\eta_N}\phi_N^2\phi_S\phi_C\chi_{C0}\eta_N(K^{\ast0}_0\overline{K}^0
+\overline{K}^{\ast0}_0 K^0+K^{\ast-}_0K^+
+K^{\ast+}_0K^-)\nonumber\\&+\frac{\sqrt{2}}{4}cZ_K\,Z_{K^\ast_0}\,Z_{\eta_S}\phi_N^3\phi_C\chi_{C0}\eta_S(K^{\ast0}_0\overline{K}^0
+\overline{K}^{\ast0}_0 K^0+K^{\ast-}_0K^+ +K^{\ast+}_0K^-)\,.\label{chiKetKs}
\end{align}
Using Eqs. (4.101, 4.102), the interaction Lagrangian (\ref{chiKetKs}) can be written as
\begin{align}
\mathcal{L}_{\chi_{C0}K^\ast_0K\eta,\eta'}=\frac{1}{\sqrt{2}}c\phi_C\phi_N^2 Z_K Z_{K^\ast_0}\chi_{co}\bigg[&(\phi_S Z_{\eta_N} \cos\varphi_\eta+\frac{1}{2}\phi_N Z_{\eta_S} \sin\varphi_\eta)\eta\nonumber\\&
-(\phi_S Z_{\eta_N} \sin\varphi_\eta-\frac{1}{2}\phi_N Z_{\eta_S} \cos\varphi_\eta)\eta'\bigg]\nonumber\\&
\times\big[K^{\ast0}_0\overline{K}^0
+\overline{K}^{\ast0}_0 K^0+K^{\ast-}_0K^+ +K^{\ast+}_0K^-\big]\,.\label{chiKetKs1}
\end{align}
Consequently, the amplitude decay for the decay channels $\chi_{C0} \rightarrow K^\ast_0K\eta$ and $\chi_{C0}\rightarrow K^\ast_0K\eta'$ can be obtain as
\begin{equation}
-i\mathcal{M}_{\chi_{C0}\rightarrow K^\ast_0K\eta}=-i\frac{1}{\sqrt{2}}c\phi_C\phi_N^2 Z_K Z_{K^\ast_0}(\phi_S Z_{\eta_N} \cos\varphi_\eta+\frac{1}{2}\phi_N Z_{\eta_S} \sin\varphi_\eta)\,,
\end{equation}
and 
\begin{equation}
-i\mathcal{M}_{\chi_{C0}\rightarrow K^\ast_0K\eta'}=\frac{1}{\sqrt{2}}c\phi_C\phi_N^2 Z_K Z_{K^\ast_0}(-\phi_S Z_{\eta_N} \sin\varphi_\eta+\frac{1}{2}\phi_N Z_{\eta_S} \cos\varphi_\eta)\,,
\end{equation}
which are used to compute $\Gamma_{\chi_{C0}\rightarrow K^\ast_0K\eta}$ and  $\Gamma_{\chi_{C0}\rightarrow K^\ast_0K\eta'}$ by Eq.(\ref{3chidec}).\\

\textbf{Decay channel $\chi_{C0}\rightarrow f_0\eta,\eta'$}\\

The corresponding interaction Lagrangian is extracted from the Lagrangian (\ref{lag}) and given by
\begin{align}
\mathcal{L}_{\chi_{C0}\sigma_{N,S}\eta_{NS}}=&-\frac{3}{2}c\,Z_{\eta_N}Z_{\eta_S}\phi_N^2\phi_S\phi_C\chi_{C0}\sigma_N\eta_N\eta_S-c\,Z^2_{\eta_N}\phi_N^2\phi_S\phi_C\chi_{C0}\sigma_S\eta_N^2\nonumber\\&-c\,Z^2_{\eta_N}\phi_N^2\phi^2_S\phi_C\chi_{C0}\sigma_N\eta^2_N-\frac{1}{2}c\,Z^2_{\eta_S}\phi_N^3\phi_C\chi_{C0}\sigma_N\eta^2_S\nonumber\\&-\frac{1}{2}cZ_{\eta_N},Z_{\eta_S}\phi_N^3\phi_C\chi_{C0}\sigma_S\eta_S\eta_N\,.\label{chif0eta}
\end{align}
 Substituting Eqs.(4.101, 4.102, 4.113, 4.114) and the relations (8.11) and (8.12), we get the decay amplitudes for several channels, which are used in Eq.(\ref{3chidec}) to compute the decay widths, as follows:\\
$$\\$$ 
$$\\$$ 

\textbf{Decay channel $\chi_{C0}\rightarrow  f_0(1370)\eta\eta$}\\

\begin{align}
\mathcal{L}_{\chi_{C0}f_0(1370)\eta\eta}=& c\phi_N\phi_C\,\big[(0.085 \phi_N-1.41\phi_S)\cos\varphi_\eta \sin\varphi_\eta Z_{\eta_N}Z_{\eta_S}\phi_N\nonumber\\& -0.47 \sin^2\varphi_\eta Z^2_{\eta_S}\phi_N^2
+(0.17\phi_N-0.94\phi_S) \phi_S  \cos^2\varphi_\eta  Z^2_{\eta_N}\big]\chi_{C0}\eta^2 f_0(1370)\,.\label{chif0eta1}
\end{align}
Thus, the decay amplitude for the decay width $\Gamma_{\chi_{C0}\rightarrow f_0(1370)\eta\eta}$ reads
\begin{align}
-i\mathcal{M}_{\chi_{C0}\rightarrow f_0(1370)\eta\eta}=&c\phi_N\phi_C\,\big[(0.085 \phi_N-1.41\phi_S)\cos\varphi_\eta \sin\varphi_\eta Z_{\eta_N}Z_{\eta_S}\phi_N\nonumber\\& -0.47 \sin^2\varphi_\eta Z^2_{\eta_S}\phi_N^2
+(0.17\phi_N-0.94\phi_S) \phi_S  \cos^2\varphi_\eta  Z^2_{\eta_N}\big]\,.
\end{align}
$$\\$$ 
 \textbf{Decay channel $\chi_{C0}\rightarrow  f_0(1370)\eta'\eta'$}\\
 \begin{align}
\mathcal{L}_{\chi_{C0}f_0(1370)\eta'\eta'}=& c\phi_N\phi_C\,\big[(-0.085 \phi_N+1.41\phi_S)\cos\varphi_\eta \sin\varphi_\eta Z_{\eta_N}Z_{\eta_S}\phi_N\nonumber\\& -0.47 \cos^2\varphi_\eta Z^2_{\eta_S}\phi_N^2
+(0.17\phi_N-0.94\phi_S) \phi_S  \sin^2\varphi_\eta  Z^2_{\eta_N}\big]\chi_{C0}\,\eta'^2 f_0(1370)\,.\label{chif0eta2}
\end{align}
Thus, the decay amplitude for the decay width $\Gamma_{\chi_{C0}\rightarrow f_0(1370)\eta'\eta'}$ reads
\begin{align}
-i\mathcal{M}_{\chi_{C0}\rightarrow f_0(1370)\eta'\eta'}=&c\phi_N\phi_C\,\big[(-0.085 \phi_N+1.41\phi_S)cos\varphi_\eta \sin\varphi_\eta Z_{\eta_N}Z_{\eta_S}\phi_N\nonumber\\& -0.47 \cos^2\varphi_\eta Z^2_{\eta_S}\phi_N^2
+(0.17\phi_N-0.94\phi_S) \phi_S  \sin^2\varphi_\eta  Z^2_{\eta_N}\big]\,.
\end{align}

$$\\$$ 
 \textbf{Decay channel $\chi_{C0}\rightarrow  f_0(1370)\eta\eta'$}\\

 \begin{align}
\mathcal{L}_{\chi_{C0}f_0(1370)\eta\eta'}=& c\phi_N\phi_C\,\big[(0.085 \phi_N-1.41\phi_S)\cos^2\varphi_\eta Z_{\eta_N}Z_{\eta_S}\phi_N\nonumber\\& +(-0.085 \phi_N+1.41\phi_S) \sin^2\varphi_\eta Z_{\eta_N}Z_{\eta_S}\phi_N\nonumber\\&
+{-0.94 Z^2_{\eta_S}\phi^2_N+(-0.34\phi_N+1.88\phi_S )\phi_S Z^2_{\eta_N}}\sin\varphi_\eta \cos\varphi_\eta \big]\chi_{C0}\eta\eta' f_0(1370)\,.\label{chif0eta3}
\end{align}
Thus, the decay amplitude for the decay width $\Gamma_{\chi_{C0}\rightarrow f_0(1370)\eta\eta'}$ reads
\begin{align}
-i\mathcal{M}_{\chi_{C0}\rightarrow f_0(1370)\eta\eta'}=& c\phi_N\phi_C\,\big[(0.085 \phi_N-1.41\phi_S)\cos^2\varphi_\eta Z_{\eta_N}Z_{\eta_S}\phi_N\nonumber\\& +(-0.085 \phi_N+1.41\phi_S) \sin^2\varphi_\eta Z_{\eta_N}Z_{\eta_S}\phi_N\nonumber\\&
+{-0.94 Z^2_{\eta_S}\phi^2_N+(-0.34\phi_N+1.88\phi_S )\phi_S Z^2_{\eta_N}}\sin\varphi_\eta \cos\varphi_\eta \big]\,.
\end{align}
 $$\\$$ 
 \textbf{Decay channel $\chi_{C0}\rightarrow  f_0(1500)\eta\eta$}\\
 
 \begin{align}
\mathcal{L}_{\chi_{C0}f_0(1500)\eta\eta}=& c\phi_N\phi_C\,\big[(-0.485\phi_N-0.3151\phi_S)\cos\varphi_\eta \sin\varphi_\eta Z_{\eta_N}Z_{\eta_S}\phi_N\nonumber\\& -0.105 \sin^2\varphi_\eta Z^2_{\eta_S}\phi_N^2
+(-0.97\phi_N-0.21\phi_S) \phi_S \cos^2\varphi_\eta  Z^2_{\eta_N}\big]\chi_{C0}\eta^2 f_0(1500)\,.\label{chif0eta4}
\end{align}
Thus, the decay amplitude for the decay width $\Gamma_{\chi_{C0}\rightarrow f_0(1500)\eta\eta}$ reads
\begin{align}
-i\mathcal{M}_{\chi_{C0}\rightarrow f_0(1500)\eta\eta}=&c\phi_N\phi_C\,\big[(-0.485\phi_N-0.3151.41\phi_S)\cos\varphi_\eta \sin\varphi_\eta Z_{\eta_N}Z_{\eta_S}\phi_N\nonumber\\& -0.105 \sin^2\varphi_\eta Z^2_{\eta_S}\phi_N^2
+(-0.97\phi_N-0.21\phi_S) \phi_S \cos^2\varphi_\eta  Z^2_{\eta_N}\big]\,.
\end{align}

 $$\\$$ 
 \textbf{Decay channel $\chi_{C0}\rightarrow  f_0(1500)\eta\eta'$}\\

\begin{align}
\mathcal{L}_{\chi_{C0}f_0(1500)\eta\eta'}=& c\phi_N\phi_C\,\big[(-0.485\phi_N-0.3151\phi_S)\cos^2\varphi_\eta Z_{\eta_N}Z_{\eta_S}\phi_N\nonumber\\& +(0.48 \phi_N+0.315\phi_S) \sin^2\varphi_\eta Z_{\eta_N}Z_{\eta_S}\phi_N\nonumber\\&
+{-0.21 Z^2_{\eta_S}\phi^2_N+(1.94\phi_N+0.42\phi_S )\phi_S Z^2_{\eta_N}}\sin\varphi_\eta \cos\varphi_\eta \big]\chi_{C0}\eta\eta' f_0(1500)\,.\label{chif0eta5}
\end{align}
Thus, the decay amplitude for the decay width $\Gamma_{\chi_{C0}\rightarrow f_0(1500)\eta\eta'}$ reads
\begin{align}
-i\mathcal{M}_{\chi_{C0}\rightarrow f_0(1500)\eta\eta'}=&c\phi_N\phi_C\,\big[(-0.485\phi_N-0.3151.41\phi_S)\cos^2\varphi_\eta Z_{\eta_N}Z_{\eta_S}\phi_N\nonumber\\& +(0.48 \phi_N+0.315\phi_S) \sin^2\varphi_\eta Z_{\eta_N}Z_{\eta_S}\phi_N\nonumber\\&
+{-0.21 Z^2_{\eta_S}\phi^2_N+(1.94\phi_N+0.42\phi_S )\phi_S Z^2_{\eta_N}}\sin\varphi_\eta \cos\varphi_\eta \big]\,.
\end{align}

 $$\\$$ 
 \textbf{Decay channel $\chi_{C0}\rightarrow  f_0(1710)\eta\eta$}\\
 
 \begin{align}
\mathcal{L}_{\chi_{C0}f_0(1710)\eta\eta}=& c\phi_N\phi_C\,\big[(-0.09\phi_N-0.39\phi_S)\cos\varphi_\eta \sin\varphi_\eta Z_{\eta_N}Z_{\eta_S}\phi_N\nonumber\\& +0.13 \sin^2\varphi_\eta Z^2_{\eta_S}\phi_N^2
+(-0.18\phi_N+0.26\phi_S) \phi_S \cos^2\varphi_\eta  Z^2_{\eta_N}\big]\chi_{C0}\eta^2 f_0(1710)\,.\label{chif0eta6}
\end{align}
Thus, the decay amplitude for the decay width $\Gamma_{\chi_{C0}\rightarrow f_0(1710)\eta\eta}$ reads
\begin{align}
-i\mathcal{M}_{\chi_{C0}\rightarrow f_0(1710)\eta\eta}=&c\phi_N\phi_C\,\big[(-0.09\phi_N-0.39\phi_S)\cos\varphi_\eta \sin\varphi_\eta Z_{\eta_N}Z_{\eta_S}\phi_N\nonumber\\& +0.13 \sin^2\varphi_\eta Z^2_{\eta_S}\phi_N^2
+(-0.18\phi_N+0.26\phi_S) \phi_S \cos^2\varphi_\eta  Z^2_{\eta_N}\big]\,.
\end{align}

 \textbf{Decay channel $\chi_{C0}\rightarrow  f_0(1710)\eta\eta'$}\\

\begin{align}
\mathcal{L}_{\chi_{C0}f_0(1710)\eta\eta'}=& c\phi_N\phi_C\,\big[(-0.09\phi_N-0.39\phi_S)\sin^2\varphi_\eta Z_{\eta_N}Z_{\eta_S}\phi_N\nonumber\\& +(-0.09 \phi_N+0.39\phi_S) \cos^2\varphi_\eta Z_{\eta_N}Z_{\eta_S}\phi_N\nonumber\\&
+{-0.26 Z^2_{\eta_S}\phi^2_N+(0.36\phi_N-0.52\phi_S )\phi_S Z^2_{\eta_N}}\sin\varphi_\eta \cos\varphi_\eta \big]\chi_{C0}\eta\eta' f_0(1710)\,.\label{chif0eta5}
\end{align}
Thus, the decay amplitude for the decay width $\Gamma_{\chi_{C0}\rightarrow f_0(1710)\eta\eta'}$ reads
\begin{align}
-i\mathcal{M}_{\chi_{C0}\rightarrow f_0(1710)\eta\eta'}=&c\phi_N\phi_C\,c\phi_N\phi_C\,\big[(-0.09\phi_N-0.39\phi_S)\sin^2\varphi_\eta Z_{\eta_N}Z_{\eta_S}\phi_N\nonumber\\& +(-0.09 \phi_N+0.39\phi_S) \cos^2\varphi_\eta Z_{\eta_N}Z_{\eta_S}\phi_N\nonumber\\&
+{-0.26 Z^2_{\eta_S}\phi^2_N+(0.36\phi_N-0.52\phi_S )\phi_S Z^2_{\eta_N}}\sin\varphi_\eta \cos\varphi_\eta \big]\,.
\end{align}

\chapter{Decay rates for $\eta_C$ \label{Sec3}}

We present the explicit expressions for the two- and three-body decay
rates for the pseudoscalar hidden-charmed meson $\eta_C$,
which are listed in Table 8.5 in Sec.8.3.

\section{Two-body decay expressions for $\eta_C$}

The explicit expressions for the two-body decay
widths of $\eta_C$ are given by \\

\textbf{Decay channel $\eta_C\rightarrow\overline{K}^\ast_0K$}\\

The corresponding interaction Lagrangian can be obtained from the Lagrangian ( \ref{intLagofetadecay}) as
\begin{align}
\mathcal{L}_{\eta_{C}}=&\frac{c\sqrt{2}}{8}\phi_N^3\phi_S \phi_C\,Z_K
Z_{K^\ast_0}\,Z_{\eta_C}\eta_C(K^{\ast0}_0\overline{K}^0+\overline{K}^{\ast0}_0 K^0+K^{\ast-}_0 K^{+}+K^{\ast+}_0 K^-)\,.
\end{align}
The decay width is obtained as 
\begin{equation}
\Gamma_{\eta_C\rightarrow
\overline{K}^\ast_0K}=\frac{|\overrightarrow{k}_1|}{68\pi\,m_{\eta_C}^2}c^2\,\phi_N^6\,\phi_S^2\,\phi_C^2\,Z_K^2\,Z_{K^\ast_0}^2\,Z_{\eta_C}^2\,,
\end{equation}
with
\begin{equation}
|\overrightarrow{k}_1|=\frac{1}{2m_{\eta_C}}\bigg[m_{\eta_C}^4+(m_K^2-m_{K_0^\ast}^2)^2-2(m_{K^\ast_0}^2+m_K^2)m_{\eta_C}^2\bigg]^{1/2}\,.
\end{equation}
$$\\$$

\textbf{Decay channel $\eta_C\rightarrow a_0\,\pi$}\\

The corresponding interaction Lagrangian is extracted as 
\begin{align}
\mathcal{L}_{\eta_{C}}=&\frac{c}{4}\phi_N^2\phi_S^2\phi_C\,Z_\pi\,Z_{\eta_C}\eta_C\,(a_0^{0} \pi^{0 }+ a_0^{+} \pi^{-} + a_0^{-}\pi^{+})\,.
\end{align}
The decay width is obtained as

\begin{equation}
\Gamma_{\eta_C\rightarrow
a_0\,\pi}=3\frac{|\overrightarrow{k}_1|}{128\pi\,m_{\eta_C}^2}c^2\,\phi_N^4\,\phi_S^4\,\phi_C^2\,Z_\pi^2\,Z_{\eta_C}^2\,,
\end{equation}
with
\begin{equation}
|\overrightarrow{k}_1|=\frac{1}{2m_{\eta_C}}\bigg[m_{\eta_C}^4+(m_\pi^2-m_{a_0}^2)^2-2(m_{\pi}^2+m_{a_0}^2)m_{\eta_C}^2\bigg]^{1/2}\,.
\end{equation}
$$\\$$

\textbf{Decay channel $\eta_C\rightarrow f_0\eta,\,\eta'$}\\

The corresponding interaction Lagrangian can be obtained from the Lagrangian ( \ref{intLagofetadecay}) as

\begin{align}
\mathcal{L}_{\eta_{C}\eta \sigma }=&\frac{c}{8}\phi_N^2\phi_C\,Z_{\eta_C}\eta_C\{-4 \phi_N \phi_S (Z_{\eta_S} \eta_S \sigma_N + Z_{\eta_N} \eta_N \sigma_S)\nonumber\\
&\,\,\,\,\,\,\,\,\,\,\,\,\,\,\,\,\,\,\,\,\,\,\,\,\,\,\,\,\,\,\,\,\,\,\,-6\phi_S^2\,Z_{\eta_N}\eta_N\,\sigma_N-\phi_N^2\eta_S\sigma_S\,Z_{\eta_S}\}\,.\label{etacetasigmaint} 
\end{align}
 Substituting Eqs.(4.102, 4.103) and the relations (8.11) and (8.12), the interaction Lagrangian (\ref{etacetasigmaint}) can be written as\\
 
\begin{align}
\mathcal{L}_{\eta_{C}f_0\eta,\eta' }=\frac{c}{8}\phi_N^2\phi_C\,Z_{\eta_C}&\bigg[\{(3.76\phi_S-0.17\phi_N)\phi_N\,Z_{\eta_S}\sin\varphi_\eta\nonumber\\
&+(5.64\phi_S-0.68\phi_N)\phi_S\,Z_{\eta_N}\cos\varphi_\eta\}\eta_C\,f_0(1370)\,\eta\nonumber\\
&+\{(0.84\phi_S+0.97\phi_N)\phi_N\,Z_{\eta_S}\sin\varphi_\eta\nonumber\\
&+(1.26\phi_S+3.88\phi_N)\phi_S\,Z_{\eta_N}\cos\varphi_\eta\}\eta_C\,f_0(1500)\eta\nonumber\\
&+ \{(0.18\phi_N-1.04\phi_S)\phi_N\,Z_{\eta_S}\sin\varphi_\eta\nonumber\\
&+(0.72\phi_N-1.56\phi_S)\phi_S\,Z_{\eta_N}\cos\varphi_\eta\}\eta_C f_0(1710)\eta\nonumber\\
& + \{(3.76\phi_S-0.17\phi_N)\phi_N\,Z_{\eta_S}\cos\varphi_\eta\nonumber\\
&-(5.64\phi_S+0.68\phi_N)\phi_S\,Z_{\eta_N}\sin\varphi_\eta\}\eta_C f_0(1370)\eta'\nonumber\\
&+\{(0.84\phi_S+0.97\phi_N)\phi_N\,Z_{\eta_S}\cos\varphi_\eta\nonumber\\
&-(1.26\phi_S+3.88\phi_N)\phi_S\,Z_{\eta_N}\sin\varphi_\eta\}\eta_C f_0(1500)\eta'\nonumber\\
&+\{(0.18\phi_N-1.04\phi_S)\phi_N\,Z_{\eta_S}\cos\varphi_\eta\nonumber\\
&-(1.56\phi_S-0.72\phi_N)\phi_S\,Z_{\eta_N}\sin\varphi_\eta\}\eta_C f_0(1710)\eta'\bigg]\,. 
\end{align}

We then obtain the following decay amplitudes: \\

\textbf{Decay channel $\eta_C\rightarrow f_0(1370)\eta$}
\begin{align}
\Gamma_{\eta_C\rightarrow
f_0(1370)\eta}=\frac{|\overrightarrow{k}_1|}{512\pi\,m_{\eta_C}^2}c^2\,\phi_N^4\,\phi_C^2\,Z_{\eta_C}^2\bigg[&(3.76\phi_S-0.17\phi_N)\phi_N\,Z_{\eta_S}\sin\varphi_\eta\nonumber\\
&+(5.64\phi_S-0.68\phi_N)\phi_S\,Z_{\eta_N}\cos\varphi_\eta\bigg]^2\,,
\end{align}
with
\begin{equation}
|\overrightarrow{k}_1|=\frac{1}{2m_{\eta_C}}\bigg[m_{\eta_C}^4+(m_{f_0(1370)}^2-m_{\eta}^2)^2-2(m_{f_0(1370)}^2+m_{\eta}^2)m_{\eta_C}^2\bigg]^{1/2}\,.
\end{equation}
$$\\$$
\textbf{ Decay channel $\eta_C\rightarrow f_0(1500)\eta$}
\begin{align}
\Gamma_{\eta_C\rightarrow
f_0(1500)\eta}=\frac{|\overrightarrow{k}_1|}{512\pi\,m_{\eta_C}^2}c^2\,\phi_N^4\,\phi_C^2\,Z_{\eta_C}^2\bigg[&(0.84\phi_S+0.97\phi_N)\phi_N\,Z_{\eta_S}\sin\varphi_\eta\nonumber\\
&+(1.26\phi_S+3.88\phi_N)\phi_S\,Z_{\eta_N}\cos\varphi_\eta\bigg]^2\,,
\end{align}
with
\begin{equation}
|\overrightarrow{k}_1|=\frac{1}{2m_{\eta_C}}\bigg[m_{\eta_C}^4+(m_{f_0(1500)}^2-m_{\eta}^2)^2-2(m_{f_0(1500)}^2+m_{\eta}^2)m_{\eta_C}^2\bigg]^{1/2}\,.
\end{equation}
$$\\$$
\textbf{Decay channel $\eta_C\rightarrow f_0(1710)\eta$}
\begin{align}
\Gamma_{\eta_C\rightarrow
f_0(1710)\eta}=\frac{|\overrightarrow{k}_1|}{512\pi\,m_{\eta_C}^2}c^2\,\phi_N^4\,\phi_C^2\,Z_{\eta_C}^2\bigg[&(0.18\phi_N-1.04\phi_S)\phi_N\,Z_{\eta_S}\sin\varphi_\eta\nonumber\\
&+(0.72\phi_N-1.56\phi_S)\phi_S\,Z_{\eta_N}cos\varphi_\eta\bigg]^2\,,
\end{align}
with
\begin{equation}
|\overrightarrow{k}_1|=\frac{1}{2m_{\eta_C}}\bigg[m_{\eta_C}^4+(m_{f_0(1710)}^2-m_{\eta}^2)^2-2(m_{f_0(1710)}^2+m_{\eta}^2)m_{\eta_C}^2\bigg]^{1/2}\,.
\end{equation}
$$\\$$
\textbf{Decay channel $\eta_C\rightarrow f_0(1370)\eta'$}
\begin{align}
\Gamma_{\eta_C\rightarrow
f_0(1370)\eta'}=\frac{|\overrightarrow{k}_1|}{512\pi\,m_{\eta_C}^2}c^2\,\phi_N^4\,\phi_C^2\,Z_{\eta_C}^2\bigg[&(3.76\phi_S-0.17\phi_N)\phi_N\,Z_{\eta_S}\cos\varphi_\eta\nonumber\\
&-(5.64\phi_S+0.68\phi_N)\phi_S\,Z_{\eta_N}\sin\varphi_\eta\bigg]^2\,,
\end{align}
with
\begin{equation}
|\overrightarrow{k}_1|=\frac{1}{2m_{\eta_C}}\bigg[m_{\eta_C}^4+(m_{f_0(1370)}^2-m_{\eta'}^2)^2-2(m_{f_0(1370)}^2+m_{\eta'}^2)m_{\eta_C}^2\bigg]^{1/2}\,.
\end{equation}
$$\\$$
\textbf{Decay channel $\eta_C\rightarrow f_0(1500)\eta'$}
\begin{align}
 \Gamma_{\eta_C\rightarrow
f_0(1500)\eta'}=\frac{|\overrightarrow{k}_1|}{512\pi\,m_{\eta_C}^2}c^2\,\phi_N^4\,\phi_C^2\,Z_{\eta_C}^2\bigg[&(0.84\phi_S+0.97\phi_N)\phi_N\,Z_{\eta_S}\cos\varphi_\eta\nonumber\\
&-(1.26\phi_S+3.88\phi_N)\phi_S\,Z_{\eta_N}\sin\varphi_\eta\bigg]^2\,,
\end{align}
with
\begin{equation}
|\overrightarrow{k}_1|=\frac{1}{2m_{\eta_C}}\bigg[m_{\eta_C}^4+(m_{f_0(1500)}^2-m_{\eta'}^2)^2-2(m_{f_0(1500)}^2+m_{\eta'}^2)m_{\eta_C}^2\bigg]^{1/2}\,.
\end{equation}
$$\\$$
\textbf{Decay channel $\eta_C\rightarrow f_0(1710)\eta'$}
\begin{align}
 \Gamma_{\eta_C\rightarrow
f_0(1710)\eta'}=\frac{|\overrightarrow{k}_1|}{512\pi\,m_{\eta_C}^2}c^2\,\phi_N^4\,\phi_C^2\,Z_{\eta_C}^2\bigg[&(0.18\phi_N-1.04\phi_S)\phi_N\,Z_{\eta_S}\cos\varphi_\eta\nonumber\\
&-(1.56\phi_S-0.72\phi_N)\phi_S\,Z_{\eta_N}\sin\varphi_\eta\bigg]^2\,,
\end{align}
with
\begin{equation}
|\overrightarrow{k}_1|=\frac{1}{2m_{\eta_C}}\bigg[m_{\eta_C}^4+(m_{f_0(1710)}^2-m_{\eta'}^2)^2-2(m_{f_0(1710)}^2+m_{\eta'}^2)m_{\eta_C}^2\bigg]^{1/2}\,.
\end{equation}
$$\\$$

\section{Three-body decay expressions for $\eta_C$}

The corresponding interaction Lagrangian, contains the three-body decay rates for the $\eta_C$ meson, is extracted as
\begin{align}
\mathcal{L}_{\eta_{C}}=&\frac{c}{8}\phi_N^2\phi_C\,Z_{\eta_C}\eta_C\{2\phi_N\,Z^2_{\eta_S}\,Z_{\eta_N}\eta_S^2\eta_N +6\phi_S\,Z_{\eta_S}\,Z_{\eta_N}^2\eta_N^2\eta_S \nonumber\\
&-\sqrt{2}\phi_N\,Z_{\eta_S}\,Z_K^2(\overline{K}^0K^0+K^-K^+)\,\eta_S
-3\sqrt{2}\phi_S\,Z_{\eta_N}Z_K^2(\overline{K}^0K^0+K^-K^+)\eta_N\nonumber\\
&+\sqrt{2}\phi_S\,Z_\pi\,Z_K^2\bigg[\sqrt{2}(\overline{K}^0 K^+ \pi^- + K^0 K^- \pi^+)-(K^0 \overline{K}^0 -K^{-}K^{+})\pi^0\bigg]\nonumber\\
&-2\phi_S\eta_S\,Z_{\eta_S}Z_\pi^2({\pi^0}^2+2\pi^-\pi^+)\}\,.\label{intLageta3decay}
\end{align}

The general formula for the three-body decay
width for $\eta_C$, which is proved in chapter 5, takes the following form

\begin{multline}
\Gamma_{A\rightarrow B_{1}B_{2}B_{3}}=\frac{S}{32(2\pi)^{3}m_{\eta_C}^{3}}\int_{(m_{1}+m_{2})^{2}%
}^{(m_{\eta_C}-m_{3})^{2}}|-i\mathcal{M}_{A\rightarrow B_{1}B_{2}B_{3}}%
|^{2}\nonumber\\
\sqrt{\frac{(-m_{1}+m_{12}-m_{2})(m_{1}+m_{12}-m_{2})(-m_{1}%
+m_{12}+m_{2})(m_{1}+m_{12}+m_{2})}{m_{12}^{2}}}\nonumber\\
\sqrt{\frac{(-m_{\eta_C}+m_{12}-m_{3})(m_{\eta_C}+m_{12}-m_{3}%
)(-m_{\eta_C}+m_{12}+m_{3})(m_{\eta_C}+m_{12}+m_{3})}{m_{12}^{2}}%
}\,dm_{12}^{2}\text{ .}%
\end{multline}

Now let us list the corresponding tree-level decay amplitudes for $\eta_C$ which are obtained from the Lagrangian (\ref{intLageta3decay}) and are presented in Table 8.5 as follows:\\

\textbf{Decay channel $\eta_C\rightarrow \eta^3$:}
$m_1=m_2=m_3=m_{\eta}$ and $S=6$
\begin{equation}
|\overline{-iM_{\eta_C\rightarrow
\eta^3}}|^2=\bigg[\frac{1}{4}c\phi_N^2\phi_C\, \sin\varphi_\eta
\cos\varphi_\eta(Z_{\eta_S}\phi_N
\sin\varphi_\eta+3Z_{\eta_N}\phi_S\,\cos\varphi_{\eta})Z_{\eta_C}Z_{\eta_N}Z_{\eta_S}\bigg]^2\,.
\end{equation}
$$\\$$

\textbf{Decay channel $\eta_C\rightarrow \eta'^3$:}
$m_1=m_2=m_3=m_{\eta'}$ and $S=6$
\begin{equation}
|\overline{-iM_{\eta_C\rightarrow
\eta'^3}}|^2=\bigg[\frac{1}{4}c\phi_N^2\phi_C\, \sin\varphi_\eta
\cos\varphi_\eta(Z_{\eta_S}\phi_N
\cos\varphi_\eta-3Z_{\eta_N}\phi_S\,\sin\varphi_{\eta})Z_{\eta_C}Z_{\eta_N}Z_{\eta_S}\bigg]^2\,.
\end{equation}
$$\\$$

\textbf{Decay channel $\eta_C\rightarrow \eta'\eta^2$:}
$m_1=m_{\eta'},\,\, m_2=m_3=m_{\eta}$ and $S=2$
\begin{align}
|\overline{-iM_{\eta_C\rightarrow
\eta'\eta^2}}|^2=\frac{1}{16}c^2\phi_N^4\phi_C^2\,\bigg[&\phi_N\,Z_{\eta_S}\,\sin\varphi_\eta
(2\cos^2\varphi_\eta-\sin^2\varphi_\eta)\nonumber\\&+3Z_{\eta_N}\phi_S
\cos\varphi_\eta(\cos^2\varphi_\eta-2\,\sin^2\varphi_{\eta})\bigg]^2\,.
\end{align}
$$\\$$

\textbf{Decay channel $\eta_C\rightarrow \eta'^2\eta$:}
$m_1=m_2=m_{\eta'},\,\, m_3=m_{\eta}$ and $S=2$
\begin{align}
|\overline{-iM_{\eta_C\rightarrow
\eta'^2\eta}}|^2=\frac{1}{16}c^2\phi_N^4\phi_C^2\,Z^2_{\eta_C}\,Z^2_{\eta_N}\,Z^2_{\eta_S}\bigg[&Z_{\eta_S}\phi_N
\cos\varphi_\eta(\cos^2\varphi_\eta-2
\sin^2\varphi_\eta)\nonumber\\&+3Z_{\eta_N}\phi_S
\sin\varphi_\eta(\cos^2\varphi_\eta-2\,\cos^2\varphi_{\eta})\bigg]^2\,.
\end{align}
$$\\$$ 
 
\textbf{Decay channel $\eta K \overline{K}$:}
$m_1=K^0,\,m_2=m_{\overline{K}^0},\,\, m_3=m_{\eta}$
\begin{align}
\Gamma_{\eta_C\rightarrow \eta K
\overline{K}}&=\Gamma_{\eta_C\rightarrow \eta K^+
K^-}+\Gamma_{\eta_C\rightarrow \eta K^0
\overline{K}^0}\nonumber\\&=2\Gamma_{\eta_C\rightarrow \eta K^0
\overline{K}^0}.
\end{align}
with the average modulus squared decay amplitude
\begin{equation}
|\overline{-iM_{\eta_C\rightarrow \eta
K\overline{K}}}|^2=\frac{1}{32}c^2\phi_N^4\phi_C^2\,Z^2_{\eta_C}\,Z^4_{K}\,(\phi_N\,Z_{\eta_S}\,\sin\varphi_\eta+3\phi_S\,Z_{\eta_N}\cos\varphi_\eta)^2\,.
\end{equation}
$$\\$$

\textbf{Decay channel $\eta_C\rightarrow \eta' K \overline{K}$:}
$m_1=K^0,\,m_2=m_{\overline{K}^0},\,\, m_3=m_{\eta'}$
\begin{equation}
\Gamma_{\eta_C\rightarrow \eta' K
\overline{K}}=2\Gamma_{\eta_C\rightarrow \eta' K^0
\overline{K}^0}.
\end{equation}
The average modulus squared decay amplitude for this process reads
\begin{equation}
|\overline{-iM_{\eta_C\rightarrow \eta'
K\overline{K}}}|^2=\frac{1}{32}c^2\phi_N^4\phi^2_C\,Z^2_{\eta_C}\,Z^4_{K}\,(\phi_N\,Z_{\eta_S}\,\cos\varphi_\eta-3\phi_S\,Z_{\eta_N}\sin\varphi_\eta)^2\,.
\end{equation}
$$\\$$

\textbf{Decay channel $\eta_C\rightarrow \eta \pi\pi$:}
$m_1=\eta,\,m_2=m_3=m_{\pi^0}$ and $S=2$
\begin{equation}
\Gamma_{\eta_C\rightarrow \eta
\pi\pi}=3\Gamma_{\eta_C\rightarrow \eta \pi^0 \pi^0}\,,
\end{equation}
where the average modulus squared decay amplitude for this process
 is obtained from the Lagrangian (\ref{intLageta3decay}) as
\begin{equation}
|\overline{-iM_{\eta_C\rightarrow \eta \pi\pi
}}|^2=\frac{1}{16}c^2\phi_N^4\phi_S^2\,\phi_C^2\,Z^2_{\eta_C}\,Z^2_{\eta_S}\,Z^4_{\pi}\,\sin^2\varphi_\eta\,.
\end{equation}
$$\\$$

\textbf{Decay channel $\eta_C\rightarrow \eta' \pi\pi$:}
$m_1=\eta',\,m_2=m_3=m_{\pi^0}$ and $S=2$
\begin{equation}
\Gamma_{\eta_C\rightarrow \eta'
\pi\pi}=3\Gamma_{\eta_C\rightarrow \eta' \pi^0 \pi^0}\,,
\end{equation}
where the average modulus squared decay amplitude for this process is
\begin{equation}
|\overline{-iM_{\eta_C\rightarrow \eta' \pi\pi
}}|^2=\frac{1}{16}c^2\phi_N^4\phi_S^2\,\phi_C^2\,Z_{\eta_C}^2\,Z_{\eta_S}^2\,Z^4_{\pi}\,\cos^2\varphi_\eta\,.
\end{equation}
$$\\$$

\textbf{Decay channel $\eta_C\rightarrow KK\pi$:}
$m_1=K^+,\,m_2=K^-,\,m_3=m_{\pi^0}$ and $S=2$
\begin{align}
\Gamma_{\eta_C\rightarrow KK\pi}&=\Gamma_{\eta_C\rightarrow
K^+K^-\pi^0}+\Gamma_{\eta_C\rightarrow
K^0\overline{K}^0\pi^0}+\Gamma_{\eta_C\rightarrow
\overline{K}^0K^+\pi^-}+\Gamma_{\eta_C\rightarrow K^0K^-\pi^+}\nonumber\\&
=4\Gamma_{\eta_C\rightarrow K^+K^-\pi^0}.
\end{align}

with the average modulus squared decay amplitude
\begin{equation}
|\overline{-iM_{\eta_C\rightarrow K^+K^-\pi^0
}}|^2=\frac{1}{32}c^2\phi_N^4\phi_S^2\,\phi_C^2\,Z^2_{\eta_C}\,Z_{\eta_K}^4\,Z_{\pi}^2\,.
\end{equation}


  \cleardoublepage
  \phantomsection
  \addcontentsline{toc}{chapter}{Bibliography}
 \bibliography{mybib}

\begin{thebibliography}{100}

\bibitem{Povh:2008jx}
B.~Povh, K.~Rith, C.~Scholz, and F.~Zetsche.
\newblock {Particles and nuclei: An introduction to the physical concepts}.
\newblock 2008.

\bibitem{Maxwell:1865zz}
J.~Clerk Maxwell.
\newblock {A dynamical theory of the electromagnetic field}.
\newblock {\em Phil.Trans.Roy.Soc.Lond.}, 155:459--512, 1865.

\bibitem{Aad:2012tfa}
G.~Aad et~al.
\newblock {Observation of a new particle in the search for the Standard Model
  Higgs boson with the ATLAS detector at the LHC}.
\newblock {\em Phys.Lett.}, B716:1--29, 2012.

\bibitem{Chatrchyan:2012ufa}
S.~Chatrchyan et~al.
\newblock {Observation of a new boson at a mass of 125 GeV with the CMS
  experiment at the LHC}.
\newblock {\em Phys.Lett.}, B716:30--61, 2012.

\bibitem{Bilenky:1982ms}
S.~M. Bilenky and J.~Hosek.
\newblock {Glashow-Weinberg-Salam theory of electroweak interactions and the
  neutral currents}.
\newblock {\em Phys.Rept.}, 90:73--157, 1982.

\bibitem{Marciano:1977su}
W.~J. Marciano and H.~Pagels.
\newblock {Quantum chromodynamics: A review}.
\newblock {\em Phys.Rept.}, 36:137, 1978.

\bibitem{Muta:1987Pert}
T.~Muta.
\newblock {Foundation of quantum chromodynamics: An introduction to
  perturbative methods in gauge theories}.
\newblock 1998.

\bibitem{Hagiwara:1984jk}
K.~Hagiwara, K.~Harada, M.~Haruyama, K.~Kato, T.~Kubota, et~al.
\newblock {Quantum chromodynamics at short distances}.
\newblock {\em Prog.Theor.Phys.Suppl.}, 77:1--334, 1983.

\bibitem{Amsler:2004ps}
C.~Amsler and N.A. Tornqvist.
\newblock {Mesons beyond the naive quark model}.
\newblock {\em Phys.Rept.}, 389:61--117, 2004.

\bibitem{Klempt:2007cp}
E.~Klempt and A.~Zaitsev.
\newblock {Glueballs, Hybrids, Multiquarks. Experimental facts versus QCD
  inspired concepts}.
\newblock {\em Phys.Rept.}, 454:1--202, 2007.

\bibitem{Gasiorowicz}
S.~Gasiorowicz and D.A. Geffen.
\newblock {Effective Lagrangians and field algebras with chiral symmetry}.
\newblock {\em Rev.Mod.Phys.}, 41:531--573, 1969.

\bibitem{Meissner:1987ge}
U.~G. Meissner.
\newblock {Low-Energy hadron physics from effective chiral Lagrangians with
  vector mesons}.
\newblock {\em Phys.Rept.}, 161:213, 1988.

\bibitem{Vafa:1983tf}
C.~Vafa and E.~Witten.
\newblock {Restrictions on symmetry breaking in vector-like gauge theories}.
\newblock {\em Nucl.Phys.}, B234:173, 1984.

\bibitem{Giusti:2007cn}
L.~Giusti and S.~Necco.
\newblock {Spontaneous chiral symmetry breaking in QCD: A Finite-size scaling
  study on the lattice}.
\newblock {\em JHEP}, 0704:090, 2007.

\bibitem{Pisarski:1983ms}
R.~D. Pisarski and F.~Wilczek.
\newblock {Remarks on the chiral phase transition in chromodynamics}.
\newblock {\em Phys.Rev.}, D29:338--341, 1984.

\bibitem{'tHooft:1976fv}
G.~'t~Hooft.
\newblock {Computation of the quantum effects due to a four-dimensional
  pseudoparticle}.
\newblock {\em Phys.Rev.}, D14:3432--3450, 1976.

\bibitem{'tHooft:1976up}
G.~'t~Hooft.
\newblock {Symmetry breaking through Bell-Jackiw anomalies}.
\newblock {\em Phys.Rev.Lett.}, 37:8--11, 1976.

\bibitem{'tHooft:1986nc}
G.~'t~Hooft.
\newblock {How instantons solve the U(1) Problem}.
\newblock {\em Phys.Rept.}, 142:357--387, 1986.

\bibitem{Gross:1973id}
D.~J. Gross and F.~Wilczek.
\newblock {Ultraviolet Behavior of Nonabelian Gauge Theories}.
\newblock {\em Phys.Rev.Lett.}, 30:1343--1346, 1973.

\bibitem{Politzer:1973fx}
H.~D. Politzer.
\newblock {Reliable perturbative results for strong interactions?}
\newblock {\em Phys.Rev.Lett.}, 30:1346--1349, 1973.

\bibitem{Kapusta:1979fh}
J.~I. Kapusta.
\newblock {Quantum chromodynamics at high temperature}.
\newblock {\em Nucl.Phys.}, B148:461--498, 1979.

\bibitem{Aitchison:1982kj}
I.J.R. Aitchison and A.J.G. Hey.
\newblock {Gauge Theories in particle physics. A practical introduction}.
\newblock 1982.

\bibitem{Nakamura:2010zzi}
K.~Nakamura et~al.
\newblock {Review of particle physics}.
\newblock {\em J.Phys.}, G37:075021, 2010.

\bibitem{Gross:1973ju}
D.J. Gross and Frank Wilczek.
\newblock {Asymptotically free gauge theories. 1}.
\newblock {\em Phys.Rev.}, D8:3633--3652, 1973.

\bibitem{Politzer:1974sm}
H.~D. Politzer.
\newblock {Setting the scale for predictions of asymptotic freedom}.
\newblock {\em Phys.Rev.}, D9:2174--2175, 1974.

\bibitem{Politzer:1974fr}
H.~D. Politzer.
\newblock {Asymptotic freedom: An approach to strong interactions}.
\newblock {\em Phys.Rept.}, 14:129--180, 1974.

\bibitem{Hartmann:1999}
Stephan Hartmann.
\newblock {Models and Stories in Hadron Physics}.
\newblock {\em Morgan and Morrison}, pages 326--346, 1999.

\bibitem{Wilets:1996kr}
L.~Wilets, S.~Hartmann, and P.~Tang.
\newblock {The chromodielectric soliton model: Quark selfenergy and hadron
  bags}.
\newblock {\em Phys.Rev.}, C55:2067--2077, 1997.

\bibitem{Pisarski:1994yp}
R.~D. Pisarski.
\newblock {Applications of chiral symmetry}.
\newblock 1994.

\bibitem{Pich:2005mk}
A.~Pich.
\newblock {The Standard model of electroweak interactions}.
\newblock pages 1--48, 2005.

\bibitem{Eidelman:2004wy}
S.~Eidelman et~al.
\newblock {Review of particle physics. Particle Data Group}.
\newblock {\em Phys.Lett.}, B592:1--1109, 2004.

\bibitem{Veltmana1}
M.~Veltman.
\newblock {Theoretical aspects of high energy neutrino interactions}.
\newblock {\em Proc. Roy. Soc.}, A301:107--112, 1967.

\bibitem{Adler:1969gk}
S.~L. Adler.
\newblock {Axial vector vertex in spinor electrodynamics}.
\newblock {\em Phys.Rev.}, 177:2426--2438, 1969.

\bibitem{Bell:1969ts}
J.S. Bell and R.~Jackiw.
\newblock {A PCAC puzzle: pi0 --> gamma gamma in the sigma model}.
\newblock {\em Nuovo Cim.}, A60:47--61, 1969.

\bibitem{Roper:1964zza}
L.~D. Roper.
\newblock {Evidence for a P-11 pion-nucleon resonance at 556 MeV}.
\newblock {\em Phys.Rev.Lett.}, 12:340--342, 1964.

\bibitem{Olsson:1966zza}
M.G. Olsson and G.B. Yodh.
\newblock {Analysis of single-pion production reactions pi+N -->
  pi1+pi2+N-prime below 1 BeV}.
\newblock {\em Phys.Rev.}, 145:1309--1326, 1966.

\bibitem{Pauli:1940zz}
W.~Pauli.
\newblock {The connection between spin and statistics}.
\newblock {\em Phys.Rev.}, 58:716--722, 1940.

\bibitem{Lattes:1947mw}
C.M.G. Lattes, H.~Muirhead, G.P.S. Occhialini, and C.F. Powell.
\newblock {Processes involving charged involving charged mesons}.
\newblock {\em Nature}, 159:694--697, 1947.

\bibitem{Lattes:1947mx}
C.M.G. Lattes, G.P.S. Occhialini, and C.F. Powell.
\newblock {Observations on the tracks of slow mesons in photographic emulsions.
  1}.
\newblock {\em Nature}, 160:453--456, 1947.

\bibitem{Lattes:1947my}
C.M.G. Lattes, G.P.S. Occhialini, and C.F. Powell.
\newblock {Observations on the tracks of slow mesons in photographic emulsions.
  2}.
\newblock {\em Nature}, 160:486--492, 1947.

\bibitem{Augustin:1974xw}
J.E. Augustin et~al.
\newblock {Discovery of a narrow resonance in e+ e- annihilation}.
\newblock {\em Phys.Rev.Lett.}, 33:1406--1408, 1974.

\bibitem{Aubert:1974js}
J.J. Aubert et~al.
\newblock {Experimental observation of a heavy particle J}.
\newblock {\em Phys.Rev.Lett.}, 33:1404--1406, 1974.

\bibitem{Li:2012tr}
H.B. Li.
\newblock {Recent results on charm physics at BESIII}.
\newblock {\em Nucl.Phys.Proc.Suppl.}, 233:185--191, 2012.

\bibitem{Albrecht:1993jn}
H.~Albrecht et~al.
\newblock {A Partial wave analysis of the decay D0 ---> K0(s) pi+ pi-}.
\newblock {\em Phys.Lett.}, B308:435--443, 1993.

\bibitem{Anastassov:2001cw}
A.~Anastassov et~al.
\newblock {First measurement of Gamma(D*+) and precision measurement of m(D*+)
  - m(D0)}.
\newblock {\em Phys.Rev.}, D65:032003, 2002.

\bibitem{Bartelt:1995rq}
J.~E. Bartelt and S.~Shukla.
\newblock {Charmed meson spectroscopy}.
\newblock {\em Ann.Rev.Nucl.Part.Sci.}, 45:133--161, 1995.

\bibitem{Godfrey:1985xj}
S.~Godfrey and N.~Isgur.
\newblock {Mesons in a relativized quark model with chromodynamics}.
\newblock {\em Phys.Rev.}, D32:189--231, 1985.

\bibitem{Godfrey:1986wj}
S.~Godfrey and R.~Kokoski.
\newblock {The properties of p wave mesons with one heavy quark}.
\newblock {\em Phys.Rev.}, D43:1679--1687, 1991.

\bibitem{Capstick:1986bm}
S.~Capstick and N.~Isgur.
\newblock {Baryons in a relativized quark model with chromodynamics}.
\newblock {\em Phys.Rev.}, D34:2809, 1986.

\bibitem{Neubert:1993mb}
M.~Neubert.
\newblock {Heavy quark symmetry}.
\newblock {\em Phys.Rept.}, 245:259--396, 1994.

\bibitem{Beringer:1900zz}
J.~Beringer et~al.
\newblock {Review of particle physics (RPP)}.
\newblock {\em Phys.Rev.}, D86:010001, 2012.

\bibitem{Guo:2009id}
F.K. Guo, C.~Hanhart, and U.G. Meissner.
\newblock {Implications of heavy quark spin symmetry on heavy meson hadronic
  molecules}.
\newblock {\em Phys.Rev.Lett.}, 102:242004, 2009.

\bibitem{Liu:2012zya}
L.~Liu, K.~Orginos, F.K. Guo, C.~Hanhart, and U.G. Meissner.
\newblock {Interactions of charmed mesons with light pseudoscalar mesons from
  lattice QCD and implications on the nature of the $D_{s0}^*(2317)$}.
\newblock {\em Phys.Rev.}, D87:014508, 2013.

\bibitem{Gutsche:2010zz}
T.~Gutsche and V.~E. Lyubovitskij.
\newblock {Heavy flavor hadron molecules}.
\newblock {\em AIP Conf.Proc.}, 1322:289--297, 2010.

\bibitem{Gutsche:2010zza}
T.~Gutsche and V.~E. Lyubovitskij.
\newblock {Heavy hadron molecules}.
\newblock {\em AIP Conf.Proc.}, 1257:385--389, 2010.

\bibitem{Guo:2006rp}
F.K. Guo, P.N. Shen, and H.C. Chiang.
\newblock {Dynamically generated 1+ heavy mesons}.
\newblock {\em Phys.Lett.}, B647:133--139, 2007.

\bibitem{Cleven:2010aw}
M.~Cleven, F.K. Guo, C.~Hanhart, and U.G. Meissner.
\newblock {Light meson mass dependence of the positive parity heavy-strange
  mesons}.
\newblock {\em Eur.Phys.J.}, A47:19, 2011.

\bibitem{Guo:2009ct}
F.K. Guo, C.~Hanhart, and U.G. Meissner.
\newblock {Interactions between heavy mesons and Goldstone bosons from chiral
  dynamics}.
\newblock {\em Eur.Phys.J.}, A40:171--179, 2009.

\bibitem{Brambilla:2010cs}
N.~Brambilla, S.~Eidelman, B.K. Heltsley, R.~Vogt, G.T. Bodwin, et~al.
\newblock {Heavy quarkonium: progress, puzzles, and opportunities}.
\newblock {\em Eur.Phys.J.}, C71:1534, 2011.

\bibitem{Jaffe:1975fd}
R.L. Jaffe and K.~Johnson.
\newblock {Unconventional States of Confined Quarks and Gluons}.
\newblock {\em Phys.Lett.}, B60:201, 1976.

\bibitem{Konoplich:1981ed}
R.~Konoplich and M.~Shchepkin.
\newblock {Glueballs' Masses in the Bag Model}.
\newblock {\em Nuovo Cim.}, A67:211, 1982.

\bibitem{Jezabek:1982ic}
M.~Jezabek and J.~Szwed.
\newblock {Glueballs in the bag models}.
\newblock {\em Acta Phys.Polon.}, B14:599, 1983.

\bibitem{StrohmeierPresicek:1999yv}
M.~Strohmeier-Presicek, T.~Gutsche, R.~V.~Mau, and A.~Faessler.
\newblock {Glueball quarkonia content and decay of scalar - isoscalar mesons}.
\newblock {\em Phys.Rev.}, D60:054010, 1999.

\bibitem{Jaffe:1985qp}
R.L. Jaffe, K.~Johnson, and Z.~Ryzak.
\newblock {Qualitative features of the glueball spectrum}.
\newblock {\em Annals Phys.}, 168:344, 1986.

\bibitem{Morningstar:1999rf}
C.~J. Morningstar and M.~J. Peardon.
\newblock {The Glueball spectrum from an anisotropic lattice study}.
\newblock {\em Phys.Rev.}, D60:034509, 1999.

\bibitem{Loan:2005ff}
M.~Loan, X.Q. Luo, and Z.H. Luo.
\newblock {Monte Carlo study of glueball masses in the Hamiltonian limit of
  SU(3) lattice gauge theory}.
\newblock {\em Int.J.Mod.Phys.}, A21:2905--2936, 2006.

\bibitem{Chen:2005mg}
Y.~Chen, A.~Alexandru, S.J. Dong, Terrence Draper, I.~Horvath, et~al.
\newblock {Glueball spectrum and matrix elements on anisotropic lattices}.
\newblock {\em Phys.Rev.}, D73:014516, 2006.

\bibitem{Close:1987er}
F.E. Close.
\newblock {Gluonic hadrons}.
\newblock {\em Rept.Prog.Phys.}, 51:833, 1988.

\bibitem{Godfrey:1998pd}
S.~Godfrey and J.~Napolitano.
\newblock {Light meson spectroscopy}.
\newblock {\em Rev.Mod.Phys.}, 71:1411--1462, 1999.

\bibitem{Giacosa:2009bj}
F.~Giacosa.
\newblock {Dynamical generation and dynamical reconstruction}.
\newblock {\em Phys.Rev.}, D80:074028, 2009.

\bibitem{Gregory:2012hu}
E.~Gregory, A.~Irving, B.~Lucini, C.~McNeile, A.~Rago, et~al.
\newblock {Towards the glueball spectrum from unquenched lattice QCD}.
\newblock {\em JHEP}, 1210:170, 2012.

\bibitem{Amsler:1995td}
C.~Amsler and F.~E. Close.
\newblock {Is f0 (1500) a scalar glueball?}
\newblock {\em Phys.Rev.}, D53:295--311, 1996.

\bibitem{Lee:1999kv}
W.J. Lee and D.~Weingarten.
\newblock {Scalar quarkonium masses and mixing with the lightest scalar
  glueball}.
\newblock {\em Phys.Rev.}, D61:014015, 2000.

\bibitem{Close:2001ga}
F.~E. Close and A.~Kirk.
\newblock {Scalar glueball q anti-q mixing above 1-GeV and implications for
  lattice QCD}.
\newblock {\em Eur.Phys.J.}, C21:531--543, 2001.

\bibitem{Giacosa:2005qr}
F.~Giacosa, T.~Gutsche, V.E. Lyubovitskij, and A.~Faessler.
\newblock {Scalar meson and glueball decays within a effective chiral
  approach}.
\newblock {\em Phys.Lett.}, B622:277--285, 2005.

\bibitem{Giacosa:2004ug}
F.~Giacosa, Th. Gutsche, and A.~Faessler.
\newblock {A Covariant constituent quark / gluon model for the
  glueball-quarkonia content of scalar - isoscalar mesons}.
\newblock {\em Phys.Rev.}, C71:025202, 2005.

\bibitem{Mathieu:2008me}
V.~Mathieu, N.~Kochelev, and V.~Vento.
\newblock {The physics of glueballs}.
\newblock {\em Int.J.Mod.Phys.}, E18:1--49, 2009.

\bibitem{Janowski:2011gt}
S.~Janowski, D.~Parganlija, F.~Giacosa, and D.~H. Rischke.
\newblock {The glueball in a chiral linear sigma model with vector mesons}.
\newblock {\em Phys.Rev.}, D84:054007, 2011.

\bibitem{Giacosa:2005zt}
F.~Giacosa, Th. Gutsche, V.E. Lyubovitskij, and A.~Faessler.
\newblock {Scalar nonet quarkonia and the scalar glueball: Mixing and decays in
  an effective chiral approach}.
\newblock {\em Phys.Rev.}, D72:094006, 2005.

\bibitem{Cheng:2006hu}
H.Y. Cheng, C.K. Chua, and K.F. Liu.
\newblock {Scalar glueball, scalar quarkonia, and their mixing}.
\newblock {\em Phys.Rev.}, D74:094005, 2006.

\bibitem{Chatzis:2011qz}
P.~Chatzis, A.~Faessler, Th. Gutsche, and V.~E. Lyubovitskij.
\newblock {Hadronic and radiative three-body decays of J/psi involving the
  scalars f0(1370), f0(1500) and f0(1710)}.
\newblock {\em Phys.Rev.}, D84:034027, 2011.

\bibitem{Gutsche:2012zz}
Th. Gutsche.
\newblock {Exotic mesons}.
\newblock {\em Prog.Part.Nucl.Phys.}, 67:380--389, 2012.

\bibitem{Janowski:2014ppa}
S.~Janowski, F.~Giacosa, and D.~H. Rischke.
\newblock {Is $f_0$(1710) a glueball?}
\newblock {\em Phys.Rev.}, D90(11):114005, 2014.

\bibitem{Giacosa:2005bw}
F.~Giacosa, Th. Gutsche, V.E. Lyubovitskij, and A.~Faessler.
\newblock {Decays of tensor mesons and the tensor glueball in an effective
  field approach}.
\newblock {\em Phys.Rev.}, D72:114021, 2005.

\bibitem{Burakovsky:1997ci}
L.~Burakovsky and J.~Terrance Goldman.
\newblock {Towards resolution of the enigmas of P wave meson spectroscopy}.
\newblock {\em Phys.Rev.}, D57:2879--2888, 1998.

\bibitem{Masoni:2006rz}
A.~Masoni, C.~Cicalo, and G.L. Usai.
\newblock {The case of the pseudoscalar glueball}.
\newblock {\em J.Phys.}, G32:R293--R335, 2006.

\bibitem{Gutsche:2009jh}
Th. Gutsche, V.~E. Lyubovitskij, and M.~C. Tichy.
\newblock {eta(1405) in a chiral approach based on mixing of the pseudoscalar
  glueball with the first radial excitations of eta and eta-prime}.
\newblock {\em Phys.Rev.}, D80:014014, 2009.

\bibitem{Cheng:2008ss}
H.Y. Cheng, H.n. Li, and K.F. Liu.
\newblock {Pseudoscalar glueball mass from eta - eta-prime - G mixing}.
\newblock {\em Phys.Rev.}, D79:014024, 2009.

\bibitem{Mathieu:2009sg}
V.~Mathieu and V.~Vento.
\newblock {Pseudoscalar glueball and eta - eta-prime mixing}.
\newblock {\em Phys.Rev.}, D81:034004, 2010.

\bibitem{DiDonato:2011kr}
C.~Di~Donato, G.~Ricciardi, and I.~Bigi.
\newblock {$\eta - \eta'$ Mixing - From electromagnetic transitions to weak
  decays of charm and beauty hadrons}.
\newblock {\em Phys.Rev.}, D85:013016, 2012.

\bibitem{Li:2009rk}
B.~A. Li.
\newblock {Chiral field theory of 0-+ glueball}.
\newblock {\em Phys.Rev.}, D81:114002, 2010.

\bibitem{Ambrosino:2009sc}
F.~Ambrosino, A.~Antonelli, M.~Antonelli, F.~Archilli, P.~Beltrame, et~al.
\newblock {A Global fit to determine the pseudoscalar mixing angle and the
  gluonium content of the eta-prime meson}.
\newblock {\em JHEP}, 0907:105, 2009.

\bibitem{Escribano:2007cd}
R.~Escribano and J.~Nadal.
\newblock {On the gluon content of the eta and eta-prime mesons}.
\newblock {\em JHEP}, 0705:006, 2007.

\bibitem{Gasiorowicz:1969kn}
S.~Gasiorowicz and D.A. Geffen.
\newblock {Effective Lagrangians and field algebras with chiral symmetry}.
\newblock {\em Rev.Mod.Phys.}, 41:531--573, 1969.

\bibitem{Schwinger:1957em}
J.~S. Schwinger.
\newblock {A theory of the fundamental interactions}.
\newblock {\em Annals Phys.}, 2:407--434, 1957.

\bibitem{GellMann:1960np}
M.~Gell-Mann and M~Levy.
\newblock {The axial vector current in beta decay}.
\newblock {\em Nuovo Cim.}, 16:705, 1960.

\bibitem{Weinberg:1966fm}
S.~Weinberg.
\newblock {Dynamical approach to current algebra}.
\newblock {\em Phys.Rev.Lett.}, 18:188--191, 1967.

\bibitem{Gasser:1983yg}
J.~Gasser and H.~Leutwyler.
\newblock {Chiral Perturbation Theory to One Loop}.
\newblock {\em Annals Phys.}, 158:142, 1984.

\bibitem{Scherer:2002tk}
S.~Scherer.
\newblock {Introduction to chiral perturbation theory}.
\newblock {\em Adv.Nucl.Phys.}, 27:277, 2003.

\bibitem{Bando:1987br}
M.~Bando, T.~Kugo, and K.~Yamawaki.
\newblock {Nonlinear Realization and Hidden Local Symmetries}.
\newblock {\em Phys.Rept.}, 164:217--314, 1988.

\bibitem{Ecker:1988te}
G.~Ecker, J.~Gasser, A.~Pich, and E.~de~Rafael.
\newblock {The role of resonances in chiral perturbation theory}.
\newblock {\em Nucl.Phys.}, B321:311, 1989.

\bibitem{Jenkins:1995vb}
E.~E. Jenkins, A.~V. Manohar, and M.~B. Wise.
\newblock {Chiral perturbation theory for vector mesons}.
\newblock {\em Phys.Rev.Lett.}, 75:2272--2275, 1995.

\bibitem{Terschlusen:2011pm}
C.~Terschlusen and S.~Leupold.
\newblock {Electromagnetic transition form factors of mesons}.
\newblock {\em Prog.Part.Nucl.Phys.}, 67:401--405, 2012.

\bibitem{Schwinger:1967tc}
J.~S. Schwinger.
\newblock {Chiral dynamics}.
\newblock {\em Phys.Lett.}, B24:473--476, 1967.

\bibitem{Weinberg:1968de}
S.~Weinberg.
\newblock {Nonlinear realizations of chiral symmetry}.
\newblock {\em Phys.Rev.}, 166:1568--1577, 1968.

\bibitem{Ko:1994en}
P.~Ko and S.~Rudaz.
\newblock {Phenomenology of scalar and vector mesons in the linear sigma
  model}.
\newblock {\em Phys.Rev.}, D50:6877--6894, 1994.

\bibitem{Urban:2001ru}
M.~Urban, M.~Buballa, and J.~Wambach.
\newblock {Vector and axial vector correlators in a chirally symmetric model}.
\newblock {\em Nucl.Phys.}, A697:338--371, 2002.

\bibitem{Parganlija:2010fz}
D.~Parganlija, F.~Giacosa, and D.~H. Rischke.
\newblock {Vacuum properties of mesons in a linear sigma model with vector
  mesons and global chiral invariance}.
\newblock {\em Phys.Rev.}, D82:054024, 2010.

\bibitem{Gallas:2009qp}
S.~Gallas, F.~Giacosa, and D.~H. Rischke.
\newblock {Vacuum phenomenology of the chiral partner of the nucleon in a
  linear sigma model with vector mesons}.
\newblock {\em Phys.Rev.}, D82:014004, 2010.

\bibitem{Parganlija:2012fy}
D.~Parganlija, P.~Kovacs, G.~Wolf, F.~Giacosa, and D.~H. Rischke.
\newblock {Meson vacuum phenomenology in a three-flavor linear sigma model with
  (axial-)vector mesons}.
\newblock {\em Phys.Rev.}, D87(1):014011, 2013.

\bibitem{Rosenzweig:1981cu}
C.~Rosenzweig, A.~Salomone, and J.~Schechter.
\newblock {A pseudoscalar glueball, the Axial anomaly and the mixing problem
  for pseudoscalar mesons}.
\newblock {\em Phys.Rev.}, D24:2545--2548, 1981.

\bibitem{Salomone:1980sp}
A.~Salomone, J.~Schechter, and T.~Tudron.
\newblock {Properties of scalar gluonium}.
\newblock {\em Phys.Rev.}, D23:1143, 1981.

\bibitem{Rosenzweig:1982cb}
C.~Rosenzweig, A.~Salomone, and J.~Schechter.
\newblock {How does a pseudoscalar glueball come unglued?}
\newblock {\em Nucl.Phys.}, B206:12, 1982.

\bibitem{Gomm:1984zq}
H.~Gomm and J.~Schechter.
\newblock {Goldstone bosons and scalar gluonium}.
\newblock {\em Phys.Lett.}, B158:449, 1985.

\bibitem{Gomm:1985ut}
R.~Gomm, P.~Jain, R.~Johnson, and J.~Schechter.
\newblock {Scale anomaly and the scalars}.
\newblock {\em Phys.Rev.}, D33:801, 1986.

\bibitem{Parganlija:2012xj}
D.~Parganlija.
\newblock {Quarkonium Phenomenology in Vacuum}.
\newblock 2012.

\bibitem{Lenaghan:2000ey}
J.~T. Lenaghan, D.~H. Rischke, and J.~Schaffner-Bielich.
\newblock {Chiral symmetry restoration at nonzero temperature in the SU(3)(r) x
  SU(3)(l) linear sigma model}.
\newblock {\em Phys.Rev.}, D62:085008, 2000.

\bibitem{Eshraim:2014afa}
W.~I. Eshraim.
\newblock {Masses of light and heavy mesons in a $U(4)_r \times U(4)_l$ linear
  sigma model}.
\newblock {\em PoS}, QCD-TNT-III:049, 2013.

\bibitem{Eshraim:2014eka}
Walaa~I. Eshraim, Francesco Giacosa, and Dirk~H. Rischke.
\newblock {Phenomenology of charmed mesons in the extended Linear Sigma Model}.
\newblock {\em Eur. Phys. J.}, A51(9):112, 2015.

\bibitem{Eshraim:2014vfa}
W.~I. Eshraim and F.~Giacosa.
\newblock {Decays of open charmed mesons in the extended Linear Sigma Model}.
\newblock {\em EPJ Web Conf.}, 81:05009, 2014.

\bibitem{Eshraim:2014gya}
Walaa~I. Eshraim.
\newblock {A pseudoscalar glueball and charmed mesons in the extended Linear
  Sigma Model}.
\newblock {\em EPJ Web Conf.}, 95:04018, 2015.

\bibitem{Eshraim:2014tla}
Walaa~I. Eshraim.
\newblock {Vacuum properties of open charmed mesons in a chiral symmetric
  model}.
\newblock {\em J. Phys. Conf. Ser.}, 599(1):012009, 2015.

\bibitem{Gallas:2013ipa}
S.~Gallas and F.~Giacosa.
\newblock {Mirror versus naive assignment in chiral models for the nucleon}.
\newblock {\em Int.J.Mod.Phys.}, A29(17):1450098, 2014.

\bibitem{Eshraim:2012jv}
W.~I. Eshraim, S.~Janowski, F.~Giacosa, and D.~H. Rischke.
\newblock {Decay of the pseudoscalar glueball into scalar and pseudoscalar
  mesons}.
\newblock {\em Phys.Rev.}, D87(5):054036, 2013.

\bibitem{Eshraim:2012ju}
W.~I. Eshraim and S.~Janowski.
\newblock {Phenomenology of the pseudoscalar glueball with a mass of 2.6 GeV}.
\newblock {\em J.Phys.Conf.Ser.}, 426:012018, 2013.

\bibitem{Eshraim:2012rb}
W.~I. Eshraim, S.~Janowski, A.~Peters, K.~Neuschwander, and F.~Giacosa.
\newblock {Interaction of the pseudoscalar glueball with (pseudo)scalar mesons
  and nucleons}.
\newblock {\em Acta Phys.Polon.Supp.}, 5:1101--1108, 2012.

\bibitem{Eshraim:2013dn}
W.~I. Eshraim and S.~Janowski.
\newblock {Branching ratios of the pseudoscalar glueball with a mass of 2.6
  GeV}.
\newblock {\em PoS}, ConfinementX:118, 2012.

\bibitem{Giacosa:2006tf}
F.~Giacosa.
\newblock {Mixing of scalar tetraquark and quarkonia states in a chiral
  approach}.
\newblock {\em Phys.Rev.}, D75:054007, 2007.

\bibitem{Kovacs:2006ym}
P.~Kovacs and Zs. Szep.
\newblock {The critical surface of the SU(3)(L) x SU(3)(R) chiral quark model
  at non-zero baryon density}.
\newblock {\em Phys.Rev.}, D75:025015, 2007.

\bibitem{Kovacs:2007sy}
P.~Kovacs and Zs. Szep.
\newblock {Influence of the isospin and hypercharge chemical potentials on the
  location of the CEP in the mu(B) - T phase diagram of the SU(3)(L) x SU(3)(R)
  chiral quark model}.
\newblock {\em Phys.Rev.}, D77:065016, 2008.

\bibitem{Heinz:2008cv}
A.~Heinz, S.~Struber, F.~Giacosa, and D.~H. Rischke.
\newblock {Role of the tetraquark in the chiral phase transition}.
\newblock {\em Phys.Rev.}, D79:037502, 2009.

\bibitem{Han:1965pf}
M.Y. Han and Y.~Nambu.
\newblock {Three triplet model with double SU(3) symmetry}.
\newblock {\em Phys.Rev.}, 139:B1006--B1010, 1965.

\bibitem{Yang:1954ek}
Chen-Ning Yang and Robert~L. Mills.
\newblock {Conservation of Isotopic Spin and Isotopic Gauge Invariance}.
\newblock {\em Phys.Rev.}, 96:191--195, 1954.

\bibitem{Koch:1997ei}
V.~Koch.
\newblock {Aspects of chiral symmetry}.
\newblock {\em Int.J.Mod.Phys.}, E6:203--250, 1997.

\bibitem{Halzen}
F.~Halzen and Martin~A. D.
\newblock {Quarks and Leptons: an introductory course in modern particle
  physics.}

\bibitem{Hung:1994eq}
C.M. Hung and E.~V. Shuryak.
\newblock {Hydrodynamics near the QCD phase transition: Looking for the longest
  lived fireball}.
\newblock {\em Phys.Rev.Lett.}, 75:4003--4006, 1995.

\bibitem{Peskin:1995ev}
M.~E. Peskin and D.~V. Schroeder.
\newblock {An Introduction to quantum field theory}.
\newblock 1995.

\bibitem{Noether:1918zz}
E.~Noether.
\newblock {Invariant variation problems}.
\newblock {\em Gott.Nachr.}, 1918:235--257, 1918.

\bibitem{Koch:1995vp}
V.~Koch.
\newblock {Introduction to chiral symmetry}.
\newblock 1995.

\bibitem{Muta:1998vi}
T.~Muta.
\newblock {Foundations of quantum chromodynamics. Second edition}.
\newblock {\em World Sci.Lect.Notes Phys.}, 57:1--409, 1998.

\bibitem{Goldstone:1961eq}
J.~Goldstone.
\newblock {Field theories with superconductor solutions}.
\newblock {\em Nuovo Cim.}, 19:154--164, 1961.

\bibitem{FGhabilit}
F.~Giacosa.
\newblock {Ein effektives chirales Modell der QCD mit Vektormesonen, Dilaton
  und Tetraquarks: Physik im Vakuum und bei nichtverschwindender Dichte und
  Temperatur}.
\newblock 2012.

\bibitem{Boguta:1982wr}
J.~Boguta.
\newblock {A saturating chiral field theory of nuclear matter}.
\newblock {\em Phys.Lett.}, B120:34--38, 1983.

\bibitem{Kaymakcalan:1984bz}
O.~Kaymakcalan and J.~Schechter.
\newblock {Chiral lagrangian of pseudoscalars and vectors}.
\newblock {\em Phys.Rev.}, D31:1109, 1985.

\bibitem{St.Di.Th}
S.~Janowski.
\newblock {Phänomenologie des Dilatons in einem chiralen Modell mit (Axial-)
  Vektormesonen}.
\newblock {\em Diploma, Thesis, Faculty of Physics at Johann Wolfgang
  Goethe–Universität Frankfurt am Main}, 2010.

\bibitem{Lebed:1998st}
R.~F. Lebed.
\newblock {Phenomenology of large N(c) QCD}.
\newblock {\em Czech.J.Phys.}, 49:1273--1306, 1999.

\bibitem{'tHooft:1973jz}
G.~'t~Hooft.
\newblock {A planar diagram theory for strong interactions}.
\newblock {\em Nucl.Phys.}, B72:461, 1974.

\bibitem{Witten:1979kh}
E.~Witten.
\newblock {Baryons in the 1/n expansion}.
\newblock {\em Nucl.Phys.}, B160:57, 1979.

\bibitem{Lashin:2003jv}
E.I. Lashin.
\newblock {CP conserved nonleptonic K --> pi pi pi decays in the chiral quark
  model}.
\newblock {\em Int.J.Mod.Phys.}, A21:3699--3726, 2006.

\bibitem{Shifman:1978rh}
M.~A. Shifman, A.I. Vainshtein, and V.~I. Zakharov.
\newblock {Resonance properties in quantum chromodynamics}.
\newblock {\em Phys.Rev.Lett.}, 42:297, 1979.

\bibitem{Migdal:1982jp}
A.~A. Migdal and M.~A. Shifman.
\newblock {Dilaton effective Lagrangian in gluodynamics}.
\newblock {\em Phys.Lett.}, B114:445, 1982.

\bibitem{DiGiacomo:2000va}
A.~Di~Giacomo, H.~G. Dosch, V.I. Shevchenko, and Yu.A. Simonov.
\newblock {Field correlators in QCD: Theory and applications}.
\newblock {\em Phys.Rept.}, 372:319--368, 2002.

\bibitem{Struber:2007bm}
S.~Struber and D.~H. Rischke.
\newblock {Vector and axialvector mesons at nonzero temperature within a gauged
  linear sigma model}.
\newblock {\em Phys.Rev.}, D77:085004, 2008.

\bibitem{Jido:1998av}
D.~Jido, Y.~Nemoto, M.~Oka, and A.~Hosaka.
\newblock {Chiral symmetry for positive and negative parity nucleons}.
\newblock {\em Nucl.Phys.}, A671:471--480, 2000.

\bibitem{Zschiesche:2006zj}
D.~Zschiesche, L.~Tolos, J.~Schaffner-Bielich, and R.~D. Pisarski.
\newblock {Cold, dense nuclear matter in a SU(2) parity doublet model}.
\newblock {\em Phys.Rev.}, C75:055202, 2007.

\bibitem{Detar:1988kn}
C.~E. Detar and T.~Kunihiro.
\newblock {Linear $\sigma$ Model with parity doubling}.
\newblock {\em Phys.Rev.}, D39:2805, 1989.

\bibitem{Lee}
B.W. Lee.
\newblock {Chiral dynamics}.
\newblock {\em Gordon and breach}, New York:1972.

\bibitem{Gallas:2011qp}
S.~Gallas, F.~Giacosa, and G.~Pagliara.
\newblock {Nuclear matter within a dilatation-invariant parity doublet model:
  the role of the tetraquark at nonzero density}.
\newblock {\em Nucl.Phys.}, A872:13--24, 2011.

\bibitem{Giacosa:2007up}
F.~Giacosa.
\newblock {Two-photon decay of light scalars: A Comparison of tetraquark and
  quarkonium assignments}.
\newblock 2007.

\bibitem{Isgur:1991wq}
N.~Isgur and M.~B. Wise.
\newblock {Spectroscopy with heavy quark symmetry}.
\newblock {\em Phys.Rev.Lett.}, 66:1130--1133, 1991.

\bibitem{Eichten:1993ub}
E.~J. Eichten, C.~T. Hill, and C.~Quigg.
\newblock {Properties of orbitally excited heavy - light mesons}.
\newblock {\em Phys.Rev.Lett.}, 71:4116--4119, 1993.

\bibitem{Maki:1977ri}
Z.~Maki and I.~Umemura.
\newblock {Masses and decay constants of charmed mesons}.
\newblock {\em Prog.Theor.Phys.}, 59:507, 1978.

\bibitem{Segovia:2013kg}
J.~Segovia, D.R. Entem, and F.~Fernandez.
\newblock {Strong charmonium decays in a microscopic model}.
\newblock {\em Nucl.Phys.}, A915:125--141, 2013.

\bibitem{Cao:2012du}
L.~Cao, Y.-C. Yang, and H.~Chen.
\newblock {Charmonium states in QCD-inspired quark potential model using
  Gaussian expansion method}.
\newblock {\em Few Body Syst.}, 53:327--342, 2012.

\bibitem{Bali:2006xt}
G.~S. Bali.
\newblock {Charmonia from lattice QCD}.
\newblock {\em Int.J.Mod.Phys.}, A21:5610--5617, 2006.

\bibitem{Donald:2012ga}
G.C. Donald, C.T.H. Davies, R.J. Dowdall, E.~Follana, K.~Hornbostel, et~al.
\newblock {Precision tests of the $J/{\psi}$ from full lattice QCD: mass,
  leptonic width and radiative decay rate to ${\eta}_c$}.
\newblock {\em Phys.Rev.}, D86:094501, 2012.

\bibitem{Kalinowski:2012re}
M.~Kalinowski and M.~Wagner.
\newblock {Strange and charm meson masses from twisted mass lattice QCD}.
\newblock {\em PoS}, ConfinementX:303, 2012.

\bibitem{Casalbuoni:1996pg}
R.~Casalbuoni, A.~Deandrea, N.~Di~Bartolomeo, R.~Gatto, F.~Feruglio, et~al.
\newblock {Phenomenology of heavy meson chiral Lagrangians}.
\newblock {\em Phys.Rept.}, 281:145--238, 1997.

\bibitem{Georgi:1990um}
H.~Georgi.
\newblock {An Effective field theory for heavy quarks at low-energies}.
\newblock {\em Phys.Lett.}, B240:447--450, 1990.

\bibitem{Georgi:1991mr}
H.~Georgi.
\newblock {Heavy quark effective field theory}.
\newblock 1991.

\bibitem{DeFazio:2000up}
F.~De~Fazio.
\newblock {Weak decays of heavy quarks}.
\newblock 2000.

\bibitem{Wise:1992hn}
M.~B. Wise.
\newblock {Chiral perturbation theory for hadrons containing a heavy quark}.
\newblock {\em Phys.Rev.}, D45:2188--2191, 1992.

\bibitem{Kolomeitsev:2003ac}
E.E. Kolomeitsev and M.F.M. Lutz.
\newblock {On Heavy light meson resonances and chiral symmetry}.
\newblock {\em Phys.Lett.}, B582:39--48, 2004.

\bibitem{Bardeen:1993ae}
W.~A. Bardeen and C.~T. Hill.
\newblock {Chiral dynamics and heavy quark symmetry in a solvable toy field
  theoretic model}.
\newblock {\em Phys.Rev.}, D49:409--425, 1994.

\bibitem{Bardeen:2003kt}
W.~A. Bardeen, E.~J. Eichten, and C.~T. Hill.
\newblock {Chiral multiplets of heavy - light mesons}.
\newblock {\em Phys.Rev.}, D68:054024, 2003.

\bibitem{Nowak:1992um}
M.~A. Nowak, M.~Rho, and I.~Zahed.
\newblock {Chiral effective action with heavy quark symmetry}.
\newblock {\em Phys.Rev.}, D48:4370--4374, 1993.

\bibitem{Nowak:2003ra}
M.~A. Nowak, M.~Rho, and I.~Zahed.
\newblock {Chiral doubling of heavy light hadrons: BABAR 2317-MeV/c**2 and CLEO
  2463-MeV/c**2 discoveries}.
\newblock {\em Acta Phys.Polon.}, B35:2377--2392, 2004.

\bibitem{Sasaki:2014asa}
C.~Sasaki.
\newblock {Fate of charmed mesons near chiral symmetry restoration in hot
  matter}.
\newblock {\em Phys.Rev.}, D90(11):114007, 2014.

\bibitem{Lutz:2007sk}
M.~F.M. Lutz and M.~Soyeur.
\newblock {Radiative and isospin-violating decays of D(s)-mesons in the
  hadrogenesis conjecture}.
\newblock {\em Nucl.Phys.}, A813:14--95, 2008.

\bibitem{Caprini:2005zr}
I.~Caprini, G.~Colangelo, and H.~Leutwyler.
\newblock {Mass and width of the lowest resonance in QCD}.
\newblock {\em Phys.Rev.Lett.}, 96:132001, 2006.

\bibitem{Yndurain:2007qm}
F.J. Yndurain, R.~Garcia-Martin, and J.R. Pelaez.
\newblock {Experimental status of the pi pi isoscalar S wave at low energy:
  f(0)(600) pole and scattering length}.
\newblock {\em Phys.Rev.}, D76:074034, 2007.

\bibitem{GarciaMartin:2011jx}
R.~Garcia-Martin, R.~Kaminski, J.R. Pelaez, and J.~Ruiz~de Elvira.
\newblock {Precise determination of the f0(600) and f0(980) pole parameters
  from a dispersive data analysis}.
\newblock {\em Phys.Rev.Lett.}, 107:072001, 2011.

\bibitem{Bugg:2007ja}
D.V. Bugg.
\newblock {A Study in Depth of f0(1370)}.
\newblock {\em Eur.Phys.J.}, C52:55--74, 2007.

\bibitem{Black:1999dx}
D.~Black, A.~H. Fariborz, and J.~Schechter.
\newblock {Chiral Lagrangian treatment of pi eta scattering}.
\newblock {\em Phys.Rev.}, D61:074030, 2000.

\bibitem{Fariborz:2007ai}
A.~H. Fariborz, R.~Jora, and J.~Schechter.
\newblock {Two chiral nonet model with massless quarks}.
\newblock {\em Phys.Rev.}, D77:034006, 2008.

\bibitem{Fariborz:2009cq}
A.~H. Fariborz, R.~Jora, and J.~Schechter.
\newblock {Global aspects of the scalar meson puzzle}.
\newblock {\em Phys.Rev.}, D79:074014, 2009.

\bibitem{Fariborz:2011in}
A.~H. Fariborz, R.~Jora, J.~Schechter, and M.~N. Shahid.
\newblock {Chiral nonet mixing in pi pi scattering}.
\newblock {\em Phys.Rev.}, D84:113004, 2011.

\bibitem{Mukherjee:2012xn}
T.~K. Mukherjee, M.~Huang, and Q.S. Yan.
\newblock {Low-lying scalars in an extended Linear $\sigma$ Model}.
\newblock {\em Phys.Rev.}, D86:114022, 2012.

\bibitem{Mohler:2011ke}
D.~Mohler and R.M. Woloshyn.
\newblock {$D$ and $D_s$ meson spectroscopy}.
\newblock {\em Phys.Rev.}, D84:054505, 2011.

\bibitem{Moir:2013ub}
G.~Moir, M.~Peardon, S.~M. Ryan, C.~E. Thomas, and L.~Liu.
\newblock {Excited spectroscopy of charmed mesons from lattice QCD}.
\newblock {\em JHEP}, 1305:021, 2013.

\bibitem{Ebert:2009ua}
D.~Ebert, R.N. Faustov, and V.O. Galkin.
\newblock {Heavy-light meson spectroscopy and Regge trajectories in the
  relativistic quark model}.
\newblock {\em Eur.Phys.J.}, C66:197--206, 2010.

\bibitem{Klempt:2004yz}
E.~Klempt.
\newblock {Glueballs, hybrids, pentaquarks: Introduction to hadron spectroscopy
  and review of selected topics}.
\newblock 2004.

\bibitem{Divotgey:2013jba}
F.~Divotgey, L.~Olbrich, and F.~Giacosa.
\newblock {Phenomenology of axial-vector and pseudovector mesons: decays and
  mixing in the kaonic sector}.
\newblock {\em Eur.Phys.J.}, A49:135, 2013.

\bibitem{Cheng:2003bn}
H.Y. Cheng.
\newblock {Hadronic charmed meson decays involving axial vector mesons}.
\newblock {\em Phys.Rev.}, D67:094007, 2003.

\bibitem{Hatanaka:2008gu}
H.~Hatanaka and K.C. Yang.
\newblock {K(1)(1270)-K(1)(1400) mixing angle and new-physics effects in B --->
  K(1) l+ l- decays}.
\newblock {\em Phys.Rev.}, D78:074007, 2008.

\bibitem{Ahmed:2011vr}
A.~Ahmed, I.~Ahmed, M.~Ali~Paracha, and A.~Rehman.
\newblock {$K_{1}(1270)-K_{1}(1400)$ mixing and the fourth generation SM
  effects in $B \to K_{1}\ell^{+}\ell^{-}$ decays}.
\newblock {\em Phys.Rev.}, D84:033010, 2011.

\bibitem{Liu:2014dxa}
X.~Liu, Z.T. Zou, and Z.J. Xiao.
\newblock {Penguin-dominated B-->Phi K1(1270) and Phi K1(1400) decays in the
  perturbative QCD approach}.
\newblock {\em Phys.Rev.}, D90(9):094019, 2014.

\bibitem{Gutsche:2010jf}
Th. Gutsche, T.~Branz, A.~Faessler, I.~W. Lee, and V.~E. Lyubovitskij.
\newblock {Hadron molecules}.
\newblock 2010.

\bibitem{GellMann:1968rz}
M.~Gell-Mann, R.J. Oakes, and B.~Renner.
\newblock {Behavior of current divergences under SU(3) x SU(3)}.
\newblock {\em Phys.Rev.}, 175:2195--2199, 1968.

\bibitem{Achasov:2004uq}
N.N. Achasov and A.V. Kiselev.
\newblock {Propagators of light scalar mesons}.
\newblock {\em Phys.Rev.}, D70:111901, 2004.

\bibitem{Giacosa:2007bn}
F.~Giacosa and G.~Pagliara.
\newblock {On the spectral functions of scalar mesons}.
\newblock {\em Phys.Rev.}, C76:065204, 2007.

\bibitem{Giacosa:2012de}
F.~Giacosa and Th. Wolkanowski.
\newblock {Propagator poles and an emergent stable state below threshold:
  general discussion and the E(38) state}.
\newblock {\em Mod.Phys.Lett.}, A27:1250229, 2012.

\bibitem{Mishra:2003se}
A.~Mishra, E.~L. Bratkovskaya, J.~Schaffner-Bielich, S.~Schramm, and Horst
  Stoecker.
\newblock {Mass modification of D meson in hot hadronic matter}.
\newblock {\em Phys. Rev.}, C69:015202, 2004.

\bibitem{Gottfried:1992bz}
F.~O. Gottfried and S.~P. Klevansky.
\newblock {Thermodynamics of open and hidden charmed mesons within the NJL
  model}.
\newblock {\em Phys. Lett.}, B286:221--224, 1992.

\bibitem{Okubo:1963fa}
S.~Okubo.
\newblock {Phi meson and unitary symmetry model}.
\newblock {\em Phys.Lett.}, 5:165--168, 1963.

\bibitem{Zweig:1964jf}
G.~Zweig.
\newblock {An SU(3) model for strong interaction symmetry and its breaking.
  Version 2}.
\newblock pages 22--101, 1964.

\bibitem{Iizuka:1966fk}
J.~Iizuka.
\newblock {Systematics and phenomenology of meson family}.
\newblock {\em Prog.Theor.Phys.Suppl.}, 37:21--34, 1966.

\bibitem{FGThesis}
F.~Giacosa.
\newblock {Glueball phenomenology within a nonlocal approach.}
\newblock {\em PhD, Thesis, Faculty of Physics at Eberhard Karls University
  Tübingen}, 2005.

\bibitem{ONachtmann}
O.~Nachtmann.
\newblock {Elementary particle physics: concepts and phenomena.}

\bibitem{Rosenzweig:1979ay}
C.~Rosenzweig, J.~Schechter, and C.G. Trahern.
\newblock {Is the Effective Lagrangian for QCD a Sigma Model?}
\newblock {\em Phys.Rev.}, D21:3388, 1980.

\bibitem{Kawarabayashi:1980dp}
K.~Kawarabayashi and N.~Ohta.
\newblock {The problem of $\eta$ in the large $N$ limit: Effective Lagrangian
  approach}.
\newblock {\em Nucl.Phys.}, B175:477, 1980.

\bibitem{Maiani:2004uc}
L.~Maiani, F.~Piccinini, A.D. Polosa, and V.~Riquer.
\newblock {A New look at scalar mesons}.
\newblock {\em Phys.Rev.Lett.}, 93:212002, 2004.

\bibitem{Giacosa:2006rg}
F.~Giacosa.
\newblock {Strong and electromagnetic decays of the light scalar mesons
  interpreted as tetraquark states}.
\newblock {\em Phys.Rev.}, D74:014028, 2006.

\bibitem{Fariborz:2005gm}
A.~H. Fariborz, R.~Jora, and J.~Schechter.
\newblock {Toy model for two chiral nonets}.
\newblock {\em Phys.Rev.}, D72:034001, 2005.

\bibitem{Fariborz:2003uj}
A.~H. Fariborz.
\newblock {Isosinglet scalar mesons below 2-GeV and the scalar glueball mass}.
\newblock {\em Int.J.Mod.Phys.}, A19:2095--2112, 2004.

\bibitem{Napsuciale:2004au}
M.~Napsuciale and S.~Rodriguez.
\newblock {A Chiral model for anti-q q and anti-qq qq mesons}.
\newblock {\em Phys.Rev.}, D70:094043, 2004.

\bibitem{Giacosa:2009qh}
F.~Giacosa and G.~Pagliara.
\newblock {Decay of light scalar mesons into vector-photon and into
  pseudoscalar mesons}.
\newblock {\em Nucl.Phys.}, A833:138--155, 2010.

\bibitem{Lutz:2009ff}
M.F.M. Lutz et~al.
\newblock {Physics performance report for PANDA: Strong interaction studies
  with antiprotons}.
\newblock 2009.

\bibitem{Ablikim:2005um}
M.~Ablikim et~al.
\newblock {Observation of a resonance X(1835) in J / psi -> gamma pi+ pi-
  eta-prime}.
\newblock {\em Phys.Rev.Lett.}, 95:262001, 2005.

\bibitem{Kochelev:2005vd}
N.~Kochelev and D.-P. Min.
\newblock {X(1835) as the lowest mass pseudoscalar glueball and proton spin
  problem}.
\newblock {\em Phys.Lett.}, B633:283--288, 2006.

\bibitem{Ablikim:2010au}
M.~Ablikim et~al.
\newblock {Confirmation of the $X(1835)$ and observation of the resonances
  $X(2120)$ and $X(2370)$ in $J/\psi\to \gamma \pi^+\pi^-\eta^\prime$}.
\newblock {\em Phys.Rev.Lett.}, 106:072002, 2011.

\bibitem{Jaffe:1976ig}
R.~L. Jaffe.
\newblock {Multi-quark hadrons. 1. the phenomenology of (2 quark 2 anti-quark)
  mesons}.
\newblock {\em Phys.Rev.}, D15:267, 1977.

\bibitem{vanBeveren:1986ea}
E.~van Beveren, T.A. Rijken, K.~Metzger, C.~Dullemond, G.~Rupp, et~al.
\newblock {A Low lying scalar meson nonet in a unitarized meson model}.
\newblock {\em Z.Phys.}, C30:615--620, 1986.

\bibitem{Tornqvist:1995kr}
N.~A. Tornqvist.
\newblock {Understanding the scalar meson q anti-q nonet}.
\newblock {\em Z.Phys.}, C68:647--660, 1995.

\bibitem{Boglione:2002vv}
M.~Boglione and M.R. Pennington.
\newblock {Dynamical generation of scalar mesons}.
\newblock {\em Phys.Rev.}, D65:114010, 2002.

\bibitem{vanBeveren:2006ua}
E.~van Beveren, D.V. Bugg, F.~Kleefeld, and G.~Rupp.
\newblock {The Nature of sigma, kappa, a(0)(980) and f(0)(980)}.
\newblock {\em Phys.Lett.}, B641:265--271, 2006.

\bibitem{Pelaez:2003dy}
J.R. Pelaez.
\newblock {On the nature of light scalar mesons from their large N(c)
  behavior}.
\newblock {\em Phys.Rev.Lett.}, 92:102001, 2004.

\bibitem{Oller:1997ti}
J.A. Oller and E.~Oset.
\newblock {Chiral symmetry amplitudes in the S wave isoscalar and isovector
  channels and the sigma, f0(980), a0(980) scalar mesons}.
\newblock {\em Nucl.Phys.}, A620:438--456, 1997.

\bibitem{Bugg:2012ys}
D.V. Bugg.
\newblock {An alternative interpretation of Belle data on $\gamma-\gamma \to
  \eta'-\pi^\pm\pi^-$}.
\newblock {\em Phys.Rev.}, D86:114006, 2012.

\bibitem{Antje}
A.~Peters.
\newblock {Baryonische Zweikörp erzerfälle im erweiterten Linearen
  Sigma-Modell}.
\newblock {\em Bachelor, Thesis, Faculty of Physics at Johann Wolfgang
  Goethe–Universität Frankfurt am Main}, 2012.

\bibitem{Brambilla:2004jw}
N.~Brambilla, A.~Pineda, J.~Soto, and A.~Vairo.
\newblock {Effective field theories for heavy quarkonium}.
\newblock {\em Rev.Mod.Phys.}, 77:1423, 2005.

\bibitem{Edwards:2000bb}
K.W. Edwards et~al.
\newblock {Study of B decays to charmonium states B --> eta(c) K and B --->
  chi(c0) K}.
\newblock {\em Phys.Rev.Lett.}, 86:30--34, 2001.

\bibitem{Deshpande:1994mk}
N.G. Deshpande and J.~Trampetic.
\newblock {Exclusive and semiinclusive B decays based on b --> s eta(c)
  transition}.
\newblock {\em Phys.Lett.}, B339:270--274, 1994.

\bibitem{Coito:2011qn}
S.~Coito, G.~Rupp, and E.~van Beveren.
\newblock {Quasi-bound states in the continuum: a dynamical coupled-channel
  calculation of axial-vector charmed mesons}.
\newblock {\em Phys.Rev.}, D84:094020, 2011.

\bibitem{Rupp:2012py}
G.~Rupp, S.~Coito, and E.~van Beveren.
\newblock {Meson spectroscopy: too much excitement and too few excitations}.
\newblock {\em Acta Phys.Polon.Supp.}, 5:1007--1014, 2012.

\bibitem{Novikov:1977dq}
V.A. Novikov, L.B. Okun, Mikhail~A. Shifman, A.I. Vainshtein, M.B. Voloshin,
  et~al.
\newblock {Charmonium and gluons: Basic experimental facts and theoretical
  introduction}.
\newblock {\em Phys.Rept.}, 41:1--133, 1978.

\bibitem{Choi:2002na}
S.K. Choi et~al.
\newblock {Observation of the eta(c)(2S) in exclusive B --> K K(S) K- pi+
  decays}.
\newblock {\em Phys.Rev.Lett.}, 89:102001, 2002.

\bibitem{Vinokurova:2011dy}
A.~Vinokurova et~al.
\newblock {Study of B{+-} --> K{+-}(KS K pi)0 decay and determination of etac
  and etac(2S) parameters}.
\newblock {\em Phys.Lett.}, B706:139--149, 2011.

\bibitem{Uehara:2007vb}
S.~Uehara et~al.
\newblock {Study of charmonia in four-meson final states produced in two-photon
  collisions}.
\newblock {\em Eur.Phys.J.}, C53:1--14, 2008.

\bibitem{Liventsev:2011ks}
D.~Liventsev.
\newblock {Exotic/charmonium hadron spectroscopy at Belle and BaBar}.
\newblock 2011.

\bibitem{Lees:2010de}
J.P. Lees et~al.
\newblock {Measurement of the gamma gamma* --> etac transition form factor}.
\newblock {\em Phys.Rev.}, D81:052010, 2010.

\bibitem{Sun:2014gca}
L.P. Sun, H.~Han, and K.T. Chao.
\newblock {Impact of $J/\psi$ pair production at the LHC and predictions in
  nonrelativistic QCD}.
\newblock 2014.

\bibitem{Khan:2013ixa}
M.~S. Khan.
\newblock {$J/\psi$ Production within the framework of nonrelativistic QCD}.

\bibitem{Kniehl:2012ffa}
B.~Kniehl.
\newblock {Testing nonrelativistic-QCD factorization in charmonium production
  at next-to-leading order}.
\newblock {\em PoS}, LL2012:009, 2012.

\bibitem{Mannel:1997ky}
T.~Mannel.
\newblock {Heavy quark effective field theory}.
\newblock {\em Rept.Prog.Phys.}, 60:1113--1172, 1997.

\bibitem{Kawanai:2011jt}
T.~Kawanai and S.~Sasaki.
\newblock {Charmonium potential from full lattice QCD}.
\newblock {\em Phys.Rev.}, D85:091503, 2012.

\bibitem{DeTar:2011nn}
C.~DeTar.
\newblock {Charmonium spectroscopy from Lattice QCD}.
\newblock {\em Int.J.Mod.Phys.Conf.Ser.}, 02:31--35, 2011.

\bibitem{Liu:2012ze}
L.~Liu et~al.
\newblock {Excited and exotic charmonium spectroscopy from lattice QCD}.
\newblock {\em JHEP}, 1207:126, 2012.

\bibitem{Peters:2007zza}
K.~Peters.
\newblock {Charmonium and exotic hadrons at PANDA}.
\newblock {\em Int.J.Mod.Phys.}, E16:919--924, 2007.

\bibitem{Close:2005iz}
F.~E. Close and P.~R. Page.
\newblock {Gluonic charmonium resonances at BaBar and BELLE?}
\newblock {\em Phys.Lett.}, B628:215--222, 2005.

\bibitem{Bali:1993fb}
G.S. Bali et~al.
\newblock {A Comprehensive lattice study of SU(3) glueballs}.
\newblock {\em Phys.Lett.}, B309:378--384, 1993.

\bibitem{Bali:2000vr}
G.~S. Bali et~al.
\newblock {Static potentials and glueball masses from QCD simulations with
  Wilson sea quarks}.
\newblock {\em Phys.Rev.}, D62:054503, 2000.

\bibitem{Morningstar:2003ew}
C.~Morningstar and M.~J. Peardon.
\newblock {Simulating the scalar glueball on the lattice}.
\newblock {\em AIP Conf.Proc.}, 688:220--230, 2004.

\bibitem{Suzuki:2002bz}
M.~Suzuki.
\newblock {Elusive vector glueball}.
\newblock {\em Phys.Rev.}, D65:097507, 2002.

\bibitem{Chan:1999px}
C.T. Chan and W.S. Hou.
\newblock {On the mixing amplitude of J/psi and vector glueball O}.
\newblock {\em Nucl.Phys.}, A675:367C--370C, 2000.

\bibitem{Gregory:2005yr}
E.~B. Gregory, A.~C. Irving, C.~C. McNeile, S.~Miller, and Z.~Sroczynski.
\newblock {Scalar glueball and meson spectroscopy in unquenched lattice QCD
  with improved staggered quarks}.
\newblock {\em PoS}, LAT2005:027, 2006.

\bibitem{Bettoni:2005bb}
D.~Bettoni and R.~Calabrese.
\newblock {Charmonium spectroscopy}.
\newblock {\em Prog.Part.Nucl.Phys.}, 54:615--651, 2005.

\bibitem{Asner:1999kj}
D.M. Asner et~al.
\newblock {Hadronic structure in the decay tau- --> tau-neutrino pi- pi0 pi0
  and the sign of the tau-neutrino helicity}.
\newblock {\em Phys.Rev.}, D61:012002, 2000.

\bibitem{Bromberg:1980bk}
C.~Bromberg, J.~Dickey, G.~Fox, R.~Gomez, W.~Kropac, et~al.
\newblock {Observations of the $D$ and $E$ mesons and possible three kaon
  enhancements in $\pi^- p \to K^0 K^\pm \pi^\mp X$, $K^0 K^+ K^- X$ at
  50-{GeV}/$c$ and 100-{GeV}/$c$}.
\newblock {\em Phys.Rev.}, D22:1513, 1980.

\bibitem{Dionisi:1980hi}
C.~Dionisi et~al.
\newblock {Observation and quantum numbers determination of the E(1420) meson
  in pi- p interactions at 3.95-GeV/c}.
\newblock {\em Nucl.Phys.}, B169:1, 1980.

\bibitem{Harada:2003jx}
M.~Harada and K.~Yamawaki.
\newblock {Hidden local symmetry at loop: A New perspective of composite gauge
  boson and chiral phase transition}.
\newblock {\em Phys.Rept.}, 381:1--233, 2003.

\bibitem{Tilma:2002kf}
T.~E. Tilma, M.~Byrd, and G.~Sudarshan.
\newblock {A parametrization of bipartite systems based on SU(4) Euler angles}.
\newblock {\em J.Phys.}, A35:10445--10465, 2002.

\bibitem{WGreiner}
W.~Greiner and Müller B.
\newblock {Quantum mechanics - symmetries}.
\newblock {\em Springer Verlag}, 2nd. ed, 1994.

\end{thebibliography}
 \cleardoublepage



 \chapter*{Curriculum vitae\markboth{Curriculum vitae}{Curriculum vitae}}
\label{Curriculum vitae}
\renewcommand{\labelenumi}{\arabic{enumi}.}

\textbf{Personal Information}\\

Name: Ms. Walaa I. Eshraim\\
                           
Languages: English (very good), German (good), Arabic (native)\\
  
Web page: http://th.physik.uni-frankfurt.de/~weshraim/\\

\textbf{Academic Qualifications}\\

1- Master of Science (M. Sc. in Physics), the Islamic University of Gaza (IUG), Gaza, Gaza Strip, 18 Feb., 2007.\\

2- Bachelor of Science (B.Sc. in Physics), the Islamic University of Gaza (IUG), Gaza, Gaza Strip, 2003.\\

3- Diploma of Education, Al-Quds Open University, Gaza, Gaza Strip 2008.\\

\textbf{Awards and Honors}\\

1- Special Prize for the Best Poster, at the 13th Workshop on Mesons Production, Properties and Interaction MESON2014, Cracow, Poland, June 2, 2014.\\

2- DAAD (Deutscher Akademischer Austausch Dienst-German Academic Exchange Service) PhD scholarship, Germany, 2010.\\ 
	
3- Majorana Prize for the best research published in the Electronic Journal of Theoretical Physics in 2008, Italy.\\

4- The Islamic University Prize for the best scientific research in Science Faculty at the Islamic University of Gaza in 2007.\\

\textbf{Books}\\

1. “Hamilton-Jacobi Treatment of Fields with Constraints”, LAP Lambert Academic Publishing, Germany, 2012. \\

$$\\$$
$$\\$$

\textbf{Professional Experience}\\

04.2015 - 07.2015 Tutorial for Quantum Mechanics II, J. W. Goethe  
                                University, Frankfurt am Main, Germany.\\

10.2012 - 03.2013 Tutorial for Electrodynamics, J. W. Goethe  
                                University, Frankfurt am Main, Germany.\\
                                
09.2009 - 02.2010 Teacher for Medical Physics in Dental Medicine
                              College, University of Palestine, Gaza, Gaza strip.\\
                              
09.2006 – 06.2010 Teacher for Physics for the Abettor, High School,   
                                Gaza, Gaza Strip.\\

\textbf{Technical and Skills Courses}\\

Attended and obtained certificates for the following technical courses:\\
\\
1- Training course in Latex-Duration.\\
2- Training course in ICDL –Duration.\\
3- Training course in Mathematica Program. \\
4- Training courses in soft skills program in Helmholtz Graduate School for Hadron and Ion Research for FAIR at Germany:\\
\indent a) ‘Making an Impact as an Effective Researcher’.\\
\indent b) Leading Teams in a Research Environment.\\
\indent c) Leadership and Career Development.\\

\textbf{Activities}\\ 

1. Participant of the HGS-HIRe Lecture Week on Hadron Physics, presentation of “Tracking techniques for high energy physics experiments”, Ebsdorfergrund, Germany, 15-20 July, 2012.\\

2. Participant of the 52nd Winter School on Theoretical Physics in Schladming, which focussed on “Physics Beyond the Higgs” and presentation of “A $U(4)_r\times U(4)_l$ linear sigma model with (axial-)vector mesons”, Austria, March 1-8, 2014.\\

\textbf{Posters}\\

1. Poster entitled “Charmed mesons in the extended Linear Sigma Model”, at the 13th Workshop on Mesons Production, Properties and Interaction MESON2014, Cracow, Poland, June 2, 2014.\\

$$\\$$
$$\\$$

\textbf{Presentations}\\

\textbf{\textit{A. Presentations at Conferences:}}\\

1. “Phenomenology of a pseudoscalar glueball and charmed mesons in the extended linear sigma model”, at the 3rd International Conference on New Frontiers in Physics, Chania, Crete, Greece, July 28 -August 6, 2014.\\

2. “Phenomenology of Charmed Mesons in the extended Linear Sigma Model”, at the 78th Annual Meeting of the DPG (DPG Spring Meeting), Frankfurt am Main, Germany, March 17-21, 2014.\\

3. “A $U(4)_r\times U(4)_l$ linear sigma model with (axial-)vector mesons”, at the 77th Annual Meeting of the DPG (DPG Spring Meeting), Dresden, March 4 - 8, 2013.\\

4. “Decay of the pseudoscalar Glueball into scalar and Pseudoscalar mesons”, at the Xth International Conference on  Quark Confinement and the Hadron Spectrum, München, Germany, October 11, 2012.\\

\textbf{\textit{B. Presentations at Workshops:}}\\

1. ``Phenomenology of a pseudoscalar glueball and charmed mesons in a chiral symmetric model'', at Bound states in QCD and beyond workshop, Schlosshotel Rheinfels, St. Goar, Germany, March 24-27, 2015.\\

2. “Phenomenology of (open and hidden) charmed mesons in a chiral symmetric model”, at FAIRNESS workshop, Vietri sul Mare, Italy, September 22-27, 2014.\\

3. “Phenomenology of charmed mesons in a chiral symmetric model”, at the HICforFAIR workshop: Heavy flavor physics with CBM, FIAS, Frankfurt/Main, Germany, May 26-28, 2014.\\

4. “A $U(4)_r\times U(4)_l$ linear sigma model with (axial-)vector mesons”, at ‘QCD-TNT-III, From quarks and gluons to hadronic matter: A bridge too far?’ workshop ECT* Trento, Italy, Sep. 2-6, 2013.\\

5. “Decay of the pseudoscalar Glueball into scalar and Pseudoscalar mesons”, at the FAIRNESS workshop, Hersonissos, Greece, September 3-8, 2012.\\

6. “Decay of the pseudoscalar Glueball into scalar and Pseudoscalar mesons”, at the ‘Excited QCD 2012’ workshop, Peniche, Portugal, May 6-12, 2012.\\

$$\\$$
$$\\$$

\textbf{\textit{C. Presentations in Seminars: }}\\

1. “Phenomenology of (open and hidden) charmed mesons in Chiral Symmetric Model”, in the institute seminar at JW Goethe University, Frankfurt am Main, Germany, May 5, 2014.\\

2. “A $U(4)_r\times U(4)_l$ linear sigma model with vector and axial-vector mesons”, in the institute seminar at JW Goethe University, Frankfurt am Main, Germany, June 3, 2013.\\

3. “Decay of the pseudoscalar glueball into scalar and pseudoscalar mesons”, in the institute seminar at JW Goethe University, Frankfurt am Main, Germany, October 25, 2012.\\

4. “A Chiral Lagrangian for the Pseudoscalar Glueball”, in the chiral group seminar at JW Goethe University, Frankfurt am Main, Germany, June 25, 2012.\\

5. “Including the charm quark into the Linear Sigma Model”, in chiral group seminar at JW Goethe University, Frankfurt am Main, Germany, December 19, 2011.\\

\chapter*{Publications\markboth{Publications}{Publications}}
\renewcommand{\labelenumi}{\arabic{enumi}.}

\textbf{A. Publications in Journals}\\

1. W. I. Eshraim and D. H. Rischke, “Decays of charmonium states in the extended Linear Sigma Model”, in preparation.\\

2. W. I. Eshraim, F. Giacosa, and D. H. Rischke, “Phenomenology of charmed mesons in the extended Linear Sigma Model”, Eur.\ Phys.\ J.\ A 51, no. 9, 112 (2015)
  [arXiv:1405.5861 [hep-ph]].\\

3. W. I. Eshraim, S. Janowski, F. Giacosa and D. H. Rischke, “Decay of the pseudoscalar glueball into scalar and pseudoscalar mesons”, Phys. Rev. D87, 054036 (2013) [arXiv:1208.6474 [hep-ph]].\\

4. W. I. Eshraim, ``Path Integral Quantization of Landau-Ginzburg Theory'', Islamic University Journal, 18, 42 (2010) [arXiv: 1301.2478 [physics.gen-ph]].\\

5. W. I. Eshraim, and N. I. Farahat, “Path Integral Quantization of The Electromagnetic Field Coupled to A Spinor”, Electronic Journal of Theoretical Physics, 22, (2009) 189.\\

6. W. I. Eshraim, and N. I. Farahat, “Hamilton-Jacobi formulation of a non-abelian Yang-Mills theories”, Electronic Journal of theoretical Physics, 17, (2008) 69.\\

7. W. I. Eshraim, and N. I. Farahat, “Hamilton-Jacobi treatment of Lagrangian with fermionic and scalar field”, Romanian Journal of Physics, 53, (2008).\\

8. W. I. Eshraim, and N. I. Farahat, “Hamilton-Jacobi formulation of the scalar field coupled to two flavours Fermionic through Yukawa couplings”, Islamic University Journal, 15, (2007), 151.\\

9. W. I. Eshraim, and N. I. Farahat, “Quantization of the scalar field coupled minimally to the vector potential”  Electronic Journal of Theoretical Physics, 14, (2007), 61.\\

10. W. I. Eshraim, and N. I. Farahat, “Hamilton-Jacobi Approach to the Relativistic Local Free Field with Linear velocity of Dimension D” Hadronic Journal, 29, (2006), 553.\\

\textbf{B. Conference Proceedings}\\

1. W. I. Eshraim “Vacuum properties of open charmed mesons in a chiral symmetric model”,  J.\ Phys.\ Conf.\ Ser.\ 599, no. 1, 012009 (2015)
  [arXiv:1411.4749 [hep-ph]]. \\
 
2. W. I. Eshraim “A pseudoscalar glueball and charmed mesons in the extended linear sigma model”, EPJ Web Conf.\  95, 04018 (2015)
  [arXiv:1411.2218 [hep-ph]].\\

3. W. I. Eshraim and F. Giacosa “Decay of open charmed mesons in the extended linear sigma model”, EPJ Web Conf. 81, 05009 (2014) [arXiv:1409.5082 [hep-ph]].\\

4. W. I. Eshraim, “Masses of light and heavy mesons in a $U(4)_R \times U(4)_L$ linear sigma model”, PoS QCD-TNT-III 049 (2014) [arXiv:1401.3260 [hep-ph]].\\

5. W. I. Eshraim and S. Janowski, “Phenomenology of the pseudoscalar glueball with a mass of 2.6 GeV”, J. Phys. Conf. Ser. 426, 012018 (2013) [arXiv:1211.7323 [hep-ph]].\\

6. W. I. Eshraim and S. Janowski, “Branching ratios of the pseudoscalar glueball with a mass of 2.6 GeV”, PoS ConfinementX, 118 (2012) [arXiv:1301.3345 [hep-ph]].\\

7. W. I. Eshraim, S. Janowski, A. Peters, K. Neuschwander and F. Giacosa, “Interaction of the pseudoscalar glueball with (pseudo)scalar mesons and nucleons”, Acta Phys. Polon. Supp. 5, 1101(2012) [arXiv:1209.3976 [hep-ph]].\\

8. W. I. Eshraim, “On the Lagrangian Formalism of Landau-Ginzburg Theory”. The Third International Conference for Science and Development, The Islamic University of Gaza (IUG),  Faculty of Science, Gaza, Palestine, March 7-8, 2009.\\

9. W. I. Eshraim and N. I. Farahat, “Quantization of the Relativistic Local Free Field with Linear Velocity of Dimension D”. The Third International Conference for Science and Development, The Islamic University of Gaza (IUG), Faculty of Science, Gaza, Palestine, March 7-8, 2009. \\

10. W. I. Eshraim and N. I. Farahat, “Hamilton-Jacobi formulation of the scalar field coupled to two flavours Fermionic through Yukawa couplings”. The Second International Conference for Science and Development, The Islamic University of Gaza (IUG), Faculty of Science, Gaza, Palestine, March 6-7, 2007. \\

  \cleardoublepage
  \phantomsection
  \addcontentsline{toc}{chapter}{Index}
  \printindex

\end{document}